\documentclass[
a4paper, 11pt, twoside,
headsepline, 
footsepline, 
numbers=noenddot, 
bibliography=totoc, 
listof=totoc, 
headings=optiontotoc 
]{scrbook} 

\usepackage{typearea}
\areaset[28mm]{156mm}{248mm}
\usepackage[utf8]{inputenc}
\usepackage[T1]{fontenc}

\usepackage[main=english]{babel}

\usepackage[
backend=bibtex,
style=phys, 
biblabel=brackets,
backref=true
]{biblatex}

\DefineBibliographyStrings{english}{%
backrefpage = {Cited on page},
backrefpages = {Cited on pages},
}

\renewbibmacro*{pageref}{\addperiod%
\iflistundef{pageref}
{}
{\newline\footnotesize\printtext[parens]{%
\ifnumgreater{\value{pageref}}{1}
{\bibstring{backrefpages}\ppspace}
{\bibstring{backrefpage}\ppspace}%
\printlist[pageref][-\value{listtotal}]{pageref}\addperiod}}}
\usepackage{pdflscape} 
\usepackage{multirow} 
\usepackage{longtable} 

\usepackage{graphicx} 
\graphicspath{{Figures/}}
\usepackage{subcaption} 
\usepackage{siunitx}[=v2]

\usepackage[version=4]{mhchem} 

\usepackage[
cal=euler, calscaled=1.01, 
scr=boondoxo, scrscaled=1.03, 
frak=pxtx, frakscaled=1.05, 
bb=ams, bbscaled=1.00 
]{mathalpha}

\usepackage{bm} 
\usepackage{upgreek} 

\usepackage{amsmath, amssymb, amsthm}

\usepackage{mleftright}
\usepackage{colorprofiles}
\usepackage[a-2b, mathxmp]{pdfx}[2018/12/22]

\hypersetup{
colorlinks=true,
allcolors=blue,
breaklinks=true,
pdfstartview=
}

\input{Macros/layout} 

\makeatletter
\newcommand*{\defeq}{\mathrel{\rlap{%
\raisebox{0.3ex}{$\m@th\cdot$}}%
\raisebox{-0.3ex}{$\m@th\cdot$}}=}
\makeatother

\newcommand{\one}{\text{\usefont{U}{bbold}{m}{n}1}}
\MakeRobust{\one}

\newcommand*{\iu}{\mathrm{i}} 
\newcommand*{\Elr}{\mathrm{e}} 
\newcommand*{\Pauli}{\upsigma} 
\newcommand*{\LCs}{\upepsilon} 
\newcommand*{\Kd}{\updelta} 
\DeclareMathOperator{\HTh}{\Uptheta} 
\DeclareMathOperator{\Dd}{\updelta} 

\newcommand*{\abs}[1]{\mleft\lvert {#1} \mright\rvert} 
\newcommand*{\norm}[1]{\mleft\lVert {#1} \mright\rVert} 

\newcommand*{\dd}[2][]{\mathop{}\!\mathrm{d}^{#1} {#2}} 
\newcommand*{\dv}[2]{\frac{\mathrm{d} #1}{\mathrm{d} #2}} 
\newcommand*{\pdv}[2]{\frac{\partial #1}{\partial #2}} 
\newcommand*{\pdvc}[3]{\mleft.\frac{\partial #1}{\partial #2}\mright|_{#3}} 
\newcommand*{\DD}[1]{\mleft[\dd{#1}\mright]} 
\newcommand*{\var}[2][]{\mathop{}\!\delta_{#1} {#2}} 
\newcommand*{\fdv}[2]{\frac{\delta #1}{\delta #2}} 

\newcommand*{\vdot}{\bm{\cdot}} 
\newcommand*{\vcross}{\bm{\times}} 
\newcommand*{\grad}{\bm{\nabla}} 
\newcommand*{\vb}[1]{\bm{#1}} 
\newcommand*{\vu}[1]{\bm{\hat{#1}}} 

\newcommand*{\bra}[1]{\mleft\langle {#1} \mright\rvert} 
\newcommand*{\ket}[1]{\mleft\lvert {#1} \mright\rangle} 
\newcommand*{\dyad}[1]{\mleft\lvert {#1} \middle\rangle\middle\langle {#1} \mright\rvert} 
\newcommand*{\braket}[2]{\mleft\langle {#1} \middle| {#2} \mright\rangle} 
\newcommand*{\mel}[3]{\mleft\langle {#1} \middle| {#2} \middle| {#3} \mright\rangle} 
\newcommand*{\ev}[1]{\mleft\langle {#1} \mright\rangle} 

\DeclareMathOperator{\Conj}{\mathscr{K}}
\newcommand*{\const}{\mathrm{const.}}
\newcommand*{\cc}{\mathrm{c.c.}}
\newcommand*{\Hc}{\mathrm{H.c.}}
\DeclareMathOperator{\tr}{tr}
\DeclareMathOperator{\Tr}{Tr}
\DeclareMathOperator{\sgn}{sgn}
\DeclareMathOperator{\diag}{diag}
\DeclareMathOperator{\erf}{erf}
\DeclareMathOperator{\Cl}{Cl}
\DeclareMathOperator{\arccot}{arccot}
\DeclareMathOperator{\arcsinh}{arcsinh}
\DeclareMathOperator{\bigO}{\mathscr{O}} 

\newcommand*{\N}{\mathbb{N}}
\newcommand*{\Z}{\mathbb{Z}}

\newcommand*{\R}{\mathbb{R}}
\let\C\relax
\newcommand*{\C}{\mathbb{C}}

\DeclareMathOperator{\GL}{GL}
\DeclareMathOperator{\SL}{SL}
\DeclareMathOperator{\Ugp}{U}
\DeclareMathOperator{\SU}{SU}
\DeclareMathOperator{\Ogp}{O}
\DeclareMathOperator{\SO}{SO}

\theoremstyle{definition}
\newtheorem*{definition}{Definition}
\newtheorem*{theorem}{Theorem}

\newcommand*{\action}{\mathcal{S}}
\newcommand*{\Haml}{\mathcal{H}}

\newcommand*{\PintW}{\overline{\mathcal{W}}}
\newcommand*{\Pintw}{W}
\newcommand*{\PintV}{\overline{\mathcal{V}}}
\newcommand*{\Pintv}{V}
\newcommand*{\PintF}{\mathscr{F}}
\newcommand*{\Pintf}{\mathscr{f}}

\newcommand*{\SymU}{\hat{\mathcal{U}}}
\newcommand*{\MatU}{U}
\newcommand*{\TRop}{\Theta}
\newcommand*{\SymTR}{\hat{\Theta}}
\newcommand*{\MatTR}{\Theta}
\newcommand*{\RepM}{\mathcal{M}}

\newcommand*{\DipD}{\mathcal{D}}

\newcommand*{\DimK}{\mathscr{k}}
\newcommand*{\DimQ}{\mathscr{q}}
\newcommand*{\DimP}{\mathscr{p}}
\newcommand*{\DOSg}{\mathrm{g}}

\makeatletter
\newcommand{\thickhline}{%
    \noalign {\ifnum 0=`}\fi \hrule height 1.25pt
    \futurelet \reserved@a \@xhline
}
\newcolumntype{"}{@{\hskip\tabcolsep\vrule width 1.25pt\hskip\tabcolsep}}
\makeatother
\newcolumntype{N}[0]{>{\renewcommand{\arraystretch}{1.0}}c}
\newcolumntype{L}[1]{>{\renewcommand{\arraystretch}{1.0}\raggedright\let\newline\\\arraybackslash\hspace{0pt}}m{#1}}
\newcolumntype{C}[1]{>{\renewcommand{\arraystretch}{1.0}\centering\let\newline\\\arraybackslash\hspace{0pt}}m{#1}}
\newcolumntype{R}[1]{>{\renewcommand{\arraystretch}{1.0}\raggedleft\let\newline\\\arraybackslash\hspace{0pt}}m{#1}}

\begin{document}

\frontmatter

\selectlanguage{english}
\thispagestyle{empty}
\pdfbookmark{Cover}{Cover}
\begin{center}
\sffamily
{
  \LARGE
  \hrule
  \vspace{1em}
  \begin{center}
  \begin{minipage}{0.30\columnwidth}
    \raggedright
    \includegraphics[width=4cm]{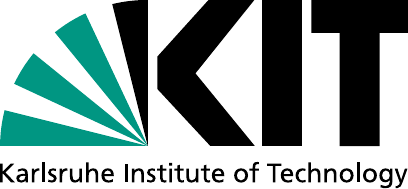}
  \end{minipage}\begin{minipage}{0.70\columnwidth}
    Department of Physics\\
    Institute for Theory of Condensed Matter
  \end{minipage}
  \end{center}
  \vspace{1em}
  \hrule
}
\vspace*{\stretch{7}}
{ 
  \Huge\bfseries
  Unconventional Superconductivity\\
  in Correlated, Multiband, and\\
  Topological Systems\\
}
\vspace*{\stretch{7}}
{
  \Large
  Dissertation of\\
}
\vspace*{\stretch{5}}
{
  \LARGE
  Grgur Palle\\
}
\vspace*{\stretch{5}}
{
  \Large
  6\textsuperscript{th} of December, 2024\\
}
\vspace*{\stretch{5}}
{
  \Large
  Advisor: Prof.~Dr.~Jörg Schmalian
}
\end{center}
\cleardoublepage

\selectlanguage{ngerman}
\pdfbookmark{Title}{Title}
\begin{titlepage}
\begin{center}
\vspace*{\stretch{3}}
{ 
  \LARGE\bfseries
  Unconventional Superconductivity in Correlated,\\
  Multiband, and Topological Systems\\
}
\vspace*{\stretch{6}}
{
  \large
  Zur Erlangung des akademischen Grades eines\\[8pt]
  \Large
  DOKTORS DER NATURWISSENSCHAFTEN (Dr.~rer.~nat.)\\
}
\vspace*{\stretch{4}}
{
  \large
  von der KIT-Fakultät für Physik des\\[4pt]
  Karlsruher Instituts für Technologie (KIT)\\[4pt]
  angenommene\\
}
\vspace*{\stretch{4}}
{
  \Large
  Dissertation\\
}
\vspace*{\stretch{4}}
{
  \large
  von\\[3pt]
  mag.~phys.~Grgur Palle\\[4pt]
  aus Zagreb\\
}
\vspace*{\stretch{5}}
{
  \large
  \begin{tabular}{rl}
  Tag der mündlichen Prüfung: & 25.~10.~2024.\\[2pt]
  Referent: & Prof.~Dr.~Jörg Schmalian\\[2pt]
  Korreferent: & Prof.~Dr.~Markus Garst
  \end{tabular}
}
\vspace*{\stretch{1}}
\end{center}
\end{titlepage}
\cleardoublepage
\thispagestyle{empty}
\vspace*{\stretch{1}}
\pdfbookmark{Erklärung}{Erklarung}
\section*{Erklärung:}
{
  \large
  Ich versichere wahrheitsgemäß, die Arbeit selbstständig verfasst, alle benutzten Hilfsmittel vollständig und genau angegeben und alles kenntlich gemacht zu haben, was aus Arbeiten anderer unverändert oder mit Abänderungen entnommen wurde sowie die Satzung des KIT zur Sicherung guter wissenschaftlicher Praxis in der Fassung vom 30.~9.~2021 beachtet zu haben.\\[64pt]
  Ort, Datum \hfill Unterschrift \hfill
}
\vspace*{\stretch{2}}
\clearpage

\selectlanguage{english}
\thispagestyle{empty}
\vspace*{\fill}
\pdfbookmark{Copyright}{Copyright}
{
  \large\sffamily
  \noindent\includegraphics[width=3cm]{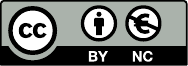}\\[4pt]
  This document, with the exception of reprinted figures whose copyright is held\\
  by the respective journals, is licensed under a Creative Commons Attribution-\\
  NonCommercial 4.0 International License (CC BY-NC 4.0).\\
  To view a copy of this license, visit \href{https://creativecommons.org/licenses/by-nc/4.0/}{https://creativecommons.org/licenses/by-nc/4.0/}.
}
\clearpage
\thispagestyle{empty}
\vspace*{\fill}
\pdfbookmark{Dedication}{Dedication}
\begin{center}
\itshape
To my family, \\[2pt]
For their (almost too) abundant support

\vspace{24pt}

To Dario Mičić, my high-school professor, \\[2pt]
For bringing me to physics
\end{center}
\vspace*{\fill}
\cleardoublepage

\chapter{Summary and Overview}
\label{chap:summary}

Despite the tremendous amount of research that has been devoted to superconductivity, a complete understanding of the phenomenon still eludes the scientific community.
In the first seventy years following its discovery in 1911~\cite{Onnes1911, Onnes1911-p2, Onnes1911-p3}, a great many advancements have been made~\cite{Meissner1933, London1935, London1935-p2, London1935-p3, Ginzburg1950, Abrikosov1952, Pippard1953, Cooper1956, Bogoljubov1958, Gorkov1959, Nambu1960, Eliashberg1960, Eliashberg1961, Morel1962, Josephson1962}, the grandest of which is, no doubt, the development of Bardeen-Cooper-Schrieffer theory in 1957~\cite{BCSshort, BCS, Cooper2010}.
Yet it is only with the discovery of high-temperature superconductivity in 1986~\cite{Bednorz1986} that we realized how incomplete our understanding truly is.
In the wake of this breakthrough, many other families of unconventional superconductors have been discovered~\cite{Norman2011, Stewart2017}, each one with its own set of challenges.
An outstanding problem is to theoretically explain -- and, one would hope, even predict -- the rich phenomenology displayed by the superconducting phases of these fascinating compounds.
Although there is no lack of theories, discerning which one, if any of the currently proposed ones, is the correct one has proven to be a formidable task.

In this thesis, we theoretically examine the unconventional superconductivity of systems whose complexity has one of three origins: strong correlations, multiple Fermi surfaces, or band structure topology.
We chiefly study two aspects of their superconductivity: the pairing mechanism and the pairing symmetry.
With regard to the former, we investigate whether a much-discussed loop-current-based pairing mechanism is viable (Chaps.~\ref{chap:loop_currents} and~\ref{chap:cuprates}), as well as propose a new electronic one (Chap.~\ref{chap:el_dip_SC}).
Regarding the latter, in the last Chap.~\ref{chap:Sr2RuO4} we take advantage of recent experiments performed on \ce{Sr2RuO4} to considerably narrow down the viable candidates for its superconducting state.

The thesis is organized into four chapters, as summarized below.
Each chapter is self-contained and can be read independently of the others.
Although we have attempted to make this work broadly accessible by providing extensive introductions within each chapter, some background knowledge is nonetheless necessary.
Namely, familiarity with quantum mechanics and second quantization~\cite{Sakurai2017}, as well as elementary knowledge of solid state physics~\cite{Simon2013} and superconductivity~\cite{Leggett2011}, are prerequisites.
More advanced knowledge of many-body theory~\cite{Altland2010} is only needed in parts of Chap.~\ref{chap:el_dip_SC}.
Symmetries play an important role throughout the thesis and we use their natural language, group theory, extensively.
We have thus supplemented the main text with a group theory introduction in Appx.~\ref{app:group_theory}, which is followed by \nameref{app:conventions} and a \nameref{app:abbreviations} which the reader may also want to peruse.
This thesis builds on Refs.~\cite{Palle2024-LC, Palle2024-el-dip, Palle2023-ECE, Jerzembeck2024} whose text has been recycled in many places~\cite{TRRP2021}.
Results that go beyond these references are pointed out at the start of each chapter.
\\[-4pt]

\textbf{Chapters 1 and 2.}
Magnetism most commonly originates in the spin sector.
However, it can also, in principle, originate in the orbital sector through the formation of loop currents (LCs).
When present, the fluctuations of these LCs mediate interactions among electrons which can potentially cause Cooper pairing.
These interactions are particularly strong when the fluctuations are strong, as is the case near quantum-critical points (QCPs) associated with LC order.
According to a prominent proposal~\cite{Varma2012, Varma2016, Varma2020}, quantum-critical LC fluctuations which are intra-unit-cell (i.e., order at $\vb{Q} = \vb{0}$) are not only an effective pairing glue, but the main explanation for the high-temperature superconductivity of cuprates.
In the first and second chapters, which are based on Ref.~\cite{Palle2024-LC}, we investigate both aspects of this proposal.

Our main finding is that, among all intra-unit-cell orders, which we also analyze, loop currents are uniquely ineffective at driving superconductivity.
The pairing that even-parity LCs mediate does not become enhanced as one approaches the assumed LC QCP, while odd-parity LCs strongly suppress any tendency towards Cooper pair formation in the vicinity of their QCP (Fig.~\ref{fig:QCP-general-results}).
In the case of cuprates, we systematically classify the possible intra-unit-cell LC orders and find that neither of the two proposed~\cite{Varma2012, Varma2016, Varma2020} to occur in the cuprates give the correct $d_{x^2-y^2}$ pairing symmetry (Figs.~\ref{fig:cuprate-A2g-results} and~\ref{fig:cuprate-Eu-results}).
Moreover, of the two proposed LCs, the odd-parity one which has been invoked~\cite{Varma1997, Simon2002, Simon2003, Varma2006} to explain the pseudogap strongly suppresses superconductivity, instead of enhancing it.

The strategy that we employ to tackle the problem of quantum-critical pairing is phenomenological and has two steps.
First, we classify the possible LC orders and assume that the system orders in one of the LC channels that we found.
Second, we use BCS theory to analyze the pairing tendency as the putative LC QCP is approached from the disordered side (Fig.~\ref{fig:QCP-calc-strategy}), where the normal state is a Fermi liquid.
Although we focus on loop currents, this strategy is clearly applicable to other orders, which we also analyze.
As we demonstrate in Sec.~\ref{sec:IUC-order-results}, for 2D systems with weak spin-orbit coupling, pairing mediated by nematic, ferromagnetic, and altermagnetic fluctuations becomes strongly enhanced as one approaches the QCP.
In contrast, the superconductivity driven by ferroelectric and spin-nematic fluctuations does not, just like in the case of even-parity LCs.
The latter result is a consequence of a suppression of forward-scattering that follows from parity and time-reversal symmetry.

In addition, we derive a number of subsidiary results.
Bloch's theorem is generalized to loop currents in Sec.~\ref{sec:Bloch-tm-Kirchhoff}.
In Sec.~\ref{sec:TR-positivity}, we prove that the exchange of bosons which are even under time-reversal favors $s$-wave pairing and that time-reversal-odd bosons robustly mediate unconventional pairing.
An extensive literature review on loop currents and symmetry-breaking in the pseudogap of cuprates has been compiled in Sec.~\ref{sec:cuprate-LC-lit-review}.
In Sec.~\ref{sec:cuprate-bilinear-class} we not only classify LC orders, but all possible orders which can arise in the widely-used three-orbital (Emery) model of cuprates.
The idea of introducing an extended basis, that is key to our classification, may prove to be useful in other multiorbital models.
Finally, in Sec.~\ref{sec:comparison-w-Aji2010} we compare our work to a previous analysis of quantum-critical intra-unit-cell LC pairing~\cite{Aji2010}.
\\[-4pt]

\textbf{Chapter 3.}
The origin of the superconductivity of doped bismuth selenide \ce{Bi2Se3} is a mystery.
On the one hand, there is considerable evidence that it is highly unconventional~\cite{Venderbos2016, Yonezawa2019, Cho2020}.
On the other hand, the normal state from which it springs is an utterly conventional low-density Fermi liquid that appears to neither be in the proximity of any competing orders nor have strong electronic correlations.
Yet there is one notable thing about \ce{Bi2Se3}: its topological band structure~\cite{Zhang2009, Ando2013}.
This raises the question:
Can parity-mixing and spin-orbit coupling that are responsible for the topological non-triviality of so many systems also be responsible for unconventional superconductivity?
In the third chapter, which draws from Ref.~\cite{Palle2024-el-dip}, we propose precisely such a pairing mechanism.

Our mechanism is based on electron-electron Coulomb interactions.
As we show in Sec.~\ref{sec:el-dip-sc-dipole-theory}, whenever the conduction band strongly mixes parity and has appreciable spin-orbit coupling, its Fermi surface attains a finite electric dipole density (Fig.~\ref{fig:el-dip-sc-FS-dip-density}).
Consequently, the dipolar contributions to the effective electron-electron interaction become particularly large.
In Sec.~\ref{sec:el-dip-sc-pairing} we then demonstrate that, with sufficient screening, these dipolar interaction may result in superconductivity, for which we prove that it is always unconventional.
Although we estimate a low-temperature $T_c$ that does not exceed a few Kelvins, the more interesting aspect is that the pairing is unconventional, even though no strong local electron correlations or quantum-critical fluctuations have been assumed.

Dirac metals are a natural platform for our theory, given that they are the effective model of spin-orbit-coupled parity-inverted bands (Sec.~\ref{sec:el-dip-sc-Dirac-model-band}), and we study them at length.
Quasi-2D Dirac systems turn out to be particularly promising for our mechanism, as we show in Sec.~\ref{sec:el-dip-sc-Dirac}.
This is because their out-of-plane ($z$-axis) dipole coupling is marginally relevant, in contrast to the monopole coupling which is marginally irrelevant, as our large-$N$ renormalization group calculation reveals.
For physically realistic parameters, we find that the effective dipole moments can get significantly enhanced (Fig.~\ref{fig:el-dip-sc-RG-etaz-flow}).
In addition, we establish that the proposed pairing glue is directly measurable in the $z$-axis optical conductivity.
Regarding the pairing symmetry, in Sec.~\ref{sec:el-dip-sc-Dirac-pairing} we find that $z$-axis electric dipole fluctuations favor unconventional odd-parity superconductivity of pseudoscalar symmetry, which is similar to the Balian-Werthamer state of \ce{^3He-B}.

There have been many mechanisms that in some aspect resemble our work, whether electronic~\cite{Kohn1965, Luttinger1966, Maiti2013, Kagan2014}, ferroelectric~\cite{Edge2015, KoziiBiRuhman2019, Klein2023}, or other, and in the last Sec.~\ref{sec:el-dip-sc-final-discussion} we compare and contrast them to the proposed electric dipole mechanism.
\\[-4pt]

\textbf{Chapter 4.}
Strontium ruthenate \ce{Sr2RuO4} is one of the most studied unconventional superconductors~\cite{Mackenzie2017, Maeno2024} whose normal Fermi-liquid state is characterized in exquisite detail.
Yet the fundamental question of what is its pairing symmetry remains unanswered.
This question gained a new life with recent NMR Knight shift experiments~\cite{Pustogow2019, Ishida2020, Chronister2021} that ruled out odd-parity pairing, including the chiral $p$-wave state which was until then considered the most likely pairing state.
In the aftermath of these landmark NMR studies, many interesting experiments have been conducted which clarify (or sometimes add to the puzzle of) the superconductivity of strontium ruthenate.
In the fourth and last chapter, which is based on Refs.~\cite{Palle2023-ECE, Jerzembeck2024}, we theoretically analyze the implications of two such experiments.

The first experiment~\cite{Li2022} is a measurement of the elastocaloric effect under $[100]$ uniaxial stress.
The elastocaloric effect is, let us recall, the effect of changes in the strain inducing changes in the temperature.
It is a measure of the strain derivative of the entropy, as follows from a thermodynamic identity.
From the measurements of Ref.~\cite{Li2022}, one clearly sees that the normal state attains an entropy maximum as a function of strain precisely when the $\gamma$ Fermi sheet crosses a Van Hove line, as expected.
However, as one enters the superconducting state, the data of Ref.~\cite{Li2022} reveal that the entropy maximum becomes a minimum (Fig.~\ref{fig:SRO-elasto}).
Our analysis of the gapping of Van Hove lines (Sec.~\ref{sec:SRO-elasto}) establishes that this can only be accounted for if there are no vertical line nodes at the Van Hove lines responsible for the normal-state maximum.
A detailed symmetry analysis (Sec.~\ref{sec:SRO-behavior-vH}, Tab.~\ref{tab:SRO-main-result}) moreover shows that only three even-parity states are consistent with this observation: $s$-wave, $d_{x^2-y^2}$-wave, and a body-centered $d_{xz} + \iu \, d_{yz}$ state that has horizontal line nodes.
The pairing state of strontium ruthenate therefore must include admixtures from at least one of these states.

The second experiment~\cite{Jerzembeck2024} are $T_c$ and elastocaloric effect measurements performed under $[110]$ uniaxial stress.
They were motivated by recently reported jumps in the $c_{66}$ elastic modulus~\cite{Benhabib2021, Ghosh2021}.
If interpreted in terms of a homogeneous superconducting state, a Ginzburg-Landau analysis shows that $c_{66}$ jumps imply cusps in $T_c$ and a second transition at a temperature $T_2 < T_c$ as $\langle 110 \rangle$ pressure is applied (Tab.~\ref{fig:schematicPhaseDiagrams}).
However, neither were observed in Ref.~\cite{Jerzembeck2024} (Figs.~\ref{fig:SRO-J24-Tc} and~\ref{fig:SRO-J24-ECE}).
As we show in Sec.~\ref{sec:SRO-110-implications}, a very large degree of fine-tuning is necessary if we are to accept both experimental results at face value.
This poses a serious challenge to any theory of bulk two-component superconductivity, whether symmetry-enforced or accidental.

The superconductivity of strontium ruthenate thus remains as puzzling as ever.
Yet some aspects of it are coming into focus.
In Sec.~\ref{sec:SRO-lit-review} we have summarized what is currently known from all the (100+) available experimental investigations of its pairing state, including the two experiments mentioned above.

{
\hypersetup{hidelinks}
\tableofcontents
}

\mainmatter

\chapter{The limitations of loop-current fluctuations as a pairing glue}
\label{chap:loop_currents}

Ordered states characterized by patterns of persistent spontaneously circulating charge currents, that is loop currents (LC), have been proposed to emerge in many systems~\cite{Varma2020, Bourges2021}, including cuprates, iridates, and kagome superconductors.
In their most general setting, such ordered states are best understood as instances of time-reversal symmetry-breaking (TRSB) that takes place in the orbital sector.
This orbital magnetism we shall refer to interchangeably as LC order throughout the thesis.

Quantum-critical LC fluctuations have been put forward as a possible source of Cooper pairing in general~\cite{Varma2012, Varma2016}, and in the case of cuprates in particular~\cite{Varma1997, Aji2010, Bok2016, Varma2020, Chakravarty2001, Laughlin2014, Laughlin2014-details}.
In this chapter, we study the pairing due to fluctuating LCs in general systems, focusing on systems with weak spin-orbit coupling (SOC) and on intra-unit-cell (IUC) loop currents which have been the most discussed as a pairing glue.
Our main finding is that quantum-critical IUC LC fluctuations are not an effective pairing glue, contrary to previous suggestions~\cite{Varma2020, Varma2016, Aji2010}.
In the next chapter, we study IUC LCs in the cuprates and what role, if any, they could have in driving the high-temperature superconductivity of the cuprates.
Both chapters are based on Ref.~\cite{Palle2024-LC}.
Although the text of Ref.~\cite{Palle2024-LC} has been reused in this and the next chapter, here we have taken the opportunity to elaborate in more detail upon the theoretical analysis of Ref.~\cite{Palle2024-LC}, including discussions and results that have not ended up in the published article.
In particular, we prove a generalized Bloch-Kirchhoff theorem (Sec.~\ref{sec:Bloch-tm-Kirchhoff}) and we derive results for general IUC order (Secs.~\ref{sec:IUC-order-results} and~\ref{sec:TR-positivity}), of which the results concerning loop currents (Fig.~\ref{fig:QCP-general-results}) are a special case.

The chapter is organized as follows.
We start by discussing the notion of orbital magnetism.
We explain why spontaneously forming patterns of charge currents must be made of loops in Sec.~\ref{sec:Bloch-tm}, after which we review previous theoretical and experimental work on LC order in systems other than the cuprates.
After that, in Sec.~\ref{sec:QCP-paradigm}, we introduce the paradigm of pairing driven by quantum-critical order-parameter fluctuations.
The studied LC pairing mechanism falls into this paradigm.
We also recall some basic facts on continuous quantum (i.e., $T = 0$) phase transitions and on the behavior of itinerant electronic systems near quantum-critical points (QCPs).
In Sec.~\ref{sec:gen-sys-LC-analysis}, we present the theoretical analysis leading up to the results of Ref.~\cite{Palle2024-LC}, summarized in Fig.~\ref{fig:QCP-general-results}.
In short, we find that even-parity IUC LCs are inefficient and odd-parity IUC LCs are detrimental to superconductivity (SC) near their QCP in 2D systems without SOC.
In Sec.~\ref{sec:QCP-pairing-model} we introduce a general model which allows us to make material-independent statements for general IUC orders other than LCs and in Sec.~\ref{sec:QCP-model-lin-gap-eq} we study the properties of the Cooper-channel interaction within this model.
Using these properties, in Sec.~\ref{sec:QCP-genera-final-analysis} we explain the strategy that we use to analyze quantum-critical pairing, summarized in Fig.~\ref{fig:QCP-calc-strategy}, and we derive the results of Fig.~\ref{fig:QCP-general-results}, but for general IUC order.
The main results are that (i) nematic, ferromagnetic, and altermagnetic fluctuations drive parametrically strong quantum-critical pairing, (ii) even-parity LC, ferroelectric, and spin-nematic fluctuations give parametrically weak SC near their QCP, while (iii) odd-parity LC fluctuations, unique among all orders, give rise to parametrically strong suppression of SC near their QCP.
Parametric strength or weakness refers to whether the pairing eigenvalue $\lambda$ ($T_c \propto \Elr^{-1/\lambda}$) diverges or stays finite, respectively, as the parameter $r$ controlling the distance from the QCP vanishes (Fig.~\ref{fig:QCP-general-results}).
These results apply to IUC orders in 2D systems with weak SOC.
In addition, in Sec.~\ref{sec:TR-positivity} we clarify the interplay between the time-reversal sign of the quantum-critical modes and the symmetry of the pairing state.

\section{Orbital magnetism and loop currents} \label{sec:orbital-mag}
\begin{figure}[t!]
\centering
\includegraphics[width=0.95\textwidth]{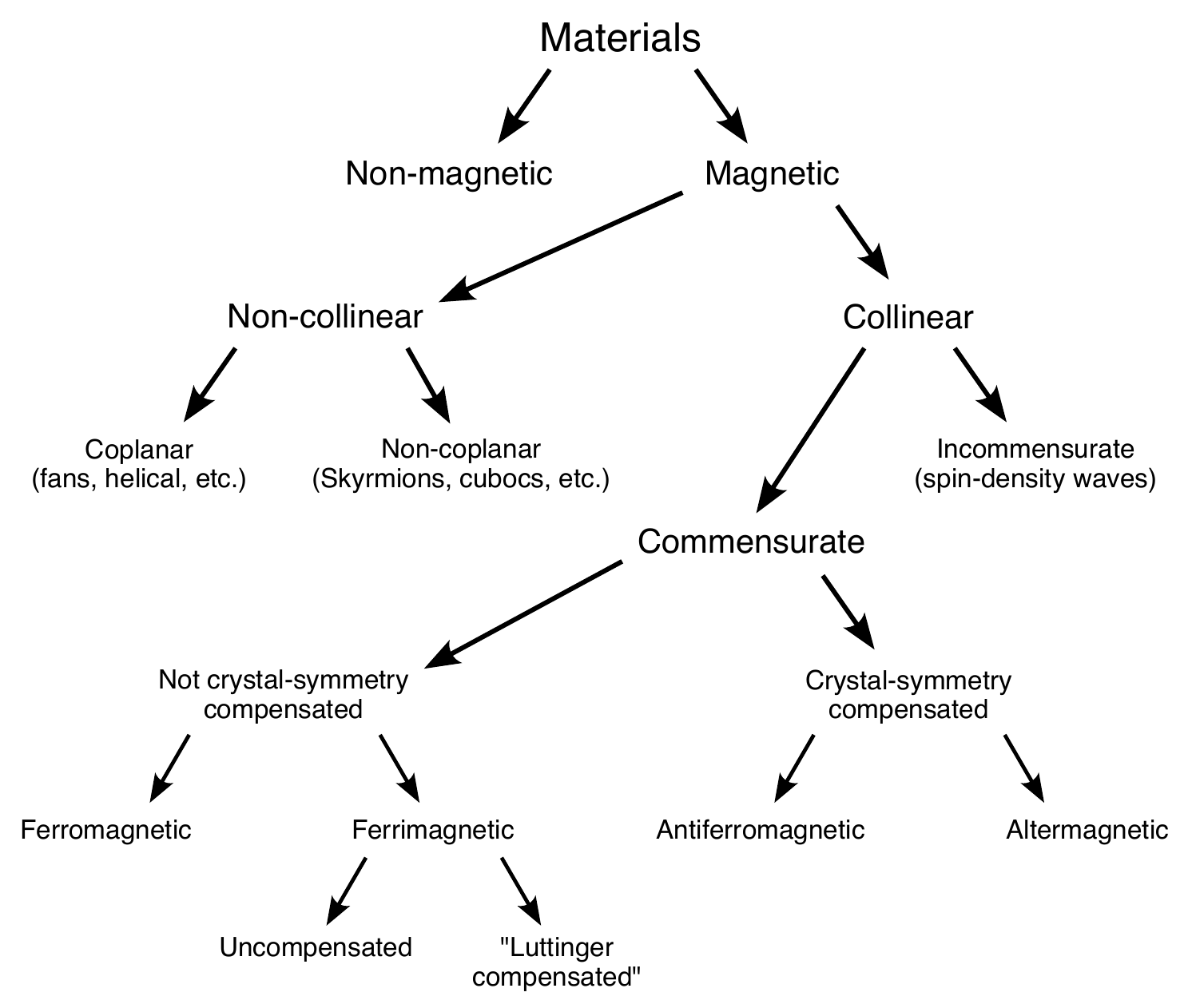}
\captionbelow[Classification of spin-magnetic orders according to symmetry and orientation of the local spin moments~\cite{Mazin2022}.]{\textbf{Classification of spin-magnetic orders according to symmetry and orientation of the local spin moments}~\cite{Mazin2022}.
Magnetism is characterized by the breaking of time-reversal (TR) symmetry $\TRop$.
The local spins can either be collinear (aligned along one axis), coplanar (orthogonal to one axis), or non-coplanar.
Furthermore, the pattern of the local spins and magnetic moments can either be commensurate (have the same periodicity as the underlying lattice) or incommensurate (break additional translation symmetries).
Even though TR symmetry is broken, composing TR with a crystal symmetry may leave the spin pattern invariant and thus compensate for TR symmetry-breaking.
Antiferromagnets are compensated by translations, whereas altermagnets are compensated by point group operations such as reflections or rotations.
Spatial-inversion symmetry is always preserved by spin-magnetic order.}
\label{fig:spin-mag-class}
\end{figure}

Magnetism most commonly arises from interactions related to the spin degrees of freedom~\cite{Kittel2005, Simon2013}.
The corresponding order parameter $S$ has a non-trivial structure in spin space and is odd under time reversal (TR):
\begin{align}
\SymTR^{-1} S \, \SymTR = - S,
\end{align}
where $\SymTR$ is the antiunitary many-body TR operator.
The simplest $S$ is, of course, spin itself, which is the appropriate order parameter for a ferromagnet.
However, there is a wide variety of orbital and spin structures that the order parameter $S$ may acquire, depending on the type of spin magnetism.
A comprehensive classification of possible spin-magnetic orders is provided in Fig.~\ref{fig:spin-mag-class}, reproduced from Ref.~\cite{Mazin2022}.
This classification is based on the local orientation of the spin moments and the symmetry of the overall spin pattern.

In correlated systems, magnetism may develop in the orbital sector as well~\cite{Varma2020, Bourges2021}.
Such orbital magnetism is characterized by ``generalized orbital angular momentum'' or ``flux'' operators $L$ which are odd under time reversal,
\begin{align}
\SymTR^{-1} L \, \SymTR = - L,
\end{align}
but have trivial spin structures.
As we shall later in Sec.~\ref{sec:gen-sys-LC-analysis} see, this difference in the spin structure has far-reaching consequences, especially when SOC is weak, and it is the main reason for why ferromagnetic and IUC LC fluctuations behave so differently near their QCPs.
The range of possible LC orders is, in principle, as varied as that of spin-magnetic orders classified in Fig.~\ref{fig:spin-mag-class}.
However, LC order is less common than spin magnetism, which is why its phases have not been explored as extensively.

\subsection{Bloch and generalized Bloch-Kirchhoff theorems on persistent currents} \label{sec:Bloch-tm}
The only way TR symmetry can be broken in the charge or orbital sector is through the formation of some sort of charge currents.
This pattern of spontaneously flowing currents, moreover, must be made of closed loops to avoid a global current, which is forbidden because of a theorem first proved by Bloch~\cite{Bohm1949, Ohashi1996, Yamamoto2015, Zhang2019, Watanabe2019, Watanabe2022}.
Here we in addition prove a generalized Bloch-Kirchhoff theorem according to which the pattern of spontaneously flowing currents must be divergenceless, i.e., respect Kirchhoff's law and not lead to an accumulation of charge in some parts of the system.
The latter was previously invoked in Ref.~\cite{Palle2024-LC} in the form of Kirchhoff constraints (see Sec.~\ref{sec:Bloch-Kirch-constr} of the next chapter), but not rigorously proved in the same manner as Bloch's theorem.
These two theorems give fundamental constraints on the possible thermodynamically stable LC orders which may arise in nature.

\subsubsection{Proof of Bloch's theorem} \label{sec:Bloch-tm-Bloch}
\begin{figure}[t]
\centering
\begin{subfigure}[t]{0.5\textwidth}
\raggedright
\includegraphics[width=0.95\textwidth]{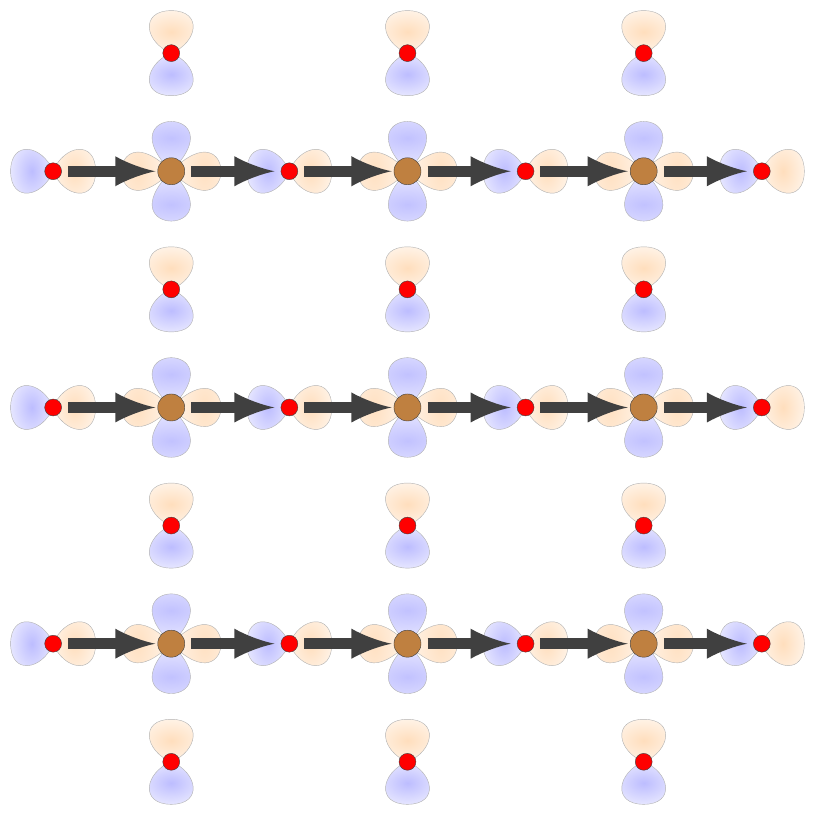}
\subcaption{}
\end{subfigure}%
\begin{subfigure}[t]{0.5\textwidth}
\raggedleft
\includegraphics[width=0.95\textwidth]{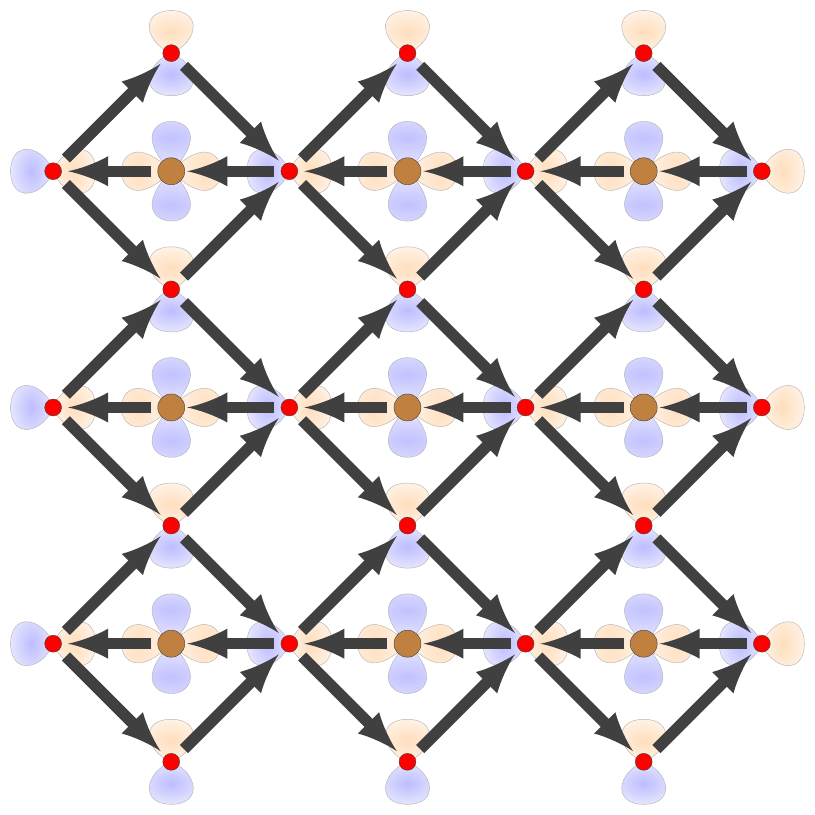}
\subcaption{}
\end{subfigure}
\captionbelow[Examples of loop-current (LC) patterns that have a finite (a) and vanishing (b) global charge current.]{\textbf{Examples of loop-current (LC) patterns that have a finite (a) and vanishing (b) global charge current.}
Arrows indicate the direction of the local currents, i.e., the flow of charge between the orbitals.
The underlying model is the three-orbital tight-binding model of the copper planes of the cuprates which we study in the next chapter (Sec.~\ref{sec:cuprate-3band-model}).
The LC pattern shown under (a) is forbidden by Bloch's theorem~\cite{Bohm1949, Ohashi1996, Yamamoto2015, Zhang2019, Watanabe2019, Watanabe2022}, as discussed in the text.
As for the currents under (b), they must not lead to any local accumulation of charge, as proved in the text.}
\label{fig:Bloch-examples}
\end{figure}

The proof of Bloch's theorem proceeds by contradiction.
Let us assume that a ground state $\ket{\Uppsi_0}$ of energy $E_0 = \mel{\Uppsi_0}{\Haml}{\Uppsi_0}$ has a finite global electron charge current $\vb{J} = \int_{\vb{r}} \mel{\Uppsi_0}{\vb{j}_e}{\Uppsi_0}$, where $\int_{\vb{r}} = \int \dd[d]{r}$.
The corresponding charge is locally conserved in the sense that $\partial_t \rho_e + \grad \vdot \vb{j}_e = 0$, where $\rho_e$ is the local charge density operator of the electrons only and $\partial_t \rho_e = \iu [\Haml, \rho_e]/ \hbar$.
We use Heisenberg's picture throughout.
Now consider the state
\begin{align}
\ket{\Uppsi_0'} = \exp\mleft(\iu \int_{\vb{r}} \vb{k} \vdot \vb{r} \, \rho_e\mright) \ket{\Uppsi_0}
\end{align}
for small $\vb{k}$.
This corresponds to a state in which every electron has been given an additional momentum $\hbar \vb{k}$.
To linear order in $\vb{k}$, its energy equals
\begin{align}
\begin{aligned}
E_0' &= \mel{\Uppsi_0'}{\Haml}{\Uppsi_0'} \\
&= E_0 + \iu \int_{\vb{r}} \vb{k} \vdot \vb{r} \mel{\Uppsi_0}{[\Haml, \rho_e]}{\Uppsi_0} + \bigO(\vb{k}^2) \\
&= E_0 + \hbar \int_{\vb{r}} \vb{k} \vdot \vb{r} \mel{\Uppsi_0}{\partial_t \rho_e}{\Uppsi_0} + \bigO(\vb{k}^2) \\
&= E_0 - \hbar \int_{\vb{r}} \vb{k} \vdot \vb{r} \, \grad \vdot \mel{\Uppsi_0}{\vb{j}_e}{\Uppsi_0} + \bigO(\vb{k}^2) \\
&= E_0 + \hbar \vb{k} \vdot \int_{\vb{r}} \mel{\Uppsi_0}{\vb{j}_e}{\Uppsi_0} + \bigO(\vb{k}^2) \\
&= E_0 + \hbar \vb{k} \vdot \vb{J} + \bigO(\vb{k}^2).
\end{aligned}
\end{align}
Hence we may always lower the energy relative to $E_0$ by orienting $\vb{k}$ in the opposite direction of $\vb{J}$.
This implies that the true ground state cannot have a finite global electron charge current~\cite{Bohm1949}.
Notice how the coupling to the ions, which proceeds via the density $\rho_e$, drops out in the above manipulations.
Examples of LC patterns which do and do not respect Bloch's theorem are provided in Fig.~\ref{fig:Bloch-examples}.

This proof applies to continuum models at zero temperature subject to open boundary conditions.
All three of these assumptions can be lifted~\cite{Ohashi1996, Yamamoto2015, Zhang2019, Watanabe2019, Watanabe2022}.
In the case of lattice models, the local charge conservation law takes the form
\begin{align}
\partial_t \rho_{\vb{R} \alpha} + \sum_{\vb{R}' \alpha'} j_{\vb{R} \alpha; \vb{R}' \alpha'} = 0,
\end{align}
where $j_{\vb{R} \alpha; \vb{R}' \alpha'} = j_{\vb{R} \alpha; \vb{R}' \alpha'}^{\dag} = - j_{\vb{R}' \alpha'; \vb{R} \alpha}$ is the current flowing from the lattice site $\vb{R}$ and orbital $\alpha$ to the lattice site $\vb{R}'$ and orbital $\alpha'$.
The variation
\begin{align}
\ket{\Uppsi_0'} = \exp\mleft(\iu \sum_{\vb{R} \alpha} \vb{k} \vdot \vb{R} \, \rho_{\vb{R} \alpha}\mright) \ket{\Uppsi_0}
\end{align}
results in an identical $E_0' = E_0 + \hbar \vb{k} \vdot \vb{J} + \bigO(\vb{k}^2)$, where the global current is defined as
\begin{align}
\vb{J} = \frac{1}{2} \sum_{\vb{R} \alpha \vb{R}' \alpha'} \mleft(\vb{R}' - \vb{R}\mright) \mel{\Uppsi_0}{j_{\vb{R} \alpha; \vb{R}' \alpha'}}{\Uppsi_0}.
\end{align}
In the proof for a finite-temperature ensemble described by the density matrix $\uprho_0 = \mathcal{Z}^{-1} \Elr^{- \upbeta \Haml}$, where $\mathcal{Z} = \Tr \Elr^{- \upbeta \Haml}$, one considers the variation
\begin{align}
\uprho_0 \to \uprho_0' = \exp\mleft(\iu \int_{\vb{r}} \vb{k} \vdot \vb{r} \, \rho_e\mright) \uprho_0 \exp\mleft(- \iu \int_{\vb{r}} \vb{k} \vdot \vb{r} \, \rho_e\mright)
\end{align}
and by completely analogous manipulations obtains that the free energy $F' = F + \hbar \vb{k} \vdot \vb{J} + \bigO(\vb{k}^2)$ given by $F = \Tr \uprho_0 \mleft(\Haml + k_B T \log \uprho_0\mright)$ is not minimal.
Finally, for periodic boundary conditions the variational $\vb{k}$ is on the order of $L^{-1}$, where $L$ is the length of the system.
Because $\vb{k}$ is not infinitesimal, the quadratic term in the expansion of $E_0$ is non-negligible and, because it is positive, it can compensate for the linear term and allow for a global charge current density which is on the order of $L^{-1}$.
The global charge current density thus vanishes in the thermodynamic limit.

Because the proof only relies on the local conservation of charge, Bloch's theorem quite generally applies to $\Ugp(1)$ symmetries and their Noether currents~\cite{Watanabe2022}.
In particular, in the absence of spin-orbit coupling global spin currents are forbidden in the ground state as well.

\subsubsection{Proof of a generalized Bloch-Kirchhoff theorem} \label{sec:Bloch-tm-Kirchhoff}
Here we prove a generalization of Bloch's theorem to local currents.
Consider the order parameter of a generic LC state:
\begin{align}
\Phi &= \int_{\vb{r}} \vb{v} \vdot \mel{\Uppsi_0}{\vb{j}_e}{\Uppsi_0},
\end{align}
where the vector field $\vb{v}$ specifies its structure.
We shall now show that $\Phi = 0$ whenever $\vb{v}$ is curl-free, $\grad \vcross \vb{v} = \vb{0}$.
Assume that we are given a ground state $\ket{\Uppsi_0}$.
By Helmholtz's theorem, curl-free $\vb{v}$ can always be written in the form $\vb{v} = \grad \vartheta$.
Hence the state
\begin{align}
\ket{\Uppsi_0'} = \exp\mleft(\iu k \int_{\vb{r}} \vartheta \, \rho_e\mright) \ket{\Uppsi_0}
\end{align}
has the energy
\begin{align}
\begin{aligned}
E_0' &= E_0 + \iu k \int_{\vb{r}} \vartheta \mel{\Uppsi_0}{[H, \rho_e]}{\Uppsi_0} + \bigO(k^2) \\
&= E_0 - \hbar k \int_{\vb{r}} \vartheta \, \grad \vdot \mel{\Uppsi_0}{\vb{j}_e}{\Uppsi_0} + \bigO(k^2) \\
&= E_0 + \hbar k \Phi + \bigO(k^2)
\end{aligned}
\end{align}
which can always be made smaller than that of the presumed ground state if $\Phi$ is finite.
Thus $\Phi$ always vanishes in the ground state for arbitrary curl-free $\vb{v}$.
In turn, this implies that the longitudinal component of $\mel{\Uppsi_0}{\vb{j}_e}{\Uppsi_0}$ vanishes, which is equivalent to stating that $\grad \vdot \mel{\Uppsi_0}{\vb{j}_e}{\Uppsi_0} = 0$.\footnote{Recall that in general $\int_{\vb{r}} \vb{v} \vdot \vb{w} = \int_{\vb{r}} \vb{v}_L \vdot \vb{w}_L + \int_{\vb{r}} \vb{v}_T \vdot \vb{w}_T$, where $\grad \vcross \vb{v}_L = \grad \vcross \vb{w}_L = \vb{0}$ and $\grad \vdot \vb{v}_T = \grad \vdot \vb{w}_T = 0$.}
On a lattice, this gives the following Kirchhoff constraint on the pattern of local currents:
\begin{align}
\sum_{\vb{R}' \alpha'} \mel{\Uppsi_0}{j_{\vb{R} \alpha; \vb{R}' \alpha'}}{\Uppsi_0} = 0.
\end{align}
Clearly, the proof of this generalized Bloch-Kirchhoff theorem proceeds with minimal modifications of the original proof.
It is straightforward to generalize it to finite temperatures, lattice models, and periodic boundary conditions.

A physical interpretation of both theorems is that if a current were initially present, it would lead to an accumulation of a charge (in the bulk or on the boundary) whose electric fields would then counteract to remove the initial current.
Hence there cannot be any longitudinal charge current in equilibrium.

\subsection{Previous theoretical and experimental work on loop currents} \label{sec:general-LC-work-review}
Although the direct spin-spin (magnetic dipole-dipole) interactions are weak, because of Pauli's exclusion principle the much stronger Coulomb interactions among electrons, as well as the Coulomb interaction between electrons and ions, may acquire spin-dependence, inducing an effective spin-spin interaction which is strong~\cite{Kittel2005, Simon2013}.
Magnetic ordering in the spin sector is therefore relatively common.

There is no analogous simple argument for why orbital angular momentum should have as large an influence on the electronic interactions in crystalline systems.
Indeed, the surrounding crystal environment lifts the angular momentum degeneracy because it breaks the full $\SO(3)$ rotational symmetry group down to a discrete subgroup.
As a result, on-site orbital angular momentum $\vb{L}$ is not a good quantum number and its average vanishes, rendering $\vb{L}$ inactive at low energies.
This is the so-called quenching of orbital angular momentum~\cite{Kittel2005, Simon2013}.

Nevertheless, if the systems has strong correlations and multiorbital physics, or if the quenching of the on-site orbital angular momentum is sufficiently weak, orbital magnetism may still arise~\cite{Varma2020, Bourges2021}.
Multiorbital physics is expected to be favorable for orbital magnetism because it enables intersite generalized orbital angular momentum operators, as we shall see in the next chapter on cuprates (see Fig.~\ref{fig:our-ASV-flux-operators}, for instance).
Even though we shall shy away from attempting to derive LC order from microscopic models in this thesis, which is an interesting but challenging problem in itself, let us briefly discuss previous theoretical work along this direction, as well as experimental evidence for LC order in systems other than the cuprates.

One of the earliest mentions of orbital magnetism is in a chapter by Halperin and Rice from 1968~\cite{Halperin1968} in which, using the screened Hartree-Fock approximation, orbital antiferromagnetism was found in a model of itinerant electrons as a possible instability alongside charge-density waves and spin-density waves.
This orbital antiferromagnetism was not studied in much detail in the chapter, however.
In the pioneering works by Kugel' and Khomskii~\cite{Kugel1973} and Inagaki~\cite{Inagaki1975}, they theoretically investigated the possibility of orbital magnetism due to exchange interactions in systems where the crystal fields do not completely quench the orbital angular momentum of the localized electronic states.
One way of understanding this orbital ordering is as a purely electronic analog of the cooperative Jahn-Teller effect~\cite{Khomskii2021}.
In a more concrete setting, Ohkawa~\cite{Ohkawa1985} argued that \ce{CeB6} has local orbital moments which order into an orbitally antiferromagnetic state.
Apart from these references, for many years orbital magnetism was not the subject of much study.
This changed with the discovery of the high-temperature superconductors in 1986~\cite{Bednorz1986}.
After the discovery, a number of researchers have found LC order in models of relevance to cuprates~\cite{Affleck1988, Marston1989, Schulz1989, Varma1997}.
In particular, Chandra M.\ Varma has suggested that intra-unit-cell LC order is, in fact, the key to understanding the phase diagram of the cuprates~\cite{Varma1997, Varma2020}.
Other researchers have made similar proposals~\cite{Chakravarty2001, Laughlin2014, Laughlin2014-details}.
This and related work we shall review in the next chapter in Sec.~\ref{sec:LC-cuprate-micro}.

Sun and Fradkin~\cite{Sun2008} have theoretically investigated orbital magnetism in general Fermi liquids without SOC.\footnote{Note that they call ``magnetic'' what we call spin-magnetic and their ``nonmangetic with TRSB'' means orbital-magnetic according to us.}
They considered two types of orbital magnetism:
type I, which are odd-parity LCs invariant under a reflection symmetry, and
type II, which are even-parity LCs odd under a reflection symmetry.
Note that the parity of a LC state refers to the behavior of the corresponding LC pattern under space inversion; see Fig.~\ref{fig:our-ASV-flux-operators} of the next chapter for examples of even- and odd-parity LCs.
In modern terms~\cite{Mazin2022, Smjekal2022}, type II states can be understood as orbital altermagnets since composing TR with a reflection leaves the system invariant, i.e., reflections compensate for TRSB; cf.\ Fig.~\ref{fig:spin-mag-class}.
What makes type II states so interesting is that, unlike type I states, they spontaneously exhibit Kerr and anomalous Hall effects even in the absence of magnetic fields and disorder.
In related work, using mean-field theory~\cite{Sun2008, Castro2011} and renormalization group methods~\cite{Sur2018, Tazai2021}, it was found that LC states can be stabilized in several microscopic models of itinerant electrons.\footnote{For Hubbard, $t$-$J$, and other models of relevance to cuprates, see Sec.~\ref{sec:LC-cuprate-micro}.}
Loop currents may also accompany other orders, as has been explored in \ce{AV3Sb5} kagome metals~\cite{Christensen2022}, TRSB superconductors on a honeycomb lattice~\cite{Brydon2019}, spin liquids in cuprate superconductors~\cite{Scheurer2018}, as well as iron-based superconductors~\cite{Klug2018} in which conventional spin magnetism induces orbital magnetism through SOC.
Spin-orbit coupling may also act in the opposite direction, inducing spin order which accompanies LC order~\cite{Aji2007}.

LC order can be experimentally probed using a number of methods~\cite{Bourges2021}.
Although TRSB takes place in the orbital sectors, external spins are still sensitive to the local magnetic fields, irrespective of origin.
Thus spin-polarized neutrons and muons, as used in polarized neutron diffraction (PND) and muon spin relaxation ($\mu$SR), can probe LC order~\cite{Bourges2021}.
PND has the additional advantage of momentum resolution, allowing it to tell apart LCs which do and do not break translation symmetry.
The latter we shall refer to as homogeneous, $\vb{q} = \vb{0}$, or intra-unit-cell (IUC) LCs.
TRSB can also be experimentally observed through the magneto-optic Kerr effect~\cite{Kapitulnik2009}.
If the LC order is odd under parity, it contributes to the third-rank optical susceptibility tensor which is measurable via optical second-harmonic generation experiments~\cite{Fiebig2005}.

Although LC order has been most extensively studied in the cuprates~\cite{Varma2020, Bourges2021}, as we shall discuss in Sec.~\ref{sec:cuprate-LC-lit-review} of the next chapter, there is evidence for LC order in other systems as well.
A state consistent with LC order has been inferred from optical second-harmonic generation measurements~\cite{Zhao2016} and PND~\cite{Jeong2017} in the iridate Mott insulator \ce{Sr2IrO4} which displays an unusual gap upon doping~\cite{Yan2015, Kim2016}.
A LC pattern that breaks translation symmetry is one of the main candidates for explaining why the charge-density wave displayed by the recently discovered kagome superconductors seemingly breaks TR symmetry~\cite{Mielke2022}.
Orbital magnetism is generically induced by spin magnetism thorough SOC, hence experimental evidence for stripe spin-magnetic order in iron-based superconductors~\cite{Johnston2010, Paglione2010, Stewart2011, Avci2014} can be taken as indirect evidence for (subleading) LC order in those compounds~\cite{Klug2018}.

\section{Paradigm of pairing driven by quantum-critical order-parameter fluctuations} \label{sec:QCP-paradigm}
The idea that bosonic modes coupled to electrons can induce an attractive interaction among electrons, resulting in Cooper pairing~\cite{Cooper1956}, has a long history, going all the way back to the landmark articles by Bardeen, Cooper, and Schrieffer (BCS)~\cite{BCSshort, BCS} in which the bosonic modes are the phonons.
For a lucid review of BCS theory, see the book by Leggett~\cite{Leggett2006}; the book by Schrieffer~\cite{Schrieffer1999} is also excellent.
Since then, boson-exchange pairing mechanisms have been extensively studied~\cite{Grabowski1984, Carbotte1990, Monthoux2007}, mostly within the framework of Migdal-Eliashberg theory~\cite{Migdal1957, Eliashberg1960, Eliashberg1961, Allen1983, Ummarino2013, Marsiglio2020}.

Besides phonons, many other collective modes have been proposed as the pairing glue, including ferromagnetic~\cite{Millis2001, Chubukov2003} and antiferromagnetic~\cite{Scalapino1995, Mathur1998, Scalapino1999, Moriya2000, Abanov2001, Metlitski2010, YuxuanWang2014, Scalapino2012} spin waves, nematic fluctuations~\cite{Yamase2013, Lederer2015, Lederer2017, Klein2018}, ferroelectric modes~\cite{Kozii2015, KoziiBiRuhman2019, Klein2023}, and orbital loop currents~\cite{Varma1997, Aji2010, Varma2016, Varma2020, Chakravarty2001, Laughlin2014, Laughlin2014-details}.
With few exceptions~\cite{Takada1978, Ruvalds1987}, the collective modes whose potential role in driving superconductivity (SC) has been investigated the most are modes associated with ordered states.
We may think of these collective modes as fluctuating order parameters.
As we shall see in Sec.~\ref{sec:QCP-pairing-model}, within the action formalism fluctuating order parameters can be rigorously introduced through a Hubbard-Stratonovich transformation~\cite{Hertz1976, Altland2010}.
Of particular interest to our work are ``purely orbital'' collective modes, that is, collective modes which are trivial in the spin sector.
Loop currents are an example, but there are other modes as well, as summarized in Tab.~\ref{tab:orbital-modes}.

\begin{table}[t]
\centering
\captionabove[A selected list of bulk orders, classified according to whether they break translation (red) or time-reversal (TR) symmetry and whether they are trivial or not in the spin sector.]{\textbf{A selected list of bulk orders, classified according to whether they break translation (\textcolor{red}{red}) or time-reversal (TR) symmetry and whether they are trivial or not in the spin sector.}
Orbital/charge orders are trivial in spin space.
Highlighted in red are orders which break translation invariance.
Nematic orders are electronic orders which spontaneously break point group symmetries such as rotations or reflections without breaking parity or translation symmetry~\cite{Fradkin2010, Hecker2024}.
Spin-nematics do the same in the spin sector, but without TR symmetry-breaking~\cite{Andreev1984, Fernandes2019}.
Ferroelectrics have spontaneous electric polarizations and their modes are usually soft polar phonons~\cite{Cochran1960}.
Alterelectrics I dub orders which break parity in the orbital sector, just like ferroelectrics, but whose order parameter does not transform like a vector.
Hence, alterelectrics do not have a net electric dipole moment.
Charge- and spin-density waves are spontaneously forming periodic modulations of the charge or spin density which are incommensurate with the underlying lattice~\cite{Gruner1988, Gruner1994}.
Loop-current order is orbital magnetism, as discussed in Sec.~\ref{sec:orbital-mag}.
Spin loop currents are loop currents which carry spin.
Regarding the spin-magnetic orders, see Fig.~\ref{fig:spin-mag-class} for an explanation.
For a more systematic classification of orders, see Refs.~\cite{Watanabe2018, Hayami2018, Winkler2023}.}
{\renewcommand{\arraystretch}{1.3}
\renewcommand{\tabcolsep}{7.2pt}
\begin{tabular}{|c"c|c|} \hline
& {orbital or charge} & {spin} \\ \thickhline
{TR-even} & \begin{tabular}{c}
nematic \\
ferroelectric \\
alterelectric \\
\textcolor{red}{charge-density waves}
\end{tabular} & \begin{tabular}{c}
spin-nematic \\
intra-unit-cell spin loop currents \\
\textcolor{red}{staggered spin loop currents}
\end{tabular} \\ \hline
{TR-odd} & \begin{tabular}{c}
intra-unit-cell (orbital) loop currents \\
\textcolor{red}{staggered (orbital) loop currents}
\end{tabular} & \begin{tabular}{c}
ferromagnetic \\
altermagnetic \\
\textcolor{red}{antiferromagnetic} \\
\textcolor{red}{spin-density waves}
\end{tabular}
\\ \hline
\end{tabular}}
\label{tab:orbital-modes}
\end{table}

When bosonic collective modes couple to electrons, they generate interactions among electrons, as depicted in Fig.~\ref{fig:boson-exchange}.
As long as this interaction is attractive in some pairing channel and sufficiently strong to overcome Coulomb and other repulsive interactions in that channel, SC will emerge at low enough temperatures, assuming a normal Fermi-liquid state~\cite{Schrieffer1999, Sigrist2005, Leggett2006, Leggett2011}.
This is a consequence of the fact that the Fermi sea is unstable against pairing in arbitrarily weak attractive Cooper channels~\cite{Schrieffer1999, Sigrist2005, Leggett2006, Leggett2011}, albeit at exponentially small temperatures $T_c \propto \omega_c \Elr^{-1/\lambda}$, where $\omega_c$ is a characteristic frequency of the modes and $\lambda$ is the pairing eigenvalue.
The appearance of SC and its strength, as reflected in the transition temperature $T_c$, is then a matter of the detailed properties of the material under study.
This is, broadly speaking, the situation for Cooper pairing due to electron-phonon coupling~\cite{Marsiglio2008, Marsiglio2020}.

In the case of coupling to order-parameter fluctuations, however, there is a well-defined regime where one expects the boson-mediated electron-electron interaction to be especially strong.
This is the regime near the quantum-critical point (QCP) of the associated ordered state, where the collective modes become soft and the associated order-parameter fluctuations are particularly strong.\footnote{Soft modes are not necessarily good at driving SC if they couple weakly to electrons, as is the case for Nambu-Goldstone modes~\cite{Watanabe2014}. For an explanation of why soft modes mediate strong interactions between fermions, see Sec.~\ref{sec:QCP-pairing-model} and the discussion surrounding Eq.~\eqref{eq:effective-multi-exch-int}.}
Such continuous quantum phase transitions have been the subject of extensive study~\cite{Sondhi1997, Vojta2000, Vojta2003, Lohneysen2007, Sachdev2011}, as has the corresponding pairing due to quantum-critical fluctuations that takes places in the vicinity of QCPs~\cite{Varma2016, She2009, Moon2010, Metlitski2015, YuxuanWang2016-inv, YuxuanWang2016-QCP, WangRaghu2017, ChubukovI_2020, ChubukovII_2020}.
Although we shall not attempt to review this vast field, below we recapitulate some fundamental notions which are relevant to our work.

\begin{figure}[t]
\centering
\includegraphics[width=\textwidth]{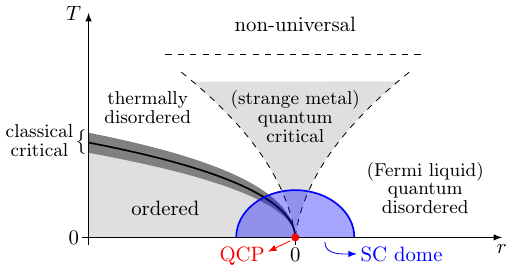}
\captionbelow[Generic phase diagram in the vicinity of a continuous quantum phase transition~\cite{Vojta2003, Lohneysen2007}.]{\textbf{Generic phase diagram in the vicinity of a continuous quantum phase transition}~\cite{Vojta2003, Lohneysen2007}.
$T$ is the temperature and $r$ is a tuning parameters, such as doping, pressure, or uniaxial stress.
At $r = T = 0$, the system passes through a quantum-critical point (QCP), which may or may not be surrounded by a superconducting (SC) dome.
See text for further discussion.}
\label{fig:QCP-pairing-paradigm}
\end{figure}

Consider a system that, as a function of some control parameter $r$ such as doping, pressure, or uniaxial stress, experiences a continuous quantum phase transition from a disordered state into an ordered state of broken symmetry, as depicted in Fig.~\ref{fig:QCP-pairing-paradigm}.
This ordered state may extend to finite temperatures, as assumed in Fig.~\ref{fig:QCP-pairing-paradigm}, or it can formally only arise at $T = 0$ because of Hohenberg-Mermin-Wagner's theorem~\cite{Hohenberg1967, MerminWagner1966, Halperin2019}.
This theorem forbids the spontaneous breaking of continuous symmetries in two or less (spatial) dimensions at finite temperatures.\footnote{In two dimensions, however, finite-size and disorder effects can render Hohenberg-Mermin-Wagner's theorem physically irrelevant~\cite{Palle2021-HMW}.}
Either way, this ordered phase can become disordered in two ways: by thermal fluctuations or by quantum fluctuations.

If we are on the ordered side of the phase diagram, $r < 0$, and approach the transition temperature $T_c$ with a constant $r$, then for a continuous phase transition the correlation length $\xi$ and the correlation (or equilibration) time $\xi_{\tau}$ diverge as~\cite{Vojta2003, Lohneysen2007}
\begin{align}
\xi &\propto \abs{t}^{- \nu}, &
\xi_{\tau} &\propto \xi^z \propto \abs{t}^{- \nu z},
\end{align}
where $\nu > 0$ is the correlation length critical exponent of the finite temperature transition, $z > 0$ is the dynamic exponent, and $t = (T - T_c) / T_c$ is a dimensionless measure of the distance from $T_c$.
From the divergence of $\xi_{\tau}$, it follows that the order parameter fluctuates coherently over increasingly large time-scales as $T \to T_c$.
This means that the characteristic frequency $\omega_c$ characterizing the large-scale order-parameter fluctuations becomes softer and vanishes at the critical point like~\cite{Vojta2003}
\begin{align}
\omega_c \propto \xi_{\tau}^{-1} \propto \abs{t}^{\nu z}.
\end{align}
Hence $\hbar \omega_c \ll k_B T$ near $T_c$ and the critical fluctuations behave in a classical way~\cite{Vojta2003, Lohneysen2007}.
The corresponding region is labeled ``classical critical'' in Fig.~\ref{fig:QCP-pairing-paradigm}.
Microscopically, quantum effects can still be important near $T_c$, but at large scales at least the transition is essentially classical.
As the temperature is further increased, we enter a ``thermally disordered'' phase.

Let us now examine the $T = 0$, or quantum, transition as $r$ is varied.
For continuous quantum phase transition, $\xi$ and $\xi_{\tau}$ again diverge, but this time as a function of the tuning parameter $r$~\cite{Vojta2003, Lohneysen2007}:
\begin{align}
\xi &\propto \abs{r}^{- \nu}, &
\xi_{\tau} &\propto \xi^z \propto \abs{r}^{- \nu z}.
\end{align}
Here we have assumed that $r$ is defined so that it is dimensionless and vanishes at the QCP.
Note that the critical exponents $\nu$ and $z$ are different from those of the finite temperature transition~\cite{Lohneysen2007}.
The characteristic frequency $\omega_c \propto \xi_{\tau}^{-1} \propto \abs{r}^{\nu z}$ of the quantum-critical fluctuations yet again vanishes as the QCP is approached, but is finite elsewhere.
Thus for finite $r > 0$ and small $T$, $\hbar \omega_c \gg k_B T$ and the systems has essentially the same behavior as the $T = 0$ ground state~\cite{Vojta2003}.
Since the disorder of the $r > 0$ ground state is driven by quantum fluctuations, the same is true for the ``quantum disordered'' region of Fig.~\ref{fig:QCP-pairing-paradigm}.
In between the quantum and thermally disordered regions, there is a ``quantum critical'' cone where both thermal and quantum fluctuations are important~\cite{Vojta2003, Lohneysen2007}.
This cone widens with increasing temperature because the continuum of quantum-critical excitations associated with the QCP is more efficiently excited at large $T$.
At very large temperatures, $k_B T$ becomes comparable to the microscopic energy scales of the system and the minute microscopic features of the system become relevant to its behavior, driving ``non-universal'' physics~\cite{Vojta2003, Lohneysen2007}, as indicated in Fig.~\ref{fig:QCP-pairing-paradigm}.

In the case of itinerant electronic systems, the quantum disordered phase is a Fermi liquid and the quantum-critical phase is called a quantum-critical metal, strange metal, or non-Fermi liquid~\cite{Vojta2003, Lohneysen2007}.
The key feature that sets strange metals apart from Fermi liquids is the absence well-defined fermionic quasi-particles near the Fermi surface.
Instead, one finds a critical continuum of excitations which drives various power-law behavior~\cite{Vojta2003, Lohneysen2007}.
Following the pioneering work by Hertz~\cite{Hertz1976}, later extended by Millis~\cite{Millis1993}, there has been much work on non-Fermi liquids driven by quantum-critical fluctuations.
Although we shall not attempt to review this fascinating field here, let us note that the broad qualitative picture of QCPs sketched above holds for itinerant electronic systems in particular.
The interested reader we refer to the many excellent reviews on the topic~\cite{Vojta2000, Vojta2003, Lohneysen2007, Sachdev2011, Schofield1999, Stewart2001, Lee2018, Chowdhury2022}.

More relevant to our work is the paradigm of Cooper pairing due to quantum-critical order-parameter fluctuations~\cite{Varma2016, She2009, Moon2010, Metlitski2015, YuxuanWang2016-inv, YuxuanWang2016-QCP, WangRaghu2017, ChubukovI_2020, ChubukovII_2020}.
According to Ref.~\cite{She2009}, some of the earliest seeds of this paradigm are the marginal Fermi liquid ideas of Varma~\cite{Varma1989}, motivated by the cuprates, and the notion of SC driven by critical spin fluctuation~\cite{Miyake1986, Mathur1998}, relevant to heavy fermion compounds.
Although there has been much work that broadly falls into this paradigm (see \cite{Varma2016, She2009, Moon2010, Metlitski2015, YuxuanWang2016-inv, YuxuanWang2016-QCP, WangRaghu2017, ChubukovI_2020, ChubukovII_2020} and references cited therein), most reviews to date cover quantum phase transitions in general, discussing Cooper pairing in the passing~\cite{Lohneysen2007}, or focus on pairing near QCPs of only one type of order~\cite{Scalapino2012, Varma2020}.
A comparative review of QCP-based pairing mechanisms would be very interesting, but is currently unavailable.
That said, I will not attempt to fill in this gap in the literature here, but rather only sketch the broad physical picture.

The most well-understood part of the QCP phase diagram of Fig.~\ref{fig:QCP-pairing-paradigm} is the Fermi-liquid region, so let us start from there.
Let us assume that we have a Fermi liquid which is weakly coupled to heavy order-parameter fluctuations.
As the QCP is approached, these bosonic order-parameter fluctuations become soft and the electron-electron interactions that they mediate become stronger.
Because of the coupling to electrons, however, the order-parameter bosons in addition become damped, which changes the effective space-time dimensionality of the bosonic theory~\cite{She2009}.
The order-parameter fluctuations, in turn, affect the electrons in two ways~\cite{Moon2010, ChubukovI_2020}.
On the one hand, they make the electrons less coherent, which tends to weaken the logarithmic Cooper divergence associated with pairing.
On the other hand, the Cooper-channel interaction that they mediate is more singular at low frequencies than it would be away from the QCP.
Depending on which of these two competing tendencies prevail, the QCP may or may not be surrounded by a SC dome~\cite{Moon2010, ChubukovI_2020}, as shown in Fig.~\ref{fig:QCP-pairing-paradigm}.
Of course, it may be the case that the interaction mediated by the order-parameter modes is repulsive in all Cooper channels, in which case SC will never take place.
It is also possible that the quantum critical region is completely hidden inside the SC dome~\cite{Metlitski2015}, in contrast to what is shown in Fig.~\ref{fig:QCP-pairing-paradigm}.

\section{Analysis of pairing due to quantum-critical loop-current and other fluctuations} \label{sec:gen-sys-LC-analysis}
Having introduced LC order and the paradigm of pairing due to quantum-critical fluctuations, we are now in a position to study the pairing when the quantum-critical fluctuations originate from an underlying LC order.
As already mentioned at the beginning of this chapter, the forthcoming analysis borrows heavily (sometimes verbatim) from Ref.~\cite{Palle2024-LC}.

Given their potential realization in a diverse set of systems, as reviewed in Sec.~\ref{sec:general-LC-work-review} and Sec.~\ref{sec:cuprate-LC-lit-review} of the next chapter, it is important to elucidate whether fluctuating loop currents can give rise to superconductivity (SC).
Related, equally important, matters are the strength of this pairing tendency and the symmetry of the resulting SC state.
In this context, intra-unit-cell (IUC) LCs have been prominently discussed as the pairing glue of the cuprates~\cite{Varma1997, Aji2010, Varma2016, Varma2020} which makes them particularly interesting, notwithstanding the difficulties in detecting them.\footnote{Note that by IUC order we mean homogeneous $\vb{q} = \vb{0}$ order which preserves the translation symmetries of the underlying lattice.}
For comparison, in the case of fluctuations deriving from IUC orders that preserve time-reversal symmetry, such as nematic~\cite{Yamase2013, Lederer2015, Lederer2017, Klein2018} and ferroelectric~\cite{Kozii2015, KoziiBiRuhman2019, Klein2023} ones, it is well established that $s$-wave pairing generally emerges with a number of attractive subleading channels.
Pairing is promoted by ferromagnetic spin fluctuations~\cite{Millis2001, Chubukov2003} as well, the main difference being the $p$-wave nature of the leading pairing state.
Furthermore, SC in all of these cases is strongly enhanced in two (spatial) dimensions as the quantum-critical point (QCP) is approached, thus establishing a robust regime in which pairing is dominated by the corresponding fluctuations.
However, the case of pure orbital magnetism is different, not only because LCs do not directly couple to the spin, but also because they usually break additional symmetries besides time reversal.
This raises the question of whether there are general conditions, independent of the details of a given material, under which pairing is dominated by quantum-critical IUC LC fluctuations.

\begin{figure}[t!]
\centering
\includegraphics[width=\textwidth]{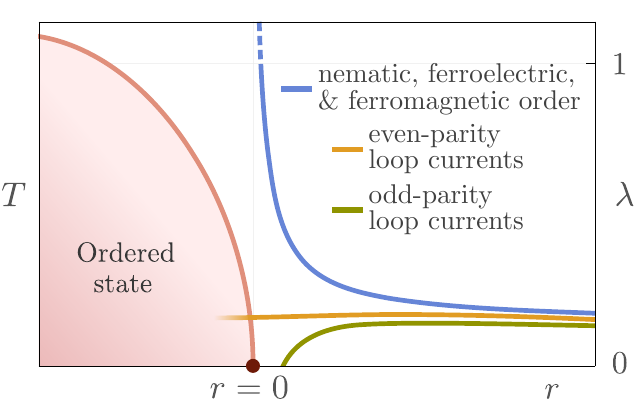}
\captionbelow[Schematic behavior of the leading pairing eigenvalue $\lambda$ as a quantum-critical point (QCP) is approached from the disordered (Fermi liquid) side, depending on the type of intra-unit-cell (IUC) order~\cite{Palle2024-LC}.]{\textbf{Schematic behavior of the leading pairing eigenvalue $\lambda$ as a quantum-critical point (QCP) is approached from the disordered (Fermi liquid) side, depending on the type of intra-unit-cell (IUC) order}~\cite{Palle2024-LC}.
The tuning parameter $r$ (horizontal scale) controls the QCP $r = T = 0$.
The vertical temperature scale on the left is for the ordered state, whereas the vertical $\lambda$ scale on the right is for the curves.
The superconducting transition temperature grows with $\lambda$ according to $k_B T_c = 1.134 \hbar \omega_c \Elr^{-1/\lambda}$, where $\hbar \omega_c$ is an energy cutoff.
Pairing mediated by time-reversal-even IUC charge-fluctuations or time-reversal-odd IUC spin-fluctuations (blue) is enhanced near the QCP, where weak-coupling theory breaks down (dashed line).
In contrast, we find that the pairing mediated by even-parity IUC loop currents (orange) is not enhanced at the QCP, whereas pairing mediated by odd-parity IUC loop currents (green) becomes strongly repulsive near the QCP.}
\label{fig:QCP-general-results}
\end{figure}

In the remainder of this chapter, we address this question.
We show that IUC LC fluctuations do not give rise to an enhanced pairing near the QCP, as shown schematically in Fig.~\ref{fig:QCP-general-results}.
Even-parity LCs, such as orbital ferromagnets or orbital altermagnets, may cause unconventional pairing.
However, they are as likely or unlikely to do so as any other degree of freedom far from its critical point.
This is because the pairing promoted by these fluctuations is not enhanced as the QCP is approached (orange line in Fig.~\ref{fig:QCP-general-results}), in sharp contrast to the cases of ferromagnetic spin fluctuations or TR-even charge fluctuations, such as nematic or ferroelectric\footnote{Here we are comparing ferroelectrics with SOC to LCs without SOC. In the absence of SOC, pairing driven by quantum-critical ferroelectric fluctuations does not become enhanced near the corresponding QCP, just like in the case of even-parity LCs. See Sec.~\ref{sec:IUC-order-results} for further discussions.} ones (blue line in Fig.~\ref{fig:QCP-general-results}).
LCs that break parity, i.e.\ states of magneto-electric order, are repulsive for all pairing symmetries as one approaches the QCP (green line in Fig.~\ref{fig:QCP-general-results}).
Hence they weaken pairing caused by other pairing mechanisms.
Such odd-parity LC states can at best support SC when their fluctuations are sufficiently weak.

These conclusions hold for general systems in which spin-orbit coupling (SOC) is weak.
In particular, they hold for the cuprates, seemingly challenging previous work on the topic which suggested that IUC loop currents are effective at driving pairing~\cite{Varma1997, Aji2010, Varma2016, Varma2020}.
However, the questions of what type of LC orders can arise in the cuprates, how can one probe and tell apart these LC orders experimentally, and whether the fluctuations of these LC orders induce the correct $d_{x^2-y^2}$-wave symmetry still need to be addressed.
To this task, we devote the next chapter.

Although until now we have focused on LC order, the analysis leading up to our results applies with minimal modifications to other IUC orders.
However, many of these were previously already studied~\cite{Lederer2015} and were not the subject of much controversy, which is why in the article itself~\cite{Palle2024-LC} we focused on LC order.
That said, comparing and contrasting with other IUC orders is enlightening, not only because we cover additional orders which are of interest, but also because it highlights how uniquely ineffective LCs are at driving Cooper pairing near their QCPs.
As we shall see in Sec.~\ref{sec:IUC-order-results}, the pair-breaking tendency of odd-parity LCs is, in fact, unique among all conceivable IUC orders.

For the above reasons, we have framed the whole analysis in the most general fashion possible, starting with an arbitrary fluctuating order parameter which we narrow down to loop currents only at the end.
As a side-effect of this approach, the formulas of this section will have many indices.
To aid comprehension, here we list the notation conventions which we shall consistently use through the rest of this chapter (see also \nameref{app:conventions}):
\begin{itemize}
\item $\vb{R}, \vb{\delta}$ are real-space lattice vectors.
\item $\vb{k}, \vb{p}, \vb{q}$ are crystal momenta and they are always within the first Brillouin zone.
\item $\alpha, \beta \in \{1, \ldots, 2M\}$ are fermion component indices, covering both orbital and spin degrees of freedom.
\item $s \in \{\uparrow, \downarrow\}$ are spin or pseudospin indices.
\item $n, m \in \{1, \ldots, M\}$ are band indices.
\item $\vb{k}_n$ is on the Fermi surface of the $n$-th band, i.e., it satisfies $\varepsilon_{\vb{k} n} = 0$, where $\varepsilon_{\vb{k} n}$ is the dispersion of the $n$-th band displaced by the chemical potential. Likewise, $\vb{p}_m \iff \varepsilon_{\vb{p} m} = 0$.
\item $A, B \in \{0, 1, 2, 3\}$ are Pauli matrix indices. Within the Balian-Werthamer $\vb{d}$-vector notation~\cite{Balian1963}, $A = 0$ is the even-parity singlet and $\{1, 2, 3\}$ are the odd-parity triplet channel components.
\item $a, b \in \{1, \ldots, \dim \Phi\}$ are the order parameter component indices.
\item $g$ are point group elements.
\end{itemize}

The rest of this section is organized as follows.
We start by defining the model of itinerant fermions coupled to order-parameter fluctuations which we use to analyze quantum-critical pairing.
Its symmetry transformation rules are provided in Sec.~\ref{sec:LC-gen-model-sym}.
This section requires some knowledge of group and representation theory, which is reviewed in Appx.~\ref{app:group_theory} for the reader's convenience.
Afterwards, in Sec.~\ref{sec:QCP-model-lin-gap-eq}, we introduce the linearized gap equation (derived in Appx.~\ref{app:lin_gap_eq}) and we prove a number of properties of its Cooper-channel interaction.
In particular, in Sec.~\ref{sec:Cp-channel-gen-sym-constr} we prove general symmetry constraints on the Cooper-channel interaction which are key to our arguments.
The results for the generic behavior near a IUC QCP, summarized in Fig.~\ref{fig:QCP-general-results} for the case of LC, are derived in Sec.~\ref{sec:QCP-genera-final-analysis}.
We end with a discussion of the effects of SOC on our results and a comparison with pairing driven by fluctuations of staggered (finite-$\vb{q}$) orders.

\subsection{Model of itinerant fermions coupled to fluctuating order parameters} \label{sec:QCP-pairing-model}
Here we introduce a general model which allows us to draw conclusions that are independent of material details.
The model is made of itinerant fermions coupled to a soft order-parameter field.
Other interactions, such as the repulsive Coulomb interaction, are not included, so any strong pairing tendency that we find near the QCP is indicative of SC, but the obtained transition temperature $T_c$ is likely overestimated.

Let us consider a centrosymmetric system with $M$ orbitals per primitive unit cell.
Introduce the fermionic spinors $\psi_{\vb{k}} \equiv \mleft(\psi_{\vb{k}, 1, \uparrow}, \psi_{\vb{k}, 1, \downarrow}, \ldots, \psi_{\vb{k}, M, \uparrow}, \psi_{\vb{k}, M, \downarrow}\mright)^{\intercal}$, where $s \in \{\uparrow, \downarrow\}$ is the physical spin.
In terms of these spinors, the one-particle Hamiltonian equals
\begin{align}
\Haml_0 = \sum_{\vb{k}} \psi_{\vb{k}}^{\dag} H_{\vb{k}} \psi_{\vb{k}}.
\end{align}
The corresponding non-interacting Euclidean action is~\cite{Altland2010, Coleman2015}:
\begin{align}
\action_0[\psi] = \int_0^{\upbeta} \dd{\tau} \sum_{\vb{k}} \psi_{\vb{k}}^{\dag}(\tau) \mleft[\partial_{\tau} + H_{\vb{k}}\mright] \psi_{\vb{k}}(\tau),
\end{align}
where $\tau$ is imaginary time, $\upbeta = 1/(k_B T)$, and all momentum summations here and elsewhere go over the first Brillouin zone only.
For $\Haml_{0}$, we assume that it respects parity and time-reversal symmetry.

The interactions we treat phenomenologically and assume from the outset that they give rise to some kind of ordered state, which we later fix to LC order.
For comparison, a microscopic treatment would start from a physically motivated microscopic interaction, such as an extended Hubbard interaction, and then decompose it into different channels (cf.\ Sec.~\ref{sec:CuO2-Hubbard-decompose}).
For each of these channels, one may then use a Hubbard-Stratonovich transformation to introduce a corresponding bosonic field which can be interpreted as a fluctuating order-parameter field~\cite{Hertz1976, Altland2010}.
The question of which of these channels prevails and orders first is a challenging one and we shall not attempt to answer it.
Instead, we focus on what happens afterwards: Are the given order-parameter fluctuations effective at driving SC near their QCP?
If the answer is negative, then the whole problem of whether this or that order arises in a given compound becomes moot, at lest with regard to quantum-critical Cooper pairing.
Within the model, we shall only include one type of fluctuating order parameter, implicitly assuming that the system orders in that channel.

To introduce the interaction, let us suppose that there is a real collective mode $\Phi_{a}(\vb{R}) = \Phi_{a}^{*}(\vb{R})$ present in the system that transforms according to an irreducible representation (irrep) of the point group and that has a well-defined sign under time reversal (TR).
This mode we couple to the fermions through a Yukawa term of the form:
\begin{align}
\begin{aligned}
\Haml_c &= g \sum_{a \vb{R}} \Phi_{a}(\vb{R}) \phi_{a}(\vb{R}) \\
&= g \sum_{a \vb{q}} \Phi_{a, -\vb{q}} \phi_{a \vb{q}},
\end{aligned} \label{eq:Yukawa-coupling}
\end{align}
where $\phi_{a}(\vb{R}) = \phi_{a}^{\dag}(\vb{R})$ is a Hermitian fermionic bilinear.
This coupling preserves all symmetries only when $\phi_{a}(\vb{R})$ belongs to the same irrep and has the same TR-sign as $\Phi_{a}(\vb{R})$.
Given such a $\phi_{a}(\vb{R})$, symmetries only break upon the condensation of the order-parameter field $\Phi_{a \vb{q}} = \Phi_{a, -\vb{q}}^{*}$.
Here, $g$ is the coupling constant, $a, b, \ldots$ are irrep component indices (they go from $1$ to the dimension of the irrep $= \dim \Phi$), $\vb{R}$ goes over the real-space lattice, and $\vb{k}, \vb{p}, \vb{q}, \ldots$ are wavevectors.
The fermionic bilinears $\phi_{a}(\vb{R})$ we shall specify a bit later.

\begin{figure}[t]
\centering
\includegraphics[width=0.95\textwidth]{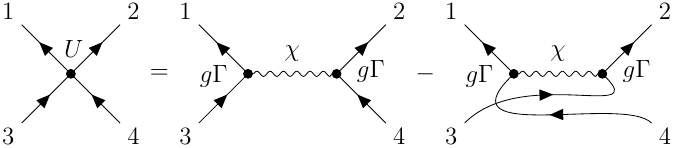}
\captionbelow[The diagram of the four-fermion interaction that is mediated by a bosonic collective mode.]{\textbf{The diagram of the four-fermion interaction that is mediated by a bosonic collective mode.}
Solid dots represent the Yukawa coupling between fermions and collective modes which is given in Eq.~\eqref{eq:Yukawa-coupling}.
Solid lines are fermion propagators, while wavy lines are boson propagators.}
\label{fig:boson-exchange}
\end{figure}

In the presence of Yukawa coupling, the collective modes mediate an interaction between the fermions, as shown in Fig.~\ref{fig:boson-exchange}.
Within the Euclidean path-integral formalism~\cite{Altland2010}, the interacting part of the action is
\begin{align}
\action_{\text{int}}[\Phi, \psi] &= \frac{1}{2} \int_{x_1 x_2} \sum_a \Phi_a(x_1) \chi^{-1}(x_1 - x_2) \Phi_a(x_2) + g \int_{x} \sum_a \Phi_{a}(x) \phi_{a}(x),
\end{align}
where $x \defeq (\vb{R}, \tau)$, $\int_x = \int_0^{\upbeta} \dd{\tau} \sum_{\vb{R}}$, and $\chi(x) = \chi(-x) = \chi^{*}(x)$ is the bosonic propagator or susceptibility.
Integrating out the collective modes yields the four-fermion interaction:
\begin{align}
\begin{aligned}
\action_{\text{int}}[\psi] &= - \frac{1}{2} g^2 \int_{x_1 x_2} \sum_a \phi_a(x_1) \chi(x_1 - x_2) \phi_a(x_2) \\
&= - \frac{1}{2} g^2 \sum_{\vb{q} \omega_{\ell} a} \phi_{a, -\vb{q}}(- \iu \omega_{\ell}) \chi(\vb{q}, \iu \omega_{\ell}) \phi_{a \vb{q}}(\iu \omega_{\ell}),
\end{aligned}
\end{align}
where $\omega_{\ell} = 2 \pi \ell / \upbeta$ are bosonic Matsubara frequencies and $\chi(\vb{q}, \iu \omega_{\ell}) = \chi(-\vb{q}, -\iu \omega_{\ell}) = \chi^*(\vb{q}, \iu \omega_{\ell})$.
Notice that
\begin{align}
\begin{aligned}
\ev{\Phi_{a \vb{q}}(\iu \omega_{\ell}) \Phi_{b, -\vb{q}'}(-\iu \omega_{\ell'})} &= \Kd_{ab} \Kd_{\vb{q} - \vb{q}'} \Kd_{\ell \ell'} \chi(\vb{q}, \iu \omega_{\ell}) \\
\implies \chi(\vb{q}, \iu \omega_{\ell}) &= \ev{\abs{\Phi_{a \vb{q}}(\iu \omega_{\ell})}^2} \geq 0
\end{aligned}
\end{align}
implies that the susceptibility is strictly non-negative.
The divergence of $\chi(\vb{Q}, 0)$, or equivalently the softening of $\chi^{-1}(\vb{Q}, 0)$, indicates condensation at $\vb{q} = \vb{Q}$.

Although this interaction is in general retarded, we may treat it as instantaneous if we renormalize the theory down to an energy cutoff $\hbar \omega_c$ which is sufficiently small to freeze the boson dynamics.
This is appropriate as long as we are far enough from the QCP~\cite{Moon2010}.
At the QCP, the non-trivial frequency dependence of the Cooper-channel interaction extends down to $\omega = 0$, precluding a BCS treatment of the low-frequency sector~\cite{Moon2010}.
Nonetheless, the behavior of the pairing as one approaches the QCP is still indicative of the pairing tendency at the QCP~\cite{Moon2010, Aji2010, Lederer2015}.
The effective interaction away from the QCP thus acquires the form:
\begin{align}
\Haml_{\text{int}} &= - \frac{1}{2} g^2 \sum_{a \vb{q}} \chi(\vb{q},0) \phi_{a, -\vb{q}} \phi_{a \vb{q}}. \label{eq:effective-multi-exch-int}
\end{align}
On the mean-field level, the static ($\omega_{\ell} = 0$) bosonic propagator (susceptibility) is given by the mode frequencies:
\begin{align}
\chi(\vb{q},0) &= \frac{1}{\hbar \Omega_{\vb{q}}}. \label{eq:effective-multi-exch-int-mode-freq}
\end{align}
Here we directly see what we alluded to many times in Sec.~\ref{sec:QCP-paradigm}: that the softening of collective modes, which is the defining feature of QCPs, implies that the electron-electron interactions which they mediate become stronger.

The appeal of constructing four-fermion interactions via collective order-parameter fields is that the motivation for their form is physically transparent.
The interaction~\eqref{eq:effective-multi-exch-int} could be expressed in terms of local and non-local Hubbard interactions as well.
Except for common practice and custom, such an approach is not any better justified than ours in the effective low-energy regime discussed here.
That said, in Sec.~\ref{sec:CuO2-Hubbard-decompose} of the next chapter we shall explore what are the ordering channels which are intrinsic to extended Hubbard interactions of the three-orbital model of the cuprates (Sec.~\ref{sec:cuprate-3band-model}).

\subsubsection{Specification of the order-parameter field and its coupling} \label{sec:LC-gen-ph-bilinears}
We still need to specify the fermionic bilinears $\phi_{a}(\vb{R})$.
As it turns out, there is an ambiguity in how one should properly define the bilinears which derives from the fact that $\tfrac{1}{2} (\vb{R}_1 + \vb{R}_2)$ is not necessarily on the same lattice as $\vb{R}_1$ and $\vb{R}_2$.
For instance, for lattice neighbors $\tfrac{1}{2} (\vb{R}_1 + \vb{R}_2)$ is always in between lattice points.
The most general possible definition of a fermionic bilinear in the particle-hole sector is
\begin{align}
\phi_{a}(\vb{R}) &= \sum_{\vb{\delta}_1 \vb{\delta}_2} \psi^{\dag}(\vb{R}+\vb{\delta}_1) \Gamma_{a}(\vb{\delta}_1, \vb{\delta}_2) \psi(\vb{R}+\vb{\delta}_2), \label{eq:general-coordinate-bilinear}
\end{align}
where $\vb{\delta}_1, \vb{\delta}_2, \ldots$ go over lattice neighbors and $\Gamma_{a}(\vb{\delta}_1, \vb{\delta}_2) = \Gamma_{a}^{\dag}(\vb{\delta}_2, \vb{\delta}_1)$ are $2M \times 2M$ matrices in spin and orbital space.
These $\Gamma$ matrices encode the information on the symmetry of the ordering channel.
After a Fourier transform with the conventions
\begin{align}
\psi(\vb{R}) &= \frac{1}{\sqrt{\mathcal{N}}} \sum_{\vb{k}} \Elr^{\iu \vb{k} \vdot \vb{R}} \psi_{\vb{k}}, &
\phi_{a}(\vb{R}) &= \frac{1}{\sqrt{\mathcal{N}}} \sum_{\vb{k}} \Elr^{\iu \vb{k} \vdot \vb{R}} \phi_{a \vb{k}}, \\
\Phi_{a}(\vb{R}) &= \frac{1}{\sqrt{\mathcal{N}}} \sum_{\vb{k}} \Elr^{\iu \vb{k} \vdot \vb{R}} \Phi_{a \vb{k}}, &
\Gamma_{a}(\vb{\delta}_1, \vb{\delta}_2) &= \frac{1}{\mathcal{N}^2} \sum_{\vb{k} \vb{p}} \Elr^{\iu (\vb{k} \vdot \vb{\delta}_1 - \vb{p} \vdot \vb{\delta}_2)} \Gamma_{a \vb{k}, \vb{p}},
\end{align}
where $\mathcal{N}$ is the number of unit cells, one finds that
\begin{align}
\phi_{a \vb{q}} &= \frac{1}{\sqrt{\mathcal{N}}} \sum_{\vb{k}} \psi_{\vb{k}}^{\dag} \Gamma_{a \vb{k}, \vb{k}+\vb{q}} \psi_{\vb{k}+\vb{q}}. \label{eq:general-momentum-bilinear}
\end{align}
This most general definition is the one that we shall use.
Notice that the displaced bilinear $\phi_{a}'(\vb{R}) = \phi_{a}(\vb{R}+\vb{\delta}')$ whose $\Gamma_{a}'(\vb{\delta}_1, \vb{\delta}_2) = \Gamma_{a}(\vb{\delta}_1+\vb{\delta}', \vb{\delta}_2+\vb{\delta}')$ is equally valid.
This arbitrariness we shall eliminate by localizing $\phi_{a}(\vb{R})$ around $\vb{R}$, i.e., making $\Gamma_{a}(\vb{\delta}_1, \vb{\delta}_2)$ finite for small $\vb{\delta}_{1,2}$ only.

It is instructive to see what goes wrong with alternative definitions.
One alternative definition is
\begin{align}
\phi_{a}^{(\text{alt}1)}(\vb{R}) &= \sum_{\vb{\delta}} \psi^{\dag}(\vb{R} + \vb{\delta}) \Gamma_{a}^{(\text{alt}1)}(\vb{\delta}) \psi(\vb{R} - \vb{\delta}),
\end{align}
which is more symmetric.
However, it is not general enough because it cannot include coupling between closest neighbors.
Another alternative definition is
\begin{align}
\phi_{a}^{(\text{alt}2)}(\vb{R}) &= \sum_{\vb{\delta}} \psi^{\dag}(\vb{R} + \vb{\delta}) \Gamma_{a}^{(\text{alt}2)}(\vb{\delta}) \psi(\vb{R}) + \Hc, \label{eq:order-param-alt2} \\
\phi_{a \vb{q}}^{(\text{alt}2)} &= \frac{1}{\sqrt{\mathcal{N}}} \sum_{\vb{k}} \psi^{\dag}_{\vb{k}} \mleft(\Gamma_{a \vb{k}}^{(\text{alt}2)} + \mleft[\Gamma_{a, \vb{k}+\vb{q}}^{(\text{alt}2)}\mright]^{\dag}\mright) \psi_{\vb{k} + \vb{q}}, \label{eq:general-momentum-bilinear-alt2}
\end{align}
where
\begin{align}
\Gamma_{a}^{(\text{alt}2)}(\vb{\delta}) &= \frac{1}{\mathcal{N}} \sum_{\vb{k}} \Elr^{\iu \vb{k} \vdot \vb{\delta}} \Gamma_{a \vb{k}}^{(\text{alt}2)}.
\end{align}
This is manifestly Hermitian and can include coupling between neighbors of all distances $\vb{\delta}$.
It is used in a number of references in the literature (e.g.\ Ref.~\cite{Kozii2015}) and is related to the previous definition through
\begin{align}
\Gamma_{a}(\vb{\delta}_1, \vb{\delta}_2) &= \Kd_{\vb{\delta}_1,\vb{0}} \Gamma_{a}^{(\text{alt}2)}(\vb{\delta}_2) + \mleft[\Gamma_{a}^{(\text{alt}2)}(-\vb{\delta}_1)\mright]^{\dag} \Kd_{\vb{\delta}_2,\vb{0}}, \\
\Gamma_{a \vb{k}, \vb{p}} &= \Gamma_{a \vb{k}}^{(\text{alt}2)} + \mleft[\Gamma_{a \vb{p}}^{(\text{alt}2)}\mright]^{\dag}.
\end{align}
In the simplest case when all the orbital are centered at the lattice points, i.e.\ have trivial Wychoff positions, this definition is completely appropriate.
However, when this is not the case, symmetry operations map $\psi(\vb{R})$ not only to $\psi(\vb{R}')$, but also to its neighbors $\psi(\vb{R}'+\vb{\delta})$.
Thus with the above definition one cannot ensure that $\phi_{a}^{(\text{alt}2)}(\vb{R})$ has well-defined symmetry transformation rules in a sense that we shall explain in the next section (see Eqs.~\eqref{eq:phi-transf-rule1-bilinear} and~\eqref{eq:phi-transf-rule2-bilinear} in particular).
In our study of a tight-binding model of cuprates (Sec.~\ref{sec:cuprate-bilinear-class}), we will face precisely such a circumstance where some orbitals have non-trivial Wychoff positions.
One can circumvent this obstacle by using in Eq.~\eqref{eq:order-param-alt2}, instead of $\psi(\vb{R})$, an extended spinor $\Psi(\vb{R})$ which transforms into itself under all symmetry operations.
This is how we will treat the cuprate model in Sec.~\ref{sec:cuprate-bilinear-class}.

Loop-current order is purely orbital order.
Hence the corresponding $\Gamma$ matrices are purely orbital and have the form
\begin{align}
\Gamma_{a \vb{k},\vb{p}} = \gamma_{a \vb{k}, \vb{p}} \otimes \Pauli_0,
\end{align}
where $\Pauli_0$ is the $2 \times 2$ identity matrix in spin space.
LCs are also odd under TR.
For the Yukawa coupling of Eq.~\eqref{eq:Yukawa-coupling} to respect TR and point group symmetries, the orbital $\gamma$ matrices must therefore also be odd under TR,
\begin{align}
\gamma_{a \vb{k}, \vb{p}}^{*} = - \gamma_{a, -\vb{k}, -\vb{p}},
\end{align}
as well as belong to the same irrep as the LC order parameter.

\subsubsection{Symmetry transformation rules} \label{sec:LC-gen-model-sym}
Here we state how the fields transform under TR, whose many-body TR operator is $\SymTR$, and under point group operations of the crystalline system, whose many-body operators are $\SymU(g)$ with $g$ denoting the point group elements.

The most general fermion transformation rules are:
\begin{align}
\SymU^{\dag}(g) \psi_{\vb{k}} \SymU(g) &= \MatU_{\vb{k}}(g) \psi_{R(g^{-1}) \vb{k}}, \label{eq:fermi-tranf-point-group} \\
\SymTR^{-1} \psi_{\vb{k}} \SymTR &= \MatTR_{\vb{k}}^{*} \psi_{-\vb{k}}, \label{eq:fermi-tranf-TR}
\end{align}
where $\MatU_{\vb{k}}(g)$ and $\MatTR_{\vb{k}}$ are unitary matrices which act on the spinors and $R(g)$ are the usual $3 \times 3$ orthogonal matrices, defined in detail in Sec.~\ref{sec:SU2-SO3-conventions} of the appendix, which act on vectors.
They satisfy:
\begin{align}
\MatU_{\vb{k}}^{-1}(g) &= \MatU_{\vb{k}}^{\dag}(g), &
\MatTR_{\vb{k}}^{-1}(g) &= \MatTR_{\vb{k}}^{\dag}(g), &
R^{-1}(g) &= R^{\intercal}(g).
\end{align}
Notice that the former two matrices depend on momentum.
These momentum-dependencies are necessary for $\MatU_{\vb{k}}(g)$ and $\MatTR_{\vb{k}}$ whenever the fermionic basis functions (i.e., orbitals) have non-trivial Wychoff positions and spin-orbit mixing, respectively.

The composition law is slightly modified for $\MatU_{\vb{k}}(g)$:
\begin{align}
\MatU_{\vb{k}}(g) \MatU_{R(g^{-1}) \vb{k}}(g') &= \MatU_{\vb{k}}(g g').
\end{align}
As a consequence, $\MatU_{\vb{k}}(g^{-1}) = \MatU_{R(g) \vb{k}}^{\dag}(g)$.
Regarding $R(g)$, it satisfies the usual $R(g) R(g') = R(g g')$ and $R(g^{-1}) = R^{\intercal}(g)$ relations.
Let us observe that the reason why $g$ acts on the spinor and $g^{-1}$ on the momentum in Eq.~\eqref{eq:fermi-tranf-point-group} is to ensure that
\begin{align}
\SymU^{\dag}(g') \mleft[\SymU^{\dag}(g) \psi_{\vb{k}} \SymU(g)\mright] \SymU(g') &= \SymU^{\dag}(g g') \psi_{\vb{k}} \SymU(g g')
\end{align}
is respected.
For fermionic systems $\SymTR^2 = - \hat{\one}$ holds which implies that
\begin{align}
\MatTR_{\vb{k}} \MatTR_{-\vb{k}}^{*} = - \one.
\end{align}
Moreover, TR commutes with all point group operations:
\begin{align}
\MatU_{\vb{k}}(g) \MatTR_{R(g^{-1}) \vb{k}} &= \MatTR_{\vb{k}} \MatU_{-\vb{k}}^{*}(g).
\end{align}

Next, let us consider the eigenvectors of the band Hamiltonian:
\begin{align}
H_0 u_{\vb{k} n s} &= \varepsilon_{\vb{k} n} u_{\vb{k} n s}, &
u_{\vb{k} n s}^{\dag} u_{\vb{k} m s'} &= \Kd_{nm} \Kd_{ss'}.
\end{align}
Here $s \in \{\uparrow, \downarrow\}$ is the pseudospin or Kramers' degeneracy index.
In the absence of SOC, $s$ reduces to the physical spin index.
We may always choose a gauge in which:
\begin{align}
\MatU_{\vb{p}}(P) \MatTR_{-\vb{p}} u_{\vb{p} m s}^{*} &= \sum_{s'} u_{\vb{p} m s'} (\iu \Pauli_y)_{s's}. \label{eq:band-eigenvector-sym-TR-rule}
\end{align}
In simpler notation, this is just a matter of defining the up-arrow pseudospin as $\ket{\uparrow} = P \TRop \ket{\downarrow}$ at each $\vb{k}$.
In this gauge:
\begin{align}
\begin{aligned}
\MatU_{\vb{p}}^{\dag}(g) u_{\vb{p} m s} &= \sum_{s'} [S_{\vb{p} m}(g)]_{ss'}^{*} u_{R(g^{-1}) \vb{p} m s'}, \\
\MatU_{R(g) \vb{p}}(g) u_{\vb{p} m s} &= \sum_{s'} u_{R(g) \vb{p} m s'} [S_{\vb{p} m}(g)]_{s's},
\end{aligned} \label{eq:band-eigenvector-sym-transf-rule}
\end{align}
where
\begin{align}
S_{\vb{k} n}(g) S_{R(g^{-1}) \vb{k} n}(g') &= S_{\vb{k} n}(g g'), \\
(\iu \Pauli_y) S_{\vb{k} n}(g) &= S_{\vb{k} n}^{*}(g) (\iu \Pauli_y).
\end{align}
The latter property of $S_{\vb{k} n}(g)$ follows from the fact that parity and TR both commute with all point group operations $g$.
It is also useful to consider the band eigen-projector
\begin{align}
\mathcal{P}_{\vb{k} n}^{A} \defeq \sum_{s s'} u_{\vb{k} n s} (\Pauli_A)_{ss'} u_{\vb{k} n s'}^{\dag}
\end{align}
weighted by the $\Pauli_A$ Pauli matrices, where $A \in \{0, 1, 2, 3\}$.
It transforms according to:
\begin{align}
\MatU_{\vb{k}}^{\dag}(g) \mathcal{P}_{\vb{k} n}^{0} \MatU_{\vb{k}}(g) &= \mathcal{P}_{R(g^{-1}) \vb{k} n}^{0}, \\
\MatU_{\vb{k}}^{\dag}(g) \mathcal{P}_{\vb{k} n}^{A} \MatU_{\vb{k}}(g) &= \sum_{B=1}^{3} \mleft[R_{\vb{k} n}(g)\mright]_{AB} \mathcal{P}_{R(g^{-1}) \vb{k} n}^{B},
\end{align}
where $R_{\vb{k} n}(g)$ are orthogonal $3 \times 3$ matrices which satisfy
\begin{align}
R_{\vb{k} n}(g) R_{R(g^{-1}) \vb{k} n}(g') &= R_{\vb{k} n}(g g')
\end{align}
and are related to $S_{\vb{k} n}(g)$ via
\begin{align}
S_{\vb{k} n}^{\dag}(g) \Pauli_A S_{\vb{k} n}(g) &= \sum_{B=1}^{3} \mleft[R_{\vb{k} n}(g)\mright]_{AB} \Pauli_B.
\end{align}
In the absence of SOC, the eigenvectors and eigen-projectors factorize into spin and orbital parts:
\begin{align}
u_{\vb{k} n s} &= u_{\vb{k} n} \otimes \ket{s}, &
\mathcal{P}_{\vb{k} n}^{A} &= u_{\vb{k} n} u_{\vb{k} n}^{\dag} \otimes \Pauli_A.
\end{align}
Moreover, $S_{\vb{k} n}(g)$ equals the product of the band-orbital and spin representations of the point group,
\begin{align}
S_{\vb{k} n}(g) &= \Elr^{\iu \varkappa_{\vb{k} n}(g)} S(g), \label{eq:band-eigenvector-sym-no-SOC}
\end{align}
and $R_{\vb{k} n}(g)$ reduces to the vector representation of the point group,
\begin{align}
R_{\vb{k} n}(g) &= R(g).
\end{align}
The spin $S(g)$ and vector $R(g)$ representations are the usual ones and they are defined in Appx.~\ref{app:group_theory}, Sec.~\ref{sec:SU2-SO3-conventions}.

The transformation rules of the bosonic order-parameter field are:
\begin{align}
\SymU^{\dag}(g) \Phi_{a \vb{q}} \SymU(g) &= \sum_{b=1}^{\dim \Phi} \RepM_{ab}^{\Phi}(g) \Phi_{b, R(g^{-1}) \vb{q}}, \\
\SymTR^{-1} \Phi_{a \vb{q}} \SymTR &= p_{\TRop} \Phi_{a, -\vb{q}}, \label{eq:ord-param-field-TR-sym}
\end{align}
where $\RepM^{\Phi}$ is a real orthogonal irreducible representation of the point group, $\dim \Phi$ is the number of components of $\Phi$, and $p_{\TRop} = \pm 1$ is the sign under TR.
In real space, these transformation rules take the form
\begin{align}
\SymU^{\dag}(g) \Phi_{a}(\vb{R}) \SymU(g) &= \sum_{b=1}^{\dim \Phi} \RepM_{ab}^{\Phi}(g) \Phi_{b}\mleft(R(g^{-1}) \vb{R}\mright), \\
\SymTR^{-1} \Phi_{a}(\vb{R}) \SymTR &= p_{\TRop} \Phi_{a}(\vb{R}).
\end{align}
The matrices $\RepM_{ab}^{\Phi}(g)$ must be real to be consistent with the reality of $\Phi_{a \vb{q}} = \Phi_{a, -\vb{q}}^{*}$.
Recall that every irreducible representation of a finite group can be made unitary (Ref.~\cite{Dresselhaus2007}, see also Sec.~\ref{sec:rep-theory-basics}), which for real representations means orthogonal.

The fermionic bilinears of the preceding section [Eq.~\eqref{eq:general-momentum-bilinear}] must transform in the same way as $\Phi$ to ensure that the Yukawa coupling [Eq.~\eqref{eq:Yukawa-coupling}] between the two preserves all symmetries:
\begin{align}
\SymU^{\dag}(g) \phi_{a \vb{q}} \SymU(g) &= \sum_{b=1}^{\dim \Phi} \RepM_{ab}^{\Phi}(g) \phi_{b, R(g^{-1}) \vb{q}}, \label{eq:phi-transf-rule1-bilinear} \\
\SymTR^{-1} \phi_{a \vb{q}} \SymTR &= p_{\TRop} \phi_{a, -\vb{q}}.\label{eq:phi-transf-rule2-bilinear}
\end{align}
In turn, these transformation laws constrain the $\Gamma_{a \vb{k}, \vb{p}}$ matrices:
\begin{align}
\MatU_{\vb{k}}^{\dag}(g) \Gamma_{a \vb{k}, \vb{p}} \MatU_{\vb{p}}(g) &= \sum_{b=1}^{\dim \Phi} \RepM_{ab}^{\Phi}(g) \Gamma_{b, R(g^{-1}) \vb{k}, R(g^{-1}) \vb{p}}, \label{eq:Gamma-transf-rule1-bilinear} \\
\MatTR_{\vb{k}}^{\dag} \Gamma_{a \vb{k}, \vb{p}} \MatTR_{\vb{p}} &= p_{\TRop} \Gamma_{a, -\vb{k}, -\vb{p}}^{*}. \label{eq:Gamma-transf-rule2-bilinear}
\end{align}
Reality entails $\phi_{a \vb{q}}^{\dag} = \phi_{a, -\vb{q}}$ and $\Gamma_{a \vb{k}, \vb{p}}^{\dag} = \Gamma_{a \vb{p}, \vb{k}}$.

\subsection{The linearized BCS gap equation} \label{sec:QCP-model-lin-gap-eq}
In this section, we introduce the linearized BCS gap equation that we shall use to study the pairing problem and we analyze the Cooper-channel interaction which enters it.

In Appx.~\ref{app:lin_gap_eq}, we have derived the linearized gap equation for a general Fermi liquid with SOC that has parity and TR symmetry and whose Fermi surfaces do not touch each other or have Van Hove singularities on them.
For the following general non-retarded four-fermion interaction
\begin{align}
\Haml_{\text{int}} = \frac{1}{4 L^d} \sum_{1234} \Kd_{\vb{k}_1 + \vb{k}_2 - \vb{k}_3 - \vb{k}_4} U_{1234} \psi_1^{\dag} \psi_2^{\dag} \psi_4 \psi_3, \label{eq:LC-lin-gap-eq-entry-Hint}
\end{align}
where $L^d$ is the volume in $d$ spatial dimensions and $1 = (\vb{k}_1, \alpha_1)$, $2 = (\vb{k}_2, \alpha_2)$, etc., are shorthands for all the quantum numbers carried by the fermions (momentum, orbital content, spin), the linearized gap equation is [Eq.~\eqref{eq:final-lin-gap-eq}]:
\begin{align}
\sum_n \int\limits_{\varepsilon_{\vb{k} n} = 0} \frac{\dd{S_{\vb{k}}}}{(2 \pi)^d} \sum_{A=0}^{3} \PintW_{BA}(\vb{p}_m, \vb{k}_n) \, d_{A}(\vb{k}_n) = \lambda \, d_{B}(\vb{p}_m), \label{eq:final-lin-gap-eq-LC-again}
\end{align}
where the momentum integrals go over the Fermi surface(s), $\varepsilon_{\vb{k} n}$ are the dispersions of the $u_{\vb{k} n s}$ band eigenvectors,
\begin{align}
\mathcal{P}_{\vb{p} m}^{B} = \sum_{s s'} u_{\vb{p} m s} (\Pauli_B)_{ss'} u_{\vb{p} m s'}^{\dag}.
\end{align}
are the projectors, and the pairing interaction is given by:
\begin{align}
\PintW_{BA}(\vb{p}_m, \vb{k}_n) = - \sum_{\alpha_1 \alpha_2 \alpha_3 \alpha_4} \frac{\mleft[\MatTR_{-\vb{p}}^{*} \mathcal{P}_{\vb{p} m}^{B}\mright]_{\alpha_2 \alpha_1} \mleft[\mathcal{P}_{\vb{k} n}^{A} \MatTR_{-\vb{k}}^{\intercal}\mright]_{\alpha_3 \alpha_4}}{4 \abs{\grad_{\vb{p}} \varepsilon_{\vb{p} m}}^{1/2} \abs{\grad_{\vb{k}} \varepsilon_{\vb{k} n}}^{1/2}} U_{\alpha_1 \alpha_2 \alpha_3 \alpha_4}(\vb{p}, -\vb{p}, \vb{k}, -\vb{k}). \label{eq:final-lin-gap-eq-LC-PintW-BA-again}
\end{align}
Positive pairing eigenvalues $\lambda$ correspond to pairing instabilities with transition temperatures
\begin{align}
k_B T_c &= \frac{2 \Elr^{\upgamma_E}}{\pi} \hbar \omega_c \, \Elr^{- 1 / \lambda} \approx 1.134 \, \hbar \omega_c \, \Elr^{- 1 / \lambda},
\end{align}
where $\hbar \omega_c$ is the energy cutoff which defines the theory.
See Appx.~\ref{app:lin_gap_eq} for further details.
$\PintW_{BA}(\vb{p}_m, \vb{k}_n)$ is the Cooper-channel interaction which we shall now analyze.

When the interactions are mediated by a bosonic mode as in Eq.~\eqref{eq:effective-multi-exch-int}, we can use Eq.~\eqref{eq:general-momentum-bilinear} to read off the interaction amplitudes:
\begin{gather}
U_{1234} = \tilde{U}_{1234} - \tilde{U}_{1243}, \\
\tilde{U}_{1234} = - g^2 \frac{L^d}{\mathcal{N}} \sum_a \chi(\vb{k}_1-\vb{k}_3,0) \mleft[\Gamma_{a \vb{k}_1, \vb{k}_3}\mright]_{\alpha_1 \alpha_3} \mleft[\Gamma_{a \vb{k}_2, \vb{k}_4}\mright]_{\alpha_2 \alpha_4}.
\end{gather}
The corresponding diagram is shown in Fig.~\ref{fig:boson-exchange}.
Assuming that the order-parameter field has a well-defined eigenvalue under TR $p_{\TRop} = \pm 1$, as defined in Eq.~\eqref{eq:ord-param-field-TR-sym} of Sec.~\ref{sec:LC-gen-model-sym}, we find that
\begin{align}
\PintW_{BA}(\vb{p}_m, \vb{k}_n) &= p_{\TRop} \, g^2 \frac{\Pintw_{BA}(\vb{p}_m, \vb{k}_n) + \Pintw_{BA}(\vb{p}_m, -\vb{k}_n) p_A}{4 \abs{\grad_{\vb{p}} \varepsilon_{\vb{p} m}}^{1/2} \abs{\grad_{\vb{k}} \varepsilon_{\vb{k} n}}^{1/2}}, \label{eq:lin-gap-eq-boson-exchange-W-mat} \\
\Pintw_{BA}(\vb{p}_m, \vb{k}_n) &= (L^d / \mathcal{N}) \chi(\vb{p} - \vb{k},0) \sum_a \Tr \mathcal{P}_{\vb{p} m}^{B} \Gamma_{a \vb{p}, \vb{k}} \mathcal{P}_{\vb{k} n}^{A} \Gamma_{a \vb{p}, \vb{k}}^{\dag}, \label{eq:lin-gap-eq-boson-exchange-M-mat}
\end{align}
where the trace $\Tr$ goes over spin and orbital indices and
\begin{align}
p_{A=0} &= +1 \qquad\text{for even-parity singlet pairing, whereas} \\
p_{A=1,2,3} &= -1 \qquad\text{for odd-parity triplet pairing.}
\end{align}
The general physical implications of this important expression for $\PintW_{BA}(\vb{p}_m, \vb{k}_n)$ we shall discuss in the next section.
Below, we derive some formal properties first.

\subsubsection{Properties of the pairing form factor} \label{sec:LC-prop-Cp-ch-int}
Given how it enters $\Pintw_{BA}(\vb{p}_m, \vb{k}_n)$, it is of prime interest to explore the properties of the pairing form factor:
\begin{align}
\begin{aligned}
\PintF_{BA}(\vb{p}_m, \vb{k}_n) &\defeq \sum_a \Tr \mathcal{P}_{\vb{p} m}^{B} \Gamma_{a \vb{p}, \vb{k}} \mathcal{P}_{\vb{k} n}^{A} \Gamma_{a \vb{p}, \vb{k}}^{\dag} \\
&= \sum_a \tr_s \Pauli_B \Pintf_{a}(\vb{p}_m, \vb{k}_n) \Pauli_A \Pintf_{a}^{\dag}(\vb{p}_m, \vb{k}_n),
\end{aligned} \label{eq:F_BA-expr}
\end{align}
where
\begin{align}
[\Pintf_{a}(\vb{p}_m, \vb{k}_n)]_{s_1 s_2} &\defeq u_{\vb{p} m s_1}^{\dag} \Gamma_{a \vb{p}, \vb{k}} u_{\vb{k} n s_2}  \label{eq:fa-definition-form-f}
\end{align}
is a matrix in pseudospin space and $\tr_s$ is a trace over pseudospins.
This pairing form factor contains information on the symmetry of the order parameter and band structure.
Physically, it measures the interference amplitude between Cooper pairs going from momenta $(\vb{k}_n, -\vb{k}_n)$ to momenta $(\vb{p}_m, -\vb{p}_m)$, and from pseudospin singlet/triplet $A$ to pseudospin singlet/triplet $B$.
Both $\PintF_{BA}(\vb{p}_m, \vb{k}_n)$ and $\Pintf_{a}(\vb{p}_m, \vb{k}_n)$ we shall call pairing form factors.

Because of $\Gamma_{a \vb{k}, \vb{p}}^{\dag} = \Gamma_{a \vb{p}, \vb{k}}$, the following reality relations hold:
\begin{align}
\Pintf_{a}^{\dag}(\vb{p}_m, \vb{k}_n) &= \Pintf_{a}(\vb{k}_n, \vb{p}_m), \label{eq:f-real-cond} \\
\PintF_{BA}^{*}(\vb{p}_m, \vb{k}_n) &= \PintF_{BA}(\vb{p}_m, \vb{k}_n) = \PintF_{AB}(\vb{k}_n, \vb{p}_m). \label{eq:F-real-cond}
\end{align}

One may now show that:
\begin{align}
p_{\TRop} p_{P} \Pintf_{a}(\vb{p}_m, \vb{k}_n) &= (\iu \Pauli_y)^{\dag} \Pintf_{a}^{*}(\vb{p}_m, \vb{k}_n) (\iu \Pauli_y), \label{eq:f-sym-cond1} \\
\sum_{b=1}^{\dim \Phi} \RepM_{ab}^{\Phi}(g) \Pintf_{b}(R(g^{-1}) \vb{p}_m, R(g^{-1}) \vb{k}_n) &= S_{\vb{p} m}^{\dag}(g) \Pintf_{a}(\vb{p}_m, \vb{k}_n) S_{\vb{k} n}(g), \label{eq:f-sym-cond2}
\end{align}
where $\RepM_{ab}^{\Phi}(P) = p_P \Kd_{ab}$, i.e., $p_P = \pm 1$ is the parity of the order-parameter field $\Phi$.

One consequence of Eq.~\eqref{eq:f-sym-cond1} is that $\PintF_{BA}(\vb{p}_m, \vb{k}_n) = p_B p_A \PintF_{BA}(\vb{p}_m, \vb{k}_n)$, where $p_{B=0} = +1$ for the singlet and $p_{B=1,2,3} = -1$ for the triplet channel.
Hence even-parity pseudospin-singlet and odd-parity pseudospin-triplet channels do not mix, as expected.
As we noted in Sec.~\ref{app:lin_gap_eq-fin-form} of Appx.~\ref{app:lin_gap_eq}, one can quite generally show that $\PintW_{BA}(\vb{p}_m, \vb{k}_n) = p_B p_A \PintW_{BA}(\vb{p}_m, \vb{k}_n)$ for $\PintW_{BA}(\vb{p}_m, \vb{k}_n)$ given by Eq.~\eqref{eq:final-lin-gap-eq-LC-PintW-BA-again}.

\subsubsection{General symmetry constraints} \label{sec:Cp-channel-gen-sym-constr}
Equations~\eqref{eq:f-real-cond} and~\eqref{eq:f-sym-cond2}, when combined, give:
\begin{align}
\sum_{b=1}^{\dim \Phi} \RepM_{ab}^{\Phi}(g) \mleft[\Pintf_{b}(R(g^{-1}) \vb{k}_n, \vb{k}_n) S_{\vb{k} n}(g^{-1})\mright]^{\dag} &= \Pintf_{a}(R(g) \vb{k}_n, \vb{k}_n) S_{\vb{k} n}(g).
\end{align}
This motivates the introduction of
\begin{align}
\tilde{\Pintf}_a(g, \vb{k}_n) \defeq \Pintf_{a}(R(g) \vb{k}_n, \vb{k}_n) S_{\vb{k} n}(g),
\end{align}
in terms of which
\begin{align}
\sum_{b=1}^{\dim \Phi} \RepM_{ab}^{\Phi}(g) \tilde{\Pintf}_b^{\dag}(g^{-1}, \vb{k}_n) &= \tilde{\Pintf}_a(g, \vb{k}_n). \label{eq:f-symmetry-constraint-general}
\end{align}

Eq.~\eqref{eq:f-sym-cond1} implies that
\begin{align}
p_{\TRop} p_{P} \tilde{\Pintf}_a(g, \vb{k}_n) &= (\iu \Pauli_y)^{\dag} \tilde{\Pintf}_a^{*}(g, \vb{k}_n) (\iu \Pauli_y).
\end{align}
Let us decompose $\tilde{\Pintf}$ into Pauli matrices:
\begin{align}
\tilde{\Pintf}_a(g, \vb{k}_n) &= \sum_{A=0}^{3} \tilde{\Pintf}_a^A(g, \vb{k}_n) \Pauli_A.
\end{align}
Because of $(\iu \Pauli_y)^{\dag} \Pauli_A^{*} (\iu \Pauli_y) = p_A \Pauli_A$, the following now holds:
\begin{align}
p_{\TRop} p_{P} = +1 &\iff \tilde{\Pintf}_a^0(g, \vb{k}_n), \iu \tilde{\Pintf}_a^1(g, \vb{k}_n), \iu \tilde{\Pintf}_a^2(g, \vb{k}_n), \iu \tilde{\Pintf}_a^3(g, \vb{k}_n) \in \R, \\
p_{\TRop} p_{P} = -1 &\iff \iu \tilde{\Pintf}_a^0(g, \vb{k}_n), \tilde{\Pintf}_a^1(g, \vb{k}_n), \tilde{\Pintf}_a^2(g, \vb{k}_n), \tilde{\Pintf}_a^3(g, \vb{k}_n) \in \R.
\end{align}

Notice that $\tilde{\Pintf}_a(g, \vb{k}_n)$ is not necessarily Hermitian or anti-Hermitian.
However, if we consider an operation $g$ whose $R(g^{-1}) = R(g)$, then Eq.~\eqref{eq:f-symmetry-constraint-general} severely constrains $\tilde{\Pintf}_a(g, \vb{k}_n)$.
Because $g$ belongs to the double group of the point group, $R(g^{-1}) = R(g)$ does not imply that $g^{-1} = g$ so we cannot assume that $S_{\vb{k} n}(g^{-1}) = S_{\vb{k} n}(g)$.\footnote{Recall that $2 \pi$ rotations act on fermions like minus identity (Sec.~\ref{sec:SU2-SO3-conventions}).}
Instead, we have $S_{\vb{k} n}(g^{-1}) = (\pm)_g S_{\vb{k} n}(g)$, where
\begin{align}
(\pm)_g &= \begin{cases}
+1, & \text{when $g$ is the identity or parity,} \\
-1, & \text{when $g$ is a \SI{180}{\degree} rotation or reflection.} \\
\end{cases}
\end{align}
If we in addition assume that $\RepM_{ab}^{\Phi}(g)$ is diagonal, we obtain the following symmetry constraint:
\begin{align}
\tilde{\Pintf}_a^A(g, \vb{k}_n) &= p_A p_{\TRop} p_{P} (\pm)_g \RepM_{aa}^{\Phi}(g) \, \tilde{\Pintf}_a^A(g, \vb{k}_n).
\end{align}
This constraint forces the $\tilde{\Pintf}_a^A$ to either have $A=0$ or $A=1,2,3$ components, but never both.
This is as far as we can go for general systems with spin-orbit coupling.

In the absence of SOC, $S_{\vb{k} n}(g)$ equals the product of the band-orbital and spin representations of the point group [Eq.~\eqref{eq:band-eigenvector-sym-no-SOC}],
\begin{align}
S_{\vb{k} n}(g) &= \Elr^{\iu \varkappa_{\vb{k} n}(g)} S(g),
\end{align}
and $\Pintf_{a}(\vb{p}_m, \vb{k}_n)$ inherits the spin structure from the $\Gamma_{a \vb{p}, \vb{k}}$ matrices.
Thus, depending on whether the matrices are purely orbital or spin, $\tilde{\Pintf}_a(g, \vb{k}_n)$ may vanish completely.
For $g \in \{\one, P\}$, $S(g) = \Pauli_0$, whereas for \SI{180}{\degree} rotations around $\vu{n}$, $S(g) = - \iu \sum_{A=1}^{3} \hat{n}_A \Pauli_A$.
Reflections are compositions of \SI{180}{\degree} rotations with parity.
By analyzing the various possible cases, for systems without SOC we find that:
\begin{itemize}
\item For $P \TRop$-odd purely orbital $\Gamma$ matrices, the forward scattering form factor vanishes, \linebreak $\Pintf_{a}(\vb{k}_n, \vb{k}_n) = 0$.
\item For $P \TRop$-even purely spin $\Gamma$ matrices, the forward scattering form factor vanishes, \linebreak $\Pintf_{a}(\vb{k}_n, \vb{k}_n) = 0$.
\item For $\TRop$-odd purely orbital $\Gamma$ matrices, the backward scattering form factor vanishes, \linebreak $\Pintf_{a}(-\vb{k}_n, \vb{k}_n) = 0$.
\item For $\TRop$-even purely spin $\Gamma$ matrices, the backward scattering form factor vanishes, \linebreak $\Pintf_{a}(-\vb{k}_n, \vb{k}_n) = 0$.
\item For purely orbital $\Gamma$ matrices belonging to 1D irreps, $\Pintf_{a}(R(g)\vb{k}_n, \vb{k}_n) = 0$ for $g$ that are \SI{180}{\degree} rotations whenever $p_{\TRop} p_{P} \RepM^{\Phi}(g) = - 1$. In the 2D irrep case, both components never vanish at the same time so the corresponding $\PintF_{BA}(R(g)\vb{k}_n, \vb{k}_n)$ never vanishes.
\item For purely spin $\Gamma$ matrices proportional to $\Pauli_3$ and belonging to 1D irreps, $\Pintf_{a}(R(g)\vb{k}_n, \linebreak \vb{k}_n) = 0$ follows from either (i) $g$ that are \SI{180}{\degree} rotations around $\vu{n} \parallel \vu{e}_3$ with $p_{\TRop} p_{P} \RepM^{\Phi}(g) = + 1$, or (ii) $g$ that are \SI{180}{\degree} rotations around $\vu{n} \perp \vu{e}_3$ with $p_{\TRop} p_{P} \RepM^{\Phi}(g) = - 1$. The reflection symmetry constraints are analogous with $\vu{n}$ pointing along the reflection normal.
\end{itemize}
Note that by ``purely orbital $\Gamma$ matrices'' we mean $\Gamma_{a \vb{k},\vb{p}} = \gamma_{a \vb{k},\vb{p}} \otimes \Pauli_0$, whereas ``purely spin $\Gamma$ matrices'' entails $\Gamma_{a \vb{k},\vb{p}} = \gamma_{a \vb{k},\vb{p}}^1 \otimes \Pauli_1 + \gamma_{a \vb{k},\vb{p}}^2 \otimes \Pauli_2 + \gamma_{a \vb{k},\vb{p}}^3 \otimes \Pauli_3$.

\subsection{Generic pairing behavior near the quantum-critical point} \label{sec:QCP-genera-final-analysis}
Here we derive the results of Ref.~\cite{Palle2024-LC} for LC pairing in general systems.
These results we already reviewed at the start of this section (Sec.~\ref{sec:gen-sys-LC-analysis}) and they are summarized in Fig.~\ref{fig:QCP-general-results}.
We also cover some additional results and discussions concerning general IUC orders which follow from our analysis, but which did not end up in the final published article.

\begin{figure}[t]
\centering
\includegraphics[width=0.95\textwidth]{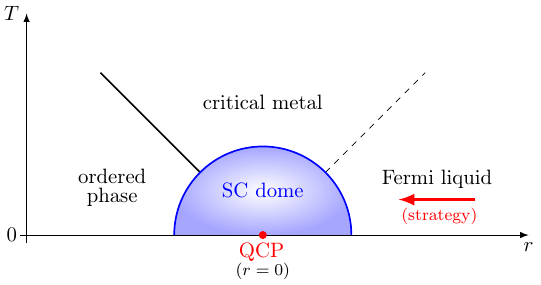}
\captionbelow[Strategy for analyzing the pairing promoted by quantum-critical fluctuations.]{\textbf{Strategy for analyzing the pairing promoted by quantum-critical fluctuations.}
The sketched phase diagram is a simplified version of Fig.~\ref{fig:QCP-pairing-paradigm}.
$T$ is the temperature and $r$ tunes the system through a continuous quantum phase transition at $r = T = 0$.
The hypothetical superconducting (SC) dome surrounding the quantum-critical point (QCP) we analyze by approaching it from the disordered (Fermi liquid) side, as indicated by the red arrow.}
\label{fig:QCP-calc-strategy}
\end{figure}

In Sec.~\ref{sec:QCP-pairing-model}, we have introduce a general model of a fluctuating order parameter coupled to itinerant fermions.
We want to study the pairing tendencies of this model as the order parameter approaches its QCP.
Solving this coupled many-body problem is a formidable challenge.
In order to make progress, we follow the strategy of Refs.~\cite{Aji2010, Lederer2015} and approach the QCP from the disordered side of the phase diagram, as depicted in Fig.~\ref{fig:QCP-calc-strategy}.
On the the disordered side, far enough from the QCP, the normal state is a well-understood Fermi liquid whose pairing can be investigated using an effective BCS description with a non-retarded interaction~\cite{Moon2010}.
Moreover, far enough from the QCP, the collective fluctuations are sufficiently weak to enable us to analyze the pairing instability to leading-order in perturbation theory using the linearized gap equation of Sec.~\ref{sec:QCP-model-lin-gap-eq}.
This approach is controlled by the coupling $g$ of the fermions to order-parameter fluctuations.
Although it is formally valid only for small $g$, any apparent breakdown of the weak-coupling theory is a compelling indicator for strong quantum-critical pairing, as evidenced by complementary analytical~\cite{She2009, Moon2010, Metlitski2015, YuxuanWang2016-inv, YuxuanWang2016-QCP, WangRaghu2017, ChubukovI_2020, ChubukovII_2020} and numerical~\cite{Berg2019, Lederer2017, Wang2017, Xu2017, Xu2020} methods which find strong quantum-critical pairing only when weak-coupling theory indicates it.

Notice that this strategy can be applied not only to LC order, but more broadly to any order in the particle-hole sector.
Let us also emphasize that this strategy is phenomenological in the sense that we assume from the outset that the system orders in a certain channel; we make no attempts at deriving the ordered phase from a microscopic model.
Instead, we focus on what happens afterwards: if we take for granted a certain ordered phase, are the corresponding quantum-critical fluctuation effective at driving pairing near their QCP?
This is the question that the just discussed strategy allows us to answer; see also the discussion of Sec.~\ref{sec:QCP-pairing-model}.

The key task in front of us is to analyze how the solutions of the linearized gap equation (Sec.~\ref{sec:QCP-model-lin-gap-eq})
\begin{align}
\sum_n \int\limits_{\varepsilon_{\vb{k} n} = 0} \frac{\dd{S_{\vb{k}}}}{(2 \pi)^d} \sum_{A=0}^{3} \PintW_{BA}(\vb{p}_m, \vb{k}_n) \, d_{A}(\vb{k}_n) = \lambda \, d_{B}(\vb{p}_m) \label{eq:final-lin-gap-eq-exchange-model}
\end{align}
depend on the distance from the QCP when the pairing interaction equals [Eqs.~(\ref{eq:lin-gap-eq-boson-exchange-W-mat},~\ref{eq:lin-gap-eq-boson-exchange-M-mat})]:
\begin{align}
\PintW_{BA}(\vb{p}_m, \vb{k}_n) &= p_{\TRop} \, g^2 \frac{\Pintw_{BA}(\vb{p}_m, \vb{k}_n) + \Pintw_{BA}(\vb{p}_m, -\vb{k}_n) p_A}{4 \abs{\grad_{\vb{p}} \varepsilon_{\vb{p} m}}^{1/2} \abs{\grad_{\vb{k}} \varepsilon_{\vb{k} n}}^{1/2}}, \label{eq:lin-gap-eq-boson-exchange-W-mat-again} \\
\Pintw_{BA}(\vb{p}_m, \vb{k}_n) &= (L^d / \mathcal{N}) \chi(\vb{p} - \vb{k},0) \sum_a \Tr \mathcal{P}_{\vb{p} m}^{B} \Gamma_{a \vb{p}, \vb{k}} \mathcal{P}_{\vb{k} n}^{A} \Gamma_{a \vb{p}, \vb{k}}^{\dag}.
\end{align}

\subsubsection{Symmetries of the pairing interaction and eigenvectors} \label{sec:lin-gap-eq-spectral-symmetries}
Let us start from the fact that
\begin{align}
\PintW_{BA}^{*}(\vb{p}_m, \vb{k}_n) &= \PintW_{AB}(\vb{k}_n, \vb{p}_m)
\end{align}
is Hermitian [Eq.~\eqref{eq:F-real-cond}], from which it follows that the pairing interaction can be completely diagonalized to obtain an eigen-expansion of the form
\begin{align}
\PintW_{BA}(\vb{p}_m, \vb{k}_n) &= \sum_a \lambda_a \, d_{B; a}(\vb{p}_m) \mleft[d_{A; a}(\vb{k}_n)\mright]^{*},
\end{align}
where $a$ is a general index here (not necessarily just the irrep index) and
\begin{align}
\sum_n \int\limits_{\varepsilon_{\vb{k} n} = 0} \frac{\dd{S_{\vb{k}}}}{(2 \pi)^d} \sum_{A=0}^{3} \mleft[d_{A; a}(\vb{k}_n)\mright]^{*} d_{A; b}(\vb{k}_n) &= \Kd_{ab}.
\end{align}
Moreover,
\begin{align}
\PintW_{BA}^{*}(\vb{p}_m, \vb{k}_n) = \PintW_{BA}(\vb{p}_m, \vb{k}_n)
\end{align}
is in addition real [Eq.~\eqref{eq:F-real-cond}] so the eigenvectors can be chosen to be real:
\begin{align}
\mleft[d_{A; a}(\vb{k}_n)\mright]^{*} &= d_{A; a}(\vb{k}_n).
\end{align}

Next, we prove the fundamental fact that the pairing interaction respects the symmetries of the system.
In Sec.~\ref{sec:QCP-model-lin-gap-eq}, we already showed that composed parity and TR symmetry imply
\begin{align}
\PintW_{BA}(\vb{p}_m, \vb{k}_n) &= p_B p_A \PintW_{BA}(\vb{p}_m, \vb{k}_n),
\end{align}
where $p_{A=0} = - p_{A=1,2,3} = 1$, and therefore
\begin{align}
\PintW_{0A'}(\vb{p}_m, \vb{k}_n) &= \PintW_{A'0}(\vb{p}_m, \vb{k}_n) = 0 \qquad\text{for $A' \in \{1, 2, 3\}$.}
\end{align}
Hence pseudospin-singlet ($A = 0$) and pseudospin-triplet ($A = 1, 2, 3$) channels do not mix.
The two channels correspond to even-parity and odd-parity, respectively, as can be seen from the relation
\begin{align}
\PintW_{BA}(\vb{p}_m, -\vb{k}_n) &= p_A \PintW_{BA}(\vb{p}_m, \vb{k}_n)
\end{align}
which follows from Pauli's principle.
Formally, Pauli's principle is reflected by the way $\PintW_{BA}(\vb{p}_m, \vb{k}_n)$ is an (anti-)symmetric average of $\Pintw_{BA}(\vb{p}_m, \vb{k}_n)$ [Eq.~\eqref{eq:lin-gap-eq-boson-exchange-W-mat-again}].
Thus in the singlet (triplet) channel only even-parity (odd-parity) eigenvectors can contract with $\PintW_{BA}$ to give non-zero eigenvalues.
The pairing eigenvectors are therefore either even-parity pseudospin-singlets with an $A = 0$ component or odd-parity pseudospin-triplets with $A = 1, 2, 3$ components.
Using the symmetry rules of Sec.~\ref{sec:LC-gen-model-sym}, one may also show that point group symmetries yield the following constraints:
\begin{gather}
\PintW_{00}\mleft(R(g^{-1}) \vb{p}_m, R(g^{-1}) \vb{k}_n\mright) = \PintW_{00}(\vb{p}_m, \vb{k}_n), \\
\sum_{A',B'=1}^{3} \mleft[R_{\vb{p} m}(g)\mright]_{BB'} \mleft[R_{\vb{k} n}(g)\mright]_{AA'} \PintW_{B'A'}\mleft(R(g^{-1}) \vb{p}_m, R(g^{-1}) \vb{k}_n\mright) = \PintW_{BA}(\vb{p}_m, \vb{k}_n).
\end{gather}
These constraints are satisfied as long as the initial interaction of Eq.~\eqref{eq:effective-multi-exch-int-mode-freq} or~\eqref{eq:LC-lin-gap-eq-entry-Hint} does not break any point group symmetries.

Because the pairing interaction respects the symmetries of the system, the solutions of the linearized gap equation fall into irreps of the point group,\footnote{Mathematically, if we reinterpret the linearized gap equation as a diagonalization problem on a infinite-dimension space of pairing states, this is equivalent to the statement that if symmetry operators commute with the pairing interaction, then we may simultaneously diagonalize the two. Since the symmetry operators do not, in general, commute among themselves, we cannot simultaneously diagonalize all the symmetry operators. Instead, the pairing eigenvectors $d_A(\vb{k}_n)$ fall into irreps, which is the closest thing to simultaneously diagonalizing all point group symmetry operators. See also Sec.~\ref{sec:rep-theory-basics} of Appx.~\ref{app:group_theory} on the basics of representation theory.} as is well-known~\cite{Schrieffer1999, Sigrist2005, Leggett2006}.
In particular, they transform according to
\begin{align}
d_{0;a}\mleft(R(g^{-1}) \vb{p}_m\mright) &= \sum_{b=1}^{\dim \zeta_g} \RepM_{ab}^{\zeta_g}(g^{-1}) d_{0;b}(\vb{p}_m)
\end{align}
in the case of pseudospin-singlet solutions and according to
\begin{align}
\sum_{B'=1}^{3} \mleft[R_{\vb{p} m}(g)\mright]_{BB'} d_{B';a}\mleft(R(g^{-1}) \vb{p}_m\mright) &= \sum_{b=1}^{\dim \zeta_u} \RepM_{ab}^{\zeta_u}(g^{-1}) d_{B;b}(\vb{p}_m) \label{eq:fundamental-triplet-dvec-sym-transf-rule}
\end{align}
in the case of pseudospin-triplet solutions.
Here $\zeta_g$ and $\zeta_u$ are even- and odd-parity irreps of the point group, respectively, and $\RepM_{ab}^{\zeta}(g)$ are the corresponding representation matrices.

Conversely, every momentum-dependent scalar function $f(\vb{k})$ can be decomposed into these same irreps by exploiting the group-theoretical identity~\cite[Chap.~4]{Dresselhaus2007}:
\begin{align}
f(\vb{k}) &= \sum_{\zeta} \sum_{a=1}^{\dim \zeta} f^{\zeta}_a(\vb{k}), \\
f^{\zeta}_a(\vb{k}) &= \frac{\dim \zeta}{\dim G} \sum_{g \in G} \RepM^{\zeta}_{aa}(g) f(R(g^{-1}) \vb{k}), \label{eq:irrep-group-decomp}
\end{align}
where $G$ is the point group and $\zeta$ goes over all irreps.
Note that, for multicomponent irreps, it is not in general true that $f^{\zeta}_a\mleft(R(g^{-1}) \vb{k}\mright) = \sum_b \RepM^{\zeta}_{ab}(g^{-1}) f^{\zeta}_b(\vb{k})$ because the various $\zeta$ irrep components contained in $f(\vb{k})$ are not necessarily related.
For instance, relative to the $D_{4h}$ tetragonal group (defined in Sec.~\ref{sec:tetragonal-group-D4h} of Appx.~\ref{app:group_theory}), $f(\vb{k}) = k_x + 2 k_y$ has $f^{E_u}_x(\vb{k}) = k_x$, but $f^{E_u}_y(\vb{k}) = 2 k_y \neq f^{E_u}_x(R(C_{4z}^{-1}) \vb{k}) = k_y$.
If we had replaced $\RepM^{\zeta}_{aa}(g)$ in Eq.~\eqref{eq:irrep-group-decomp} with $\RepM^{\zeta}_{ac}(g)$, where $c$ is fixed, then $f^{\zeta}_a(R(g^{-1}) \vb{k}) = \sum_b \RepM^{\zeta}_{ab}(g^{-1}) f^{\zeta}_b(\vb{k})$ would indeed hold.
However, the corresponding $f^{\zeta}_a(\vb{k})$ now depends on $c$.
For instance, $f(\vb{k}) = - k_x + 3 k_y$ has $(f^{E_u}_x, f^{E_u}_y) = (- k_x, - k_y)$ when $c = 1$, but $(f^{E_u}_x, f^{E_u}_y) = (3 k_x, 3 k_y)$ when $c = 2$.
In the case of momentum-dependent vector-valued functions $\vb{f}(\vb{k})$, in Eq.~\eqref{eq:irrep-group-decomp} one replaces $f(R(g^{-1}) \vb{k})$ with $R_{\vb{k} n}(g) \vb{f}(R(g^{-1}) \vb{k}_n)$.

The leading pairing instability has the largest positive pairing eigenvalue $\lambda_a$ and its $T_c \propto \omega_c \Elr^{-1/\lambda}$ is the largest.
Its symmetry is determined by the corresponding eigenvector $d_{A; a}(\vb{k}_n)$.
When the eigenvector irrep is 1D, there is no ambiguity in the resulting pairing symmetry.
However, for multidimensional irreps, there is a continuum of possible pairing states which are all degenerate on the linearized-gap-equation level.
This degeneracy is lifted by high-order terms, such as the quartic coefficients in the Ginzburg-Landau expansion.
We shall see an example of this phenomenon in Chap.~\ref{chap:Sr2RuO4} during our Ginzburg-Landau analysis of \ce{Sr2RuO4}.

\subsubsection[Pairing symmetry and the time-reversal sign of the order \\ parameter]{Pairing symmetry and the time-reversal sign of the order parameter} \label{sec:TR-positivity}
In the expression~\eqref{eq:lin-gap-eq-boson-exchange-M-mat} for $\Pintw_{BA}(\vb{p}_m, \vb{k}_n)$,
\begin{align}
\Pintw_{BA}(\vb{p}_m, \vb{k}_n) &= (L^d / \mathcal{N}) \chi(\vb{p} - \vb{k},0) \sum_a \Tr \mathcal{P}_{\vb{p} m}^{B} \Gamma_{a \vb{p}, \vb{k}} \mathcal{P}_{\vb{k} n}^{A} \Gamma_{a \vb{p}, \vb{k}}^{\dag},
\end{align}
the susceptibility is strictly non-negative:
\begin{align}
\chi(\vb{q}, \iu \omega_{\ell}) = \ev{\abs{\Phi_{a \vb{q}}(\iu \omega_{\ell})}^2} \geq 0.
\end{align}
Furthermore, the pairing form factor $\PintF_{BA}(\vb{p}_m, \vb{k}_n) = \sum_a \Tr \mathcal{P}_{\vb{p} m}^{B} \Gamma_{a \vb{p}, \vb{k}} \mathcal{P}_{\vb{k} n}^{A} \Gamma_{a \vb{p}, \vb{k}}^{\dag}$ which we studied in Sec.~\ref{sec:QCP-model-lin-gap-eq} can be written as [Eq.~\eqref{eq:F_BA-expr}]:
\begin{align}
\PintF_{BA}(\vb{p}_m, \vb{k}_n) = \sum_a \tr_s \Pauli_B \Pintf_{a}(\vb{p}_m, \vb{k}_n) \Pauli_A \Pintf_{a}^{\dag}(\vb{p}_m, \vb{k}_n).
\end{align}
Hence in the singlet channel, the pairing form factor is also strictly non-negative:
\begin{align}
\begin{aligned}
\PintF_{00}(\vb{p}_m, \vb{k}_n) &= \sum_a \tr_s \Pintf_{a}(\vb{p}_m, \vb{k}_n) \Pintf_{a}^{\dag}(\vb{p}_m, \vb{k}_n) \\
&= \sum_{a s s'} \abs{\mleft[\Pintf_{a}(\vb{p}_m, \vb{k}_n)\mright]_{ss'}}^2 \geq 0.
\end{aligned}
\end{align}
Consequently
\begin{align}
\Pintw_{00}(\vb{p}_m, \vb{k}_n) \geq 0,
\end{align}
and the singlet Cooper-channel interaction
\begin{align}
\PintW_{00}(\vb{p}_m, \vb{k}_n) &= p_{\TRop} \, g^2 \frac{\Pintw_{00}(\vb{p}_m, \vb{k}_n) + \Pintw_{00}(\vb{p}_m, -\vb{k}_n)}{4 \abs{\grad_{\vb{p}} \varepsilon_{\vb{p} m}}^{1/2} \abs{\grad_{\vb{k}} \varepsilon_{\vb{k} n}}^{1/2}}
\end{align}
can be either positive or negative-semidefinite, depending on the TR sign of the order parameter $p_{\TRop} = \pm 1$.

By the Perron-Frobenius theorem~\cite{Berman1994}, the definiteness of the pairing interaction implies that:
\begin{enumerate}
\item The largest-in-magnitude pairing eigenvalue $\lambda$ arising in Eq.~\eqref{eq:final-lin-gap-eq-exchange-model} is positive (negative) for TR-even (TR-odd) order-parameter fields $\Phi_{a \vb{q}}$.
\item The corresponding eigenvector has components which are all positive, $d_{A=0}(\vb{k}_n) > 0$.
\item This leading singlet channel is non-degenerate.
\end{enumerate}
Hence the leading singlet pairing state of TR-even collective modes is always a conventional $s$-wave pairing state.
This leading singlet pairing state is, in fact, the preferred pairing channel overall for TR-even order parameters.
To prove this, consider the triplet eigenvector $\vb{d}_{A=1,2,3}(\vb{k}_n)$ with the largest triplet eigenvalue.
We may always switch to a gauge in which this $\vb{d}$-vector is parallel to $\vu{e}_3$, i.e., in which only the $\vb{d}_{A=3}(\vb{k}_n)$ component is finite.
Since in this gauge
\begin{align}
\begin{aligned}
\PintF_{33}(\vb{p}_m, \vb{k}_n) &= \sum_a \tr_s \Pauli_3 \Pintf_{a}(\vb{p}_m, \vb{k}_n) \Pauli_3 \Pintf_{a}^{\dag}(\vb{p}_m, \vb{k}_n) \\
&= \sum_{a s s'} (\pm)_s (\pm)_{s'} \abs{\mleft[\Pintf_{a}(\vb{p}_m, \vb{k}_n)\mright]_{ss'}}^2
\end{aligned}
\end{align}
is bounded from above by $\PintF_{00}(\vb{p}_m, \vb{k}_n)$, a corollary of the Perron-Frobenius theorem tells us that:
\begin{enumerate}\setcounter{enumi}{3}
\item The largest-in-magnitude triplet eigenvalue is strictly smaller than the largest singlet eigenvalue.
\end{enumerate}

This result is a generalization of a result due to Brydon et al.~\cite{Brydon2014} concerning phonon exchange to generic TR-even bosons, such as nematic, ferroelectric, alterelectric, spin-nematic, or charge-density wave modes.
This result holds quite generally for systems with parity and TR symmetry.
It even holds for multiband systems with topological band structures and spin-orbit coupling.
In the multiband case, $s^{+-}$-wave pairing that has different signs on different Fermi surfaces is not precluded by this result.
It is also possible that interactions not included in the model, such as Coulomb repulsion, suppress the leading $s$-wave state more than the subleading state so that we get unconventional pairing in the end~\cite{Brydon2014}.

In the case of TR-odd collective modes, the singlet interaction is overall repulsive.
This does not, however, preclude the appearance of singlet pairing~\cite{Maiti2013}.
But it does mean that the existence of an attractive channel and its symmetry both highly depend on the form factor $\Pintf_{a}(\vb{p}_m, \vb{k}_n)$ and on the $\vb{q}$-dependence of the susceptibility $\chi(\vb{q}, 0)$.
The conceptual aspects of this we discuss in Sec.~\ref{sec:cuprate-sym-choice-mechanism} of the next chapter.
Moreover, the Perron-Frobenius theorem implies that any attractive singlet channel, if it exists, must be unconventional in the sense that $d_{A=0}(\vb{k}_n)$ changes sign as a function of $\vb{k}_n$.
This is necessary to ensure that it is orthogonal to the most-repulsive (Perron-Frobenius) eigenvector which does not have sign-changes.
The overall most attractive channel can be either singlet or triplet in the case of exchange of TR-odd modes, as we shall see in Sec.~\ref{sec:pairing-cuprate-actual-analysis-results} of the next chapter when we study pairing due to loop-current and spin-magnetic modes in cuprates.

\subsubsection{The case of intra-unit-cell order with weak spin-orbit coupling} \label{sec:IUC-order-results}
In Sec.~\ref{sec:Cp-channel-gen-sym-constr}, we have seen that, for systems without SOC, symmetries sometimes force the pairing form factors to vanish at certain momenta.
Here we show that this has far-reaching implications for quantum-critical pairing driven by intra-unit-cell (IUC) order.

For IUC order, the static order-parameter correlation function or susceptibility $\chi(\vb{q}, 0)$ is peaked at $\vb{q} = \vb{0}$ in momentum space.
This peak is characterized by a correlation length
\begin{align}
\xi = a_0 r^{-\nu}
\end{align}
which diverges as the QCP is approached.
Here, $a_0$ is a microscopic length scale on the order of the lattice constant and $r$ is a dimensionless parameter that, by definition, vanishes at the QCP; see Fig.~\ref{fig:QCP-calc-strategy}.
For $r \sim 1$, the correlation function is structureless in momentum space.
For the static correlation function we shall use the following critical scaling expression
\begin{align}
\chi(\vb{q}, 0) &= \frac{\mathcal{F}\mleft(\xi \abs{\vb{q}}\mright)}{\abs{\vb{q}}^{2-\eta}}, \label{eq:chi-scaling-expression}
\end{align}
with critical exponents $\nu$ and $\eta$ and scaling function $\mathcal{F}(y)$ that has the usual asymptotic behavior:
\begin{align}
\mathcal{F}(y) \sim \begin{cases}
\const, & \text{for $y \gg 1$,} \\
y^{2-\eta}, & \text{for $y \ll 1$.}
\end{cases}
\end{align}
This critical scaling expression for $\chi(\vb{q}, 0)$ allows us to phenomenologically treat a general critical boson sector, beyond the mean-field level ($\eta = 0$).

Before we present our analysis, let us briefly comment on previous results for pairing mediated by critical fluctuations.
For order parameters that are nematic or spin-ferromagnetic, as the QCP is approached ($r \to 0$) one finds that the largest eigenvalue of the linearized gap equation diverges like
\begin{align}
\lambda &\propto r^{-\psi}
\end{align}
with $\psi > 0$~\cite{Abanov2001, Varma2012, Lederer2015}, as schematically shown by the blue line in Fig.~\ref{fig:QCP-general-results}.
While this corresponds to a breakdown of the weak-coupling analysis, it also signals the emergence of a strong pairing tendency near the QCP.
Weak-coupling theory alone cannot determine the precise behavior in the immediate vicinity of the QCP, yet numerous computational approaches show that $T_c$ is largest at or near the QCP~\cite{Berg2019, Lederer2017, Wang2017, Xu2017, Xu2020}.
This is the much-discussed efficiency of quantum-critical pairing~\cite{She2009, Moon2010, Metlitski2015, YuxuanWang2016-inv, YuxuanWang2016-QCP, WangRaghu2017, ChubukovI_2020, ChubukovII_2020}.

Following Ref.~\cite{Lederer2015}, the divergence of $\lambda$ in the case of IUC order is based on the assumption that the forward-scattering form factor $\mleft.\PintF_{BA}(\vb{p}_m,\vb{k}_n)\mright|_{\vb{p} \to \vb{k}}$ is attractive and varies smoothly as a function of the exchanged momentum $\vb{q} = \vb{p} - \vb{k}$.
Under these circumstances, the largest eigenvalue of the gap equation is given by
\begin{equation}
\lambda \approx \lambda_0 \int d^{d-1}q_{\parallel} \, \chi(\vb{q}_{\parallel}, 0), \label{eq:Lederer2015-expr}
\end{equation}
where $\vb{q}_{\parallel}$ are the components of the transferred momentum $\vb{q}$ tangential to the Fermi surface and
\begin{align}
\lambda_0 &= g^{2} \ev{\frac{p_{\TRop} \PintF_{AB}(\vb{k}_n,\vb{k}_n)}{\abs{\grad_{\vb{k}} \varepsilon_{\vb{k} n}}}}_{\text{FS}}
\end{align}
is a suitably defined average over the Fermi surface(s) of the pairing form factor at forward scattering.
The precise components ($A, B$ indices) of $\PintF_{AB}(\vb{k}_n,\vb{k}_n)$ which enter the above average vary and depend on whether the leading state is a singlet or triplet.
Using the scaling form for $\chi(\vb{q}_{\parallel}, 0)$ of Eq.~\eqref{eq:chi-scaling-expression}, the integral in Eq.~\eqref{eq:Lederer2015-expr} gives
\begin{align}
\psi = (3 - d - \eta) \nu \qquad\text{in spatial dimensions $d < 3 - \eta$.}
\end{align}
Hence QCPs in $d = 2$ with $\eta < 1$ yield strong pairing.
Since most studies find $\eta$ which are smaller than $\tfrac{1}{2}$, QCPs in two dimensions almost always give strong pairing.
The origin of this pairing enhancement is the $\vb{q} = \vb{0}$ divergence of the susceptibility at the QCP.
In $d=3$ the enhancement is logarithmic, $\lambda \propto \log r$, provided that $\eta = 0$, and it can never become stronger than this because $\eta$ is never negative.
In the remaining cases (when we obtain a negative $\psi$), $\lambda$ goes to a constant at the QCP (and not to zero as the negative $\psi$ suggests) and there is no enhancement.

The enhancement of pairing near IUC QCPs in 2D rests on two assumptions:
\begin{enumerate}
\item That the forward-scattering pairing form factor $\PintF_{AB}(\vb{k}_n,\vb{k}_n)$ has at least one singlet or triplet component which has the same sign as the TR-sign of the order parameter, thus rendering $p_{\TRop} \PintF_{AB}(\vb{k}_n,\vb{k}_n) > 0$ attractive in that channel.
\item That the pairing form factor $\PintF_{BA}(\vb{p}_m,\vb{k}_n)$ in this same channel does not vanish at forward scattering ($\vb{p} \to \vb{k}$). Note that if it did vanish, then the forward-scattering divergence of the susceptibility would be suppressed.
\end{enumerate}
In the presence of SOC, both the singlet and triplet components of $\PintF_{AB}(\vb{k}_n,\vb{k}_n)$ are generically finite, with different components having opposite signs when SOC is strong enough.
Hence both assumptions can be true in the presence SOC, i.e., IUC order can be effective at driving pairing near its QCP in 2D.
That said, when the pairing is not effective in the absence of SOC, the strength of SOC has to be sufficiently large to overcome this tendency.

Without SOC, both assumptions depend sensitively on the type of order.
In the case of purely orbital order, the coupling $\Gamma$ matrix introduced in Eq.~\eqref{eq:general-momentum-bilinear} has the form $\Gamma_{a \vb{k}, \vb{p}} = \gamma_{a \vb{k}, \vb{p}} \otimes \Pauli_0$ which implies that the pairing form factor of Eq.~\eqref{eq:fa-definition-form-f} equals
\begin{align}
\Pintf_{a}(\vb{p}_m,\vb{k}_n) &= \mleft(u_{\vb{p} m}^{\dag} \gamma_{a \vb{p}, \vb{k}} u_{\vb{k} n}\mright) \Pauli_0,
\end{align}
where $u_{\vb{k} n s} = u_{\vb{k} n} \otimes \ket{s}$ are the band eigenvectors.
Thus
\begin{align}
\begin{aligned}
\PintF_{BA}(\vb{p}_m,\vb{k}_n) &= \sum_a \tr_s \Pauli_B \Pintf_{a}(\vb{p}_m, \vb{k}_n) \Pauli_A \Pintf_{a}^{\dag}(\vb{p}_m, \vb{k}_n) \\
&= \Kd_{AB} \sum_a \abs{u_{\vb{p} m}^{\dag} \gamma_{a \vb{p}, \vb{k}} u_{\vb{k} n}}^2 \geq 0
\end{aligned} \label{eq:IUC-tab-underlined-cond}
\end{align}
has only positive components and the first assumption holds only for TR-even purely orbital order.
In the case of purely spin order, the spin trace has both positive and negative components and the first assumption is always valid.
Regarding the second assumption, the composition of parity $P$ with time-reversal $\TRop$ is the only symmetry operation which maps generic $\vb{k}$ to themselves.
In Sec.~\ref{sec:Cp-channel-gen-sym-constr}, we have seen that $\PintF_{AB}(\vb{k}_n,\vb{k}_n) = 0$ for $P \TRop$-odd purely orbital order and for $P \TRop$-even purely spin order.
As a result, the forward-scattering form factor is suppressed by two powers of $\vb{q} = \vb{p} - \vb{k}$,
\begin{align}
\mleft.\PintF_{BA}(\vb{p}_n,\vb{k}_n)\mright|_{\vb{p} \to \vb{k}} \propto \mleft(\vb{p} - \vb{k}\mright)^2, \label{eq:IUC-tab-red-cond}
\end{align}
which changes the $\psi$ value obtained from Eq.~\eqref{eq:Lederer2015-expr} into:
\begin{align}
\psi = (1 - d - \eta) \nu \qquad\text{in spatial dimensions $d < 1 - \eta$.}
\end{align}
Thus quantum-critical pairing can only be logarithmically enhanced in $d = 1$ if $\eta = 0$.
Even a small deviation away from the mean-field value of $\eta = 0$ eliminates this logarithmic enhancement.
In all other cases, there is no enhancement under any circumstance.
This is the main result of our analysis.

In the derivation of Eq.~\eqref{eq:IUC-tab-red-cond}, we have assumed that the form factor $\PintF_{BA}(\vb{p}_n,\vb{k}_n)$ is analytic near forward-scattering.
However, even if the vanishing was slightly non-analytic, i.e.\ if Eq.~\eqref{eq:IUC-tab-red-cond} had $\abs{\vb{p} - \vb{k}}^{2+\kappa}$ instead of $\mleft(\vb{p} - \vb{k}\mright)^2$ for a small $\abs{\kappa} < 1$, in 2D the exponent $\psi = (1 - d - \eta - \kappa) \nu$ would yet again be negative, indicating the absence of an enhancement of $\lambda$ near the QCP.
Thus our main result is robust against both the non-analyticity of the vertex and the non-mean-field scaling of the susceptibility.

Based on this understanding of IUC QCPs, we may now state the behavior of quantum-critical Cooper pairing in the absence of SOC.
We focus on two-dimensional systems.
The QCP pairing behavior of the various IUC orders, classified in Tab.~\ref{tab:orbital-spin-IUC-orders}, is:
\begin{itemize}
\item Nematic, ferromagnetic, altermagnetic, and odd-parity spin LC fluctuations drive \emph{parametrically strong} quantum-critical pairing.
\item Even-parity LC, ferroelectric, alterelectric, spin-nematic, even-parity spin LC, toroidal ferromagnetic, and toroidal altermagnetic fluctuations gives \emph{parametrically weak} pairing near their QCP.
\item Odd-parity LC fluctuations are \emph{parametrically strong pair-breakers} near the QCP, i.e., they completely suppress the pairing driven by other mechanisms.
\end{itemize}

By ``parametrically strong'' pairing we mean that the leading $\lambda \propto r^{-\psi}$ is attractive and diverges as $r \to 0$ with a $\psi > 0$, as shown by the blue line in Fig.~\ref{fig:QCP-general-results}.
Although this divergence indicates a breakdown of our weak-coupling treatment at the QCP, complementary numeric~\cite{Berg2019, Lederer2017, Wang2017, Xu2017, Xu2020} and analytic~\cite{She2009, Moon2010, Metlitski2015, YuxuanWang2016-inv, YuxuanWang2016-QCP, WangRaghu2017, ChubukovI_2020, ChubukovII_2020} calculations indicate a robust SC dome surrounding the QCP.

\begin{table}[t!]
\centering
\captionabove[Orbital and spin intra-unit-cell orders, classified according to parity (P) and time reversal (TR).]{\textbf{Orbital and spin intra-unit-cell orders, classified according to parity (P) and time reversal (TR).}
Colored red are orders which, due to combine parity and TR symmetry, have suppressed forward-scattering pairing form factors in the absence of spin-orbit coupling [Eq.~\eqref{eq:IUC-tab-red-cond}].
Underlined are the TR-odd orbital orders which are repulsive in all channels for weak spin-orbit coupling [Eq.~\eqref{eq:IUC-tab-underlined-cond}].
LC stands for loop currents.
By our conventions, nematic and altermagnetic orders do not break space-inversion symmetry.
Even though spin-nematic and even-parity spin LC orders cannot be distinguished using symmetries, I have included them as separate entries due to their different microscopic origins.
For descriptions of the various ordered phases, see captions of Fig.~\ref{fig:spin-mag-class} and Tab.~\ref{tab:orbital-modes}.}
{\renewcommand{\arraystretch}{1.3}
\renewcommand{\tabcolsep}{10pt}
\begin{tabular}{|c"c|c|}
\multicolumn{3}{c}{\large Orbital intra-unit-cell orders:} \\ \hline
& P-even & P-odd \\ \thickhline
TR-even & nematic & \begin{tabular}{c}
\textcolor{red}{ferroelectric} \\
\textcolor{red}{alterelectric}
\end{tabular} \\ \hline
TR-odd & \textcolor{red}{\underline{even-parity LC}} & \underline{odd-parity LC}
\\[2pt] \hline 
\multicolumn{3}{c}{} \\[-2pt]
\multicolumn{3}{c}{\large Spin intra-unit-cell orders:} \\ \hline
& P-even & P-odd \\ \thickhline
TR-even & \begin{tabular}{c}
\textcolor{red}{spin-nematic} \\
\textcolor{red}{even-parity spin LC}
\end{tabular} & odd-parity spin LC \\ \hline
TR-odd & \begin{tabular}{c}
ferromagnetic \\
altermagnetic
\end{tabular} & \begin{tabular}{c}
\textcolor{red}{toroidal ferromagnetic} \\
\textcolor{red}{toroidal altermagnetic}
\end{tabular}
\\ \hline
\end{tabular}}
\label{tab:orbital-spin-IUC-orders}
\end{table}

For ``parametrically weak'' pairing, in contrast, $\lambda \to \const$ as $r \to 0$, as depicted by the orange line in Fig.~\ref{fig:QCP-general-results}.
Hence superconductivity may result, but there is no particular reason for why $T_c$ should be larger near the QCP.
Indeed, with regard to pairing, these ``parametrically weak'' orders near QCPs behave as any other degree of freedom far from its critical point.

Finally, odd-parity LCs which are ``parametrically strong pair-breakers'' require special elaboration.
On the one hand, because they are TR-odd in the orbital sector, their Cooper-channel interaction is repulsive between all momenta for both singlet and triplet channels; this follows from Eq.~\eqref{eq:IUC-tab-underlined-cond}.
On the other hand, this repulsive interaction is strongly peaked at $\vb{q} = \vb{0}$ near the QCP.
These two properties imply that the interaction behaves essentially the same as the unscreened Coulomb repulsion ($\sim 1/\vb{q}^2$), which is known to strongly suppress pairing.
The largest attractive pairing eigenvalue $\lambda$ therefore must vanish as we approach the QCP, as shown schematically by the green line in Fig.~\ref{fig:QCP-general-results}.
Only away from the QCP can finite-$\vb{q}$ features of the pairing form factor $\PintF_{BA}(\vb{p}_m,\vb{k}_n)$ or the Fermi velocity $\abs{\grad_{\vb{k}} \varepsilon_{\vb{k n}}}$ result in an attractive pairing channel that, however, is parametrically weak.
For even-parity LCs, note that the $\vb{q} = \vb{0}$ repulsion is suppressed because of $P \TRop$ symmetry [Eq.~\eqref{eq:IUC-tab-red-cond}] so that finite-$\vb{q}$ features can give pairing even precisely at the QCP (orange line in Fig.~\ref{fig:QCP-general-results}).

Spin-orbit coupling has the effect of eliminating the symmetry-based suppression of forward scattering [Eq.~\eqref{eq:IUC-tab-red-cond}].
Thus parametrically weak pairing becomes parametrically strong.
In the case of even-parity LCs, it is worth noting that the forward-scattering amplitude behaves like a pseudospin triplet and changes signs so we get parametrically strong pairing instead of parametrically strong pair-breaking.
Pairing and pair-breaking which is parametrically strong in the absence of SOC continues to be so with SOC.

In conclusion, focusing on loop currents in two-dimensional systems without SOC, even-parity IUC LCs are inefficient at driving pairing near their QCP, whereas odd-parity IUC LCs are detrimental to pairing near their QCP, as summarized in Fig.~\ref{fig:QCP-general-results}.
Note that the absence of a strong attractive pairing interaction at the QCP justifies \textit{a posteriori} the weak-coupling analysis employed in our analysis. 

Let us end with a comparison to finite-$\vb{q}$ order.
In the case of staggered order, the static susceptibility $\chi(\vb{q}, 0)$ peaks at a finite ordering vector $\vb{q} = \vb{Q}$, as well as at symmetry-related $\vb{Q}' = R(g) \vb{Q}$, where $g$ are point group operations.
Finite-momentum Cooper pair scattering, however, only takes place at certain ``hot spots'' on the Fermi surface where both $\vb{k}$ and $\vb{k}+\vb{Q}$ reside on some Fermi surface.
Geometrically, these hot spots are found by translating the Fermi surface(s) by $\vb{Q}$ and then looking at intersections.
As long as the Fermi surface geometry is such that hot spots exist, the largest pairing eigenvalue $\lambda$ will be essentially given by the same Eq.~\eqref{eq:Lederer2015-expr}, albeit with a modified $\lambda_0 \sim \PintF_{AB}(\vb{k}_n, (\vb{k}+\vb{Q})_m)$ which is an average over hot spots.
Thus $\psi = (3 - d - \eta) \nu$ yet again and we find parametrically strong enhancement as we approach QCPs in 2D.
As in the IUC ($\vb{Q} = \vb{0}$) case, at the QCP this weak-coupling treatment breaks down and complementary methods which include the effects of retardation, damping, etc., are needed to confirm that we get a SC dome around the QCP.
The big difference from IUC order is that there are no generic symmetries which map arbitrary $\vb{k}$ to $\vb{k} + \vb{Q}$.
Hence, there are no symmetry constraints which would suppress the $\vb{q} = \vb{Q}$ divergence of the susceptibility and give results analogous to the ones we just derived for IUC order.
Only when the hot spots reside on special high-symmetry points of the Brillouin zone can some similar statements be made.
This remains as an interesting venue for further study.

\chapter{Intra-unit-cell loop currents in~cuprates}
\label{chap:cuprates}

The physics of cuprates has attracted an enormous amount of attention since the discovery of high-$T_c$ superconductivity in these compounds in 1986~\cite{Bednorz1986}, almost four decades ago.
As of the time of writing, more than 30000 papers have been published that deal with cuprates and their superconductivity~\cite{WebofScience}.
Yet questions concerning the proper pairing mechanism, the role of competing orders, the origin of the pseudogap and strange metal regimes, and the remnants of Mott localization phenomena near optimal doping continue to be debated~\cite{Keimer2015}.

In a prominent proposal~\cite{Varma2020, Varma2016}, Chandra M.\ Varma has suggested that loop currents (LCs) are the key to understanding the phase diagram of the cuprates.
The phase diagram of hole-doped cuprates is shown in Fig.~\ref{fig:cuprate-phase-diagram}.
Within this proposal, the underlying order of the pseudogap phase is a LC order which preserves translation symmetry, but breaks space-inversion and time-reversal symmetry in the orbital sector~\cite{Varma1997, Simon2002, Simon2003, Varma2006}.
As the doping is increased, this odd-parity intra-unit-cell (IUC) LC order vanishes at a quantum-critical point (QCP) around which LC fluctuations are especially strong.
According to the proposal, it is precisely these LC fluctuations that are responsible for both the $d_{x^{2}-y^{2}}$-wave superconductivity (SC) surrounding the QCP~\cite{Aji2010} and the strange metal behavior above the QCP, which is described using marginal Fermi liquid theory~\cite{Varma1989, Varma1999}.
The final proposal is summarized in Refs.~\cite{Varma2020, Varma2016}.

In this chapter, we explore to what degree are quantum-critical IUC LC fluctuations a viable source of Cooper pairing in the cuprates.
We do so by employing the same strategy we used in the previous chapter (Sec.~\ref{sec:QCP-genera-final-analysis}, Fig.~\ref{fig:QCP-calc-strategy}) to analyze the quantum-critical pairing of general IUC orders.
Given the strong evidence that cuprates are a Fermi liquid at overdoping with a negligible spin-orbit coupling (SOC)~\cite{Keimer2015, Proust2019, Kramer2019, Lee-Hone2020}, the results of Sec.~\ref{sec:IUC-order-results}, summarized in Fig.~\ref{fig:QCP-general-results}, apply to the current case and essentially answer our question.

The answer is that even-parity LCs are not effective at driving SC near their QCP, whereas odd-parity LCs are pair breaks which strongly suppress any pairing tendency near their QCP.
Since pairing driven by even-parity LCs is parametrically weak, there is no particular reason to think that even-parity LCs can drive SC at such high temperatures as in the cuprates.
Although even-parity LCs can support unconventional pairing driven by other mechanisms, they are not likely to be the main source of pairing in the cuprates.
Regarding odd-parity LCs, our results give a compelling reason to believe that the pseudogap phase is not associated with an odd-parity IUC LC order, contrary to Varma's suggestion~\cite{Varma2020, Varma2016}.
At best, such odd-parity IUC LCs may arise as a subsidiary order, as proposed in Refs.~\cite{Agterberg2015, Scheurer2018, Sarkar2019}.
Note that, according to our analysis, quantum-critical staggered LCs (whose $\vb{q} \neq \vb{0}$), considered in Refs.~\cite{Chakravarty2001, Nayak2002, Dimov2008, Hsu2011, Chakravarty2011, Laughlin2014, Laughlin2014-details, Makhfudz2016}, constitute an effective pairing glue, unlike IUC LCs (whose $\vb{q} = \vb{0}$).

Even though the main question has thus already been answered in the negative in Chap.~\ref{chap:loop_currents}, a number of related questions still remain.
What types of LC orders are possible in the cuprates?
How can we experimentally distinguish these LC orders?
And for which of these LC orders do we get the correct $d_{x^2-y^2}$ symmetry?
These are the questions that we answer in the current chapter.
Just like the previous chapter, the current chapter is based on Ref.~\cite{Palle2024-LC} and in a number of places the text of Ref.~\cite{Palle2024-LC} has been recycled.
Additional material not covered by Ref.~\cite{Palle2024-LC} includes the literature review of Sec.~\ref{sec:cuprate-LC-lit-review}, details of how fermionic bilinears are classified (Sec.~\ref{sec:cuprate-bilinear-class}), analytic solutions of the linearized gap equation (Sec.~\ref{sec:cuprate-analytic-res}), and an extended comparison with the work by Aji, Shekhter, and Varma~\cite{Aji2010} which also studied LC-driven pairing with a similar strategy, but came to very different conclusions (Sec.~\ref{sec:comparison-w-Aji2010}).

This chapter is organized as follows.
We start with Sec.~\ref{sec:cuprate-basics} in which we recall the basics of cuprates: their composition, crystal structure (Fig.~\ref{fig:cuprate-crystal-structure}), and phase diagram as a function of hole doping and temperature (Fig.~\ref{fig:cuprate-phase-diagram}).
After that, in Sec.~\ref{sec:exp-sym-break-pseudogap} we survey the experimental evidence on symmetry-breaking in the pseudogap regime, which is overall mixed, in Sec.~\ref{sec:LC-proposals} we review the proposals which put loop currents forth as the hidden order of the pseudogap regime, the most prominent of which are those by Varma~\cite{Varma2020, Varma2016} and Chakravarty, Laughlin, et al.~\cite{Chakravarty2001, Laughlin2014, Laughlin2014-details}, and in Sec.~\ref{sec:LC-cuprate-micro} we review microscopic theoretical investigations of loop currents, some of which find them to be competitive for realistic parameters.
In the next Sec.~\ref{sec:cuprate-3band-model}, we discuss the electronic structure of the cuprates and we introduce the three-orbital model of the \ce{CuO2} planes, summarized in Fig.~\ref{fig:CuO2-sketch}, which we use in the rest of the chapter.
The fermionic bilinears of the three-orbital model are classified in Sec.~\ref{sec:cuprate-bilinear-class}, the main result being Tab.~\ref{tab:orbital-Lambda-mats} (Sec.~\ref{sec:orbital-Lambda-mats}) which lists all possible local orbital orders.
How this classification is put to practice is explained in Sec.~\ref{sec:cup-ord-param-constr}.
As an interesting application of potentially broader interest, in Sec.~\ref{sec:CuO2-Hubbard-decompose} we decompose extended Hubbard interactions into symmetry channels and derive Fierz identities which reflect ambiguities in the decomposition.

In Sec.~\ref{sec:pairing-cuprate-actual-analysis}, we finally turn to the analysis of pairing due to IUC LC fluctuations in cuprates.
First, we establish that the strategy of Sec.~\ref{sec:QCP-genera-final-analysis} (Fig.~\ref{fig:QCP-calc-strategy}) applies to cuprates.
Then, in Sec.~\ref{sec:pairing-cuprate-formalism}, we explain how the formalism of the previous chapter gets simplified for purely orbitals orders in the absence of SOC.
The allowed IUC LC orders are determined in Sec.~\ref{sec:Bloch-Kirch-constr} by taking into account the Bloch and generalized Bloch-Kirchhoff theorems of Sec.~\ref{sec:Bloch-tm} of the previous chapter.
We find three local LC orders, with $g_{xy(x^2-y^2)}$, $d_{x^2-y^2}$, and $(p_x|p_y)$ symmetry, which we study in the remainder of the chapter.
In Sec.~\ref{sec:pairing-cuprate-actual-analysis-VH}, we explore how efficiently LC fluctuations scatter Cooper pairs off Van Hove points, depending on the LC symmetry and band structure.
In Sec.~\ref{sec:pairing-cuprate-actual-analysis-results} we present the main results: numerical solutions of the linearized gap equation, together with discussions of how to experimentally probe the LCs and of the influence of SOC.
We find that $g$-wave LCs robustly favor $d_{xy}$ pairing (Fig.~\ref{fig:cuprate-A2g-results}), while $d$-wave LCs robustly favor $d_{x^2-y^2}$ pairing (Fig.~\ref{fig:cuprate-B1g-results}).
As for $p$-wave LCs, they suppress pairing near the QCP, while away from it they support extended $s$-wave pairing (Fig.~\ref{fig:cuprate-Eu-results}).
A conceptual understanding of our results is outlined in Sec.~\ref{sec:cuprate-sym-choice-mechanism}.
In the penultimate Sec.~\ref{sec:cuprate-analytic-res}, we derive analytic solutions of the linearized gap equation, which supplement the numerics.
In the final Sec.~\ref{sec:comparison-w-Aji2010}, we provide an extended comparison to the work by Aji, Shekhter, and Varma~\cite{Aji2010} which claimed that $d_{x^2-y^2}$ superconductivity robustly appears in a model of $p$-wave and $g$-wave LCs appropriate for cuprates.
We argue to the contrary.

\section{Basics of cuprate superconductors} \label{sec:cuprate-basics}
The cuprate superconductors are a family of copper oxides materials which are famous for their high-temperature superconductivity.
Superconductivity in these materials was first discovered in the copper oxide perovskite \ce{La_{2-x}Ba_{x}CuO4} with a $T_c \sim \SI{30}{\kelvin}$ in 1986 by Bednorz and Müller~\cite{Bednorz1986}, for which they were soon awarded a Nobel prize.

A $T_c$ of \SI{-240}{\celsius} might not seem impressive, but it greatly surprised the community~\cite{Blundell2009}.
On the one hand, its $T_c \sim \SI{30}{\kelvin}$ exceeded what was though to be possible based on BCS theory~\cite{Cohen1972}, which was by then well-established.
The intuition behind this expectation is that the relatively weak electron-electron attraction mediated by phonons can, because of retardation, only overcome the strongly repulsive Coulomb interaction at low energies, making $T_c$ small.
On the other hand, the material itself defied what were conventionally understood to be favorable properties for SC, as summarized in the empirical Matthias' rules~\cite{Matthias1957}, for instance.
The simplified Matthias' rules are~\cite{Mazin2010}: (i) high symmetry is good and cubic symmetry is best, (ii) high density of electronic states is good, (iii) stay away from oxygen, (iv) stay away from magnetism, and (v) stay away from insulators.
In contrast, the first cuprate superconductor \ce{La_{2-x}Ba_{x}CuO4} is strongly anisotropic and a poor conductor which, after a slight change of doping, becomes a strongly insulating antiferromagnet.
Since magnetism mostly arises from repulsive electron-electron interactions, whereas Cooper pairing needs attraction, its vicinity to a SC phase of such high $T_c$ seemed perplexing.

Very soon after the discovery, a flurry of scientific activity ensued in which many other cuprate superconductors were discovered, with the highest ambient-pressure $T_c$ reaching \SI{138}{\kelvin} in \ce{Hg_{0.8}Tl_{0.2}Ba2Ca2Cu3O_{8+\delta}}~\cite{Sun1994, Stewart2017}, which is larger than in any other compound discovered to date.
Currently, there are more than 200 compounds that fall into the family of cuprate superconductors~\cite{Stewart2017}.

In the intervening four decades, a vast literature has grown on cuprate superconductors, spanning more than 30000 articles~\cite{WebofScience}.
Reviewing this vast field would require a whole book of its own, and indeed many have been written~\cite{Anderson1997, Schrieffer2007, Waldram1996, Kamimura2010, Plakida2010}.
In this section, we take up the more modest task of reviewing the basics regarding the composition, crystal structure, and phase diagram of the cuprates.
These basics provide the background that is necessary for understanding the current work, which is based on Ref.~\cite{Palle2024-LC}.
The main references for this section are the chapter on cuprate superconductivity in Leggett's book~\cite{Leggett2006} and the review article by Keimer et al.~\cite{Keimer2015}.
Both references are an excellent entry into the field, and we refer the interested reader to these references (as well as the literature cited therein) for further reading.

\subsection{Composition and crystal structure} \label{sec:cuprate-structure}
\begin{figure}[t]
\centering
\begin{subfigure}[t]{0.42\textwidth}
\raggedright
\includegraphics[width=0.95\textwidth]{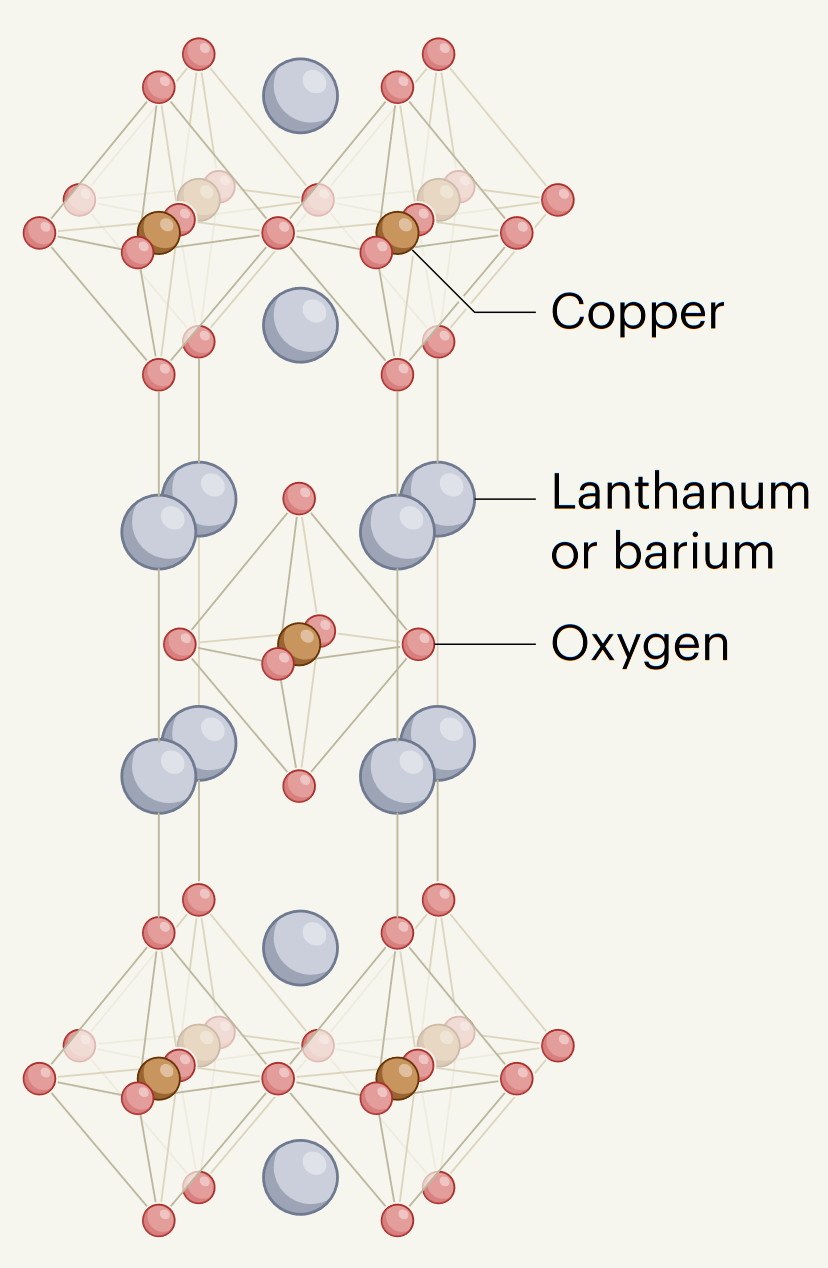}
\subcaption{}
\end{subfigure}%
\begin{subfigure}[t]{0.58\textwidth}
\raggedleft
\includegraphics[width=0.95\textwidth]{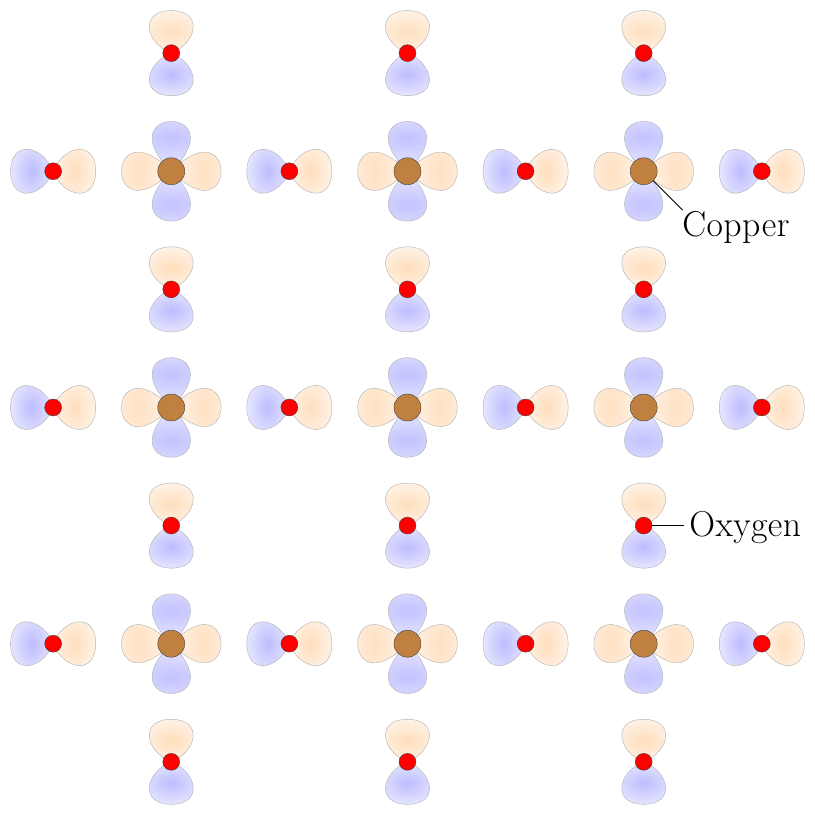}
\subcaption{}
\end{subfigure}
\captionbelow[Crystal structure of lanthanum barium copper oxide \ce{La_{2-x}Ba_{x}CuO4}~\cite{Armitage2019} (a) and of the copper oxide planes \ce{CuO2} (b).]{\textbf{Crystal structure of lanthanum barium copper oxide \ce{La_{2-x}Ba_{x}CuO4}}~\cite{Armitage2019} \textbf{(a) and of the copper oxide planes \ce{CuO2} (b).}
Lanthanum barium copper oxide was the first discovered cuprate superconductor~\cite{Bednorz1986} and it has one of the simpler crystal structures among the cuprates, which is that of a perovskite.
Its horizontal planes, which are made of copper and oxygen atoms, are shown from above under (b), together with the chemically most active orbitals which are, namely, \ce{Cu}:$3d_{x^2-y^2}$ and \ce{O}:$2p_{x,y}$ (Sec.~\ref{sec:cuprate-3band-model}).
Orange (blue) are positive (negative) lobes of the orbitals.
Figure (a) is reproduced from Ref.~\cite{Armitage2019}, with permission from Springer Nature.}
\label{fig:cuprate-crystal-structure}
\end{figure}

The cuprates are layered compounds whose key feature are the \ce{CuO2} copper oxide planes, which are depicted in Fig.~\ref{fig:cuprate-crystal-structure}(b).
Conceptually, the chemical composition of a generic cuprate material can be written in the following way~\cite{Leggett2006}:
\begin{align}
\ce{(CuO2)_n A_{n-1} X}, \label{eq:Leggett-chem-formula}
\end{align}
where \ce{CuO2} stands for the copper oxide planes, \ce{A} is an alkaline earth element, a rare earth element, or yttrium, and \ce{X} can be an arbitrary collection of elements, possibly including copper and/or oxygen.
The appeal of writing the chemical formula in this way is that it now reflects the crystal structure characteristic of cuprates~\cite{Leggett2006}, which is the following.
The \ce{CuO2} planes come in groups made of $n$ layers which are intercalated with $(n-1)$ \ce{A} elements; together we shall call this groupation an ``$n$-fold multilayer.''
In between these $n$-fold multilayers are the \ce{X} groups which act as a charge reservoir for the \ce{CuO2} planes.
Whereas the distance between the \ce{CuO2} planes is small within the $n$-fold multilayers (assuming $n \geq 2$), the distance between the multilayers is generally larger, although how much depends on \ce{X}.
Moreover, the \ce{CuO2} planes are stacked on top of each other within the $n$-fold multilayers, but are usually staggered relative to one another between multilayers.

The cuprate crystal structure can also be understood as a variation of the perovskite structure in which (for $n = 1$) \ce{Cu} is surrounded by an octahedron of oxygen atoms, while \ce{A} reside at the corners of the cube surrounding \ce{Cu}~\cite{Leggett2006}.
Within this picture, the in-plane oxygen atoms (which are called ``ligand'' oxygens) are shared between the copper elements, but the out-of-plane oxygen atoms (the so-called ``apical'' oxygens) are staggered relative to each other.
For $n \geq 2$, the oxygen octahedra become vertically elongated.
An example of a cuprate with a $n = 1$ perovskite structure is shown in Fig.~\ref{fig:cuprate-crystal-structure}(a); its formula can be recast into~\eqref{eq:Leggett-chem-formula} using $n = 1$ and $\ce{X} = \ce{La_{2-x}Ba_{x}CuO2}$.
Although the $\ce{A} = \ce{La_{1-x/2}Ba_{x/2}}$ intercalant is present in this example, it gets grouped with \ce{X} due to the staggering of the \ce{CuO2} layers.
Let us also note that some cuprate superconductors, such as \ce{Sr_{x}Ca_{1-x}CuO2}, are ``infinitely layered'' in the sense that they do not have a charge reservoir group \ce{X}~\cite{Leggett2006}.

Given how the \ce{CuO2} planes are the motif that is common to all cuprates, it is almost universally believed that SC originates in these planes~\cite{Leggett2006, Keimer2015}.
This does not, however, mean that the surrounding is unimportant.
The surrounding must play some role (beyond doping) if we are to explain the large variations of $T_c$\footnote{For some cuprates $T_c = 0$ at all doping levels, which might be an important clue regarding the origin of their superconductivity~\cite{Leggett2006}.} and other properties with $n$, \ce{A}, and \ce{X} entering Eq.~\eqref{eq:Leggett-chem-formula}~\cite{Leggett2006}.
Rather, the belief is that understanding the physics of the \ce{CuO2} planes is the correct first step (``zeroth-order approximation'') in understanding the broad qualitative features of the cuprate phase diagram~\cite{Leggett2006}.

Although intuitively plausible, one may wonder if there is hard experimental evidence supporting this belief.
The answer is affirmative.
The main experimental evidence comes from studies of atomically-thin two-dimensional superconductors~\cite{Uchihashi2017} and from x-ray absorption spectroscopy measurements~\cite{Yano2009}.
In Refs.~\cite{Gozar2008, Logvenov2009}, layered heterostructures were fabricated in which a single cuprate layer situated between insulating layers was found to be superconducting, without any apparent suppression of $T_c$ relative to the bulk crystal.
However, the neighboring insulating layers could, in principle, play a role in stabilizing this SC, making the interpretation of this finding not completely clear-cut~\cite{Palle2021-HMW}.
More recently, undiminished superconductivity was measured in exfoliated monolayer crystals of \ce{Bi2Sr2CaCu2O_{8+\delta}}~\cite{Yu2019}, which has conclusively shown that SC essentially originates from one layer in isolation.
The \ce{CuO2} planes are the central parts of these layers, and complementary x-ray absorption studies reveal that the low-energy electronic states primarily derive from the in-plane \ce{CuO2} orbitals~\cite{Chen1991, Chen1992, Pellegrin1993, Nucker1995, Peets2009}, in agreement with theoretical considerations~\cite{Pickett1989, Dagotto1994, Hybertsen1992}.

For the above reasons, the majority of theoretical work on cuprates has focused on the copper oxide planes~\cite{Leggett2006}; see Refs.~\cite{Anderson1987resonating, Zhang1988, Eskes1988, Varma1987, Emery1987, Emery1988, Littlewood1989, Gaididei1988, Scalettar1991, Hirsch1989, Marsiglio1990} for early examples.\footnote{There are innumerable later references as well.}
In this chapter, we shall do the same and study LC-driven pairing within a model of the \ce{CuO2} plane.

The \ce{CuO2} plane is, to a first approximation, made of a square lattice of copper atoms, with (ligand) oxygen atoms situated midway between the copper atoms, as shown under Fig.~\ref{fig:cuprate-crystal-structure}(b).
Hence it has a tetragonal point group.
For $n \geq 2$ in Eq.~\eqref{eq:Leggett-chem-formula}, the neighboring layers are stacked on top, with the same orientation and with copper atoms residing precisely above copper atoms.
In reality, there are a number of deviations from this idealized picture~\cite{Leggett2006}.
Many cuprates have a slightly rectangular lattice, instead of a square one, rendering the point group orthorhombic.
When there are multiple neighboring layers, the \ce{Cu-O} bonds often buckle away from the crystalline $a$ and $b$ directions.
Finally, in some compounds, such as \ce{Bi2Sr2CaCu2O8}, slight distortions with a large period appear within the \ce{CuO2} planes~\cite{Leggett2006}.
These inessential features, which are not common to all cuprates, we shall not include in our analysis.
As for the chemical structure, we shall discuss later, in Sec.~\ref{sec:cuprate-3band-model}, when we introduce the three-orbital tight-binding model of the \ce{CuO2} planes~\cite{Varma1987, Emery1987, Emery1988, Littlewood1989, Gaididei1988, Scalettar1991} that we employ in our calculation.

\begin{figure}[p!]
\centering
\includegraphics[width=\textwidth]{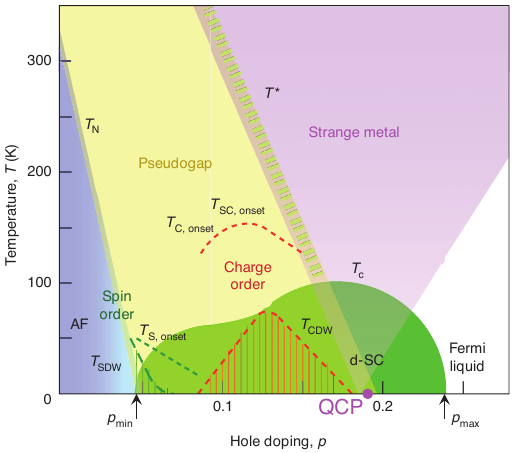}
\captionbelow[Schematic phase diagram of the hole-doped cuprates, as a function of hole doping $p$ and temperature $T$~\cite{Keimer2015}.]{\textbf{Schematic phase diagram of the hole-doped cuprates, as a function of hole doping $p$ and temperature $T$}~\cite{Keimer2015}.
The blue region indicates antiferromagnetic (AF) order, which onsets below the Néel temperature $T_N$, whereas the green region indicates superconducting order of $d_{x^2-y^2}$-wave symmetry (d-SC), which onsets below $T_c$.
There is a smooth crossover from Fermi liquid behavior at high doping (white) to strange metal behavior above optimal doping (purple).
The pseudogap phase (yellow-brown) develops below $T^{*}$ in a fairly sharp crossover.
Within the pseudogap region, there are competing orders which are, namely, charge-density waves (CDW) and spin-density wave (SDW).
They become fully developed below $T_{\text{CDW}}$ (red stripes) and $T_{\text{SDW}}$ (green stripes), respectively.
$T_{\text{SC, onset}}$, $T_{\text{C, onset}}$, and $T_{\text{S, onset}}$ next to the red and green dashed lines indicate the onset of superconducting, charge, and spin fluctuations, respectively.
Indications of a quantum-critical point (QCP), denoted with a purple dot, at which the pseudogap phase becomes a Fermi liquid are experimentally seen by suppressing the superconducting state with a strong magnetic field~\cite{Proust2019}.
See text for further discussion.
Reproduced with editing from Ref.~\cite{Keimer2015}, with permission from Springer Nature.}
\label{fig:cuprate-phase-diagram}
\end{figure}

\subsection{Phase diagram of hole-doped cuprates}
There are many experimental knobs one may use to tune the cuprates: pressure~\cite{Mark2022}, uniaxial or shear stress, magnetic fields~\cite{Proust2019}, temperature, and chemical composition or doping.
The two most important ones are temperature and doping~\cite{Leggett2006, Keimer2015}.
A schematic temperature vs.\ doping phase diagram of the hole-doped cuprates is shown in Fig.~\ref{fig:cuprate-phase-diagram}, reproduced from Ref.~\cite{Keimer2015}.
The hole doping $p$ is defined so that $1 + p$ is the number of holes in the \ce{CuO2} planes per copper atom.
Electron-doped cuprates will not be discussed here or studied later on.
Although closely related to the stoichiometric doping levels (usually denoted with $x$ or $\delta$), the relation to $p$ is not necessarily the simple linear one that follows from the naive chemical valencies of \ce{A} and \ce{X} because the injected holes can end up elsewhere in the system, and not only in the \ce{CuO2} planes~\cite{Leggett2006}.
Once rescaled to $p$, there is considerable evidence for the universality of the corresponding cuprate phase diagram~\cite{Leggett2006, Keimer2015}, as shown in Fig.~\ref{fig:cuprate-phase-diagram}.
That said, one should keep in mind that, in practice, this diagram is reconstructed from multiple cuprate compounds since no one compound is chemically and structurally stable over the whole range of doping which is of interest~\cite{Leggett2006}.
One should also bare in mind that the precise quantitative features, such as the value of $T_c$, significantly vary with chemical composition in a way that is not completely encapsulated by $p$~\cite{Leggett2006}.
Below we briefly discuss the various phases appearing in Fig.~\ref{fig:cuprate-phase-diagram}.

In the absence of doping ($p = 0$), that is ``at stoichiometry,'' cuprates are magnetically ordered insulators~\cite{Leggett2006, Keimer2015} (blue region in Fig.~\ref{fig:cuprate-phase-diagram}).
The obstruction to electric conduction comes not from a gap at the Fermi level in the band structure, which is not expected for an odd number of electrons per unit cell anyway, but from the strong on-site Coulomb repulsion which localizes the holes to the copper atoms.
The result is a Mott insulating phase~\cite{Leggett2006, Keimer2015, Sachdev2003, Lee2006}.
Although localized, the spins of the holes are still active degrees of freedom and they interact via virtual hopping processes (``superexchange'').
Since Pauli's principle forbids parallel spin exchange, whereas anti-parallel spin exchange is only suppressed by the Coulomb repulsion, the effective spin-spin interaction favors opposite spins and is antiferromagnetic.
In agreement with these considerations, neutron diffraction and other studies show that at $p = 0$ cuprates have antiferromagnetic Néel long-range order in which neighboring copper elements have oppositely oriented spins~\cite{Leggett2006, Keimer2015, Sachdev2003, Lee2006}.
This antiferromagnetic phase persists up to room temperature, with little variation in the Néel transition temperature $T_N \sim \SI{300}{\kelvin}$ between compounds~\cite{Leggett2006}.

With the injection of holes into the \ce{CuO2} planes via doping, the holes become more mobile and through their interactions a variety of other phases appear~\cite{Keimer2015}.
At low temperatures, we have the prominent SC phase (green region in Fig.~\ref{fig:cuprate-phase-diagram}) which spans a SC dome ranging from $p_{\text{min}}$ to $p_{\text{max}}$.
In between somewhat closer to $p_{\text{max}}$, the SC transition temperature $T_c$ attains a maximum at the so-called optimal doping $p_{\text{opt}}$ which is then used to orient oneself within the diagram.
Samples with $p < p_{\text{opt}}$ are called ``underdoped,'' those with $p \sim p_{\text{opt}}$ are called ``optimally doped,'' and those with $p > p_{\text{opt}}$ ``overdoped.''
As we now increase the temperature, underdoped samples enter the pseudogap phase (yellow-brown region in Fig.~\ref{fig:cuprate-phase-diagram}), optimally doped samples enter the strange metal phase (purpler region in Fig.~\ref{fig:cuprate-phase-diagram}), and overdoped samples which do not exceed $p_{\text{max}}$ enter a Fermi liquid phase (white region in Fig.~\ref{fig:cuprate-phase-diagram}).
Although the transitions at $T_N$ and $T_c$ are sharp, the change from the Fermi liquid to the strange metal is a broad crossover, whereas the change from the strange metal to the pseudogap phase at $T^{*}$ is a sharp crossover~\cite{Keimer2015, Tallon2001, Proust2019} which some interpret as a transition~\cite{Chakravarty2001, Varma2020}.

Even though the origin of cuprate SC is hotly debated, a few of its properties are agreed upon~\cite{Leggett2006, Keimer2015}.
There is a large body of experimental evidence which shows that the SC state is an even-parity singlet state which in the case of tetragonal systems has $d_{x^2-y^2}$ symmetry~\cite{Wollman1993, VanHarlingen1995, Tsuei2000} and which for weakly orthorhombic systems is dominated by the $d_{x^2-y^2}$ component~\cite{Kirtley2006, Tsuei2000}.
Its phenomenology is well-described using Ginzburg-Landau theory~\cite{Leggett2006}, albeit with the noted difference that, unlike for classical SC, the Cooper pairs are a lot smaller and phase fluctuations play a much more important role~\cite{Emery1995, Carlson2008}.
Finally, there is considerable evidence~\cite{Campuzano1996, Damascelli2003, Sobota2021, Wei1998, Petrov2007, Zhou2019} that well-defined Bogoliubov quasi-particles are present in the SC state.
The normal-state quasi-particles, however, may or may not be well-defined, depending on the doping.

At overdoping, numerous experiments indicate a conventional Fermi liquid normal state~\cite{Leggett2006, Keimer2015, Proust2019}.
Thermodynamic and transport measurements exhibit the expected Fermi liquid behavior~\cite{Kubo1991, Manako1992, Mackenzie1996, Proust2002, Nakamae2003, Hussey2003}, as do magneto-oscillation experiments~\cite{Vignolle2008, Bangura2010} and angle-resolved photoemission spectroscopy (ARPES)~\cite{Damascelli2003, Plate2005, Yoshida2006, Peets2007, Horio2018}.
Moreover, the overdoped normal state is well-described by density functional theory~\cite{Kramer2019}.

Near optimal doping, the normal state is a ``strange metal'' whose most fundamental feature is the absence of quasi-particles~\cite{Keimer2015}.
This has far-reaching implications for its phenomenology, perhaps the most striking of which is the linear in temperature resistivity which spans from as low to as high temperatures as one can measure~\cite{Proust2019, Phillips2022}.
At low $T$, very general phase space considerations of electron-electron scattering processes show that the resistivity should increase $\propto T^2$ for a system with well-defined quasi-particles, as in a Fermi liquid~\cite{Kittel2005, Behnia2022}.
At $T$ comparable to the Debye temperature, the dominant contribution to the resistivity is electron-phonon scattering which is linear in $T$~\cite{Kittel2005}.
However, the resistivity should saturate at high $T$ once the electron's mean free path becomes comparable in magnitude to the de~Broglie wavelength; this is the so-called Mott-Ioffe-Regel limiting~\cite{Hussey2004}.
The linear in $T$ resistivity, sometimes stretching up to three orders of magnitude in temperature, is thus one of the big mysteries that has attracted much scholarly attention~\cite{Varma2016, Varma2020, Phillips2022, Hartnoll2022}, for it not only arises in the cuprates.

In this context, high magnetic fields have been used to suppress the SC phase and uncover the low-$T$ normal state near optimal doping to great effect, as nicely reviewed in Ref.~\cite{Proust2019}.
The picture that is emerging is that at $T = 0$ the pseudogap phase ends at a near-optimal $p = p^{*}$ in a sharp transition~\cite{Tallon2001, Proust2019}.
In the absence of a magnetic field this transition is shielded by a SC dome~\cite{Tallon2001, Proust2019} and for finite $T = T^{*}$ it turns into a sharp crossover~\cite{Proust2019}.
The main evidence supporting that the $T = 0$ phase transition is second-order is the divergence of the heat capacity~\cite{Proust2019}.\footnote{It would be desirable if this heat capacity divergence as a function of doping could be reproduced in additional cuprate compounds which do not have a Van Hove singularity near optimal doping.}
However, the observation of a diverging length is missing and the identification of the order which underlies the pseudogap phase has not been settled~\cite{Proust2019}.

In addition to the proposal, mentioned at the beginning of this chapter, that intra-unit-cell loop currents underlie the pseudogap phase~\cite{Varma1997, Varma2020}, $d$-density wave~\cite{Chakravarty2001, Laughlin2014, Laughlin2014-details},\footnote{$d$-density waves are the same thing as staggered loop currents, usually with a $\vb{Q} = (\pi, \pi)$. Sometimes they are also called staggered flux states~\cite{Affleck1988, Marston1989} or orbital antiferromagnets~\cite{Schulz1989}.} oxygen orbital moment~\cite{Moskvin2012}, spin magnetoelectric~\cite{Orenstein2011, Lovesey2015, Fechner2016}, nematic~\cite{Nie2014}, topological spin liquid~\cite{Sachdev2003}, and many other orders~\cite{Kivelson2003, Chowdhury2015, Kivelson2019} have been suggested as well.
For a concise review, see the introduction of Ref.~\cite{Volkov2018}; see also Sec.~\ref{sec:LC-proposals}.

The pseudogap phase is characterized by three main experimental signatures~\cite{Timusk1999, Sadovskii2001, Tallon2001, Proust2019}: the opening of a gap at the Fermi surface near the anti-nodal regions $(k_x, k_y) \approx (\pi, 0)$ and $(0, \pi)$, the retention of a well-defined Fermi surface (``Fermi arcs'') at the nodal diagonal regions intersecting $k_x = \pm k_y$, and a carrier density $n = p$ which abruptly changes to $n = 1 + p$ at higher doping.
The first two are most directly seen in ARPES~\cite{Damascelli2001, Damascelli2003, Vishik2018, Sobota2021} and the name ``pseudogap'' derives from this partial opening of a gap, while the last is inferred from Hall effect measurements~\cite{Proust2019}.
The pseudogap is also seen in optical conductivity measurements, scanning tunneling microscopy, and electronic Raman spectroscopy~\cite{Timusk1999}.
The pseudogap regime is also marked by various competing orders whose interplay is not completely understood~\cite{Keimer2015}.
The most prominent among these orders are charge-density waves~\cite{Chen2016, Frano2020}, which are denoted with red stripes in Fig.~\ref{fig:cuprate-phase-diagram}.

\section{Previous experimental and theoretical work} \label{sec:cuprate-LC-lit-review}
In Sec.~\ref{sec:general-LC-work-review} of the previous chapter, we have reviewed the theoretical and experimental literature on orbital magnetism and loop currents in general systems.
Here we continue this discussion, focusing on the cuprate superconductors in which loop currents have arguably drawn more attention~\cite{Chakravarty2001, Varma2020, Bourges2021} than in all other systems combined.

We start by reviewing the experimental literature on symmetry-breaking in the pseudogap regime.
After that, in Sec.~\ref{sec:LC-proposals}, we discuss theoretical proposals in which some type of LC order is put forward as the hidden order underlying the pseudogap regime.
We end with an overview of microscopic theoretical calculations which investigated whether LC order is a competitive instability that could arise in a realistic model of the cuprates.

\subsection{Experiments on symmetry-breaking in the pseudogap regime} \label{sec:exp-sym-break-pseudogap}
As already noted in Sec.~\ref{sec:general-LC-work-review}, LC order, although it takes place in the orbital sector, is measurable using spin probes because they are sensitive to local magnetic fields and time-reversal symmetry-breaking (TRSB), regardless of origin~\cite{Bourges2021}.

The earliest evidence supporting TRSB in the pseudogap phase of the hole-doped cuprates comes from a spin-polarized neutron diffraction (PND) study of \ce{YBa2Cu3O_{6+x}} performed in the group of Philippe Bourges~\cite{Fauque2006}.
Their main finding is that the rate of neutron spin-flipping changes at the $\vb{Q} = (0,1,1)$ Bragg peak below $T^{*}$, indicating a TRSB state which preserves the translation symmetries of the lattice.
Later PND studies performed on a number of other cuprate compounds~\cite{Bourges2008, Bourges2011, Bourges2015, Bourges2017, Greven2008, Greven2010, Greven2011, Bourges2018, Bourges2010, Bourges2012, Laffez-Monot2014}, mostly carried out by the groups of Bourges and Greven using similar methodology, have reproduced and extended this finding.
However, an independent group led by Stephen M.\ Hayden using different methodology has found no change in the spin-flipping rate at the same $\vb{Q} = (0, 1, 1)$ Bragg peak below $T^{*}$~\cite{Croft2017}.
In the ensuing dispute~\cite{Croft2018-comment, Croft2018-reply, Bourges2019} two notable points of contention are whether the data of Ref.~\cite{Croft2017} has the necessary statistics to resolve the effect and whether the measurement protocol and background subtraction procedure of Ref.~\cite{Fauque2006} results in spurious temperature-dependent drifts in the signal.
Although all the just cited studies reported no evidence of translation symmetry-breaking, very recently two PND studies have found a short-ranged in-plane TRSB order with a commensurate $\vb{q} = (\pi, 0) \cong (0, \pi)$ below $T^{*}$~\cite{Bounoua2022, Bounoua2023}.

If taken at face value, the PND measurements (excluding Ref.~\cite{Croft2017}) suggest local magnetic moments per unit cell that are on the order of $0.1$ Bohr magnetons~\cite{Bourges2011-review}.
Local magnetic moments of this size should be measurable using nuclear magnetic resonance (NMR)~\cite{Lederer2012}, as well as muon spin spectroscopy~\cite{Zhang2018}.
Numerous NMR and nuclear quadruple resonance experiments have been carried out through the years~\cite{Strassle2008, Kawasaki2010, Wu2011, Strassle2011, Mounce2013, Wu2015} and they do not find any evidence for such local moments.
A possible interpretation is that these local magnetic moments fluctuate slowly enough in time to appear static when probed by neutrons, but that they average out to zero on the longer time-scales which are probed by NMR~\cite{Wu2015, Zhu2021}.
If true, this interpretation would suggest that the putative order underlying the pseudogap phase is not a genuine static order.

Muon spin relaxation ($\mu$SR) measurements with zero and longitudinal magnetic fields have also been used to probe TRSB, with mixed results.
One $\mu$SR experiment does find a signal at the expected $T^{*}$~\cite{Zhang2018}, indicating TRSB in agreement with PND, another finds a signal at a different temperature~\cite{Sonier2009}, and the remaining~\cite{MacDougall2008, Huang2012, Pal2016} do not find any indications of TRSB.
Concerns were raised~\cite{Zhang2018-comment} and responded to~\cite{Zhang2018-reply} regarding the positive results of Ref.~\cite{Zhang2018}, and a later study~\cite{Gheidi2020} investigated whether certain assumptions that enter the analysis of Ref.~\cite{Zhang2018} hold at overdoping, as a reference.
In particular, they find that the nuclear-dipole field is temperature-dependent~\cite{Gheidi2020}, which casts some doubt on the conclusions of Ref.~\cite{Zhang2018}, although the longitudinal magnetic field might be strong enough for the nuclear dipolar relaxation to decouple~\cite{Zhu2021}.
The same group thus used stronger magnetic fields in a recent experiment~\cite{Zhu2021} in which they reproduced the finding of Ref.~\cite{Zhang2018} that there are slow magnetic fluctuations present in \ce{YBa2Cu3O_y}, albeit with relaxation rates whose $T$-dependence below $T^{*}$ is unusual when compared to conventional magnetic orders.
We refer the interested reader to Ref.~\cite{Gheidi2022} for an accessible discussion of $\mu$SR technicalities and further discussion of the just-mentioned $\mu$SR experiments of cuprates.

The magneto-optic Kerr effect is another experimental probe capable of observing TRSB~\cite{Kapitulnik2009}.
A series of experiments have been performed on cuprates which all found a non-zero Kerr rotation~\cite{Xia2008, Kapitulnik2009, He2011, Karapetyan2012, Karapetyan2014}, indicating TRSB.
However, the signal onsets at a distinct temperature $T_K$ below $T^{*}$ which is close to the charge ordering temperature.
The observed Kerr effect is also unusual because a magnetic field cannot be used to flip the sign of the Kerr rotation angle and because opposite surfaces of the same crystal have the same Kerr rotation sign~\cite{Karapetyan2014}.
These unusual features motivated a number of proposals that a novel gyrotropic order which preserves time-reversal symmetry is the explanation~\cite{Mineev2010, Pershoguba2013, Hosur2013}, but later work established that TRSB is necessary~\cite{Armitage2014, Fried2014, Kapitulnik2015, Cho2016} which lead to retractions~\cite{Mineev2010-E, Pershoguba2013-E, Hosur2013-E}.
Intra-unit-cell loop currents~\cite{Aji2013, Yakovenko2015} are a possible explanation, but so are other TRSB orders~\cite{Orenstein2011, Orenstein2013, YuxuanWang2014-Kerr, Gradhand2015, Lovesey2015, Sarkar2019}.
Let us also mention an early observation of circular dichroism in the pseudogap regime~\cite{Kaminski2002}.
This is indicative of TRSB~\cite{Kaminski2002}, although alternative interpretations exist~\cite{Armitage2004, Borisenko2004, Arpiainen2009, Arpiainen2009-reply} which were disputed~\cite{Kaminski2002-reply, Campuzano2004, Arpiainen2009-comment}.

In addition to time reversal, many studies have investigated whether any point group symmetries are broken at $T^{*}$ as one enters the pseudogap regime.\footnote{Since the experiments that I list here are sometimes used to define $T^{*}$ proper, let me just note that the approximate value of $T^{*}$ can be independently determined from the crossovers seen in ARPES and tunnelling spectroscopy~\cite{Timusk1999, Tallon2001}, for instance.}
A sharp feature indicating a transition was observed in resonant ultrasound spectroscopy~\cite{Shekhter2013}.\footnote{They were unable to deduce any information on symmetry-breaking in Ref.~\cite{Shekhter2013}, however.}
Torque magnetometry measurements found a cusp in the $T$-dependence of the in-plane anisotropy~\cite{Sato2017-cuprate, Murayama2019}, strongly suggesting a nematic transition at $T^{*}$ which breaks the four-fold in-plane rotation symmetry $C_4$.
Multiple studies observed that a large in-plane anisotropy develops in the Nernst effect at $T^{*}$~\cite{Daou2010, Chang2011, Doiron-Leyraud2013, Cyr-Choiniere2018}, supporting a $C_4$-nematic transition.
A terahertz polarimetry experiment found evidence for the breaking of both $C_4$ and mirror symmetries at $T^{*}$~\cite{Lubashevsky2014}, while an optical second-harmonic generation experiment reported evidence for the breaking of spatial inversion and two-fold rotation symmetries~\cite{Zhao2017}.\footnote{There have also been second-harmonic generation measurements at zero doping which found evidence for mirror symmetry-breaking above the Néel temperature~\cite{Torre2021}.}
Finally, elastoresistance measurements also corroborate two-fold rotation symmetry-breaking at $T^{*}$~\cite{Ishida2020-cuprate}.
Thus most studies support nematicity in the pseudogap phase, with the exception of one study which found no signatures of in-plane anisotropy in the resistivity or the Seeback coefficient at $T^{*}$~\cite{Grissonnanche2023}.
One point of debate~\cite{Cooper2014, Tallon2020} is whether the transition at $T^{*}$ is a genuine mean-field thermodynamic one, as suggested in Refs.~\cite{He2011, Shekhter2013, Sato2017-cuprate}, because this would normally entail a specific heat anomaly which, despite an intensive search, has not been observed~\cite{Loram1993, Loram2001, Cooper2014, Tallon2020}.
The Ashkin-Teller/XY model nature of the transition to intra-unit-cell LC order has been suggested~\cite{Gronsleth2009, Varma2015} to explain why the transition is more easily observable in ultrasound than in heat capacity measurements.

In summary, the experimental evidence on TRSB in the pseudogap phase is mixed.
PND finds it, NMR does not, $\mu$SR is mixed, and polar Kerr measurements find something, but at a $T_K < T^{*}$ and with a number of strange properties~\cite{Karapetyan2014}.
Regarding nematicity, there is strong evidence that it sets in below $T^{*}$, while spatial-inversion symmetry-breaking is supported primarily by one study~\cite{Zhao2017}.

\begin{figure}[t]
\centering
\includegraphics[width=\textwidth]{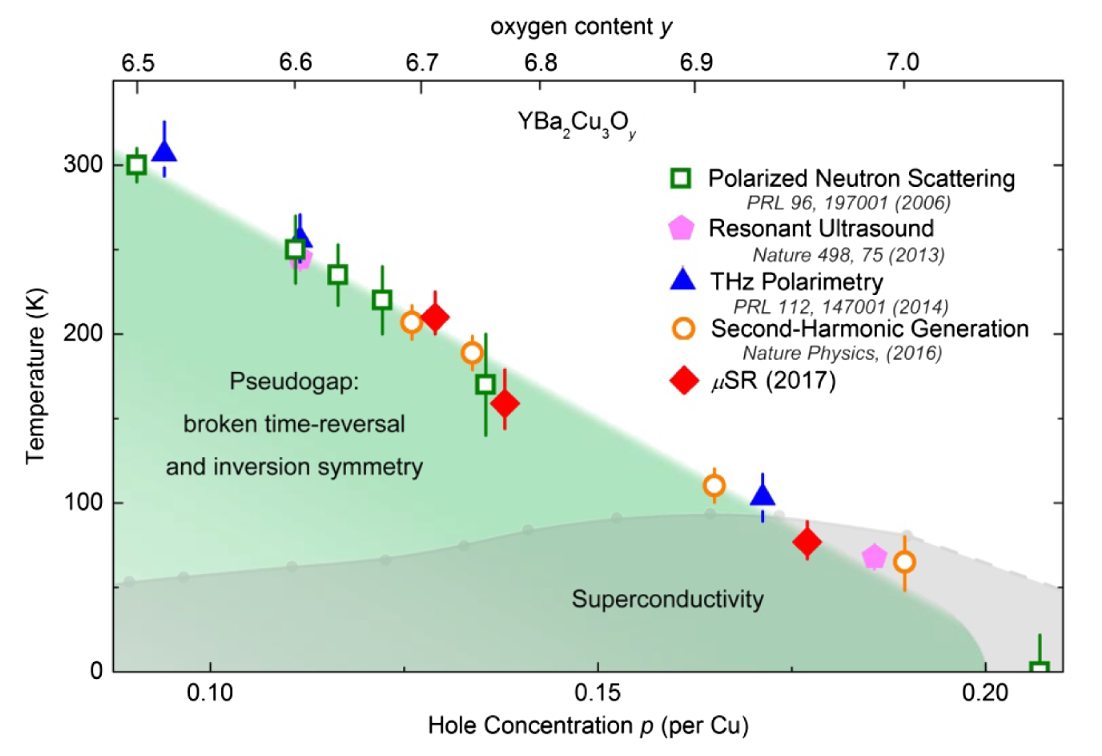}
\captionbelow[Selected experimental evidence on the nature of the pseudogap phase of hole-doped cuprates~\cite{Varma2020}.]{\textbf{Selected experimental evidence on the nature of the pseudogap phase of hole-doped cuprates}~\cite{Varma2020}.
Polarized neutron scattering points are from Ref.~\cite{Fauque2006}, resonant ultrasound measurements are from Ref.~\cite{Shekhter2013}, terahertz polarimetry data is from Ref.~\cite{Lubashevsky2014}, second-harmonic generation points are from Ref.~\cite{Zhao2017}, and muon spin relaxation ($\mu$SR) data is from Ref.~\cite{Zhang2018}.
See also Fig.~3 of Ref.~\cite{Zhang2018} for a similar plot and the discussion in the text regarding evidence (not shown) which is at odds or complicates the interpretation of the pseudogap as a state of broken time-reversal and space-inversion symmetry.
Reprinted with permission from Ref.~\cite{Varma2020}. Copyright (2020) by the American Physical Society.}
\label{fig:Varma-evidence}
\end{figure}

\subsection{Loop-current proposals for the pseudogap phase} \label{sec:LC-proposals}
One possible interpretation of this state of affairs, advocated by Varma~\cite{Varma1997, Simon2002, Simon2003, Varma2006}, is that the pseudogap regime is an ordered LC phase of broken time-reversal and space-inversion symmetries that preserves lattice translation symmetry.
Later this was revised~\cite{Varma2014} to include order parameter fluctuations on time-scales shorter than those measured by NMR and zero-field $\mu$SR, to avoid conflict with the latter two null-results.
More recently, a small degree of lattice translation symmetry-breaking was also included in the proposal to explain the pseudogap and Fermi arcs observed in ARPES~\cite{Varma2019}.
The remaining evidence can then be fit reasonably well~\cite{Simon2003, Shekhter2009, Yakovenko2015}, with the phase diagram as shown in Fig.~\ref{fig:Varma-evidence}, reproduced from Ref.~\cite{Varma2020}.
The final proposal is reviewed in Ref.~\cite{Varma2020}.

In a conceptually similar proposal, Chakravarty, Laughlin, et al.~\cite{Chakravarty2001} have also put forward odd-parity loop currents as the underlying order of the pseudogap phase, with the main difference being that they are staggered with a finite ordering wavevector $\vb{Q} = (\pi, \pi)$.
This staggered LC order can be understood as a $d_{x^2-y^2}$ density wave~\cite{Nayak2000}, which is the original and most common way of referring to it.
In light of even earlier work~\cite{Affleck1988, Marston1989, Schulz1989} which found such a state in the Hubbard and Heisenberg-Hubbard models, $d$-density waves are sometimes also called orbital antiferromagnets or staggered flux order.
This proposal has been further developed in a number of later articles~\cite{Nayak2002, Dimov2008, Hsu2011, Chakravarty2011, Laughlin2014, Laughlin2014-details, Makhfudz2016} of which I would like to highlight the two by Laughlin~\cite{Laughlin2014, Laughlin2014-details} as good reviews.

For a long time, the main evidence supporting intra-unit-cell LCs over staggered ones was the absence of translation symmetry-breaking observed in PND (see previous section).
Recently, however, two PND studies found $\vb{Q} = (\pi, 0) \cong (0, \pi)$ TRSB order~\cite{Bounoua2022, Bounoua2023}, which could be taken as evidence for staggered LCs.
Although most microscopic studies find staggered LCs with $\vb{Q} = (\pi, \pi)$, if they find any (see next section), one study following~\cite{Laughlin2014, Laughlin2014-details} actually found $\vb{Q} = (\pi, 0) \cong (0, \pi)$ as the preferred ordering~\cite{Makhfudz2016}.
In any case, the difficulty in observing a $\vb{Q} = (\pi, \pi)$ Bragg peak could be due to pseudogap glassiness~\cite{Laughlin2014, Laughlin2014-details} and the agreement of the proposal with experiment is as reasonable, broadly speaking, as that of intra-unit-cell LCs.

Finally, let us also mention one unrelated paper~\cite{Moskvin2012} in which oxygen orbital moments were considered as an alternative to loop currents between orbitals.
Chiral nematics~\cite{YuxuanWang2014-Kerr} and imaginary charge-density waves~\cite{Gradhand2015} have also been discussed in the context of pseudogap physics.
Both are orbital orders which are symmetry-wise similar to LCs, but microscopically different.

Although these proposals are interesting and consistent with a significant portion of experiments, it is worth pointing out that, even if we take for granted that TRSB takes place, there is currently no experimental evidence which could tells us whether orbital magnetism (LC order) is preferred as an explanation over some type of conventional spin-magnetism, or torroidal spin-magnetism if we accept space-inversion symmetry-breaking.
Moreover, even if LC order is present, it could simply be a ``passenger'' accompanying other orders, instead of a ``driver'' generating the superconductivity and strange metal behavior, as proposed by Varma~\cite{Varma2020} and Chakravarty, Laughlin, et al.~\cite{Chakravarty2001, Laughlin2014, Laughlin2014-details}.
The fluctuations of both pair-density waves~\cite{Agterberg2015} and bond-density waves~\cite{Sarkar2019} have been shown to induce subsidiary orbital loop currents, for example.
Orbital LCs have also been found to emerge in models where the pseudogap parent phase is a spin liquid~\cite{Scheurer2018}, and even for simple spin-magnetic orders spin-orbit coupling is expected to induce loop currents, as has been found in iron-based superconductors~\cite{Klug2018}.
Needless to say, there are numerous other interpretations of the experimental evidence.
For instance, both spin magnetoelectric order (torroidal spin-magnetism)~\cite{Orenstein2011, Lovesey2015, Fechner2016} and (fluctuating) pair-density waves~\cite{Lee2014, Agterberg2015, Chakraborty2019, Agterberg2020} are similar to the proposed LC orders in terms of symmetries and are therefore difficult to tell apart experimentally.
Microscopically, however, the two are very different from LCs, with the former taking place in the spin sector and the latter in the particle-particle sector.

\subsection{Loop currents in microscopic models of cuprates} \label{sec:LC-cuprate-micro}
The cuprates are microscopically most often modeled using repulsive Hubbard models~\cite{Dagotto1994, Keimer2015}, possibly including extended interactions, hoppings, and multiple orbitals.
These models have been the subject of extensive research~\cite{Arovas2022, Qin2022}, in no small part because of their potential relevance to high-temperature SC.
Moreover, for the reasons discussed in Sec.~\ref{sec:cuprate-structure}, the vast majority of theoretical work has focused on the copper oxide planes.
The \ce{CuO2} planes are most commonly modeled using the one-band Hubbard model~\cite{Zhang1988, Eskes1988}, which is based on the \ce{Cu}:$3d_{x^2-y^2}$ orbital, and the three-band extended Hubbard model~\cite{Varma1987, Emery1987, Emery1988, Littlewood1989, Gaididei1988, Scalettar1991}, which in addition includes the ligand \ce{O}:$2p_{x,y}$ orbitals; see also Sec.~\ref{sec:cuprate-3band-model} and Sec.~\ref{sec:CuO2-Hubbard-decompose}.
The $t$-$J$~\cite{Chao1978} and Hubbard-Heisenberg~\cite{Affleck1988, Marston1989} models, obtained from a strong-coupling expansion of the one-band Hubbard model, are also often employed.
When the two-dimensional models of interest are out of reach, in the hope of gleaning some insight into their physics one frequently resorts to studying one-dimensional chains, ladders, cylinders, etc., which are amenable to Bethe Ansatz, bosonization, and other controlled techniques~\cite{Carlson2008, Arovas2022}.
Below we review the work on these models in which loop currents (LCs) have been in some way addressed.

In an early study from 1988, Affleck and Marston~\cite{Affleck1988, Marston1989} investigated the one-band Heisenberg-Hubbard model within a large-$N$ expansion, where $N$ is the number of spin components.
Among the possible orders, they found a competitive staggered flux order with a $\vb{Q} = (\pi, \pi)$, also called an orbital antiferromagnet, $d$-density wave, or staggered LC order.
Within a weak-coupling treatment of the one-band Hubbard model at half-filling, Schulz~\cite{Schulz1989} also found a competitive staggered LC order whose consequences were explored on the mean-field level in Ref.~\cite{Nersesyan1989}.
The appearance of this staggered LC order, later proposed by Chakravarty et al.~\cite{Chakravarty2001} for the pseudogap, has in fact been investigated in many more theoretical studies than the intra-unit-cell LC order proposed by Varma~\cite{Varma2020}.

It is worth recalling that during these early years it was not immediately clear that cuprates are spin antiferromagnets at stoichiometry, as opposed to something more exotic like a resonating valence bond state~\cite{Anderson1987resonating} or orbital antiferromagnet.
The early work on staggered LCs~\cite{Kotliar1988, Lederer1989, Zhang1990, Poilblanc1990, Liang1990, Lee1990, Hsu1991, Ogata1991, Ubbens1992} thus chiefly focused on whether they are the ground state of the (one-band) $t$-$J$ model at underdoping.
Unsurprisingly, for the ground state they by and large found spin antiferromagnetism and $d_{x^2-y^2}$-wave superconductivity, although the numerical studies were significantly limited by the then-available technology.
Later studies of the $t$-$J$ model, both numerical and analytic, explored whether it could be the normal state of the pseudogap~\cite{Yokoyama1996, Wen1996, Lee1998, Cappelluti1999, Hamada2003, Ivanov2003, Weber2006}, mostly finding that staggers LCs are competitive at underdoping for realistic parameters.
Motivated by STM experiments, related work~\cite{Lee2001, Kishine2001, Wang2001, Ivanov2003, Tsuchiura2003, Kuribayashi2003, Weber2006} examined whether these subleading staggered LCs could emerge in vortex cores of the mixed SC state.

The majority of studies on the one-band two-dimensional Hubbard model find that staggered LCs with $\vb{Q} = (\pi, \pi)$ are competitive and sometimes prevail for realistic parameter values at underdoping.
Staggered LCs were found using a variety of methods, including weak-coupling mean-field~\cite{Nayak2002, Hankevych2002, Hankevych2003, Normand2004, Hirose2019}, perturbative renormalization group~\cite{Honerkamp2002}, and strong-coupling~\cite{Stanescu2000, Stanescu2001} analytic approaches, but also numeric approaches based on the Hartree-Fock approximation augmented with Gutzwiller projection factors~\cite{Normand2004}, Gutzwiller-projected variational minimization~\cite{Allais2014}, variational cluster approximation~\cite{Lu2012-Hub}, and variational Monte Carlo~\cite{Yokoyama2016}.
A tendency towards staggered LC order was also found using dynamical cluster approximation~\cite{Macridin2004} and dynamical mean-field theory~\cite{Otsuki2014}.
However, the last two studies found no divergence in the susceptibility, suggesting short-range order.
For a complementary literature review on staggered LCs in one-band models, we refer the reader to Yokoyama et al.~\cite{Yokoyama2016}.

Staggered LCs do not require oxygen orbitals, which is why there have not been many studies searching for them in the three-band Hubbard model.
Mean-field Hartree-Fock~\cite{Atkinson2016} and generalized random phase approximation~\cite{Bulut2015} calculations including three bands both find staggered LCs to be competitive at underdoping, while variational cluster approximation~\cite{Lu2012-Hub} and variational Monte Carlo~\cite{Weber2014} calculations find the opposite.

When it comes to intra-unit-cell (IUC) LCs, there have been only a few studies, all of them dealing with the three-band Hubbard model.
Early mean-field studies by Varma et al.~\cite{Simon2002, Weber2009} showed that the LC order they dubbed $\Theta_{\text{II}}$ can be stabilized for strong enough \ce{Cu-O} nearest-neighbor repulsion.
This $\Theta_{\text{II}}$ LC order, which is odd under parity and transforms like an in-plane vector, is the same one Varma proposes for the pseudogap~\cite{Varma2020}.
Below, by IUC LC order we always mean $\Theta_{\text{II}}$ LC order, unless explicitly stated otherwise.
In our terminology, we would call $\Theta_{\text{II}}$ LC order $p$-wave IUC LC order.
Later the mean-field phase diagram was more thoroughly investigated by Fischer and Kim~\cite{Fischer2011}, who again find that IUC LC order is viable alongside nematic and spin-nematic IUC orders.
However, staggered LCs were found to always prevail over IUC LCs in a generalized random phase approximation study~\cite{Bulut2015}.

Exact diagonalization of \ce{Cu8O16} and \ce{Cu8O24} clusters (the latter includes apical oxygens) with periodic boundary conditions at zero temperature~\cite{Greiter2007, Thomale2008, Kung2014} and determintental quantum Monte Carlo performed on \ce{Cu16O32} and \ce{Cu36O72} clusters at high temperatures ($\sim \SI{1000}{\kelvin}$)~\cite{Kung2014} found no tendency towards $\Theta_{\text{II}}$ IUC LC order in the current-current correlation functions at underdoping.
This remained the case even when the model parameters were tuned to be favorable for IUC LC order, as suggested by mean-field theory.
Spin-antiferromagnetism is clearly observed near stoichiometry in these numerical studies~\cite{Greiter2007, Thomale2008, Kung2014}, thus confirming their sensitivity to ordering.
In contrast, variational Monte Carlo studies on clusters ranging from \ce{Cu16O32} to \ce{Cu64O128} with open~\cite{Weber2009} and periodic~\cite{Weber2014} boundary conditions reported that IUC LC order is very competitive and that it is the ground state in a significant portion of the phase diagram.
The appropriateness of their variational procedure was checked by comparing against exact diagonalization of a \ce{Cu8O16} cluster~\cite{Weber2009, Weber2014}.
Notably, the stabilization of IUC LCs required physically-motivated modifications of the standard three-band Hubbard model.
The stabilization of IUC LCs was aided by apical oxygens in the first study~\cite{Weber2009}, while in the second study the next-nearest oxygen hopping $t_{pp}'$, mediated by the \ce{Cu}:$4s$ orbital, played this role~\cite{Weber2014}.

One-dimensional chains and ladders~\cite{Nersesyan1991, Orignac1997, Tsuchiizu2002, Fjaerestad2002, Schollwock2003, Fjarrestad2006, Chudzinski2008, Nishimoto2009, Nishimoto2010}, sometimes including the oxygen orbitals, have also been studied in this context, with mixed results regarding the competitiveness of intra-unit-cell and staggered LCs.

In summary, staggered LCs are competitive in both the $t$-$J$ and one-band Hubbard models in two dimensions, but it is not clear whether they set in as a long-range order.
In the three-band extended Hubbard model, the results are mixed regarding both staggered and intra-unit-cell LCs.

While these microscopic investigations are important and interesting, it is worthwhile to ponder what can they actually tell us about the correct theory of cuprates.
If one robustly finds an order which agrees with experiment, as is the case for antiferromagnetism near $p = 0$, then this supports the notion that the one-band and three-band Hubbard models capture at least some of the essential physics of cuprates.
Going from there, there is a broad range of effective interaction parameters that are physically reasonable and that give antiferromagnetism at half-filling.
Within this broad range, one can apparently stabilize many ordered states away from half-filling, including LC states.
Given how no single order is robustly favored away from stoichiometry (apart from $d_{x^2-y^2}$-wave superconductivity), and how this sensitively depends on what we put into the model (and what we \emph{should} put into the model is hotly debated), it seems that the main thing one can deduce from these studies is whether a certain order is a reasonable candidate for the pseudogap.
In this regard, loop currents, both staggered and intra-unit-cell, are viable candidates.
Beyond this, it is difficult to see how further theoretical work along these directions could settle whether LC order arises at underdoping.
There are simply too many free tuning parameters, not to mention the lack of reliable methods for intermediate coupling.
At best, it could help clarify the microscopic mechanism which underlies LC order and whether the same (or related) mechanisms could play a role in driving SC or strange metal behavior.
Addressing this last question is quite challenging within microscopic approaches, however.

More phenomenological approaches, in which one starts from a higher-level description, hold the promise of being able to shed more light on these issues.
Such approaches have been fruitfully applied to the problem of Cooper pairing due to quantum-critical order-parameter fluctuations (Sec.~\ref{sec:QCP-paradigm}, Chap.~\ref{chap:loop_currents}), and the current work, based on Ref.~\cite{Palle2024-LC}, continues this tradition.
As was explained in Sec.~\ref{sec:QCP-genera-final-analysis} of the previous chapter, in our analysis we have not attempted to derive LC order microscopically.
Instead, we have adopted a phenomenological approach and identified from the start the pseudogap phase with LC order.
This allowed us to clarify whether LC fluctuations are an effective pairing glue near their quantum-critical point.
Within a weak-coupling analysis coming from the Fermi liquid regime (Fig.~\ref{fig:QCP-calc-strategy}), we found that even-parity IUC LCs are ineffective at driving pairing, while odd-parity IUC LCs act as strong pair-breakers near the QCP (Fig.~\ref{fig:QCP-general-results}).
Thus, without making any assumptions about their microscopic origin, we have shown that odd-parity intra-unit-cell loop currents are not a good candidate for the hidden order of the pseudogap~\cite{Palle2024-LC}.
Staggered loop currents are still viable, as discussed at the end of Sec.~\ref{sec:IUC-order-results}.

There have been a few previous works on loop currents which are similar in spirit to our own work~\cite{Palle2024-LC}.
In particular, the paper by Aji, Shekhter, and Varma~\cite{Aji2010} addressed the same question and came to very different conclusions.
As we explain in detail in Sec.~\ref{sec:comparison-w-Aji2010}, certain assumptions were made in that paper which a careful analysis reveals to be incorrect.
The same limitations apply to later work by Varma~\cite{Varma2012, Bok2016, Varma2016, Varma2020} concerning the pairing due to IUC LCs.
When it comes to staggered LCs, the two papers by Laughlin~\cite{Laughlin2014, Laughlin2014-details} are conceptually similar in the sense that one approaches the phase diagram from the overdoped Fermi-liquid regime, just like in our own work~\cite{Palle2024-LC}, but with the different goal of attempting to reproduce as much of cuprate physics at under- and optimal doping as possible.
Methodologically, the strategy we adopted is the same as the one employed by Lederer et al.~\cite{Lederer2015} to investigate pairing near nematic QCPs.

\section{Electronic structure and the three-orbital model of the copper oxide planes} \label{sec:cuprate-3band-model}
The copper oxide planes are the structural component that is shared by all cuprates (Sec.~\ref{sec:cuprate-structure}) and there are good reasons to think that the essential physics of cuprate SC is contained in these planes, as is commonly believed~\cite{Leggett2006, Keimer2015}.
X-ray absorption studies show that the low-energy electronic states of the cuprates derive from the orbitals which are within the \ce{CuO2} planes~\cite{Chen1991, Chen1992, Pellegrin1993, Nucker1995, Peets2009}, while studies of atomically-thin cuprate monolayers observe undiminished superconductivity~\cite{Uchihashi2017, Gozar2008, Logvenov2009, Yu2019}.
For these reasons, in the remainder of this chapter we study the copper oxide planes.
We do so using the three-orbital tight-binding model~\cite{Varma1987, Emery1987, Emery1988, Littlewood1989, Gaididei1988, Scalettar1991} which is introduced in this section.

\begin{figure}[t]
\centering
\includegraphics[width=\linewidth]{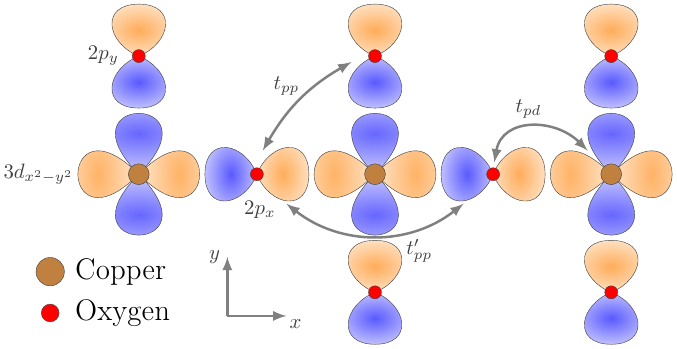}
\captionbelow[The \ce{CuO2} plane of the cuprates and its chemically most active \ce{Cu}:$3d_{x^2-y^2}$ and \ce{O}:$2p_{x,y}$ orbitals.]{\textbf{The \ce{CuO2} plane of the cuprates and its chemically most active \ce{Cu}:$3d_{x^2-y^2}$ and \ce{O}:$2p_{x,y}$ orbitals.}
Orange (blue) are positive (negative) lobes of the orbitals.
Arrows indicate hopping amplitudes we include in the three-orbital tight-binding model~\cite{Varma1987, Emery1987, Emery1988, Littlewood1989, Gaididei1988, Scalettar1991}.}
\label{fig:CuO2-sketch}
\end{figure}

To start, let us recall that the atomic electron configuration of \ce{Cu} is [\ce{Ar}]$4s^{1}3d^{10}$ and of \ce{O} is [\ce{He}]$2s^{2}2p^{4}$.
Keeping in mind that at stoichiometry two electrons are donated to each \ce{CuO2}, this means that (as a first approximation) the two oxygen atoms have filled shells, while the copper atom has the electronic configuration [\ce{Ar}]$4s^{0}3d^{9}$ with a singly-occupied $d_{x^2-y^2}$ orbital~\cite{Leggett2006}.
More accurately, and more generally in the presence of doping, the states closest to the Fermi level primarily derive from anti-bonding hybridization between \ce{Cu}:$3d_{x^2-y^2}$ orbitals and \ce{O}:$2p_{x,y}$ orbitals oriented along the ligands~\cite{Pickett1989, Dagotto1994, Pellegrin1993, Varma1987, Emery1987}.
These orbitals are the basis of the three-band tight-binding model that was first introduced in Refs.~\cite{Varma1987, Emery1987, Emery1988, Littlewood1989, Gaididei1988} and which is shown in Fig.~\ref{fig:CuO2-sketch}.
Between the partially filled anti-bonding band and the filled $3d_{x^2-y^2}$--$2p_{x,y}$ bonding bands, there are additional states coming from the remaining \ce{Cu}:$3d$, \ce{Cu}:$4s$, as well as in-plane and apical \ce{O}:$2p$ orbitals~\cite{Pickett1989, Dagotto1994, Photopoulos2019, Andersen1995, Andersen2001, Pavarini2001, Jiang2020}.
Integrating these states out strongly renormalizes the $t_{pp}$ and $t_{pp}'$ hopping amplitudes, mostly through the $2p_x$--$4s$--$2p_y$ and $2p_x$--$4s$--$2p_x$ virtual processes~\cite{Andersen1995, Andersen2001, Pavarini2001, Kent2008}.
Most of the variation in the tight-binding parameters between the various cuprate compounds comes from $t_{pp}$ and $t_{pp}'$~\cite{Andersen2001, Pavarini2001, Kent2008}.
Upon downfolding, the apical \ce{O}:$2p_z$ orbitals generate interlayer hopping~\cite{Photopoulos2019}, which we neglect.
We also neglect spin-orbit coupling.
Further downfolding to a one-band model~\cite{Zhang1988, Eskes1988} is possible, but at the expense of greatly delocalizing the effective \ce{Cu}:$3d_{x^2-y^2}$ orbital~\cite{Kent2008} and limiting the number of possible intra-unit-cell orders.

We use the following orbital basis
\begin{align}
\psi(\vb{R}) &\defeq \begin{pmatrix}
\ce{Cu}\colon 3d_{x^2-y^2}(\vb{R}) \\[2pt]
\ce{O}\colon 2p_x(\vb{R} + \tfrac{1}{2} \vu{e}_x) \\[2pt]
\ce{O}\colon 2p_y(\vb{R} + \tfrac{1}{2} \vu{e}_y)
\end{pmatrix} = \begin{pmatrix}
\ce{Cu}\colon 3d_{x^2-y^2}(\vb{R}) \,\otimes \ket{\uparrow} \\[2pt]
\ce{Cu}\colon 3d_{x^2-y^2}(\vb{R}) \,\otimes \ket{\downarrow} \\[2pt]
\ce{O}\colon 2p_x(\vb{R} + \tfrac{1}{2} \vu{e}_x) \,\otimes \ket{\uparrow} \\[2pt]
\ce{O}\colon 2p_x(\vb{R} + \tfrac{1}{2} \vu{e}_x) \,\otimes \ket{\downarrow} \\[2pt]
\ce{O}\colon 2p_y(\vb{R} + \tfrac{1}{2} \vu{e}_y) \,\otimes \ket{\uparrow} \\[2pt]
\ce{O}\colon 2p_y(\vb{R} + \tfrac{1}{2} \vu{e}_y) \,\otimes \ket{\downarrow}
\end{pmatrix} \label{eq:primitive-basis}
\end{align}
with the Fourier convention
\begin{align}
\psi_{\vb{k}} &= \frac{1}{\sqrt{\mathcal{N}}} \sum_{\vb{R}} \Elr^{- \iu \vb{k} \vdot \vb{R}} \psi(\vb{R}), \label{eq:first-psi-gauge-choice}
\end{align}
where the lattice constant is set to unity so the Cartesian unit vectors $\vu{e}_{x,y}$ are the primitive lattice vectors (which connect the neighboring copper atoms and are oriented along the $x, y$ axes denoted in Fig.~\ref{fig:CuO2-sketch}), $\vu{e}_i \vdot \vu{e}_j = \Kd_{ij}$, $\vb{R} \in \Z \vu{e}_x + \Z \vu{e}_y$ goes over the real-space square lattice (on which the copper atoms are positioned), $\vb{k} = (k_x, k_y)$ are crystal momenta which always go over the first Brillouin zone only, and $\mathcal{N}$ is the number of unit cells.
The spins $\uparrow, \downarrow$ and tensor products with the $2 \times 2$ identity $\Pauli_0 = \one$ in spin space shall be suppressed when obvious.

\begin{figure}[t]
\centering
\includegraphics[width=\textwidth]{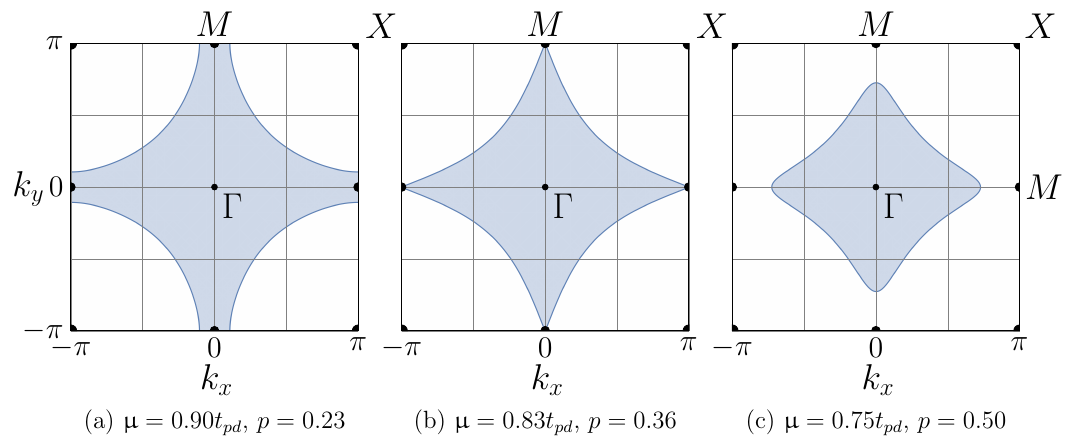}
\captionbelow[Evolution of the Fermi surface of the three-orbital \ce{CuO2} model as the hole doping is increased, ranging from slight overdoping (a), across the Lifshitz transition (b) into the far-overdoped regime (c).]{\textbf{Evolution of the Fermi surface of the three-orbital \ce{CuO2} model as the hole doping is increased, ranging from slight overdoping (a), across the Lifshitz transition (b) into the far-overdoped regime (c).}
The parameters used in these plots are $\epsilon_d - \epsilon_p = 3 t_{pd}$, $t_{pp} = 0.6 t_{pd}$, and $t_{pp}' = 0.5 t_{pd}$ [Eq.~\eqref{eq:standard-CuO2-parameter-set}] with $\epsilon_d = 0$ and $\upmu$ as given in the subcaption.
The hole doping $p$, as determined by Eq.~\eqref{eq:CuO2-model-dopin}, is also provided in the subcaption.}
\label{fig:cuprate-Fermi-surfaces}
\end{figure}

With the orbital orientation conventions as depicted in Fig.~\ref{fig:CuO2-sketch},
the three-band Hamiltonian takes the form
\begin{align}
H_{\vb{k}} &= \begin{pmatrix}
h_d(k_x, k_y) & h_{pd}(k_x, k_y) & - h_{pd}(k_y, k_x) \\
& h_p(k_x, k_y) & h_{pp}(k_x, k_y) \\
\cc & & h_p(k_y, k_x)
\end{pmatrix}, \label{eq:3band-Haml}
\end{align}
where
\begin{align}
h_d(k_x, k_y) &= \epsilon_{d} - \upmu, \\
h_p(k_x, k_y) &= \epsilon_{p} + 2 t_{pp}' \cos k_x - \upmu, \\
h_{pd}(k_x, k_y) &= t_{pd} (1 - \Elr^{- \iu k_x}), \\
h_{pp}(k_x, k_y) &= - t_{pp} (1 - \Elr^{\iu k_x}) (1 - \Elr^{- \iu k_y}).
\end{align}
Here $\upmu$ is the chemical potential, $\epsilon_d - \epsilon_p$ is the charge-transfer gap, and $t_{pd}$, $t_{pp}$, and $t_{pp}'$ are the hopping amplitudes depicted in Fig.~\ref{fig:CuO2-sketch}.
$t_{pd}$ is the the largest one and we shall use it to set the overall energy scale.
Typical values for the tight-binding parameters used in the literature are~\cite{Photopoulos2019}: $(\epsilon_{d} - \epsilon_{p}) / t_{pd} \in [2.5, 3.5]$, $t_{pp} / t_{pd} \in [0.5, 0.6]$, and $t_{pp}' / t_{pd} \approx 0$ with $t_{pd} \in [1.2, 1.5]~\si{\electronvolt}$.
$t_{pp}'$ is not really negligible~\cite{Andersen1995, Andersen2001, Pavarini2001, Kent2008}, although it is often assumed to be.
The importance of $t_{pp}'$ for stabilizing loop currents has been emphasized in Ref.~\cite{Weber2014}.

\begin{figure}[p!]
\centering
\includegraphics[width=0.90\textwidth]{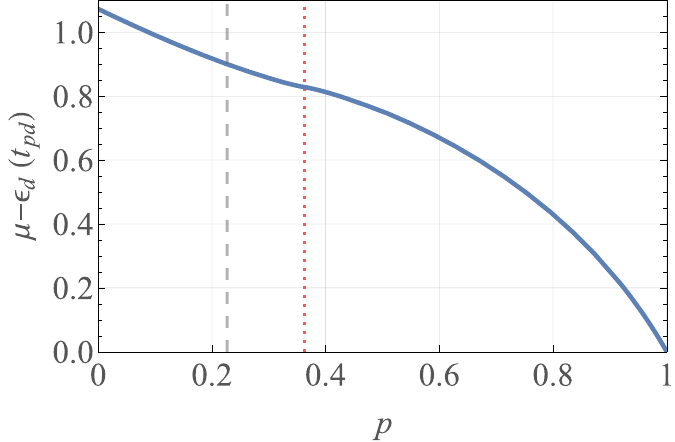}
\captionbelow[The chemical potential $\upmu$ relative to $\epsilon_d$ in units of $t_{pd}$ as a function of hole doping $p$, as determined by Eq.~\eqref{eq:CuO2-model-dopin}.]{\textbf{The chemical potential $\upmu$ relative to $\epsilon_d$ in units of $t_{pd}$ as a function of hole doping $p$, as determined by Eq.~\eqref{eq:CuO2-model-dopin}.}
The parameters used are $\epsilon_d - \epsilon_p = 3 t_{pd}$, $t_{pp} = 0.6 t_{pd}$, and $t_{pp}' = 0.5 t_{pd}$ [Eq.~\eqref{eq:standard-CuO2-parameter-set}].
The dashed vertical line corresponds to $\upmu - \epsilon_d = 0.9 t_{pd}$ and $p = 0.23$ [Fig.~\ref{fig:cuprate-Fermi-surfaces}(a)], while the red dotted vertical line is at $\upmu_{\text{VH}} - \epsilon_d = 0.83 t_{pd}$ and $p_{\text{VH}} = 0.36$ for which the Fermi surface crosses the Van Hove singularities at the high-symmetry points $M$ [Fig.~\ref{fig:cuprate-Fermi-surfaces}(b)].}
\label{fig:chem-pot-vs-doping}
\end{figure}

\begin{table}[p!]
\centering
\captionabove[Tight-binding parameter sets of the three-orbital \ce{CuO2} model that we considered in our calculations.]{\textbf{Tight-binding parameter sets of the three-orbital \ce{CuO2} model that we considered in our calculations.}
$t_{pd}$ sets the overall energy scale and is in between \num{1.2} and \SI{1.5}{\electronvolt}~\cite{Photopoulos2019}.
The hopping amplitudes are defined in Eq.~\eqref{eq:3band-Haml} and Fig.~\ref{fig:CuO2-sketch}.
DFT stands for density functional theory.}
{\renewcommand{\arraystretch}{1.3}
\renewcommand{\tabcolsep}{9.2pt}
\begin{tabular}{ccccc} \hline\hline \\[-15pt]
No. & $\displaystyle \frac{\epsilon_{d} - \epsilon_{p}}{t_{pd}}$ & $\displaystyle \frac{t_{pp}}{t_{pd}}$ & $\displaystyle \frac{t_{pp}'}{t_{pd}}$ & Comment \\[8pt] \hline
1. & $3$ & $0.6$ & $0$ & One of the most conventional parameter choices~\cite{Photopoulos2019}. \\
2. & $1$ & $0.6$ & $0$ & Reduced charge transfer gap. \\
3. & $3$ & $0.6$ & $0.5$ & Includes $t_{pp}'$. \\
4. & $1.5$ & $0.6$ & $0.5$ & Reduced charge transfer gap and includes $t_{pp}'$. \\
5. & $0.5$ & $0.45$ & $0$ & Based on DFT~\cite{Kent2008}. \\
6. & $0$ & $-0.1$ & $0.3$ & Based on DFT~\cite{Kent2008}. \\
7. & $0.7$ & $-0.2$ & $0.7$ & Employed in Ref.~\cite{Weber2014}. \\
8. & $1.4$ & $-0.35$ & $0.7$ & Employed in Ref.~\cite{Weber2014}.
\\ \hline\hline
\end{tabular}}
\label{tab:CuO2-model-parameters}
\end{table}

The chemical potential $\upmu$ is set to intersect the band whose dispersion $\varepsilon_{\vb{k} n}$ has the highest energy among the three bands.
We shall order the band index in ascending ordering of energy,
\begin{align}
\varepsilon_{\vb{k} 1} < \varepsilon_{\vb{k} 2} < \varepsilon_{\vb{k} 3}
\end{align}
so that $\varepsilon_{\vb{k} 3}$ is the conduction band.
At zero temperature, ignoring interactions, $\upmu$ is related to the hole doping $p$ through
\begin{align}
1 + p &= 2 \int_{\text{1\textsuperscript{st}BZ}} \frac{\dd{k_x} \dd{k_y}}{(2 \pi)^2} \, \HTh(\varepsilon_{\vb{k} 3}), \label{eq:CuO2-model-dopin}
\end{align}
i.e., $1 + p$ is the total number of holes per \ce{CuO2}.
Note that both $H_{\vb{k}}$ and its eigenvalues $\varepsilon_{\vb{k} n}$ are displaced by $\upmu$.
Give that the lattice constant is set to unity, the first Brillouin zone (1\textsuperscript{st}BZ) spans $[-\pi,\pi]_{k_x} \times [-\pi,\pi]_{k_y}$.
The evolution of the Fermi surface at overdoping is drawn in Fig.~\ref{fig:cuprate-Fermi-surfaces}.
The chemical potential as a function of hole doping is plotted in Fig.~\ref{fig:chem-pot-vs-doping}.

As we shall explain in Sec.~\ref{sec:pairing-cuprate-actual-analysis}, within out calculation the one-particle Hamiltonian describes Fermi liquid quasi-particles of the overdoped regime.
Hubbard interactions are known to be strong in these compounds and they drastically change the orbital character of the conduction band depending on the doping~\cite{Chen1991, Pellegrin1993, Peets2009, Pickett1989}.
To account for this, we have considered eight different parameter sets that cover a wide range of the physically reasonable possibilities.
They are listed in Tab.~\ref{tab:CuO2-model-parameters}.
We have ensured that all eight parameter sets reproduce the ARPES Fermi surface shapes~\cite{Plate2005, Yoshida2006, Peets2007, Horio2018} and, relatedly, that the Lifshitz transition occurs at a hole doping $p_L > 0.15$, as found in experiment~\cite{Sobota2021}.
In the end, our results have turned out to be insensitive to these changes in the one-particle Hamiltonian.
All results which we show or quote in the remained of this chapter are for the representative parameter set (No.~3 in Tab.~\ref{tab:CuO2-model-parameters}):
\begin{align}
\epsilon_d - \epsilon_p &= 3 t_{pd}, &
t_{pp} &= 0.6 t_{pd}, &
t_{pp}' &= 0.5 t_{pd}, \label{eq:standard-CuO2-parameter-set}
\end{align}
with the reference energy and chemical potential:
\begin{align}
\epsilon_d &= 0, &
\upmu &= 0.9 t_{pd},
\end{align}
unless stated otherwise.
The corresponding Fermi surface is shown in Fig.~\ref{fig:cuprate-Fermi-surfaces}(a).

When comparing to the work of others, one should keep in mind that there are multiple possible orbital orientation conventions and momentum-space gauges that one may use.
Ours are given in Fig.~\ref{fig:CuO2-sketch} and Eq.~\eqref{eq:first-psi-gauge-choice}.
Alternative choices are discussed in Sec.~\ref{sec:ASV-conventions}.

\section{Classification of particle-hole bilinears in the three-orbital model} \label{sec:cuprate-bilinear-class}
Although the three-orbital model of the previous section has been known for almost forty years~\cite{Varma1987, Emery1987, Emery1988, Littlewood1989, Gaididei1988, Scalettar1991}, a systematic classification of all possible particle-hole fermionic bilinears which one can construct within it is absent in the literature.
Partial classifications are available in Refs.~\cite{Aji2010, Fischer2011}.
Here we provide such a classification by exploiting a certain redundant ``extended basis'' which has particularly simple symmetry transformation rules.
Physically, fermionic bilinears are interesting because their expectation values can be taken to represent order parameters, or alternatively they can be used to construct the symmetry-allowed Yukawa couplings to fluctuating order parameter fields, as we discussed in Sec.~\ref{sec:QCP-pairing-model} of the previous chapter.
As a simple application of the classification, in Sec.~\ref{sec:CuO2-Hubbard-decompose} we use it to decompose Hubbard interactions into symmetry channels.
This classification has already been presented in Ref.~\cite{Palle2024-LC}, but without its derivation or the listing of TR-even bilinears.
Note that some of the TR-odd matrices are defined with an additional minus sign compared to Ref.~\cite{Palle2024-LC}.

As was discussed in Sec.~\ref{sec:LC-gen-ph-bilinears}, the most general form of a Hermitian fermionic bilinear in the particle-hole sector is [Eq.~\eqref{eq:general-coordinate-bilinear}]:
\begin{align}
\phi_{a}(\vb{R}) &= \sum_{\vb{\delta}_1 \vb{\delta}_2} \psi^{\dag}(\vb{R}+\vb{\delta}_1) \Gamma_{a}(\vb{\delta}_1, \vb{\delta}_2) \psi(\vb{R}+\vb{\delta}_2), \label{eq:general-coordinate-bilinear-again}
\end{align}
where $\psi = \mleft(\psi_{1, \uparrow}, \psi_{1, \downarrow}, \ldots, \psi_{M, \uparrow}, \psi_{M, \downarrow}\mright)^{\intercal}$ are the fermionic field operators, assuming $M$ orbitals per unit cell, and $\vb{\delta}_1, \vb{\delta}_2, \ldots$ go over lattice neighbors.
The $2M \times 2M$ matrices $\Gamma_{a}(\vb{\delta}_1, \vb{\delta}_2) = \Gamma_{a}^{\dag}(\vb{\delta}_2, \vb{\delta}_1)$, which are in general non-trivial in both spin and orbital space, determine the symmetry properties of $\phi_{a}(\vb{R})$ under time reversal (TR) and under crystalline operations, as specified by the irreducible representation (irrep) of the point group under which it transforms.
The subscript $a$ denotes different irrep components and is relevant only in the case of multidimensional irreps.
The classification of possible fermionic bilinears thus amounts to the classification of the $2 M \times 2 M$ matrices $\Gamma_{a}(\vb{\delta}_1, \vb{\delta}_2)$.

\begin{table}[t!]
\centering
\captionabove[The character table of the tetragonal point group $D_{4h}$~\cite{Dresselhaus2007}.]{\textbf{The character table of the tetragonal point group $D_{4h}$}~\cite{Dresselhaus2007}.
The irreps are divided according to parity into even (subscript $g$) and odd ($u$) ones.
To the left of the irreps are the simplest polynomials constructed from the coordinates $\vb{r} = (x, y, z)$ that transform according to them.
$C_4$ are \SI{90}{\degree} rotations around $\vu{e}_z$.
$C_2$, $C_2'$, and $C_2''$ are \SI{180}{\degree} rotations around $\vu{e}_z$, $\vu{e}_x$ or $\vu{e}_y$, and the diagonals $\vu{e}_x \pm \vu{e}_y$, respectively.
$P$ is space inversion or parity.
Improper rotations $S_4$ and mirror reflections $\Sigma_h$, $\Sigma_v'$, and $\Sigma_d''$ are obtained by composing $C_4$, $C_2$, $C_2'$, and $C_2''$ with $P$, respectively.}
{\renewcommand{\arraystretch}{1.3}
\renewcommand{\tabcolsep}{8.4pt}
\begin{tabular}{cc|rrrrr|rrrrr} \hline\hline
\multicolumn{2}{c|}{$D_{4h}$} & $E$ & $2 C_4$ & $C_2$ & $2 C_2'$ & $2 C_2''$ & $P$ & $2 S_4$ & $\Sigma_h$ & $2 \Sigma_v'$ & $2 \Sigma_d''$
\\ \hline
$1$, $x^2+y^2$, $z^2$ & $A_{1g}$ & $1$ & $1$ & $1$ & $1$ & $1$ & $1$ & $1$ & $1$ & $1$ & $1$
\\
$xy(x^2-y^2)$ & $A_{2g}$ & $1$ & $1$ & $1$ & $-1$ & $-1$ & $1$ & $1$ & $1$ & $-1$ & $-1$
\\
$x^2-y^2$ & $B_{1g}$ & $1$ & $-1$ & $1$ & $1$ & $-1$ & $1$ & $-1$ & $1$ & $1$ & $-1$
\\
$xy$ & $B_{2g}$ & $1$ & $-1$ & $1$ & $-1$ & $1$ & $1$ & $-1$ & $1$ & $-1$ & $1$
\\
$(yz|-xz)$ & $E_g$ & $2$ & $0$ & $-2$ & $0$ & $0$ & $2$ & $0$ & $-2$ & $0$ & $0$
\\ \hline
$xyz(x^2-y^2)$ & $A_{1u}$ & $1$ & $1$ & $1$ & $1$ & $1$ & $-1$ & $-1$ & $-1$ & $-1$ & $-1$
\\
$z$ & $A_{2u}$ & $1$ & $1$ & $1$ & $-1$ & $-1$ & $-1$ & $-1$ & $-1$ & $1$ & $1$
\\
$xyz$ & $B_{1u}$ & $1$ & $-1$ & $1$ & $1$ & $-1$ & $-1$ & $1$ & $-1$ & $-1$ & $1$
\\
$(x^2-y^2)z$ & $B_{2u}$ & $1$ & $-1$ & $1$ & $-1$ & $1$ & $-1$ & $1$ & $-1$ & $1$ & $-1$
\\
$(x|y)$ & $E_u$ & $2$ & $0$ & $-2$ & $0$ & $0$ & $-2$ & $0$ & $2$ & $0$ & $0$
\\ \hline\hline
\end{tabular}}
\label{tab:D4h-char-tab-again}
\end{table}

In the three-orbital \ce{CuO2} model under consideration $M = 3$ and the fermionic spinor is the one given in Eq.~\eqref{eq:primitive-basis}.
The orbitals and their orientation are shown in Figs.~\ref{fig:CuO2-sketch} and \ref{fig:extended-unit-cell}.
As for the symmetries, the point group of the copper oxide plane is the tetragonal point group $D_{4h}$.
The structure of this point group is worked out in detail in Sec.~\ref{sec:tetragonal-group-D4h} of Appx.~\ref{app:group_theory}, where one may also find its character table (repeater here in Tab.~\ref{tab:D4h-char-tab-again} for the reader's convenience) and irrep product table (Tab.~\ref{tab:D4h-irrep-prod-tab}).
Here we shall just note that $D_{4h}$ is generated by four-fold rotations around the $z$ axis $C_{4z}$, two-fold rotations around the $x$ axis $C_{2x}$ and $d_+ = x + y$ diagonal $C_{2d_+}$, and parity $P$.
These symmetries are evident from Fig.~\ref{fig:extended-unit-cell}.
The center of rotation and inversion we always take to be at the center of a copper atom.\footnote{The center of rotation and inversion can also be chosen to lie at $\vb{R} + \tfrac{1}{2} (\vu{e}_x + \vu{e}_y)$, which is in the middle of four copper atoms, instead of $\vb{R}$. Point group operations $g$ which leave $\vb{R} + \tfrac{1}{2} (\vu{e}_x + \vu{e}_y)$ fixed are related to those that keep $\vb{R}$ fixed through a commensurate translation by $R(g) \tfrac{1}{2} (\vu{e}_x + \vu{e}_y) - \tfrac{1}{2} (\vu{e}_x + \vu{e}_y) = \vb{\delta} \in \Z \vu{e}_x + \Z \vu{e}_y$.}

\subsection{Extended basis and the simplification of symmetry transformation rules} \label{sec:extended-basis-def}
\begin{figure}[t]
\centering
\includegraphics[width=0.95\linewidth]{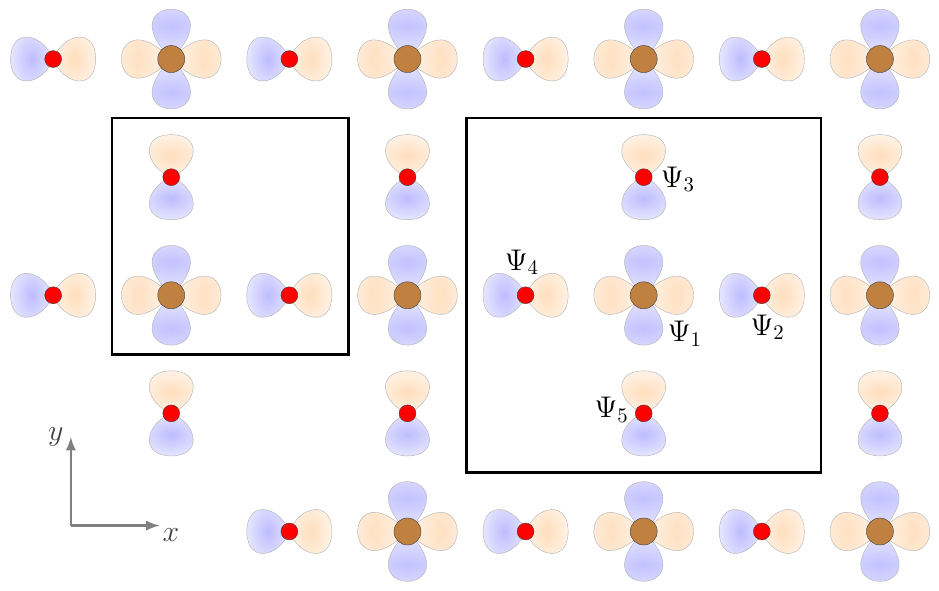}
\captionbelow[The primitive (left) and extended (right) unit cells of the \ce{CuO2} plane.]{\textbf{The primitive (left) and extended (right) unit cells of the \ce{CuO2} plane.}
The components of the extended fermionic field operator $\Psi$ [Eq.~\eqref{eq:extended-basis}] are designed within the extended unit cell.
The underlying three-orbital copper oxide model is described in Sec.~\ref{sec:cuprate-3band-model} and Fig.~\ref{fig:CuO2-sketch}.}
\label{fig:extended-unit-cell}
\end{figure}

Because of the non-trivial Wyckoff positions of the oxygen atoms,\footnote{More simply stated, the oxygen atoms do not lie on the (Bravais) lattice points like cooper, but are instead displaced away from them.} some point group operations (e.g., \SI{90}{\degree} rotations around the $z$ axis $C_{4z}$ and parity $P$) map orbitals between different primitive unit cells, as one can convince oneself by examining Fig.~\ref{fig:extended-unit-cell}.
To be more precise, for some point group operations $g \in D_{4h}$ the orbitals of the unit cell at $\vb{R}$ get mapped not only to the orbitals of the unit cell at $R(g) \vb{R}$, where $R(g)$ is the vector rotation matrix, but also to the orbitals of neighboring unit cells which are at $R(g) \vb{R} + \vb{\delta}$.
This remains true irrespective of which primitive unit cell one chooses.

In momentum space, the corresponding unitary matrices therefore acquire $\vb{k}$-dependent phases.
This we have already seen in Sec.~\ref{sec:LC-gen-model-sym} of the previous chapter when we wrote down the most general possible fermionic transformation rules [Eqs.~\eqref{eq:fermi-tranf-point-group} and~\eqref{eq:fermi-tranf-TR}]:
\begin{align}
\SymU^{\dag}(g) \psi_{\vb{k}} \SymU(g) &= \MatU_{\vb{k}}(g) \psi_{R(g^{-1}) \vb{k}}, \label{eq:fermi-tranf-point-group-again} \\
\SymTR^{-1} \psi_{\vb{k}} \SymTR &= \MatTR^{*} \psi_{-\vb{k}}, \label{eq:fermi-tranf-TR-again}
\end{align}
where $\SymU(g)$ are the many-body symmetry operators and $\SymTR$ is the many-body TR operator.
Notice how the $2M \times 2M$ unitary transformation matrices $\MatU_{\vb{k}}(g)$ depend on $\vb{k}$.
Thus the change in the orbital structure of $\psi_{\vb{k}}$ depends not only on the point group transformation $g$, but also on the momentum $\vb{k}$.

In the current model, the TR symmetry matrix $\MatTR$ of Eq.~\eqref{eq:fermi-tranf-TR-again} has no $\vb{k}$-dependence because there is no spin-orbit mixing in the basis.
One may always choose a gauge for the spins in which
\begin{align}
\MatTR &= \one \otimes \iu \Pauli_y,
\end{align}
as we henceforth assume.
$\Pauli_y \equiv \Pauli_2$ is the second Pauli matrix.

For classification purposes, it is much more convenient if the orbital and momentum dependencies of the point group matrices $\MatU_{\vb{k}}(g)$ do not mix.
This can be accomplished by employing the extended basis
\begin{align}
\Psi(\vb{R}) &\defeq \begin{pmatrix}
\ce{Cu}\colon 3d_{x^2-y^2}(\vb{R}) \\[2pt]
\ce{O}\colon 2p_x(\vb{R} + \tfrac{1}{2} \vu{e}_x) \\[2pt]
\ce{O}\colon 2p_y(\vb{R} + \tfrac{1}{2} \vu{e}_y) \\[2pt]
\ce{O}\colon 2p_x(\vb{R} - \tfrac{1}{2} \vu{e}_x) \\[2pt]
\ce{O}\colon 2p_y(\vb{R} - \tfrac{1}{2} \vu{e}_y)
\end{pmatrix} \label{eq:extended-basis}
\end{align}
instead of the primitive basis $\psi(\vb{R})$ that we introduced in Eq.~\eqref{eq:primitive-basis}.
The corresponding extended unit cell is shown in Fig.~\ref{fig:extended-unit-cell}.
If we use the same Fourier transform convention as in Eq.~\eqref{eq:first-psi-gauge-choice}, namely
\begin{align}
\Psi_{\vb{k}} &= \frac{1}{\sqrt{\mathcal{N}}} \sum_{\vb{R}} \Elr^{- \iu \vb{k} \vdot \vb{R}} \Psi(\vb{R}),
\end{align}
then this new basis is related to the primitive basis through:
\begin{align}
\Psi_{\vb{k}} &= \mathcal{K}_{\vb{k}} \psi_{\vb{k}},
\end{align}
where ($\Pauli_0 = \one$ is the $2 \times 2$ identity):
\begin{align}
\mathcal{K}_{\vb{k}} &\defeq \begin{pmatrix}
1 & 0 & 0 \\
0 & 1 & 0 \\
0 & 0 & 1 \\
0 & \Elr^{- \iu k_x} & 0 \\
0 & 0 & \Elr^{- \iu k_y}
\end{pmatrix} \otimes \Pauli_0. \label{eq:K-proj-mat}
\end{align}
Conversely, the primitive basis is related to the extended basis through
\begin{align}
\psi_{\vb{k}} &= \mathcal{K}^{-1} \Psi_{\vb{k}},
\end{align}
where $\mathcal{K}^{-1}$ is the pseudo-inverse of $\mathcal{K}_{\vb{k}}$:
\begin{align}
\mathcal{K}^{-1} &\defeq \begin{pmatrix}
1 & 0 & 0 & 0 & 0 \\
0 & 1 & 0 & 0 & 0 \\
0 & 0 & 1 & 0 & 0
\end{pmatrix} \otimes \Pauli_0. \label{eq:K-proj-mat-inv}
\end{align}
The two matrices multiply to give an identity only for one ordering:
\begin{align}
\mathcal{K}^{-1} \mathcal{K}_{\vb{k}} &= \begin{pmatrix}
1 & 0 & 0 \\
0 & 1 & 0 \\
0 & 0 & 1
\end{pmatrix}, &
\mathcal{K}_{\vb{k}} \mathcal{K}^{-1} &= \begin{pmatrix}
1 & 0 & 0 & 0 & 0 \\
0 & 1 & 0 & 0 & 0 \\
0 & 0 & 1 & 0 & 0  \\
0 & \Elr^{- \iu k_x} & 0 & 0 & 0 \\
0 & 0 & \Elr^{- \iu k_y} & 0 & 0
\end{pmatrix}.
\end{align}
The reason why $\mathcal{K}_{\vb{k}} \mathcal{K}^{-1} \neq \one$ lies in the fact that generic extended-basis vectors $\vb{v} = (v_1, v_2, v_3, \linebreak v_4, v_5)^{\intercal}$ do not satisfy $v_{4,5} = \Elr^{- \iu k_{x,y}} v_{2,3}$, as every output of $\mathcal{K}_{\vb{k}}$ must.
In the case of $\Psi_{\vb{k}}$, $\mathcal{K}_{\vb{k}} \mathcal{K}^{-1} \Psi_{\vb{k}} = \Psi_{\vb{k}}$.

\begin{table}[t!]
\centering
\captionabove[The symmetry transformation matrices of the four generators $g$ of the point group $D_{4h}$ in the extended basis $\Psi$.]{\textbf{The symmetry transformation matrices of the four generators $g$ of the point group $D_{4h}$ in the extended basis $\Psi$.}
$C_{4z}$ is a \SI{90}{\degree} rotation around $\vu{e}_z$, $C_{2x}$ is a \SI{180}{\degree} rotation around $\vu{e}_x$, $C_{2d_+}$ is a \SI{180}{\degree} rotation around $\vu{e}_x + \vu{e}_y$, and $P$ is parity.
$R(g)$ and $S(g)$ are vector and spin transformation matrices, respectively.
$\Pauli_A$ are Pauli matrices.
The extended basis $\Psi$ is defined in Eq.~\eqref{eq:extended-basis} and $O(g)$ are its orbital transformation matrices which are easily deduced from Fig.~\ref{fig:extended-unit-cell}.}
{\renewcommand{\arraystretch}{1.3}
\renewcommand{\tabcolsep}{10pt}
\begin{tabular}{cNNN} \hline\hline
$g$ & $R(g)$ & $O(g)$ & $S(g)$
\\ \hline \\[-14pt]
$C_{4z}$ & $\begin{pmatrix} 0 & -1 & 0 \\ 1 & 0 & 0 \\ 0 & 0 & 1 \end{pmatrix}$ &
$\begin{pmatrix}
-1 & 0 & 0 & 0 & 0 \\
0 & 0 & 0 & 0 & -1 \\
0 & 1 & 0 & 0 & 0 \\
0 & 0 & -1 & 0 & 0 \\
0 & 0 & 0 & 1 & 0
\end{pmatrix}$
& $\displaystyle \frac{\Pauli_0 - \iu \Pauli_z}{\sqrt{2}}$ \\[30pt]
$C_{2x}$ & $\begin{pmatrix} 1 & 0 & 0 \\ 0 & -1 & 0 \\ 0 & 0 & -1 \end{pmatrix}$ & $\begin{pmatrix}
1 & 0 & 0 & 0 & 0 \\
0 & 1 & 0 & 0 & 0 \\
0 & 0 & 0 & 0 & -1 \\
0 & 0 & 0 & 1 & 0 \\
0 & 0 & -1 & 0 & 0
\end{pmatrix}$
& $\displaystyle - \iu \Pauli_x$ \\[30pt]
$C_{2d_+}$ & $\begin{pmatrix} 0 & 1 & 0 \\ 1 & 0 & 0 \\ 0 & 0 & -1 \end{pmatrix}$ &
$\begin{pmatrix}
-1 & 0 & 0 & 0 & 0 \\
0 & 0 & 1 & 0 & 0 \\
0 & 1 & 0 & 0 & 0 \\
0 & 0 & 0 & 0 & 1 \\
0 & 0 & 0 & 1 & 0
\end{pmatrix}$
& $\displaystyle - \iu \frac{\Pauli_x + \Pauli_y}{\sqrt{2}}$ \\[30pt]
$P$ & $\begin{pmatrix} -1 & 0 & 0 \\ 0 & -1 & 0 \\ 0 & 0 & -1 \end{pmatrix}$ &
$\begin{pmatrix}
1 & 0 & 0 & 0 & 0 \\
0 & 0 & 0 & -1 & 0 \\
0 & 0 & 0 & 0 & -1 \\
0 & -1 & 0 & 0 & 0 \\
0 & 0 & -1 & 0 & 0
\end{pmatrix}$
& $\displaystyle \Pauli_0$
\\ \\[-14pt] \hline\hline
\end{tabular}}
\label{tab:CuO2-model-D4h-generators}
\end{table}

In the extended basis the symmetry transformation matrices do not depend on crystal momentum:
\begin{align}
\SymU^{\dag}(g) \Psi_{\vb{k}} \SymU(g) &= O(g) \otimes S(g) \Psi_{R(g^{-1}) \vb{k}}, \label{eq:fermi-tranf-point-group-extended} \\
\SymTR^{-1} \Psi_{\vb{k}} \SymTR &= (\one \otimes \iu \Pauli_y) \Psi_{- \vb{k}},
\end{align}
in contrast to what we found in Eq.~\eqref{eq:fermi-tranf-point-group-again} (or Eqs.~\eqref{eq:fermi-tranf-point-group} and~\eqref{eq:fermi-tranf-TR} of the previous chapter).
Here, $R(g)$ and $S(g)$ are the usual vector and spin transformation matrices which are precisely defined in Sec.~\ref{sec:SU2-SO3-conventions} of Appx.~\ref{app:group_theory}.\footnote{In short, for a rotation by $\vartheta$ around $\vu{n}$, $R(g) = \exp(- \iu \vartheta \vu{n} \vdot \vb{L})$ with $(L_i)_{jk} = - \iu \LCs_{ijk}$ and $S(g) = \exp(- \iu \vartheta \vu{n} \vdot \vb{S})$ with $S_i = \tfrac{1}{2} \Pauli_i$, while for parity $P$, $R(P) = - \one$ and $S(P) = \Pauli_0$.}
Because we are dealing with a fermionic field, $g$ belongs to the double group of the tetragonal point group $D_{4h}$.
The orbital transformation matrices $O(g)$ encode the detailed orbital structure of the model and they are readily deduced from Fig.~\ref{fig:extended-unit-cell}.
The symmetry matrices corresponding to the four generators of the point group $D_{4h}$ are given in Tab.~\ref{tab:CuO2-model-D4h-generators}.
They are related to the $\vb{k}$-dependent matrices of Eq.~\eqref{eq:fermi-tranf-point-group-again} via:
\begin{align}
\mathcal{K}_{\vb{k}} \MatU_{\vb{k}}(g) &= \mleft[O(g) \otimes S(g)\mright] \mathcal{K}_{R(g^{-1}) \vb{k}}, \\
\MatU_{\vb{k}}(g) &= \mathcal{K}^{-1} \mleft[O(g) \otimes S(g)\mright] \mathcal{K}_{R(g^{-1}) \vb{k}}.
\end{align}
Even though the $5 \times 5$ $O(g)$ matrices are slightly larger than the corresponding $3 \times 3$ matrices in the primitive basis $\psi$, their momentum-independence greatly simplifies the symmetry classification, as we shall see in the next section.

\subsection{Symmetry analysis and classification}
In momentum space, fermionic bilinears of the general form given in Eq.~\eqref{eq:general-coordinate-bilinear-again} become [Eq.~\eqref{eq:general-momentum-bilinear}]:
\begin{align}
\phi_{a \vb{q}} &= \frac{1}{\sqrt{\mathcal{N}}} \sum_{\vb{k}} \psi_{\vb{k}}^{\dag} \Gamma_{a \vb{k}, \vb{k}+\vb{q}} \psi_{\vb{k}+\vb{q}}, \label{eq:general-momentum-bilinear-again}
\end{align}
where $\Gamma_{a \vb{k}, \vb{p}} = \sum_{\vb{\delta}_1 \vb{\delta}_2} \Elr^{- \iu (\vb{k} \vdot \vb{\delta}_1 - \vb{p} \vdot \vb{\delta}_2)} \Gamma_{a}(\vb{\delta}_1, \vb{\delta}_2)$.

Our goal is to classify the possible fermionic bilinears according to how they transform under point group operations and TR.
The point group and TR transformation rules are [Eqs.~\eqref{eq:phi-transf-rule1-bilinear} and~\eqref{eq:phi-transf-rule2-bilinear}]:
\begin{align}
\SymU^{\dag}(g) \phi_{a \vb{q}} \SymU(g) &= \sum_{b=1}^{\dim \zeta} \RepM_{ab}^{\zeta}(g) \phi_{b, R(g^{-1}) \vb{q}}, \label{eq:phi-transf-rule1-bilinear-again} \\
\SymTR^{-1} \phi_{a \vb{q}} \SymTR &= p_{\TRop} \phi_{a, -\vb{q}}, \label{eq:phi-transf-rule2-bilinear-again}
\end{align}
where $\zeta$ is an irrep of the point group $D_{4h}$, $g \in D_{4h}$, and $p_{\TRop} = \pm 1$ is the TR sign.
Irreps of $D_{4h}$ are listed in Tab.~\ref{tab:D4h-char-tab-again}.
Given that $\phi_{a \vb{q}}^{\dag} = \phi_{a, -\vb{q}}$ is real, the irrep transformation matrices $\RepM_{ab}^{\zeta}(g)$ must be real as well, which can be made true for all irreps $\zeta$ of $D_{4h}$.
These transformation rules are satisfied if and only if the $\Gamma_{a \vb{k}, \vb{p}}$ matrices satisfy [cf.\ Eqs.~\eqref{eq:Gamma-transf-rule1-bilinear} and~\eqref{eq:Gamma-transf-rule2-bilinear}]:
\begin{align}
S^{\dag}(g) O^{\dag}(g) \Gamma_{a, R(g) \vb{k}, R(g) \vb{p}} O(g) S(g) &= \sum_{b=1}^{\dim \phi} \RepM_{ab}^{\zeta}(g) \Gamma_{b \vb{k}, \vb{p}}, \label{eq:Gamma-transf-rule1-bilinear-again} \\
(\iu \Pauli_y)^{\dag} \Gamma_{a, -\vb{k}, -\vb{p}}^{*} (\iu \Pauli_y) &= p_{\TRop} \Gamma_{a \vb{k}, \vb{p}}. \label{eq:Gamma-transf-rule2-bilinear-again}
\end{align}
Constructing $6 \times 6$ $\Gamma_{a \vb{k}, \vb{p}}$ matrices which combine their dependence on $\vb{k}$, $\vb{p}$, spin, and orbital indices in just the right way so that they transform under an \emph{irreducible} representation is a non-trivial task to which we devote the current section.

To make progress, we introduce extended-basis ($10 \times 10$) matrices $\Gamma_{a \vb{k}}$ that depend on only one momentum and that transform according to:
\begin{align}
S^{\dag}(g) O^{\dag}(g) \Gamma_{a, R(g) \vb{k}} O(g) S(g) &= \sum_{b=1}^{\dim \zeta} \RepM_{ab}^{\zeta}(g) \Gamma_{b \vb{k}}, \label{eq:Gamma-transf-rule1-bilinear-again-extended} \\
(\iu \Pauli_y)^{\dag} \Gamma_{a, -\vb{k}}^{*} (\iu \Pauli_y) &= p_{\TRop} \Gamma_{a \vb{k}}. \label{eq:Gamma-transf-rule2-bilinear-again-extended}
\end{align}
If we now construct the primitive-basis matrices by projecting the extended-basis matrices like so [cf.\ Eq.~\eqref{eq:general-momentum-bilinear-alt2}]:
\begin{align}
\Gamma_{a \vb{k}, \vb{p}} &= \mathcal{K}_{\vb{k}}^{\dag} \mleft(\Gamma_{a \vb{k}} + \Gamma_{a \vb{p}}^{\dag}\mright) \mathcal{K}_{\vb{p}}, \label{eq:momentum-bilinear-v2}
\end{align}
then Eqs.~\eqref{eq:Gamma-transf-rule1-bilinear-again} and~\eqref{eq:Gamma-transf-rule2-bilinear-again} are automatically satisfied.
Note that reality $\Gamma_{a \vb{k}, \vb{p}}^{\dag} = \Gamma_{a \vb{p}, \vb{k}}$ is also automatically satisfied.
$\mathcal{K}_{\vb{k}}$ is defined in Eq.~\eqref{eq:K-proj-mat}.
However, to construct $\Gamma_{a \vb{k}}$ with proper transformation properties, we first need to separately classify orbital matrices, spin matrices, and momentum-dependent functions into irreps.
After that, we use the irrep multiplication Tab.~\ref{tab:D4h-irrep-prod-tab}, provided in Appx.~\ref{app:group_theory}, to construct the $\Gamma_{a \vb{k}}$.
We will give a number of examples in Sec.~\ref{sec:cup-ord-param-constr}.

Let us emphasize that the reason why we can assume dependence on only one momentum in the first place is because the extended basis transformation rules [Eq.~\eqref{eq:fermi-tranf-point-group-extended}] do not mix orbital and momentum transformations.
Thus dependence on only one momentum is sufficiently general to cover all possible bilinears, as was explained in Sec.~\ref{sec:LC-gen-ph-bilinears} of the previous chapter.

A collection of momentum-dependent scalar functions $f_{a}(\vb{k})$, indexed by $a$, is classified according to:
\begin{align}
f_{a}\mleft(R(g) \vb{k}\mright) &= \sum_{b=1}^{\dim \zeta} \RepM_{ab}^{\zeta}(g) f_{b}(\vb{k}), \label{eq:scalar-func-transf-rule1} \\
f_{a}^{*}(-\vb{k}) &= p_{\TRop} f_{a}(\vb{k}). \label{eq:scalar-func-transf-rule2}
\end{align}
By going through the lattice harmonics, one readily retrieves the well-known result that:
\begin{align}
\begin{aligned}
1, \cos k_x + \cos k_y &\in A_{1g}^{+}, &\hspace{50pt}
(\cos k_x - \cos k_y) \sin k_x \sin k_y &\in A_{2g}^{+}, \\
\cos k_x - \cos k_y &\in B_{1g}^{+}, &\hspace{50pt}
\sin k_x \sin k_y &\in B_{2g}^{+}, \\
(\sin k_x | \sin k_y) &\in E_u^{-},
\end{aligned} \label{eq:classification-of-fk-func}
\end{align}
where the irrep superscripts indicate the TR sign $p_{\TRop}$.
Notice that their Taylor expansions agree with Tab.~\ref{tab:D4h-example-tab} of Appx.~\ref{app:group_theory}.
As long as the functions are real, the parity and TR sign will be equal.

A feature specific to two dimensions is that \SI{180}{\degree} rotations around the $z$ axis $C_{2z}$ act in the same way as parity $P$.
Hence the same must hold for scalar functions $f_{a}(\vb{k})$ that do not depend on $k_z$ and for orbital matrices $\Lambda$ constructed from in-plane orbitals.
As long as there is no $k_z$-dependence or out-of-plane orbitals, respectively, $f_{a}(\vb{k})$ and $\Lambda$ cannot transform under the irreps $A_{1u}$, $A_{2u}$, $B_{1u}$, $B_{2u}$, and $E_g$ since their $\RepM^{\zeta}(C_{2z}) \neq \RepM^{\zeta}(P)$.

Spin matrices are classified according to:
\begin{align}
S^{\dag}(g) \Pauli_a S(g) &= \sum_{b=1}^{\dim \zeta} \RepM_{ab}^{\zeta}(g) \Pauli_b, \\
(\iu \Pauli_y)^{\dag} \Pauli_a^{*} (\iu \Pauli_y) &= p_{\TRop} \Pauli_a,
\end{align}
and one readily finds that (Sec.~\ref{sec:examples-convention-D4h}):
\begin{align}
\Pauli_0 &\in A_{1g}^{+}, &
(\Pauli_1 | \Pauli_2) &\in E_g^{-}, &
\Pauli_3 &\in A_{2g}^{-}. \label{eq:classification-of-Pauli-mat}
\end{align}

As for the orbital matrices, we denote them with capital $\Lambda$-s and we classify them according to:
\begin{align}
O^{\intercal}(g) \Lambda_a O(g) &= \sum_{b=1}^{\dim \zeta} \RepM_{ab}^{\zeta}(g) \Lambda_b, \label{eq:Lambda-transf-rule1} \\
\Lambda_a^{*} &= p_{\TRop} \Lambda_a, \label{eq:Lambda-transf-rule2}
\end{align}
where $O^{\dag}(g) = O^{\intercal}(g) = O^{-1}(g) = O(g^{-1})$ because all $O(g)$ are real and orthogonal (Tab.~\ref{tab:CuO2-model-D4h-generators}).
We shall always chose them so that they are Hermitian, $\Lambda_a^{\dag} = \Lambda$.
Component-wise, $(\Lambda_a)_{\alpha \beta}$ transforms under the direct product representation $O \otimes O$ which we can decompose using representation characters:
\begin{align}
\vec{\chi}_{O \otimes O} &= \begin{pmatrix}
5, & -1, & 1, & 3, & -1, & 1
\end{pmatrix} = \vec{\chi}_{A_{1g}} + 2 \vec{\chi}_{B_{1g}} + \vec{\chi}_{E_u}.
\end{align}
This is explained in Sec.~\ref{sec:multid-irrep-product}, Appx.~\ref{app:group_theory}.
After the change of basis
\begin{align}
\mathcal{B} &= \begin{pmatrix}
1 & 0 & 0 & 0 & 0 \\[2pt]
0 & \tfrac{1}{2} & \tfrac{1}{2} & -\tfrac{1}{2} & -\tfrac{1}{2} \\[2pt]
0 & \tfrac{1}{2} & -\tfrac{1}{2} & -\tfrac{1}{2} & \tfrac{1}{2} \\[2pt]
0 & \frac{1}{\sqrt{2}} & 0 & \frac{1}{\sqrt{2}} & 0 \\[2pt]
0 & 0 & \frac{1}{\sqrt{2}} & 0 & \frac{1}{\sqrt{2}} \label{eq:mathcalB-class-mat}
\end{pmatrix},
\end{align}
the representation $O$ takes the block-diagonal form of a direct sum of irreps:
\begin{align}
\overline{O}(g) = \mathcal{B} O(g) \mathcal{B}^{\intercal} &= \begin{pmatrix}
\RepM^{B_{1g}}(g) & & & \\
& \RepM^{A_{1g}}(g) & & \\
& & \RepM^{B_{1g}}(g) & \\
& & & \RepM^{E_u}(g)
\end{pmatrix}.
\end{align}
We shall use overlines to designate matrices in the rotated basis.
Notice that $\mathcal{B}^{*} = \mathcal{B}$ is real and orthogonal, $\mathcal{B}^{\intercal} \mathcal{B} = \one$.
For the 2D $E_g$ and $E_u$ representations we always use the matrices (Eqs.~\eqref{eq:Egu-irrep-mat-conventions1} and~\eqref{eq:Egu-irrep-mat-conventions2}, Sec.~\ref{sec:examples-convention-D4h}):
\begin{align}
\RepM^{E}(C_{4z}) &= \begin{pmatrix}
 0 & -1 \\
 1 & 0 \\
\end{pmatrix}, &
\RepM^{E}(C_{2x}) &= \begin{pmatrix}
 1 & 0 \\
 0 & -1 \\
\end{pmatrix}, &
\RepM^{E}(C_{2d_+}) &= \begin{pmatrix}
 0 & 1 \\
 1 & 0 \\
\end{pmatrix}.
\end{align}

\begin{table}[t]
\centering
\captionabove[Statistics of the classification of extended-basis orbital $\Lambda$ matrices belonging to the three-orbital \ce{CuO2} model.]{\textbf{Statistics of the classification of extended-basis orbital $\Lambda$ matrices belonging to the three-orbital \ce{CuO2} model.}
Table entries indicate the number of Hermitian $5 \times 5$ $\Lambda$ matrices which transform according to Eqs.~\eqref{eq:Lambda-transf-rule1} and~\eqref{eq:Lambda-transf-rule2}.
The $D_{4h}$ irrep $\zeta$ is specified by the corresponding row, while the time-reversal (TR) sign $p_{\TRop}$ is specified by the corresponding column.
The irrep $E_u$ is two-dimensional.
The last row is the net number of TR-even and TR-odd matrices, which coincides with the number of symmetric and antisymmetric Hermitian $5 \times 5$ matrices.}
{\renewcommand{\arraystretch}{1.3}
\renewcommand{\tabcolsep}{10pt}
\begin{tabular}{c|cc} \hline\hline
& TR-even & TR-odd \\ \hline
$A_{1g}$ & $5$ & $1$ \\
$A_{2g}$ & $0$ & $1$ \\
$B_{1g}$ & $3$ & $2$ \\
$B_{2g}$ & $1$ & $0$ \\
$E_u$ & $3 \times 2$ & $3 \times 2$ \\ \hline
$\sum$ & $15 = \frac{5 \times 6}{2}$ & $10 = \frac{5 \times 4}{2}$
\\[2pt] \hline\hline
\end{tabular}}
\label{tab:bilin-classification-stats}
\end{table}

With the help of the irrep product Tab.~\ref{tab:D4h-irrep-prod-tab}, the orbital matrices can now be classified in a straightforward way.
Schematically, we may write
\begin{align}
\overline{\Lambda} = \mathcal{B} \Lambda \mathcal{B}^{\intercal} \sim \overline{O} \otimes \overline{O} \sim \begin{pmatrix}
A_{1g} & B_{1g} & A_{1g} & E_u & E_u \\
B_{1g} & A_{1g} & B_{1g} & \textcolor{red}{E_u} & \textcolor{red}{E_u} \\
A_{1g} & B_{1g} & A_{1g} & \textcolor{green!70!black}{E_u} & \textcolor{green!70!black}{E_u} \\
E_u & \textcolor{red}{E_u} & \textcolor{green!70!black}{E_u} & \textcolor{blue}{\{A_{1g},} & \textcolor{blue}{A_{2g},} \\
E_u & \textcolor{red}{E_u} & \textcolor{green!70!black}{E_u} & \textcolor{blue}{B_{1g},} & \textcolor{blue}{B_{2g}\}}
\end{pmatrix},
\end{align}
where for $E_u$, components of the same color go together.
The arbitrariness in the definition of the various $\Lambda$-s we partially eliminate by making them Hermitian, $\Lambda^{\dag} = \Lambda$,  as well as orthogonal and normalized according to:
\begin{align}
\Tr \Lambda^{\zeta}_{n,a} \Lambda^{\xi}_{m,b} &= 2 \Kd_{\zeta \xi} \Kd_{n m} \Kd_{ab}, \label{eq:Lambda-orthonormalization}
\end{align}
where $\zeta, \xi$ are irreps of $D_{4h}$, $n, m \in \{1, 2, \ldots\}$ enumerate the orbital matrices belonging to each irrep, and $a \in \{1, \ldots, \dim \zeta\}$, $b \in \{1, \ldots, \dim \xi\}$ are irrep component indices which are only relevant for the 2D irrep $E_u$.

In total, there are $5 \times 5 = 25$ orbital matrices, of which $15$ are symmetric and TR-even and $10$ are antisymmetric and TR-odd.
How they fall into the various irreps is summarized in Tab.~\ref{tab:bilin-classification-stats}.
The final results are shown in Tab.~\ref{tab:orbital-Lambda-mats} of the next section.
In the text, the orbital matrices of Tab.~\ref{tab:orbital-Lambda-mats} we shall denote $\Lambda^{\zeta^{p_{\TRop}}}_{n,a}$, like for instance:
\begin{align}
\Lambda^{B_{1g}^{-}}_{2} &= \frac{1}{2} \begin{pmatrix}
 0 & 0 & 0 & 0 & 0 \\
 0 & 0 & \iu & 0 & -\iu \\
 0 & -\iu & 0 & \iu & 0 \\
 0 & 0 & -\iu & 0 & \iu \\
 0 & \iu & 0 & -\iu & 0 \\
\end{pmatrix}, &
\Lambda^{E_{u}^{+}}_{1,y} &= \frac{1}{\sqrt{2}} \begin{pmatrix}
 0 & 0 & -1 & 0 & -1 \\
 0 & 0 & 0 & 0 & 0 \\
 -1 & 0 & 0 & 0 & 0 \\
 0 & 0 & 0 & 0 & 0 \\
 -1 & 0 & 0 & 0 & 0 \\
\end{pmatrix}.
\end{align}

\newpage
\subsubsection{Table of orbital matrices, classified according to symmetry} \label{sec:orbital-Lambda-mats}

{\small
\renewcommand{\arraystretch}{1.3}
\renewcommand{\tabcolsep}{5pt}
\begin{longtable}[c]{lcC{0.26\textwidth}C{0.26\textwidth}C{0.24\textwidth}}
\caption[The symmetry classification of extended-basis orbital $\Lambda$ matrices belonging to the three-orbital \ce{CuO2} model.]{\textbf{The symmetry classification of extended-basis orbital $\Lambda$ matrices belonging to the three-orbital \ce{CuO2} model.}
The symmetry transformation rule are given in Eqs.~\eqref{eq:Lambda-transf-rule1} and~\eqref{eq:Lambda-transf-rule2}.
The $D_{4h}$ irrep $\zeta$ (Tab.~\ref{tab:D4h-char-tab-again}) is specified in the first column, with the superscript denoting the time-reversal sign (TRs) $p_{\TRop}$.
$n$ enumerates the matrices belonging to each irrep, while the irrep index $a$, relevant only to the 2D irrep $E_u$, indicates to which component a given matrix corresponds to.
$\Lambda$ and $\overline{\Lambda}$ are the orbital matrices in the original and rotated basis, respectively.
Since $\mathcal{B}^{*} = \mathcal{B}$ is real [Eq.~\eqref{eq:mathcalB-class-mat}], real matrices are TR-even and imaginary matrices are TR-odd.
All matrices are Hermitian.
The last column is a graphical representation of $\Lambda$.
TR-even $\Lambda$ represent densities, with yellow (cyan) denoting positive (negative) superpositions, while TR-odd $\Lambda$ represent currents, denoted with arrows.
To ensure that the schematics are physical and orbital convention-invariant (cf.\ Sec.~\ref{sec:ASV-conventions}), bond densities ($\sim \Psi_i^{\dag} \Psi_j + \Psi_j^{\dag} \Psi_i$) and currents ($\sim \iu \Psi_i^{\dag} \Psi_j - \iu \Psi_j^{\dag} \Psi_i$) have been consistently multiplied with the sign of the overlap (hopping $t_{ij}$) between the $i$ and $j$ orbitals.
Note that the $\Lambda^{A_{2g}^{-}}_1$, $\Lambda^{E_{u}^{-}}_{2,x/y}$, and $\Lambda^{E_{u}^{-}}_{3,x/y}$ matrices are minus those of Ref.~\cite{Palle2024-LC}.} \label{tab:orbital-Lambda-mats} \\[208pt] \hline\hline
{\normalsize irrep\textsuperscript{TRs}} & {\large $n, a$} & \raisebox{-2pt}{\large $\overline{\Lambda} = \mathcal{B} \Lambda \mathcal{B}^{\intercal}$} & \raisebox{-2pt}{\large $\Lambda$} & {\normalsize schematic} \endfirsthead
\caption[]{(continued)} \\[20pt] \hline\hline
{\normalsize irrep\textsuperscript{TRs}} & {\large $n, a$} & \raisebox{-2pt}{\large $\overline{\Lambda} = \mathcal{B} \Lambda \mathcal{B}^{\intercal}$} & \raisebox{-2pt}{\large $\Lambda$} & {\normalsize schematic} \endhead
\hline
{\LARGE $A_{1g}^{+}$} & {\large $1$} & $\begin{pmatrix}
 \sqrt{2} & 0 & 0 & 0 & 0 \\
 0 & 0 & 0 & 0 & 0 \\
 0 & 0 & 0 & 0 & 0 \\
 0 & 0 & 0 & 0 & 0 \\
 0 & 0 & 0 & 0 & 0 \\
\end{pmatrix}$ & $\begin{pmatrix}
 \sqrt{2} & 0 & 0 & 0 & 0 \\
 0 & 0 & 0 & 0 & 0 \\
 0 & 0 & 0 & 0 & 0 \\
 0 & 0 & 0 & 0 & 0 \\
 0 & 0 & 0 & 0 & 0 \\
\end{pmatrix}$ & \raisebox{-28pt}{\includegraphics[width=0.23\textwidth]{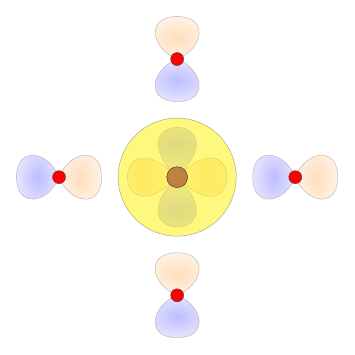}} \\
& {\large $2$} & $\displaystyle \frac{1}{\sqrt{2}} \begin{pmatrix}
 0 & 0 & 0 & 0 & 0 \\
 0 & 1 & 0 & 0 & 0 \\
 0 & 0 & 1 & 0 & 0 \\
 0 & 0 & 0 & 1 & 0 \\
 0 & 0 & 0 & 0 & 1 \\
\end{pmatrix}$ & $\displaystyle \frac{1}{\sqrt{2}} \begin{pmatrix}
 0 & 0 & 0 & 0 & 0 \\
 0 & 1 & 0 & 0 & 0 \\
 0 & 0 & 1 & 0 & 0 \\
 0 & 0 & 0 & 1 & 0 \\
 0 & 0 & 0 & 0 & 1 \\
\end{pmatrix}$ & \raisebox{-28pt}{\includegraphics[width=0.23\textwidth]{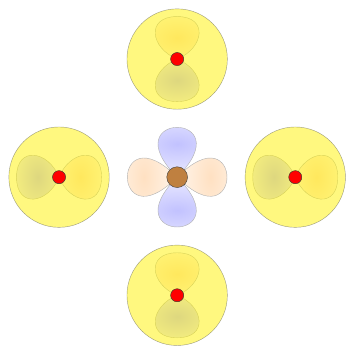}} \\
& {\large $3$} & $\begin{pmatrix}
 0 & 0 & 1 & 0 & 0 \\
 0 & 0 & 0 & 0 & 0 \\
 1 & 0 & 0 & 0 & 0 \\
 0 & 0 & 0 & 0 & 0 \\
 0 & 0 & 0 & 0 & 0 \\
\end{pmatrix}$ & $\displaystyle \frac{1}{2} \begin{pmatrix}
 0 & 1 & -1 & -1 & 1 \\
 1 & 0 & 0 & 0 & 0 \\
 -1 & 0 & 0 & 0 & 0 \\
 -1 & 0 & 0 & 0 & 0 \\
 1 & 0 & 0 & 0 & 0 \\
\end{pmatrix}$ & \raisebox{-28pt}{\includegraphics[width=0.23\textwidth]{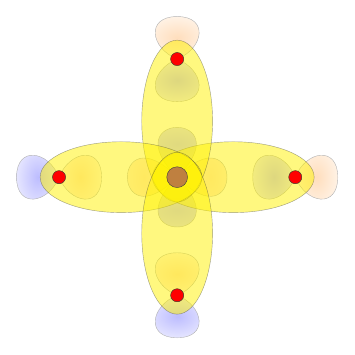}} \\ \hline\hline
\pagebreak \hline
{\LARGE $A_{1g}^{+}$} & {\large $4$} & $\begin{pmatrix}
 0 & 0 & 0 & 0 & 0 \\
 0 & -1 & 0 & 0 & 0 \\
 0 & 0 & 1 & 0 & 0 \\
 0 & 0 & 0 & 0 & 0 \\
 0 & 0 & 0 & 0 & 0 \\
\end{pmatrix}$ & $\displaystyle \frac{1}{2} \begin{pmatrix}
 0 & 0 & 0 & 0 & 0 \\
 0 & 0 & -1 & 0 & 1 \\
 0 & -1 & 0 & 1 & 0 \\
 0 & 0 & 1 & 0 & -1 \\
 0 & 1 & 0 & -1 & 0 \\
\end{pmatrix}$ & \raisebox{-28pt}{\includegraphics[width=0.23\textwidth]{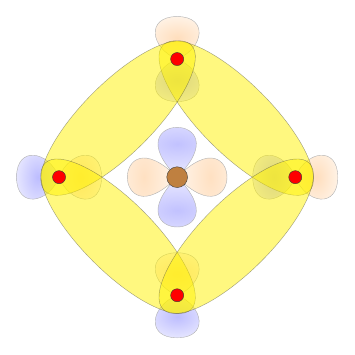}} \\
& {\large $5$} & $\displaystyle \frac{1}{\sqrt{2}} \begin{pmatrix}
 0 & 0 & 0 & 0 & 0 \\
 0 & -1 & 0 & 0 & 0 \\
 0 & 0 & -1 & 0 & 0 \\
 0 & 0 & 0 & 1 & 0 \\
 0 & 0 & 0 & 0 & 1 \\
\end{pmatrix}$ & $\displaystyle \frac{1}{\sqrt{2}} \begin{pmatrix}
 0 & 0 & 0 & 0 & 0 \\
 0 & 0 & 0 & 1 & 0 \\
 0 & 0 & 0 & 0 & 1 \\
 0 & 1 & 0 & 0 & 0 \\
 0 & 0 & 1 & 0 & 0 \\
\end{pmatrix}$ & \raisebox{-28pt}{\includegraphics[width=0.23\textwidth]{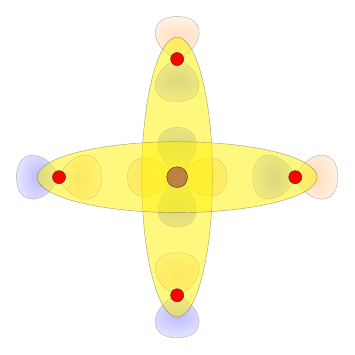}} \\ \hline
{\LARGE $A_{1g}^{-}$} & {\large $1$} & $\begin{pmatrix}
 0 & 0 & -\iu & 0 & 0 \\
 0 & 0 & 0 & 0 & 0 \\
 \iu & 0 & 0 & 0 & 0 \\
 0 & 0 & 0 & 0 & 0 \\
 0 & 0 & 0 & 0 & 0 \\
\end{pmatrix}$ & $\displaystyle \frac{1}{2} \begin{pmatrix}
 0 & -\iu & \iu & \iu & -\iu \\
 \iu & 0 & 0 & 0 & 0 \\
 -\iu & 0 & 0 & 0 & 0 \\
 -\iu & 0 & 0 & 0 & 0 \\
 \iu & 0 & 0 & 0 & 0 \\
\end{pmatrix}$ & \raisebox{-28pt}{\includegraphics[width=0.23\textwidth]{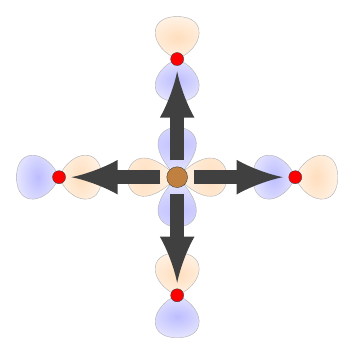}} \\ \hline
{\LARGE $A_{2g}^{-}$} & {\large $1$} & $\begin{pmatrix}
 0 & 0 & 0 & 0 & 0 \\
 0 & 0 & 0 & 0 & 0 \\
 0 & 0 & 0 & 0 & 0 \\
 0 & 0 & 0 & 0 & \iu \\
 0 & 0 & 0 & -\iu & 0 \\
\end{pmatrix}$ & $\displaystyle \frac{1}{2} \begin{pmatrix}
 0 & 0 & 0 & 0 & 0 \\
 0 & 0 & \iu & 0 & \iu \\
 0 & -\iu & 0 & -\iu & 0 \\
 0 & 0 & \iu & 0 & \iu \\
 0 & -\iu & 0 & -\iu & 0 \\
\end{pmatrix}$ & \raisebox{-28pt}{\includegraphics[width=0.23\textwidth]{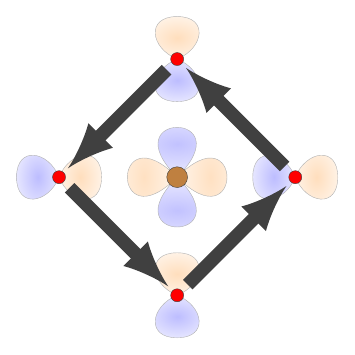}} \\  \hline\hline
\pagebreak \hline
{\LARGE $B_{1g}^{+}$} & {\large $1$} & $\begin{pmatrix}
 0 & 1 & 0 & 0 & 0 \\
 1 & 0 & 0 & 0 & 0 \\
 0 & 0 & 0 & 0 & 0 \\
 0 & 0 & 0 & 0 & 0 \\
 0 & 0 & 0 & 0 & 0 \\
\end{pmatrix}$ & $\displaystyle \frac{1}{2} \begin{pmatrix}
 0 & 1 & 1 & -1 & -1 \\
 1 & 0 & 0 & 0 & 0 \\
 1 & 0 & 0 & 0 & 0 \\
 -1 & 0 & 0 & 0 & 0 \\
 -1 & 0 & 0 & 0 & 0 \\
\end{pmatrix}$ & \raisebox{-28pt}{\includegraphics[width=0.23\textwidth]{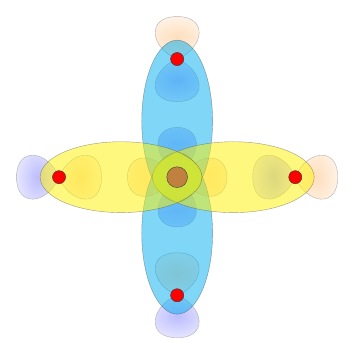}} \\
& {\large $2$} & $\displaystyle \frac{1}{\sqrt{2}} \begin{pmatrix}
 0 & 0 & 0 & 0 & 0 \\
 0 & 0 & 1 & 0 & 0 \\
 0 & 1 & 0 & 0 & 0 \\
 0 & 0 & 0 & 1 & 0 \\
 0 & 0 & 0 & 0 & -1 \\
\end{pmatrix}$ & $\displaystyle \frac{1}{\sqrt{2}} \begin{pmatrix}
 0 & 0 & 0 & 0 & 0 \\
 0 & 1 & 0 & 0 & 0 \\
 0 & 0 & -1 & 0 & 0 \\
 0 & 0 & 0 & 1 & 0 \\
 0 & 0 & 0 & 0 & -1 \\
\end{pmatrix}$ & \raisebox{-28pt}{\includegraphics[width=0.23\textwidth]{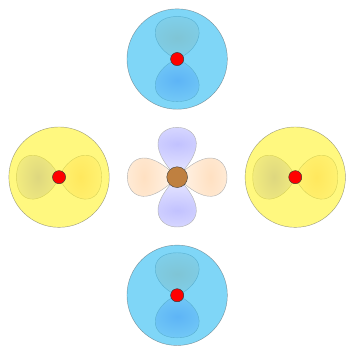}} \\
& {\large $3$} & $\displaystyle \frac{1}{\sqrt{2}} \begin{pmatrix}
 0 & 0 & 0 & 0 & 0 \\
 0 & 0 & -1 & 0 & 0 \\
 0 & -1 & 0 & 0 & 0 \\
 0 & 0 & 0 & 1 & 0 \\
 0 & 0 & 0 & 0 & -1 \\
\end{pmatrix}$ & $\displaystyle \frac{1}{\sqrt{2}} \begin{pmatrix}
 0 & 0 & 0 & 0 & 0 \\
 0 & 0 & 0 & 1 & 0 \\
 0 & 0 & 0 & 0 & -1 \\
 0 & 1 & 0 & 0 & 0 \\
 0 & 0 & -1 & 0 & 0 \\
\end{pmatrix}$ & \raisebox{-28pt}{\includegraphics[width=0.23\textwidth]{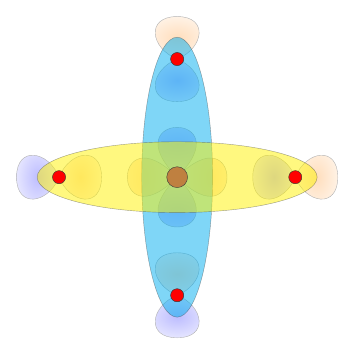}} \\ \hline
{\LARGE $B_{1g}^{-}$} & {\large $1$} & $\begin{pmatrix}
 0 & -\iu & 0 & 0 & 0 \\
 \iu & 0 & 0 & 0 & 0 \\
 0 & 0 & 0 & 0 & 0 \\
 0 & 0 & 0 & 0 & 0 \\
 0 & 0 & 0 & 0 & 0 \\
\end{pmatrix}$ & $\displaystyle \frac{1}{2} \begin{pmatrix}
 0 & -\iu & -\iu & \iu & \iu \\
 \iu & 0 & 0 & 0 & 0 \\
 \iu & 0 & 0 & 0 & 0 \\
 -\iu & 0 & 0 & 0 & 0 \\
 -\iu & 0 & 0 & 0 & 0 \\
\end{pmatrix}$ & \raisebox{-28pt}{\includegraphics[width=0.23\textwidth]{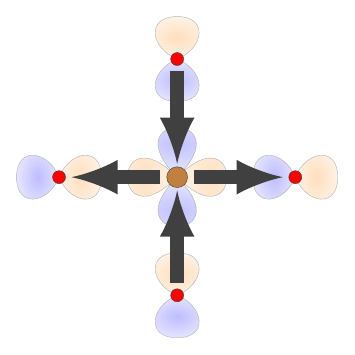}} \\
& {\large $2$} & $\begin{pmatrix}
 0 & 0 & 0 & 0 & 0 \\
 0 & 0 & -\iu & 0 & 0 \\
 0 & \iu & 0 & 0 & 0 \\
 0 & 0 & 0 & 0 & 0 \\
 0 & 0 & 0 & 0 & 0 \\
\end{pmatrix}$ & $\displaystyle \frac{1}{2} \begin{pmatrix}
 0 & 0 & 0 & 0 & 0 \\
 0 & 0 & \iu & 0 & -\iu \\
 0 & -\iu & 0 & \iu & 0 \\
 0 & 0 & -\iu & 0 & \iu \\
 0 & \iu & 0 & -\iu & 0 \\
\end{pmatrix}$ & \raisebox{-28pt}{\includegraphics[width=0.23\textwidth]{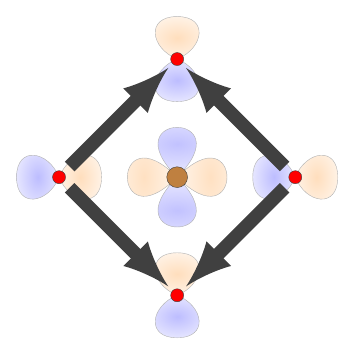}} \\  \hline\hline
\pagebreak \hline
{\LARGE $B_{2g}^{+}$} & {\large $1$} & $\begin{pmatrix}
 0 & 0 & 0 & 0 & 0 \\
 0 & 0 & 0 & 0 & 0 \\
 0 & 0 & 0 & 0 & 0 \\
 0 & 0 & 0 & 0 & 1 \\
 0 & 0 & 0 & 1 & 0 \\
\end{pmatrix}$ & $\displaystyle \frac{1}{2} \begin{pmatrix}
 0 & 0 & 0 & 0 & 0 \\
 0 & 0 & 1 & 0 & 1 \\
 0 & 1 & 0 & 1 & 0 \\
 0 & 0 & 1 & 0 & 1 \\
 0 & 1 & 0 & 1 & 0 \\
\end{pmatrix}$ & \raisebox{-28pt}{\includegraphics[width=0.23\textwidth]{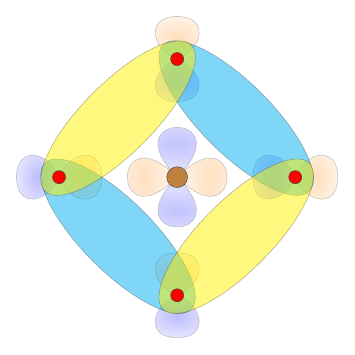}} \\ \hline
{\LARGE $E_{u}^{+}$} & {\large $1,x$} & $\begin{pmatrix}
 0 & 0 & 0 & 1 & 0 \\
 0 & 0 & 0 & 0 & 0 \\
 0 & 0 & 0 & 0 & 0 \\
 1 & 0 & 0 & 0 & 0 \\
 0 & 0 & 0 & 0 & 0 \\
\end{pmatrix}$ & $\displaystyle \frac{1}{\sqrt{2}} \begin{pmatrix}
 0 & 1 & 0 & 1 & 0 \\
 1 & 0 & 0 & 0 & 0 \\
 0 & 0 & 0 & 0 & 0 \\
 1 & 0 & 0 & 0 & 0 \\
 0 & 0 & 0 & 0 & 0 \\
\end{pmatrix}$ & \raisebox{-28pt}{\includegraphics[width=0.23\textwidth]{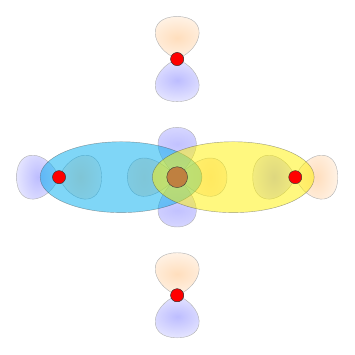}} \\
& {\large $1,y$} & $\begin{pmatrix}
 0 & 0 & 0 & 0 & -1 \\
 0 & 0 & 0 & 0 & 0 \\
 0 & 0 & 0 & 0 & 0 \\
 0 & 0 & 0 & 0 & 0 \\
 -1 & 0 & 0 & 0 & 0 \\
\end{pmatrix}$ & $\displaystyle \frac{1}{\sqrt{2}} \begin{pmatrix}
 0 & 0 & -1 & 0 & -1 \\
 0 & 0 & 0 & 0 & 0 \\
 -1 & 0 & 0 & 0 & 0 \\
 0 & 0 & 0 & 0 & 0 \\
 -1 & 0 & 0 & 0 & 0 \\
\end{pmatrix}$ & \raisebox{-28pt}{\includegraphics[width=0.23\textwidth]{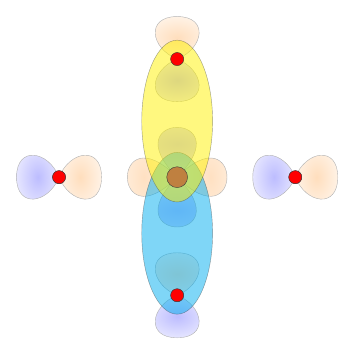}} \\
& {\large $2,x$} & $\displaystyle \frac{1}{\sqrt{2}} \begin{pmatrix}
 0 & 0 & 0 & 0 & 0 \\
 0 & 0 & 0 & 1 & 0 \\
 0 & 0 & 0 & 1 & 0 \\
 0 & 1 & 1 & 0 & 0 \\
 0 & 0 & 0 & 0 & 0 \\
\end{pmatrix}$ & $\begin{pmatrix}
 0 & 0 & 0 & 0 & 0 \\
 0 & 1 & 0 & 0 & 0 \\
 0 & 0 & 0 & 0 & 0 \\
 0 & 0 & 0 & -1 & 0 \\
 0 & 0 & 0 & 0 & 0 \\
\end{pmatrix}$ & \raisebox{-28pt}{\includegraphics[width=0.23\textwidth]{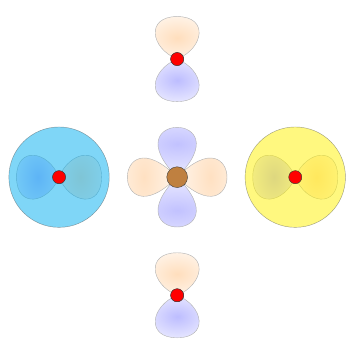}} \\
& {\large $2,y$} & $\displaystyle \frac{1}{\sqrt{2}} \begin{pmatrix}
 0 & 0 & 0 & 0 & 0 \\
 0 & 0 & 0 & 0 & 1 \\
 0 & 0 & 0 & 0 & -1 \\
 0 & 0 & 0 & 0 & 0 \\
 0 & 1 & -1 & 0 & 0 \\
\end{pmatrix}$ & $\begin{pmatrix}
 0 & 0 & 0 & 0 & 0 \\
 0 & 0 & 0 & 0 & 0 \\
 0 & 0 & 1 & 0 & 0 \\
 0 & 0 & 0 & 0 & 0 \\
 0 & 0 & 0 & 0 & -1 \\
\end{pmatrix}$ & \raisebox{-28pt}{\includegraphics[width=0.23\textwidth]{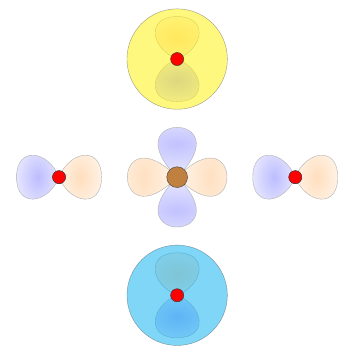}} \\ \hline\hline
\pagebreak \hline
{\LARGE $E_{u}^{+}$} & {\large $3,x$} & $\displaystyle \frac{1}{\sqrt{2}} \begin{pmatrix}
 0 & 0 & 0 & 0 & 0 \\
 0 & 0 & 0 & 1 & 0 \\
 0 & 0 & 0 & -1 & 0 \\
 0 & 1 & -1 & 0 & 0 \\
 0 & 0 & 0 & 0 & 0 \\
\end{pmatrix}$ & $\displaystyle \frac{1}{2} \begin{pmatrix}
 0 & 0 & 0 & 0 & 0 \\
 0 & 0 & 1 & 0 & -1 \\
 0 & 1 & 0 & 1 & 0 \\
 0 & 0 & 1 & 0 & -1 \\
 0 & -1 & 0 & -1 & 0 \\
\end{pmatrix}$ & \raisebox{-28pt}{\includegraphics[width=0.23\textwidth]{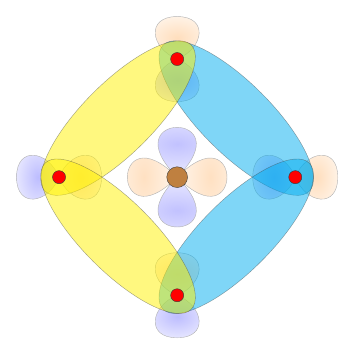}} \\
& {\large $3,y$} & $\displaystyle \frac{1}{\sqrt{2}} \begin{pmatrix}
 0 & 0 & 0 & 0 & 0 \\
 0 & 0 & 0 & 0 & 1 \\
 0 & 0 & 0 & 0 & 1 \\
 0 & 0 & 0 & 0 & 0 \\
 0 & 1 & 1 & 0 & 0 \\
\end{pmatrix}$ & $\displaystyle \frac{1}{2} \begin{pmatrix}
 0 & 0 & 0 & 0 & 0 \\
 0 & 0 & 1 & 0 & 1 \\
 0 & 1 & 0 & -1 & 0 \\
 0 & 0 & -1 & 0 & -1 \\
 0 & 1 & 0 & -1 & 0 \\
\end{pmatrix}$ & \raisebox{-28pt}{\includegraphics[width=0.23\textwidth]{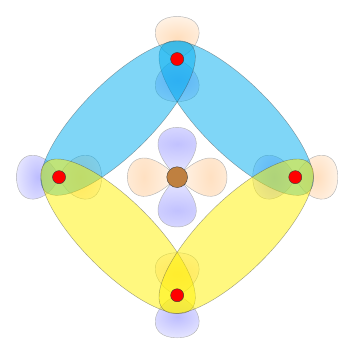}} \\ \hline
{\LARGE $E_{u}^{-}$} & {\large $1,x$} & $\begin{pmatrix}
 0 & 0 & 0 & -\iu & 0 \\
 0 & 0 & 0 & 0 & 0 \\
 0 & 0 & 0 & 0 & 0 \\
 \iu & 0 & 0 & 0 & 0 \\
 0 & 0 & 0 & 0 & 0 \\
\end{pmatrix}$ & $\displaystyle \frac{1}{\sqrt{2}} \begin{pmatrix}
 0 & -\iu & 0 & -\iu & 0 \\
 \iu & 0 & 0 & 0 & 0 \\
 0 & 0 & 0 & 0 & 0 \\
 \iu & 0 & 0 & 0 & 0 \\
 0 & 0 & 0 & 0 & 0 \\
\end{pmatrix}$ & \raisebox{-28pt}{\includegraphics[width=0.23\textwidth]{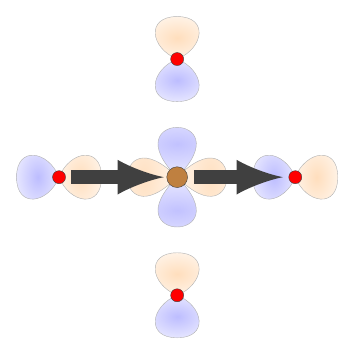}} \\
& {\large $1,y$} & $\begin{pmatrix}
 0 & 0 & 0 & 0 & \iu \\
 0 & 0 & 0 & 0 & 0 \\
 0 & 0 & 0 & 0 & 0 \\
 0 & 0 & 0 & 0 & 0 \\
 -\iu & 0 & 0 & 0 & 0 \\
\end{pmatrix}$ & $\displaystyle \frac{1}{\sqrt{2}} \begin{pmatrix}
 0 & 0 & \iu & 0 & \iu \\
 0 & 0 & 0 & 0 & 0 \\
 -\iu & 0 & 0 & 0 & 0 \\
 0 & 0 & 0 & 0 & 0 \\
 -\iu & 0 & 0 & 0 & 0 \\
\end{pmatrix}$ & \raisebox{-28pt}{\includegraphics[width=0.23\textwidth]{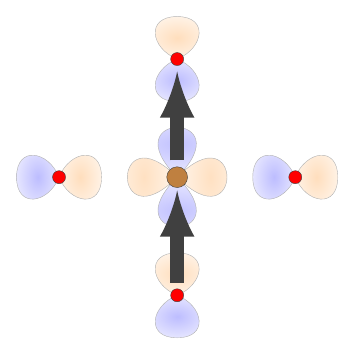}} \\ \hline\hline
\pagebreak \hline
{\LARGE $E_{u}^{-}$} & {\large $2,x$} & $\displaystyle \frac{1}{\sqrt{2}} \begin{pmatrix}
 0 & 0 & 0 & 0 & 0 \\
 0 & 0 & 0 & \iu & 0 \\
 0 & 0 & 0 & \iu & 0 \\
 0 & -\iu & -\iu & 0 & 0 \\
 0 & 0 & 0 & 0 & 0 \\
\end{pmatrix}$ & $\begin{pmatrix}
 0 & 0 & 0 & 0 & 0 \\
 0 & 0 & 0 & \iu & 0 \\
 0 & 0 & 0 & 0 & 0 \\
 0 & -\iu & 0 & 0 & 0 \\
 0 & 0 & 0 & 0 & 0 \\
\end{pmatrix}$ & \raisebox{-28pt}{\includegraphics[width=0.23\textwidth]{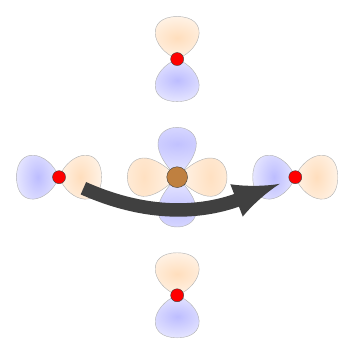}} \\
& {\large $2,y$} & $\displaystyle \frac{1}{\sqrt{2}} \begin{pmatrix}
 0 & 0 & 0 & 0 & 0 \\
 0 & 0 & 0 & 0 & \iu \\
 0 & 0 & 0 & 0 & -\iu \\
 0 & 0 & 0 & 0 & 0 \\
 0 & -\iu & \iu & 0 & 0 \\
\end{pmatrix}$ & $\begin{pmatrix}
 0 & 0 & 0 & 0 & 0 \\
 0 & 0 & 0 & 0 & 0 \\
 0 & 0 & 0 & 0 & \iu \\
 0 & 0 & 0 & 0 & 0 \\
 0 & 0 & -\iu & 0 & 0 \\
\end{pmatrix}$ & \raisebox{-28pt}{\includegraphics[width=0.23\textwidth]{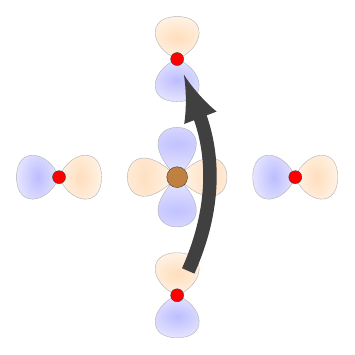}} \\
& {\large $3,x$} & $\displaystyle \frac{1}{\sqrt{2}} \begin{pmatrix}
 0 & 0 & 0 & 0 & 0 \\
 0 & 0 & 0 & \iu & 0 \\
 0 & 0 & 0 & -\iu & 0 \\
 0 & -\iu & \iu & 0 & 0 \\
 0 & 0 & 0 & 0 & 0 \\
\end{pmatrix}$ & $\displaystyle \frac{1}{2} \begin{pmatrix}
 0 & 0 & 0 & 0 & 0 \\
 0 & 0 & -\iu & 0 & \iu \\
 0 & \iu & 0 & \iu & 0 \\
 0 & 0 & -\iu & 0 & \iu \\
 0 & -\iu & 0 & -\iu & 0 \\
\end{pmatrix}$ & \raisebox{-28pt}{\includegraphics[width=0.23\textwidth]{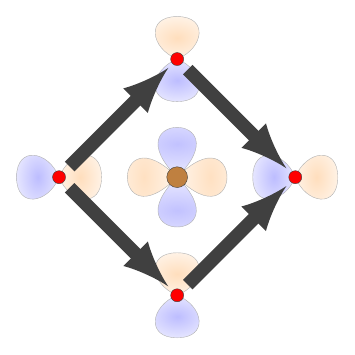}} \\
& {\large $3,y$} & $\displaystyle \frac{1}{\sqrt{2}} \begin{pmatrix}
 0 & 0 & 0 & 0 & 0 \\
 0 & 0 & 0 & 0 & \iu \\
 0 & 0 & 0 & 0 & \iu \\
 0 & 0 & 0 & 0 & 0 \\
 0 & -\iu & -\iu & 0 & 0 \\
\end{pmatrix}$ & $\displaystyle \frac{1}{2} \begin{pmatrix}
 0 & 0 & 0 & 0 & 0 \\
 0 & 0 & \iu & 0 & \iu \\
 0 & -\iu & 0 & \iu & 0 \\
 0 & 0 & -\iu & 0 & -\iu \\
 0 & -\iu & 0 & \iu & 0 \\
\end{pmatrix}$ & \raisebox{-28pt}{\includegraphics[width=0.23\textwidth]{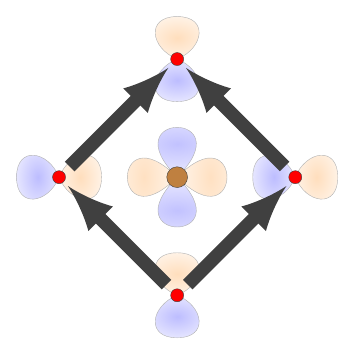}} \\ \hline\hline
\end{longtable}}

\newpage

\subsection{Construction of particle-hole fermionic bilinears} \label{sec:cup-ord-param-constr}
Having separately classified momentum functions, spin matrices, and orbital matrices in the previous two sections, we can now combine them to systematically construct particle-hole fermionic bilinears of any type.
The procedure for doing so is explained here.

Suppose we have a collection of scalar functions $f_{a}(\vb{k})$, orbital matrices $\Lambda_b$, and spin matrices $\Pauli_c$ which transform under the irreps $\RepM^f$, $\RepM^{\Lambda}$, and $\RepM^{\Pauli}$, respectively.
Then the collection of extended-basis $\Gamma$ matrices
\begin{align}
\Gamma_{a b c \vb{k}} &= f_{a}(\vb{k}) \Lambda_b \otimes \Pauli_c
\end{align}
transforms according to [Eqs.~\eqref{eq:Gamma-transf-rule1-bilinear-again-extended} and~\eqref{eq:Gamma-transf-rule2-bilinear-again-extended}]:
\begin{align}
S^{\dag}(g) O^{\dag}(g) \Gamma_{a b c, R(g) \vb{k}} O(g) S(g) &= \sum_{a'=1}^{\dim f} \sum_{b'=1}^{\dim \Lambda} \sum_{c'=1}^{\dim \Pauli} \RepM_{aa'}^{f}(g) \RepM_{bb'}^{\Lambda}(g) \RepM_{cc'}^{\Pauli}(g) \Gamma_{a'b'c' \vb{k}}, \\
(\iu \Pauli_y)^{\dag} \Gamma_{a b c, -\vb{k}}^{*} (\iu \Pauli_y) &= p_{\TRop}^{f} p_{\TRop}^{\Lambda} p_{\TRop}^{\Pauli} \Gamma_{a \vb{k}}.
\end{align}
In other words, $\Gamma_{a b c \vb{k}}$ transforms under the direct product representation $\RepM^f \otimes \RepM^{\Lambda} \otimes \RepM^{\Pauli}$ and it has the TR sign $p_{\TRop}^{f} p_{\TRop}^{\Lambda} p_{\TRop}^{\Pauli}$.
Note that a reality condition, such as $\Gamma_{a b c \vb{k}}^{\dag} = \Gamma_{a b c \vb{k}}$, does not need to be imposed because $\Gamma_{a \vb{k}, \vb{p}}$ as given by Eq.~\eqref{eq:momentum-bilinear-v2} automatically satisfies $\Gamma_{a \vb{k}, \vb{p}}^{\dag} = \Gamma_{a \vb{p}, \vb{k}}$.
That said, if one looks at the $\Gamma_{a \vb{k}, \vb{k}+\vb{q}}$ which enter the $\phi_{a \vb{q}}$ bilinear [Eq.~\eqref{eq:general-momentum-bilinear-again}], one notices that
\begin{align}
\Gamma_{a \vb{k}, \vb{k}+\vb{q}} &= \mathcal{K}_{\vb{k}}^{\dag} \mleft(\Gamma_{a \vb{k}} + \Gamma_{a, \vb{k}+\vb{q}}^{\dag}\mright) \mathcal{K}_{\vb{k}+\vb{q}}
\end{align}
vanishes in the $\vb{q} \to \vb{0}$ limit when $\Gamma_{a \vb{k}}^{\dag} = - \Gamma_{a \vb{k}}$.
Hence for intra-unit-cell orders, only Hermitian $\Gamma_{a \vb{k}}^{\dag} = \Gamma_{a \vb{k}}$ do not vanish at the condensation momentum $\vb{q} = \vb{0}$.

In Sec.~\ref{sec:multid-irrep-product} of Appx.~\ref{app:group_theory}, we have worked out how to decompose composite objects such as $\Gamma_{a b c \vb{k}}$ into irreps.
The results are summarized in the irrep product Tab.~\ref{tab:D4h-irrep-prod-tab}.
The idea is to first decompose $\RepM^f$ and $\RepM^{\Lambda} \otimes \RepM^{\Pauli}$ into irreps and only afterwards decompose $\RepM^f \otimes \mleft(\RepM^{\Lambda} \otimes \RepM^{\Pauli}\mright)$.
Here we give a few examples of how this is done with the help of Tab.~\ref{tab:D4h-irrep-prod-tab}.

Let us start with a purely orbital order, such as orbital current order.
Then $\Pauli_c = \Pauli_0 \in A_{1g}^{+}$ transforms trivially and we can focus on $f_{a}(\vb{k})$ and $\Lambda_b$.
For purely local or contact bilinears all the coupling takes places within the extended unit cell (Fig.~\ref{fig:extended-unit-cell}).
Hence  $f_{a}(\vb{k}) = 1 \in A_{1g}^{+}$ and Tab.~\ref{tab:orbital-Lambda-mats} tells us that there are four possible orbital current bilinears.
Their extended-basis $\Gamma_{a \vb{k}}$ matrices are given by:
\begin{align}
\Gamma_{\vb{k}} &= \Lambda^{A_{1g}^{-}}_1 &\in A_{1g}^{-}, \\
\Gamma_{\vb{k}} &= \Lambda^{A_{2g}^{-}}_1 &\in A_{2g}^{-}, \\
\Gamma_{\vb{k}} &= c_1 \Lambda^{B_{1g}^{-}}_1 + c_2 \Lambda^{B_{1g}^{-}}_2 &\in B_{1g}^{-}, \\
\begin{pmatrix}
\Gamma_{x, \vb{k}} \\
\Gamma_{y, \vb{k}}
\end{pmatrix} &= c_1' \begin{pmatrix}
\Lambda^{E_{u}^{-}}_{1,x} \\
\Lambda^{E_{u}^{-}}_{1,y}
\end{pmatrix} + c_2' \begin{pmatrix}
\Lambda^{E_{u}^{-}}_{2,x} \\
\Lambda^{E_{u}^{-}}_{2,y}
\end{pmatrix} + c_3' \begin{pmatrix}
\Lambda^{E_{u}^{-}}_{3,x} \\
\Lambda^{E_{u}^{-}}_{3,y}
\end{pmatrix} &\in E_{u}^{-}.
\end{align}
Here $c_i$ and $c_j'$ are real coefficients which express the freedom to superimpose bilinears belonging to the same irrep.
The tensor product with $\Pauli_0$ has been suppressed.
If we imagine expanding some Yukawa coupling in powers of momentum, the above would represent the lowest order terms in the expansion.
Let us note that the TR sign cannot be changed by using a purely imaginary $f_{a}(\vb{k}) = \iu \in A_{1g}^{-}$ because the corresponding $\Gamma_{a \vb{k}, \vb{p}}$ then vanish identically.
For instance, plugging
\begin{align}
\Gamma_{\vb{k}} &= \iu \Lambda^{B_{2g}^{+}}_1 &\in B_{2g}^{-}
\end{align}
into Eq.~\eqref{eq:momentum-bilinear-v2} gives
\begin{align}
\Gamma_{\vb{k}, \vb{p}} &= \mathcal{K}_{\vb{k}}^{\dag} \mleft(\mleft[\iu \Lambda^{B_{2g}^{+}}_1\mright] + \mleft[\iu \Lambda^{B_{2g}^{+}}_1\mright]^{\dag}\mright) \mathcal{K}_{\vb{p}} = 0.
\end{align}

To construct orbital current bilinears belonging to $B_{2g}^{-}$, we need to allow for momentum dependence.
From Eq.~\eqref{eq:classification-of-fk-func}, we see that the lowest order lattice functions are $\cos k_x + \cos k_y \in A_{1g}^{+}$, $\cos k_x - \cos k_y \in B_{1g}^{+}$, and $(\sin k_x | \sin k_y) \in E_u^{-}$.
These can be multiplied with the imaginary unit to flip the TR-sign.
Because there is no $A_{2g}^{+}$ orbital matrix (Tab.~\ref{tab:orbital-Lambda-mats}) which we could multiply with $\iu (\cos k_y - \cos k_y) \in B_{1g}^{-}$ to get $B_{2g}^{-}$, the only option which uses the 1D irrep momentum functions is:
\begin{align}
\iu (\cos k_x + \cos k_y) \Lambda^{B_{2g}^{+}}_1 &\in A_{1g}^{-} \otimes B_{2g}^{+} = B_{2g}^{-}.
\end{align}
There are three $E_{u}^{+}$ orbital matrices.
Recalling that (Tab.~\ref{tab:D4h-irrep-prod-tab})
\begin{align}
\begin{aligned}
E_u(f_x | f_y) \otimes E_u(\Lambda_x | \Lambda_y) &= A_{1g}(f_x \Lambda_x + f_y \Lambda_y) \oplus A_{2g}(f_x \Lambda_y - f_y \Lambda_x) \\
&\qquad {} \oplus B_{1g}(f_x \Lambda_x - f_y \Lambda_y) \oplus B_{2g}(f_x \Lambda_y + f_y \Lambda_x),
\end{aligned}
\end{align}
we find another option:
\begin{align}
\sin k_x \mleft(c_1 \Lambda^{E_{u}^{+}}_{1,x} + c_2 \Lambda^{E_{u}^{+}}_{2,x} + c_3 \Lambda^{E_{u}^{+}}_{3,x}\mright) + \sin k_y \mleft(c_1 \Lambda^{E_{u}^{+}}_{1,y} + c_2 \Lambda^{E_{u}^{+}}_{2,y} + c_3 \Lambda^{E_{u}^{+}}_{3,y}\mright) \in B_{2g}^{-}.
\end{align}
This one uses 2D irrep momentum functions.
Altogether:
\begin{align}
\begin{aligned}
\Gamma_{\vb{k}} &= c_0 \, \iu (\cos k_x + \cos k_y) \Lambda^{B_{2g}^{+}}_1 + \sin k_x \mleft(c_1 \Lambda^{E_{u}^{+}}_{1,x} + c_2 \Lambda^{E_{u}^{+}}_{2,x} + c_3 \Lambda^{E_{u}^{+}}_{3,x}\mright) \\
&\qquad {} + \sin k_y \mleft(c_1 \Lambda^{E_{u}^{+}}_{1,y} + c_2 \Lambda^{E_{u}^{+}}_{2,y} + c_3 \Lambda^{E_{u}^{+}}_{3,y}\mright) \hspace{120pt} \in B_{2g}^{-},
\end{aligned}
\end{align}
where $c_i \in \R$.
In agreement with what we previously said, in the homogeneous limit $\vb{q} \to \vb{0}$ the first term which is non-Hermitian vanishes:
\begin{align}
\Gamma_{\vb{k}, \vb{k}+\vb{q}} &= c_0 \, \iu \mleft(\cos k_x + \cos k_y - \cos(k_x+q_x) - \cos(k_y+q_y)\mright) \mathcal{K}_{\vb{k}}^{\dag} \Lambda^{B_{2g}^{+}}_1 \mathcal{K}_{\vb{k}+\vb{q}} + \cdots \, .
\end{align}

We can also ask what type of orbital current bilinears are possible within the one-orbital model of the \ce{CuO2} planes~\cite{Zhang1988, Eskes1988}.
This model is based on the \ce{Cu}:$3d_{x^2-y^2}$ orbital and within it the only possible orbital matrix is
\begin{align}
\Lambda^{A_{1g}^{+}}_1 &= \begin{pmatrix}
 \sqrt{2} & 0 & 0 & 0 & 0 \\
 0 & 0 & 0 & 0 & 0 \\
 0 & 0 & 0 & 0 & 0 \\
 0 & 0 & 0 & 0 & 0 \\
 0 & 0 & 0 & 0 & 0 \\
\end{pmatrix}.
\end{align}
The source of TRSB therefore must lie in the momentum dependence.
The simplest options are:
\begin{align}
\Gamma_{\vb{k}} &= \iu (\cos k_x + \cos k_y) \Lambda^{A_{1g}^{+}}_1 &\in A_{1g}^{-}, \\
\Gamma_{\vb{k}} &= \iu (\cos k_x - \cos k_y) \Lambda^{A_{1g}^{+}}_1 &\in B_{1g}^{-}, \\
\begin{pmatrix}
\Gamma_{x, \vb{k}} \\
\Gamma_{y, \vb{k}}
\end{pmatrix} &= \begin{pmatrix}
\sin k_x \, \Lambda^{A_{1g}^{+}}_1\\[2pt]
\sin k_y \, \Lambda^{A_{1g}^{+}}_1
\end{pmatrix} &\in E_{u}^{-}.
\end{align}
The second $B_{1g}^{-}$ option corresponds to $d$-density waves, which are also known as orbital antiferromagnets or staggered flux states.
For $A_{1g}^{-}$ and $B_{1g}^{-}$ orders of this kind, the ordering must take place at a finite $\vb{q}$, which is usually taken to be $\vb{Q} = (\pi, \pi)$, for $\ev{\phi_{\vb{q}}}$ to be finite.

Up to now, we have simply listed the possible orbital current $\Gamma_{a \vb{k}}$.
These extended-basis matrices define the fermionic bilinears $\phi_{a \vb{q}}$ through Eqs.~\eqref{eq:momentum-bilinear-v2} and~\eqref{eq:general-momentum-bilinear-again}.
If we want to use the expectation value of $\phi_{a \vb{q}}$ as an order parameter, then we have to ensure that $\ev{\phi_{a \vb{q}}}$ is allowed to be finite.
For orbital current orders, in Sec.~\ref{sec:Bloch-tm} of the previous chapter we have seen that the Bloch and Bloch-Kirchhoff theorems fundamentally constrain the orbital current patterns to not have net currents or induce net accumulations of charge.
Within our phenomenological treatment, these constraints on the allowed $\Gamma_{a \vb{k}}$ will have to be enforced by hand, as will be explained in Sec.~\ref{sec:Bloch-Kirch-constr}.
The $\phi_{a \vb{q}}$ correspond to a proper orbital \emph{loop}-current orders only once this is done.

There is a host of other purely orbital bilinears which one can construct.
The possible orbital orders were systematically listed in Tabs.~\ref{tab:orbital-modes} and~\ref{tab:orbital-spin-IUC-orders} of the previous chapter and for each one of them one can construct a bilinear.
For example, here are two extended-basis $\Gamma$ matrices which correspond to nematic and ferroelectric order, respectively:
\begin{align}
\Gamma_{\vb{k}} &= c_1 \Lambda^{B_{1g}^{+}}_1 + c_2 \Lambda^{B_{1g}^{+}}_2 + c_3 \Lambda^{B_{1g}^{+}}_3 + \cdots &\in B_{1g}^{+}, \\
\begin{pmatrix}
\Gamma_{x, \vb{k}} \\
\Gamma_{y, \vb{k}}
\end{pmatrix} &= \begin{pmatrix}
\sin k_x \, \Lambda^{A_{1g}^{-}}_1 \\[2pt]
\sin k_y \, \Lambda^{A_{1g}^{-}}_1
\end{pmatrix} + \cdots &\in E_{u}^{+}.
\end{align}
There are many more.
We shall not pursue this any further since the construction is analogous to the construction of orbital current bilinears.

\begin{table}[t]
\centering
\captionabove[Statistics of the classification of extended-basis spin-orbital matrices $\Gamma = \Lambda \otimes \Pauli$ belonging to the three-orbital \ce{CuO2} model.]{\textbf{Statistics of the classification of extended-basis spin-orbital matrices $\Gamma = \Lambda \otimes \Pauli$ belonging to the three-orbital \ce{CuO2} model.}
Table entries indicate the number of Hermitian $10 \times 10$ momentum-independent $\Gamma$ matrices which transform according to Eqs.~\eqref{eq:Gamma-transf-rule1-bilinear-again-extended} and~\eqref{eq:Gamma-transf-rule2-bilinear-again-extended}.
The $D_{4h}$ irrep $\zeta$ is specified by the corresponding row, while the time-reversal (TR) sign $p_{\TRop}$ is specified by the corresponding column.
The irreps $E_g$ and $E_u$ are two-dimensional.
The last row is the net number of TR-even and TR-odd matrices.}
{\renewcommand{\arraystretch}{1.3}
\renewcommand{\tabcolsep}{10pt}
\hspace*{\stretch{1}} \begin{tabular}{c|cc} \hline\hline
\multicolumn{3}{c}{out-of-plane spin ($\otimes \, \Pauli_z$)} \\[2pt]
& TR-even & TR-odd \\ \hline
$A_{1g}$ & $1$ & $0$ \\
$A_{2g}$ & $1$ & $5$ \\
$B_{1g}$ & $0$ & $1$ \\
$B_{2g}$ & $2$ & $3$ \\
$E_u$ & $3 \times 2$ & $3 \times 2$ \\ \hline
$\sum$ & $10$ & $15$
\\[2pt] \hline\hline
\end{tabular} \hspace*{\stretch{1}} \begin{tabular}{c|cc} \hline\hline
\multicolumn{3}{c}{in-plane spin ($\otimes \, \Pauli_{x,y}$)} \\[2pt]
& TR-even & TR-odd \\ \hline
$A_{1u}$ & $3$ & $3$ \\
$A_{2u}$ & $3$ & $3$ \\
$B_{1u}$ & $3$ & $3$ \\
$B_{2u}$ & $3$ & $3$ \\
$E_g$ & $4 \times 2$ & $9 \times 2$ \\ \hline
$\sum$ & $20$ & $30$
\\[2pt]  \hline\hline
\end{tabular} \hspace*{\stretch{1}}}
\label{tab:SOC-bilin-classification-stats}
\end{table}

Regarding spin orders, there are three possible Pauli matrices which combine with the orbital matrices to give a net of $3 \times 25 = 75$ possible spin-orbit matrices, which we shall denote with a $\Gamma$.
Given that we know the irreps and TR signs of the orbital $\Lambda$ matrices (Tab.~\ref{tab:orbital-Lambda-mats}) and of the spin $\Pauli$ matrices [Eq.~\eqref{eq:classification-of-Pauli-mat}] and that we also know how to decompose their direct products (Tab.~\ref{tab:D4h-irrep-prod-tab}), working out the irreps and TR sign of the $75$ spin-orbit matrices is a straightforward task.
We shall not go through all the matrices, however.
Instead, we shall simply list how many spin-orbit matrices belong to each irrep in Tab.~\ref{tab:SOC-bilin-classification-stats} and go through a few examples below.

For example, let us see in how many ways can one construct a $A_{2g}^{-}$ spin-orbit matrix.
Tab.~\ref{tab:D4h-irrep-prod-tab} informs us that $A_{2g}^{-}$ can arise only by multiplying $\Lambda \in A_{1g}^{+}$ with $\Pauli_z \in A_{2g}^{-}$ or $(\Lambda_x | \Lambda_y) \in E_{g}^{+}$ with $(\Pauli_x | \Pauli_y) \in E_{g}^{-}$.
Since $\Lambda$ matrices transforming under  $E_g$ do not exist, we find that the most general spin-orbit $A_{2g}^{-}$ matrix is:
\begin{align}
\Gamma &= \mleft(c_1 \Lambda^{A_{1g}^{+}}_1 + c_2 \Lambda^{A_{1g}^{+}}_2 + c_3 \Lambda^{A_{1g}^{+}}_3 + c_4 \Lambda^{A_{1g}^{+}}_4 + c_5 \Lambda^{A_{1g}^{+}}_5\mright) \otimes \Pauli_z &\in A_{2g}^{-}, \label{eq:Gamma-along-z-1}
\end{align}
where $c_i \in \R$.
Similarly for $B_{1g}^{-}$ and $E_u^{-}$ we find that:
\begin{align}
\Gamma &= \Lambda^{B_{2g}^{+}}_1 \otimes \Pauli_z &\in B_{1g}^{-}, \label{eq:Gamma-along-z-2} \\
\begin{pmatrix}
\Gamma_x \\
\Gamma_y
\end{pmatrix} &= c_1' \begin{pmatrix}
\Lambda^{E_{u}^{-}}_{1,y} \otimes \Pauli_z \\
- \Lambda^{E_{u}^{-}}_{1,x} \otimes \Pauli_z
\end{pmatrix} + c_2' \begin{pmatrix}
\Lambda^{E_{u}^{-}}_{2,y} \otimes \Pauli_z \\
- \Lambda^{E_{u}^{-}}_{2,x} \otimes \Pauli_z
\end{pmatrix} + c_3' \begin{pmatrix}
\Lambda^{E_{u}^{-}}_{3,y} \otimes \Pauli_z \\
- \Lambda^{E_{u}^{-}}_{3,x} \otimes \Pauli_z
\end{pmatrix} &\in E_u^{-}. \label{eq:Gamma-along-z-3}
\end{align}
Clearly, allowing for momentum dependence in the $\Gamma_{a \vb{k}}$ matrices opens up even more possibilities.
Here is a non-trivial example where care needs to be taken to ensure the proper ordering of the $E_{g}^{-}$ components:
\begin{align}
\begin{pmatrix}
\Gamma_x \\
\Gamma_y
\end{pmatrix} &= \begin{pmatrix}
\Lambda^{B_{2g}^{+}}_{1} \otimes \Pauli_y \\
\Lambda^{B_{2g}^{+}}_{1} \otimes \Pauli_x
\end{pmatrix} &\in E_g^{-}, \\
\Gamma_{\vb{k}} &= \sin k_x \Lambda^{B_{2g}^{+}}_{1} \otimes \Pauli_y - \sin k_y \Lambda^{B_{2g}^{+}}_{1} \otimes \Pauli_x &\in B_{1u}^{+}.
\end{align}
In this example, one could have also first constructed $\sin k_x \, \Pauli_y - \sin k_y \, \Pauli_x \in A_{2u}^{+}$ and then multiplied it with $\Lambda^{B_{2g}^{+}}_{1}$.

\subsection{Simple applications of the classification}
Here we give two simple examples of how the classification of bilinears can be used to analyze the three-orbital model of Sec.~\ref{sec:cuprate-3band-model} and the associated three-band Hubbard model.

\subsubsection{Rewriting the one-particle Hamiltonian}
All the hopping amplitudes which are included in the three-band model, depicted in Fig.~\ref{fig:CuO2-sketch}, are between orbitals that are within an extended unit cell, shown in Fig.~\ref{fig:extended-unit-cell}.
The hopping amplitudes can therefore be collected into the following extended-basis matrix:
\begin{align}
\mathcal{T} &\defeq \begin{pmatrix}
\epsilon_{d} - \upmu & t_{pd} & - t_{pd} & - t_{pd} & t_{pd} \\[2pt]
& \tfrac{1}{2} (\epsilon_{p} - \upmu) & - t_{pp} & t_{pp}' & t_{pp} \\[2pt]
&& \tfrac{1}{2} (\epsilon_{p} - \upmu) & t_{pp} & t_{pp}' \\[2pt]
&&& \tfrac{1}{2} (\epsilon_{p} - \upmu) & - t_{pp} \\[2pt]
\cc &&&& \tfrac{1}{2} (\epsilon_{p} - \upmu)
\end{pmatrix}, \label{eq:hopping_matrix} 
\end{align}
where a factor of $\tfrac{1}{2}$ has been added to $\epsilon_{p} - \upmu$ to avoid double counting.
Notice how $\mathcal{T}$ can be expressed in terms of the $A_{1g}^{+}$ matrices of Tab.~\ref{tab:orbital-Lambda-mats}:
\begin{align}
\mathcal{T} &= \frac{1}{\sqrt{2}} (\epsilon_{d} - \upmu) \Lambda^{A_{1g}^{+}}_1 + \frac{1}{\sqrt{2}} (\epsilon_{p} - \upmu) \Lambda^{A_{1g}^{+}}_2 + 2 t_{pd} \Lambda^{A_{1g}^{+}}_3 + 2 t_{pp} \Lambda^{A_{1g}^{+}}_4 + \sqrt{2} t_{pp}' \Lambda^{A_{1g}^{+}}_5,
\end{align}
as expected from symmetry.

The three-band Hamiltonian [Eq.~\eqref{eq:3band-Haml}] is recovered by projecting $\mathcal{T}$ down to the non-redundant basis $\psi$ [Eq.~\eqref{eq:primitive-basis}] with the aid of $\mathcal{K}_{\vb{k}}$ [Eq.~\eqref{eq:K-proj-mat}]:
\begin{align}
H_{\vb{k}} &= \mathcal{K}_{\vb{k}}^{\dag} \mathcal{T} \mathcal{K}_{\vb{k}} = \begin{pmatrix}
\epsilon_{d} - \upmu & t_{pd} (1 - \Elr^{- \iu k_x}) & - t_{pd} (1 - \Elr^{- \iu k_y}) \\
& \epsilon_{p} + 2 t_{pp}' \cos k_x - \upmu & - t_{pp} (1 - \Elr^{\iu k_x}) (1 - \Elr^{- \iu k_y}) \\
\cc & & \epsilon_{p} + 2 t_{pp}' \cos k_y - \upmu
\end{pmatrix}.
\end{align}
The corresponding second quantized one-particle Hamiltonian can be written in a number of equivalent ways:
\begin{align}
\Haml_0 &= \sum_{\vb{R}} \Psi^{\dag}(\vb{R}) \mathcal{T} \Psi(\vb{R}) = \sum_{\vb{k}} \Psi_{\vb{k}}^{\dag} \mathcal{T} \Psi_{\vb{k}} = \sum_{\vb{k}} \psi_{\vb{k}}^{\dag} H_{\vb{k}} \psi_{\vb{k}}.
\end{align}

\subsubsection{Decomposition of Hubbard interactions} \label{sec:CuO2-Hubbard-decompose}
Conventionally, the interactions that are added to the three-band \ce{CuO2} model have the form of (possibly extended) Hubbard interactions.
Here we decompose these Hubbard interactions into symmetry channels.
Compare with the mean-field analyses of Refs.~\cite{Ovchinnikov1990, Fischer2011, Atkinson2016}.

In the algebra below, we treat the extended-basis fields $\Psi(\vb{R})$ [Eq.~\eqref{eq:extended-basis}] as Grassmann variables, neglecting any one-particle terms that would otherwise appear.
Given that we only deal with operators within one extended unit cell, in this subsection we suppress the argument $\vb{R}$.
Introduce the following densities:
\begin{align}
n_d &= \Psi^{\dag} \diag(1,0,0,0,0) \Psi, \\
n_{p1} &= \Psi^{\dag} \diag(0,1,0,0,0) \Psi \equiv n_{p5}, \\
n_{p2} &= \Psi^{\dag} \diag(0,0,1,0,0) \Psi, \\
n_{p3} &= \Psi^{\dag} \diag(0,0,0,1,0) \Psi, \\
n_{p4} &= \Psi^{\dag} \diag(0,0,0,0,1) \Psi,
\end{align}
and for each $10 \times 10$ spin-orbit matrix $\Gamma$ let us denote the corresponding operator:
\begin{align}
\mathcal{O}(\Gamma) &= \Psi^{\dag} \Gamma \Psi.
\end{align}
Traditionally, the following four Hubbard interactions are considered~\cite{Fischer2011}:
\begin{align}
\Haml' &= \frac{U_d}{2} n_d^2 + \frac{U_p}{4} \sum_{\ell=1}^{4} n_{p\ell}^2 + V_{pd} n_d \sum_{\ell=1}^{4} n_{p\ell} + V_{pp} \sum_{\ell=1}^{4} n_{p\ell} n_{p,\ell+1},
\end{align}
where $n_{p5} \equiv n_{p1}$.

The $U_d$ Hubbard interaction can be written as:
\begin{align}
n_d^2 &= \frac{1}{2} \Big[\mathcal{O}\big(\Lambda^{A_{1g}^{+}}_1\big)\Big]^2.
\end{align}
However, this decomposition is not unique due to the Fierz identities:
\begin{align}
\Big[\mathcal{O}\big(\Lambda^{A_{1g}^{+}}_1\big)\Big]^2 + \Big[\mathcal{O}\big(\Lambda^{A_{1g}^{+}}_1 \Pauli_A\big)\Big]^2 &= 0, \label{eq:Fierz-id-Ud}
\end{align}
where $A \in \{1, 2, 3\}$ is fixed.
These identities follow from the Pauli exclusion principle ($\Psi_n^{\dag} \Psi_n^{\dag} = \Psi_m \Psi_m = 0$).
They do not arise for interactions between distinct unit cells.

After some algebra aided by \texttt{Mathematica}, the $U_p$, $V_{pd}$, and $V_{pp}$ Hubbard interactions can be rewritten as well:
\begin{align}
\sum_{\ell=1}^{4} n_{p\ell}^2 &= \frac{1}{2} \sum_{\Lambda \in \mathcal{L}_p^{+}} \big[\mathcal{O}(\Lambda)\big]^2, \\
n_d \sum_{\ell=1}^{4} n_{p\ell} &= - \frac{1}{2} \sum_{\Lambda \in \mathcal{L}_{pd}^{-}} \big[\mathcal{O}(\Lambda)\big]^2 - \frac{1}{4} \sum_{\Lambda \in \mathcal{L}_{pd}^{-}} \sum_{A=1}^{3} \big[\mathcal{O}(\Lambda \Pauli_A)\big]^2 + \frac{1}{4} \sum_{\Lambda \in \mathcal{L}_{pd}^{+}} \big[\mathcal{O}(\Lambda)\big]^2, \label{eq:Hubbard-Vpd-decomp} \\
\sum_{\ell=1}^{4} n_{p\ell} n_{p,\ell+1} &= - \frac{1}{2} \sum_{\Lambda \in \mathcal{L}_{pp}^{-}} \big[\mathcal{O}(\Lambda)\big]^2 - \frac{1}{4} \sum_{\Lambda \in \mathcal{L}_{pp}^{-}} \sum_{A=1}^{3} \big[\mathcal{O}(\Lambda \Pauli_A)\big]^2 + \frac{1}{4} \sum_{\Lambda \in \mathcal{L}_{pp}^{+}} \big[\mathcal{O}(\Lambda)\big]^2,
\end{align}
where:
\begin{align}
\mathcal{L}_p^{+} &= \Big(\Lambda^{A_{1g}^{+}}_2, \quad\Lambda^{B_{1g}^{+}}_2, \quad\Lambda^{E_u^{+}}_{2,x}, \quad\Lambda^{E_u^{+}}_{2,y}\Big), \\
\mathcal{L}_{pd}^{-} &= \Big(\Lambda^{A_{1g}^{-}}_1, \quad\Lambda^{B_{1g}^{-}}_1, \quad\Lambda^{E_u^{-}}_{1,x}, \quad\Lambda^{E_u^{-}}_{1,y}\Big), \\
\mathcal{L}_{pd}^{+} &= \Big(\Lambda^{A_{1g}^{+}}_3, \quad\Lambda^{B_{1g}^{+}}_1, \quad\Lambda^{E_u^{+}}_{1,x}, \quad\Lambda^{E_u^{+}}_{1,y}\Big), \\
\mathcal{L}_{pp}^{-} &= \Big(\Lambda^{A_{2g}^{-}}_1, \quad\Lambda^{B_{1g}^{-}}_2, \quad\Lambda^{E_u^{-}}_{3,x}, \quad\Lambda^{E_u^{-}}_{3,y}\Big), \\
\mathcal{L}_{pp}^{+} &= \Big(\Lambda^{B_{2g}^{+}}_1, \quad\Lambda^{A_{1g}^{+}}_4, \quad\Lambda^{E_u^{+}}_{3,x}, \quad\Lambda^{E_u^{+}}_{3,y}\Big).
\end{align}
The $V_{pd}$ decomposition is explicitly derived by hand in Sec.~\ref{sec:ASV-eff-Haml}.
These decompositions are ambiguous too, as there are three Fierz identities pertaining to $U_p$:
\begin{align}
\sum_{\Lambda \in \mathcal{L}_p^{+}} \big[\mathcal{O}(\Lambda)\big]^2 + \sum_{\Lambda \in \mathcal{L}_p^{+}} \big[\mathcal{O}(\Lambda \Pauli_A)\big]^2 &= 0, \label{eq:Fierz-id-Up}
\end{align}
where $A \in \{1, 2, 3\}$ is fixed.
There are twelve Fierz identities relevant for $V_{pd}$:
\begin{align}
\big[\mathcal{O}(\Lambda^{-})\big]^2 + \big[\mathcal{O}(\Lambda^{-} \Pauli_A)\big]^2 - \big[\mathcal{O}(\Lambda^{+})\big]^2 - \big[\mathcal{O}(\Lambda^{+} \Pauli_A)\big]^2 &= 0, \label{eq:Fierz-Vpd-Vpp}
\end{align}
where $A \in \{1, 2, 3\}$ is fixed and $\Lambda^{\pm}$ are the first, second, third, or fourth matrices appearing in $\mathcal{L}_{pd}^{\pm}$.
For instance, one possible choice is $\Lambda^{-} = \Lambda^{E_u^{-}}_{1,x}$, $\Lambda^{+} = \Lambda^{E_u^{+}}_{1,x}$.
Note that if we use, e.g., the first matrix of $\mathcal{L}_{pd}^{-}$, then we must also use the first matrix of $\mathcal{L}_{pd}^{+}$.
There are twelve Fierz identities relevant for $V_{pp}$ that have the same form as Eq.~\eqref{eq:Fierz-Vpd-Vpp}, except that now $\Lambda^{\pm}$ are the first, second, third, or fourth matrices appearing in $\mathcal{L}_{pp}^{\pm}$.

Having decomposed the Hubbard interactions, let us now discuss their interpretation.
Microscopically, Hubbard interactions derive from Coulomb repulsion so $U_d$, $U_p$, $V_{pd}$, and $V_{pp}$ are all positive.
In the decompositions, however, some terms are attractive and negative.
For instance, the $U_d$ and $U_p$ interactions result in terms that are attractive in the spin channels:
\begin{align}
n_d^2 &= - \frac{1}{6} \sum_{A=1}^{3} \Big[\mathcal{O}\big(\Lambda^{A_{1g}^{+}}_1 \Pauli_A\big)\Big]^2, \\
\sum_{\ell=1}^{4} n_{p\ell}^2 &= - \frac{1}{6} \sum_{\Lambda \in \mathcal{L}_p^{+}} \sum_{A=1}^{3} \big[\mathcal{O}(\Lambda \Pauli_A)\big]^2,
\end{align}
as follows from the Fierz identities~\eqref{eq:Fierz-id-Ud} and~\eqref{eq:Fierz-id-Up}.
The same is true for the $V_{pd}$ and $V_{pp}$ interactions.
Recalling how integrating out a fluctuating order parameter always gives a negative interaction (Sec.~\ref{sec:QCP-pairing-model}), the negative terms in the decompositions can be interpreted as being indicative of a possible instability towards condensation in the corresponding channel.
That said, in the $V_{pd}$ and $V_{pp}$ interactions there is an ambiguity in which channels are attractive since, by employing the Fierz identity~\eqref{eq:Fierz-Vpd-Vpp}, one can also write:
\begin{align}
n_d \sum_{\ell=1}^{4} n_{p\ell} &= - \frac{1}{2} \sum_{\Lambda \in \mathcal{L}_{pd}^{+}} \big[\mathcal{O}(\Lambda)\big]^2 - \frac{1}{4} \sum_{\Lambda \in \mathcal{L}_{pd}^{+}} \sum_{A=1}^{3} \big[\mathcal{O}(\Lambda \Pauli_A)\big]^2 + \frac{1}{4} \sum_{\Lambda \in \mathcal{L}_{pd}^{-}} \big[\mathcal{O}(\Lambda)\big]^2, \label{eq:Hubbard-Vpd-decomp2} \\
\sum_{\ell=1}^{4} n_{p\ell} n_{p,\ell+1} &= - \frac{1}{2} \sum_{\Lambda \in \mathcal{L}_{pp}^{+}} \big[\mathcal{O}(\Lambda)\big]^2 - \frac{1}{4} \sum_{\Lambda \in \mathcal{L}_{pp}^{+}} \sum_{A=1}^{3} \big[\mathcal{O}(\Lambda \Pauli_A)\big]^2 + \frac{1}{4} \sum_{\Lambda \in \mathcal{L}_{pp}^{-}} \big[\mathcal{O}(\Lambda)\big]^2.
\end{align}
In any case, finding out in which channel the system condenses is a non-trivial task that is not the focus of the current work.
In Sec.~\ref{sec:LC-cuprate-micro} we reviewed previous theoretical work that dealt with this task.

\section{Pairing due to intra-unit-cell loop-current fluctuations in cuprates} \label{sec:pairing-cuprate-actual-analysis}
The pairing due to quantum-critical intra-unit-cell (IUC) loop-current (LC) fluctuations has been analyzed in Sec.~\ref{sec:gen-sys-LC-analysis} of the previous chapter.
The main result of the analysis, summarized in Fig.~\ref{fig:QCP-general-results}, is that IUC LCs are uniquely incapable of driving strong pairing near their quantum-critical point (QCP).
Even-parity IUC LCs are an ineffective pairing glue, while odd-parity IUC LCs are parametrically strong pair breakers.
This result holds for general two-dimensional systems without SOC.
In this section, we apply the analysis of Sec.~\ref{sec:gen-sys-LC-analysis} to the cuprates.

Let us recall that the strategy we used in Sec.~\ref{sec:gen-sys-LC-analysis} is a phenomenological strategy in which we assume LC order from the outset and then explore whether there is an enhancement in the pairing tendency as we approach the QCP from the disordered, Fermi liquid side (Fig.~\ref{fig:QCP-calc-strategy}).
The first question that we need to address is whether this strategy is applicable to cuprates.

Although much of cuprate physics is hotly debated, there are several well-established facts about these materials that are agreed upon~\cite{Keimer2015}, as already discussed in Sec.~\ref{sec:cuprate-basics}:
\begin{itemize}
\item The pairing state for tetragonal systems is an even-parity spin-singlet state with $d_{x^2-y^2}$ symmetry~\cite{Wollman1993, VanHarlingen1995, Tsuei2000}, whereas for weakly orthorhombic systems it is dominated by this pairing state~\cite{Kirtley2006, Tsuei2000}.
\item In the SC state there are well-defined Bogoliubov quasi-particles, as evidenced by angle-resolved photoemission spectroscopy~\cite{Campuzano1996, Damascelli2003, Sobota2021}, Andreev reflection experiments~\cite{Wei1998, Petrov2007}, and shot noise measurements~\cite{Zhou2019}.
\item Superconductivity originates in the \ce{CuO2} planes, as explicitly seen in atomically-thin cuprate monolayers~\cite{Uchihashi2017, Gozar2008, Logvenov2009, Yu2019}, and the predominant orbitals of the \ce{CuO2} planes are \ce{Cu}:$3d_{x^2-y^2}$ and \ce{O}:$2p_{x,y}$, as deduced from x-ray absorption studies~\cite{Chen1991, Chen1992, Pellegrin1993, Nucker1995, Peets2009} and theoretical considerations~\cite{Pickett1989, Dagotto1994, Hybertsen1992}.
\item The overdoped normal state is a Fermi liquid~\cite{Keimer2015, Proust2019}, as evidenced by thermodynamic and transport measurements~\cite{Kubo1991, Manako1992, Mackenzie1996, Proust2002, Nakamae2003, Hussey2003}, angular-resolved photoemission spectroscopy~\cite{Damascelli2003, Plate2005, Yoshida2006, Peets2007, Horio2018}, and magneto-oscillation experiments~\cite{Vignolle2008, Bangura2010}.
Moreover, the overdoped normal and SC states are well-described by density functional theory~\cite{Kramer2019} and dirty $d$-wave BCS theory~\cite{Lee-Hone2020}, respectively.
\end{itemize}
Let us also remark that there is some evidence supporting that a QCP near optimal doping lies beneath the SC dome~\cite{Tallon2001, Proust2019}.
Clearly, these established findings justify the use of our strategy to cuprates, but with the additional point that a viable pairing glue must reproduce the correct $d_{x^2-y^2}$ pairing symmetry.

The idea is thus to focus on the far-overdoped regime and assume a Fermi liquid normal state.
Starting from this well-understood normal state, we shall then phenomenologically analyze within weak-coupling theory the pairing due to various types of LC fluctuations and explore which ones yield the observed singlet $d_{x^2-y^2}$-wave state.
Which ones become enhanced as the putative LC QCP is approached we already know from the results of the previous chapter.
As there is no experimental indication that the pairing symmetry changes upon doping~\cite{VanHarlingen1995, Tsuei2000, Keimer2015}, this approach should allow us to draw conclusions for optimally doped materials, even though all the complications of the Mott state, the pseudogap, etc., have been ignored.
As as long as there is sufficient continuity within the SC phase itself, the crucial pairing interactions should be closely related across the phase diagram.
This is certainly true for LC-based proposals (Sec.~\ref{sec:LC-proposals}) which we are currently examining.
That said, the scenario of two different, but complementary, mechanisms acting on the under- and overdoped sides of the phase diagram cannot be excluded.

The rest of this section, which is based on Ref.~\cite{Palle2024-LC}, is organized as follows.
First, we set up the formalism.
We state the band structure, the precise form of the interaction, and write down the simplified linearized gap equation appropriate to the current problem.
Then, in Sec.~\ref{sec:Bloch-Kirch-constr}, we discuss how the Bloch and Bloch-Kirchhoff theorems of Sec.~\ref{sec:Bloch-tm} (Chap.~\ref{chap:loop_currents}) constrain the viable LC patterns in cuprates down to three options.
In Sec.~\ref{sec:pairing-cuprate-actual-analysis-VH}, we investigate how efficiently LC fluctuations couple Van Hove points to the rest of the Fermi surface, depending on the LC symmetry and band structure.
The numerical solutions of the linearized gap equation are presented in Sec.~\ref{sec:pairing-cuprate-actual-analysis-results}.
There are three possible LC orders with $g_{xy(x^2-y^2)}$-wave, $d_{x^2-y^2}$-wave, and $(p_x|p_y)$-wave symmetry.
We find that their leading SC states have, respectively, $d_{xy}$-wave, $d_{x^2-y^2}$-wave, and extended $s$-wave symmetry.
Hence only $d_{x^2-y^2}$-wave LCs yield the correct pairing symmetry.
However, since they have even parity, their pairing tendency does not become enhanced near the QCP (Sec.~\ref{sec:gen-sys-LC-analysis}, Fig.~\ref{fig:QCP-general-results}).
Moreover, if we include weak SOC, then it induces subsidiary $d_{x^2-y^2}$-wave spin fluctuations whose pairing does become enhanced near the QCP, but with the incorrect $p$-wave symmetry.
These are the main results of Ref.~\cite{Palle2024-LC} concerning cuprates.
In Sec.~\ref{sec:pairing-cuprate-actual-analysis-results}, we also discuss how to experimentally measure these LC orders.
In the Sec.~\ref{sec:cuprate-sym-choice-mechanism} thereafter, we explain how the pairing symmetry gets chosen in boson exchange mechanisms based on IUC orders, as opposed to those based on finite-$\vb{q}$ instabilities.
We supplement our numerics with analytic solutions of the linearized gap equation in Sec.~\ref{sec:cuprate-analytic-res}.
We conclude with an extended comparison with the work by Aji, Shekhter, and Varma~\cite{Aji2010} which, in contrast to our results, suggested that $p$-wave LCs and their conjugate momentum, $g$-wave LCs, give strong pairing near the QCP with the correct $d_{x^2-y^2}$ symmetry.

\subsection{Formalism} \label{sec:pairing-cuprate-formalism}
Having established the applicability of the formalism of Chap.~\ref{chap:loop_currents}, Sec.~\ref{sec:gen-sys-LC-analysis}, we now discuss its application to IUC LCs in cuprates.

In the general model of Sec.~\ref{sec:QCP-pairing-model} that we considered in the previous chapter, we assumed a general band Hamiltonian which respects parity and time reversal.
Here, for the one-particle Hamiltonian we use the three-band Hamiltonian of Sec.~\ref{sec:cuprate-3band-model}:
\begin{align}
\Haml_0 &= \sum_{\vb{k}} \psi_{\vb{k}}^{\dag} H_{\vb{k}} \psi_{\vb{k}},
\end{align}
where
\begin{align}
H_{\vb{k}} &= \begin{pmatrix}
\epsilon_{d} - \upmu & t_{pd} (1 - \Elr^{- \iu k_x}) & - t_{pd} (1 - \Elr^{- \iu k_y}) \\
& \epsilon_{p} + 2 t_{pp}' \cos k_x - \upmu & - t_{pp} (1 - \Elr^{\iu k_x}) (1 - \Elr^{- \iu k_y}) \\
\cc & & \epsilon_{p} + 2 t_{pp}' \cos k_y - \upmu
\end{pmatrix} \otimes \Pauli_0. \label{eq:3band-Haml-again}
\end{align}
This Hamiltonian we diagonalize into:
\begin{align}
H_{\vb{k}} &= \sum_{n=1}^{3} \varepsilon_{\vb{k} n} \mathcal{P}_{\vb{k} n}, \label{eq:CuO2-disp-proj-def}
\end{align}
where $n \in \{1, 2, 3\}$ is the band index, $\varepsilon_{\vb{k} n}$ are the band dispersions, sorted so that $\varepsilon_{\vb{k} 1} < \varepsilon_{\vb{k} 2} < \varepsilon_{\vb{k} 3}$, and $\mathcal{P}_{\vb{k} n}$ are the corresponding band projectors.
Given that there is no spin-orbit coupling:
\begin{align}
\mathcal{P}_{\vb{k} n} &= u_{\vb{k} n} u_{\vb{k} n}^{\dag} \otimes \Pauli_0,
\end{align}
where $u_{\vb{k} n}$ is the normalized (orbital part of the) band eigenvector.
Even though $H_{\vb{k}}$ is just a $3 \times 3$ matrix, its $\varepsilon_{\vb{k} n}$ and $u_{\vb{k} n}$ cannot be found in closed form for general parameters.
As we shall discuss in Sec.~\ref{sec:ASV-fermion-coupling}, one can diagonalize $H_{\vb{k}}$ analytically for $t_{pp} = t_{pp}' = 0$ [Eqs.~(\ref{eq:tpp-tpp'-0-closed-form-1},~\ref{eq:tpp-tpp'-0-closed-form-2})], but this is clearly too restrictive.
Examples of Fermi surfaces are shown in Fig.~\ref{fig:cuprate-Fermi-surfaces}.
It is worth noting that the band states of this one-particle Hamiltonian are suppose to describe the already dressed Fermi liquid quasi-particles of the overdoped regime, since no additional Hubbard or similar interactions will be included in the model, apart from the effective interaction mediated by LC fluctuations.

The effective interaction between fermions has the form [Eq.~\eqref{eq:effective-multi-exch-int}]:
\begin{align}
\Haml_{\text{int}} &= - \frac{1}{2} g^2 \sum_{a \vb{q}} \chi(\vb{q}) \phi_{a, -\vb{q}} \phi_{a \vb{q}}.
\end{align}
Instead of the critical scaling expression of Sec.~\ref{sec:IUC-order-results}, for the susceptibility we shall use the following mean-field expression:
\begin{align}
\chi(\vb{q}) &= \frac{\chi_{0}}{\displaystyle \frac{1+r}{2} - \frac{1-r}{4} \mleft(\cos q_{x} + \cos q_{y}\mright)}, \label{eq:chi-mean-field-expr-r}
\end{align}
where $\chi_0 > 0$ and the lattice constant has been set to unity.
For $r = 1$, $\chi(\vb{q}) = \chi_0$ is a constant.
As $r \to 0$, $\chi(\vb{q})$ becomes increasingly strongly peaked at $\vb{q} = \vb{0}$ and diverges like $1/(8 r + \vb{q}^2)$ near the QCP $r = 0$.
Hence the critical exponents of Sec.~\ref{sec:IUC-order-results} are $\nu = \tfrac{1}{2}$ and $\eta = 0$.
In light of Eq.~\eqref{eq:effective-multi-exch-int-mode-freq}, this divergence is equivalent to a softening of the order parameter modes at $\vb{q} = \vb{0}$.
For $r < 0$, $\chi(\vb{q} = \vb{0})$ becomes negative, indicating condensation to a homogeneous intra-unit-cell order.

Regarding the fermionic bilinears $\phi_{a \vb{q}}$, we have classified them at length in Sec.~\ref{sec:cuprate-bilinear-class} and now we take full advantage of this classification.
LC orders are, by definition, TR-odd and orbital (Sec.~\ref{sec:orbital-mag}).
Hence their bilinears have the form [Eq.~\eqref{eq:general-momentum-bilinear-again}]
\begin{align}
\phi_{a \vb{q}} &= \frac{1}{\sqrt{\mathcal{N}}} \sum_{\vb{k}} \psi_{\vb{k}}^{\dag} \mleft(\gamma_{a \vb{k}, \vb{k}+\vb{q}} \otimes \Pauli_0\mright) \psi_{\vb{k}+\vb{q}}, \label{eq:general-momentum-bilinear-again2}
\end{align}
where
\begin{align}
\gamma_{a \vb{k}, \vb{p}}^{*} = - \gamma_{a, -\vb{k}, -\vb{p}}.
\end{align}
In principle, there are infinitely many LC bilinears which one could consider.
These bilinears describe the Yukawa coupling to the fermions $\Haml_c = g \sum_{a \vb{q}} \Phi_{a, -\vb{q}} \phi_{a \vb{q}}$ [Eq.~\eqref{eq:Yukawa-coupling}] and, at least for small Fermi surfaces, a renormalization group argument can be made that the non-local terms in the Yukawa coupling are irrelevant.
However, even for large Fermi surfaces it is expected, although by no means necessary, that for a given LC channel the most local Yukawa couplings in real space, or equivalently the lowest order harmonics in momentum space, are the largest.
We shall therefore restrict ourselves to only those LC bilinears that can be constructed from one extended unit cell.
In particular, this covers the most-discussed IUC LC proposal put forward by Varma~\cite{Varma2020, Varma2016, Aji2010}.
The $\gamma$ matrices we thus write as [Eq.~\eqref{eq:momentum-bilinear-v2}]:
\begin{align}
\gamma_{a \vb{k}, \vb{p}} &= \mathcal{K}_{\vb{k}}^{\dag} \Lambda_a \mathcal{K}_{\vb{p}}, \label{eq:momentum-bilinear-v3}
\end{align}
where the orbital extended-basis $\Lambda$ matrices are the TR-odd ones from Tab.~\ref{tab:orbital-Lambda-mats} (Sec.~\ref{sec:orbital-Lambda-mats}).
Since $\mathcal{K}_{\vb{k}}^{*} = \mathcal{K}_{-\vb{k}}$ [Eq.~\eqref{eq:K-proj-mat}], the condition $\gamma_{a \vb{k}, \vb{p}}^{*} = - \gamma_{a, -\vb{k}, -\vb{p}}$ implies that LCs have purely imaginary $\Lambda$.
The purely imaginary nature of the orbital matrix $\Lambda$ can be interpreted as introducing phase shifts in the bare hopping parameters of $\Haml_0$.
Via a reverse Peierls substitution, these phase shifts correspond to magnetic fluxes generated by orbital currents (cf.\ Sec.~\ref{sec:cup-g-wave-res-discus}).
However, this construction is not yet finished since not all $\Phi_{a \vb{q}}$ are able to condensed due to Bloch and Kirchhoff constraints, as we shall discuss in the next Sec.~\ref{sec:Bloch-Kirch-constr}.

The linearized gap equation that we wrote down in Sec.~\ref{sec:QCP-model-lin-gap-eq} applies to arbitrary orders in general systems with multiple Fermi surfaces and spin-orbit coupling.
For the system under consideration, however, the order is purely orbital and there is only one Fermi surface and no SOC.
The linearized gap equation~\eqref{eq:final-lin-gap-eq-LC-again} thus simplifies to:
\begin{align}
\oint\limits_{\varepsilon_{\vb{k}} = 0} \frac{\dd{\ell_{\vb{k}}}}{(2\pi)^2 v_{\vb{k}}} \PintV_p(\vb{p}, \vb{k}) \Delta_p(\vb{k}) &= \lambda \, \Delta_p(\vb{p}), \label{eq:final-lin-gap-eq-cup-simplified}
\end{align}
where the integral goes over the Fermi surface (line),
\begin{align}
\varepsilon_{\vb{k}} &\equiv \varepsilon_{\vb{k} 3}, &
v_{\vb{k}} &\defeq \abs{\grad_{\vb{k}} \varepsilon_{\vb{k}}}
\end{align}
are the conduction band dispersion and Fermi velocity, and $p = +1$ ($-1$) stands for singlet (triplet) pairing.
The largest eigenvalue $\lambda$ determines the superconducting transition temperature through $k_{B} T_c = \tfrac{2\Elr^{\upgamma_E}}{\pi} \hbar \omega_{c} \Elr^{-1/\lambda}$, where $\omega_{c}$ is the characteristic cutoff for LC fluctuations and $\upgamma_E$ is the Euler-Mascheroni constant.
The eigenvector $\Delta_{p}(\vb{p})$ determines the symmetry of the pairing and is related to the SC gap function of the Bogoliubov-de~Gennes Hamiltonian $\Delta_{ss'}(\vb{p})$ through
\begin{align}
\Delta_{ss'}(\vb{p}) &= \begin{cases}
\Delta_{+}(\vb{p}) (\iu \Pauli_y)_{ss'}, & \text{for singlet pairing ($p = +1$),} \\
\Delta_{-}(\vb{p}) \mleft(\Pauli_{A'} \iu \Pauli_y\mright)_{ss'}, & \text{for triplet pairing ($p = -1$).}
\end{cases}
\end{align}
Here all triplet orientations $A' = 1, 2, 3$ are degenerate because, on the one hand, there is no SOC, while, on the other hand, LCs are purely orbital.
Hence nothing breaks the spin rotation symmetry.

The linearized gap equation~\eqref{eq:final-lin-gap-eq-cup-simplified} is unsymmetrized, i.e., $v_{\vb{k}}$ has not been absorbed into the Cooper-channel interaction (cf.\ Eq.~\eqref{eq:final-lin-gap-eq-unsym} of Appx.~\ref{app:lin_gap_eq}).
The Cooper-channel interaction is thus given by
\begin{align}
\PintV_{\pm}(\vb{p}, \vb{k}) &= - g^2 \frac{1}{2} \mleft[\Pintv_0(\vb{p}, \vb{k}) \pm \Pintv_0(\vb{p}, -\vb{k})\mright], \label{eq:final-lin-gap-eq-cup-int}
\end{align}
where the overall minus sign arises because LCs are odd under TR.
\begin{align}
\Pintv_0(\vb{p}, \vb{k}) &= \chi(\vb{p}-\vb{k}) \sum_a \abs{\Pintf_{a}(\vb{p}, \vb{k})}^2 > 0
\end{align}
is a combination of the LC correlation function $\chi(\vb{q})$ and the pairing form factor
\begin{align}
\Pintf_{a}(\vb{p}, \vb{k}) &\defeq u_{\vb{p} 3}^{\dag} \gamma_{a \vb{p}, \vb{k}} u_{\vb{k} 3}. \label{eq:final-lin-gap-eq-cup-pairing-form-factor}
\end{align}
This pairing form factors contains information about the nature and symmetry of the LC state via the coupling matrix $\gamma_{a \vb{p}, \vb{k}}$.
For the coupling constant $g$, we assume a value that yields sufficiently small dimensionless eigenvalues $\lambda$ to justify a weak-coupling treatment.

We have studied the symmetry properties of $\Pintf_{a}(\vb{p}, \vb{k})$ in full generality in Sec.~\ref{sec:Cp-channel-gen-sym-constr}.
The most important finding was that the pairing form factor vanishes at forward-scattering,
\begin{align}
\lim_{\vb{p} \to \vb{k}} \Pintf_{a}(\vb{p}, \vb{k}) &= 0, \label{eq:cup-fwd-suppression}
\end{align}
for order parameters that are odd under the composed parity and TR operation $P \TRop$, $p_P p_{\TRop} = -1$.
For LC order, which is always odd under TR, this implies that even-parity LCs have a suppressed forward-scattering Cooper-channel interaction.
Moreover, from a Taylor expansion it follows that
\begin{align}
\sum_a \abs{\Pintf_{a}(\vb{p}, \vb{k})}^2 &\propto \mleft(\vb{p} - \vb{k}\mright)^2 \qquad\text{as $\vb{p} \to \vb{k}$.}
\end{align}
The $1/\vb{q}^2$ divergence of the susceptibility near the QCP is thus completely eliminated in $\Pintv_0(\vb{p}, \vb{k})$ for intra-unit-cell ($\vb{q} = \vb{0}$) orders.
Hence no pairing enhancement takes place, as was demonstrated in Sec.~\ref{sec:IUC-order-results}.
For odd-parity LCs, $\PintV_{\pm}(\vb{p}, \vb{k})$ is uniformly repulsive with an unchecked divergence near $\vb{q} = \vb{0}$ as $r \to 0$, imply that they act as strong pair breakers.

The general symmetry formalism of Sec.~\ref{sec:gen-sys-LC-analysis} can be imposing and at times difficult to follow so it is instructive to prove Eq.~\eqref{eq:cup-fwd-suppression} directly once more.
Under spatial inversion $\gamma_{a \vb{p},\vb{k}} \overset{P}{\to} p_{P} \gamma_{a, -\vb{p},-\vb{k}}$, where $p_{P}$ is the parity of the LC order parameter $\Phi_{a \vb{q}}$.
Since LCs are odd under time reversal, $\gamma_{a \vb{p},\vb{k}} \overset{\TRop}{\to} - \gamma_{a, -\vb{p},-\vb{k}}^{*}$.
If we further use the transformation properties of orbital Bloch states $u_{\vb{k} 3} \overset{P}{\to} u_{-\vb{k}, 3} \overset{\TRop}{=} u_{\vb{k} 3}^{*}$ under these same symmetries, we obtain $u_{\vb{k} 3}^{\dag} \gamma_{a \vb{k},\vb{p}} u_{\vb{p} 3} = - p_{P} u_{\vb{p} 3}^{\dag} \gamma_{a \vb{p},\vb{k}} u_{\vb{k} 3}$ from which Eq.~\eqref{eq:cup-fwd-suppression} follows.

In the current model, the precise orbital structure of the conduction band eigenvectors $u_{\vb{k} 3}$ and LC coupling matrices $\gamma_{a \vb{p}, \vb{k}}$ can make the pairing form factor $\Pintf_{a}(\vb{p}, \vb{k})$ vanish when one or both of the momenta are at high-symmetry points.
These additional constraints, specific to the model, are important for understanding some of our results and we discuss them in Sec.~\ref{sec:pairing-cuprate-actual-analysis-VH}.

\subsection{Bloch and Kirchhoff constraints on intra-unit-cell current patterns} \label{sec:Bloch-Kirch-constr}
\begin{figure}[t]
\centering
\begin{subfigure}[t]{0.25\textwidth}
\centering
\includegraphics[width=0.9\textwidth]{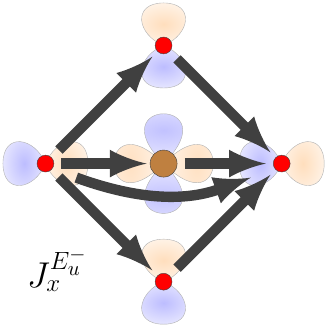}
\subcaption{}
\end{subfigure}%
\begin{subfigure}[t]{0.25\textwidth}
\centering
\includegraphics[width=0.9\textwidth]{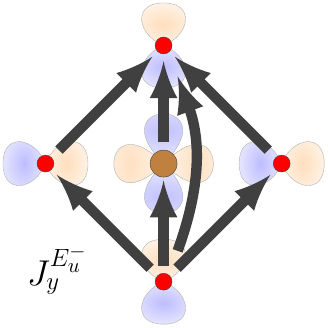}
\subcaption{}
\end{subfigure}%
\begin{subfigure}[t]{0.25\textwidth}
\centering
\includegraphics[width=0.9\textwidth]{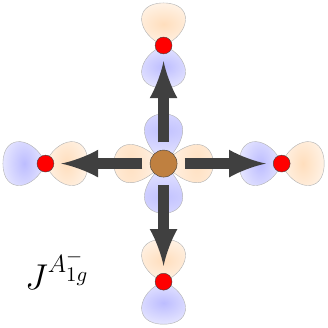}
\subcaption{}
\end{subfigure}%
\begin{subfigure}[t]{0.25\textwidth}
\centering
\includegraphics[width=0.9\textwidth]{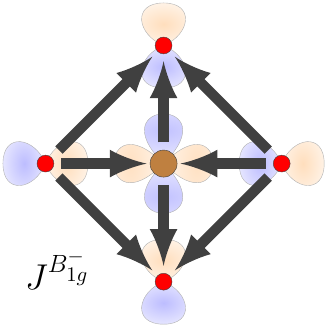}
\subcaption{}
\end{subfigure}
\captionbelow[Current patterns which violate Bloch's theorem (a \& b) and which violate the generalized Bloch-Kirchhoff theorem (c  \& d).]{\textbf{Current patterns which violate Bloch's theorem (a \& b) and which violate the generalized Bloch-Kirchhoff theorem (c  \& d).}
Both theorems are proved in Sec.~\ref{sec:Bloch-tm} of Chap.~\ref{chap:loop_currents}.
The current patterns under (a) and (b) result in a global current, while those under (c) and (d) result in a continuous accumulation of charge around some of the orbitals.}
\label{fig:Bloch-Kirchhoff-patterns}
\end{figure}

A bosonic mode $\Phi_a(\vb{R})$ is a viable candidate for a quantum-critical mode only if it can condense, in our case to a homogeneous state with $\vb{q} = \vb{0}$.
If the mode acquires a finite expectation value, we may expand it around its mean value:
\begin{align}
\Phi_a(\vb{R}) = \ev{\Phi_a} + \var{\Phi_a(\vb{R})}.
\end{align}
Neglecting the fluctuations $\var{\Phi_a(\vb{R})}$, the Yukawa coupling to the electrons [Eq.~\eqref{eq:Yukawa-coupling}] becomes:
\begin{align}
\Haml_c &= g \sum_{a \vb{R}} \ev{\Phi_a} \phi_{a}(\vb{R}). \label{eq:Yukawa-coupling-condensed}
\end{align}
However, such a term in the Hamiltonian can have aphysical consequences, such as those illustrated in Fig~\ref{fig:Bloch-Kirchhoff-patterns}.
For TR-odd $\phi_a(\vb{R})$, $\Haml_c$ may induce global currents, in violation of Bloch's theorem~\cite{Bohm1949, Ohashi1996, Yamamoto2015, Zhang2019, Watanabe2019, Watanabe2022}, or it may induce local currents that violate Kirchhoff's current law, resulting in an ever-increasing accumulation of charge on some of the orbitals, in violation of the generalized Bloch-Kirchhoff theorem of Sec.~\ref{sec:Bloch-tm-Kirchhoff}.
In the extended basis, $\vb{q} = \vb{0}$ condensation may also not be possible if $\phi_{a, \vb{q} = \vb{0}}$ vanishes due to a cancellation between overlapping extended unit cells.

Below, we analyze these constraints under the assumption that $g \ev{\Phi_a}$ is small.
This enables the use of linear response theory with current operators derived from only the kinetic part of the Hamiltonian.
Although the interacting part of the Hamiltonian also contributes to the current operators, these corrections are of higher order and can thus be neglected.

\subsubsection{Bloch constraints} \label{sec:Bloch-Bloch-constr}
In the extended basis of Sec.~\ref{sec:cuprate-bilinear-class}, the global current operator can be written as
\begin{align}
\vb{j} &= \frac{\iu}{\mathcal{N}} \sum_{\vb{R} \alpha \beta} \mleft(\vb{x}_{\beta} - \vb{x}_{\alpha}\mright) \mathcal{T}_{\alpha \beta} \Psi^{\dag}_{\alpha}(\vb{R}) \Psi_{\beta}(\vb{R}), \label{eq:CuO2-global_current_operator}
\end{align}
where $\mathcal{N}$ is the number of copper atoms, $\alpha, \beta$ are orbital indices, $\mathcal{T}_{\alpha \beta}$ is the hopping matrix of Eq.~\eqref{eq:hopping_matrix}, and the basis vectors of the atoms are (see Fig.~\ref{fig:CuO2-sketch} or \ref{fig:extended-unit-cell}):
\begin{align}
\vb{x}_{\alpha} = \begin{pmatrix}
\vb{0} \\[2pt]
\tfrac{1}{2} \vu{e}_x \\[2pt]
\tfrac{1}{2} \vu{e}_y \\[2pt]
- \tfrac{1}{2} \vu{e}_x \\[2pt]
- \tfrac{1}{2} \vu{e}_y
\end{pmatrix}. \label{eq:cup-r-alpha-pos}
\end{align}
By introducing the matrices
\begin{align}
\big(J_a^{E_u^{-}}\big)_{\alpha \beta} &= \iu \, \vu{e}_a \vdot \mleft(\vb{x}_{\beta} - \vb{x}_{\alpha}\mright) \mathcal{T}_{\alpha \beta},
\end{align}
taking the expectation value of Eq.~\eqref{eq:CuO2-global_current_operator},
and assuming that the translation symmetry is not broken, we find that the currents along the $x$ and $y$ directions equal:
\begin{align}
j_x &= \ev{\Psi^{\dag}(\vb{R}) J_x^{E_u^{-}} \Psi(\vb{R})},\\
j_y &= \ev{\Psi^{\dag}(\vb{R}) J_y^{E_u^{-}} \Psi(\vb{R})}.
\end{align}
Global currents are odd under parity and time reversal.
Hence the $J_a^{E_u^{-}}$ matrices belong to the $E_u^{-}$ irrep and can be expressed in terms of the $E_u^{-}$ matrices of Tab.~\ref{tab:orbital-Lambda-mats}:
\begin{align}
\big(J_a^{E_u^{-}}\big)_{\alpha \beta} &= - \frac{1}{\sqrt{2}} t_{pd} \Lambda^{E_u^{-}}_{1,a} - t_{pp}' \Lambda^{E_u^{-}}_{2,a} - t_{pp} \Lambda^{E_u^{-}}_{3,a}.
\end{align}
The condensation of a $E_u^{-}$ mode will therefore generically induce global currents, in violation of Bloch's theorem (Sec.~\ref{sec:Bloch-tm}), unless we fine-tune the bilinears to cancel the global current.

If we restrict ourselves to bilinears that are localized within only one extended unit cell, we are left with only three options which can induce global currents along the $\vu{e}_a$ directions, which are namely (Tab.~\ref{tab:orbital-Lambda-mats}):
\begin{align}
\gamma_{a \vb{k}, \vb{p}}^{E_u^{-}} &= \mathcal{K}_{\vb{k}}^{\dag} \mleft(c_1 \Lambda^{E_u^{-}}_{1,a} + c_2 \Lambda^{E_u^{-}}_{2,a} + c_3 \Lambda^{E_u^{-}}_{3,a}\mright) \mathcal{K}_{\vb{p}}. \label{eq:gamma-Eum-bilinear}
\end{align}
Within linear response theory at zero temperature, adding
\begin{align}
\Haml_c &= \frac{g}{\sqrt{\mathcal{N}}} \sum_{a \vb{k}} \ev{\Phi_a} \psi_{\vb{k}}^{\dag} \mleft(\gamma_{a \vb{k}, \vb{k}}^{E_u^{-}} \otimes \Pauli_0\mright) \psi_{\vb{k}}
\end{align}
to the Hamiltonian induces a global current
\begin{align}
j_a &= - \frac{g}{\sqrt{\mathcal{N}}} \ev{\Phi_a}
\int\limits_{\text{1\textsuperscript{st}BZ}} \frac{\dd[2]{k}}{(2 \pi)^2} \Bigg[\sum_n \Dd(\varepsilon_{\vb{k} n}) \Tr\mleft(\mathcal{K}_{\vb{k}}^{\dag} J_a^{E_u^{-}} \mathcal{K}_{\vb{k}} \mathcal{P}_{\vb{k} n} \gamma_{a \vb{k}, \vb{k}}^{E_u^{-}} \mathcal{P}_{\vb{k} n}\mright) \label{eq:cuprate-Eum-lin-response} \\[-8pt]
&\hspace{130pt} + \sum_{n \neq m} \frac{\HTh(- \varepsilon_{\vb{k} m}) - \HTh(- \varepsilon_{\vb{k} n})}{\varepsilon_{\vb{k} n} - \varepsilon_{\vb{k} m}} \Tr\mleft(\mathcal{K}_{\vb{k}}^{\dag} J_a^{E_u^{-}} \mathcal{K}_{\vb{k}} \mathcal{P}_{\vb{k} n} \gamma_{a \vb{k}, \vb{k}}^{E_u^{-}} \mathcal{P}_{\vb{k} m}\mright)\Bigg] \notag \\
&= - \frac{g}{\sqrt{\mathcal{N}}} \ev{\Phi_a} \, \vb{h} \vdot \vb{c},
\end{align}
where $\vb{h} = (h_1, h_2, h_3)$ are the linear response coefficients obtained by evaluating the above integral and $\vb{c} = (c_1, c_2, c_3)$ specify the $\gamma$ matrix of Eq.~\eqref{eq:gamma-Eum-bilinear} which determines the fermionic bilinear through Eq.~\eqref{eq:general-momentum-bilinear-again2}.
$\varepsilon_{\vb{k} n}$ and $\mathcal{P}_{\vb{k} n}$ are the dispersions and band projectors introduced in Eq.~\eqref{eq:CuO2-disp-proj-def}.
Note that only $\gamma_{b \vb{k}, \vb{k}}^{E_u^{-}}$ with the same $b = a$ arises above because the trace with the other component $b \neq a$ vanishes identically by symmetry.
Moreover, $\vb{h}$ has the same value for both $a = x$ and $a = y$, again due to symmetry.
The $\sqrt{\mathcal{N}}$ appears because $\ev{\Phi_a} \equiv \ev{\Phi_a(\vb{R})} = \sqrt{\mathcal{N}} \ev{\Phi_{a, \vb{q} = \vb{0}}}$.

The direction of $\vb{h}$ depends weakly on chemical potential and for the parameter set of Eq.~\eqref{eq:standard-CuO2-parameter-set} with $\upmu = \epsilon_d + 0.9 t_{pd}$ it equals
\begin{align}
\frac{\vb{h}}{\abs{\vb{h}}} &= \begin{pmatrix}
0.85 \\ -0.29 \\ 0.44
\end{pmatrix}. \label{eq:Bloch-h-vector-value}
\end{align}
Naively, if in Eq.~\eqref{eq:cuprate-Eum-lin-response} we dropped the $\mathcal{P}_{\vb{k}n}$ projectors, integral weights, etc., the trace identity~\eqref{eq:Lambda-orthonormalization} would suggest that $\vb{h}$ approximately points along $(t_{pd}/\sqrt{2}, t_{pp}', t_{pp}) = (0.707..., 0.5, \linebreak 0.6) t_{pd}$.
However, from the numerical result we see that the next-nearest hopping $\propto c_2$ actually reduces the net current, even though $t_{pp}'$ is positive just like $t_{pd}$ and $t_{pp}$.

Bloch's theorem gives the linear constraint:
\begin{align}
\vb{h} \vdot \vb{c} &= 0.
\end{align}
If we further normalize the coefficients to $\vb{c} \vdot \vb{c} = 1$, this leaves a one-parameter family of $E_u^{-}$ bilinears that we parameterize with an angle $\upalpha$:
\begin{align}
\vb{c} &= \vu{h}_c \cos \upalpha + \vu{h}_s \sin \upalpha.
\end{align}
Here $\vb{h}_c = (0,1,0) \vcross \vb{h}$, $\vu{h}_c = \vb{h}_c / \abs{\vb{h}_c}$, $\vb{h}_s = \vb{h} \vcross \vb{h}_c$, and $\vu{h}_s = \vb{h}_s / \abs{\vb{h}_s}$.
Explicitly, for the $\vb{h}$ from above:
\begin{align}
\vu{h}_c &= \begin{pmatrix}
0.46 \\ 0 \\ -0.89
\end{pmatrix}, &
\vu{h}_s &= \begin{pmatrix}
0.25 \\ 0.96 \\ 0.13
\end{pmatrix}.
\end{align}
The dependence of $\vb{c}$ on $\upalpha$ is plotted in Fig.~\ref{fig:cuprate-Eu-results-alpha}(b).

\subsubsection{Kirchhoff constraints} \label{sec:Kirch-Kirch-constr}
Local charge conservation entails that for each site $\alpha$:
\begin{align}
\dot{n}_{\alpha} + \sum_{\beta} j_{\alpha \beta} &= 0,
\end{align}
where $n_{\alpha}$ is the charge on site $\alpha$ and $j_{\alpha \beta} = j_{\alpha \beta}^{\dag} = - j_{\beta \alpha}$ is the charge current flowing from the site $\alpha$ to some other site $\beta$.
When $\Haml = \sum_{\alpha \beta} \mathcal{T}_{\alpha \beta} \psi_{\alpha}^{\dag} \psi_{\beta}$, Heisenberg's equations of motion give
\begin{align}
j_{\alpha \beta} &= \iu \, \mathcal{T}_{\alpha \beta} \psi_{\alpha}^{\dag} \psi_{\beta} + \Hc
\end{align}
For steady phases of matter $\dot{n}_{\alpha} = 0$, which in turn implies that any currents that may appear due to breaking of TR symmetry must obey Kirchhoff's law:
\begin{align}
\sum_{\beta} j_{\alpha \beta} = 0.
\end{align}
A TR-odd bosonic mode can be quantum-critical only if, after condensation, it satisfies the above constraint.
Indeed, in Sec.~\ref{sec:Bloch-tm-Kirchhoff} we have have adapted the proof of Bloch's theorem to show that any state of matter that does not satisfy Kirchhoff's law is unstable against charge relaxation.

The global charges located on the various orbitals are given by:
\begin{align}
n_d &= \sum_{\vb{R}} \Psi^{\dag}(\vb{R}) \diag(1,0,0,0,0) \Psi(\vb{R}), \\
n_{p_x} &= \sum_{\vb{R}} \Psi^{\dag}(\vb{R}) \diag(0,\tfrac{1}{2},0,\tfrac{1}{2},0) \Psi(\vb{R}), \\
n_{p_y} &= \sum_{\vb{R}} \Psi^{\dag}(\vb{R}) \diag(0,0,\tfrac{1}{2},0,\tfrac{1}{2}) \Psi(\vb{R}).
\end{align}
With respect to the non-interacting three-band Hamiltonian of Sec.~\ref{sec:cuprate-3band-model}, their time derivatives equal
\begin{align}
\dot{n}_d = - \dot{n}_{p_x} - \dot{n}_{p_y} &= \sum_{\vb{R}} \Psi^{\dag}(\vb{R}) J^{A_{1g}^{-}} \Psi(\vb{R}), \\
\dot{n}_{p_x} - \dot{n}_{p_y} &= \sum_{\vb{R}} \Psi^{\dag}(\vb{R}) J^{B_{1g}^{-}} \Psi(\vb{R}),
\end{align}
where
\begin{align}
J^{A_{1g}^{-}} &= 2 t_{pd} \Lambda^{A_{1g}^{-}}_1, \\
J^{B_{1g}^{-}} &= - 2 t_{pd} \Lambda^{B_{1g}^{-}}_1 + 4 t_{pp} \Lambda^{B_{1g}^{-}}_2.
\end{align}
Hence the $A_{1g}^{-}$ state described by the $\gamma$ matrix [Eq.~\eqref{eq:general-momentum-bilinear-again2}]
\begin{align}
\gamma_{\vb{k}, \vb{p}}^{A_{1g}^{-}} &= \mathcal{K}_{\vb{k}}^{\dag} \Lambda^{A_{1g}^{-}}_1 \mathcal{K}_{\vb{p}}
\end{align}
is forbidden because it would cause charge accumulation on the $d$ orbitals.
The $A_{2g}^{-}$ state
\begin{align}
\gamma_{\vb{k}, \vb{p}}^{A_{2g}^{-}} &= \mathcal{K}_{\vb{k}}^{\dag} \Lambda^{A_{2g}^{-}}_1 \mathcal{K}_{\vb{p}} \label{eq:gamma-A2g-bilinear}
\end{align}
satisfies Kirchhoff's law identically since all the orbitals are located on mirror planes over which the irrep changes sign.
As for the local $B_{1g}^{-}$ state
\begin{align}
\gamma_{\vb{k}, \vb{p}}^{B_{1g}^{-}} &= \mathcal{K}_{\vb{k}}^{\dag} \mleft(c_1 \Lambda^{B_{1g}^{-}}_1 + c_2 \Lambda^{B_{1g}^{-}}_2\mright) \mathcal{K}_{\vb{p}}, \label{eq:gamma-B1g-bilinear}
\end{align}
linear response theory yields the Kirchhoff constraint
\begin{align}
j^{B_{1g}^{-}} &= \ev{\Psi^{\dag}(\vb{R}) J^{B_{1g}^{-}} \Psi(\vb{R})} = - \frac{g}{\sqrt{\mathcal{N}}} \ev{\Phi} \, \vb{h} \vdot \vb{c},
\end{align}
where $\vb{h} = (h_1, h_2)$ are obtained from Eq.~\eqref{eq:cuprate-Eum-lin-response} by replacing $J_a^{E_u^{-}}$ with $J^{B_{1g}^{-}}$ and $\gamma_{a \vb{k}, \vb{k}}^{E_u^{-}}$ with the $\gamma_{\vb{k}, \vb{k}}^{B_{1g}^{-}}$ of Eq.~\eqref{eq:gamma-B1g-bilinear}. 
After normalization, we are left with only one viable LC $B_{1g}^{-}$ state:
\begin{align}
\begin{pmatrix}
c_1 \\ c_2
\end{pmatrix} &= \frac{1}{\sqrt{h_1^2 + h_2^2}}
\begin{pmatrix} h_2 \\ - h_1 \end{pmatrix} \approx \begin{pmatrix}
0.59 \\ 0.81
\end{pmatrix}.
\end{align}
The numerical value is for $\upmu = \epsilon_d + 0.9t_{pd}$ and the standard parameter set of Eq.~\eqref{eq:standard-CuO2-parameter-set}.
The $\vb{c} = (c_1, c_2)$ coefficients do not depend strongly on chemical potential.
For the $E_u^{-}$ state of Eq.~\eqref{eq:gamma-Eum-bilinear}, Kirchhoff's law is enforced by symmetry at each orbital site and does not give any additional constraints.

\subsubsection{No constraints for spin-magnetic orders}
In the next section, we shall also considered spin-dependent bilinears belonging to $E_u^{-}$ that have the form:
\begin{align}
\Gamma_{a \vb{k}, \vb{p}} &= \mathcal{K}_{\vb{k}}^{\dag} \mleft(c_1 \Lambda^{E_u^{+}}_{1,a}+ c_2 \Lambda^{E_u^{+}}_{2,a} + c_3 \Lambda^{E_u^{+}}_{3,a}\mright) \mathcal{K}_{\vb{p}} \otimes \Pauli_z.
\end{align}
As was explained in Sec.~\ref{sec:cup-ord-param-constr}, there are only three pairs of local spin-dependent $E_u^{-}$ bilinears (Tab.~\ref{tab:SOC-bilin-classification-stats}), which are precisely those given above.
In this case, $\mathcal{K}_{\vb{k}}^{\dag} \Lambda^{E_u^{+}}_{2,a} \mathcal{K}_{\vb{k}} = 0$ identically due to exact cancellation deriving from translation invariance (see the schematic of Tab.~\ref{tab:orbital-Lambda-mats}) so we are again left with a 1D parameter space:
\begin{align}
\vb{c} &= \begin{pmatrix}
\cos \upalpha \\ 0 \\ \sin \upalpha
\end{pmatrix}.
\end{align}

Although these bilinears cannot induce global charge currents, perhaps they can induce global spin currents described by the matrix $J_a^{E_u^{-}} \otimes \Pauli_z$.
However, given the absence of SOC, one readily observes that the spin parts of the traces factor out in Eq.~\eqref{eq:cuprate-Eum-lin-response}, leaving orbital parts that vanish because they couple $E_u$ matrices of opposite TR signs.
Thus there is no Bloch constraint on the spin $E_u^{-}$ bilinears.
For similar reasons, local spin $A_{2g}^{-}$, spin $B_{1g}^{-}$, and spin $B_{2g}^{-}$ bilinears have no Kirchhoff constraints.
Orbital $B_{2g}^{-}$ and spin $A_{1g}^{-}$ bilinears that are localized within one extended unit cell do not exist for the three-orbital \ce{CuO2} model.
The physical explanation for the absence of Bloch and Kirchhoff constraints is that TR-odd spin order is fundamentally about spin densities, not currents.
Spin loop currents, which \emph{are} subject to these constraints, are even under TR, as noted in Tabs.~\ref{tab:orbital-modes} and~\ref{tab:orbital-spin-IUC-orders} of the previous chapter.

\subsection{Cooper pair scattering off Van Hove points} \label{sec:pairing-cuprate-actual-analysis-VH}
Van Hove points are points in crystal momentum space where the Fermi velocity $\vb{v}_{\vb{k}} = \grad_{\vb{k}} \varepsilon_{\vb{k}}$ vanishes.
This, in turn, implies that the density of states (DOS), which is $\propto \int_{\text{FS}} \dd{S_{\vb{k}}} / \abs{\vb{v}_{\vb{k}}}$, receives singular contributions from these points when the Fermi surface crosses them.
Since the Cooper pairing strength is proportional to the DOS, it is important to elucidate how Van Hove points affect the pairing mediated by LC fluctuations.

\begin{figure}[t]
\centering
\begin{subfigure}[t]{0.3\textwidth}
\raggedright
\includegraphics[width=0.95\textwidth]{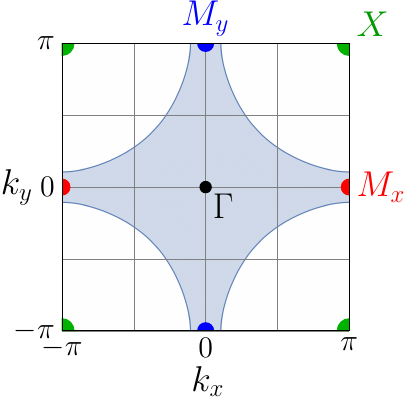}
\subcaption{}
\end{subfigure}%
\begin{subfigure}[t]{0.7\textwidth}
\centering
\includegraphics[width=0.95\textwidth]{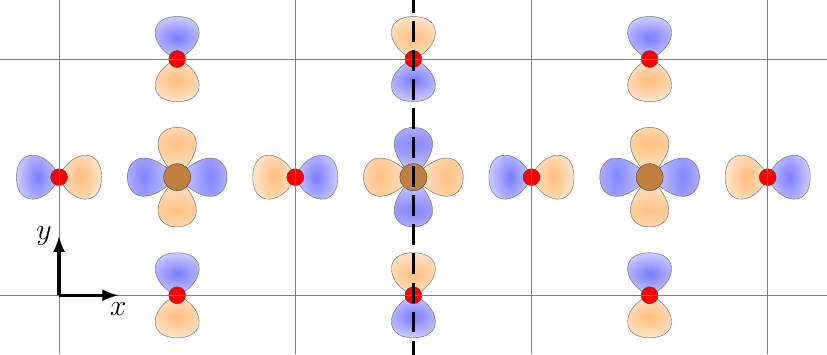}
\subcaption{}
\end{subfigure}
\captionbelow[The first Brillouin zone and its high-symmetry points (a) and the orbitals of the \ce{CuO2} plane modulated by the Van Hove wavevector $\vb{k}_{M_x} = (\pi, 0)$ (b).]{\textbf{The first Brillouin zone and its high-symmetry points (a) and the orbitals of the \ce{CuO2} plane modulated by the Van Hove wavevector $\vb{k}_{M_x} = (\pi, 0)$ (b).}
Under (a), shaded in blue is a typical Fermi sea at overdoping.
Under (b), solid lines outline the unit cells, while the dashed line denotes the $yz$-plane of reflection.
At the $M_x$ momentum, all orbitals are even under mirroring $\Sigma_x$ across the $yz$-plane (Tab.~\ref{tab:CuO2-orbital-sym-eigenvalues}).
However, the $p_y$ orbital belongs to a different irrep from $d_{x^2-y^2}$ and $p_x$ (Tab.~\ref{tab:CuO2-band-irreps}), as can be seen from the fact that it is odd under parity (spatial inversion across the copper site), unlike $d_{x^2-y^2}$ and $p_x$.}
\label{fig:cup-VH-reflection}
\end{figure}

At generic momenta in $d$ spatial dimensions, all $d$ components of $\vb{v}_{\vb{k}}$ are finite.
The equation $\vb{v}_{\vb{k}} = \vb{0}$ thus has solutions only when symmetries force some or all of the components of $\vb{v}_{\vb{k}}$ to vanish identically.
Van Hove points therefore reside on high-symmetry points and lines of the Brillouin zone.
In the three-orbital \ce{CuO2} model, there are four high-symmetry points:
\begin{align}
\vb{k}_{\Gamma} &= \begin{pmatrix}
0 \\ 0
\end{pmatrix}, &
\vb{k}_{M_x} &= \begin{pmatrix}
\pi \\ 0
\end{pmatrix}, &
\vb{k}_{M_y} &= \begin{pmatrix}
0 \\ \pi
\end{pmatrix}, &
\vb{k}_{X} &= \begin{pmatrix}
\pi \\ \pi
\end{pmatrix}, \label{eq:high-symmetry-points}
\end{align}
shown in Fig.~\ref{fig:cup-VH-reflection}(a).
All are, up to a reciprocal lattice vector, invariant with respect to the vertical reflections $\Sigma_{x}$ and $\Sigma_{y}$ whose normals are $\vu{e}_x$ and $\vu{e}_y$, respectively.\footnote{The corresponding planes of reflection are $yz$ and $xz$, respectively.}
Given that $\varepsilon_{\Sigma_{x} \vb{k} + \vb{G}} = \varepsilon_{\vb{k}} = \varepsilon_{\Sigma_{y} \vb{k} + \vb{G}}$, differentiating this identity tells us that the $x$ and $y$ components of $\vb{v}_{\vb{k}}$ both vanish.
Hence these four high-symmetry points are Van Hove points.
In principle, additional accidental Van Hove points are possible, e.g., along the $\Gamma$--$M_x$ high-symmetry line, but for the model at hand they are not present.
The $\vb{k}_{\Gamma}$ and $\vb{k}_{X}$ points are associated with the minimum and maximum of the conduction band dispersion, respectively.
More interesting to us are the $\vb{k}_{M_x}$ and $\vb{k}_{M_y}$ Van Hove points which are associated with saddle points of the conduction band and near which the DOS gets logarithmically enhanced.\footnote{The DOS and gapping of a dispersion saddle point we study in a different context in Chap.~\ref{chap:Sr2RuO4} on \ce{Sr2RuO4}.}

\subsubsection{Symmetries and the little group of the Van Hove points}
Let us consider the Van Hove point $\vb{k}_{M_x}$.
The full point group of the system $D_{4h}$, which is review in Sec.~\ref{sec:tetragonal-group-D4h}, is generated by four-fold rotations around $z$ $C_{4z}$, two-fold rotations around $x$ $C_{2x}$, two-fold rotations around the diagonal $d_{+} = x+y$ $C_{2d_{+}}$, and parity $P$.
The subgroup of $D_{4h}$ which keeps $\vb{k}_{M_x}$ invariant up to a reciprocal lattice vector $\vb{G}$ is called the little group of $\vb{k}_{M_x}$, or sometimes also the point group of the wavevector $\vb{k}_{M_x}$.
Formally we may write it as $\{g \in D_{4h} \mid \exists \vb{G}\colon R(g) \vb{k}_{M_x} = \vb{k}_{M_x} + \vb{G}\}$.
For $\vb{k}_{M_x}$, its little group equals the orthorhombic point group $D_{2h}$ which is generated by two-fold rotations around $x$, $y$, $z$, and parity.
Its character table is given in Tab.~\ref{tab:D2h-char-tab}.
Irreps of the little group $D_{2h}$ we shall denote with primes to avoid confusion (e.g., the $B_{1g}$ irrep of $D_{4h}$ is even under $C_{2x}$ and $C_{2y}$, while the $B_{1g}'$ irrep of $D_{2h}$ is odd under these two \SI{180}{\degree} rotations).
The $C_{2d_{+}}$ and $C_{4z}$ rotations map the two high-symmetry points $\vb{k}_{M_x}$ and $\vb{k}_{M_y}$.

\begin{table}[t]
\centering
\captionabove[The character table of the orthorhombic point group $D_{2h}$~\cite{Dresselhaus2007}.]{\textbf{The character table of the orthorhombic point group $D_{2h}$}~\cite{Dresselhaus2007}.
This point group is the little group of ${M_x} = (\pi, 0)$ and ${M_y} = (0, \pi)$.
The irreps are divided according to parity into even (subscript $g$) and odd ($u$) ones.
To the left of the irreps are the simplest polynomials constructed from the coordinates $\vb{r} = (x, y, z)$ that transform according to them.
Primes have been added on the irreps to distinguish them from $D_{4h}$ irreps.
$C_{2z}$, $C_{2y}$, and $C_{2x}$ are \SI{180}{\degree} rotations around $\vu{e}_z$, $\vu{e}_y$, and $\vu{e}_x$, respectively.
$P$ is space inversion or parity.
Mirror reflections $\Sigma_z$, $\Sigma_y$, and $\Sigma_x$ are obtained by composing $C_{2z}$, $C_{2y}$, and $C_{2x}$ with $P$, respectively.}
{\renewcommand{\arraystretch}{1.3}
\renewcommand{\tabcolsep}{10pt}
\begin{tabular}{cc|rrrr|rrrr} \hline\hline
\multicolumn{2}{c|}{$D_{2h}$} & $E$ & $C_{2z}$ & $C_{2y}$ & $C_{2x}$ & $P$ & $\Sigma_z$ & $\Sigma_y$ & $\Sigma_x$
\\ \hline
$1$, $x^2$, $y^2$, $z^2$ & $A_{1g}'$ & $1$ & $1$ & $1$ & $1$ & $1$ & $1$ & $1$ & $1$
\\
$xy$ & $B_{1g}'$ & $1$ & $1$ & $-1$ & $-1$ & $1$ & $1$ & $-1$ & $-1$
\\
$xz$ & $B_{2g}'$ & $1$ & $-1$ & $1$ & $-1$ & $1$ & $-1$ & $1$ & $-1$
\\
$yz$ & $B_{3g}'$ & $1$ & $-1$ & $-1$ & $1$ & $1$ & $-1$ & $-1$ & $1$
\\ \hline
$xyz$ & $A_{1u}'$ & $1$ & $1$ & $1$ & $1$ & $-1$ & $-1$ & $-1$ & $-1$
\\
$z$ & $B_{1u}'$ & $1$ & $1$ & $-1$ & $-1$ & $-1$ & $-1$ & $1$ & $1$
\\
$y$ & $B_{2u}'$ & $1$ & $-1$ & $1$ & $-1$ & $-1$ & $1$ & $-1$ & $1$
\\
$x$ & $B_{3u}'$ & $1$ & $-1$ & $-1$ & $1$ & $-1$ & $1$ & $1$ & $-1$
\\ \hline\hline
\end{tabular}}
\label{tab:D2h-char-tab}
\end{table}

The band Hamiltonian at $\vb{k}_{M_x}$ commutes with all elements of the little group of $\vb{k}_{M_x}$.
Since there is no SOC, the orbital parts of the band eigenvectors fall into irreps of the little group, as opposed to irreps of the double group of the little group which would allow for \SI{360}{\degree} rotations $\mathscr{C}$ equal to minus unity.
In two dimensions, $C_{2z} = P$ so only $A_{1g}'$, $B_{1g}'$, $B_{2u}'$, and $B_{3u}'$ are possible irreps.
The Hamiltonian~\eqref{eq:3band-Haml-again} is easily diagonalized and by exploiting the symmetry matrices of Tab.~\ref{tab:CuO2-model-D4h-generators}, one readily finds the irreps of the bands given in Tab.~\ref{tab:CuO2-band-irreps}.
The irreps are robust against variations of the model parameters and we indeed find the same result for all eight parameter sets of Tab.~\ref{tab:CuO2-model-parameters}.
Moreover, even if we add strong Hubbard or other interactions, as long as they respect the point group symmetries, the symmetry and orbital content of the band states at the Van Hove points will remain intact.

\begin{table}[t]
\centering
\captionabove[Irreps and orbital contents of the bands of the three-orbital \ce{CuO2} model at the high-symmetry point $\vb{k}_{M_x} = (\pi, 0)$.]{\textbf{Irreps and orbital contents of the bands of the three-orbital \ce{CuO2} model at the high-symmetry point $\vb{k}_{M_x} = (\pi, 0)$.}
The model is defined in Sec.~\ref{sec:cuprate-3band-model} and Fig.~\ref{fig:CuO2-sketch}.
The band energies are ordered according to $\varepsilon_{\vb{k} 1} < \varepsilon_{\vb{k} 2} < \varepsilon_{\vb{k} 3}$, i.e., $n = 3$ is the conduction band.
The irreps are those of the little group $D_{2h}$ (Tab.~\ref{tab:D2h-char-tab}).
These results hold for all eight parameter sets of Tab.~\ref{tab:CuO2-model-parameters}.}
{\renewcommand{\arraystretch}{1.3}
\renewcommand{\tabcolsep}{10pt}
\begin{tabular}{ccc} \hline\hline
band index & irrep at $M_x$ & orbital content at $M_x$
\\ \hline
$n = 3$ & $A_{1g}'$ & \ce{Cu}:$3d_{x^2-y^2}$ and \ce{O}:$2p_x$
\\
$n = 2$ & $B_{2u}'$ & \ce{O}:$2p_y$
\\
$n = 1$ & $A_{1g}'$ & \ce{Cu}:$3d_{x^2-y^2}$ and \ce{O}:$2p_x$
\\ \hline\hline
\end{tabular}}
\label{tab:CuO2-band-irreps}
\end{table}

Apart from the band eigenvectors $u_{\vb{k} n}$, it is worthwhile to contemplate whether one can sensibly speak of the symmetry properties of orbitals for high-symmetry points, or even for generic momenta $\vb{k}$.
For comparison, the band eigenvectors express the orbital content of the band at a given $\vb{k}$.
Symmetries tell us that the orbital contents at different momenta are related [cf.\ Eq.~\eqref{eq:band-eigenvector-sym-transf-rule}]:
\begin{align}
\begin{aligned}
\mathcal{K}^{-1} O(g) \mathcal{K}_{\vb{k}} u_{\vb{k} n} &= \Elr^{- \iu \varkappa_{\vb{k} n}(g^{-1})} u_{R(g) \vb{k} n}, \\
O(g) \mathcal{K}_{\vb{k}} u_{\vb{k} n} &= \Elr^{- \iu \varkappa_{\vb{k} n}(g^{-1})} \mathcal{K}_{R(g) \vb{k}} u_{R(g) \vb{k} n},
\end{aligned}
\end{align}
where $\mathcal{K}_{\vb{k}}$ and $\mathcal{K}^{-1}$ are defined in Eqs.~\eqref{eq:K-proj-mat} and~\eqref{eq:K-proj-mat-inv} and $\varkappa_{\vb{k} n}(g)$ is a global phase, here made consistent with Eq.~\eqref{eq:band-eigenvector-sym-no-SOC}.
Clearly, $R(g) \vb{k}$ and $\vb{k}$ need to be commensurate for one to be able to say that $u_{\vb{k} n}$ are definite under a certain symmetry.
Otherwise, there is no reason why $u_{\vb{k} n}$ and $u_{R(g) \vb{k} n}$ should be proportional\footnote{When $R(g) \vb{k} = \vb{k} + \vb{G}$ for inverse lattice $\vb{G}$, the two eigenvectors have the same energy. Since they are also non-degenerate, they must be proportional.} and whatever proportionality one finds is dependent on the global phase (gauge) chosen for $u_{\vb{k} n}$.
In the case when the vectors are made of only one orbital, as in
\begin{align}
u_{d_{x^2-y^2}} &= \begin{pmatrix}
1 \\ 0 \\ 0
\end{pmatrix}, &
u_{p_x} &= \begin{pmatrix}
0 \\ 1 \\ 0
\end{pmatrix}, &
u_{p_y} &= \begin{pmatrix}
0 \\ 0 \\ 1
\end{pmatrix},
\end{align}
it turns out that even for generic $\vb{k}$ one can define their symmetry eigenvalue according to
\begin{align}
\mathcal{K}^{-1} O(g) \mathcal{K}_{\vb{k}} u_{\alpha} &= \RepM_{\vb{k}}^{\alpha}(g) u_{\alpha} \label{eq:orbital-sym-eigenvalues}
\end{align}
as long as the group operation $g$ maps the orbital into itself.
Here $\alpha \in \{d_{x^2-y^2}, p_x, p_y\}$ and $\RepM_{\vb{k}}^{\alpha}(g)$ is the $\vb{k}$-dependent symmetry eigenvalue.
This requirement excludes diagonal rotations $C_{2d_{+}}$, four-fold rotations $C_{4z}$, and related elements of $D_{4h}$, but it still includes all the operations of the little group $D_{2h}$.
The corresponding symmetry eigenvalues are provided in Tab.~\ref{tab:CuO2-orbital-sym-eigenvalues}.
From the table one notices that, at the $\vb{k}_{M_x} = (\pi, 0)$ momentum, \ce{Cu}:$3d_{x^2-y^2}$ and \ce{O}:$2p_x$ transform under $A_{1g}'$, while \ce{O}:$2p_y$ transforms under $B_{2u}'$, in agreement with Tab.~\ref{tab:CuO2-band-irreps}.

\begin{table}[t]
\centering
\captionabove[Symmetry eigenvalues of the \ce{CuO2} orbitals under $D_{2h}$ transformations.]{\textbf{Symmetry eigenvalues of the \ce{CuO2} orbitals under $D_{2h}$ transformations.}
The symmetry eigenvalues are defined in Eqs.~\eqref{eq:orbital-sym-eigenvalues} and ~\eqref{eq:orbital-sym-eigenvalues-v2}.
Underlined are those eigenvalues for which $g$ does not satisfy the additional relation~\eqref{eq:orbital-sym-eigenvalues-constr} for generic $\vb{k}$.
At the high-symmetry momenta $\Gamma$, $M_{x,y}$, and $X$, shown in Fig.~\ref{fig:cup-VH-reflection}(a), the $g$ of the underlined eigenvalues always satisfy Eq.~\eqref{eq:orbital-sym-eigenvalues-constr}.}
{\renewcommand{\arraystretch}{1.3}
\renewcommand{\tabcolsep}{10pt}
\begin{tabular}{c|cccc|cccc} \hline\hline
& $\one$ & $C_{2z}$ & $C_{2y}$ & $C_{2x}$ & $P$ & $\Sigma_z$ & $\Sigma_y$ & $\Sigma_x$
\\ \hline
\raisebox{-3pt}{$\RepM_{\vb{k}}^{d_{x^2-y^2}}(g)$} & $1$ & $1$ & $1$ & $1$ & $1$ & $1$ & $1$ & $1$
\\
$\RepM_{\vb{k}}^{p_x}(g)$ & $1$ & \underline{$- \Elr^{- \iu k_x}$} & \underline{$- \Elr^{- \iu k_x}$} & $1$ & \underline{$- \Elr^{- \iu k_x}$} & $1$ & $1$ & \underline{$- \Elr^{- \iu k_x}$}
\\
$\RepM_{\vb{k}}^{p_y}(g)$ & $1$ & \underline{$- \Elr^{- \iu k_y}$} & $1$ & \underline{$- \Elr^{- \iu k_y}$} & \underline{$- \Elr^{- \iu k_y}$} & $1$ & \underline{$- \Elr^{- \iu k_y}$} & $1$
\\ \hline\hline
\end{tabular}}
\label{tab:CuO2-orbital-sym-eigenvalues}
\end{table}

In the extended basis, Eq.~\eqref{eq:orbital-sym-eigenvalues} is equivalent to
\begin{align}
O(g) \mathcal{K}_{\vb{k}} u_{\alpha} &= \RepM_{\vb{k}}^{\alpha}(g) \mathcal{K}_{R(g) \vb{k}} u_{\alpha}, \label{eq:orbital-sym-eigenvalues-v2}
\end{align}
i.e., the extended-basis vectors on the left-hand and right-hand side have different momenta.
This means that the corresponding symmetries apply to generic momenta, and give rise to constraints for generic $\vb{k}$, only when
\begin{align}
\mathcal{K}_{R(g) \vb{k}} u_{\alpha} = \mathcal{K}_{\vb{k}} u_{\alpha}. \label{eq:orbital-sym-eigenvalues-constr}
\end{align}
This relation always holds for the $d_{x^2-y^2}$ orbital, but it does not always hold for the $p_{x,y}$ orbitals because of their non-trivial Wyckoff positions.
The reason is that some $g$ map the $p_{x, y}$ orbitals at $\vb{R}$ to those at $R(g) \vb{R} + \vb{\delta}$ (Sec.~\ref{sec:extended-basis-def}) so additional $\Elr^{\pm \iu \vb{k} \vdot \vb{\delta}}$ phase factors appear which make $\mathcal{K}_{R(g) \vb{k}} u_{\alpha}$ different from $\mathcal{K}_{\vb{k}} u_{\alpha}$, unless $\vb{k}$ is one of the high-symmetry wavevector listed in Eq.~\eqref{eq:high-symmetry-points}.
As we shall see below, it is the extended-basis relation
\begin{align}
O(g) \mathcal{K}_{\vb{k}} u_{\alpha} &= \RepM_{\vb{k}}^{\alpha}(g) \mathcal{K}_{\vb{k}} u_{\alpha}
\end{align}
that will constrain the pairing form factor for generic $\vb{k}$.

\subsubsection{Constraints on pairing form factors and Van Hove decoupling} \label{sec:pairing-cuprate-actual-analysis-VH-details}
With this understanding of the orbitals and bands at the Van Hove point, we can now analyze the pairing form factor of Eq.~\eqref{eq:final-lin-gap-eq-cup-pairing-form-factor}:
\begin{align}
\Pintf_{a}(\vb{p}, \vb{k}) &= u_{\vb{p} 3}^{\dag} \mathcal{K}_{\vb{p}}^{\dag} \Lambda_a \mathcal{K}_{\vb{k}} u_{\vb{k} 3}.
\end{align}
We want to see whether it vanishes for generic $\vb{p}$ and $\vb{k} = \vb{k}_{M_x} = (\pi, 0)$ at the $M_x$ Van Hove point.

Let us consider $A_{2g}^{-}$ LCs.
The first thing to notice is that $A_{2g}^{-}$ LCs only couple $p_x$ and $p_y$ orbitals, as depicted in the schematic of Tab.~\ref{tab:orbital-Lambda-mats}.
Since we know that the conduction band at $M_x$ is made of $d_{x^2-y^2}$ and $p_x$ orbitals (Tab.~\ref{tab:CuO2-band-irreps}), it follows that
\begin{align}
u_{\vb{p} 3}^{\dag} \mathcal{K}_{\vb{p}}^{\dag} \Lambda^{A_{2g}^{-}}_1 \mathcal{K}_{\vb{k}_{M_x}} u_{\vb{k}_{M_x} 3} = \const \times u_{p_y}^{\dag} \mathcal{K}_{\vb{p}}^{\dag} \Lambda^{A_{2g}^{-}}_1 \mathcal{K}_{\vb{k}_{M_x}} u_{p_x}.
\end{align}
There are only three symmetry operations of the generic extended-basis vector $\mathcal{K}_{\vb{p}} u_{p_y}$: $C_{2y}$, $\Sigma_z$, and $\Sigma_x$ (Tab.~\ref{tab:CuO2-orbital-sym-eigenvalues}).
Both $\mathcal{K}_{\vb{p}} u_{p_y}$ and $\mathcal{K}_{\vb{k}_{M_x}} u_{p_x}$ are even under them, so the minus sign must come from the LC $\Lambda$ matrix.
And indeed
\begin{align}
O^{\dag}(g) \Lambda^{A_{2g}^{-}}_1 O(g) &= - \Lambda^{A_{2g}^{-}}_1
\end{align}
for $g = C_{2y}$ or $\Sigma_x$.
The $\Sigma_x$ mirroring operation is depicted in Fig.~\ref{fig:cup-VH-reflection}(b).
Hence
\begin{align}
\Pintf(\vb{p}, \vb{k}_{M_x}) &= - \Pintf(\vb{p}, \vb{k}_{M_x}) = 0 \qquad\text{for $A_{2g}^{-}$ loop currents.}
\end{align}
An analogous argument applies to the $M_y$ Van Hove point.
Thus $A_{2g}^{-}$ LCs are unable to scatter Cooper pairs away from the Van Hove points.

Although not our focus, for $B_{2g}^{+}$ nematic order the Van Hove points also decouple from the rest of the Fermi surface.
The argument is the same as for $A_{2g}^{-}$ LCs.
The corresponding $\Lambda^{B_{2g}^{+}}_1$ matrix only couples $p_x$ and $p_y$ orbitals (see Tab.~\ref{tab:orbital-Lambda-mats}) and it is odd under $C_{2y}$ and $\Sigma_x$.

Among the other LC orders, the contributions
\begin{align}
u_{\vb{p} 3}^{\dag} \mathcal{K}_{\vb{p}}^{\dag} \Lambda^{E_u^{-}}_{2,y} \mathcal{K}_{\vb{k}_{M_x}} u_{\vb{k}_{M_x} 3} = u_{\vb{p} 3}^{\dag} \mathcal{K}_{\vb{p}}^{\dag} \Lambda^{E_u^{-}}_{3,x} \mathcal{K}_{\vb{k}_{M_x}} u_{\vb{k}_{M_x} 3} = 0
\end{align}
to the $E_u^{-}$ pairing form factor vanish.
The former $\Lambda^{E_u^{-}}_{2,y}$ matrix couples the $p_y$ orbital to itself.
Since there is no $p_y$ component of the conduction band, its contribution vanishes.
The latter $\Lambda^{E_u^{-}}_{3,x}$ matrix couples $p_x$ and $p_y$ and is odd under $C_{2y}$ and $\Sigma_x$ so the argument proceeds analogously to $A_{2g}^{-}$ LCs.
The above constraints are not that interesting because all other contributions to the $E_u^{-}$ pairing form factor are finite.

More interesting is the observation that $B_{1g}^{-}$ LCs would not couple Van Hove points if the $n = 2$ band (Tab.~\ref{tab:CuO2-band-irreps}) were the conduction band.
In this scenario,
\begin{align}
u_{\vb{p} 2}^{\dag} \mathcal{K}_{\vb{p}}^{\dag} \Lambda^{B_{1g}^{-}}_{1} \mathcal{K}_{\vb{k}_{M_x}} u_{\vb{k}_{M_x} 2} = u_{\vb{p} 2}^{\dag} \mathcal{K}_{\vb{p}}^{\dag} \Lambda^{B_{1g}^{-}}_{2} \mathcal{K}_{\vb{k}_{M_x}} u_{\vb{k}_{M_x} 2} = 0.
\end{align}
The first $\Lambda$ matrix couples the $p_{x,y}$ and $d_{x^2-y^2}$ orbitals and its contribution is thus proportional to
\begin{align}
u_{d_{x^2-y^2}}^{\dag} \mathcal{K}_{\vb{p}}^{\dag} \Lambda^{B_{1g}^{-}}_{1} \mathcal{K}_{\vb{k}_{M_x}} u_{p_y}.
\end{align}
Even though the $B_{1g}^{-}$ irrep of $D_{4h}$ is even under all $D_{2h}$ operations, as is the $d_{x^2-y^2}$ orbital, the $p_y$ orbital is odd under $C_{2z}$, $C_{2x}$, $P$, and $\Sigma_y$ so the above vanishes.
The second $\Lambda$ matrix couples the $p_x$ and $p_y$ orbitals, which implies that its contribution is proportional to
\begin{align}
u_{p_x}^{\dag} \mathcal{K}_{\vb{p}}^{\dag} \Lambda^{B_{1g}^{-}}_{2} \mathcal{K}_{\vb{k}_{M_x}} u_{p_y},
\end{align}
which again vanishes because the $p_y$ orbital is odd.
Let us also note that $A_{2g}^{-}$ LCs would effectively couple the Van Hove points in this scenario:
\begin{align}
u_{\vb{p} 2}^{\dag} \mathcal{K}_{\vb{p}}^{\dag} \Lambda^{A_{2g}^{-}}_1 \mathcal{K}_{\vb{k}_{M_x}} u_{\vb{k}_{M_x} 2} \neq 0.
\end{align}
Although this scenario does not apply to cuprates, it illustrates the important fact that the Van Hove decoupling is a consequences of the interplay between the symmetry of the LC order, on the one hand, and the symmetry of the conduction band, on the other.

\subsection{Results: numerical solutions of the linearized gap equation} \label{sec:pairing-cuprate-actual-analysis-results}
Here we present the main results of Ref.~\cite{Palle2024-LC}: the numerical solutions of the linearized gap equation~\eqref{eq:final-lin-gap-eq-cup-simplified} for the three LC orders that we found in the previous section.
These results tell us which LC fluctuations induced superconductivity of the correct $d_{x^2-y^2}$ symmetry, as well as confirm the analytic results of the previous chapter (Sec.~\ref{sec:gen-sys-LC-analysis}) that near the QCP, pairing due to even-parity LCs does not become enhanced, while odd-parity LCs become strongly repulsive.
In addition, here we also briefly discuss the statistical mechanics of the three LC orders, how to experimentally probe them, as well as the impact of spin-orbit coupling on our conclusions.

The linearized gap equation~\eqref{eq:final-lin-gap-eq-cup-simplified}, supplemented by the definitions and formulas of Sec.~\ref{sec:pairing-cuprate-formalism} and by the $\gamma_{a \vb{k}, \vb{p}}$ matrices of Eqs.~\eqref{eq:gamma-Eum-bilinear}, \eqref{eq:gamma-A2g-bilinear}, and~\eqref{eq:gamma-B1g-bilinear}, is a numerically well-conditioned problem.
It is readily solved by discretizing the Fermi surface (line) and then diagonalizing the corresponding matrix.
The leading solutions converge already for $\sim 20$ Fermi surface points, while grids up to $\sim 300$ and more points are easily accessible numerically.
When the Fermi surface grid respects\footnote{Respects in the sense that the grid maps into itself under all point group operations.} the tetragonal lattice symmetries, the solutions fall exactly into the irreps of the $D_{4h}$ point group, as expected (Sec.~\ref{sec:lin-gap-eq-spectral-symmetries}).
For dense grids in general, this is also true to a very high degree of accuracy.
Using the group-theoretic identity~\eqref{eq:irrep-group-decomp} of Sec.~\ref{sec:lin-gap-eq-spectral-symmetries}, one can, in fact, completely automate the process of the identification of the eigenvector irreps.
Instead of discretizing the momenta, another option is to expand the Cooper-channel interaction in angle-dependent harmonics and then diagonalize the corresponding truncated matrix.
Although numerically slower, this approach gives the same results as the direct discretization of the Fermi surface.

In the preceding Sec.~\ref{sec:Bloch-Kirch-constr}, by restricting ourselves to the most-local couplings in real space, that is lowest-order harmonics in momentum space, we have found three possible LC orders: $g_{xy(x^2-y^2)}$-wave LCs belonging to the irrep $A_{2g}^{-}$, $d_{x^2-y^2}$-wave LCs transforming under the irrep $B_{1g}^{-}$, and $(p_x|p_y)$-wave LCs whose irrep is $E_{u}^{-}$.\footnote{The $D_{4h}$ point group and its irreps are reviewed in Sec.~\ref{sec:tetragonal-group-D4h} of Appx.~\ref{app:group_theory}.}
For the $(p_x|p_y)$-wave LCs, we have, in fact, uncovered a whole one-parameter family of possible LC patterns.
These LC orders we shall refer to, respectively, as $g$-wave, $d$-wave, and $p$-wave loop currents.
The symmetry class of the LC fluctuations is the single most important factor governing our results.
Hence below we present our results according to LC type.

There are many parameters that enter the linearized gap equation~\eqref{eq:final-lin-gap-eq-cup-simplified}.
The pairing eigenvalues $\lambda$ of Eq.~\eqref{eq:final-lin-gap-eq-cup-simplified} are dimensionless and they are linearly proportional to the dimensionless ratio $g^2 \chi_0 / t_{pd}$.
Physically, the reason for this is that, on the one hand, $g^2 \chi_0$ is proportional to the overall interaction strength, while, on the other hand, $t_{pd}^{-1}$ is proportional to the density of states at the Fermi surface.
The overall interaction strength is proportional to the coupling strength $g$ squared, due to the two vertexes of the diagram of Fig.~\ref{fig:boson-exchange}, and to the strength of LC fluctuations, as quantified by the LC order parameter correlation function $\chi(\vb{q}) = \ev{\abs{\Phi_{\vb{q}}}^2}$ which is $\propto \chi_0$.
This overall proportionality factor is well-understood from BCS theory and by diving $\lambda$ with $g^2 \chi_0 / t_{pd}$, that is measuring it in units of $g^2 \chi_0 / t_{pd}$, in the forthcoming we can focus on the impact of other parameters.
Using these units for $\lambda$ also has the advantage of rendering the $\lambda$ shown in the different figures comparable.

\begin{figure}[t]
\centering
\includegraphics[width=0.90\textwidth]{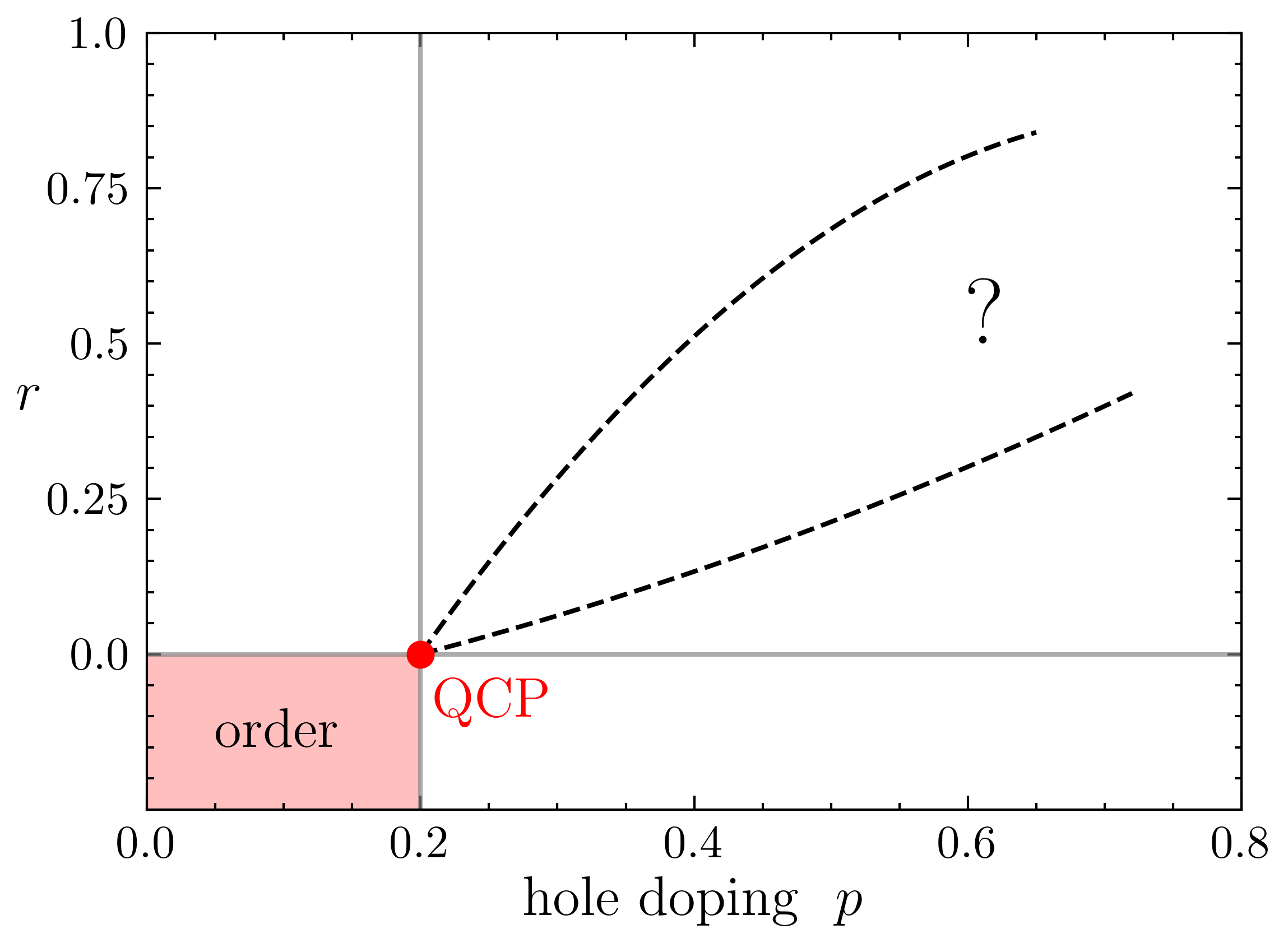}
\captionbelow[The parameter space of our pairing problem for a given loop-current order.]{\textbf{The parameter space of our pairing problem for a given loop-current order.}
The horizontal axis is the hole doping $p$.
The vertical axis is the loop-current softness parameter $r$ which determines the loop-current susceptibility $\chi(\vb{q})$ through Eq.~\eqref{eq:chi-mean-field-expr-r}.
At the quantum-critical point (QCP) it vanishes, $r = 0$, while for negative $r < 0$ the system is unstable against homogeneous loop-current ordering.
The dashed lines are possible dependencies of $r$ on $p$ as the QCP is approached from the overdoped regime $p > p_c$ with $p_c = 0.2$ for concreteness.
The question mark highlights the fact that the precise $r(p)$ dependence is not known within our phenomenological treatment.
Our numerical results reveal that the precise $r(p)$ does not matter, as discussed in the text.}
\label{fig:cuprate-r-p-model-params}
\end{figure}

How do the results depend on the tight-binding parameters $\epsilon_d - \epsilon_p$, $t_{pp}$, and $t_{pp'}$ of Sec.~\ref{sec:cuprate-3band-model}?
To test this, we have tried eight different parameter sets, listed in Tab.~\ref{tab:CuO2-model-parameters}, which cover a broad range of physically realistic values.
In the end, we have found that they affect the pairing solutions only quantitatively, but not qualitatively.
All the important features of the pairing solutions, like their symmetries or behavior near the QCP, are robust against the variations of the band Hamiltonian.
This is somewhat surprising, since the orbital content of the conduction band varies drastically between the parameter sets.
For some parameter sets of Tab.~\ref{tab:CuO2-model-parameters}, the conduction band is predominantly of \ce{Cu}:$3d_{x^2-y^2}$ orbital character, while for others it is predominantly of \ce{O}:$2p_{x,y}$ character.
As we shall explain in Sec.~\ref{sec:cuprate-sym-choice-mechanism}, the reason for this insensitivity lies in the fact that the LC coupling $\gamma_{a \vb{k}, \vb{p}}$ matrix is the one that primarily governs the pairing solutions.
All the shown results are therefore for the typical parameter set (Eq.~\eqref{eq:standard-CuO2-parameter-set}, No.~3 in Tab.~\ref{tab:CuO2-model-parameters}):
\begin{align}
\epsilon_d - \epsilon_p &= 3 t_{pd}, &
t_{pp} &= 0.6 t_{pd}, &
t_{pp}' &= 0.5 t_{pd}, \label{eq:standard-CuO2-parameter-set-again}
\end{align}
with the reference energy $\epsilon_d = 0$.

Of all the parameters, we are thus lead to the conclusion that only two are important for our pairing problem: the hole doping $p$ and the LC softness parameter $r$.
The hole doping $p$ is related to the the chemical potential $\upmu$ through Eq.~\eqref{eq:CuO2-model-dopin} and the evolution of the Fermi surface with $p$ is depicted in Fig.~\ref{fig:cuprate-phase-diagram}.
The LC softness parameter $r > 0$ measures the proximity to the QCP and specifies the gap in the susceptibility through Eq.~\eqref{eq:chi-mean-field-expr-r}.
At the QCP, $r = 0$.
In principle, if we had a microscopic model, its solution would tells us how $r$ depends on $p$ as we approach the putative LC QCP from the overdoped side of the phase diagram.
A few hypothetical $r(p)$ trajectories are illustrated in Fig.~\ref{fig:cuprate-r-p-model-params}.
Within our phenomenological approach, however, the $r(p)$ dependence is unknown.

We have therefore numerically explored the whole $r$-$p$ parameter space of Fig.~\ref{fig:cuprate-r-p-model-params} and found that there are two main features: the QCP line $(r = 0, p)$ on which the LC susceptibility $\chi(\vb{q}) \propto 1/\vb{q}^2$ diverges and the Van Hove line $(r, p = p_{\text{VH}})$ on which the Fermi velocity $v_{\vb{k}}$ vanishes at the Van Hove points $\vb{k} = (\pm \pi, 0)$ and $(0, \pm \pi)$.
For the parameter set of Eq.~\eqref{eq:standard-CuO2-parameter-set-again}, $p_{\text{VH}} = 0.36$ with $\upmu_{\text{VH}} = 0.83 t_{pd}$.
At this $p = p_{\text{VH}}$, the Fermi surface undergoes a Lifshitz transition, as shown in Fig.~\ref{fig:cuprate-Fermi-surfaces}(b).
The most important point about these two features is that they are independent: ARPES measurements of the Van Hove doping $p_{\text{VH}}$ find that $p_{\text{VH}}$ significantly varies between cuprate compounds, with apparently no relation to the critical doping $p_c$ of the QCP shown in Fig.~\ref{fig:cuprate-phase-diagram}~\cite{Sobota2021}.
The only consistent finding is that $p_{\text{VH}} > p_c$~\cite{Sobota2021}.
Thus the enhancement in the density of states expected at $p_{\text{VH}}$ does not directly play a role in the quantum-critical pairing around $p_c$.
Moreover, even tough the leading eigenvalue $\lambda$ depends strongly on $r$ and $p$, we find that the leading pairing state $\Delta(\vb{k})$ does not.
The precise trajectory $r(p)$ is therefore not important within our phenomenological treatment, for neither the enhancement (or suppression) of the pairing tendency as one approaches the QCP nor the leading pairing state and its symmetry depend on the detailed path $r(p)$.
Accordingly, in the figures it is enough to show one cross-section for a fixed $p$, and another for a fixed $r$, as we do in the following.

\subsubsection{$g_{xy(x^2-y^2)}$-wave loop currents} \label{sec:cup-g-wave-res-discus}
The first type of LCs found in Sec.~\ref{sec:Bloch-Kirch-constr} are $g_{xy(x^2-y^2)}$-wave LCs.
They are depicted in the bottom right of Fig.~\ref{fig:cuprate-A2g-results}.
We shall call them ``$g$-wave'' and denote their order parameter with $\Phi_{g}$.
Physically, $\Phi_{g}$ describes a LC order which gives rise to an orbital-magnetic dipole, i.e., an orbital ferromagnet.
It is odd under TR ($p_{\TRop} = -1$), even under parity ($p_{P} = +1$), and transforms under the $A_{2g}^{-}$ irrep of $D_{4h}$.
$\Phi_{g}$ is an Ising order parameter and its statistical mechanics is governed by the Ising model.
It can be polarized by applying an external magnetic field orientated along the $z$ direction $B_{z}$ via the coupling
\begin{align}
\Haml_c &= - \kappa \Phi_{g} B_{z},
\end{align}
where $\kappa$ is a coupling constant.
The coupling of $\Phi_{g}$ to fermions proceeds through
\begin{align}
\Haml_c &= g \Phi_{g} \sum_{\vb{k}} \psi_{\vb{k}}^{\dag} \mleft(\gamma_{\vb{k}, \vb{k}}^{A_{2g}^{-}} \otimes \Pauli_0\mright) \psi_{\vb{k}},
\end{align}
where the coupling $\gamma$ matrix was found to be [Eq.~\eqref{eq:gamma-A2g-bilinear}]:
\begin{align}
\gamma_{\vb{k}, \vb{p}}^{A_{2g}^{-}} &= \mathcal{K}_{\vb{k}}^{\dag} \Lambda^{A_{2g}^{-}}_1 \mathcal{K}_{\vb{p}}. \label{eq:gamma-A2g-bilinear-again}
\end{align}
Here $\mathcal{K}_{\vb{k}}$ is the projection matrix of Eq.~\eqref{eq:K-proj-mat} and the extended-basis orbital matrix $\Lambda^{A_{2g}^{-}}_1$ is listed in Tab.~\ref{tab:orbital-Lambda-mats} (Sec.~\ref{sec:orbital-Lambda-mats}).
The $\gamma_{\vb{k}, \vb{p}}$ of Eq.~\eqref{eq:gamma-A2g-bilinear-again} enters the Cooper-channel interaction via Eq.~\eqref{eq:final-lin-gap-eq-cup-pairing-form-factor}, with the irrep component index suppressed because the $A_{2g}^{-}$ irrep is one-dimensional.

\begin{figure}[t]
\centering
\includegraphics[width=0.45\textwidth]{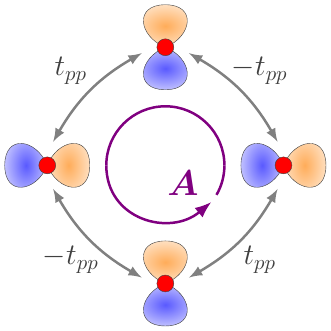}
\captionbelow[The four $p$ orbitals of an extended unit cell with a flux threaded through them.]{\textbf{The four $p$ orbitals of an extended unit cell with a flux threaded through them.}
Grey arrows indicate the hopping amplitudes, while the purple arrow indicates the direction of the vector potential $\vb{A}$.}
\label{fig:cup-Peierls}
\end{figure}

\begin{figure}[p!]
\centering
\includegraphics[width=\textwidth]{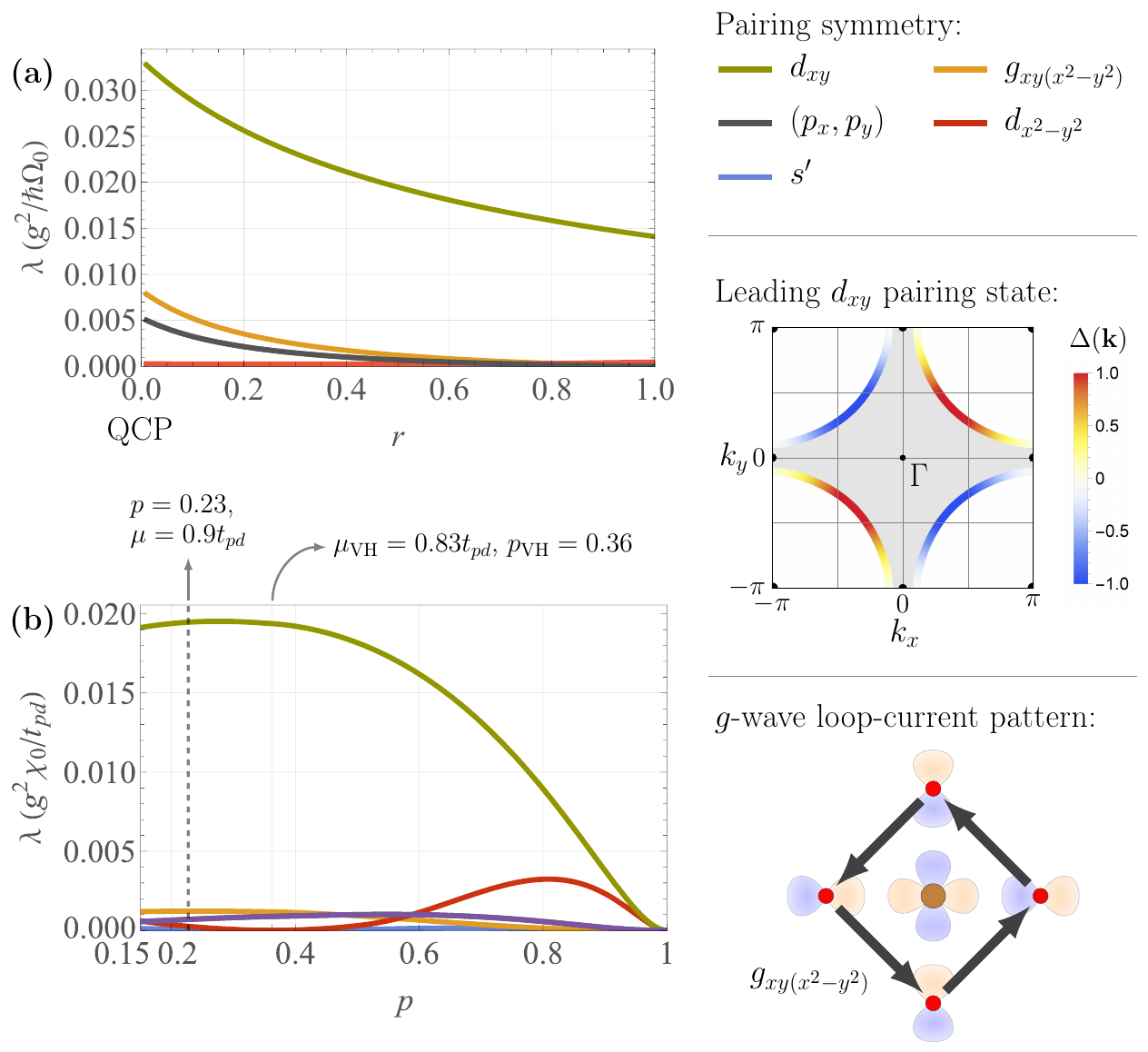}
\captionbelow[Results for the pairing mediated by $g$-wave loop-current fluctuations~\cite{Palle2024-LC}.]{\textbf{Results for the pairing mediated by $g$-wave loop-current fluctuations}~\cite{Palle2024-LC}.
The $g$-wave loop currents have $g_{xy(x^2-y^2)}$ symmetry and transform under the 1D irrep $A_{2g}^{-}$ of the $D_{4h}$ point group.
Their coupling matrix is given in Eq.~\eqref{eq:gamma-A2g-bilinear-again}, with the corresponding current pattern depicted in the bottom right.
The plots show the pairing eigenvalues $\lambda$ of Eq.~\eqref{eq:final-lin-gap-eq-cup-simplified} as a function of the tuning parameter $r$ at fixed chemical potential $\upmu = 0.9 t_{pd}$ (a) and as a function of the hole doping $p$ at fixed $r = 0.5$ (b).
The colors of the curves indicate the pairing symmetry (upper right).
$s'$ stands for extended $s$-wave.
The gap function $\Delta(\vb{k})$ of the leading pairing state, normalized to a maximum of $\pm 1$, is shown on the center right.
The tight-binding parameters used are those of Eq.~\eqref{eq:standard-CuO2-parameter-set-again}.
$r$ determines the susceptibility through Eq.~\eqref{eq:chi-mean-field-expr-r} and the putative loop-current quantum-critical point (QCP) is at $r = 0$.
$p$ is related to $\upmu$ via Eq.~\eqref{eq:CuO2-model-dopin}.
The dashed vertical line under (b) shows the $p = 0.23$ hole doping used in (a).
The additional solid vertical line under (b) corresponds to the Van Hove doping $p_{\text{VH}} = 0.36$.
The corresponding Fermi surfaces are shown in Fig.~\ref{fig:cuprate-Fermi-surfaces}(a),(b).
See Sec.~\ref{sec:cup-g-wave-res-discus} for further discussion.}
\label{fig:cuprate-A2g-results}
\end{figure}

As an alternative way of deriving the coupling to fermions, one can use the Peierls substitution.
To wit, let us consider the four oxygen $p$ orbitals of an extended unit cell, shown in Fig.~\ref{fig:cup-Peierls}.
The four orbitals form a loop and if we thread a magnetic flux through it, since both $\Phi_g$ and $B_z$ belong to the same $A_{2g}^{-}$ irrep, the flux will couple to the fermions in the same way as the $g$-wave LC order parameter $\Phi_g$.
Hence, up to a constant, $\gamma_{\vb{k}, \vb{p}}^{A_{2g}^{-}}$ follows from the Peierls substitution.
The tight-binding Hamiltonian of the $p$ orbitals is [cf.\ Eq.~\eqref{eq:hopping_matrix}]:
\begin{align}
H_{pp} &= 2 t_{pp} \Lambda^{A_{1g}^{+}}_4 = \begin{pmatrix}
0 & 0 & 0 & 0 & 0 \\
0 & 0 & - t_{pp} & 0 & t_{pp} \\
0 & - t_{pp} & 0 & t_{pp} & 0 \\
0 & 0 & t_{pp} & 0 & - t_{pp} \\
0 & t_{pp} & 0 & - t_{pp} & 0
\end{pmatrix}.
\end{align}
In the presence of a magnetic field, the hopping amplitudes get modified via the Peierls substitution:
\begin{equation}
\mathcal{T}_{\alpha \beta} \to \mathcal{T}_{\alpha \beta} \exp\mleft(-\iu \vb{A}_{\alpha \beta} \vdot (\vb{x}_{\alpha} -\vb{x}_{\beta})\mright) \approx \mathcal{T}_{\alpha \beta} \mleft(1 - \iu \vb{A}_{\alpha \beta} \vdot (\vb{x}_{\alpha} -\vb{x}_{\beta})\mright), \label{eq:cup-Peierls}
\end{equation}
where $\vb{A}_{\alpha \beta} = \vb{A}_{\beta\alpha}$ is the average vector potential along the line connecting the sites $\alpha$ and $\beta$ and $\vb{x}_{\alpha}$ is the position of the orbital $\Psi_{\alpha}$ given in Eq.~\eqref{eq:cup-r-alpha-pos}.
A magnetic flux can be represented by a circulating vector potential, shown in Fig.~\ref{fig:cup-Peierls}, that satisfies $\vb{A}_{\alpha \beta} \vdot (\vb{x}_{\alpha} - \vb{x}_{\beta}) = \Phi/4$ when $\vb{x}_{\alpha} = R(C_{4z}) \vb{x}_{\beta}$ is the neighbor in the counterclockwise direction of $\vb{x}_{\beta}$.\footnote{When $\vb{x}_{\alpha}$ is the neighbor in the clockwise direction of $\vb{x}_{\beta}$, $\vb{A}_{\alpha \beta} \vdot (\vb{x}_{\alpha} -\vb{x}_{\beta}) = - \Phi/4$.}
Here $\Phi$ is the total magnetic flux through the loop.
By enacting this substitution in $H_{pp}$, we find that
\begin{equation}
H_{pp} \to H_{pp} - \Phi \, \tfrac{1}{2} t_{pp} \Lambda^{A_{2g}^{-}}_{1},
\end{equation}
with the same
\begin{align}
\Lambda^{A_{2g}^{-}}_{1} &= \frac{1}{2} \begin{pmatrix}
 0 & 0 & 0 & 0 & 0 \\
 0 & 0 & \iu & 0 & \iu \\
 0 & -\iu & 0 & -\iu & 0 \\
 0 & 0 & \iu & 0 & \iu \\
 0 & -\iu & 0 & -\iu & 0 \\
\end{pmatrix}
\end{align}
from earlier.
Thus magnetic fields along the $z$ direction, up to a constant, couple the same way to fermions as $g$-wave LC order parameters.
Conversely, from the above we may deduce Eq.~\eqref{eq:gamma-A2g-bilinear-again}.

The results for the pairing mediated by $g$-wave LCs are given in Fig.~\ref{fig:cuprate-A2g-results}.
As shown in Fig.~\ref{fig:cuprate-A2g-results}(a), $\Phi_{g}$ fluctuations result in parametrically weak $d_{xy}$ pairing, which is parametrically weak in the sense that the pairing eigenvalue $\lambda$ does not diverge at the QCP ($r \to 0$).
This is in agreement with the general results of Chap.~\ref{chap:loop_currents}, Sec.~\ref{sec:gen-sys-LC-analysis}, which are visualized in Fig.~\ref{fig:QCP-general-results}.
Sub-leading singlet and triplet instabilities arise as well.
In Fig.~\ref{fig:cuprate-A2g-results}(b), one sees that the leading $d_{xy}$ instability is weakly enhanced near the Van Hove singularity at $p = p_{\text{VH}}$, while $d_{x^2-y^2}$ pairing is strongly suppressed in the same limit.

The reported~\cite{Aji2010} degeneracy between $d_{xy}$ and $d_{x^2-y^2}$ pairing for $\Phi_g$ is recovered in the limit of extremely overdoped systems with small Fermi surfaces, $p \to 1$.
The counter-intuitive\footnote{Counter-intuitive because, on the one hand, the SC gap function is usually largest where the density of states is largest (Fermi velocity is smallest), while, on the other hand, $d_{xy}$ pairing precisely vanishes at the Van Hove point where the Fermi velocity vanishes. See also Sec.~\ref{sec:cuprate-sym-choice-mechanism}.} result that this degeneracy is lifted in favor of $d_{xy}$ pairing by realistic $\upmu$ values follows from the fact that the pairing form factor $\Pintf(\vb{k}, \vb{p})$ vanishes when either $\vb{k}$ or $\vb{p}$ are at the high-symmetry Van Hove points $\vb{k}_{M_x} = (\pi, 0)$ or $\vb{k}_{M_y} = (0, \pi)$, as we proved in Sec.~\ref{sec:pairing-cuprate-actual-analysis-VH-details}.
Hence $\Phi_{g}$-mediated pairing cannot take advantage of the enhanced density of states due to the Van Hove singularity.

\subsubsection{$d_{x^2-y^2}$-wave loop currents} \label{sec:cup-d-wave-res-discus}
\begin{figure}[p!]
\centering
\includegraphics[width=\textwidth]{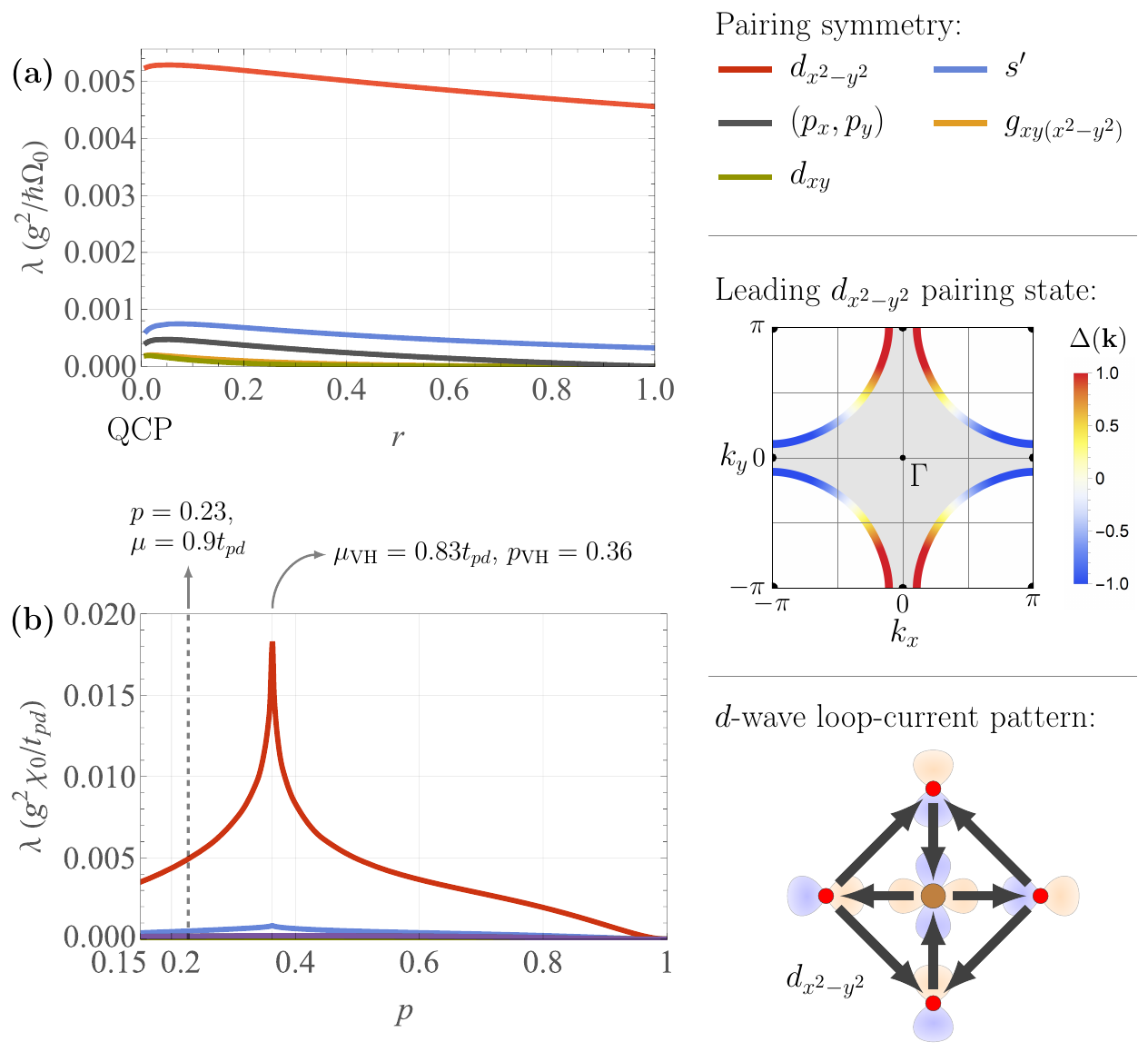}
\captionbelow[Results for the pairing mediated by $d$-wave loop-current fluctuations~\cite{Palle2024-LC}.]{\textbf{Results for the pairing mediated by $d$-wave loop-current fluctuations}~\cite{Palle2024-LC}.
The $d$-wave loop currents have $d_{x^2-y^2}$ symmetry and transform under the 1D irrep $B_{1g}^{-}$ of the $D_{4h}$ point group.
Their coupling matrix is given in Eq.~\eqref{eq:gamma-B1g-bilinear-again}, with the corresponding current pattern depicted in the bottom right.
The plots show the pairing eigenvalues $\lambda$ of Eq.~\eqref{eq:final-lin-gap-eq-cup-simplified} as a function of the tuning parameter $r$ at fixed chemical potential $\upmu = 0.9 t_{pd}$ (a) and as a function of the hole doping $p$ at fixed $r = 0.5$ (b).
The colors of the curves indicate the pairing symmetry (upper right).
$s'$ stands for extended $s$-wave.
The gap function $\Delta(\vb{k})$ of the leading pairing state, normalized to a maximum of $\pm 1$, is shown on the center right.
The tight-binding parameters used are those of Eq.~\eqref{eq:standard-CuO2-parameter-set-again}.
$r$ determines the susceptibility through Eq.~\eqref{eq:chi-mean-field-expr-r} and the putative loop-current quantum-critical point (QCP) is at $r = 0$.
$p$ is related to $\upmu$ via Eq.~\eqref{eq:CuO2-model-dopin}.
The dashed vertical line under (b) shows the $p = 0.23$ hole doping used in (a).
The additional solid vertical line under (b) corresponds to the Van Hove doping $p_{\text{VH}} = 0.36$.
The corresponding Fermi surfaces are shown in Fig.~\ref{fig:cuprate-Fermi-surfaces}(a),(b).
See Sec.~\ref{sec:cup-d-wave-res-discus} for further discussion.}
\label{fig:cuprate-B1g-results}
\end{figure}

The second type of LCs found in Sec.~\ref{sec:Bloch-Kirch-constr} are $d_{x^2-y^2}$-wave LCs.
They are depicted in the bottom right of Fig.~\ref{fig:cuprate-B1g-results}.
We shall call them ``$d$-wave'' and denote their order parameter with $\Phi_{d}$.
Physically, $\Phi_{d}$ describes a LC order which has a magnetic octupole moment.
They can be understood as an orbital altermagnet, i.e., a TR-odd state which is invariant under the combination of TR and a four-fold rotation around the $z$ axis, $\TRop C_{4z} \Phi_{d} = \Phi_{d}$; see also Fig.~\ref{fig:spin-mag-class} of Chap.~\ref{chap:loop_currents}.
$\Phi_{d}$ transforms under the $B_{1g}^{-}$ irrep of $D_{4h}$ and as such it has even parity, $p_{P} = +1$.
Like $\Phi_{g}$, $\Phi_{d}$ is an Ising order parameter, but unlike $\Phi_{g}$, it does not have a magnetic moment.
Instead, it displays piezomagnetism.
This means that it can be polarized by the combination of an external magnetic field pointing in the $z$ direction and shear strain $\epsilon_{xy}$:
\begin{align}
\Haml_{c} = - \kappa \Phi_{d} B_{z} \epsilon_{xy}.
\end{align}
Here $\kappa$ is a coupling constant.
The coupling of $\Phi_{d}$ to fermions proceeds via
\begin{align}
\Haml_c &= g \Phi_{d} \sum_{\vb{k}} \psi_{\vb{k}}^{\dag} \mleft(\gamma_{\vb{k}, \vb{k}}^{B_{1g}^{-}} \otimes \Pauli_0\mright) \psi_{\vb{k}},
\end{align}
where the coupling $\gamma$ matrix was determined to be [Eq.~\eqref{eq:gamma-B1g-bilinear}]:
\begin{align}
\gamma_{\vb{k}, \vb{p}}^{B_{1g}^{-}} &= \mathcal{K}_{\vb{k}}^{\dag} \mleft(c_1 \Lambda^{B_{1g}^{-}}_1 + c_2 \Lambda^{B_{1g}^{-}}_2\mright) \mathcal{K}_{\vb{p}}, \label{eq:gamma-B1g-bilinear-again}
\end{align}
with $\mathcal{K}_{\vb{k}}$ defined in Eq.~\eqref{eq:K-proj-mat}, the $\Lambda$ matrices given in Tab.~\ref{tab:orbital-Lambda-mats}, and $(c_1, c_2) \approx (0.58, 0.81)$ as found in Sec.~\ref{sec:Kirch-Kirch-constr}.
During the numerics the $c_{1,2}$ coefficients are recalculated for each $\upmu$.

The results for the pairing mediated by $d$-wave LCs are provided in Fig.~\ref{fig:cuprate-B1g-results}.
As is evident from Fig.~\ref{fig:cuprate-B1g-results}(a), $\Phi_{d}$ fluctuations promote pairing of the correct $d_{x^2-y^2}$ symmetry.
However, this pairing is parametrically weak in the sense that the pairing eigenvalue $\lambda$ does not diverge at the QCP ($r \to 0$), in agreement with the general results of Sec.~\ref{sec:gen-sys-LC-analysis} (Fig.~\ref{fig:QCP-general-results}).
In addition, several sub-leading singlet and one triplet pairing instabilities appear, but none of them are competitive to the leading instability.
The pairing strength of the leading $d_{x^2-y^2}$ channel is logarithmically enhanced if one tunes the chemical potential to the Van Hove singularity, as can be seen in Fig.~\ref{fig:cuprate-B1g-results}(b).

\begin{figure}[t!]
\centering
\includegraphics[width=\textwidth]{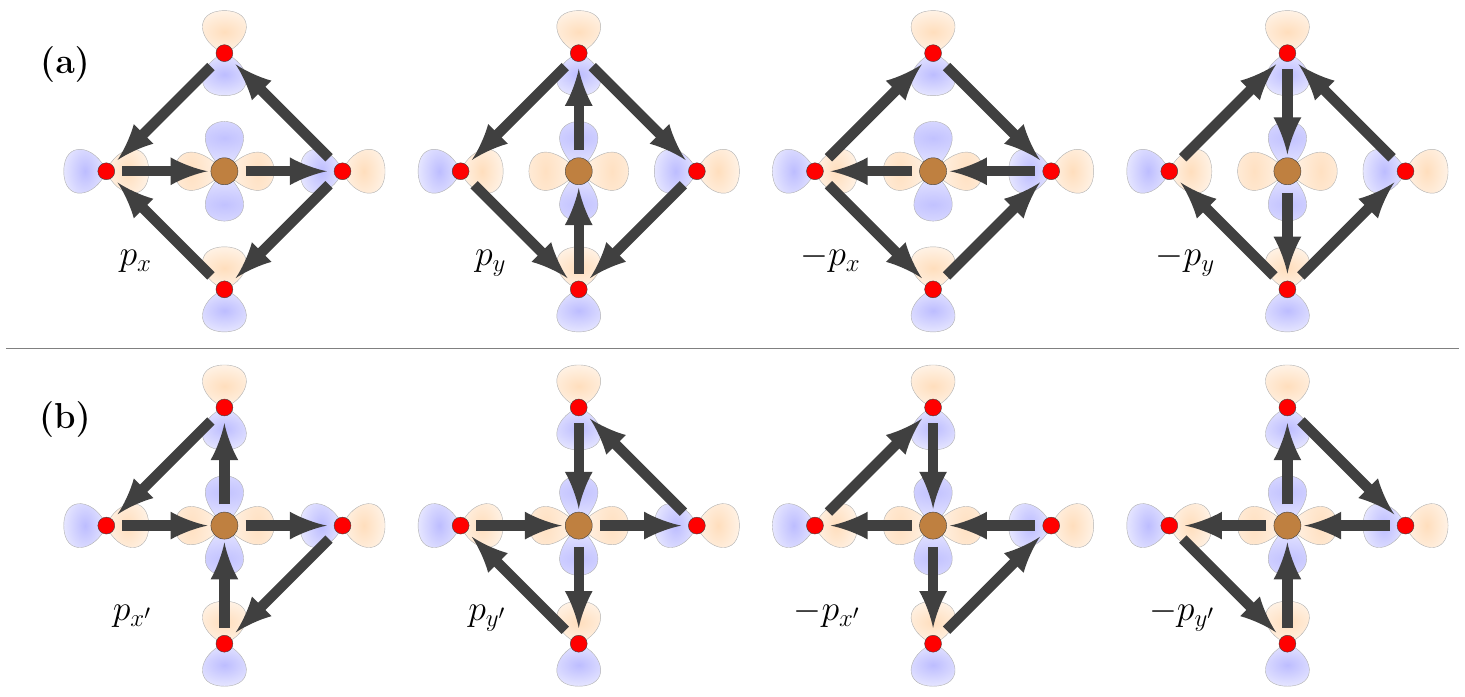}
\captionbelow[The four degenerate $p$-wave loop-current patterns when the in-plane tetragonal anisotropy favors $x$ and $y$ directions (a) vs.\ $x' = (x+y)/\sqrt{2}$ and $y' = (x-y)/\sqrt{2}$ directions (b).]{\textbf{The four degenerate $p$-wave loop-current patterns when the in-plane tetragonal anisotropy favors $x$ and $y$ directions (a) vs.\ $x' = (x+y)/\sqrt{2}$ and $y' = (x-y)/\sqrt{2}$ directions (b).}
Which ones are favored depends on the quartic coefficients of the Ginzburg-Landau expansion.
Here the $\upalpha$ of Eq.~\eqref{eq:gamma-Eum-bilinear-alpha} was set to zero.}
\label{fig:p-wave-four-LC-patterns}
\end{figure}

\begin{figure}[p!]
\centering
\includegraphics[width=\textwidth]{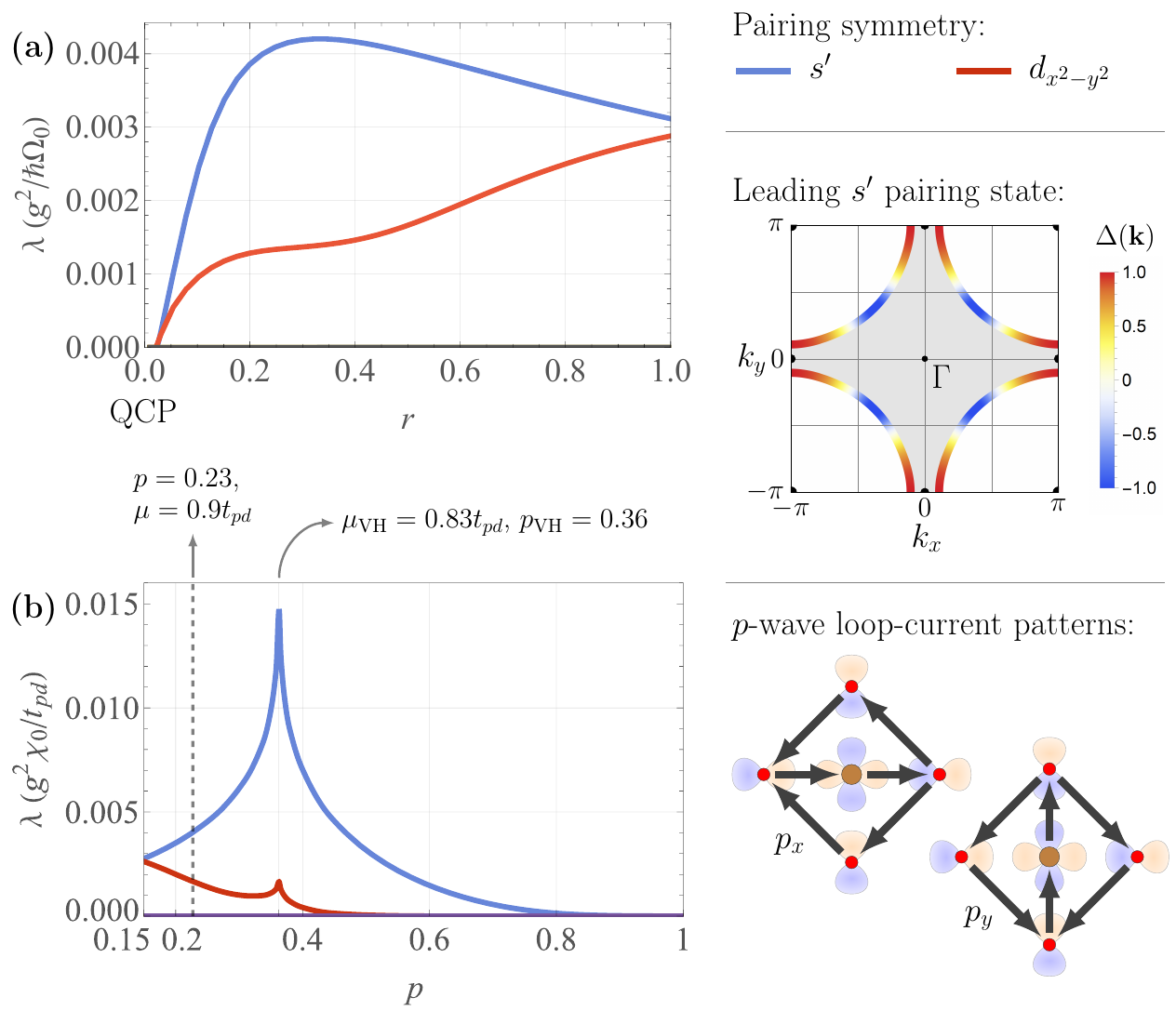}
\captionbelow[Results for the pairing mediated by $p$-wave loop-current fluctuations~\cite{Palle2024-LC}.]{\textbf{Results for the pairing mediated by $p$-wave loop-current fluctuations}~\cite{Palle2024-LC}.
The $p$-wave loop currents have $(p_x|p_y)$ symmetry and transform under the 2D irrep $E_{u}^{-}$ of the $D_{4h}$ point group.
Their coupling matrix is given in Eq.~\eqref{eq:gamma-Eum-bilinear-again} with $\upalpha = 0$ in Eq.~\eqref{eq:gamma-Eum-bilinear-alpha} and the corresponding current patterns is depicted in the bottom right.
For other $\upalpha$ values, see Fig.~\ref{fig:cuprate-Eu-results-alpha}.
The plots show the pairing eigenvalues $\lambda$ of Eq.~\eqref{eq:final-lin-gap-eq-cup-simplified} as a function of the tuning parameter $r$ at fixed chemical potential $\upmu = 0.9 t_{pd}$ (a) and as a function of the hole doping $p$ at fixed $r = 0.5$ (b).
The colors of the curves indicate the pairing symmetry (upper right).
$s'$ stands for the extended $s$-wave solution whose gap function $\Delta(\vb{k}) \approx \cos 4 \theta_k$ is shown on the center right.
The tight-binding parameters used are those of Eq.~\eqref{eq:standard-CuO2-parameter-set-again}.
$r$ determines the susceptibility through Eq.~\eqref{eq:chi-mean-field-expr-r} and the putative loop-current quantum-critical point (QCP) is at $r = 0$.
$p$ is related to $\upmu$ via Eq.~\eqref{eq:CuO2-model-dopin}.
The dashed vertical line under (b) shows the $p = 0.23$ hole doping used in (a).
The additional solid vertical line under (b) corresponds to the Van Hove doping $p_{\text{VH}} = 0.36$.
The corresponding Fermi surfaces are shown in Fig.~\ref{fig:cuprate-Fermi-surfaces}(a),(b).
See Sec.~\ref{sec:cup-p-wave-res-discus} for further discussion.}
\label{fig:cuprate-Eu-results}
\end{figure}

\begin{figure}[p!]
\centering
\includegraphics[width=\textwidth]{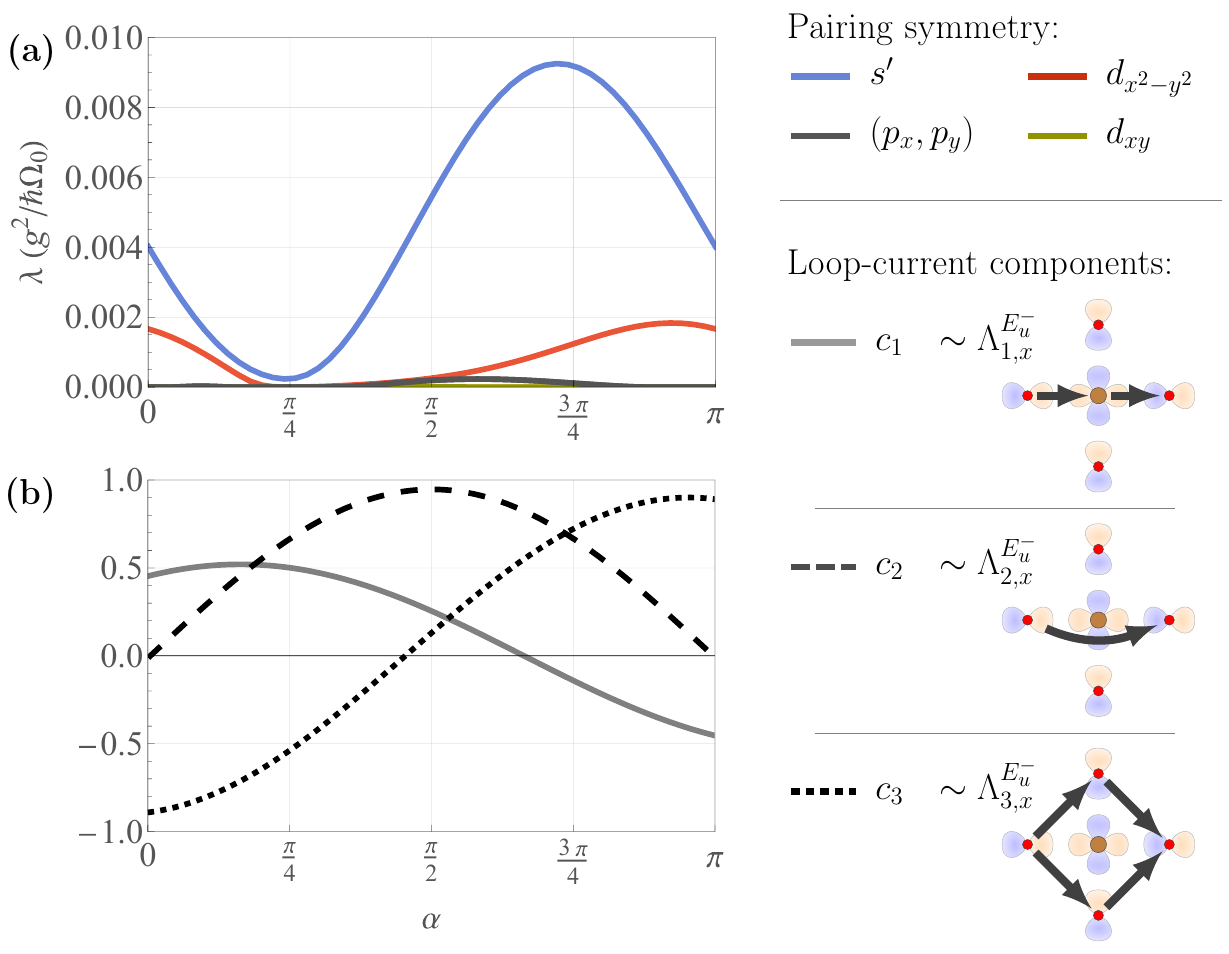}
\captionbelow[Dependence of the results for the pairing mediated by $p$-wave loop-current fluctuations on the angle $\upalpha$~\cite{Palle2024-LC}.]{\textbf{Dependence of the results for the pairing mediated by $p$-wave loop-current fluctuations on the angle $\upalpha$}~\cite{Palle2024-LC}.
The $\upalpha$ angle specifies the $c_1$, $c_2$, and $c_3$ coefficients via Eq.~\eqref{eq:gamma-Eum-bilinear-alpha}, as plotted under (b).
These coefficients correspond to the current patterns depicted on the right and they specify the coupling matrix through Eq.~\eqref{eq:gamma-Eum-bilinear-again}.
Note that $c_2$ contributes oppositely to the global current than what is shown because of band structure effects, as discussed after Eq.~\eqref{eq:Bloch-h-vector-value}.
The plot under (a) shows the pairing eigenvalues $\lambda$ of Eq.~\eqref{eq:final-lin-gap-eq-cup-simplified} as a function of $\upalpha$ at fixed $r = 0.5$ and chemical potential $\upmu = 0.9 t_{pd}$.
The colors of the curves indicate the pairing symmetry (upper right).
The gap function of the leading extended $s'$-wave pairing state can found in the center right of Fig.~\ref{fig:cuprate-Eu-results}.
The tight-binding parameters used are those of Eq.~\eqref{eq:standard-CuO2-parameter-set-again}.
$r$ determines the susceptibility through Eq.~\eqref{eq:chi-mean-field-expr-r}.
The Fermi surface corresponding to $\upmu = 0.9 t_{pd}$ is shown in Fig.~\ref{fig:cuprate-Fermi-surfaces}(a).
See Sec.~\ref{sec:cup-p-wave-res-discus} for further discussion.}
\label{fig:cuprate-Eu-results-alpha}
\end{figure}

\subsubsection{$(p_x, p_y)$-wave loop currents} \label{sec:cup-p-wave-res-discus}
Finally, the last type of LCs found in Sec.~\ref{sec:Bloch-Kirch-constr} are $(p_x|p_y)$-wave LCs.
They are depicted in Fig.~\ref{fig:p-wave-four-LC-patterns}.
We shall call them ``$p$-wave'' and denote their two-component order parameter $\vb{\Phi}_{p} = \mleft(\Phi_{p_{x}},\Phi_{p_{y}}\mright)$.
Physically, $\vb{\Phi}_{p}$ describes a LC order with a toroidal magnetic dipole moment.
It transforms under the $E_{u}^{-}$ irrep and thus has odd parity, $p_{P} = -1$, in contrast to the $g$-wave and $d$-wave LCs considered previously.
Although all directions of $\vb{\Phi}_{p}$ are degenerate on the quadratic level, quartic terms in the Landau expansion reduce the degeneracy down to four discrete directions (Fig.~\ref{fig:p-wave-four-LC-patterns}).\footnote{We shall explicitly see this in a similar context during our Ginzburg-Landau analysis of \ce{Sr2RuO4} in Sec.~\ref{sec:SRO-GL-analysis} of Chap.~\ref{chap:Sr2RuO4}.}
Its statistical mechanics is therefore governed by a four-state clock model.
$\vb{\Phi}_{p}$ has a magneto-electric response, that is to say it can be polarized by crossed electric and magnetic fields according to:
\begin{align}
\Haml_{c} &= - \kappa \mleft(\Phi_{p_x} B_{x} + \Phi_{p_y} B_{y}\mright) E_{z}.
\end{align}
A similar effect can be achieved by applying, instead of electric fields, time-varying currents along the $z$ axis.
The coupling of $\vb{\Phi}_{p}$ to fermions is given by:
\begin{align}
\Haml_c &= g \sum_{a \vb{k}} \Phi_{p_a} \psi_{\vb{k}}^{\dag} \mleft(\gamma_{a \vb{k}, \vb{k}}^{E_{u}^{-}} \otimes \Pauli_0\mright) \psi_{\vb{k}},
\end{align}
where $a \in \{x, y\}$ and [Eq.~\eqref{eq:gamma-Eum-bilinear}]:
\begin{align}
\gamma_{a \vb{k}, \vb{p}}^{E_u^{-}} &= \mathcal{K}_{\vb{k}}^{\dag} \mleft(c_1 \Lambda^{E_u^{-}}_{1,a} + c_2 \Lambda^{E_u^{-}}_{2,a} + c_3 \Lambda^{E_u^{-}}_{3,a}\mright) \mathcal{K}_{\vb{p}}. \label{eq:gamma-Eum-bilinear-again}
\end{align}
In Sec.~\ref{sec:Bloch-Bloch-constr}, we found a one-parameter family of possible $\gamma_{a \vb{k}, \vb{p}}^{E_u^{-}}$.
Its existence follows from the fact that Bloch's theorem gives one constraint, while three paths connecting opposite oxygen orbitals of the same kind are possible: an indirect path through the \ce{Cu} atom (process $c_1$ in Fig.~\ref{fig:cuprate-Eu-results-alpha}), a direct path (process $c_2$), and an indirect path through the \ce{O} atoms (process $c_3$).
In the actual cuprate structure, the second process is mediated by the \ce{Cu}:$4s$ orbital~\cite{Andersen1995, Pavarini2001, Kent2008}.
We shall parameterize this one-parameter family with an angle $\upalpha$ according to:
\begin{align}
\begin{pmatrix}
c_1 \\ c_2 \\ c_3
\end{pmatrix} &= \vu{h}_c \cos \upalpha + \vu{h}_s \sin \upalpha, \label{eq:gamma-Eum-bilinear-alpha}
\end{align}
where the $\vu{h}_{c,s}$ are defined in Sec.~\ref{sec:Bloch-Bloch-constr}.
Although the $\vu{h}_{c,s}$ are recalculated for each $\upmu$ during the numerics, they depend weakly on $\upmu$ and for $\upmu = \epsilon_d + 0.9 t_{pd}$ and the parameter set of Eq.~\eqref{eq:standard-CuO2-parameter-set-again} we find $\vu{h}_{c} = (0.45, 0, -0.89)$ and $\vu{h}_{s} = (0.28, 0.95, 0.14)$.
Since $\gamma_{a \vb{k}, \vb{p}}$ and $-\gamma_{a \vb{k}, \vb{p}}$ give the same interaction [Eq.~\eqref{eq:final-lin-gap-eq-cup-int}], it is sufficient to consider the range $\upalpha \in [0, \pi]$.
In Figs.~\ref{fig:p-wave-four-LC-patterns} and~\ref{fig:cuprate-Eu-results} we use $\upalpha = 0$.

The results for the pairing mediated by $p$-wave LCs are shown in Fig.~\ref{fig:cuprate-Eu-results}.
As can be seen from Fig.~\ref{fig:cuprate-Eu-results}(a), away from the QCP $\vb{\Phi}_{p}$ fluctuations result in weak extended $s$-wave superconductivity that is dominated by an angle-dependent gap function of the form $\Delta(\theta_k) = \Delta_{0} + \Delta_{1} \cos(4 \theta_k)$ with $\abs{\Delta_{1}} \gg \abs{\Delta_{0}}$, yielding eight vertical line nodes.
The corresponding gap function is draw on the center right of Fig.~\ref{fig:cuprate-Eu-results}.
This finding is perfectly consistent with the result of Sec.~\ref{sec:TR-positivity} of the previous chapter, where we proved that the fluctuations of TR-odd modes can never yield conventional (nodeless) $s$-wave pairing.
In addition, we find a sub-leading weak $d_{x^2-y^2}$ pairing state.
From Fig.~\ref{fig:cuprate-Eu-results}(b), we see that this $d_{x^2-y^2}$-wave SC state could only become dominant if one could approach smaller hole doping values without increasing the LC correlation length.
We also notice that the Van Hove singularity logarithmically enhances all pairing, as expected for $E_u^{-}$ LCs which effectively scatter Van Hove momenta (Sec.~\ref{sec:pairing-cuprate-actual-analysis-VH-details}).
Most importantly, and in complete agreement with the general result of Sec.~\ref{sec:gen-sys-LC-analysis} (Fig.~\ref{fig:QCP-general-results}), the pairing eigenvalues turn strongly repulsive in all symmetry channels as one approaches the QCP, as signaled by the absence of any positive eigenvalue in Fig.~\ref{fig:cuprate-Eu-results}(a) as $r \to 0$.
While the results in Fig.~\ref{fig:cuprate-Eu-results} refer to $\upalpha=0$, in Fig.~\ref{fig:cuprate-Eu-results-alpha} we show the impact of the parameter $\upalpha$ on pairing.
The impact is clearly minor, consisting of the emergence of other weak subleading SC states for a range of $\upalpha$ values and of the suppression of all SC states near $\upalpha = \pi/4$.

\subsubsection{Spin-orbit coupling and subsidiary spin-magnetic fluctuations} \label{sec:cup-spin-wave-res-discus}
Our analysis so far has considered only pure orbital magnetism in the absence of spin-orbit coupling (SOC).
Here we explore how SOC impacts our results.
There are two ways SOC can influence our results: through the band structure, or by modifying the interaction.
The two are closely related, as we show below.

In Sec.~\ref{sec:LC-prop-Cp-ch-int} of Chap.~\ref{chap:loop_currents}, we have studied the pairing form factor which for general systems with SOC is a $2 \times 2$ matrix in pseudospin space [Eq.~\eqref{eq:fa-definition-form-f}],
\begin{align}
[\Pintf_{a}(\vb{p}, \vb{k})]_{s_1 s_2} &\defeq u_{\vb{p} s_1}^{\dag} \Gamma_{a \vb{p}, \vb{k}} u_{\vb{k} s_2}. \label{eq:fa-definition-form-f-again}
\end{align}
It satisfies $\Pintf_{a}^{\dag}(\vb{p}, \vb{k}) = \Pintf_{a}(\vb{k}, \vb{p})$.
An important result from that section is Eq.~\eqref{eq:f-sym-cond1}:
\begin{align}
(\iu \Pauli_y)^{\dag} \Pintf_{a}^{*}(\vb{p}, \vb{k}) (\iu \Pauli_y) &= p_{P} p_{\TRop} \Pintf_{a}(\vb{p}, \vb{k}),
\end{align}
which follows from the combined parity and time-reversal symmetry.
Depending on the $p_{P} p_{\TRop}$ sign of the order described by $\Gamma_{a \vb{p}, \vb{k}}$, this result implies that at forward-scattering ($\vb{q} = \vb{p} - \vb{k} = \vb{0}$):
\begin{align}
\Pintf_{a}(\vb{k}, \vb{k}) &\propto \begin{cases}
\Pauli_0, & \text{when $p_{P} p_{\TRop} = +1$,} \\
\Pauli_{1,2,3}, & \text{when $p_{P} p_{\TRop} = -1$.}
\end{cases}
\end{align}
In other words, the forward-scattering pairing form factor $\Pintf_{a}(\vb{k}, \vb{k})$ is a pseudospin-singlet for $p_{P} p_{\TRop} = +1$ and a pseudospin-triplet for $p_{P} p_{\TRop} = -1$.
Therefore odd-parity LCs are pseudospin-singlets, while even-parity LCs are pseudospin-triplets.
In the absence of SOC, the trivial spin structure of the LCs is directly inherited from the $\Gamma_{a \vb{p}, \vb{k}} = \gamma_{a \vb{p}, \vb{k}} \otimes \Pauli_0$ matrices, which explains why even-parity LCs vanish at forward scattering.
Finite SOC, however, allows the even-parity LCs to be finite at forward scattering, with a Cooper pairing form factor that has the same form as for symmetry-equivalent spin orders.
Regarding odd-parity LCs, they are strongly repulsive at forward scattering irrespective of the SOC and no change is expected in their pair-breaking tendency near the QCP.

\begin{figure}[p!]
\centering
\includegraphics[width=\textwidth]{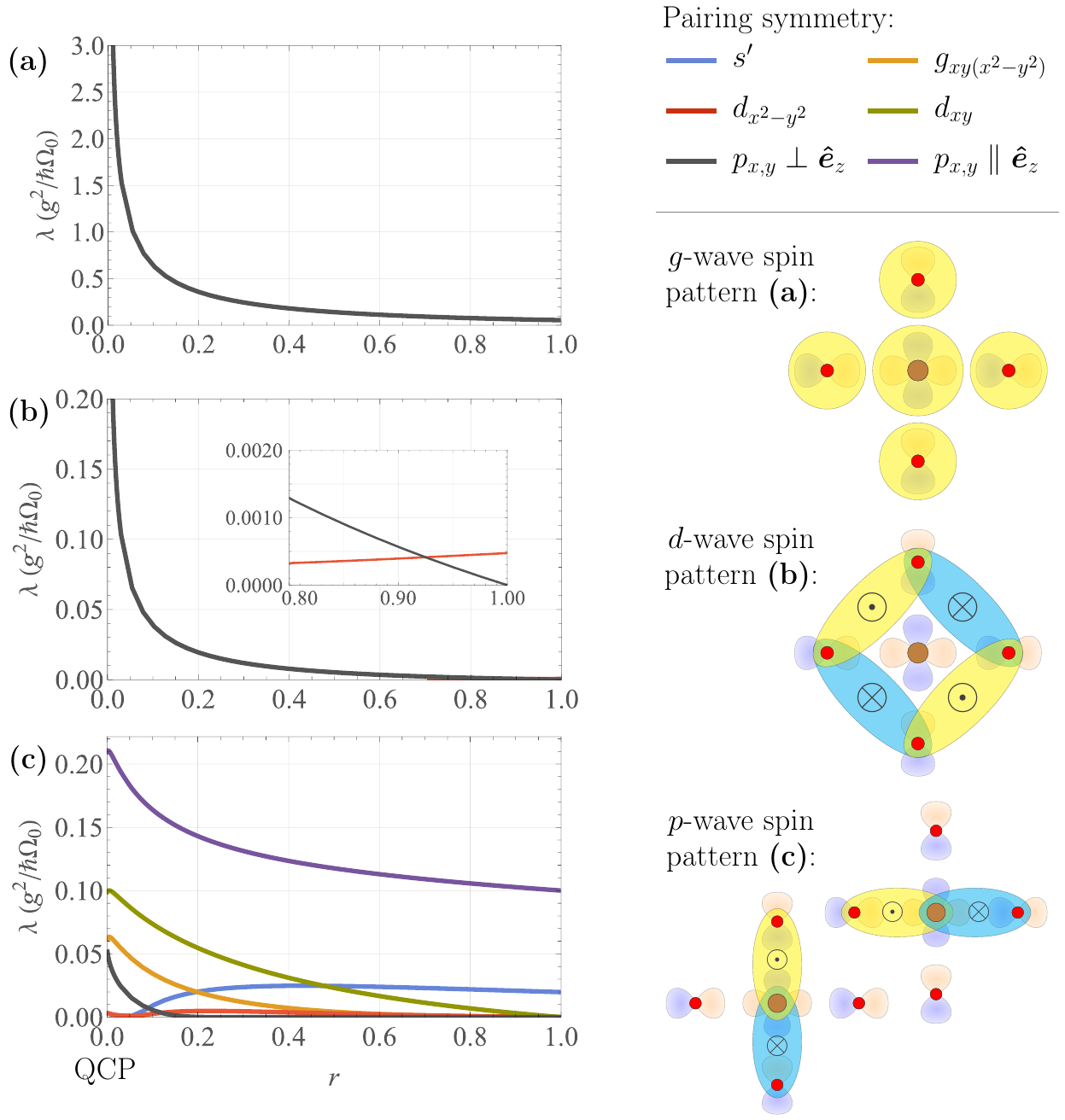}
\captionbelow[Results for the pairing mediated by subsidiary spin-magnetic fluctuations~\cite{Palle2024-LC}.]{\textbf{Results for the pairing mediated by subsidiary spin-magnetic fluctuations}~\cite{Palle2024-LC}.
The $g$-wave (a), $d$-wave (b), and $p$-wave (c) spin orders transform under $A_{2g}^{-}$, $B_{1g}^{-}$, and $E_{u}^{-}$ irreps, respectively, and their coupling matrices are given with $c_1 = \sqrt{2} c_2 = 1$ and $c_3 = c_4 = c_5 = 0$ in Eq.~\eqref{eq:gamma-A2g-bilinear-spin}, Eq.~\eqref{eq:gamma-B1g-bilinear-spin}, and $c_1' = 1$ and $c_2' = c_3' = 0$ in Eq.~\eqref{eq:gamma-Eu-bilinear-spin}.
The corresponding spin patterns are depicted on the right.
The plots show the pairing eigenvalues $\lambda$ of Eq.~\eqref{eq:final-lin-gap-eq-cup-spin} as a function of the tuning parameter $r$ at fixed chemical potential $\upmu = 0.9 t_{pd}$.
The colors of the curves indicate the pairing symmetry (upper right).
$s'$ stands for extended $s$-wave, while $p_{x,y} \perp \vu{e}_z$ and $p_{x,y} \parallel \vu{e}_z$ are triplet $p$-wave states whose Balian-Werthamer vector $\Delta_{A=1,2,3}$ is oriented along the $xy$-plane and the $z$-axis, respectively.
The tight-binding parameters used are those of Eq.~\eqref{eq:standard-CuO2-parameter-set-again}.
$r$ determines the susceptibility through Eq.~\eqref{eq:chi-mean-field-expr-r} and the putative loop-current quantum-critical point (QCP) is at $r = 0$.
See Sec.~\ref{sec:cup-spin-wave-res-discus} for further discussion.}
\label{fig:cuprate-SOC-results}
\end{figure}

Expect influence the band structure, on the interaction level SOC can also give rise to spin fluctuations which have the same symmetry as the orbital LC order~\cite{Aji2007, Klug2018, Christensen2022}.
In Sec.~\ref{sec:cup-ord-param-constr}, we have classified them and found that only those spin orders which have moments oriented along the $z$ direction can belong to the $A_{2g}^{-}$, $B_{1g}^{-}$, and $E_u^{-}$ irreps of the loop currents (Tab.~\ref{tab:SOC-bilin-classification-stats}).
This assumes, as in the case of LCs, that we only consider orders which are local in the sense that they are constructed from couplings within one extended unit cell.
Explicitly, for the possible coupling matrices we found [Eqs.~\eqref{eq:Gamma-along-z-1} and~\eqref{eq:Gamma-along-z-2})]:
\begin{align}
\Gamma_{\vb{k}, \vb{p}}^{A_{2g}^{-}} &= \mathcal{K}_{\vb{k}}^{\dag} \mleft(c_1 \Lambda^{A_{1g}^{+}}_{1} + c_2 \Lambda^{A_{1g}^{+}}_{2} + c_3 \Lambda^{A_{1g}^{+}}_{3} + c_4 \Lambda^{A_{1g}^{+}}_{4} + c_5 \Lambda^{A_{1g}^{+}}_{5}\mright) \otimes \Pauli_z \mathcal{K}_{\vb{p}}, \label{eq:gamma-A2g-bilinear-spin} \\
\Gamma_{\vb{k}, \vb{p}}^{B_{1g}^{-}} &= \mathcal{K}_{\vb{k}}^{\dag} \Lambda^{B_{2g}^{+}}_{1} \otimes \Pauli_z \mathcal{K}_{\vb{p}}, \label{eq:gamma-B1g-bilinear-spin}
\end{align}
and [Eq.~\eqref{eq:Gamma-along-z-3}]
\begin{align}
\begin{aligned}
\Gamma_{x, \vb{k}, \vb{p}}^{E_{u}^{-}} &= \mathcal{K}_{\vb{k}}^{\dag} \mleft(c_1' \Lambda^{E_{u}^{+}}_{1,y} + c_2' \Lambda^{E_{u}^{+}}_{2,y} + c_3' \Lambda^{E_{u}^{+}}_{3,y}\mright) \otimes \Pauli_z \mathcal{K}_{\vb{p}}, \\
\Gamma_{y, \vb{k}, \vb{p}}^{E_{u}^{-}} &= - \mathcal{K}_{\vb{k}}^{\dag} \mleft(c_1' \Lambda^{E_{u}^{+}}_{1,x} + c_2' \Lambda^{E_{u}^{+}}_{2,x} + c_3' \Lambda^{E_{u}^{+}}_{3,x}\mright) \otimes \Pauli_z \mathcal{K}_{\vb{p}}.
\end{aligned} \label{eq:gamma-Eu-bilinear-spin}
\end{align}
The most important point about these symmetry-equivalent spin orders is that their forward-scattering ($\vb{q} = \vb{0}$) pairing form factors in the absence of SOC behave the same as the corresponding LC form factors $\Pintf_{a}(\vb{k}, \vb{k})$ in the presence of SOC.
In particular, spin-magnetic orders in the absence of SOC satisfy the inverse of what LC orders do: they are finite at $\vb{q} = \vb{0}$ for even parity, but vanish for odd parity (Sec.~\ref{sec:Cp-channel-gen-sym-constr}).
Thus even-parity (odd-parity) subsidiary spin-magnetic fluctuations will give rise to strong (weak) quantum-critical pairing, respectively, as we already established in Sec.~\ref{sec:IUC-order-results}.
As long as the subsidiary odd-parity spin-magnetic fluctuations soften at the same QCP as the LCs, their weak quantum-critical pairing will be completely suppressed at the QCP because of the pair-breaking tendency of odd-parity LCs.

The pairing mediated by subsidiary spin-magnetic fluctuations we analyzed by solving a generalization of Eq.~\eqref{eq:final-lin-gap-eq-cup-simplified} to spin exchange [Eq.~\eqref{eq:final-lin-gap-eq-unsym}]:
\begin{align}
\oint\limits_{\varepsilon_{\vb{k}} = 0} \frac{\dd{\ell_{\vb{k}}}}{(2\pi)^2 v_{\vb{k}}} \sum_{A=0}^{3} \PintV_{BA}(\vb{p}, \vb{k}) \Delta_A(\vb{k}) &= \lambda \, \Delta_B(\vb{p}), \label{eq:final-lin-gap-eq-cup-spin}
\end{align}
where $A = 0$ is the pseudospin-singlet even-parity channel, while $A=1,2,3$ is the pseudospin-triplet odd-parity channel.
The Cooper-channel interaction is
\begin{align}
\PintV_{BA}(\vb{p}, \vb{k}) &= - g^2 \frac{1}{4} \mleft[\Pintv_{BA}(\vb{p}, \vb{k}) + p_A \Pintv_{BA}(\vb{p}, -\vb{k})\mright],
\end{align}
where the overall minus sign arises because we consider TR-odd modes, $p_{A=0} = - p_{A=1,2,3} = 1$, and
\begin{align}
\Pintv_{BA}(\vb{p}, \vb{k}) &= \chi(\vb{p}-\vb{k}) \sum_a \tr_s \Pauli_B \Pintf_{a}(\vb{p}, \vb{k}) \Pauli_A \Pintf_{a}^{\dag}(\vb{p}, \vb{k}),
\end{align}
with the pairing form factor of Eq.~\eqref{eq:fa-definition-form-f-again}.
We use the same band structure as before.
In particular, SOC has not been included at the one-particle level for the just discussed reasons.
Due to the non-trivial spin structure of the modes, the degeneracy between the in-plane and out-of-plane triplet channels is now lifted.

The results for the pairing mediated by subsidiary spin-magnetic fluctuations are shown in Fig.~\ref{fig:cuprate-SOC-results}.
For $g$-wave and $d$-wave spin fluctuations, we find strong pairing in the in-plane $p$-wave channels.
As the QCP is approached ($r \to 0$), this $p$-wave pairing will eventually surpass the weak singlet instabilities cause by LC fluctuations discussed earlier.
Conversely, for $p$-wave spin fluctuations, we find that they promote parametrically weak out-of-plane $p$-wave pairing.
Hence the strongly repulsive behavior of the pairing interaction in the orbital sector cannot be offset by the attractive contribution from subsidiary spin fluctuations.
Even when used $c_i$ and $c_j'$ coefficients in Eqs.~\eqref{eq:gamma-A2g-bilinear-spin} and~\eqref{eq:gamma-Eu-bilinear-spin} different from those shown in Fig.~\ref{fig:cuprate-SOC-results}, we never managed to get the correct leading pairing symmetry, which for cuprates is $d_{x^2-y^2}$.

\subsection{How the pairing symmetry gets chosen} \label{sec:cuprate-sym-choice-mechanism}
Broadly speaking, if one considers the linearized gap equation, which we may schematically write
\begin{align}
\int_{\text{FS}} \frac{\dd{S_{\vb{k}}}}{v_{\vb{k}}} \sum_A p_{\TRop} \chi(\vb{p}-\vb{k}) \PintF_{BA}(\vb{p}, \vb{k}) \Delta_A(\vb{k}) &= \lambda \, \Delta_B(\vb{p}), \label{eq:super-simple-lin-gap-eq}
\end{align}
there are two conceptually different ways the symmetry of the leading pairing channel can get chosen: either through the finite-momentum features of the boson susceptibility $\chi(\vb{q})$ or through the form factor $\PintF_{AB}(\vb{k}, \vb{p})$.
Here we focus on the exchange of TR-odd modes ($p_{\TRop} = -1$) because, as we proved in Sec.~\ref{sec:TR-positivity}, TR-even modes always give rise to conventional $s$-wave superconductivity, although the considerations of this section are potentially relevant to subleading channels.
These subleading instabilities could become leading due to additional interactions not included in the model, for instance.

Perhaps the simplest way of obtaining unconventional superconductivity is for the TR-odd modes to have a non-trivial spin structure, in which case the form factor $\PintF_{AB}(\vb{k}, \vb{p})$ is negative for at least some triplet components, thereby resulting in triplet pairing.
Recall that the singlet $\PintF_{00}(\vb{p}, \vb{k}) \geq 0$ is always positive (Sec.~\ref{sec:TR-positivity}).

A less obvious possibility is that of unconventional pairing due to the exchange of TR-odd orbital modes, i.e., loop currents.
Even though they have a uniformly repulsive interaction, since $\PintF_{AB}(\vb{k}, \vb{p}) = \Kd_{AB} \abs{\Pintf(\vb{k}, \vb{p})}^2 \geq 0$, unconventional pairing may arise in two different ways.

The first is by having the Cooper-channel interaction peak at a finite momentum transfer $\vb{Q}$~\cite{Scalapino1995, Maiti2013}.
The intuition behind why this would result in pairing can be gathered by simplifying Eq.~\eqref{eq:super-simple-lin-gap-eq} down to the pairing of two ``hot spots,'' in which case we essentially have:
\begin{align}
\begin{pmatrix}
- \chi_{\vb{0}} & - \chi_{\vb{Q}} \\
- \chi_{\vb{Q}} & - \chi_{\vb{0}}
\end{pmatrix} \vb{\Delta} &= \lambda \, \vb{\Delta}.
\end{align}
This $2 \times 2$ matrix is easily diagonalized:
\begin{align}
\lambda_{\pm} &= - \chi_{\vb{0}} \mp \chi_{\vb{Q}}, &
\vb{\Delta}_{\pm} &= \begin{pmatrix}
1 \\ \pm 1
\end{pmatrix}.
\end{align}
Hence, if $\chi_{\vb{Q}} > \chi_{\vb{0}}$, $\lambda_{-} > 0$ and we find an attractive Cooper instability whose gap function changes sign for points differing by $\vb{Q}$.
In the actual linearized gap equation, $\Delta(\vb{k}+\vb{Q}) = - \Delta(\vb{k})$ cannot hold everywhere.
Instead, it only holds where the gap function is weighted the most, which are the places where the Fermi velocity $v_{\vb{k}}$ is smallest and the DOS largest, as follows from  Eq.~\eqref{eq:super-simple-lin-gap-eq}.
As the QCP points is approached, $\chi_{\vb{Q}}$ diverges and the pairing eigenvalue will diverge as well in two dimensions, indicating strong quantum-critical pairing, unless some symmetry suppresses the form factor at the hot spots (Sec.~\ref{sec:IUC-order-results}).
The generic symmetry $P \TRop$ maps $\vb{k} \mapsto \vb{k}$ and can only constrain $\PintF_{AB}(\vb{k}, \vb{k}+\vb{Q})$ for $\vb{Q} = \vb{0}$.

As discussed in Sec.~\ref{sec:LC-proposals}, LC orders which break translation symmetry have been proposed for the pseudogap of the cuprates~\cite{Chakravarty2001, Laughlin2014, Laughlin2014-details, Makhfudz2016}.\footnote{Staggered LCs have also been proposed in the kagome superconductors~\cite{Mielke2022}.}
In the case of the cuprates, the Van Hove points are at $\vb{k}_{M_x} = (\pi,0)$ and $\vb{k}_{M_y} = (0,\pi)$ and a TR-odd mode ordering at $\vb{Q} = (\pi,\pi)$ is naturally expected to induce $d_{x^2-y^2}$ superconductivity~\cite{Scalapino1995}.
And indeed, if we numerically solve the linearized gap equation for $d$-wave and $p$-wave LCs with $\vb{Q} = (\pi,\pi)$ using the same band structure as before, we find that they both favor $d_{x^2-y^2}$-wave pairing which becomes strongly enhanced as the QCP is approached.
In the case of $\vb{Q} = (\pi,\pi)$ $g$-wave LCs, however, we get a different pairing symmetry because of the suppression of the form factor $\PintF_{00}(\vb{k}, \vb{p}) = \abs{\Pintf(\vb{k},\vb{p})}^2$ at the van Hove points (Sec.~\ref{sec:pairing-cuprate-actual-analysis-VH-details}).

The case of intra-unit-cell LCs is different because the susceptibility $\chi(\vb{q})$ is peaked at $\vb{q} = \vb{0}$.
Hence the pairing symmetry, which is fundamentally about the phase differences at different momenta ($\vb{q} \neq \vb{0}$), cannot be chosen by the susceptibility, especially when the IUC QCP is approached.
As we saw in the numerical results of the previous section, the leading pairing channels is always chosen away from the QCP and as the QCP is approached ($r \to 0$) it becomes enhanced (spin fluctuations), suppressed (odd-parity LCs), or stays the same (even-parity LCs), but never changes.
The actual choosing of the pairing symmetry is carried out by the form factor which we can imagine diagonalizing like so:
\begin{align}
\PintF(\vb{p}, \vb{k}) &\sim \sum \nu_n \mathrm{w}_n(\vb{p}) \mathrm{w}_n^{*}(\vb{k}).
\end{align}
For pairing due to exchange of $\vb{q} = \vb{0}$ fluctuations, it is the interplay of these form-factor eigenvectors with the DOS that selects the leading pairing instability, as we discuss in the next section.

\subsection{Analytic solutions of the linearized gap equation} \label{sec:cuprate-analytic-res}
The insights of the previous section motivate the following approach to analytically solving the linearized gap equation.
As previously, we consider IUC LCs in the absence of SOC.
Let us start by rewriting Eq.~\eqref{eq:final-lin-gap-eq-cup-simplified} in a more symmetric fashion:
\begin{align}
\oint\limits_{\varepsilon_{\vb{k}} = 0} \frac{\dd{\ell_{\vb{k}}}}{(2\pi)^2 \sqrt{v_{\vb{p}} v_{\vb{k}}}} \PintV(\vb{p}, \vb{k}) d(\vb{k}) &= \lambda \, d(\vb{p}), \label{eq:final-lin-gap-eq-cup-symmetric}
\end{align}
where $d(\vb{k}) = \sqrt{v_{\vb{k}}} \, \Delta_{\pm}(\vb{k})$ and
\begin{align}
\PintV(\vb{p}, \vb{k}) &= - g^2 \Pintv_0(\vb{p}, \vb{k}) = - g^2 \chi(\vb{p}-\vb{k}) \sum_a \abs{u_{\vb{p} 3}^{\dag} \gamma_{a \vb{p}, \vb{k}} u_{\vb{k} 3}}^2.
\end{align}
Although we did not explicitly split $\PintV(\vb{p}, \vb{k})$ into even and odd parts [$\PintV_{\pm}$ in Eq.~\eqref{eq:final-lin-gap-eq-cup-simplified}], the eigenvectors $d(\vb{k})$ are nonetheless always either even or odd, $d(-\vb{k}) = \pm d(\vb{k})$, as follows from parity and TR symmetry.

This eigenvalue problem is difficult to solve analytically because it is infinite dimensional ($\vb{k}$ is continuous).
Even finite-dimensional matrices are difficult to diagonalize in closed form.
In numerical approaches, one either discretizes the Fermi surface or expands $d(\vb{k})$ in harmonics and then truncates the expansion (Sec.~\ref{sec:pairing-cuprate-actual-analysis-results}).
Both approaches are approximate.
Here we show that this problem can be reduced to a finite-dimensional one exactly.
The essential idea is to first separately diagonalize $\chi(\vb{p}-\vb{k})$ and $u_{\vb{p} 3}^{\dag} \gamma_{a \vb{p},\vb{k}} u_{\vb{k} 3}$.
If these two operators have only a finite number of non-zero eigenvalues, then the diagonalization of $\PintV(\vb{p},\vb{k})$ can be reduced to the diagonalization of a finite matrix.

The mean-field susceptibility Ansatz of Eq.~\eqref{eq:chi-mean-field-expr-r},
\begin{align}
\chi(\vb{q}) &= \frac{\chi_0}{\displaystyle \frac{1+r}{2} - \frac{1-r}{4} \mleft(\cos q_x + \cos q_y\mright)}, \label{eq:chi-mean-field-expr-r-again}
\end{align}
has an infinite number of non-zero eigenvalues for $r \neq 1$.
However, if we instead use
\begin{align}
\chi(\vb{q}) &= \frac{1 + r}{2r} + \frac{1 - r}{4r} \mleft(\cos q_x + \cos q_y\mright), \label{eq:chi-new-simp-expr}
\end{align}
this also also has a maximum of $r^{-1}$ at $\vb{q} = \vb{0}$ and a minimum of $1$ at $\vb{q} = (\pi, \pi)$.
The $r = 0$ behavior is drastically different ($\sim 1/\vb{q}^2$ vs.\ $\sim 1/r + \bigO(\vb{q}^2)$), but, as we have seen in the numerics of Sec.~\ref{sec:pairing-cuprate-actual-analysis-results}, the pairing symmetry of our problem is always chosen away from $r = 0$ so this replacement should not matter for our purposes.
For $\vb{q} = \vb{p} - \vb{k}$:
\begin{align}
\cos q_x + \cos q_y &= \cos k_x \cos p_x + \cos k_y \cos p_y + \sin k_x \sin p_x + \sin k_y \sin p_y.
\end{align}
We may therefore write:
\begin{align}
\chi(\vb{p}-\vb{k}) &= \chi_0 \sum_{n} \mu_n \mathrm{v}_{n}(\vb{p}) \mathrm{v}_{n}^{*}(\vb{k}), \label{eq:chi-new-simp-expansion}
\end{align}
with the eigenvalues and eigenvectors listed in Tab.~\ref{tab:cuprate-chi-eigs}.

\begin{table}[t]
\centering
\captionabove[Eigenvalues $\mu_n$ and eigenvectors $\mathrm{v}_{n}(\vb{k})$ of the susceptibility of Eq.~\eqref{eq:chi-new-simp-expr}.]{\textbf{Eigenvalues $\mu_n$ and eigenvectors $\mathrm{v}_{n}(\vb{k})$ of the susceptibility of Eq.~\eqref{eq:chi-new-simp-expr}.}
The irreps of scalar functions are defined according to Eq.~\eqref{eq:scalar-func-transf-rule1}.
The eigen-expansion is given in Eq.~\eqref{eq:chi-new-simp-expansion}.
The eigenvectors are not normalized.}
{\renewcommand{\arraystretch}{1.3}
\renewcommand{\tabcolsep}{10pt}
\begin{tabular}{cccc} \hline\hline
$n$ & $\mu_n$ & $\mathrm{v}_{n}(\vb{k})$ & irrep \\ \hline \\[-13pt]
$1$ & $\displaystyle \frac{1 + r}{2r}$ & $1$ & $A_{1g}$ \\[6pt]
$2$ & $\displaystyle \frac{1 - r}{8r}$ & $\cos k_x + \cos k_y$ & $A_{1g}$ \\[6pt]
$3$ & $\displaystyle \frac{1 - r}{8r}$ & $\cos k_x - \cos k_y$ & $B_{1g}$ \\[6pt]
$4,x$ & $\displaystyle \frac{1 - r}{4r}$ & $\sin k_x$ & $E_u$ \\[6pt]
$4,y$ & $\displaystyle \frac{1 - r}{4r}$ & $\sin k_y$ & $E_u$
\\[6pt] \hline\hline
\end{tabular}}
\label{tab:cuprate-chi-eigs}
\end{table}

The coupling $\gamma$ matrices are given by [Eq.~\eqref{eq:momentum-bilinear-v3}]
\begin{align}
\gamma_{a \vb{p}, \vb{k}} &= \mathcal{K}_{\vb{p}}^{\dag} \Lambda_a \mathcal{K}_{\vb{k}}.
\end{align}
If we diagonalize the $\Lambda_a$ matrices,
\begin{align}
\Lambda_a &= \sum_{n} \nu_{a;n} w_{a;n} w_{a;n}^{\dag},
\end{align}
we find that
\begin{align}
u_{\vb{p} 3}^{\dag} \gamma_{a \vb{p},\vb{k}} u_{\vb{k}} &= \sum_{n} \nu_{a;n} \mathrm{w}_{a;n}(\vb{p}) \mathrm{w}_{a;n}^{*}(\vb{k}),
\end{align}
where
\begin{align}
\mathrm{w}_{a;n}(\vb{k}) \defeq u_{\vb{k} 3}^{\dag} \mathcal{K}_{\vb{k}}^{\dag} w_{a;n}.
\end{align}
Clearly, there are only a finite number of eigenvectors of $\Lambda_a$.
In fact, for most of the matrices we listed in Tab.~\ref{tab:orbital-Lambda-mats}, there are only two finite eigenvalues because they have only two finite components in the basis rotated by $\mathcal{B}$.

We thereby arrive at the following expansion of the Cooper-channel interaction:
\begin{align}
\PintV(\vb{p},\vb{k}) &= - g^2 \chi_0 \sum_{a n_1 n_2 n_3} \mu_{n_1} \nu_{a;n_2} \nu_{a;n_3} \cdot \mathrm{v}_{n_1}(\vb{p}) \mathrm{w}_{a;n_2;n_3}(\vb{p}) \cdot \mleft[\mathrm{v}_{n_1}(\vb{k}) \mathrm{w}_{a;n_2;n_3}(\vb{k})\mright]^{*},
\end{align}
where
\begin{align}
\mathrm{w}_{a;n_2;n_3}(\vb{k}) \defeq \mathrm{w}_{a;n_2}^{*}(\vb{k}) \mathrm{w}_{a;n_3}(\vb{k}).
\end{align}
The most notable thing about it is that, no matter what input vector $d(\vb{k})$ we multiply and integrate against $\PintV(\vb{p},\vb{k})$, the output vector will be proportional to some linear superposition of $\mathrm{v}_{n_1}(\vb{p}) \mathrm{w}_{a;n_2;n_3}(\vb{p})$.
In other words, the spectrum of $\PintV(\vb{p},\vb{k})$ is finite.
Given that $\PintV(\vb{p},\vb{k})$ is diagonalizable, it thus follows that eigenvectors with non-zero eigenvalues have the form:
\begin{align}
d(\vb{k}) &= \frac{1}{\sqrt{v_{\vb{k}}}} \sum_{a n_i} \mathscr{d}_{a n_1 n_2 n_3} \, \mathrm{v}_{n_1}(\vb{k}) \mathrm{w}_{a;n_2;n_3}(\vb{k}).
\end{align}
Let us note that
\begin{align}
\mathrm{w}_{a;n_2;n_3}(\vb{k}) &= w_{a;n_2}^{\dag} \mathcal{K}_{\vb{k}} u_{\vb{k} 3} u_{\vb{k} 3}^{\dag} \mathcal{K}_{\vb{k}}^{\dag} w_{a;n_3} = \mathrm{w}_{a;n_3;n_2}^{\dag}(\vb{k})
\end{align}
is gauge-invariant under $u_{\vb{k} 3} \mapsto \Elr^{\iu \vartheta_{\vb{k}}} u_{\vb{k} 3}$ and that it has well-defined transformation rules inherited from the $\Lambda_a$ matrices.
The latter follows from the fact that $\mathcal{K}_{\vb{k}} u_{\vb{k} 3} u_{\vb{k} 3}^{\dag} \mathcal{K}_{\vb{k}}^{\dag} \in A_{1g}$ transforms trivially.
The original eigenvalue problem of Eq.~\eqref{eq:final-lin-gap-eq-cup-symmetric} has thus been reduced to:
\begin{align}
\sum_{b m_1 m_2 m_3} \PintV_{a n_1 n_2 n_3; b m_1 m_2 m_3} \mathscr{d}_{b m_1 m_2 m_3} &= \lambda \, \mathscr{d}_{a n_1 n_2 n_3},
\end{align}
where
\begin{align}
\begin{aligned}
\PintV_{a n_1 n_2 n_3; b m_1 m_2 m_3} &= - g^2 \chi_0 \mu_{n_1} \nu_{a;n_2} \nu_{a;n_3} \\
&\hspace{22pt} \times \oint \frac{\dd{\ell_{\vb{k}}}}{(2\pi)^2 v_{\vb{k}}} \mleft[\mathrm{v}_{n_1}(\vb{k}) \mathrm{w}_{a;n_2;n_3}(\vb{k})\mright]^{*} \mathrm{v}_{m_1}(\vb{k}) \mathrm{w}_{b;m_2;m_3}(\vb{k}).
\end{aligned}
\end{align}
This is a finite matrix diagonalization problem whenever the sum over $b$, $m_1$, $m_2$, and $m_3$ is finite.

\subsubsection{$g_{xy(x^2-y^2)}$-wave loop currents}
The eigenvalues and eigenvectors of the $\Lambda^{A_{2g}^{-}}_1$ matrix (Tab.~\ref{tab:orbital-Lambda-mats}) are
\begin{align}
\nu_1 &= + 1, &
\nu_2 &= - 1,
\end{align}
and
\begin{align}
w_1 &= \frac{1}{2} \begin{pmatrix}
0 \\ 1 \\ - \iu \\ 1 \\ - \iu
\end{pmatrix}, &
w_2 &= \frac{1}{2} \begin{pmatrix}
0 \\ 1 \\ \iu \\ 1 \\ \iu
\end{pmatrix}.
\end{align}
Under the orbital matrices of Tab.~\ref{tab:CuO2-model-D4h-generators} they transform according to:
\begin{align}
O(g) \begin{pmatrix}
w_1 & w_2
\end{pmatrix} &= \begin{pmatrix}
w_1 & w_2
\end{pmatrix} \RepM(g),
\end{align}
where:
\begin{align}
\begin{aligned}
\RepM(C_{4z}) &= \begin{pmatrix}
\iu & 0 \\
0 & -\iu
\end{pmatrix}, &\hspace{50pt}
\RepM(C_{2x}) &= \begin{pmatrix}
0 & 1 \\
1 & 0
\end{pmatrix}, \\
\RepM(C_{2d_+}) &= \begin{pmatrix}
0 & \iu \\
-\iu & 0
\end{pmatrix}, &\hspace{50pt}
\RepM(P) &= \begin{pmatrix}
-1 & 0 \\
0 & -1
\end{pmatrix}.
\end{aligned}
\end{align}
Given that $O(g) \mathcal{K}_{\vb{k}} u_{\vb{k} 3} u_{\vb{k} 3}^{\dag} \mathcal{K}_{\vb{k}}^{\dag} O^{\dag}(g) = \mathcal{K}_{\vb{p}} u_{\vb{p} 3} u_{\vb{p} 3}^{\dag} \mathcal{K}_{\vb{p}}$ with $\vb{p} = R(g) \vb{k}$, the composite eigenvectors $\mathrm{w}_{n_1;n_2}(\vb{k})$ transform according to:
\begin{align}
\mathrm{w}_{n_1;n_2}\mleft(R(g) \vb{k}\mright) &= \sum_{m_1 m_2} \RepM_{n_1 m_1}(g) \RepM_{n_2 m_2}^{*}(g) \mathrm{w}_{m_1;m_2}(\vb{k}),
\end{align}
that is, under the representation $\RepM \otimes \RepM^{*}$.
This direct product representation can be decomposed as explained in Sec.~\ref{sec:multid-irrep-product} of Appx.~\ref{app:group_theory}, with the result:
\begin{align}
\mathrm{w}_{1;1}(\vb{k}) + \mathrm{w}_{2;2}(\vb{k}) &\in A_{1g}, &
\mathrm{w}_{1;1}(\vb{k}) - \mathrm{w}_{2;2}(\vb{k}) &\in A_{2g}, \\
\mathrm{w}_{1;2}(\vb{k}) + \mathrm{w}_{2;1}(\vb{k}) &\in B_{1g}, &
\mathrm{w}_{1;2}(\vb{k}) - \mathrm{w}_{2;1}(\vb{k}) &\in B_{2g}.
\end{align}
When $r = 1$, under the momentum integral the above functions are orthogonal so they directly give the pairing eigenvectors.
From $\nu_1 \nu_1 = \nu_2 \nu_2 = +1$ and $\nu_1 \nu_2 = \nu_2 \nu_1 = -1$ we see that only $B_{1g}$ and $B_{2g}$ irreps have positive eigenvalues and yield superconductivity, in agreement with Fig.~\ref{fig:cuprate-A2g-results}.

The exact pairing eigenvectors and eigenvalues for $r = 1$ are:
\begin{align}
d^{B_{1g}}(\vb{k}) &= \frac{1}{\sqrt{2 v_{\vb{k}}}} \mleft[\mathrm{w}_{1;2}(\vb{k}) + \mathrm{w}_{2;1}(\vb{k})\mright], &
\lambda^{B_{1g}} &= g^2 \chi_0 \oint \frac{\dd{\ell_{\vb{k}}}}{(2\pi)^2} \abs{d^{B_{1g}}(\vb{k})}^2, \\
d^{B_{2g}}(\vb{k}) &= \frac{1}{\sqrt{2 v_{\vb{k}}}} \mleft[\mathrm{w}_{1;2}(\vb{k}) - \mathrm{w}_{2;1}(\vb{k})\mright], &
\lambda^{B_{2g}} &= g^2 \chi_0 \oint \frac{\dd{\ell_{\vb{k}}}}{(2\pi)^2} \abs{d^{B_{2g}}(\vb{k})}^2.
\end{align}
The pairing eigenvectors are not normalized and numerically we find that the $d_{xy} \in B_{2g}$ state is significantly bigger than the $d_{x^2-y^2} \in B_{1g}$ one.
When $r \neq 1$, the exact $B_{1g}$ eigenvector has the form
\begin{align}
d^{B_{1g}} &= \frac{1}{\sqrt{2 v_{\vb{k}}}} \mleft[(\mathscr{d}_1 \mathrm{v}_1 + \mathscr{d}_2 \mathrm{v}_2) (\mathrm{w}_{1;2} + \mathrm{w}_{2;1}) + \mathscr{d}_3 \mathrm{v}_3 (\mathrm{w}_{1;1} + \mathrm{w}_{2;2})\mright].
\end{align}
The coefficients $\mathscr{d}_{1,2,3}$ are determined by diagonalizing the corresponding $3 \times 3$ matrix.
Due to the non-trivial irreps of the susceptibility eigenvectors (Tab.~\ref{tab:cuprate-chi-eigs}), pairing instabilities with symmetries other than $B_{1g}$ and $B_{2g}$ can also arise.

\subsubsection{$d_{x^2-y^2}$-wave loop currents}
For $\Lambda = c_1 \Lambda^{B_{1g}^{-}}_1 + c_2 \Lambda^{B_{1g}^{-}}_2$ with $c_1^2 + c_2^2 = 1$, we find that
\begin{align}
\nu_1 &= + 1, &
\nu_2 &= - 1,
\end{align}
and
\begin{align}
w_1 &= \frac{1}{\sqrt{8}} \begin{pmatrix}
2 c_1 \\ \iu - c_2 \\ \iu + c_2 \\ - \iu + c_2 \\ - \iu - c_2
\end{pmatrix}, &
w_2 &= \begin{pmatrix}
2 c_1 \\ - \iu - c_2 \\ - \iu + c_2 \\ \iu + c_2 \\ \iu - c_2
\end{pmatrix},
\end{align}
and that under point group transformations:
\begin{align}
O(g) \begin{pmatrix}
w_1 & w_2
\end{pmatrix} &= \begin{pmatrix}
w_1 & w_2
\end{pmatrix} \RepM(g),
\end{align}
where:
\begin{align}
\begin{aligned}
\RepM(C_{4z}) &= \begin{pmatrix}
0 & -1 \\
-1 & 0
\end{pmatrix}, &\hspace{50pt}
\RepM(C_{2x}) &= \begin{pmatrix}
1 & 0 \\
0 & 1
\end{pmatrix}, \\
\RepM(C_{2d_+}) &= \begin{pmatrix}
0 & -1 \\
-1 & 0
\end{pmatrix}, &\hspace{50pt}
\RepM(P) &= \begin{pmatrix}
1 & 0 \\
0 & 1
\end{pmatrix}.
\end{aligned}
\end{align}
Hence:
\begin{align}
\begin{aligned}
A_1(\vb{k}) &= \frac{1}{\sqrt{2}} \mleft[\mathrm{w}_{11}(\vb{k}) + \mathrm{w}_{22}(\vb{k})\mright] \in A_{1g}, &\hspace{20pt}
B_1(\vb{k}) &= \frac{1}{\sqrt{2}} \mleft[\mathrm{w}_{11}(\vb{k}) - \mathrm{w}_{22}(\vb{k})\mright] \in B_{1g}, \\
A_2(\vb{k}) &= \frac{1}{\sqrt{2}} \mleft[\mathrm{w}_{12}(\vb{k}) + \mathrm{w}_{21}(\vb{k})\mright] \in A_{1g}, &\hspace{20pt}
B_2(\vb{k}) &= \frac{1}{\sqrt{2}} \mleft[\mathrm{w}_{12}(\vb{k}) - \mathrm{w}_{21}(\vb{k})\mright] \in B_{1g}.
\end{aligned}
\end{align}
The exact eigenvectors for $r = 1$ therefore have the form:
\begin{align}
d^{A_{1g}}(\vb{k}) &= \frac{1}{\sqrt{v_{\vb{k}}}} \mleft[\mathscr{d}_1 A_1(\vb{k}) + \mathscr{d}_2 A_2(\vb{k})\mright], \\
d^{B_{1g}}(\vb{k}) &= \frac{1}{\sqrt{v_{\vb{k}}}} \mleft[\mathscr{d}_1 B_1(\vb{k}) + \mathscr{d}_2 B_2(\vb{k})\mright],
\end{align}
where the coefficients and pairing eigenvalues are found by diagonalizing
\begin{align}
- g^2 \chi_0 \oint \frac{\dd{\ell_{\vb{k}}}}{(2\pi)^2 v_{\vb{k}}} \begin{pmatrix}
A_1^{*}(\vb{k}) A_1(\vb{k}) & A_1^{*}(\vb{k}) A_2(\vb{k}) \\
- A_2^{*}(\vb{k}) A_1(\vb{k}) & - A_2^{*}(\vb{k}) A_2(\vb{k})
\end{pmatrix} \begin{pmatrix}
\mathscr{d}_1 \\ \mathscr{d}_2
\end{pmatrix} &= \lambda \begin{pmatrix}
\mathscr{d}_1 \\ \mathscr{d}_2
\end{pmatrix}
\end{align}
for the $A_{1g}$ channel, and an analogous matrix with $A$ replaced by $B$ for the $B_{1g}$ channel.
Although not Hermitian, one can show that the eigenvalues of these matrices are always real.
Given that only $A_{1g}$, $B_{1g}$, and $E_u$ irreps appear among the susceptibility eigenvectors (Tab.~\ref{tab:cuprate-chi-eigs}), from the $D_{4h}$ irrep product Tab.~\ref{tab:D4h-irrep-prod-tab} it follows that \emph{only} $s$-wave, $d_{x^2-y^2}$-wave, and $p$-wave instabilities are possible, at least if we use Eq.~\eqref{eq:chi-new-simp-expr}.
These results are consistent with Fig.~\ref{fig:cuprate-B1g-results}(a) in which we find $s'$ and $d_{x^2-y^2}$ pairing at $r = 1$ and an additional $p$-wave pairing for $r < 1$, while the ``forbidden'' $g_{xy(x^2-y^2)}$ and $d_{xy}$ channels only appear at very small $r$ for which the difference between the susceptibility expressions~\eqref{eq:chi-new-simp-expr} and~\eqref{eq:chi-mean-field-expr-r-again} is the largest.

\subsubsection{$(p_x, p_y)$-wave loop currents}
For $\Lambda_a = c_1 \Lambda^{E_u}_{1,a} + c_3 \Lambda^{E_u}_{3,a}$ with $c_1^2 + c_3^2 = 1$, which corresponds to $\upalpha = 0$ in Eq.~\eqref{eq:gamma-Eum-bilinear-alpha}, we obtain
\begin{align}
\nu_{a=x;1} &= + 1, &
\nu_{a=x;2} &= - 1, &
\nu_{a=y;1} &= + 1, &
\nu_{a=y;2} &= - 1
\end{align}
and
\begin{align}
\begin{aligned}
w_{a=x;1} &= \frac{1}{2} \begin{pmatrix}
\sqrt{2} c_1 \\ \iu \\ - c_3 \\ \iu \\ c_3
\end{pmatrix}, &\hspace{50pt}
w_{a=x;2} &= \frac{1}{2} \begin{pmatrix}
\sqrt{2} c_1 \\ -\iu \\ - c_3 \\ -\iu \\ c_3
\end{pmatrix}, \\
w_{a=y;1} &= \frac{1}{2} \begin{pmatrix}
\sqrt{2} c_1 \\ c_3 \\ -\iu \\ - c_3 \\ -\iu
\end{pmatrix}, &\hspace{50pt}
w_{a=y;2} &= \frac{1}{2} \begin{pmatrix}
\sqrt{2} c_1 \\ c_3 \\ \iu \\ - c_3 \\ \iu
\end{pmatrix}
\end{aligned}
\end{align}
as the eigenvectors and eigenvalues.
Under point group transformations they transform according to:
\begin{align}
O(g) \begin{pmatrix}
w_{x;1} & w_{x;2} & w_{y;1} & w_{y;2}
\end{pmatrix} &= \begin{pmatrix}
w_{x;1} & w_{x;2} & w_{y;1} & w_{y;2}
\end{pmatrix} \RepM(g),
\end{align}
where:
\begin{align}
\begin{aligned}
\RepM(C_{4z}) &= \begin{pmatrix}
0 & 0 & 0 & -1 \\
0 & 0 & -1 & 0 \\
-1 & 0 & 0 & 0 \\
0 & -1 & 0 & 0
\end{pmatrix}, &\hspace{50pt}
\RepM(C_{2x}) &= \begin{pmatrix}
1 & 0 & 0 & 0 \\
0 & 1 & 0 & 0 \\
0 & 0 & 0 & 1 \\
0 & 0 & 1 & 0
\end{pmatrix}, \\
\RepM(C_{2d_+}) &= \begin{pmatrix}
0 & 0 & -1 & 0 \\
0 & 0 & 0 & -1 \\
-1 & 0 & 0 & 0 \\
0 & -1 & 0 & 0
\end{pmatrix}, &\hspace{50pt}
\RepM(P) &= \begin{pmatrix}
0 & 1 & 0 & 0 \\
1 & 0 & 0 & 0 \\
0 & 0 & 0 & 1 \\
0 & 0 & 1 & 0
\end{pmatrix}.
\end{aligned}
\end{align}
Now there are $8$ possible momentum-dependent functions $\mathrm{w}_{a;n_1;n_2}(\vb{k})$ that may arise in the exact pairing eigenvectors $d(\vb{k})$.
From the representation characters (Sec.~\ref{sec:character-theory-D4h}) one may deduce that $\RepM = A_{1g} \oplus B_{1g} \oplus E_u$ and therefore $\RepM \otimes \RepM^{*} = 3 A_{1g} \oplus A_{2g} \oplus 3 B_{1g} \oplus B_{2g} \oplus 4 E_u$, as follows from the irrep product Tab.~\ref{tab:D4h-irrep-prod-tab}.
At $r = 0$, at least at first sight, the pairing eigenvectors can belong to any irrep.
By writing the most general superpositions as we did for $d$-wave LCs, one can now formulate finite-dimensional eigenvalue problems that exactly determine the pairing eigenvalues $\lambda$ at $r = 0$.
In the numerics shown in Fig.~\ref{fig:cuprate-Eu-results}, we found that only the $A_{1g}$ and $B_{1g}$ channels have positive $\lambda$ corresponding to pairing instabilities.

\subsection{Comparison to the work by Aji, Shekhter, and Varma (2010)} \label{sec:comparison-w-Aji2010}
In the context of the cuprates, the most prominent theory in which intra-unit-cell loop currents play an important role is the one proposed by Varma~\cite{Varma2020, Varma2016}.
This theory has been developed by Varma and his collaborators in many ways during the last three decades~\cite{Varma1997, Varma1999, Simon2002, Simon2003, Varma2006, Aji2007, Gronsleth2009, Shekhter2009, Weber2009, Aji2010, Aji2013, Varma2014, Weber2014, Yakovenko2015, Varma2015, Varma2019}.
Some aspects of this theory pertaining to the pseudogap regime~\cite{Varma1997, Simon2002, Simon2003, Varma2006, Varma2014, Varma2019} and to the numerical derivation of IUC LCs from microscopic models~\cite{Weber2009, Weber2014} we have already reviewed in Secs.~\ref{sec:LC-proposals} and~\ref{sec:LC-cuprate-micro}, respectively.
For other aspects not directly related to superconductivity, we refer the reader to Refs.~\cite{Varma2020, Varma2016} in which Varma has summarized the final proposal.

The main tenant of Varma's theory is that the pseudogap regime corresponds to a hidden odd-parity intra-unit-cell loop-current order~\cite{Varma2020, Varma2016}.
As the hole doping is increased, this putative IUC LC order ends at a QCP, as denoted in the phase diagram of Fig.~\ref{fig:cuprate-phase-diagram}; see also Fig.~\ref{fig:Varma-evidence}.
Within this theory, the ordering of odd-parity IUC LCs explains the phenomenology of the pseudogap regime, while their quantum-critical fluctuations drive both the strange metal behavior and the $d_{x^2-y^2}$-wave superconductivity~\cite{Varma2020, Varma2016}.
It is this last issue -- Can Varma's theory explain the high-temperature $d_{x^2-y^2}$-wave superconductivity of cuprates?\ -- that we discuss in this final part of the chapter.
Needless to say, any viable theory of cuprates must be able to account for their remarkable high-temperature superconductivity.

The main work in which Varma and collaborators have addressed cuprate superconductivity is the work by Aji, Shekhter, and Varma (ASV) from 2010~\cite{Aji2010}.
Here, we compare and contrast our own analysis of pairing due to IUC LC fluctuations in cuprates, and general systems (Chap.~\ref{chap:loop_currents}), to that of ASV~\cite{Aji2010}.
We start by noting that ASV use a different orbital orientation convention and momentum-space gauge than the current work (and Ref.~\cite{Palle2024-LC} on which the current work is based).
This should be kept in mind whenever comparing formulas between the two works.
In Sec.~\ref{sec:ASV-LC-operators} thereafter, we show that the flux operators introduced by ASV~\cite{Aji2010} agree with our classification of LC operators (Sec.~\ref{sec:cuprate-bilinear-class}).
Afterwards, in Sec.~\ref{sec:ASV-eff-Haml}, we discuss ASV's decompositions of the $V_{pd}$ and $V_{pp}$ Hubbard interactions and compare them with the results we derived in Sec.~\ref{sec:CuO2-Hubbard-decompose}.
In the last Sec.~\ref{sec:ASV-fermion-coupling}, we examine the most important point of disagreement: how the loop currents couple to fermions.
We review ASV's theory~\cite{Aji2010} and argue that the direct coupling of the main odd-parity LC order parameter to fermions cannot be neglected, as ASV have done~\cite{Aji2010}.
Since we have shown that quantum-critical odd-parity LCs are parametrically strong pair breakers (Sec.~\ref{sec:gen-sys-LC-analysis}, Fig.~\ref{fig:QCP-general-results}), this by itself strongly undermines Varma's proposal.
But even if we accept ASV's suggestion~\cite{Aji2010} that $g$-wave loop currents, as the conjugate momentum of the main $p$-wave LC order parameter, primarily drive superconductivity, due to their decoupling from the Van Hove points (Sec.~\ref{sec:pairing-cuprate-actual-analysis-VH-details}) they robustly yield the incorrect $d_{xy}$ pairing symmetry (Fig.~\ref{fig:cuprate-A2g-results}).
In fact, if it was not for a subtle mistake in the $g$-wave LC coupling ($\sin \tfrac{1}{2} k_{x,y}$ vs.\ $\sin k_{x,y}$), ASV~\cite{Aji2010} would have noticed in their own work the decoupling of the Van Hove points, as we explicitly demonstrate.
In the end, even though loop currents may be present in cuprates, given the experimental evidence for TRSB and parity-breaking in the pseudogap regime (Sec.~\ref{sec:exp-sym-break-pseudogap}), our results show that they are an unlikely candidate for the pairing glue.
We finish with a discussion of the challenges in circumventing our results.

\subsubsection{Differences in the orbital orientations and momentum-space gauge} \label{sec:ASV-conventions}
There are two possible sources of ambiguity in how one defines the three-orbital model of Sec.~\ref{sec:cuprate-3band-model}: the orientations ($\pm$ signs) of the orbital states and the precise definition (gauge) of the Fourier transform.
Our work differs in both from ASV~\cite{Aji2010}.
Of course, as long as one consistently uses a given convention, its choice does not matter, except when comparing to the work of others.

\begin{figure}[t]
\centering
\begin{subfigure}[t]{0.5\textwidth}
\raggedright
\includegraphics[width=0.95\textwidth]{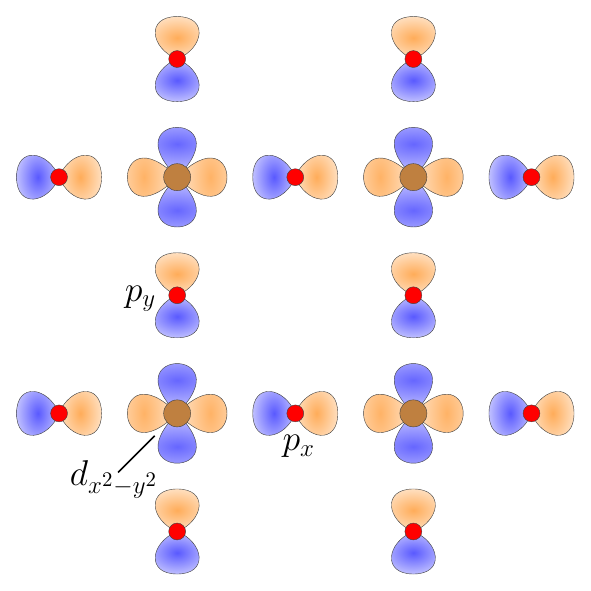}
\subcaption{}
\end{subfigure}%
\begin{subfigure}[t]{0.5\textwidth}
\raggedleft
\includegraphics[width=0.95\textwidth]{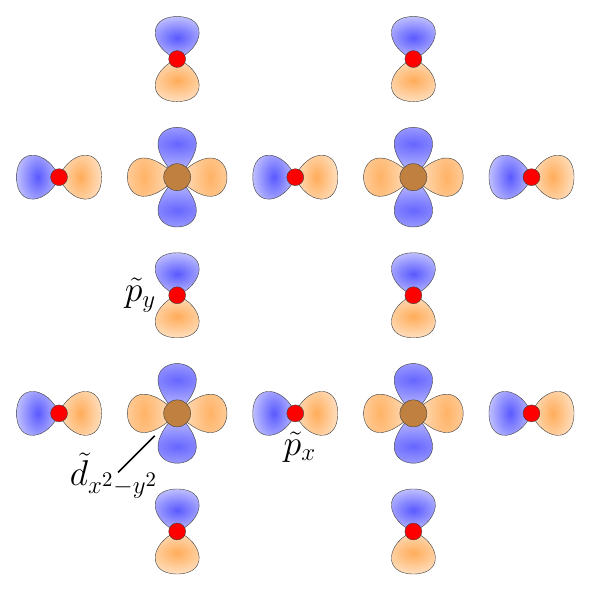}
\subcaption{}
\end{subfigure}
\captionbelow[The convention for the orientation of the \ce{Cu}:$3d_{x^2-y^2}$ and \ce{O}:$2p_{x,y}$ orbitals employed by Refs.~\cite{Varma1997, Weber2014, Photopoulos2019, Palle2024-LC} and us (a) and the convention employed by Aji, Shekhter, and Varma~\cite{Aji2010} (b).]{\textbf{The convention for the orientation of the \ce{Cu}:$3d_{x^2-y^2}$ and \ce{O}:$2p_{x,y}$ orbitals employed by Refs.}~\cite{Varma1997, Weber2014, Photopoulos2019, Palle2024-LC} \textbf{and us (a) and the convention employed by Aji, Shekhter, and Varma}~\cite{Aji2010} \textbf{(b).}
Orange (blue) are positive (negative) lobes of the orbitals.
The underlying three-orbital \ce{CuO2} model is defined in Sec.~\ref{sec:cuprate-3band-model}.
Throughout this section, we use tildes to denote the orbitals, parameters, and operators of Ref.~\cite{Aji2010}.}
\label{fig:us-vs-ASV-orb-convention}
\end{figure}

The orbital orientation conventions that we employ are transparently stated in Fig.~\ref{fig:CuO2-sketch}, which we repeat here in Fig.~\ref{fig:us-vs-ASV-orb-convention}(a) for the reader's convenience.
ASV~\cite{Aji2010}, on the other hand, use the convention shown in Fig.~\ref{fig:us-vs-ASV-orb-convention}(b), i.e., their oxygen $p_y$ orbitals have the opposite sign compared to ours.
Although the orbitals have not been drawn anywhere in Ref.~\cite{Aji2010}, the differences in conventions can be deduced by comparing the kinetic energies.
Our kinetic energy is given by [Eq.~\eqref{eq:hopping_matrix}]:
\begin{align}
\begin{aligned}
\mathrm{K.E.} &= \sum_{\vb{R}} \Psi^{\dag}(\vb{R}) \begin{pmatrix}
0 & t_{pd} & - t_{pd} & - t_{pd} & t_{pd} \\
& 0 & - t_{pp} & 0 & t_{pp} \\
&& 0 & t_{pp} & 0 \\
&&& 0 & - t_{pp} \\
\cc &&&& 0
\end{pmatrix} \Psi(\vb{R}) \\
&= t_{pd} \sum_{\vb{R}} \mleft[d^{\dag}(\vb{R}) p_x(\vb{R} + \tfrac{1}{2} \vu{e}_x) - d^{\dag}(\vb{R}) p_y(\vb{R} + \tfrac{1}{2} \vu{e}_y)\mright] + \cdots + \Hc,
\end{aligned} \label{eq:our-kinetic-en}
\end{align}
where [Eq.~\eqref{eq:extended-basis}]
\begin{align}
\Psi(\vb{R}) &= \begin{pmatrix}
d(\vb{R}) \\[2pt]
p_x(\vb{R} + \tfrac{1}{2} \vu{e}_x) \\[2pt]
p_y(\vb{R} + \tfrac{1}{2} \vu{e}_y) \\[2pt]
p_x(\vb{R} - \tfrac{1}{2} \vu{e}_x) \\[2pt]
p_y(\vb{R} - \tfrac{1}{2} \vu{e}_y)
\end{pmatrix} \label{eq:extended-spinor}
\end{align}
is the extended-basis fermionic annihilation operator.
It creates orbital states which are oriented as shown in Fig.~\ref{fig:us-vs-ASV-orb-convention}(a).
To ease the comparison to ASV's work~\cite{Aji2010}, through this last section we shall denote the components of $\Psi$ with orbitals labels ($d = d_{x^2-y^2}$ and $p_{x,y}$) instead of indices ($\Psi_{1,2,3,4,5}$ as in Fig.~\ref{fig:extended-unit-cell}).

\begin{figure}[t]
\centering
\begin{subfigure}[t]{0.5\textwidth}
\centering
\includegraphics[width=0.91\textwidth]{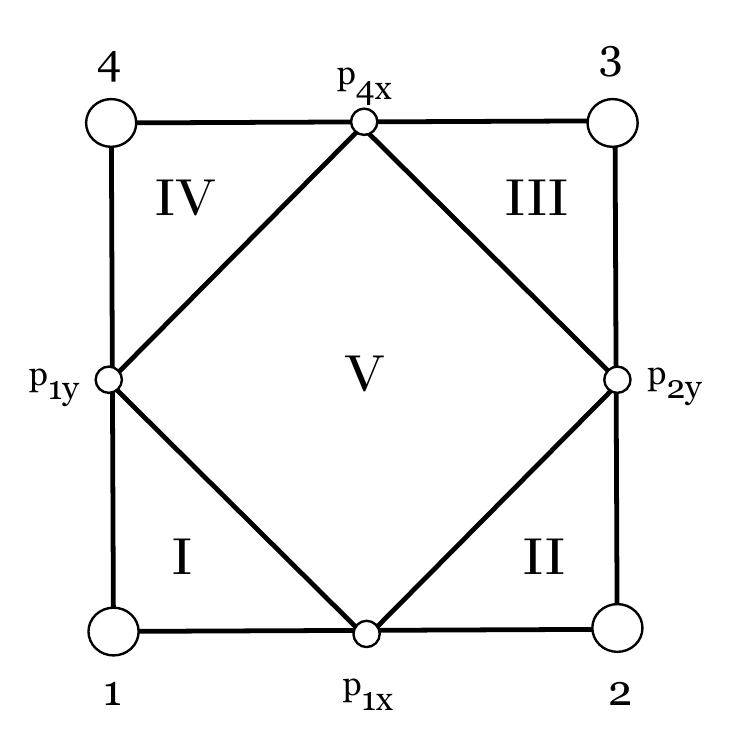}
\subcaption{}
\end{subfigure}%
\begin{subfigure}[t]{0.5\textwidth}
\centering
\includegraphics[width=0.81\textwidth]{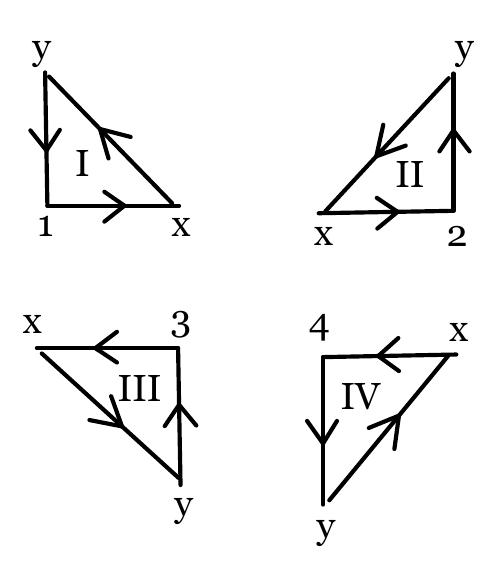}
\subcaption{}
\end{subfigure}
\captionbelow[Labeling of the eight orbitals and five unit cell areas employed by Aji, Shekhter, and Varma~\cite{Aji2010} (a) and the conventions they use for defining their triangle operators {[Eq.~\eqref{eq:ASV-triangle-ops}]} in terms of their link operators {[Eq.~\eqref{eq:ASV-link-ops}]} (b).]{\textbf{Labeling of the eight orbitals and five unit cell areas employed by Aji, Shekhter, and Varma}~\cite{Aji2010} \textbf{(a) and the conventions they use for defining their triangle operators [Eq.~\eqref{eq:ASV-triangle-ops}] in terms of their link operators [Eq.~\eqref{eq:ASV-link-ops}] (b).}
The implicit orbital orientation convention is shown in Fig.~\ref{fig:us-vs-ASV-orb-convention}(b).
Reprinted with permission from Ref.~\cite{Aji2010}. Copyright (2010) by the American Physical Society.}
\label{fig:ASV-labels}
\end{figure}

In their work~\cite{Aji2010}, ASV use the labeling for the orbitals that is shown in Fig.~\ref{fig:ASV-labels}(a).
To compare their equations to ours, we shall find it convenient to write:
\begin{align}
\begin{aligned}
d_{i1} &= \tilde{d}(\vb{R}), &\hspace{50pt}
d_{i2} &= \tilde{d}(\vb{R}+\vu{e}_x), \\
d_{i3} &= \tilde{d}(\vb{R}+\vu{e}_x+\vu{e}_y), &\hspace{50pt}
d_{i4} &= \tilde{d}(\vb{R}+\vu{e}_y), \\
p_{i1x} &= \tilde{p}_x(\vb{R}+\tfrac{1}{2}\vu{e}_x), &\hspace{50pt}
p_{i1y} &= \tilde{p}_y(\vb{R}+\tfrac{1}{2}\vu{e}_y), \\
p_{i4x} &= \tilde{p}_x(\vb{R}+\tfrac{1}{2}\vu{e}_x+\vu{e}_y), &\hspace{50pt}
p_{i2y} &= \tilde{p}_y(\vb{R}+\vu{e}_x+\tfrac{1}{2}\vu{e}_y).
\end{aligned} \label{eq:our-notation-for-ASV-orbs}
\end{align}
We shall use tildes to denote operators and variables from ASV~\cite{Aji2010}.
In this notation, the kinetic energy written in Eq.~(B1) of Ref.~\cite{Aji2010} equals
\begin{align}
\mathrm{K.E.} &= \tilde{t}_{pd} \sum_{\vb{R}} \mleft[\tilde{d}^{\dag}(\vb{R}) \tilde{p}_x(\vb{R} + \tfrac{1}{2} \vu{e}_x) + \tilde{d}^{\dag}(\vb{R}) \tilde{p}_y(\vb{R} + \tfrac{1}{2} \vu{e}_y)\mright] + \cdots + \Hc \label{eq:ASV-kinetic-en}
\end{align}
Here we have only included the $t_{pd}$ term because, as the only term that couples all three orbitals, it completely specifies the orbital conventions, up to an absolute sign.
Our kinetic energy [Eq.~\eqref{eq:our-kinetic-en}] agrees with this kinetic energy by ASV if we identify:
\begin{align}
\tilde{d}(\vb{R}) &= d(\vb{R}), &
\tilde{p}_x(\vb{R}) &= p_x(\vb{R}), &
\tilde{p}_y(\vb{R}) &= - p_y(\vb{R}), \label{eq:ASV-py-convention}
\end{align}
i.e., if we take into account that the $p_y$ orbitals are oppositely oriented, as depicted in Fig.~\ref{fig:us-vs-ASV-orb-convention}.
With this identification, all the other terms that we have not written out in the kinetic energy [ellipses in Eqs.~\eqref{eq:our-kinetic-en} and~\eqref{eq:ASV-kinetic-en}] agree as well.
Let us also note that the $t_{pp}$ parameter of ASV is by definition the opposite of ours:
\begin{align}
\tilde{t}_{pd}&= t_{pd}, &
\tilde{t}_{pp} &= - t_{pp}.
\end{align}
The convention for the signs of $t_{pd}$ and $t_{pp}$ we adopted from Ref.~\cite{Photopoulos2019}.
Of course, the convention for the hopping amplitudes does not matter as long as the correct sign and value are used in the end.

In addition, ASV~\cite{Aji2010} use a different definition of the Fourier transform.
According to our definition [Eq.~\eqref{eq:first-psi-gauge-choice}],
\begin{align}
\psi_{\vb{k}} &= \frac{1}{\sqrt{\mathcal{N}}} \sum_{\vb{R}} \Elr^{- \iu \vb{k} \vdot \vb{R}} \psi(\vb{R}), \label{eq:first-psi-gauge-choice-again}
\end{align}
i.e., the $\tilde{p}_{x,y}(\vb{R}+\tfrac{1}{2}\vu{e}_{x,y})$ orbitals use the same phase factor as the corresponding $d(\vb{R})$ orbital.
An equally viable gauge, used in Ref.~\cite{Photopoulos2019} for instance, is
\begin{align}
\psi_{\vb{k}}^{(\text{alt})} &= \begin{pmatrix}
1 & 0 & 0 \\
0 & \Elr^{- \iu k_x / 2} & 0 \\
0 & 0 & \Elr^{- \iu k_y / 2}
\end{pmatrix} \psi_{\vb{k}}
\end{align}
in which the band Hamiltonian of Eq.~\eqref{eq:3band-Haml} or~\eqref{eq:3band-Haml-again} is a bit simpler:
\begin{align}
H_{\vb{k}}^{(\text{alt})} &= \begin{pmatrix}
\epsilon_{d} - \upmu & 2 \iu t_{pd} \sin(k_x/2) & - 2 \iu t_{pd} \sin(k_y/2) \\
& \epsilon_{p} + 2 t_{pp}' \cos k_x - \upmu & - 4 t_{pp} \sin(k_x/2) \sin(k_y/2) \\
\cc & & \epsilon_{p} + 2 t_{pp}' \cos k_y - \upmu
\end{pmatrix}.
\end{align}
Note that the Fourier transform phase factors coincide with the actual positions of the oxygen atoms in this gauge:
\begin{align}
\psi_{\vb{k}, p_{x,y}}^{(\text{alt1})} = \frac{1}{\sqrt{\mathcal{N}}} \sum_{\vb{R}} \Elr^{- \iu \vb{k} \vdot (\vb{R} + \tfrac{1}{2} \vu{e}_{x,y})} p_{x,y}(\vb{R} + \tfrac{1}{2} \vu{e}_{x,y}).
\end{align}
We have avoided it because it suffers from the disadvantage that $\psi_{\vb{k}}^{(\text{alt})}$ is discontinuous at the Brillouin zone boundary, i.e., $\psi_{\vb{k} + \vb{G}}^{(\text{alt})} \neq \psi_{\vb{k}}^{(\text{alt})}$ for $k_x = - \pi$ and $\vb{G} = (2 \pi, 0)$ and analogously for the $k_y = - \pi$ boundary.
This renders $H_{\vb{k}}^{(\text{alt})}$ aperiodic, as one explicitly sees from the $\sin \tfrac{1}{2} k_{x,y}$ appearing in it.
Given that the cuprate Fermi surface intersects the Brillouin zone boundary (Fig.~\ref{fig:cuprate-Fermi-surfaces}), it is desirable to have eigenvectors which are smooth and periodic functions of $\vb{k}$, not only for the numerics but also for the various symmetry analyses.
Hence our decision to use the Fourier convention of Eq.~\eqref{eq:first-psi-gauge-choice-again}.

Combining these two differences, we find that the momentum-space field operators of ASV $\tilde{\psi}_{\vb{k}}$ are related to our field operators $\psi_{\vb{k}}$ through:
\begin{align}
\tilde{\psi}_{\vb{k}} &= \tilde{\mathcal{B}}_{\vb{k}} \psi_{\vb{k}},
\end{align}
where
\begin{align}
\tilde{\mathcal{B}}_{\vb{k}} &= \begin{pmatrix}
1 & 0 & 0 \\
0 & \Elr^{- \iu k_x / 2} & 0 \\
0 & 0 & - \Elr^{- \iu k_y / 2}
\end{pmatrix}. \label{eq:ASV-Bk-matrix}
\end{align}
Note that their real-space field operator
\begin{align}
\tilde{\psi}_i &= \begin{pmatrix}
d_{i1} \\[2pt]
p_{i1x} \\[2pt]
p_{i1y}
\end{pmatrix} = \begin{pmatrix}
\tilde{d}(\vb{R}) \\[2pt]
\tilde{p}_x(\vb{R} + \tfrac{1}{2} \vu{e}_x) \\[2pt]
\tilde{p}_y(\vb{R} + \tfrac{1}{2} \vu{e}_y)
\end{pmatrix} = \begin{pmatrix}
d(\vb{R}) \\[2pt]
p_x(\vb{R} + \tfrac{1}{2} \vu{e}_x) \\[2pt]
- p_y(\vb{R} + \tfrac{1}{2} \vu{e}_y)
\end{pmatrix}
\end{align}
includes the same orbitals as ours.
The primitive unit cell (Fig.~\ref{fig:extended-unit-cell}) is thus the same in both works.
The relation to the extended basis of Sec.~\ref{sec:extended-basis-def} is given by
\begin{align}
\Psi_{\vb{k}} &= \tilde{\mathcal{K}}_{\vb{k}} \tilde{\psi}_{\vb{k}},
\end{align}
where
\begin{align}
\tilde{\mathcal{K}}_{\vb{k}} = \mathcal{K}_{\vb{k}} \tilde{\mathcal{B}}_{\vb{k}}^{\dag} &= \begin{pmatrix}
1 & 0 & 0 \\
0 & \Elr^{\iu k_x/2} & 0 \\
0 & 0 & -\Elr^{- \iu k_y/2} \\
0 & \Elr^{- \iu k_x/2} & 0 \\
0 & 0 & -\Elr^{- \iu k_y/2}
\end{pmatrix}. \label{eq:ASV-Kk-matrix}
\end{align}
The old $\mathcal{K}_{\vb{k}}$ is defined in Eq.~\eqref{eq:K-proj-mat}.
This gauge difference has been deduced from the conduction band eigenvector ASV provided in Eq.~(17) of their article~\cite{Aji2010}, as we explain in more detail in Sec.~\ref{sec:ASV-fermion-coupling} after Eq.~\eqref{eq:3band-Haml-ASV}.

\subsubsection{Agreement between the loop-current operators} \label{sec:ASV-LC-operators}
\begin{figure}[t]
\centering
\begin{subfigure}[t]{0.43\textwidth}
\centering
\includegraphics[width=\textwidth]{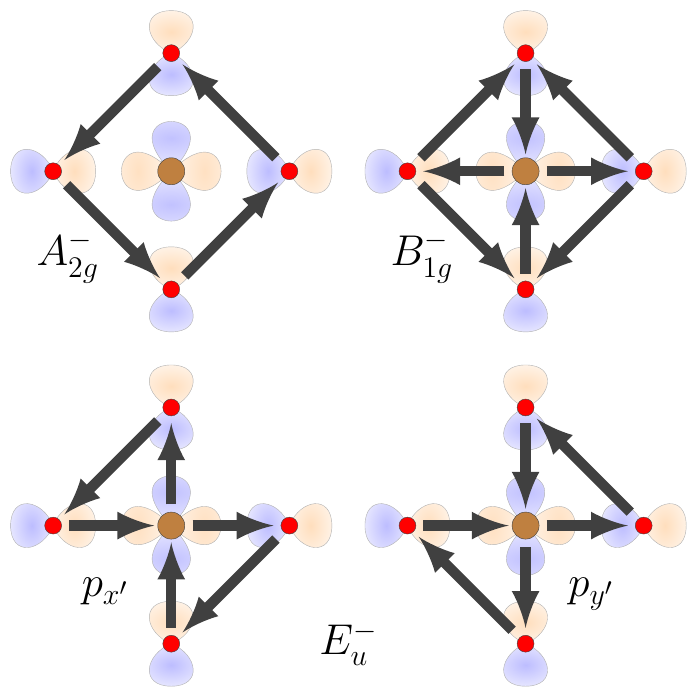}
\subcaption{}
\end{subfigure}%
\begin{subfigure}[t]{0.57\textwidth}
\centering
\includegraphics[width=0.98\textwidth]{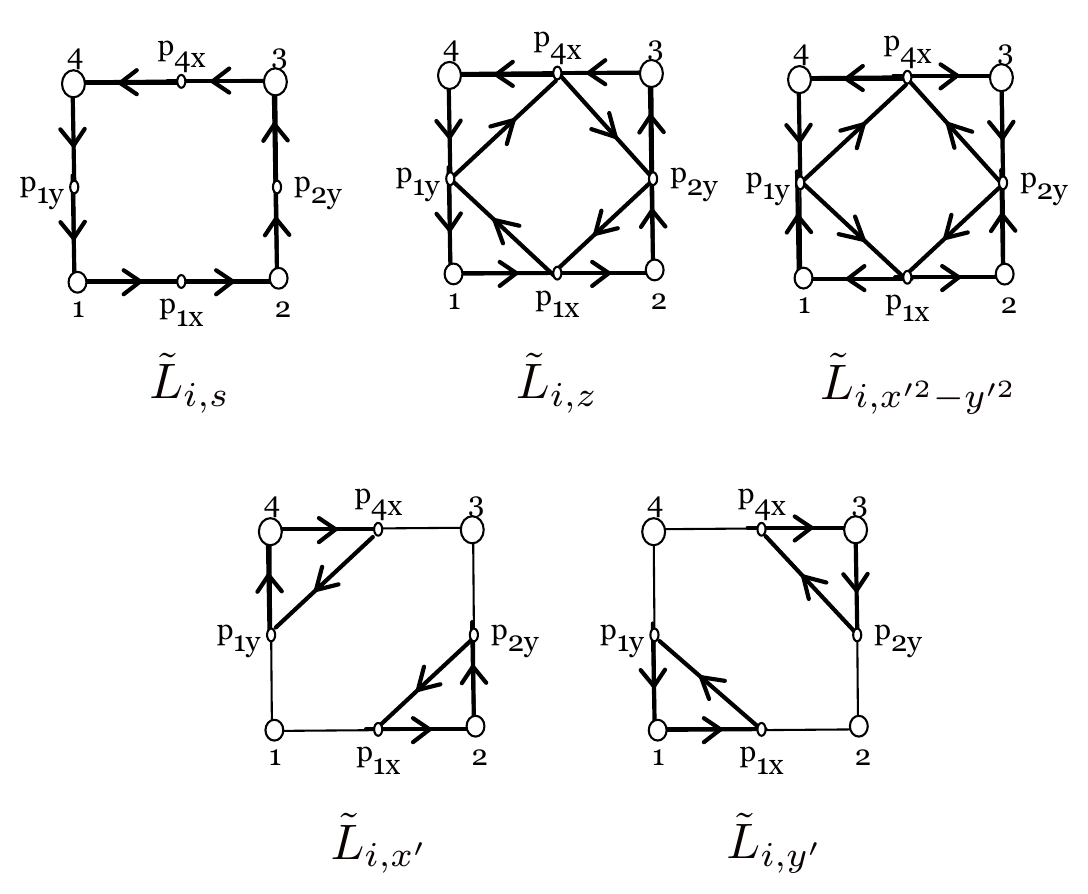}
\subcaption{}
\end{subfigure}
\captionbelow[Comparison of the loop-current operators introduced by us (a) to those introduced by Aji, Shekhter, and Varma in Ref.~\cite{Aji2010} (b).]{\textbf{Comparison of the loop-current operators introduced by us (a) to those introduced by Aji, Shekhter, and Varma in Ref.}~\cite{Aji2010} \textbf{(b).}
Our operators are constructed from the five orbitals of an extended unit cell, drawn in Fig.~\ref{fig:extended-unit-cell}, while the flux operators of Ref.~\cite{Aji2010} are constructed from the eight orbitals shown in Fig.~\ref{fig:ASV-labels}(a).
Here $x' = (x+y)/\sqrt{2}$ and $y' = (x-y)/\sqrt{2}$.
Figure (b) is reprinted with permission from Ref.~\cite{Aji2010}. Copyright (2010) by the American Physical Society.}
\label{fig:our-ASV-flux-operators}
\end{figure}

As part of their analysis, ASV~\cite{Aji2010} have introduced a number of LC or flux operators, depicted in Fig.~\ref{fig:our-ASV-flux-operators}(b).
We have gone through the effort of explicitly transcribing these flux operators and comparing them with our classification of fermionic bilinears of Sec.~\ref{sec:cuprate-bilinear-class}.
Our LC operators are made from one copper $d_{x^2-y^2}$ orbital and the four oxygen $p_{x,y}$ orbitals that surround it [Fig.~\ref{fig:our-ASV-flux-operators}(a)], while ASV's LC operators are constructed from four copper $d_{x^2-y^2}$ orbitals and the four oxygen $p_{x,y}$ orbitals that are in between them [Fig.~\ref{fig:our-ASV-flux-operators}(b)].
Here we show that the two sets of operators are in agreement, despite the different appearances.

Given that ASV's unit cell contains four copper atoms [Fig.~\ref{fig:ASV-labels}(a)], some of the LC operators that they introduce break translation symmetry.
This includes, in particular, the $\tilde{L}_{i,s}$ operator which according to Eq.~(C1) of Ref.~\cite{Aji2010} equals:
\begin{align}
\tilde{L}_{i,s} &= \iu \mleft[\tilde{d}^{\dag}(\vb{R}) \tilde{p}_x(\vb{R}+\tfrac{1}{2} \vu{e}_x) - \tilde{d}^{\dag}(\vb{R}+\vu{e}_y) \tilde{p}_x(\vb{R}+\tfrac{1}{2} \vu{e}_x+\vu{e}_y) + \cdots\mright] + \Hc
\end{align}
Thus if we look at the $\vb{q} = \vb{0}$ component
\begin{align}
\tilde{L}_{\vb{q}=\vb{0},s} &= \sum_{\vb{R}} \tilde{L}_{i,s} = 0,
\end{align}
it vanishes identically because the currents of neighboring copper atoms have opposite orientations.
This is clearly visible in Fig.~\ref{fig:our-ASV-flux-operators}(b).
Although such non-homogeneous LC operators are allowed and can condense for finite $\vb{q}$, they are not the focus of our work or of the work by ASV~\cite{Aji2010}, so we shall not discuss them further.
In addition, let us note that the $\tilde{L}_{i,z}$ operator of Ref.~\cite{Aji2010} includes a component proportional to $\tilde{L}_{i,s}$, as can be seen by examining the outer rim of $\tilde{L}_{i,z}$ in Fig.~\ref{fig:our-ASV-flux-operators}(b).
As we are interested in intra-unit-cell LCs, we shall set this part of $\tilde{L}_{i,z}$ to zero.

The remaining $\tilde{L}_{i,z}$, $\tilde{L}_{i,x'^2-y'^2}$, $\tilde{L}_{i,x'}$, and $\tilde{L}_{i,y'}$ flux operator are in direct correspondence to our LC operators, whose LC patterns are shown in Fig.~\ref{fig:our-ASV-flux-operators}(a).
These operators are precisely defined in the Appx.es~C and~D of Ref.~\cite{Aji2010}, but in a notation that is quite different from ours.
After (i) transcribing the expressions provided in Appx.es~C and~D of ASV's paper~\cite{Aji2010}, (ii) taking into account the different conventions for the $p_y$ orbitals, as in Eq.~\eqref{eq:ASV-py-convention}, and (iii) translating the operators so that they are centered around only one copper atom, one finds that:
\begin{align}
\tilde{L}_{i,z} &= \Psi^{\dag}(\vb{R}) \mleft[2 \Lambda_1^{A_{2g}^{-}}\mright] \Psi(\vb{R}), \label{eq:ASV-to-our-LC-op1} \\
\tilde{L}_{i,x'^2-y'^2} &= \Psi^{\dag}(\vb{R}) \mleft[4 \Lambda_1^{B_{1g}^{-}} - 2 \Lambda_2^{B_{1g}^{-}}\mright] \Psi(\vb{R}),
\end{align}
and
\begin{align}
\tilde{L}_{i,x'} &= \Psi^{\dag}(\vb{R}) \mleft[- \sqrt{2} \mleft(\Lambda_{1,x}^{E_u^{-}} + \Lambda_{1,y}^{E_u^{-}}\mright) - \mleft(\Lambda_{3,x}^{E_u^{-}} + \Lambda_{3,y}^{E_u^{-}}\mright)\mright] \Psi(\vb{R}), \\
\tilde{L}_{i,y'} &= \Psi^{\dag}(\vb{R}) \mleft[- \sqrt{2} \mleft(\Lambda_{1,x}^{E_u^{-}} - \Lambda_{1,y}^{E_u^{-}}\mright) - \mleft(\Lambda_{3,x}^{E_u^{-}} - \Lambda_{3,y}^{E_u^{-}}\mright)\mright] \Psi(\vb{R}), \label{eq:ASV-to-our-LC-op4}
\end{align}
i.e., the flux operators agree with our LC matrices.
Here the equality holds modulo lattice translations.
The $\Lambda$ matrices of the right-hand side are listed in Tab.~\ref{tab:orbital-Lambda-mats}.\footnote{Note that the $\Lambda_1^{A_{2g}^{-}}$ and $\Lambda^{E_{u}^{-}}_{3,x/y}$ matrices used in this thesis are the opposite of those given in Appx.~C of Ref.~\cite{Palle2024-LC}.}

The subscripts of the $\tilde{L}$ flux operators are suppose to indicate how they transform.
Thus, according to Ref.~\cite{Aji2010}, $\tilde{L}_{i,z}$ transforms under the same $D_{4h}$ point group irrep as $z$, which is $A_{2u}$, while $(\tilde{L}_{i,x'} | \tilde{L}_{i,y'})$ transform as $(x'|y') \in E_u$.
Keeping in mind that according to ASV $x' = (x+y)/\sqrt{2}$ and $y' = (x-y)/\sqrt{2}$~\cite{Aji2010}, the latter claim is in agreement with the relations~\eqref{eq:ASV-to-our-LC-op1} to~\eqref{eq:ASV-to-our-LC-op4}.
The former statement is also correct if we take that $\tilde{L}_{i,z}$ is even under parity, $A_{2u} \to A_{2g}$, as one would expect for an orbital angular momentum operator~\cite{Aji2010}.
Regarding $\tilde{L}_{i,x'^2-y'^2}$, we find that it transforms according to the $B_{1g}$ irrep.
The appropriate polynomial is $x^2-y^2$ without the primes (see Tab.~\ref{tab:D4h-char-tab-again} or Tab.~\ref{tab:D4h-example-tab} in Appx.~\ref{app:group_theory}, for instance), and not $x'^2-y'^2 = 2xy \in B_{2g}$, as suggested by ASV~\cite{Aji2010}.
That said, in their paper~\cite{Aji2010} it is also stated that $\tilde{L}_{i,x'^2-y'^2}$ has the symmetry of the so-called $\Theta_{\text{I}}$ IUC LC phase~\cite{Simon2002, Simon2003, Varma2006}, whose irrep was previously correctly identified as $B_{1g}$~\cite{Simon2002, Simon2003, Varma2006}.
The other, so-called $\Theta_{\text{II}}$ LC phase~\cite{Simon2002, Simon2003, Varma2006}  corresponds to $\tilde{L}_{i,x'}$ and $\tilde{L}_{i,y'}$~\cite{Aji2010}.
Apart from the misleading naming of one flux operator ($\tilde{L}_{i,x'^2-y'^2}$), the flux operators of ASV~\cite{Aji2010} are in agreement with our classification of fermionic bilinears (Sec.~\ref{sec:cuprate-bilinear-class}).
A point of difference between our LC operators and their flux operators is that we have determined the relative weights between the two $E_u^{-}$ and $B_{1g}^{-}$ components from the Bloch and Kirchhoff constraints (Sec.~\ref{sec:Bloch-Kirch-constr}).

Let us demonstrate how we obtained the results of Eqs.~\eqref{eq:ASV-to-our-LC-op1} to~\eqref{eq:ASV-to-our-LC-op4} using the $\tilde{L}_{i,z}$ flux operator as an example.
In light of its direct coupling to fermions, this is the most important operator within ASV's theory~\cite{Aji2010}.
The others can be analyzed in similar fashion.
We start from Eq.~(D11) of ASV~\cite{Aji2010} which defines $\tilde{L}_{i,z}$ as the sum of triangle flux operators:
\begin{equation}
\tilde{L}_{i,z} = \sum_{L=I,\ldots,IV} f_{i,L}.
\end{equation}
The triangle flux operators $f_{iL}$ are defined in Eq.~(D4) as
\begin{align}
\begin{aligned}
f_{i,I} &= \mathcal{O}_{i,1,x}-\mathcal{O}_{i,1,y}+\mathcal{O}_{i,1,xy}, \\
f_{i,II} &= \mathcal{O}_{i,2,x}+\mathcal{O}_{i,2,y}+\mathcal{O}_{i,2,xy}, \\
f_{i,III} &= -\mathcal{O}_{i,3,x}+\mathcal{O}_{i,3,y}+\mathcal{O}_{i,3,xy}, \\
f_{i,IV} &= -\mathcal{O}_{i,4,x}-\mathcal{O}_{i,4,y}+\mathcal{O}_{i,4,xy}
\end{aligned} \label{eq:ASV-triangle-ops}
\end{align}
with the conventions shown in Fig.~\ref{fig:ASV-labels}(b).
The $d_{x^2-y^2}$--$p_{x}$ link (or current) operator $\mathcal{O}_{i,\ell,x}$ is defined in Eq.~(D2), the $d_{x^2-y^2}$--$p_{y}$ link operator $\mathcal{O}_{i,\ell,y}$ follows by extension, while the $p_x$--$p_y$ link operator $\mathcal{O}_{i,\ell,xy}$ is defined in Eq.~(D3) of Ref.~\cite{Aji2010}:
\begin{align}
\begin{aligned}
\mathcal{O}_{i,\ell,x} &= \iu \sum_{s} d_{i,\ell,s}^{\dag} p_{i,\ell,x,s} + \Hc, \\
\mathcal{O}_{i,\ell,y} &= \iu \sum_{s} d_{i,\ell,s}^{\dag} p_{i,\ell,y,s} + \Hc, \\
\mathcal{O}_{i,\ell,xy} &= \iu \sum_{s} p_{i,\ell,y,s}^{\dag} p_{i,\ell,x,s} + \Hc
\end{aligned} \label{eq:ASV-link-ops}
\end{align}
As we explain in the next section, the diagonal spin summation is not the appropriate one for decomposing Hubbard interactions.
However, for the purpose of relating their flux operator to ours, the above definitions (based on Eqs.~(D2) and~(D3) of Ref.~\cite{Aji2010}) are the right ones.
Below we suppress the summation over spins $s \in \{\uparrow, \downarrow\}$.

As discussed previously, in $\tilde{L}_{i,z}$ we ignore the $\propto \tilde{L}_{i,s}$ outer rim [Fig.~\ref{fig:our-ASV-flux-operators}(b)] because it breaks translation invariance.
The $\mathcal{O}_{i,\ell,x}$ and $\mathcal{O}_{i,\ell,y}$ operators we thus eliminate, leaving:
\begin{align}
\tilde{L}_{i,z} &= \sum_{l=1}^{4}\mathcal{O}_{i,\ell,xy} = - \iu \mleft(p_{i,1,x}^{\dag}p_{i,1,y}+p_{i,1,x}^{\dag}p_{i,2,y}+p_{i,4,x}^{\dag}p_{i,2,y}+p_{i,4,x}^{\dag} p_{i,1,y}\mright) + \Hc
\end{align}
Here we used the triangle labeling shown in Fig.~\ref{fig:ASV-labels}(b).
Next, we rewrite this in terms of the notation introduced in Eq.~\eqref{eq:our-notation-for-ASV-orbs}:
\begin{align}
\tilde{L}_{i,z} = &- \iu \mleft[\tilde{p}_{x}^{\dag}\mleft(\vb{R}+\tfrac{1}{2}\vu{e}_x\mright)\tilde{p}_{y}\mleft(\vb{R}+\tfrac{1}{2}\vu{e}_y\mright)+\tilde{p}_{x}^{\dag}\mleft(\vb{R}+\tfrac{1}{2}\vu{e}_x\mright)\tilde{p}_{y}\mleft(\vb{R}+\vu{e}_x+\tfrac{1}{2}\vu{e}_y\mright)\mright] + \Hc \\[2pt]
&- \iu \mleft[\tilde{p}_{x}^{\dag}\mleft(\vb{R}+\tfrac{1}{2}\vu{e}_x+\vu{e}_y\mright)\tilde{p}_{y}\mleft(\vb{R}+\vu{e}_x+\tfrac{1}{2}\vu{e}_y\mright)+\tilde{p}_{x}^{\dag}\mleft(\vb{R}+\tfrac{1}{2}\vu{e}_x+\vu{e}_y\mright)\tilde{p}_{y}\mleft(\vb{R}+\tfrac{1}{2}\vu{e}_y\mright)\mright] + \Hc \notag
\end{align}
Finally, we exploit translation invariance to center the orbitals around $\vb{R}$ and switch to our convention for the $p_y$ orbitals (the latter only gives an overall minus sign) to obtain:
\begin{align}
\begin{aligned}
\tilde{L}_{i,z} = &+ \iu \mleft[p_{x}^{\dag}\mleft(\vb{R}+\tfrac{1}{2}\vu{e}_x\mright)p_{y}\mleft(\vb{R}+\tfrac{1}{2}\vu{e}_y\mright)+p_{x}^{\dag}\mleft(\vb{R}+\tfrac{1}{2}\vu{e}_x\mright)p_{y}\mleft(\vb{R}-\tfrac{1}{2}\vu{e}_y\mright)\mright] + \Hc \\[2pt]
&+ \iu \mleft[p_{x}^{\dag}\mleft(\vb{R}-\tfrac{1}{2}\vu{e}_x\mright)p_{y}\mleft(\vb{R}-\tfrac{1}{2}\vu{e}_y\mright)+p_{x}^{\dag}\mleft(\vb{R}-\tfrac{1}{2}\vu{e}_x\mright)p_{y}\mleft(\vb{R}+\tfrac{1}{2}\vu{e}_y\mright)\mright] + \Hc
\end{aligned}
\end{align}
The last step is to express this result in matrix notation:
\begin{align}
\begin{aligned}
\tilde{L}_{i,z} &= \begin{pmatrix}
d\mleft(\vb{R}\mright)\\
p_{x}\mleft(\vb{R}+\tfrac{1}{2}\vu{e}_x\mright)\\[2pt]
p_{y}\mleft(\vb{R}+\tfrac{1}{2}\vu{e}_y\mright)\\[2pt]
p_{x}\mleft(\vb{R}-\tfrac{1}{2}\vu{e}_x\mright)\\[2pt]
p_{y}\mleft(\vb{R}-\tfrac{1}{2}\vu{e}_y\mright)
\end{pmatrix}^{\dag} \begin{pmatrix}
0 & 0 & 0 & 0 & 0\\
0 & 0 & \iu & 0 & \iu\\
0 & -\iu & 0 & -\iu & 0\\
0 & 0 & \iu & 0 & \iu\\
0 & -\iu & 0 & -\iu & 0
\end{pmatrix} \begin{pmatrix}
d\mleft(\vb{R}\mright)\\
p_{x}\mleft(\vb{R}+\tfrac{1}{2}\vu{e}_x\mright)\\[2pt]
p_{y}\mleft(\vb{R}+\tfrac{1}{2}\vu{e}_y\mright)\\[2pt]
p_{x}\mleft(\vb{R}-\tfrac{1}{2}\vu{e}_x\mright)\\[2pt]
p_{y}\mleft(\vb{R}-\tfrac{1}{2}\vu{e}_y\mright)
\end{pmatrix} \\[2pt]
&= \Psi^{\dag}(\vb{R}) \mleft[2 \Lambda_1^{A_{2g}^{-}}\mright] \Psi(\vb{R}).
\end{aligned}
\end{align}
This is the relation stated in Eq.~\eqref{eq:ASV-to-our-LC-op1}.
Analogous manipulations give the other relations.

\begin{table}[t]
\centering
\captionabove[Statistics of the classification of the orbital matrices constructed from the eight orbitals of Aji, Shekhter, and Varma~\cite{Aji2010}, shown in Fig.~\ref{fig:ASV-labels}(a).]{\textbf{Statistics of the classification of the orbital matrices constructed from the eight orbitals of Aji, Shekhter, and Varma}~\cite{Aji2010}\textbf{, shown in Fig.~\ref{fig:ASV-labels}(a).}
Table entries indicate the number of Hermitian $8 \times 8$ orbitals matrices which transform according a given $D_{4h}$ irrep and time-reversal (TR) sign.
The last row is the net number of TR-even and TR-odd matrices, which coincides with the number of symmetric and antisymmetric Hermitian $8 \times 8$ matrices.}
{\renewcommand{\arraystretch}{1.3}
\renewcommand{\tabcolsep}{10pt}
\begin{tabular}{c|cc} \hline\hline
& TR-even & TR-odd \\ \hline
$A_{1g}$ & $8$ & $2$ \\
$A_{2g}$ & $2$ & $4$ \\
$B_{1g}$ & $5$ & $3$ \\
$B_{2g}$ & $5$ & $3$ \\
$E_u$ & $8 \times 2$ & $8 \times 2$ \\ \hline
$\sum$ & $36 = \frac{8 \times 9}{2}$ & $28 = \frac{8 \times 7}{2}$
\\[2pt] \hline\hline
\end{tabular}}
\label{tab:ASV-bilin-classification-stats}
\end{table}

As an aside, the classification procedure of Sec.~\ref{sec:cuprate-bilinear-class} can be adapted to the enlarged unit cell of ASV, shown in Fig.~\ref{fig:ASV-labels}(a), with minimal modifications.
The orbital transformation matrices analogous to the $O(g)$ of Tab.~\ref{tab:CuO2-model-D4h-generators}, call them $\tilde{O}(g)$, are now $8 \times 8$ matrices.
Using characters (Sec.~\ref{sec:character-theory-D4h}), it is easily seen that $\tilde{O} = 2 A_{2g} \oplus B_{1g} \oplus B_{2g} \oplus 2 E_u$.
With the aid of Tab.~\ref{tab:D4h-irrep-prod-tab}, one can now readily decompose $\tilde{O} \otimes \tilde{O}$, with the classification statistics as given in Tab.~\ref{tab:ASV-bilin-classification-stats}.
The additional matrices that arise, when compared to Tab.~\ref{tab:bilin-classification-stats}, are equivalent to the old $5 \times 5$ orbital $\Lambda$ matrices multiplied with momentum-dependent functions, as explained in Sec.~\ref{sec:cup-ord-param-constr}.

\subsubsection[On the decompositions of the $V_{pd}$ and $V_{pp}$ Hubbard \\ interactions]{On the decompositions of the $V_{pd}$ and $V_{pp}$ Hubbard interactions} \label{sec:ASV-eff-Haml}
The starting point of ASV's derivation of their effective LC Hamiltonian is the following exact identity (Eq.~(2) in Ref.~\cite{Aji2010}):
\begin{align}
2 a_s^{\dag} a_s b_{s'}^{\dag} b_{s'} &= - \abs{\tilde{\mathcal{J}}_{ss'}}^2 + a_s^{\dag} a_s + b_{s'}^{\dag} b_{s'}, \label{eq:ASV-exact-starting}
\end{align}
where $a$ and $b$ are fermionic annihilation operators and $\tilde{\mathcal{J}}_{ss'} = - \iu (a_s^{\dag} b_{s'} - b_{s'}^{\dag} a_s) = \tilde{\mathcal{J}}_{ss'}^{\dag}$ is a current operator.
This identity enables one to decompose density-density interactions into (spin-dependent) current channels.
Here we discuss how we obtain very different results from ASV~\cite{Aji2010} when carrying out this decomposition.

If one drops the uninteresting one-particle terms and also neglects the spin operators, ASV find that the $V_{pd}$ and $V_{pp}$ Hubbard interactions decompose into (Eqs.~(D10) and~(D12) in Ref.~\cite{Aji2010}):
\begin{align}
\Haml_i' &= - \frac{V_{pd}}{16} \mleft[\abs{\tilde{L}_{i,x'}}^2 + \abs{\tilde{L}_{i,y'}}^2 + \tfrac{1}{2} \abs{\tilde{L}_{i,x'^2-y'^2}}^2 + \abs{\tilde{L}_{i,s}}^2 + \abs{\tilde{L}_{i,z}}^2\mright] - \frac{V_{pp}}{8} \abs{\tilde{L}_{i,\bar{s}}}^2 + \cdots \, . \label{eq:ASV-H_nn}
\end{align}
On the other hand, the LC operators that we found to appear in Sec.~\ref{sec:CuO2-Hubbard-decompose} are
\begin{align}
\mathcal{L}_{pd}^{-} &= \Big(\Lambda^{A_{1g}^{-}}_1, \quad\Lambda^{B_{1g}^{-}}_1, \quad\Lambda^{E_u^{-}}_{1,x}, \quad\Lambda^{E_u^{-}}_{1,y}\Big), \\
\mathcal{L}_{pp}^{-} &= \Big(\Lambda^{A_{2g}^{-}}_1, \quad\Lambda^{B_{1g}^{-}}_2, \quad\Lambda^{E_u^{-}}_{3,x}, \quad\Lambda^{E_u^{-}}_{3,y}\Big)
\end{align}
for $V_{pd}$ and $V_{pp}$, respectively.
Keeping in mind Eqs.~\eqref{eq:ASV-to-our-LC-op1} to~\eqref{eq:ASV-to-our-LC-op4}, the two decompositions appear quite different.
Part of this difference might be due to using different unit cells (Fig.~\ref{fig:our-ASV-flux-operators}), but the symmetries and orbital contents should be the same at the very least.
The symmetries of $\tilde{L}_{i,s}$, $\tilde{L}_{i,x'^2-y'^2}$, $\tilde{L}_{i,x'}$, and $\tilde{L}_{i,y'}$ agree with the matrices of $\mathcal{L}_{pd}^{-}$, respectively.
However, $\tilde{L}_{i,x'^2-y'^2}$, $\tilde{L}_{i,x'}$, and $\tilde{L}_{i,y'}$ include $p_x$--$p_y$ currents [Fig.~\ref{fig:our-ASV-flux-operators}(b)] which are absent in all the matrices of $\mathcal{L}_{pd}^{-}$ (see the schematics of Tab.~\ref{tab:orbital-Lambda-mats}).
In light of Eq.~\eqref{eq:ASV-exact-starting}, a $d_{x^2-y^2}$--$p_{x,y}$ density-density interaction cannot result in $p_x$--$p_y$ currents.
Given its pure $p_x$--$p_y$ current character (up to translation symmetry-breaking terms), the appearance of $\tilde{L}_{i,z}$ is even more mysterious.
In the article~\cite{Aji2010}, ASV state that $\abs{\tilde{L}_{i,z}}^2$ ``is also present in the interactions'' without proof or elaboration.
Writing out ASV's decomposition does not yield exact cancellations between the $p_x$--$p_y$ currents and the present author has not managed to reproduce their $V_{pd}$ Hubbard interaction decomposition.
The same goes for $V_{pp}$ which only includes $\abs{\tilde{L}_{i,\bar{s}}}^2$, even though we found components of $\tilde{L}_{i,z}$, $\tilde{L}_{i,x'^2-y'^2}$, $\tilde{L}_{i,x'}$, and $\tilde{L}_{i,y'}$ to appear as well (see $\mathcal{L}_{pp}^{-}$).

As the $V_{pd}$ decomposition is the most pertinent one to ASV's work~\cite{Aji2010},  let us state our result once more [Eq.~\eqref{eq:Hubbard-Vpd-decomp}]:
\begin{align}
n_d \sum_{\ell=1}^{4} n_{p\ell} &= - \frac{1}{2} \sum_{\Lambda \in \mathcal{L}_{pd}^{-}} \big[\mathcal{O}(\Lambda)\big]^2 - \frac{1}{4} \sum_{\Lambda \in \mathcal{L}_{pd}^{-}} \sum_{A=1}^{3} \big[\mathcal{O}(\Lambda \Pauli_A)\big]^2 + \frac{1}{4} \sum_{\Lambda \in \mathcal{L}_{pd}^{+}} \big[\mathcal{O}(\Lambda)\big]^2, \label{eq:Hubbard-Vpd-decomp-again}
\end{align}
where:
\begin{align}
\mathcal{L}_{pd}^{+} &= \Big(\Lambda^{A_{1g}^{+}}_3, \quad\Lambda^{B_{1g}^{+}}_1, \quad\Lambda^{E_u^{+}}_{1,x}, \quad\Lambda^{E_u^{+}}_{1,y}\Big).
\end{align}
Here $\Psi = (d, p_1, p_2, p_3, p_4)^{\intercal}$, $n_d = d^{\dag} d$, $n_{p\ell} = p_{\ell}^{\dag} p_{\ell}$, and
\begin{align}
\mathcal{O}(\Gamma) &= \Psi^{\dag} \Gamma \Psi.
\end{align}
As explained in Sec.~\ref{sec:CuO2-Hubbard-decompose}, Fierz identities allow one to also write [Eq.~\eqref{eq:Hubbard-Vpd-decomp2}]:
\begin{align}
n_d \sum_{\ell=1}^{4} n_{p\ell} &= - \frac{1}{2} \sum_{\Lambda \in \mathcal{L}_{pd}^{+}} \big[\mathcal{O}(\Lambda)\big]^2 - \frac{1}{4} \sum_{\Lambda \in \mathcal{L}_{pd}^{+}} \sum_{A=1}^{3} \big[\mathcal{O}(\Lambda \Pauli_A)\big]^2 + \frac{1}{4} \sum_{\Lambda \in \mathcal{L}_{pd}^{-}} \big[\mathcal{O}(\Lambda)\big]^2. \label{eq:Hubbard-Vpd-decomp2-again}
\end{align}
The most notable thing about Eqs.~\eqref{eq:Hubbard-Vpd-decomp-again} and~\eqref{eq:Hubbard-Vpd-decomp2-again} is that nematic ($\sim \Lambda^{+} \in \mathcal{L}_{pd}^{+}$), spin-magnetic ($\sim \Lambda^{+} \Pauli_A$ for $\Lambda^{+} \in \mathcal{L}_{pd}^{+}$), and spin LC ($\sim \Lambda^{-} \Pauli_A$ for $\Lambda^{-} \in \mathcal{L}_{pd}^{-}$) instabilities at first sight appear to be as competitive as orbital LC instabilities ($\sim \Lambda^{-} \in \mathcal{L}_{pd}^{-}$).
LC operators are, in fact, repulsive in the latter form.
That said, we shall not carry out any mean-field~\cite{Ovchinnikov1990, Fischer2011, Atkinson2016} or numerical~\cite{Greiter2007, Thomale2008, Kung2014, Weber2009, Weber2014} analyses to find out which order prevails.
We reviewed such work in Sec.~\ref{sec:LC-cuprate-micro}.
The main point is that ASV~\cite{Aji2010} by dropping all other terms are essentially assuming, rather than deriving, LC order.
Conceptually, their treatment is therefore very similar to ours.
The philosophy behind our treatment was explained in the introduction of Sec.~\ref{sec:pairing-cuprate-formalism}.

To derive Eq.~\eqref{eq:Hubbard-Vpd-decomp-again}, we start from the following relation (in which we ignore quadratic terms):
\begin{align}
\begin{aligned}
n_d n_{p\ell} = \sum_{ss'} d_s^{\dag} d_s p_{\ell s'}^{\dag} p_{\ell s'} &= - \frac{1}{2} \sum_{ss'} \tilde{\mathcal{J}}_{\ell;ss'} \tilde{\mathcal{J}}_{\ell;ss'} \\
&= - \frac{1}{2} \sum_{ss'} \mathcal{J}_{\ell;ss'} \mathcal{J}_{\ell;s's} + \frac{1}{4} \sum_{ss'} (\mathcal{R}_{\ell;ss} \mathcal{R}_{\ell;s's'} - \mathcal{J}_{\ell;ss} \mathcal{J}_{\ell;s's'}),
\end{aligned}
\end{align}
where:
\begin{align}
\mathcal{R}_{\ell;ss'} &= d_s^{\dag} p_{\ell s'} + p_{\ell s}^{\dag} d_{s'}, \\
\mathcal{J}_{\ell;ss'} &= - \iu (d_s^{\dag} p_{\ell s'} - p_{\ell s}^{\dag} d_{s'}), \\
\tilde{\mathcal{J}}_{\ell;ss'} &= - \iu (d_s^{\dag} p_{\ell s'} - p_{\ell s'}^{\dag} d_s).
\end{align}
The first line follows from Eq.~\eqref{eq:ASV-exact-starting}, but it is actually the second line that more accurately reflects the channels contained in the $d$--$p$ density-density interaction.
The $\mathcal{R}_{\ell;ss'}$, $\mathcal{J}_{\ell;ss'}$, and $\tilde{\mathcal{J}}_{\ell;ss'}$ operators are not independent:
\begin{align}
\tilde{\mathcal{J}}_{\ell;ss'} + \tilde{\mathcal{J}}_{\ell;s's} &= \mathcal{J}_{\ell;ss'} + \mathcal{J}_{\ell;s's}, \\
\tilde{\mathcal{J}}_{\ell;ss'} - \tilde{\mathcal{J}}_{\ell;s's} &= - \iu \mleft(\mathcal{R}_{\ell;ss'} - \mathcal{R}_{\ell;s's}\mright).
\end{align}
In addition $(2 d_s^{\dag} p_{\ell s'})^2 = (\mathcal{R}_{\ell;ss'} + \iu \mathcal{J}_{\ell;ss'})^2 = 0$ and $(\mathcal{R}_{\ell;ss'} - \iu \mathcal{J}_{\ell;ss'})^2 = 0$.
Notice how:
\begin{align}
\mathcal{R}_{\ell;ss'}^{\dag} &= \mathcal{R}_{\ell;s's}, &
\mathcal{J}_{\ell;ss'}^{\dag} &= \mathcal{J}_{\ell;s's}, &
\tilde{\mathcal{J}}_{\ell;ss'}^{\dag} &= \tilde{\mathcal{J}}_{\ell;ss'}.
\end{align}
The fact that in $\tilde{\mathcal{J}}$ we do not interchange the spin indices complicates things when we construct spin operators from the $\tilde{\mathcal{J}}$, as we explain below.

Next, we introduce for each orbital extended-basis $\Lambda$ matrix the operators:
\begin{align}
\mathcal{O}_{ss'}(\Lambda) &= \Psi^{\dag}_s \Lambda \Psi_{s'}, \\
\tilde{\mathcal{O}}_{ss'}(\Lambda) &= \begin{pmatrix} d^{\dag}_s & p^{\dag}_{s'} \end{pmatrix} \Lambda \begin{pmatrix} d_s \\ p_{s'} \end{pmatrix},
\end{align}
where $p_s = (p_{1s}, p_{2s}, p_{3s}, p_{4s})^{\intercal}$.
A straightforward comparison to the matrices of Tab.~\ref{tab:orbital-Lambda-mats} shows that:
\begin{align}
\begin{pmatrix}
\mathcal{O}_{ss'}\big(\Lambda^{A_{1g}^{-}}_{1}\big) \\[2pt]
\mathcal{O}_{ss'}\big(\Lambda^{B_{1g}^{-}}_{1}\big) \\[2pt]
\mathcal{O}_{ss'}\big(\Lambda^{E_{u}^{-}}_{1,x}\big) \\[2pt]
\mathcal{O}_{ss'}\big(\Lambda^{E_{u}^{-}}_{1,y}\big)
\end{pmatrix} &= \mathcal{X} \begin{pmatrix}
\mathcal{J}_{1;ss'} \\
\mathcal{J}_{2;ss'} \\
\mathcal{J}_{3;ss'} \\
\mathcal{J}_{4;ss'}
\end{pmatrix}, &
\begin{pmatrix}
\tilde{\mathcal{O}}_{ss'}\big(\Lambda^{A_{1g}^{-}}_{1}\big) \\[2pt]
\tilde{\mathcal{O}}_{ss'}\big(\Lambda^{B_{1g}^{-}}_{1}\big) \\[2pt]
\tilde{\mathcal{O}}_{ss'}\big(\Lambda^{E_{u}^{-}}_{1,x}\big) \\[2pt]
\tilde{\mathcal{O}}_{ss'}\big(\Lambda^{E_{u}^{-}}_{1,y}\big)
\end{pmatrix} &= \mathcal{X} \begin{pmatrix}
\tilde{\mathcal{J}}_{1;ss'} \\
\tilde{\mathcal{J}}_{2;ss'} \\
\tilde{\mathcal{J}}_{3;ss'} \\
\tilde{\mathcal{J}}_{4;ss'}
\end{pmatrix},
\end{align}
and
\begin{align}
\begin{pmatrix}
\mathcal{O}_{ss'}\big(\Lambda^{A_{1g}^{+}}_{3}\big) \\[2pt]
\mathcal{O}_{ss'}\big(\Lambda^{B_{1g}^{+}}_{1}\big) \\[2pt]
\mathcal{O}_{ss'}\big(\Lambda^{E_{u}^{+}}_{1,x}\big) \\[2pt]
\mathcal{O}_{ss'}\big(\Lambda^{E_{u}^{+}}_{1,y}\big)
\end{pmatrix} &= \mathcal{X} \begin{pmatrix}
\mathcal{R}_{1;ss'} \\
\mathcal{R}_{2;ss'} \\
\mathcal{R}_{3;ss'} \\
\mathcal{R}_{4;ss'}
\end{pmatrix},
\end{align}
where
\begin{align}
\mathcal{X} &= \begin{pmatrix}
\tfrac{1}{2} & - \tfrac{1}{2} & - \tfrac{1}{2} & \tfrac{1}{2} \\[2pt]
\tfrac{1}{2} & \tfrac{1}{2} & - \tfrac{1}{2} & - \tfrac{1}{2} \\[2pt]
\tfrac{1}{\sqrt{2}} & 0 & \tfrac{1}{\sqrt{2}} & 0 \\[2pt]
0 & - \tfrac{1}{\sqrt{2}} & 0 & - \tfrac{1}{\sqrt{2}}
\end{pmatrix}.
\end{align}
Above, notice that the matrices that enter the columns on the left-hand side are those listed in $\mathcal{L}_{pd}^{-}$ and $\mathcal{L}_{pd}^{+}$.
Since $\mathcal{X}$ is orthogonal, $\mathcal{X}^{\intercal} \mathcal{X} = \one$, it follows that:
\begin{align}
n_d \sum_{\ell=1}^{4} n_{p\ell} &= - \frac{1}{2} \sum_{\Lambda \in \mathcal{L}_{pd}^{-}} \sum_{ss'} \tilde{\mathcal{O}}_{ss'}(\Lambda) \tilde{\mathcal{O}}_{ss'}(\Lambda) \\
&= - \frac{1}{2} \sum_{\Lambda \in \mathcal{L}_{pd}^{-}} \mathcal{O}_{ss'}(\Lambda) \mathcal{O}_{ss'}(\Lambda) - \frac{1}{4} \sum_{\Lambda \in \mathcal{L}_{pd}^{-}} \mathcal{O}_{ss}(\Lambda) \mathcal{O}_{s's'}(\Lambda) + \frac{1}{4} \sum_{\Lambda \in \mathcal{L}_{pd}^{+}} \mathcal{O}_{ss}(\Lambda) \mathcal{O}_{s's'}(\Lambda). \notag
\end{align}
Finally, we replace spins with Pauli matrices in the latter equation using
\begin{align}
\mathcal{O}(\Lambda \Pauli_A) &= \sum_{ss'} \mathcal{O}_{ss'}(\Lambda) (\Pauli_A)_{ss'}, \\
\mathcal{O}_{ss'}(\Lambda) &= \frac{1}{2} \sum_{A=0}^{3} \mathcal{O}(\Lambda \Pauli_A) (\Pauli_A)_{s's}
\end{align}
to obtain Eq.~\eqref{eq:Hubbard-Vpd-decomp-again}.
This completes the proof.

Alternatively, we could have also defined
\begin{align}
\tilde{\mathcal{O}}_A(\Lambda) &\defeq \sum_{ss'} \tilde{\mathcal{O}}_{ss'}(\Lambda) (\Pauli_A)_{ss'}, \\
\tilde{\mathcal{O}}_{ss'}(\Lambda) &= \frac{1}{2} \sum_{A=0}^{3} \tilde{\mathcal{O}}_A(\Lambda) (\Pauli_A)_{s's}
\end{align}
to obtain
\begin{align}
n_d \sum_{\ell=1}^{4} n_{p\ell} &= - \frac{1}{4} \sum_{\Lambda \in \mathcal{L}_{pd}^{-}} \sum_{A=0}^{3} \big[\tilde{\mathcal{O}}_A(\Lambda)\big]^{\dag} \big[\tilde{\mathcal{O}}_A(\Lambda)\big].
\end{align}
Although this equation looks simpler, one should keep in mind that $\tilde{\mathcal{O}}_A(\Lambda) \neq \Psi^{\dag} \Lambda \Pauli_A \Psi$ in general.
In the current case of $d$--$p$ orbital coupling, one finds that
\begin{align}
\begin{aligned}
\tilde{\mathcal{O}}_0(\Lambda^{-}) &= \mathcal{O}(\Lambda^{-}), &\hspace{50pt}
\tilde{\mathcal{O}}_1(\Lambda^{-}) &= \mathcal{O}(\Lambda^{-} \Pauli_1), \\
\tilde{\mathcal{O}}_2(\Lambda^{-}) &= - \iu \, \mathcal{O}(\Lambda^{+} \Pauli_2), &\hspace{50pt}
\tilde{\mathcal{O}}_3(\Lambda^{-}) &= \mathcal{O}(\Lambda^{-} \Pauli_3),
\end{aligned}
\end{align}
where $\Lambda^{-}$ and $\Lambda^{+}$ are the first, second, third, or fourth matrices of $\mathcal{L}_{pd}^{-}$ and $\mathcal{L}_{pd}^{+}$, respectively.
By exploiting the Fierz identity [Eq.~\eqref{eq:Fierz-Vpd-Vpp}]
\begin{align}
\big[\mathcal{O}(\Lambda^{+} \Pauli_2)\big]^2 &= \big[\mathcal{O}(\Lambda^{-})\big]^2 + \big[\mathcal{O}(\Lambda^{-} \Pauli_2)\big]^2 - \big[\mathcal{O}(\Lambda^{+})\big]^2,
\end{align}
one recovers Eq.~\eqref{eq:Hubbard-Vpd-decomp-again}.

\subsubsection[Unappreciated aspects of the coupling of loop currents to \\ electrons]{Unappreciated aspects of the coupling of loop currents to electrons} \label{sec:ASV-fermion-coupling}
Before we discuss the shortcomings of ASV's theory~\cite{Aji2010}, we first review it.
Let us call $\Phi_{p_{x'},i}$, $\Phi_{p_{y'},i}$, and $\Phi_{g,i}$ the order parameters which, through a Hubbard-Stratonovich transformation, correspond to the $\tilde{L}_{i,x'}$, $\tilde{L}_{i,y'}$, and $\tilde{L}_{i,z}$ flux operators of ASV~\cite{Aji2010}, respectively.
In ASV's notation, $\Phi_{p_{x'},i}$, $\Phi_{p_{y'},i}$, and $\Phi_{g,i}$ would be called ${L}_{i,x'}$, ${L}_{i,y'}$, and ${L}_{i,z}$, in that order.
The pair $\vb{\Phi}_{p,i} = (\Phi_{p_{x'},i}|\Phi_{p_{y'},i})$ transforms according to the $E_u^{-}$ irrep of the underlying tetragonal $D_{4h}$ point group, while $\Phi_{g,i}$ transforms according to the $A_{2g}^{-}$ irrep of $D_{4h}$ (the irrep superscripts are TR signs).

\begin{figure}[t]
\centering
\includegraphics[width=0.70\textwidth]{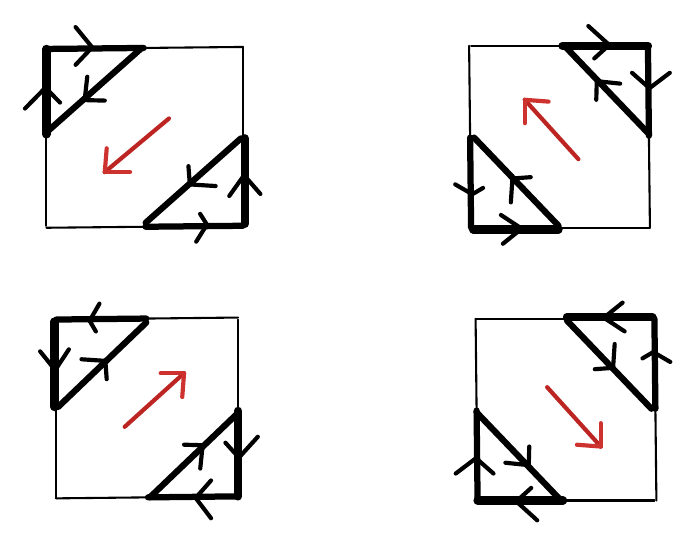}
\captionbelow[The four domains of the loop-current phase in the theory of Aji, Shekhter, and Varma~\cite{Aji2010}.]{\textbf{The four domains of the loop-current phase in the theory of Aji, Shekhter, and Varma}~\cite{Aji2010}.
The domains can be specified by the four orientations of a $p$-wave order parameter vector $\vb{\Phi}_p$ shown in red.
Compare with Fig.~\ref{fig:p-wave-four-LC-patterns}(b).
Reprinted with permission from Ref.~\cite{Aji2010}. Copyright (2010) by the American Physical Society.}
\label{fig:ASV-domains}
\end{figure}

Within the theory of ASV~\cite{Aji2010}, it is the $p$-wave LC order parameter $\vb{\Phi}_p$ that condenses.
The resulting ordered phase is the so-called $\Theta_{\text{II}}$ LC phase which was studied earlier by Varma and collaborators~\cite{Simon2002, Simon2003, Varma2006}.
However, for superconductivity the regime of interest is where $\vb{\Phi}_p$ still fluctuates~\cite{Aji2010}.
Due to in-plane tetragonal anisotropy, two easy in-plane axes are expected and, according to ASV~\cite{Aji2010}, they are oriented along the $x' = (x+y)/\sqrt{2}$ and $y' = (x-y)/\sqrt{2}$ diagonals (Fig.~\ref{fig:ASV-domains}).
Furthermore, a fluctuating $g$-wave LC order parameter $\Phi_{g}$ is also present in the theory.
According to ASV~\cite{Aji2010}, $\vb{\Phi}_p$ and $\Phi_{g}$ are conjugate momenta, with the latter acting as a generator of rotations for the former.
Because of the $A_{1g}^{+}$ contributions to the logarithm of the orbital rotation matrix $O(C_{4z})$ from Tab.~\ref{tab:CuO2-model-D4h-generators},
\begin{align}
- \iu \frac{4}{\pi} \log O(C_{4z}) &= \mleft(2 \sqrt{2} \Lambda^{A_{1g}^{+}}_1 + \sqrt{2} \Lambda^{A_{1g}^{+}}_2 + 2 \Lambda^{A_{1g}^{+}}_4 - \sqrt{2} \Lambda^{A_{1g}^{+}}_5\mright) + 2 \Lambda^{A_{2g}^{-}}_1,
\end{align}
it cannot be said that $\Phi_{g} \sim \Lambda^{A_{2g}^{-}}_1$ by itself generates fermionic rotations
However, for the $p$-wave LC fermionic bilinears one could say so based on the spin-like commutator relations:
\begin{align}
\big[\Lambda^{E_{u}^{-}}_{1,x}, \Lambda^{E_{u}^{-}}_{1,y}\big] &= \iu \Lambda^{A_{2g}^{-}}_1, &
\big[\Lambda^{E_{u}^{-}}_{1,y}, \Lambda^{A_{2g}^{-}}_1\big] &= \iu \Lambda^{E_{u}^{-}}_{1,x}, &
\big[\Lambda^{A_{2g}^{-}}_1, \Lambda^{E_{u}^{-}}_{1,x}\big] &= \iu \Lambda^{E_{u}^{-}}_{1,y}.
\end{align}
For the other $E_u^{-}$ matrices of Tab.~\ref{tab:orbital-Lambda-mats}, the commutator relations are not so neat:
\begin{align}
\begin{aligned}
\big[\Lambda^{E_{u}^{-}}_{2,x}, \Lambda^{E_{u}^{-}}_{2,y}\big] &= 0, &\hspace{60pt}
\big[\Lambda^{E_{u}^{-}}_{3,x}, \Lambda^{E_{u}^{-}}_{3,y}\big] &= 0, \\
\big[\Lambda^{A_{2g}^{-}}_1, \Lambda^{E_{u}^{-}}_{2,x}\big] &= - \iu \Lambda^{E_{u}^{-}}_{3,y}, &\hspace{60pt}
\big[\Lambda^{A_{2g}^{-}}_1, \Lambda^{E_{u}^{-}}_{2,y}\big] &= - \iu \Lambda^{E_{u}^{-}}_{3,x}, \\
\big[\Lambda^{A_{2g}^{-}}_1, \Lambda^{E_{u}^{-}}_{3,x}\big] &= \iu \Lambda^{E_{u}^{-}}_{2,y}, &\hspace{60pt}
\big[\Lambda^{A_{2g}^{-}}_1, \Lambda^{E_{u}^{-}}_{3,y}\big] &= \iu \Lambda^{E_{u}^{-}}_{2,x}.
\end{aligned}
\end{align}
In any case, from these considerations ASV have come to the conclusion that the effective model of their LC fluctuations is the quantum rotor model~\cite[Eq.~(12)]{Aji2010}:
\begin{align}
\Haml &= \sum_i \frac{\abs{L_{\theta_i}}^2}{2 I} + J \sum_{\langle ij \rangle} \cos(\theta_i - \theta_j), \label{eq:ASV-rotor}
\end{align}
where $\Phi_{g,i}$ is identified with $L_{\theta_i} = \iu \partial_{\theta_i}$ and the $\theta_i$ angles specify the in-plane directions of $\vb{\Phi}_{p,i}$~\cite{Aji2010}.
Two notable features are that the susceptibility of $\Phi_{g,i}$ has weak momentum dependence and that the amplitude fluctuations of $\vb{\Phi}_{p,i}$ are not included in the model, $\vb{\Phi}_{p,i} \vdot \vb{\Phi}_{p,j} \to \cos(\theta_i - \theta_j)$.
Let us emphasize that this effective Hamiltonian has not been rigorously derived (see previous section), but rather constitutes an educated guess, assuming loop currents as the ordering channel.

To asses the Cooper pairing instability, next ASV~\cite{Aji2010} couple the fluctuating loop currents to fermions.
They only consider the coupling of the $g$-wave LC order parameter $\Phi_{g,i}$, however.
From a Hubbard-Stratonovich transformation of the $- \tfrac{1}{16} V_{pd} \abs{\tilde{L}_{i,z}}^2$ term in Eq.~\eqref{eq:ASV-H_nn}, they obtained the $\abs{\Phi_{g,i}}^2 / (2 I)$ term of Eq.~\eqref{eq:ASV-rotor}, while the remaining $\Phi_{g,i} \tilde{L}_{i,z}$ term gives the desired coupling to fermions (Eq.~(16) in Ref.~\cite{Aji2010}):
\begin{align}
\Haml_c &= \frac{V_{pd}}{16} \sum_i \Phi_{g,i} \tilde{L}_{i,z} + \Hc \label{eq:ASV-gLC-fermion-coupling}
\end{align}
By Fourier transforming this expression and projecting it onto the conduction band states given by the approximate eigenvector~\cite[Eq.~(17)]{Aji2010}
\begin{align}
\tilde{u}_{\vb{k} 3} &= \frac{1}{\sqrt{2}} \begin{pmatrix}
1 \\[2pt]
\displaystyle - \iu \frac{s_x(\vb{k})}{s_{xy}(\vb{k})} \\[10pt]
\displaystyle - \iu \frac{s_y(\vb{k})}{s_{xy}(\vb{k})}
\end{pmatrix}, \label{eq:ASV-absence-orb-ord-eig}
\end{align}
where
\begin{align}
s_x(\vb{k}) &\equiv \sin\tfrac{1}{2} k_x, &
s_y(\vb{k}) &\equiv \sin\tfrac{1}{2} k_y, &
s_{xy}(\vb{k}) &\equiv \sqrt{s_x^2(\vb{k}) + s_y^2(\vb{k})},
\end{align}
ASV obtained the LC-fermion coupling~\cite[Eq.~(18)]{Aji2010}:
\begin{align}
\Haml_c &= - \frac{V_{pd}}{32} \sum_{\vb{k} \vb{p}} \Phi_{g,\vb{k}-\vb{p}} \, c_{\vb{k}}^{\dag} \tilde{f}(\vb{k}, \vb{p}) c_{\vb{p}},
\end{align}
where $c_{\vb{k}} = \tilde{u}_{\vb{k} 3}^{\dag} \tilde{\psi}_{\vb{k}}$ are the conduction band annihilation operators.
For the $g$-wave LC-fermion coupling matrix ASV find~\cite[Eq.~(19)]{Aji2010}:\footnote{I have replaced $s_{xy}^{-1}(\vb{k}) + s_{xy}^{-1}(\vb{p})$ with $s_{xy}^{-1}(\vb{k}) s_{xy}^{-1}(\vb{p})$ in Eq.~(19) of Ref.~\cite{Aji2010} since this is likely a typo.}
\begin{align}
\tilde{f}(\vb{k}, \vb{p}) &= - \iu s_{xy}^{-1}(\vb{k}) s_{xy}^{-1}(\vb{p}) \mleft(\sin\frac{k_x}{2} \sin\frac{p_y}{2} - \sin\frac{k_y}{2} \sin\frac{p_x}{2}\mright). \label{eq:ASV-tilde-f}
\end{align}
In the continuum this simplifies to $\tilde{f}(\vb{p},\vb{k}) \propto \vu{e}_z \vdot (\vb{k} \vcross \vb{p})$, which they then proceeded to analyze by integrating out $\Phi_{g \vb{q}}$ and solving the BCS gap equation~\cite{Aji2010}.
ASV find that the leading pairing states have $d_{x^2-y^2}$ and $d_{xy}$ symmetry~\cite{Aji2010}, as can be seen from ($k_x \to \cos \theta_k$, $k_y \to \sin \theta_k$):
\begin{align}
\abs{\iu \, \vu{e}_z \vdot (\vb{k} \vcross \vb{p})}^2 &= \sin^2(\theta_k - \theta_p) = \frac{1}{2} - \frac{\cos 2 \theta_k}{\sqrt{2}} \frac{\cos 2 \theta_p}{\sqrt{2}} - \frac{\sin 2 \theta_k}{\sqrt{2}} \frac{\sin 2 \theta_p}{\sqrt{2}}.
\end{align}
Note that $\cos 2 \theta_k = k_x^2 - k_y^2 \in B_{1g}$ and $\sin 2 \theta_k = 2 k_x k_y \in B_{2g}$.
For a circular Fermi surface in the continuum, these two pairing states are exactly degenerate (see above), as follows from the fact that \SI{45}{\degree} rotations around $\vu{e}_z$ map one into the other.
Finally, ASV conclude that~\cite{Aji2010}: ``For the actual Fermi surface of the cuprates in which the Fermi velocity is largest in the $(1,1)$ directions and the least in the $(1,0)$ or the \ce{Cu-O} bond directions, $d_{x^2-y^2}$ pairing is favored because in that case the maximum gap is in directions where the density of states is largest.''

With ASV's theory outlined, we may now discuss aspects of it that have not been sufficiently carefully treated by ASV~\cite{Aji2010}.
Let us start by discussing the coupling of $g$-wave LCs and demonstrating that Eq.~\eqref{eq:ASV-tilde-f} is incorrect for large momenta.
As we observed in Sec.~\ref{sec:ASV-conventions}, ASV use a different gauge than us.
Their Hamiltonian is given by
\begin{align}
\tilde{H}_{\vb{k}} &= \tilde{\mathcal{B}}_{\vb{k}} H_{\vb{k}} \tilde{\mathcal{B}}_{\vb{k}}^{\dag} = \begin{pmatrix}
\epsilon_{d} - \upmu & 2 \iu t_{pd} \sin(k_x/2) & 2 \iu t_{pd} \sin(k_y/2) \\
& \epsilon_{p} + 2 t_{pp}' \cos k_x - \upmu & 4 t_{pp} \sin(k_x/2) \sin(k_y/2) \\
\cc & & \epsilon_{p} + 2 t_{pp}' \cos k_y - \upmu
\end{pmatrix}, \label{eq:3band-Haml-ASV}
\end{align}
where $H_{\vb{k}}$ is the three-band Hamiltonian of Eq.~\eqref{eq:3band-Haml-again} and $\tilde{\mathcal{B}}_{\vb{k}}$ is the gauge transition matrix of Eq.~\eqref{eq:ASV-Bk-matrix}.
This Hamiltonian cannot be diagonalized in closed form.
However, if we set $t_{pp} = t_{pp}' = 0$, for the conduction band we obtain:
\begin{align}
\tilde{u}_{\vb{k} 3} &= \frac{1}{\sqrt{2}} \frac{1}{\sqrt{\mleft(\delta + S_{xy}(\vb{k})\mright) S_{xy}(\vb{k})}} \begin{pmatrix}
\delta + S_{xy}(\vb{k}) \\
\displaystyle - \iu s_x(\vb{k}) \\
\displaystyle - \iu s_y(\vb{k})
\end{pmatrix}, \label{eq:tpp-tpp'-0-closed-form-1} \\
\varepsilon_{\vb{k} 3} &= \tfrac{1}{2} (\epsilon_d + \epsilon_p) + 2 t_{pd} S_{xy}(\vb{k}) - \upmu, \label{eq:tpp-tpp'-0-closed-form-2}
\end{align}
where
\begin{align}
S_{xy}(\vb{k}) &= \sqrt{s_{xy}^2(\vb{k}) + \delta^2}, &
\delta &= \frac{\epsilon_d - \epsilon_p}{4 t_{pd}}.
\end{align}
If we further set $\delta = 0$, we recover Eq.~\eqref{eq:ASV-absence-orb-ord-eig} and what was meant by the cryptic ``absence of orbital order'' of ASV~\cite{Aji2010}.
This agreement confirms the gauge difference we claimed in Sec.~\ref{sec:ASV-conventions}.

Using Eq.~\eqref{eq:ASV-Kk-matrix}, the appropriate $g$-wave coupling matrix is now easily found to be:
\begin{align}
\tilde{f}(\vb{k}, \vb{p}) &= \tilde{u}_{\vb{k} 3}^{\dag} \tilde{\mathcal{K}}_{\vb{k}}^{\dag} \Lambda^{A_{2g}^{-}}_1 \tilde{\mathcal{K}}_{\vb{p}} \tilde{u}_{\vb{p} 3} = - \iu \frac{\sin k_x \sin p_y - \sin k_y \sin p_x}{\sqrt{\mleft(\delta + S_{xy}(\vb{k})\mright) S_{xy}(\vb{k})} \sqrt{\mleft(\delta + S_{xy}(\vb{p})\mright) S_{xy}(\vb{p})}}. \label{eq:ASV-tilde-f2}
\end{align}
At small momenta, this reduces to the $\vu{e}_z \vdot (\vb{k} \vcross \vb{p})$ from earlier.
Indeed, the continuum coupling can be guessed purely from symmetries, as ASV point out~\cite{Aji2010}.\footnote{The lowest-order coupling follows from the transformation rules $\tilde{f}\mleft(R(g)\vb{k}, R(g)\vb{p}\mright) = \RepM^{A_{2g}}(g) \tilde{f}(\vb{k}, \vb{p})$ and $\tilde{f}^{*}(-\vb{k}, -\vb{p}) = - \tilde{f}(\vb{k},\vb{p})$, supplemented by the reality condition $\tilde{f}^{*}(\vb{k}, \vb{p}) = \tilde{f}(\vb{p}, \vb{k})$ which gives the $\iu$ in Eq.~\eqref{eq:ASV-tilde-f2}.}
The continuum model is only accurate near the $\Gamma$ point, i.e., when the Fermi surface forms a small electron pocket at very large overdoping ($p \to 1$).
Using perturbation theory on the $\tilde{H}_{\vb{k}}$ of Eq.~\eqref{eq:3band-Haml-ASV} near $\vb{k}_{\Gamma} = \vb{0}$,
\begin{align}
\tilde{u}_{\vb{k} 3} &= \begin{pmatrix}
1 \\ 0 \\ 0
\end{pmatrix} + \frac{t_{pd}}{\epsilon_d - \epsilon_p - 2 t_{pp}'} \begin{pmatrix}
0 \\ - \iu k_x \\ - \iu k_y
\end{pmatrix} + \cdots \, ,
\end{align}
one may confirm that the continuum coupling has the same form for generic parameters:
\begin{align}
\tilde{f}(\vb{k}, \vb{p}) &= - \iu \frac{2 t_{pd}^2}{\mleft(\epsilon_d - \epsilon_p - 2  t'_{pp}\mright)^2} \vu{e}_z \vdot (\vb{k} \vcross \vb{p}) + \cdots \, .
\end{align}

More importantly, for large momenta $\sin k_{x,y}$ appears instead of $\sin\tfrac{1}{2} k_{x,y}$ in Eq.~\eqref{eq:ASV-tilde-f2}, which makes all the difference at the Van Hove points $\vb{k}_{M_x} = (\pi, 0)$ and $\vb{k}_{M_y} = (0, \pi)$.
The correct $g$-wave coupling therefore exactly vanishes at the Van Hove points, as we proved in general in Sec.~\ref{sec:pairing-cuprate-actual-analysis-VH}.
Although setting $t_{pp} = t_{pp}' = \epsilon_d - \epsilon_p = 0$ is clearly aphysical, which is what ASV did to get Eq.~\eqref{eq:ASV-absence-orb-ord-eig}, one may verify that the conduction band still transforms according to the correct irreps at the high-symmetry points, which explains why we still observe the effect of Sec.~\ref{sec:pairing-cuprate-actual-analysis-VH}.
Because of this, the exact degeneracy between $d_{x^2-y^2}$ and $d_{xy}$ pairing states is lifted in favor of $d_{xy}$ symmetry.
This is precisely what we found in our numerics, shown in Fig.~\ref{fig:cuprate-A2g-results}.
In our numerics, we recover the degeneracy between $d_{x^2-y^2}$ and $d_{xy}$ pairing only in the $p \to 1$ limit where the Fermi surface is a small circle surrounding the $\Gamma$ point.
In light of their effective rotor model [Eq.~\eqref{eq:ASV-rotor}], the $g$-wave susceptibility is not strongly peaked at $\vb{q} = \vb{0}$ and ASV's theory corresponds to $r \sim 1$ in our formalism.
Why should the SC dome be centered at the $p$-wave LC QCP is not entirely clear in ASV's theory~\cite{Aji2010}, nor has later work given a crisp answer to this question~\cite{Varma2012, Bok2016, Varma2016, Varma2020}.
Any potential softening of the $g$-wave LCs at $\vb{q} = \vb{0}$ cannot be the answer, as follows from the results of Sec.~\ref{sec:gen-sys-LC-analysis} (Fig.~\ref{fig:QCP-general-results}).
Including spin-orbit coupling does not help either, given that the corresponding subsidiary spin-magnetic fluctuations favor $p$-wave pairing (Fig.~\ref{fig:cuprate-SOC-results}).
In conclusion, intra-unit-cell $g$-wave loop currents cannot explain the $d_{x^2-y^2}$ superconductivity of cuprates.

But there is another difficulty with ASV's theory~\cite{Aji2010}: the direct coupling of the main $p$-wave LC order parameter $\vb{\Phi}_p$ to fermions has not been included.
This is quite surprising, since the direct coupling of the main order parameter to fermions is what normally anyone would first write down and study.
This coupling is not even commented on in Ref.~\cite{Aji2010}, or later work~\cite{Varma2012, Bok2016, Varma2016, Varma2020}, even though it was discussed earlier~\cite{Varma2006}, and one can only speculate what explains this lacuna.
However interesting the coupling of the conjugate momentum -- the $g$-wave LCs -- may be, the main order parameter itself will always couple directly to electrons, if allowed by symmetry.
In our analysis, we found a whole one-parameter family of possible direct, local couplings of $\vb{\Phi}_p$ to electrons that are consistent with Bloch's theorem (Sec.~\ref{sec:Bloch-Bloch-constr}).
Indeed, just like for $g$-wave LCs [Eq.~\eqref{eq:ASV-gLC-fermion-coupling}], the Hubbard-Stratonovich transformation employed by ASV~\cite{Aji2010}, if consistently applied to all LC operators appearing in their Hubbard interaction decomposition [Eq.~\eqref{eq:ASV-H_nn}], yields a term $\propto \sum_i \Phi_{p_{x'},i} \tilde{L}_{i,x'} + \Phi_{p_{y'},i} \tilde{L}_{i,y'}$.
In the language of the quantum rotor problem [Eq.~\eqref{eq:ASV-rotor}], this represents a coupling of the fermions to the direction vector $(\cos \theta, \sin \theta)$.
These couplings are relevant operators in the renormalization-group sense and the effective low-energy theory of $p$-wave LC fluctuations will therefore generically include them.
Most importantly, the fact that $\vb{\Phi}_p$ is even under $P \TRop$ allows it to directly couple to fermions at forward scattering ($\vb{q} = \vb{0}$).
As we showed in Sec.~\ref{sec:IUC-order-results}, this has the dramatic consequence that odd-parity IUC LC fluctuations, uniquely among all IUC orders (Tab.~\ref{tab:orbital-spin-IUC-orders}), act as parametrically strong pair breakers near their quantum-critical point.
Even if the coupling constant of $\vb{\Phi}_p$ is substantially smaller than the one of $\Phi_g$, the $\vb{q} = \vb{0}$ divergence of the susceptibility will render the pair-breaking interaction mediated by $\vb{\Phi}_p$ stronger than the attractive interaction mediated by $\Phi_g$ near the IUC $p$-wave LC QCP.
It is worth emphasizing that this result is robust to the precise details of the quantum-critical LC sector.
As long as the $\vb{\Phi}_p$ susceptibility peaks at $\vb{q} = \vb{0}$ with critical exponents that are in-line with theoretical bounds, suppression of pairing will take place near the QCP (Sec.~\ref{sec:IUC-order-results}).
At best, away from the QCP $p$-wave LC fluctuations can give rise to extended $s$-wave superconductivity (Fig.~\ref{fig:cuprate-Eu-results}).
If the pseudogap phase is to be interpreted as an intra-unit-cell loop-current order, as argued by Varma~\cite{Varma2020, Varma2016}, the experimental evidence unambiguously points towards $E_u^{-}$ or $(p_x|p_y)$ symmetry (Sec.~\ref{sec:exp-sym-break-pseudogap}, Fig.~\ref{fig:Varma-evidence}).
The strong pair-breaking of $p$-wave loop currents thus poses a serious challenge to Varma's theory~\cite{Varma2020, Varma2016}.

Are there ways our results could be circumvented?
The principal idea behind our analysis is to, in a phenomenological spirit, assume an IUC LC QCP and then to explore its pairing instabilities within a weak-coupling treatment coming from the far-overdoped regime, where complications relating to Mott physics, the pseudogap, and competing orders can be neglected~\cite{Keimer2015, Proust2019, Kramer2019, Lee-Hone2020}.
Strong-coupling physics will not change the appearance of a $\vb{q} = \vb{0}$ peak in the susceptibility, nor is it likely to change the pairing symmetry.
If there is no pairing instability at weak coupling to begin with, the experience~\cite{She2009, Moon2010, Metlitski2015, YuxuanWang2016-inv, YuxuanWang2016-QCP, WangRaghu2017, ChubukovI_2020, ChubukovII_2020} of all other quantum-critical modes suggests that nothing interesting will happen in the Cooper channel near the QCP.
For comparison, in the case of nematic~\cite{Yamase2013, Lederer2015, Lederer2017, Klein2018}, ferroelectric~\cite{Kozii2015, KoziiBiRuhman2019, Klein2023}, and ferromagnetic~\cite{Millis2001, Chubukov2003} quantum-critical IUC fluctuations a coherent picture of a SC dome surrounding the QCP emerges, whether one studies it numerically~\cite{Berg2019, Lederer2017, Wang2017, Xu2017, Xu2020} or analytically using weak-coupling or other methods.

If the bare ingredients of Varma's theory -- $p$-wave and $g$-wave LCs -- cannot reproduce cuprate superconductivity, it is difficult to see how would including additional aspects of cuprates, like Mott physics or competing orders, help out.
For spin-orbit coupling, we have established that it is of no assistance (Sec.~\ref{sec:cup-spin-wave-res-discus}, Fig.~\ref{fig:cuprate-SOC-results}).
If the symmetry of the conduction band were different at the Van Hove points, this would help because $g$-wave LCs would then efficiently couple Van Hove points (Sec.~\ref{sec:pairing-cuprate-actual-analysis-VH-details}), yet such a proposition strongly departs from the well-established understanding of the \ce{CuO2} band structure~\cite{Pickett1989, Dagotto1994, Varma1987, Emery1987, Emery1988, Littlewood1989, Gaididei1988, Scalettar1991, Andersen1995, Andersen2001, Pavarini2001, Kent2008, Photopoulos2019, Jiang2020}, reviewed in Sec.~\ref{sec:cuprate-3band-model}.
In particular, for this to work, it is not enough for the interactions to merely redistribution the weights among the orbitals: the interactions would need to fundamentally alter the symmetry of the conduction band at the Van Hove points, as we demonstrated in Sec.~\ref{sec:pairing-cuprate-actual-analysis-VH}.
A modest improvement over the theory of Aji, Shekhter, and Varma~\cite{Aji2010} can be made by replacing $g$-wave LCs with the $d$-wave LCs of Sec.~\ref{sec:cup-d-wave-res-discus}.
These $B_{1g}^{-}$ LCs were previously discussed by Varma et al.~\cite{Simon2002, Simon2003, Varma2006} under the name $\Theta_{\text{I}}$ LCs.
Even though ASV~\cite{Aji2010} found the $\tilde{L}_{i,x'^2-y'^2} \in B_{1g}^{-}$ LC operator in their decomposition [Eq.~\eqref{eq:ASV-H_nn}], this term was subsequently neglected in their analysis.
In the continuum, $d$-wave LCs couple through a pairing form factor $\tilde{f}(\vb{k}, \vb{p}) \propto \iu (k_x^2 - k_y^2 - p_x^2 + p_y^2)$ which robustly favors $d_{x^2-y^2}$ pairing, as can be seen from ($k_x \to \cos \theta_k$, $k_y \to \sin \theta_k$):
\begin{align}
\abs{\iu (k_x^2-k_y^2-p_x^2+p_y^2)}^2 &= \begin{pmatrix}
1 & \frac{\cos 4 \theta_k}{\sqrt{2}}
\end{pmatrix} \begin{pmatrix}
1 & \frac{1}{\sqrt{2}} \\[2pt]
\frac{1}{\sqrt{2}} & 0
\end{pmatrix} \begin{pmatrix}
1 \\
\frac{\cos 4 \theta_p}{\sqrt{2}}
\end{pmatrix} - 4 \frac{\cos 2 \theta_k}{\sqrt{2}} \frac{\cos 2 \theta_p}{\sqrt{2}}.
\end{align}
Diagonalizing the $2 \times 2$ matrix from above gives an extended $s$-wave instability with the eigenvalue $\tfrac{1}{2} (\sqrt{3} - 1) = 0.37$, which is much smaller than the eigenvalue $4$ characterizing the $\cos 2 \theta_k = k_x^2-k_y^2 \in B_{1g}$ pairing channel.
The more realistic numerical calculation performed in Sec.~\ref{sec:cup-d-wave-res-discus} confirms robust $d_{x^2-y^2}$ pairing (Fig.~\ref{fig:cuprate-B1g-results}).
Nonetheless, even with $d$-wave LCs, the theory suffers from the pair-breaking of $p$-wave LCs.
It is curious that ASV have not included $d$-wave LC fluctuations in their analysis~\cite{Aji2010}, even though they previously studied them~\cite{Simon2002, Simon2003, Varma2006}.

In conclusion, intra-unit-cell $p$-wave loop currents strongly suppress superconductivity near their quantum-critical point.
Their conjugate momentum -- $g$-wave loop currents -- robustly favor $d_{xy}$ symmetry.
Both conclusions follow from previously unappreciated aspects of the coupling of loop currents to fermions.
Although we focused on the original work by Aji, Shekhter, and Varma~\cite{Aji2010}, these two issues have not been addressed in later work by Varma et al.~\cite{Varma2012, Bok2016, Varma2016, Varma2020}.
It remains to be seen whether a theory based on intra-unit-cell loop currents can overcome these two obstacles and credibly explain cuprate superconductivity.

\chapter{Unconventional superconductivity from electronic dipole fluctuations}
\label{chap:el_dip_SC}

The fluctuations of electric dipole moments of electrons are fundamental to understanding a wide variety of systems, ranging from atomic gases and molecules interacting through van der Waals interactions~\cite{Margenau1939, Israelachvili1974, Langbein1974, Parsegian2005, Kaplan2006, Stone2013, Hermann2017}, to small metallic clusters and their cohesive energies~\cite{Garcia1991}, up to solids with sizable contributions to the binding energy and optical conductivity coming from interband dipole excitations~\cite{Hermann2017, Andersson1998, Dion2004, Grimme2011, Klimes2011, Klimes2012, Berland2015}.
From a microscopic point of view, all these effects are due to processes involving electromagnetic interactions among virtual or real excitations that have electric dipole moments.
The above examples usually involve high-energy processes, at least when compared to typical energy scales of collective modes in correlated electron materials.
For electrons near the Fermi level, on the other hand, the Coulomb interactions among them are crucial to facilitating phenomena such as Mott insulation~\cite{Mott1949, Mott1982}, itinerant magnetism~\cite{Moriya1979, Shimizu1981}, and unconventional superconductivity~\cite{Maiti2013}.
This raises two questions.
First, can one sensibly generalize the concept of electronic dipole excitations to states residing on or near the Fermi surface?
And second, can their Coulomb interactions give rise to non-trivial electronic phases, such as superconductivity?

In this chapter, we address both of these questions.
We develop the theory of dipole excitations of electronic states near the Fermi surface (Sec.~\ref{sec:el-dip-sc-dipole-theory}) and we use it to show that the dipolar parts of the Coulomb interaction can result in unconventional superconductivity (Sec.~\ref{sec:el-dip-sc-pairing}).
In addition, we study Dirac metals (Secs.~\ref{sec:el-dip-sc-Dirac} and~\ref{sec:el-dip-sc-Dirac-pairing}) as quintessential systems with the two key ingredients for strong Fermi-level dipole effects: parity-mixing, but also strong spin-orbit coupling (SOC), as we explain below.
This chapter is based on Ref.~\cite{Palle2024-el-dip}.
Since Ref.~\cite{Palle2024-el-dip} is written in a long-paper format already appropriate for a monograph chapter, the majority of the text and figures of this chapter have been recycled from Ref.~\cite{Palle2024-el-dip}.
Apart from the reorganizing, editing, and the inclusion of additional discussions (see Secs.~\ref{sec:el-dip-sc-Dirac-model-band}, \ref{sec:el-dip-sc-Dirac-RG-tree}, \ref{sec:el-dip-sc-Ward-id}, and~\ref{sec:el-dip-sc-comparison-OP-ex} in particular), the content of this chapter is essentially the same as that of Ref.~\cite{Palle2024-el-dip}.

\begin{figure}[t]
\centering
\begin{subfigure}[t]{0.5\textwidth}
\centering
\includegraphics[width=0.90\textwidth]{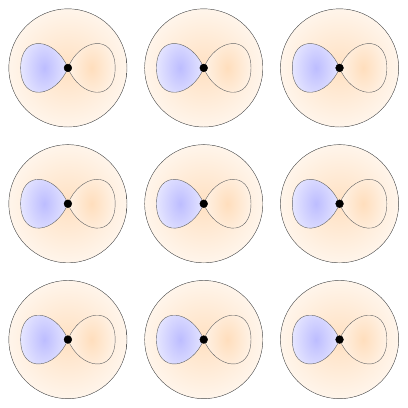}
\subcaption{}
\end{subfigure}%
\begin{subfigure}[t]{0.5\textwidth}
\raggedleft
\includegraphics[width=0.90\textwidth]{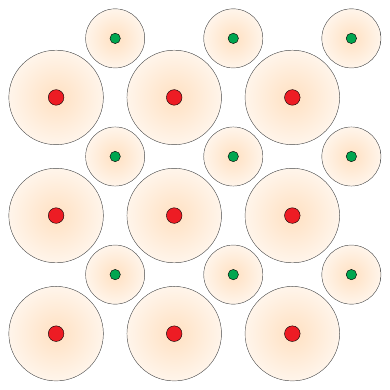}
\subcaption{}
\end{subfigure}
\captionbelow[Two simple examples of periodic systems in which local electric dipole operators can be introduced.]{\textbf{Two simple examples of periodic systems in which local electric dipole operators can be introduced.}
This is made possible by the opposite parities of the $s$ and $p_x$ orbitals under (a), and by the different inversions centers (non-trivial Wyckoff positions) of the two $s$ orbitals under (b).
The latter possibility is explained in more detail in Sec.~\ref{sec:el-dip-sc-itinerant-dipoles}.
Orange (blue) are positive (negative) lobes of the orbitals.}
\label{fig:el-dip-sc-example-lattices}
\end{figure}

Electric dipole excitations, while present in generic solids, only contribute to the Fermi surfaces of itinerant systems in the presence of SOC.
To elucidate this important fact, consider a simple lattice with orbitals of opposite parities on each site, such as the $s$ and $p_x$ orbitals shown in Fig.~\ref{fig:el-dip-sc-example-lattices}(a).
Then in the basis of these two orbitals, a local electric dipole operator $\DipD_x = \uptau_1 \otimes \Pauli_0$ exists and is perfectly well-defined.
($\uptau_{\mu}$ and $\Pauli_{\nu}$ are Pauli matrices in orbital and spin space, respectively.)
However, what matters for the description of the itinerant periodic solids is the matrix element
\begin{equation}
\mleft[\DipD_{x; \vb{k} n}\mright]_{ss'} = \mel{u_{\vb{k} n s}}{\DipD_x}{u_{\vb{k} n s'}} \label{eq:eds-dipole-intro}
\end{equation}
in the basis of the Bloch states $u_{\vb{k} n s}$.
Here $\vb{k}$, $n$, and $s$ stand for the crystal momentum, band, and spin, respectively.
In the absence of SOC, the dipole operator is trivial in spin space: $\DipD_{x; \vb{k} n} \propto \Pauli_0$.
It then follows that $\DipD_{x; \vb{k} n} = - \DipD_{x; \vb{k} n} = 0$ for systems invariant under the product $P \TRop$ of parity and time reversal (TR).
The same applies to dipole operators constructed in any other way, such as by mixing orbitals of the same parity located at different positions, like in Fig.~\ref{fig:el-dip-sc-example-lattices}(b).\footnote{How a finite displacement between the orbitals allows for dipole operators is explained in Sec.~\ref{sec:el-dip-sc-itinerant-dipoles}.}
As we will prove in Sec.~\ref{sec:el-dip-sc-itinerant-dipoles}, as long as there is no SOC, electric dipole operators vanish when projected onto the Bloch states.
The argument is essentially the same one from Sec.~\ref{sec:Cp-channel-gen-sym-constr} of Chap.~\ref{chap:loop_currents} regarding the pairing form factor at forward scattering.
In contrast, with SOC the Fermi surface may acquire a sizable electric dipole density (Fig.~\ref{fig:el-dip-sc-FS-dip-density}).

A notable feature of electronic dipole fluctuations, as opposed to polar phononic ones~\cite{Cochran1960}, is that their interactions are mediated and screened together with electric monopole, quadrupole, etc., fluctuations.
More precisely, as we will show in Secs.~\ref{sec:el-dip-sc-dipole-coupling} and~\ref{sec:el-dip-sc-plasmon-field}, the dipolar contribution to the total electronic charge density comes alongside a monopolar one, and the corresponding interactions are mediated by the same plasmon field which mediates all electrostatic interactions.

The description of electric dipole moments of insulating periodic solids in terms of Bloch states and their Berry connection played an important role in resolving the ambiguity in the definition of the polarization~\cite{KingSmith1993, Resta1992, Resta1993, Vanderbilt1993, Resta1994, Resta2000}.
This description is, in fact, closely related to our treatment of electric dipoles.
As we explain in Sec.~\ref{sec:el-dip-sc-modern-pol-teo}, the finite extent of the electronic wavefunctions used as a tight-binding basis modifies the periodicity conditions relating $\vb{k} + \vb{G}$ to $\vb{k}$ for inverse lattice vectors $\vb{G}$.
As a result, within the tight-binding basis, the dipole operator as given by the King-Smith--Vanderbilt formula~\cite{KingSmith1993} acquires an anomalous (or intrinsic) contribution
\begin{equation}  
\iu \grad_{\vb{k}} \longrightarrow \iu \grad_{\vb{k}} + \vb{\Gamma}
\end{equation}
which is determined by the same dipole matrix elements that are key to our treatment.
For quasi-2D materials in particular, the anomalous contribution can easily be the dominant one along the out-of-plane direction.

Materials featuring strong SOC and conduction bands which mix parities are therefore natural applications of our theory.
In many materials, such as the topological insulators \ce{Bi2Se3}, \ce{Bi2Te3}, \ce{Sb2Te3}, and \ce{(PbSe)_5(Bi2Se3)_6}~\cite{Ando2013} or the topological crystalline insulators \ce{SnTe} and \ce{Pb_{1-x}Sn_{x}Te}~\cite{Ando2013, Ando2015}, the parity-mixing and SOC come together through SOC-induced band inversion.
As we establish in Sec.~\ref{sec:el-dip-sc-Dirac-model}, in the vicinity of such band-inverted points, the band structure has essentially the form of a massive Dirac model.
This motivates the investigation of dipole excitations in Dirac metals that we carry out in Sec.~\ref{sec:el-dip-sc-Dirac}.
Using a large-$N$ renormalization group (RG) analysis of the Coulomb interaction (Sec.~\ref{sec:el-dip-sc-Dirac-RG}), we show that for quasi-2D Dirac systems, where the monopole coupling is known to be marginally irrelevant~\cite{DTSon2007, Kotov2012}, the $z$-axis dipole coupling becomes marginally relevant.
In Sec.~\ref{sec:el-dip-sc-Dirac-opticond} we also demonstrate that these enhanced dipole excitations are directly observable in the $z$-axis optical conductivity.

Interestingly, all the materials listed in the previous paragraph become superconductors (SC) at low temperatures when doped or pressured.\footnote{SC under pressure was found in \ce{Bi2Se3}~\cite{Kong2013, Kirshenbaum2013}, \ce{Bi2Te3}~\cite{Zhang2011}, and \ce{Sb2Te3}~\cite{Zhu2013}.
Under ambient pressure, SC was observed in the following compounds doped via intercalation: \ce{Cu_{x}Bi2Se3}~\cite{Hor2010, Kriener2011, Kriener2011-p2}, \ce{Sr_{x}Bi2Se3}~\cite{Liu2015, Shruti2015}, \ce{Nb_{x}Bi2Se3}~\cite{Qiu2015}, \ce{Pd_{x}Bi2Te3} ~\cite{Hor2011}, and \ce{Cu_{x}(PbSe)_5(Bi2Se3)_6}~\cite{Sasaki2014}.
Non-intercalated doping was found to give SC in \ce{Tl_{x}Bi2Se3}~\cite{Wang2016, Trang2016}, \ce{Sn_{1-x}In_{x}Te}~\cite{Sasaki2012, Sato2013}, and \ce{(Pb_{0.5}Sn_{0.5})_{1-x}In_{x}Te}~\cite{Zhong2014, Du2015}.}
In the case of doped \ce{Bi2Se3}, there is strong evidence that its superconductivity spontaneously breaks rotational symmetry~\cite{Yonezawa2019, Matano2016, Pan2016, Yonezawa2017, Asaba2017, Du2017, Smylie2018, Cho2020} and has nodal excitations~\cite{Yonezawa2017, Smylie2016, Smylie2017}, indicating an unconventional odd-parity state~\cite{Fu2010, Venderbos2016, Hecker2018}.
Conversely, experiments performed on \ce{In}-doped \ce{SnTe} point towards a fully gapped pairing~\cite{Novak2013, Zhong2013, Smylie2018-p2, Smylie2020} which preserves time-reversal symmetry~\cite{Saghir2014} and has a pronounced drop in the Knight shift~\cite{Maeda2017}.
Although most simply interpreted as conventional $s$-wave pairing, given the moderate change in the Knight shift, a fully-gapped odd-parity state of $A_{1u}$ symmetry is also consistent with these findings~\cite{Smylie2020}.
Because of their topological band structures, these two materials are prominent candidates for topological superconductivity~\cite{Sato2017, Mandal2023}.

When electric dipole fluctuations are present on the Fermi surface, their monopole-dipole and dipole-dipole interactions can give rise to superconductivity, as we will show in Sec.~\ref{sec:el-dip-sc-pairing}.
The resulting pairing is necessarily unconventional, as we explicitly prove in Sec.~\ref{sec:el-dip-sc-qualitative-pairing} using arguments similar to those of Sec.~\ref{sec:TR-positivity}.
It also requires substantial screening, which is true of most other pairing mechanisms.
Although we find that the dimensionless coupling constant $\lambda$ of the leading pairing channel is comparatively small and not expected to exceed $\sim 0.1$, dipole fluctuations can still be the dominant source of pairing for systems without strong local electronic correlations.
In the case of quasi-2D Dirac metals (Sec.~\ref{sec:el-dip-sc-Dirac-pairing}), the leading pairing state is an odd-parity state of pseudoscalar ($A_{1u}$) symmetry, similar to the Balian-Werthamer state of \ce{^3He-B}~\cite{Leggett1975, Vollhardt1990, Volovik2003}, while the subleading instability is a two-component $p$-wave state, as required for nematic SC.
Though the latter is the second dominant pairing channel in most cases, it could prevail if aided by a complementary pairing mechanism, such as a phononic one~\cite{Brydon2014, Wu2017}.

The chapter is organized as follows.
In Sec.~\ref{sec:el-dip-sc-dipole-theory}, we study electronic dipole excitations of Fermi-surface states in general systems.
We derive how they interact, when is their Coulomb coupling direct, and the relation of our work to the modern theory of polarization.
After that, in Sec.~\ref{sec:el-dip-sc-Dirac}, we introduce a general Dirac model with dipolar coupling and using RG show that the $z$-axis dipole moment becomes enhanced for quasi-2D systems.
In addition, we demonstrate that this $z$-axis dipole moment is directly measurable in the $z$-axis optical conductivity.
In Sec.~\ref{sec:el-dip-sc-pairing}, we study Cooper pairing due to electronic dipole fluctuations in general systems.
We write down the linearized gap equation, show that the proposed dipole mechanism can only give unconventional pairing, and derive a number of estimates on the pairing strength.
In the penultimate Sec.~\ref{sec:el-dip-sc-Dirac-pairing}, we study Cooper pairing due to electronic dipole fluctuations in the particular case of quasi-2D Dirac metals.
We solve the linearized gap equation both analytically and numerically and for the leading instability find an unconventional odd-parity pairing state with pseudoscalar symmetry.
In the last Sec.~\ref{sec:el-dip-sc-final-discussion}, we recapitulate the main results of the current chapter and compare them at length to related work.

\section{Theory of dipole excitations of electronic Fermi-surface states} \label{sec:el-dip-sc-dipole-theory}
Electric dipole moments are conventionally only associated with localized electronic states.
Here, we first show that itinerant electronic states can carry electric dipole moments as well if SOC is present.
After that, in Sec.~\ref{sec:el-dip-sc-dipole-coupling}, we derive the corresponding dipolar contributions to the electron-electron interaction.
In Sec.~\ref{sec:el-dip-sc-modern-pol-teo} our treatment is related to the modern theory of polarization.
Lastly, in Sec.~\ref{sec:el-dip-sc-plasmon-field}, we reformulate the electron-electron Coulomb interaction in terms of a plasmon field, showing that monopole-monopole, monopole-dipole, and dipole-dipole interactions are all mediated by the same plasmon field.

\subsection{Electric dipole moments of itinerant electronic states} \label{sec:el-dip-sc-itinerant-dipoles}
Itinerant electronic states are states of definite crystal momentum $\vb{k}$, which is defined through the eigenvalues $\Elr^{\iu \vb{k} \vdot \vb{R}}$ of the lattice translation operators $\mathscr{T}_{\vb{R}}$.
Crystal momentum, however, is not the same as physical momentum, the eigenvalue of the continuous translation generator $\vb{P} = - \iu \grad$.
Because of this difference, itinerant electronic states carry not only electric charge and spin, which commute with $\vb{P}$, but also the generalized charges associated with any Hermitian operator $\mathscr{Q}$ that is periodic in the lattice, i.e., that commutes with $\mathscr{T}_{\vb{R}} = \Elr^{- \iu \vb{R} \vdot \vb{P}}$.

For instance, the Bloch state (not to be confused with the fermionic fields $\psi$ or $\Psi$)
\begin{equation}
\uppsi_{\vb{k} n s}(\vb{r}) = \Elr^{\iu \vb{k} \vdot \vb{r}} u_{\vb{k} n s}(\vb{r}) \label{eq:eds-Bloch}
\end{equation}
carries the charge
\begin{equation}
\mel{u_{\vb{k} n s}}{\mathscr{Q}}{u_{\vb{k} n s}} = \int_{\vb{r}} u_{\vb{k} n s}^{\dag}(\vb{r}) Q(\vb{r}) u_{\vb{k} n s}(\vb{r})
\end{equation}
for any
\begin{equation}
\mathscr{Q}(\vb{r}) = \frac{1}{\mathcal{N}} \sum_{\vb{R}} Q(\vb{r}-\vb{R}),
\end{equation}
where $\mathcal{N} = \sum_{\vb{R}} 1$ is the number of unit cells and the $\int_{\vb{r}} = \int \dd[d]{r}$ integral goes over all space.
Within tight-binding descriptions, a possible generalized charge is the orbital composition of the Bloch waves.
However, generalized charges associated with electric or magnetic multipoles, local charge or current patterns, and more broadly collective modes in the particle-hole sector of all types are also possible.

\begin{figure}[t]
\centering
\begin{subfigure}[t]{0.5\textwidth}
\centering
\includegraphics[width=\textwidth]{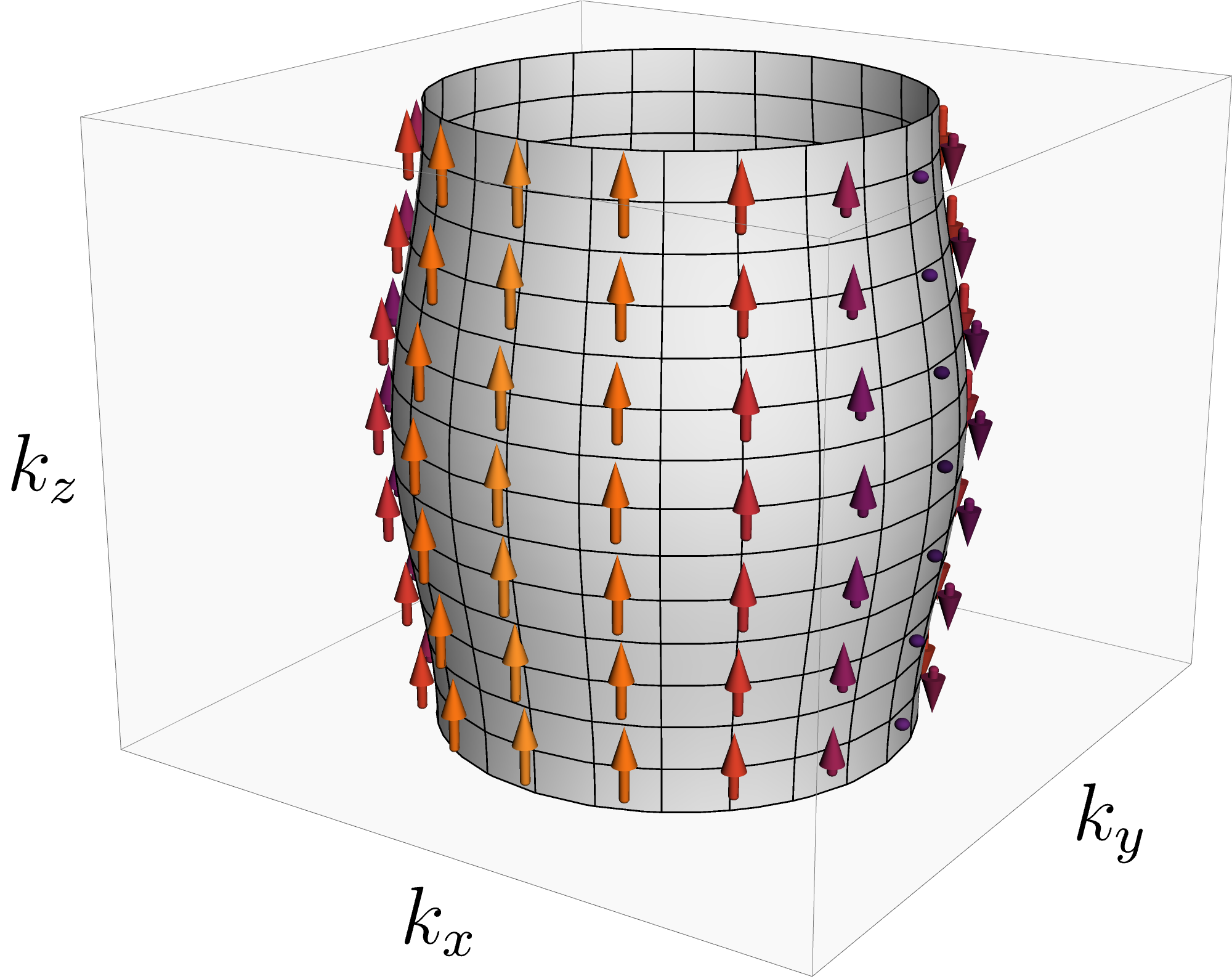}
\subcaption{}
\end{subfigure}%
\begin{subfigure}[t]{0.5\textwidth}
\centering
\includegraphics[width=\textwidth]{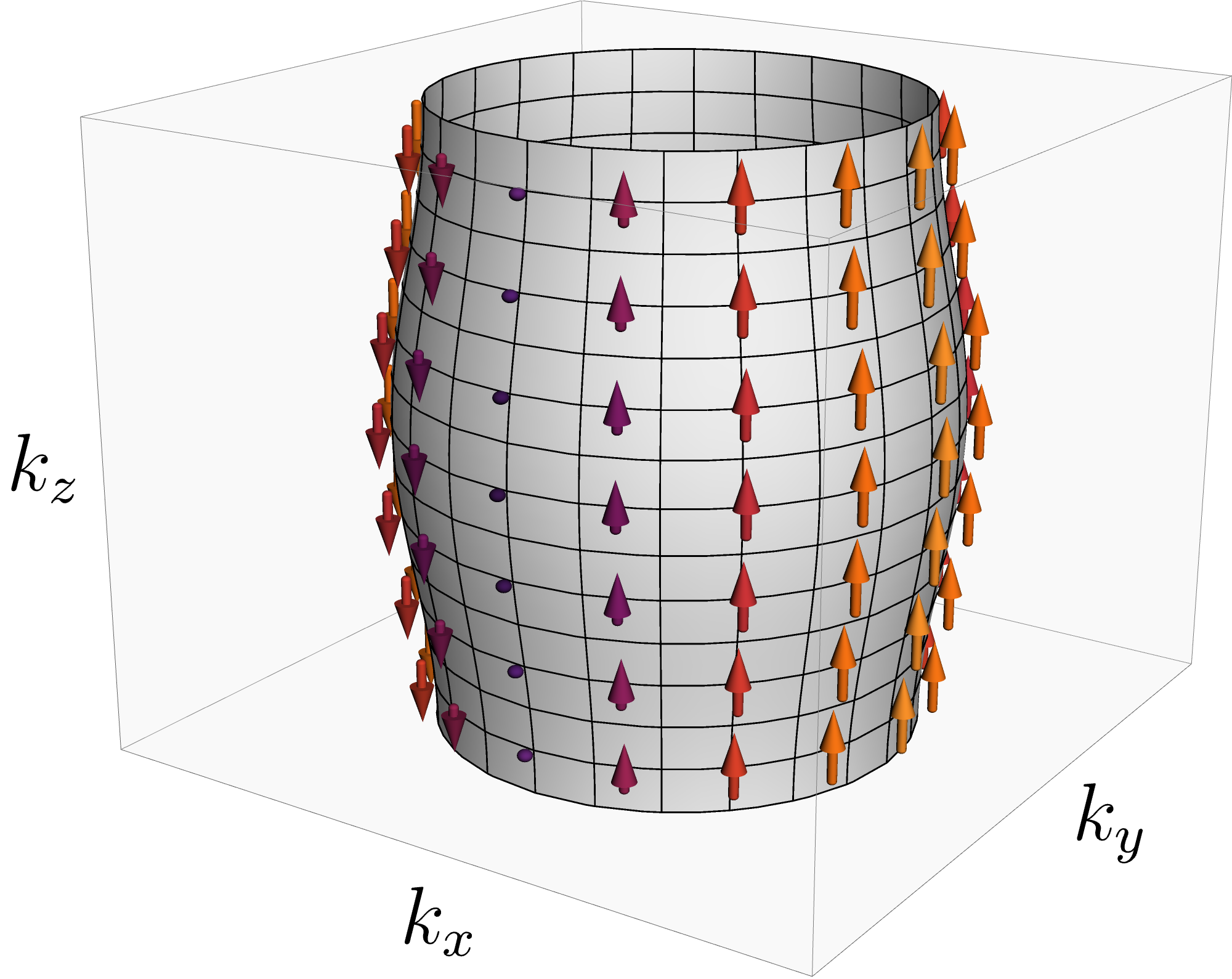}
\subcaption{}
\end{subfigure}
\captionbelow[An example of a cylindrical Fermi surface with a finite electric dipole density along the $\vu{e}_z$ direction.]{\textbf{An example of a cylindrical Fermi surface with a finite electric dipole density along the $\vu{e}_z$ direction.}
The arrows indicate the direction and strength of the electric dipole density $\mel{s}{\DipD_{z; \vb{k} n}}{s}$ for the pseudospin $s$ directed along $+\vu{e}_x$, $\ket{\uparrow}_{x} = \tfrac{1}{\sqrt{2}} \mleft(\ket{\uparrow} + \ket{\downarrow}\mright)$, under (a) and for the pseudospin $s$ directed along $+\vu{e}_y$, $\ket{\uparrow}_{y} = \tfrac{1}{\sqrt{2}} \mleft(\ket{\uparrow} + \iu \ket{\downarrow}\mright)$, under (b).
Opposite pseudospins and opposite momenta have opposite electric dipole densities.}
\label{fig:el-dip-sc-FS-dip-density}
\end{figure}

Collective modes couple to their associated generalized charges.
Because they exchange momentum with the electrons, the key matrix elements to analyze are
\begin{equation}
\mel{u_{\vb{k} n s}}{\mathscr{Q}}{u_{\vb{k}+\vb{q} n' s'}} \label{eq:eds-dipole-outro}
\end{equation}
of which the dipole element of Eq.~\eqref{eq:eds-dipole-intro} is a special case with $\vb{q} =  \vb{0}$ and $n' = n$.
At finite $\vb{q}$, or alternatively for $n' \neq n$, these matrix elements are generically finite.
However, the coupling to the Fermi-level electrons ($n' = n$) is particularly strong when they remain finite in the limit $\vb{q} \to \vb{0}$.
This is the limit we discuss in what follows.

In systems without SOC, the periodic parts of the Bloch wavefunctions $\ket{u_{\vb{k} n s}}$ decompose into an orbital and spin part:
\begin{equation}
\ket{u_{\vb{k} n s}} = \ket{u_{\vb{k} n}} \otimes \ket{s}.
\end{equation}
Since the composed space-inversion and time-reversal operation $P \TRop$ is the only symmetry that maps generic $\vb{k}$ to themselves, this is the only symmetry that limits the types of generalized charges that itinerant states can carry.
For a purely orbital charge $\mathscr{Q} = Q \otimes \Pauli_0$ that has sign $p_P \in \{\pm 1\}$ under parity and $p_{\TRop} \in \{\pm 1\}$ under TR, one readily finds the $P \TRop$ symmetry constraint to be
\begin{equation}
\mel{u_{\vb{k} n}}{Q}{u_{\vb{k} n}} = p_P p_{\TRop} \mel{u_{\vb{k} n}}{Q}{u_{\vb{k} n}}.
\end{equation}
Hence in the orbital sector only generalized charges with $p_P = p_{\TRop}$ are allowed.
In the spin sector an additional minus sign appears during time reversal so the generalized charges must satisfy $p_P = - p_{\TRop}$ to be finite.
Thus quite generically, a theory of itinerant electronic states that couple without SOC to collective modes as $\vb{q} \to \vb{0}$ is a theory of charge ($p_P = p_{\TRop} = +1$), spin ($p_P = - p_{\TRop} = +1$), and their currents.

Because their $p_P = -1 \neq p_{\TRop} = +1$, electric dipole moments cannot be carried by itinerant electronic states in the absence of SOC (cf.\ Refs.~\cite{KoziiBiRuhman2019, Volkov2020}) and, as a result, they tend to be negligible in most Fermi liquids.
The same is true for even-parity loop currents ($p_P = +1 \neq p_{\TRop} = -1$) which also decouple from electrons in the $\vb{q} \to \vb{0}$ limit, as was discussed in Sec.~\ref{sec:Cp-channel-gen-sym-constr} of Chap.~\ref{chap:loop_currents}.
In particular, notice that the pairing form factor $[\Pintf_{a}(\vb{p}_m, \vb{k}_n)]_{s_1 s_2}$ of Eq.~\eqref{eq:fa-definition-form-f} that we previously studied in Sec.~\ref{sec:QCP-model-lin-gap-eq} is the same thing, \textit{mutatis mutandis}, as the matrix element of Eq.~\eqref{eq:eds-dipole-outro}.

With spin-orbit coupling, restrictions are much less stringent and generalized charges such as electric dipoles can be carried.
The main difference from the case without SOC is that even-parity orbital operators that commute with the physical spin can acquire a non-trivial structure in pseudospin (Kramers' degeneracy) space.
Conversely, purely spin operators can have trivial pseudospin structures.
In the gauge $\ket{u_{\vb{k} n \uparrow}} = P \TRop \ket{u_{\vb{k} n \downarrow}}$, where $s \in \{\uparrow, \downarrow\}$ are pseudospins, the $P \TRop$ symmetry constraint has the form
\begin{align}
\Pauli_y Q_{\vb{k} n}^{*} \Pauli_y = p_P p_{\TRop} Q_{\vb{k} n},
\end{align}
where $\mleft[Q_{\vb{k} n}\mright]_{ss'} \defeq \mel{u_{\vb{k} n s}}{\mathscr{Q}}{u_{\vb{k} n s'}}$ and $\Pauli_y$ acts in pseudospin space.
Hence $p_P p_{\TRop}$ determines whether $Q_{\vb{k} n}$ is a pseudospin singlet or triplet.
In both cases, $Q_{\vb{k} n}$ can be finite for all charges $\mathscr{Q}$.

Electric dipoles are pseudospin triplets.
Given their purely orbital nature, this means that SOC need to be relatively strong near the Fermi surface for the electric dipole density to be large.
There is no net electric dipole moment, however.
The total electric dipole density averages to zero at each $\vb{k}$ because of the relation
\begin{align}
[\DipD_{\vu{e}; \vb{k} n}]_{\downarrow \downarrow} = - [\DipD_{\vu{e}; \vb{k} n}]_{\uparrow \uparrow}
\end{align}
which follows from TR symmetry.
Here $\DipD_{\vu{e}}$ is the electric dipole operator along the $\vu{e}$ direction.
This is also true for each pseudospin individually in the gauge $\ket{u_{-\vb{k} n s}} = P \ket{u_{\vb{k} n s}}$ since oddness under parity then implies
\begin{align}
[\DipD_{\vu{e}; -\vb{k} n}]_{s s} = - [\DipD_{\vu{e}; \vb{k} n}]_{s s}.
\end{align}
In the simplest case when the point group symmetry matrices can be made momentum-independent,\footnote{Note: contrary to what is claimed in Ref.~\cite{Fu2015}, the ``Manifestly Covariant Bloch Basis'' for which the point group symmetry matrices are momentum-independent does not exist across the whole Brillouin zone in general systems; e.g., if the parity of all time-reversal invariant momenta is not the same.} one finds that $\DipD_{\vu{e}; \vb{k} n} \propto \vu{e} \vdot (\vb{k} \vcross \vb{\Pauli})$~\cite{Fu2015}.
An example of a Fermi surface with an electric dipole density is drawn in Fig.~\ref{fig:el-dip-sc-FS-dip-density} for the case of a quasi-2D Dirac metal of the type we study in Sec.~\ref{sec:el-dip-sc-Dirac}.

Up to now, we have treated electric dipole operators $\DipD_{\vu{e}}$ abstractly as fermionic operators in the particle-hole sector which are even under TR, odd under parity, and transform like a vector under rotations and reflections.
Let us briefly comment on how such operators are constructed within tight-binding descriptions of solids.
As was already noted in the introduction of this chapter, the most straightforward way of constructing local electric dipole operators is by mixing orbitals of opposite parity centered on the same point.
For the example shown in Fig.~\ref{fig:el-dip-sc-example-lattices}(a), the local dipole operator has the form $\DipD_x(\vb{R}) = s^{\dag}(\vb{R}) p_x(\vb{R}) + p_x^{\dag}(\vb{R}) s(\vb{R})$, where $s(\vb{R})$ and $p_x(\vb{R})$ and fermionic annihilation operators of the respective orbitals at site $\vb{R}$.
Less obviously, even when orbitals have the same parity as in Fig.~\ref{fig:el-dip-sc-example-lattices}(b), local dipole operators exist whenever not all orbitals are centered at the same site.
This is made possible by the fact that the inversion centers are distinct for the different orbitals, thereby allowing for the construction of bonding and anti-bonding superpositions which do have opposite parities.
In the example of Fig.~\ref{fig:el-dip-sc-example-lattices}(b), the (anti-)bonding annihilation operators are $\tilde{s}_{\pm}(\vb{R}) = s(\vb{R}+\vb{\delta}) \pm s(\vb{R}-\vb{\delta})$ for $\vb{\delta} = \tfrac{1}{2} \vu{e}_x + \tfrac{1}{2} \vu{e}_y = \frac{1}{\sqrt{2}} \vu{e}_{d_+}$.
Hence the local dipole operator along $d_+ = x + y$ is $\DipD_{d_+}(\vb{R}) = s^{\dag}(\vb{R}) \tilde{s}_{-}(\vb{R}) + \tilde{s}_{-}^{\dag}(\vb{R}) s(\vb{R})$.
In an analogous way, non-local electric dipole operators can always be constructed because the orbitals are allowed to belong to different unit cells.

\subsection{Coulomb interactions and electronic dipole excitations} \label{sec:el-dip-sc-dipole-coupling}
Here we derive how itinerant electrons which carry electric monopole and dipole moments interact.
Our starting point is the electron-electron Coulomb interaction (in SI units):
\begin{align}
\Haml_C &= \frac{1}{2} \int_{\vb{r}, \vb{r}'} \rho_e(\vb{r}) \frac{1}{4 \pi \epsilon_0 \abs{\vb{r} - \vb{r}'}} \rho_e(\vb{r}'). \label{eq:eds-Coulomb-interaction-full}
\end{align}
The electronic charge density operator is given by
\begin{align}
\rho_e(\vb{r}) = - e \sum_s \Psi_s^{\dag}(\vb{r}) \Psi_s(\vb{r}),
\end{align}
where $e$ is the elementary charge and $s \in \{\uparrow, \downarrow\}$ are the physical spins.

Next, we expand the fermionic field operators in a complete lattice basis:
\begin{align}
\Psi_s(\vb{r}) &= \sum_{\vb{R} \alpha} \mleft[\upvarphi_{\alpha}(\vb{r} - \vb{R})\mright]_s \psi_{\alpha}(\vb{R}). \label{eq:eds-Psi-s-expansion}
\end{align}
Here, we allow the basis to depend on spin $s$.
$\alpha$ is a combined orbital and spin index.
One popular choice of basis functions are the Wannier functions~\cite{Marzari2012}.
If they are constructed from a set of bands which (i) has vanishing Chern numbers and (ii) does not touch any of the bands of the rest of the spectrum, then the corresponding Wannier functions can always be made exponentially localized~\cite{Panati2007, Brouder2007}.
Condition (i) is always satisfied in the presence of time-reversal symmetry, while the second condition can be satisfied to an adequate degree by including many bands.
Thus as long as we do not restrict ourselves to the description of low-energy bands, we may assume that our basis functions $\upvarphi_{\alpha}(\vb{r} - \vb{R})$ are exponentially localized.
Using this basis, we may now decompose the charge density into localized parts:
\begin{align}
\rho_e(\vb{r}) = \sum_{\vb{R}} \rho_{\vb{R}}(\vb{r} - \vb{R}),
\end{align}
where the $\rho_{\vb{R}}(\vb{r})$ are localized around $\vb{r} = \vb{0}$:
\begin{align}
\rho_{\vb{R}}(\vb{r}) &= - \frac{e}{2} \sum_{\vb{\delta} \alpha \beta} \upvarphi_{\alpha}^{\dag}(\vb{r}) \upvarphi_{\beta}(\vb{r} - \vb{\delta}) \, \psi_{\alpha}^{\dag}(\vb{R}) \psi_{\beta}(\vb{R} + \vb{\delta}) + \Hc
\end{align}
Here the $\vb{\delta}$ sum goes over lattice neighbors.

By expanding $\Haml_C$ to dipolar order in multipoles, we obtain
\begin{equation}
\Haml_{\text{int}} = \frac{1}{2} \sum_{\vb{R}_1 \vb{R}_2} \sum_{\mu \nu} \DipD_{\mu}(\vb{R}_1) V_{\mu \nu}(\vb{R}_1 - \vb{R}_2) \DipD_{\nu}(\vb{R}_2), \label{eq:eds-Hint-vR}
\end{equation}
where $\mu, \nu \in \{0, 1, 2, 3\}$,
\begin{equation}
\DipD_0(\vb{R}) = \int_{\vb{r}} \rho_{\vb{R}}(\vb{r})
\end{equation}
is the electric monopole moment operator, and 
\begin{equation}
\DipD_i(\vb{R}) = \int_{\vb{r}} r_i \rho_{\vb{R}}(\vb{r})  
\end{equation}
are the components of the electric dipole operator.
Here the integration $\int_{\vb{r}} = \int \dd[3]{r}$ extends over the whole space, and not merely over a unit cell.
Due to exponential localization, these integrals converge and give well-defined operators.
Because we are working with a non-periodic $\rho_{\vb{R}}(\vb{r})$, there is no ambiguity in these definitions, other than the obvious dependence on the choice of basis functions $\upvarphi_{\alpha}(\vb{r} - \vb{R})$.

The interaction matrix which follows from the multipole expansion equals
\begin{equation}
V_{\mu \nu}(\vb{R}) = \frac{1}{4 \pi \epsilon_0} \begin{pmatrix}
1 & - \partial_j \\
\partial_i & - \partial_i \partial_j
\end{pmatrix} \frac{1}{R}.
\end{equation}
Here, $R = \abs{\vb{R}}$, $i, j \in \{1, 2, 3\}$, and $\partial_i = \partial/\partial{R_i}$. 
At $\vb{R} = \vb{0}$, $V_{\mu \nu}(\vb{R})$ has an aphysical divergence that we regularize by replacing $R^{-1}$ with $R^{-1} \erf\tfrac{R}{2 a_0}$:
\begin{equation}
V_{\mu \nu}^{\text{(reg.)}}(\vb{R}) = \frac{1}{4 \pi \epsilon_0} \begin{pmatrix}
1 & - \partial_j \\
\partial_i & - \partial_i \partial_j
\end{pmatrix} \frac{1}{R} \erf\frac{R}{2 a_0}.
\end{equation}
This corresponds to an unscreened on-site Hubbard interaction $U = e^2/(4 \pi^{3/2} \epsilon_0 a_0)$.
The Fourier transform $q^{-2} \Elr^{- a_0^2 q^2}$ of $R^{-1} \erf\tfrac{R}{2 a_0}$ now decays exponentially for large $q = \abs{\vb{q}}$:
\begin{align}
V_{\mu \nu}^{\text{(reg.)}}(\vb{q}) &= \frac{L^d}{\mathcal{N}} \sum_{\vb{R}} \Elr^{- \iu \vb{q} \vdot \vb{R}} V_{\mu \nu}^{\text{(reg.)}}(\vb{R}) = \sum_{\vb{G}} \tilde{V}_{\mu \nu}^{\text{(reg.)}}(\vb{q} + \vb{G}), \\
\tilde{V}_{\mu \nu}^{\text{(reg.)}}(\vb{q}) &= \int_{\vb{r}} \Elr^{- \iu \vb{q} \vdot \vb{r}} V_{\mu \nu}^{\text{(reg.)}}(\vb{r}) = \begin{pmatrix}
1 & - \iu q_j \\
\iu q_i & q_i q_j
\end{pmatrix} \frac{\Elr^{- a_0^2 q^2}}{\epsilon_0 \vb{q}^2}.
\end{align}
Here we have exploited the Poisson summation formula to express the Fourier series sum $V_{\mu \nu}^{\text{(reg.)}}(\vb{q})$ in terms of the Fourier transform $\tilde{V}_{\mu \nu}^{\text{(reg.)}}(\vb{q})$.
The exponential decay of $\tilde{V}_{\mu \nu}^{\text{(reg.)}}(\vb{q})$ renders the Umklapp sum over inverse lattice vectors $\vb{G}$ appearing in $V_{\mu \nu}^{\text{(reg.)}}(\vb{q})$ convergent.
For $a_0$ small compared to the lattice constant, the Umklapp sum is well-approximated with just the $\vb{G} = \vb{0}$ term.
Hence in momentum space:
\begin{align}
\Haml_{\text{int}} &= \frac{1}{2 L^d} \sum_{\vb{q} \mu \nu} \DipD_{\mu,-\vb{q}} V_{\mu \nu}(\vb{q}) \DipD_{\nu \vb{q}}, \label{eq:eds-Hint-vq}
\end{align}
where $L^d$ is the total volume in $d$ spatial dimensions, $\vb{q}$ goes over the first Brillouin zone, and
\begin{align}
V_{\mu \nu}(\vb{q}) &= V_{\mu \nu}^{\text{(reg.)}}(\vb{q}) \approx \begin{pmatrix}
1 & - \iu q_j \\
\iu q_i & q_i q_j
\end{pmatrix} \frac{1}{\epsilon_0 \vb{q}^2}. \label{eq:eds-VUmklapp}
\end{align}
Keeping only the $\vb{G} = \vb{0}$ Umklapp term in $V_{\mu \nu}(\vb{q})$ can be understood as another way of regularizing the $V_{\mu \nu}(\vb{R} = \vb{0})$ divergence.
When we later consider quasi-2D systems, the Umklapp sum for the out-of-plane $\vb{G}$ will not be negligible.
Its main effect is to make $V_{\mu \nu}(\vb{q})$ periodic in the out-of-plane $q_z$, which we shall later account for by replacing all $q_z$ with $\sin q_z$.

For the $\DipD_{\mu}(\vb{R})$, we now obtain, in matrix notation,
\begin{align}
\DipD_{\mu}(\vb{R}) &= - \frac{e}{2} \psi^{\dag}(\vb{R}) \Gamma_{\mu}(\vb{\delta}) \psi(\vb{R} + \vb{\delta}) + \Hc, \label{eq:eds-dipole_real_space}
\end{align}
where
\begin{align}
\mleft[\Gamma_{0}(\vb{\delta})\mright]_{\alpha \beta} &\defeq \int_{\vb{r}} \upvarphi_{\alpha}^{\dag}(\vb{r}) \upvarphi_{\beta}(\vb{r} - \vb{\delta}), \\
\mleft[\Gamma_{i}(\vb{\delta})\mright]_{\alpha \beta} &\defeq \int_{\vb{r}} r_i \, \upvarphi_{\alpha}^{\dag}(\vb{r}) \upvarphi_{\beta}(\vb{r} - \vb{\delta}). \label{eq:eds-dipol-matrix-elem}
\end{align}
When the lattice bases $\upvarphi_{\alpha}(\vb{r} - \vb{R})$ are orthogonal and normalized $\Gamma_{0}(\vb{\delta}) = \Kd_{\vb{\delta}, \vb{0}} \one$, and when they are sufficiently localized $\Gamma_{i}(\vb{\delta}) \approx 0$ for $\vb{\delta}$ which are not $\vb{0}$ or the nearest-lattice neighbors.
Moreover, $[\Gamma_{i}(\vb{\delta}=\vb{0})]_{\alpha \beta}$ is finite for $\upvarphi_{\alpha}(\vb{r})$ and $\upvarphi_{\beta}(\vb{r})$ centered at $\vb{r} = \vb{0}$ only when they have opposite parities.
That said, substantial dipole moments can also arise from orbitals of the same parity if they belong to different neighboring atoms because of the possibility of forming anti-bonding superpositions.
This last point we discussed at the end of Sec.~\ref{sec:el-dip-sc-itinerant-dipoles}.

In the simplest case when only $\Gamma_{0}(\vb{\delta}=\vb{0}) \equiv \Gamma_0 = \one$ and $\Gamma_{i}(\vb{\delta}=\vb{0}) \equiv \Gamma_i$ are finite, in momentum space we get
\begin{align}
\begin{aligned}
\DipD_{\mu \vb{q}} &= \sum_{\vb{R}} \Elr^{- \iu \vb{q} \vdot \vb{R}} \DipD_{\mu}(\vb{R}) \\
&= - e \sum_{\vb{k}} \psi^{\dag}_{\vb{k}} \Gamma_{\mu} \psi_{\vb{k} + \vb{q}},
\end{aligned}
\end{align}
where $\vb{k}$ runs over the first Brillouin zone.
The associated matrix elements $\mel{u_{\vb{k} n s}}{\Gamma_{\mu}}{u_{\vb{k} + \vb{q} n' s'}}$ were analyzed in the previous section.
The monopole matrix elements ($\mu = 0$) become diagonal in the band index as $\vb{q} \to \vb{0}$, but are otherwise finite.
The intraband dipole matrix elements ($\mu = 1, 2, 3$), on the other hand, vanish in the $\vb{q} \to \vb{0}$ limit in the absence of SOC.
The corresponding coupling of the electric dipoles to Fermi-level electrons thus gains an additional momentum power, which makes these interactions even more irrelevant with respect to RG flow than usual, unless the system has spin-orbit coupling.

The multipole expansion employed in Eq.~\eqref{eq:eds-Hint-vR} is justified whenever two charges are localized on length scales smaller than their distance.
In the limit of strong screening that we later analyze, however, the strongest interactions come from nearby particles, indicating a breakdown of the multipole expansion.
Nonetheless, the additional dipolar terms that we identified in the effective electron-electron interaction of Eq.~\eqref{eq:eds-Hint-vR} will still be present, albeit with coefficients that are phenomenological parameters.
Although their values cannot be inferred from a direct multipole expansion when screening is strong, the momentum-dependence and form of the dipolar coupling follows from symmetry and retains the same structure as derived in this section.

It is worth noting that the exact Coulomb interaction elements are, in principle, exactly known.
They are found by simply inserting the basis expansion~\eqref{eq:eds-Psi-s-expansion} into the Coulomb interaction~\eqref{eq:eds-Coulomb-interaction-full}:
\begin{align}
\int_{\vb{r}, \vb{r}'} \sum_{ss'} \mleft[\upvarphi_{\alpha_1}(\vb{r} - \vb{R}_1)\mright]_s^{*} \mleft[\upvarphi_{\alpha_2}(\vb{r}' - \vb{R}_2)\mright]_{s'}^{*} \frac{1}{4 \pi \epsilon_0 \abs{\vb{r} - \vb{r}'}} \mleft[\upvarphi_{\alpha_3}(\vb{r} - \vb{R}_3)\mright]_s \mleft[\upvarphi_{\alpha_4}(\vb{r}' - \vb{R}_4)\mright]_{s'}.
\end{align}
In practice, however, this expression, with its four indices and three relative distances, is too complex to treat.
The most common approximation employed in theoretical studies is to include only the monopole-monopole term in Eq.~\eqref{eq:eds-Hint-vR}, perhaps even restricting it to solely the on-site Hubbard term.
The main novelty of the current work, which is based on Ref.~\cite{Palle2024-el-dip}, is thus that we include the additional dipolar terms in Eq.~\eqref{eq:eds-Hint-vR} and explore their consequences.

\subsection{Relation to the modern theory of polarization} \label{sec:el-dip-sc-modern-pol-teo}
Our theory deals with dynamical electric dipole moments and their fluctuations.
Nonetheless, it is enlightening to make contact to the modern theory of polarization~\cite{KingSmith1993, Resta1992, Resta1993, Vanderbilt1993, Resta1994, Resta2000} in which the static polarization is expressed in terms of the Berry connection via the King-Smith--Vanderbilt formula~\cite{KingSmith1993}
\begin{equation}
\ev{\vb{\DipD}} = - e \sum_{\vb{k} n s}^{\text{occ.}} \mel{u_{\vb{k} n s}}{\iu \grad_{\vb{k}}}{u_{\vb{k} n s}}, \label{eq:eds-KingSmith}
\end{equation}
where $\vb{k}$ goes over the first Brillouin zone, $n$ goes over occupied bands only, and $s \in \{\uparrow, \downarrow\}$ is the pseudospin.
The intuition behind this formula is that $\iu \grad_{\vb{k}}$ roughly represents the position operator $\vb{r}$ in momentum space so $\ev{\vb{\DipD}} \sim - e \ev{\vb{r}}$, as one would expect.
However, for systems under periodic boundary conditions a position operator $\vb{r}$ cannot be defined, which is reflected in the above formula by its apparent gauge-dependence: enacting $\ket{u_{\vb{k} n s}} \mapsto \Elr^{\iu \vartheta_{\vb{k} n s}} \ket{u_{\vb{k} n s}}$ causes a change in $\ev{\vb{\DipD}}$ proportional to the winding numbers $\frac{1}{2 \pi} \oint \dd{k_i} \, \partial_{k_i} \vartheta_{\vb{k} n s}$.
Effectively, this gauge transformation moves the weight of the charge density by a direct lattice vector $\vb{R}$, thereby changing $\ev{\vb{\DipD}}$ by $- e \vb{R}$.
Since only differences in the static polarization are physically meaningful, this ambiguity is not a problem.
Another notable feature of formula~\eqref{eq:eds-KingSmith} is that the static polarization is not only a function of the charge density, but also of the Bloch wavefunction phases, as measured by the $\mel{u_{\vb{k} n s}}{\iu \grad_{\vb{k}}}{u_{\vb{k} n s}}$ average which is precisely the Berry connection.
There is a deep connection between the static polarization and geometric phases.
For further discussion of the modern theory of polarization, we direct the interested reader toward the excellent review by Resta~\cite{Resta2000}.

As we shall now show below, the finite extent of the $\upvarphi_{\alpha}(\vb{r} - \vb{R})$ basis wavefunctions, which is crucial for the definition of the higher-order multipoles in the first place, gives rise to an anomalous contribution to the polarization of Eq.~\eqref{eq:eds-KingSmith} when expressed within a tight-binding description.

Assuming time-reversal symmetry, the Bloch wavefunctions of Eq.~\eqref{eq:eds-Bloch} can always be chosen to be periodic in $\vb{k}$,\footnote{This follows from the fact that the band energies are bounded from below, thereby precluding spectral flow in which the $n$-th band maps to a different band as loops are traversed in the Brillouin zone.
This remains true even if the bands cross, albeit with a non-analytic $\vb{k}$-dependence around the crossing.
If a (possibly degenerate) band with vanishing Chern numbers does not touch any other band, one can always choose a gauge in which $\uppsi_{\vb{k} n s}$ and $u_{\vb{k} n s}$ are analytic functions of $\vb{k}$~\cite{Panati2007}.} meaning $\uppsi_{\vb{k} + \vb{G} n s}(\vb{r}) = \uppsi_{\vb{k} n s}(\vb{r})$ for all inverse lattice vectors $\vb{G}$, where $\uppsi_{\vb{k} n s}(\vb{r})$ are continuous, but not necessarily analytic, functions of $\vb{k}$.
The real-space periodic parts $u_{\vb{k} n s}(\vb{r}) = u_{\vb{k} n s}(\vb{r} + \vb{R})$ then satisfy
\begin{equation}
u_{\vb{k} n s}(\vb{r}) = \Elr^{\iu \vb{G} \vdot \vb{r}} u_{\vb{k} + \vb{G} n s}(\vb{r}). \label{eq:eds-k-period-cond}
\end{equation}
Next, we expand the $u_{\vb{k} n s}(\vb{r})$ with respect to an orthonormal tight-binding basis:
\begin{equation}
u_{\vb{k} n s}(\vb{r}) = \sum_{\vb{R} \alpha} \upvarphi_{\alpha}(\vb{r} - \vb{R}) \mleft[v_{\vb{k} n s}\mright]_{\alpha}.
\end{equation}
The periodicity condition~\eqref{eq:eds-k-period-cond} now becomes:
\begin{equation}
v_{\vb{k} n s} = U_{\vb{G}} \, v_{\vb{k} + \vb{G} n s}, \label{eq:eds-k-period-cond2}
\end{equation}
where
\begin{equation}
\mleft[U_{\vb{G}}\mright]_{\alpha \beta} = \sum_{\vb{\delta}} \int_{\vb{r}} \upvarphi_{\alpha}^{\dag}(\vb{r}) \Elr^{\iu \vb{G} \vdot \vb{r}} \upvarphi_{\beta}(\vb{r} - \vb{\delta}).
\end{equation}
In evaluating this expression, one often argues that the wavefunctions are point-like objects such that $\Elr^{\iu \vb{G} \vdot \vb{r}} \upvarphi_{\alpha}(\vb{r}) \approx \Elr^{\iu \vb{G} \vdot \vb{x}_{\alpha}} \upvarphi_{\alpha}(\vb{r})$, where $\vb{x}_{\alpha}$ are the positions of the orbitals; see also Refs.~\cite{Fruchart2014, Simon2020}.
This would then give a diagonal $\mleft[U_{\vb{G}}\mright]_{\alpha \beta} = \Elr^{\iu \vb{G} \vdot \vb{x}_{\alpha}} \Kd_{\alpha \beta}$ with $\Ugp(1)$ phase factors which can be absorbed into the $\mleft[v_{\vb{k} n s}\mright]_{\alpha}$ through a $\Ugp(1)$ gauge transformation.
However, the spread of the $\upvarphi_{\alpha}(\vb{r})$ around $\vb{x}_{\alpha}$ also contributes significantly to $U_{\vb{G}}$ when the orbitals mix parities or overlap.
By expanding the $\Elr^{\iu \vb{G} \vdot \vb{r}}$ exponential to linear order in $\vb{r}$, one readily finds that these corrections result in
\begin{equation}
U_{\vb{G}} = \Elr^{\iu \vb{G} \vdot \sum_{\vb{\delta}} \vb{\Gamma}(\vb{\delta})}, \label{eq:eds-U-G-expression}
\end{equation}
where the $\Gamma_{i}(\vb{\delta})$ are the matrix elements of Eq.~\eqref{eq:eds-dipol-matrix-elem}.
Having found tight-binding vectors $v_{\vb{k} n s}^{(0)}$ that are periodic, $v_{\vb{k} + \vb{G} n s}^{(0)} = v_{\vb{k} n s}^{(0)}$, the periodicity condition~\eqref{eq:eds-k-period-cond2} can be accommodated by the unitary $\Ugp(N)$ transformation
\begin{equation}
v_{\vb{k} n s} = \Elr^{- \iu \vb{k} \vdot \sum_{\vb{\delta}} \vb{\Gamma}(\vb{\delta})} v_{\vb{k} n s}^{(0)}.
\end{equation}
This holds to the same order in momentum as the expression for $U_{\vb{G}}$.\footnote{Note that the $\Gamma_{i} = \sum_{\vb{\delta}} \Gamma_{i}(\vb{\delta})$ matrices do not commute so $\Elr^{\iu \vb{k} \vdot \vb{\Gamma}} \Elr^{\iu \vb{G} \vdot \vb{\Gamma}} \Elr^{- \iu (\vb{k} + \vb{G}) \vdot \vb{\Gamma}} = \Elr^{\tfrac{1}{2} [\vb{k} \vdot \vb{\Gamma}, \vb{G} \vdot \vb{\Gamma}] + \cdots} \neq \one$.}
Within the $\upvarphi_{\alpha}(\vb{r} - \vb{R})$ basis, the King-Smith--Vanderbilt formula~\eqref{eq:eds-KingSmith} therefore acquires an additional term:
\begin{equation}
\ev{\vb{\DipD}} = - e \sum_{\vb{k} n s}^{\text{occ.}} \mel{v_{\vb{k} n s}^{(0)}}{\iu \grad_{\vb{k}} + \sum\nolimits_{\vb{\delta}} \vb{\Gamma}(\vb{\delta})}{v_{\vb{k} n s}^{(0)}}. \label{eq:eds-anomal-KingSmith}
\end{equation}
This additional, or anomalous, term is determined by the same $\Gamma_{i}(\vb{\delta})$ of Eq.~\eqref{eq:eds-dipol-matrix-elem} that govern the dipolar interactions.

To illustrate the importance of this anomalous term, let us consider a system whose tight-binding Hamiltonian is independent of $k_z$.
This is often approximately true in quasi-2D systems.
The eigenvectors $v_{\vb{k} n s}^{(0)}$ are then independent of $k_z$ and a naive application of Eq.~\eqref{eq:eds-KingSmith} would suggest that the out-of-plane polarization vanishes.
However, Eq.~\eqref{eq:eds-anomal-KingSmith} reveals that this is not necessarily true:
\begin{equation}
\ev{\DipD_z} = - e \sum_{\vb{k} n s}^{\text{occ.}} \sum_{\vb{\delta}} \mel{v_{\vb{k} n s}^{(0)}}{\Gamma_z(\vb{\delta})}{v_{\vb{k} n s}^{(0)}}
\end{equation}
can be finite when the wavefunctions are spread along the $\vu{e}_z$ direction, even though there is no hopping along $z$.
In Dirac systems, this regime, which is dominated by the anomalous term, will turn out to have the strongest enhancement of dipole fluctuations, as we show in Sec.~\ref{sec:el-dip-sc-Dirac}.

\subsection{Formulation in terms of a plasmon field} \label{sec:el-dip-sc-plasmon-field}
Here we reformulate the effective interaction $\Haml_{\text{int}}$ of Eq.~\eqref{eq:eds-Hint-vR} in terms of Hubbard-Stratonovich fields~\cite{Altland2010}.
Naively, one would do this by introducing a Hubbard-Stratonovich field for each component of $\DipD_{\mu}$.
The result would then formally look like the models of ferroelectric critical fluctuations coupled to fermions that have been the subject of much recent interest~\cite{KoziiBiRuhman2019, Volkov2020, Enderlein2020, Kozii2022, Klein2023}.
Specifically, there would be a monopole Hubbard-Stratonovich field and an independent dipole Hubbard-Stratonovich field with the same symmetry and coupling to fermions as ferroelectric modes.
However, this is not correct for our $\Haml_{\text{int}}$ because the same electrostatic fields mediates all electric interactions, whether they are monopole-monopole, monopole-dipole, or dipole-dipole.
Formally, this manifests itself through the non-invertible rank~1 matrix structure of $V_{\mu \nu}(\vb{q})$ in Eq.~\eqref{eq:eds-VUmklapp}.
Within perturbation theory, one may indeed confirm that this rank~1 matrix structure stays preserved and that only $\epsilon_0 \vb{q}^2 \to \epsilon(\vb{q}) \vb{q}^2$ gets renormalized.

\begin{figure}[t]
\centering
\includegraphics[width=0.95\textwidth]{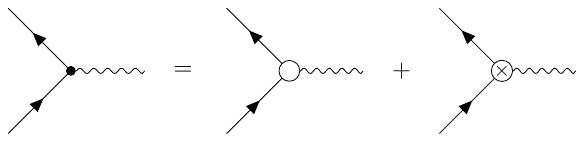}
\captionbelow[Decomposition of the total electron-plasmon vertex (solid dot) into a monopole-plasmon (open circle) and dipole-plasmon (crossed circle) contribution.]{\textbf{Decomposition of the total electron-plasmon vertex (solid dot) into a monopole-plasmon (open circle) and dipole-plasmon (crossed circle) contribution.}
This decomposition follows from the expansion of the electron density given in Eq.~\eqref{eq:eds-multipole-exp}.
Solid lines stand for electrons and wiggly lines for plasmons.}
\label{fig:el-dip-sc-plasmon-fermion-vertex}
\end{figure}

To carry out the Hubbard-Stratonovich transformations, we group all $\DipD_{\mu}$ into one effective charge density:
\begin{align}
\rho_{\vb{q}} &= \DipD_{0 \vb{q}} - \iu \sum_{j=1}^{3} q_j \DipD_{j \vb{q}}. \label{eq:eds-multipole-exp}
\end{align}
If we were not on a lattice, in real space this expression would reduce to the familiar $\rho(\vb{r}) = \DipD_0(\vb{r}) - \grad \vdot \vb{\DipD}(\vb{r})$, with $\DipD_0$ playing the role of the free charge density and $\vb{\DipD}$ the role of the polarization density.
The Euclidean action of $\Haml_{\text{int}}$ is
\begin{align}
\action_{\text{int}}[\psi] &= \frac{1}{2 \upbeta L^d} \sum_{q} \rho_{-q} V(q) \rho_{q},
\end{align}
where $q = (\omega_q, \vb{q})$, $\omega_q \in 2 \pi \Z / \upbeta$ are bosonic Matsubara frequencies, and
\begin{align}
V(q) &= V_{\mu=0,\nu=0}(\vb{q}) = \frac{1}{\epsilon_0 \vb{q}^2}.
\end{align}
After the Hubbard-Stratonovich transformation, it becomes:
\begin{align}
\action_{\text{int}}[\Phi, \psi] &= \frac{1}{2} \sum_{q} \Phi_{-q} V^{-1}(q) \Phi_{q} + \frac{\iu}{\sqrt{\upbeta L^d}} \sum_{q} \Phi_{-q} \rho_{q},  \label{eq:eds-HS-int-action}
\end{align}
where $\Phi_{q} = \Phi^{*}_{-q}$ is the electrostatic (plasmon) field.
The only difference from the usual Hubbard-Stratonovich-formulated action of plasma excitations is that the charge density has additional contributions coming from itinerant electric dipoles.
This is illustrated in Fig.~\ref{fig:el-dip-sc-plasmon-fermion-vertex}, where we show the decomposition of the total electron-plasmon vertex into monopole-plasmon and dipole-plasmon contributions, in agreement with the expansion of Eq.~\eqref{eq:eds-multipole-exp}.

\section{Dipole excitations in Dirac metals} \label{sec:el-dip-sc-Dirac}
In many systems, the electric dipole moments are relatively small, and if the spin-orbit coupling (SOC) is weak, their contribution to the interaction of Fermi surface states is even smaller.
Yet in Dirac systems which are generated through band inversion the opposite is the case.
Band inversion takes place when SOC inverts bands of opposite parities near high-symmetry points.
This large mixing of parities enables large electric dipole moments which, due to strong SOC, project onto the Fermi surface to significantly modify the electrostatic interaction.
Dirac metals therefore provide fertile ground for sizable electric dipole effects.

In the first part~\ref{sec:el-dip-sc-Dirac-model} of this section, we introduce the model which we study in the remainder of this section and whose Cooper pairing we study later in Sec.~\ref{sec:el-dip-sc-Dirac-pairing}.
In Sec.~\ref{sec:el-dip-sc-Dirac-model-band}, we show that the band Hamiltonian describing the vicinity of band-inverted points has the form of an anisotropic gapped Dirac model.
We derive how the electric dipole moments are represented within this model (Tab.~\ref{tab:eds-gamma-class}) and we introduce the corresponding electrostatic interactions of Sec.~\ref{sec:el-dip-sc-dipole-coupling} to the model in Sec.~\ref{sec:el-dip-sc-Dirac-model-int}.
In Sec.~\ref{sec:el-dip-sc-Dirac-opticond} thereafter, we turn to the study of the polarization of this model in the quasi-2D limit of weak $z$-axis dispersion.
Although it should naively vanish in this limit, we show that the additional dipole coupling renders the $z$-axis optical conductivity finite, thereby opening a route towards experimentally measuring the dipole excitations of our theory.
After that, in Sec.~\ref{sec:el-dip-sc-Dirac-RG}, we use renormalization group (RG) methods to investigate the dipole-coupled Dirac model in the regime of strong screening, as schematically shown in Fig.~\ref{fig:el-dip-sc-RG-sketch}.
This regime coincides with strong coupling and, to access it analytically, we employ a large-$N$ expansion to $1$-loop order, $N$ being the number of fermion flavors.
For generic Fermi surfaces, we find that electric dipole coupling is RG-irrelevant at the tree level (Sec.~\ref{sec:el-dip-sc-Dirac-RG-tree}) and thus becomes weaker at low energies.
However, if the dipole moments are parallel to the Fermi surface, as is the case for the out-of-plane moments in quasi-2D systems, they are marginal.
The detailed analysis of Sec.~\ref{sec:el-dip-sc-Dirac-RG-flow} furthermore shows that they are marginally relevant (Fig.~\ref{fig:el-dip-sc-RG-etaz-flow}), in contrast to the monopole coupling constant which is marginally irrelevant (Fig.~\ref{fig:el-dip-sc-RG-alpha-flow}).
Note that the dispersion along the out-of-plane direction here needs to be flat on the scale of the band gap of the semimetal because otherwise $z$-axis scaling would tend to make the $z$-axis dipole moments irrelevant.
The band gap also needs to be finite for the $z$-axis dipole moment to flow, because otherwise a chiral symmetry protects it, as we explain in the last Sec.~\ref{sec:el-dip-sc-Ward-id}.

\subsection{The model: Dirac fermions with dipolar interactions} \label{sec:el-dip-sc-Dirac-model}
The minimal model which captures the essential physics and that we shall study has the Euclidean (imaginary time) action
\begin{align}
\action[\psi, \Phi] &= \action_{\psi}[\psi] + \action_{\Phi}[\Phi] + \action_c[\psi, \Phi], \label{eq:eds-model-action}
\end{align}
where $\action_{\psi}$ and $\action_{\Phi}$ are the non-interacting fermionic and plasmonic parts, while $\action_c$ describes the electrostatic coupling between the two.
All three action parts are defined in this Sec.~\ref{sec:el-dip-sc-Dirac-model}.

\subsubsection{Effective band Hamiltonian of band-inverted points} \label{sec:el-dip-sc-Dirac-model-band}
To construct the fermionic part, we consider two bands of opposite parities in the vicinity of the $\Gamma$ point $\vb{k} = \vb{0}$ and assume that the other bands are sufficiently far away to not be important at low energies.
Within this two-band subspace, the parity and TR transformation matrices are
\begin{align}
\MatU(P) &= \uptau_3 \Pauli_0, &
\MatTR &= \uptau_3 \iu \Pauli_y, \label{eq:eds-U_P-TH-def}
\end{align}
where $\uptau_{\mu}$ and $\Pauli_{\nu}$ are Pauli matrices in band and pseudospin space, respectively.
Note that we have chosen TR to be $\MatTR = \uptau_3 \iu \Pauli_y$ so that
\begin{align}
\MatU(P) \MatTR &= \uptau_0 \iu \Pauli_y,
\end{align}
which maps $\vb{k} \mapsto \vb{k}$, becomes simpler and proportional to $\uptau_0$.
In Tab.~\ref{tab:eds-tausigma-class} we classify all the matrices according to their parity and TR signs, which are defined according to
\begin{align}
\MatU^{\dag}(P) \Gamma \MatU(P) &= p_P \Gamma, \\
\MatTR^{-1} \Gamma^{*} \MatTR &= p_{\TRop} \Gamma.
\end{align}
The only two matrices which are even under both parity and TR are $\uptau_3 \Pauli_0$ and $\uptau_0 \Pauli_0$ and they give the band gap and chemical potential displacement in the Hamiltonian, respectively.

\begin{table}[t]
\centering
\captionabove[The classification of Hermitian $4 \times 4$ spin-orbital matrices $\uptau_{\mu} \Pauli_{\nu}$ according to their eigenvalues under parity $P$ and time-reversal $\TRop$.]{\textbf{The classification of Hermitian $4 \times 4$ spin-orbital matrices $\uptau_{\mu} \Pauli_{\nu}$ according to their eigenvalues under parity $P$ and time-reversal $\TRop$.}
Here $\Conj$ is the complex conjugation operator.}
{\renewcommand{\arraystretch}{1.3}
\renewcommand{\tabcolsep}{10pt}
\begin{tabular}{cc|cc} \hline\hline
&& $\hat{P} = \uptau_3 \Pauli_0$ & $\hat{TRop} = \uptau_3 \iu \Pauli_y \Conj$ \\ \hline
$\uptau_0 \Pauli_0$, & $\uptau_3 \Pauli_0$ & $+1$ & $+1$ \\
$\uptau_0 \Pauli_{x,y,z}$, & $\uptau_3 \Pauli_{x,y,z}$ & $+1$ & $-1$ \\
$\uptau_2 \Pauli_0$, & $\uptau_1 \Pauli_{x,y,z}$ & $-1$ & $+1$ \\
$\uptau_1 \Pauli_0$, & $\uptau_2 \Pauli_{x,y,z}$ & $-1$ & $-1$
\\ \hline\hline
\end{tabular}}
\label{tab:eds-tausigma-class}
\end{table}

Because of the parity-mixing, terms linear in $\vb{k}$ also arise in the Hamiltonian.
They are constructed by combining $\vb{k}$ with three out of the four odd-parity and TR-odd matrices $\uptau_1 \Pauli_0$, $\uptau_2 \Pauli_x$, $\uptau_2 \Pauli_y$, and $\uptau_2 \Pauli_z$; which ones depends on the rotational symmetries.
When there is $n$-fold rotation symmetry around the $z$ axis, with $n \geq 3$, that has the form
\begin{align}
\MatU(C_{n z}) &= \uptau_0 \exp\mleft(- \iu \tfrac{2 \pi}{n} \tfrac{1}{2} \Pauli_z\mright), \label{eq:eds-J12-exp}
\end{align}
the pairs $\mleft(\uptau_2 \Pauli_x | \uptau_2 \Pauli_y\mright)$ and $\mleft(\uptau_2 \Pauli_y | - \uptau_2 \Pauli_x\mright)$ transform the same as $(k_x | k_y)$, giving a Rashba-like term in the Hamiltonian.
When there is twofold rotation symmetry around the $x$ axis, its form determines which of these two pairs continues to transform as $(k_x | k_y)$, as well as whether $\uptau_1 \Pauli_0$ or $\uptau_2 \Pauli_z$ transforms the same as $k_z$.
For
\begin{gather}
\begin{gathered}
\MatU(C_{2x}) = \uptau_3 (- \iu \Pauli_x), \\
\text{$\mleft(\uptau_2 \Pauli_y | - \uptau_2 \Pauli_x\mright) \sim (k_x | k_y)$ and $\uptau_1 \Pauli_0 \sim k_z$,}
\end{gathered} \label{eq:eds-J31-exp}
\end{gather}
whereas for
\begin{gather}
\begin{gathered}
\overline{\MatU}(C_{2x}) = \uptau_0 (- \iu \Pauli_x), \\
\text{$\mleft(\uptau_2 \Pauli_x | \uptau_2 \Pauli_y\mright) \sim (k_x | k_y)$ and $\uptau_2 \Pauli_z \sim k_z$.}
\end{gathered} \label{eq:eds-J31-exp-v2}
\end{gather}
For concreteness, below we assume the former [Eq.~\eqref{eq:eds-J31-exp}].
The latter choice for $\MatU(C_{2x})$ is related to the former one through the basis change $\mathcal{B}^{\dag} \overline{\MatU}(C_{2x}) \mathcal{B} = \MatU(C_{2x})$, where $\mathcal{B} = \diag(1,1,-\iu,\iu)$.
This basis change preserves the other symmetry matrices ($\mathcal{B}^{\dag} \MatU(P) \mathcal{B} = \MatU(P)$, $\mathcal{B}^{\dag} \MatU(P) \MatTR \mathcal{B}^{*} = \MatU(P) \MatTR$, and $\mathcal{B}^{\dag} \MatU(C_{n z}) \mathcal{B} = \MatU(C_{n z})$), which implies that all subsequent results are independent of which $\MatU(C_{2x})$ we use.

\begin{table}[t!]
\centering
\captionabove[The symmetry transformation matrices of the three generators $g$ of the dihedral point group of the model.]{\textbf{The symmetry transformation matrices of the three generators $g$ of the dihedral point group of the model.}
$C_{n z}$ is an $n$-fold rotation around $\vu{e}_z$, $C_{2x}$ is a \SI{180}{\degree} rotation around $\vu{e}_x$, and $P$ is parity.
$R(g)$ and $S(g)$ are vector and spin transformation matrices, respectively, $O(g)$ are orbital transformation matrices, and $\MatU(g) = O(g) \otimes S(g)$.
$\uptau_{\mu}, \Pauli_{\nu}$ are Pauli matrices.}
{\renewcommand{\arraystretch}{1.3}
\renewcommand{\tabcolsep}{10pt}
\begin{tabular}{cNNNN} \hline\hline
$g$ & $R(g)$ & $O(g)$ & $S(g)$ & $\MatU(g) = O(g) \otimes S(g)$
\\ \hline \\[-14pt]
$C_{n z}$ & $\begin{pmatrix}
\cos\tfrac{2\pi}{n} & -\sin\tfrac{2\pi}{n} & 0 \\[2pt]
\sin\tfrac{2\pi}{n} & \cos\tfrac{2\pi}{n} & 0 \\[2pt]
0 & 0 & 1 \end{pmatrix}$ &
$\uptau_0$ & $\Elr^{- \iu \pi \Pauli_z / n}$ & $\uptau_0 \Elr^{- \iu \pi \Pauli_z / n}$ \\[18pt]
$C_{2x}$ & $\begin{pmatrix}
1 & 0 & 0 \\
0 & -1 & 0 \\
0 & 0 & -1
\end{pmatrix}$ & $\uptau_3$ & $- \iu \Pauli_x$ & $\uptau_3 (- \iu \Pauli_x)$ \\[18pt]
$P$ & $\begin{pmatrix}
-1 & 0 & 0 \\
0 & -1 & 0 \\
0 & 0 & -1
\end{pmatrix}$ & $\uptau_3$ & $\Pauli_0$ & $\uptau_3 \Pauli_0$
\\ \\[-14pt] \hline\hline
\end{tabular}}
\label{tab:eds-sym-tranf-matrices}
\end{table}

In summary, the symmetry transformation rules have the following form when acting on the two-band fermionic spinors:
\begin{align}
\SymU^{\dag}(g) \psi_{\vb{k}} \SymU(g) &= \MatU(g) \psi_{R(g) \vb{k}} \equiv O(g) \otimes S(g) \psi_{R(g) \vb{k}}, \\
\SymTR^{-1} \psi_{\vb{k}} \SymTR &= \uptau_3 \iu \Pauli_y \, \psi_{-\vb{k}},
\end{align}
where $\SymU(g)$ and $\SymTR$ are the Fock-space point group and TR symmetry operators, respectively, with the corresponding $\MatU(g) = O(g) \otimes S(g)$ and $R(g)$ matrices given in Tab.~\ref{tab:eds-sym-tranf-matrices}.
The reason why we are allowed to assume that $\MatU(g)$ and $\MatTR$ do not depend on $\vb{k}$, which they do in general (see Eqs.~\eqref{eq:fermi-tranf-point-group} and~\eqref{eq:fermi-tranf-TR} in Sec.~\ref{sec:LC-gen-model-sym}), is because the $\vb{k}$ are restricted to the vicinity of the $\Gamma$ point $\vb{k} = \vb{0}$.
Using gauge transformations, one may always make $\MatU_{\vb{k}}(g)$ and $\MatTR_{\vb{k}}$ locally $\vb{k}$-independent.
All the complications we had to deal with in the previous chapter on cuprates (see Sec.~\ref{sec:extended-basis-def} in particular) are thus not relevant to the construction of the current model.
That said, when we later consider the quasi-2D limit, we shall be expanding around the $k_x = k_y = 0$ line.
A non-trivial constraint on the applicability of the model will thus be that the symmetry transformation matrices must be the same at both the $\Gamma$ point $k_z = 0$ and the $Z$ point $k_z = \pm \pi$.

The effective Hamiltonian near $\vb{k} = \vb{0}$ therefore reads
\begin{align}
H_{\vb{k}} &= m \uptau_3 \Pauli_0 + v \uptau_2 (k_x \Pauli_y - k_y \Pauli_x) + v_z k_z \uptau_1 \Pauli_0 - \upmu \uptau_0 \Pauli_0, \label{eq:eds-Dirac-Haml}
\end{align}
with the corresponding Euclidean action being:
\begin{align}
\action_{\psi}[\psi] &= \sum_k \psi^{\dag}_k \mleft[- \iu \omega_k + H_{\vb{k}}\mright] \psi_k,
\end{align}
where $k = (\omega_k, \vb{k})$ and $\omega_k \in \pi (2 \Z + 1) / \upbeta$ are fermionic Matsubara frequencies.
Because the $\vb{k}$-linear terms depend on spin, they need SOC to be large.
At quadratic order in $\vb{k}$, $m$ and $\upmu$ gain momentum dependence, as do $v$ and $v_z$ at cubic order in $\vb{k}$. 
This does not affect things qualitatively as long as the $\vb{k}$-linear terms are dominant so we shall not include this higher order $\vb{k}$-dependence in our analysis.
We shall also neglect the $\propto (3 k_x^2 k_y - k_y^3) \uptau_2 \Pauli_z$ term which arises at cubic order and breaks the emergent Dirac form.

\begin{table}[t]
\centering
\captionabove[The symmetry classification of Hermitian $4 \times 4$ matrices that can be constructed from the four $\gamma_{\mu}$ matrices.]{\textbf{The symmetry classification of Hermitian $4 \times 4$ matrices that can be constructed from the four $\gamma_{\mu}$ matrices.}
Below $\gamma_5 \defeq \gamma_0 \gamma_1 \gamma_2 \gamma_3$, $L_i \defeq - \tfrac{\iu}{4} \sum_{jk} \LCs_{ijk} \gamma_j \gamma_k$, $i, j, k \in \{1, 2, 3\}$, $\ell$ is the angular momentum under $\SO(3)$ rotations which are generated by $L_i$, and $\Conj$ is the complex conjugation operator.
Note that we are using a basis in which all five $\gamma_{\mu} = \gamma_{\mu}^{\dag}$ are Hermitian (including $\gamma_5 = \gamma_5^{\dag}$) and for which $\gamma_0^{*} = \gamma_0$, $\gamma_1^{*} = - \gamma_1$, $\gamma_2^{*} = \gamma_2$, and $\gamma_3^{*} = - \gamma_3$.}
{\renewcommand{\arraystretch}{1.3}
\renewcommand{\tabcolsep}{10pt}
\begin{tabular}{cc|ccc} \hline\hline
&& $\hat{P} = \gamma_0$ & $\SO(3)$ rotations & $\hat{\TRop} = - \gamma_1 \gamma_3 \Conj$ \\ \hline
$\one$, & $\gamma_0$ & $+1$ & $\ell = 0$ & $+1$ \\
& $\gamma_5$ & $-1$ & $\ell = 0$ & $+1$ \\
& $\iu \gamma_0 \gamma_5$ & $-1$ & $\ell = 0$ & $-1$ \\
& $\gamma_i$ & $-1$ & $\ell = 1$ & $+1$ \\
& $\iu \gamma_0 \gamma_i$ & $-1$ & $\ell = 1$ & $-1$ \\
$L_i$, & $\iu \gamma_i \gamma_5$ & $+1$ & $\ell = 1$ & $-1$
\\ \hline\hline
\end{tabular}}
\label{tab:eds-gamma-class}
\end{table}

To recast the action more closely as a Dirac model, introduce the Euclidean gamma matrices
\begin{align}
\begin{aligned}
\gamma_0 &= \uptau_3 \Pauli_0, &\hspace{80pt}
\gamma_1 &= \uptau_1 \Pauli_y, \\
\gamma_2 &= - \uptau_1 \Pauli_x, &\hspace{80pt}
\gamma_3 &= - \uptau_2 \Pauli_0.
\end{aligned}
\end{align}
The Euclidean signature we shall use not only for the gamma matrices, in the sense that
\begin{align}
\{\gamma_{\mu}, \gamma_{\nu}\} = 2 \Kd_{\mu \nu},
\end{align}
but also for raising, lowering, and contracting the indices of any four-vector.
By switching from $\psi^{\dag}$ to
\begin{align}
\bar{\psi} \defeq \psi^{\dag} \gamma_0,
\end{align}
one now readily finds that
\begin{align}
\action_{\psi}[\psi] &= \sum_k \bar{\psi}_{k} G^{-1}(k) \psi_{k}, \label{eq:eds-S-psi}
\end{align}
where
\begin{align}
G^{-1}(k) &= m \one - \iu \mleft[\omega_k \gamma_0 + v (k_x \gamma_1 + k_y \gamma_2) + v_z k_z \gamma_3\mright] - \upmu \gamma_0 \label{eq:eds-G-1-psi}
\end{align}
has a Dirac form.
Consequently, at high energies ($\abs{\omega_k} \gg \abs{\upmu}$) the symmetry of the system is enhanced to $\SO(4)$ with generators
\begin{align}
K_{\mu \nu} \defeq - \tfrac{\iu}{4} [\gamma_{\mu}, \gamma_{\nu}].
\end{align}
$K_{\mu \nu}$ generate rotations within the $k_{\mu}k_{\nu}$-plane, where $k_{\mu} = (\omega_k, \vb{k})$, and they satisfy the standard $\SO(4)$ generator commutation relations:
\begin{align}
[K_{\mu_1 \nu_1}, K_{\mu_2 \nu_2}] &= \iu \mleft(\Kd_{\mu_1 \mu_2} K_{\nu_1 \nu_2} + \Kd_{\nu_1 \nu_2} K_{\mu_1 \mu_2} - \Kd_{\mu_1 \nu_2} K_{\nu_1 \mu_2} - \Kd_{\nu_1 \mu_2} K_{\mu_1 \nu_2}\mright).
\end{align}
The chemical potential $\upmu$ breaks this symmetry down to $\SO(3)$: the group of spatial rotations which is generated by
\begin{align}
L_i \defeq \frac{1}{2} \sum_{jk} \LCs_{ijk} K_{jk}.
\end{align}
These generators satisfy the usual spin Lie algebra $[L_i, L_j] = \iu \sum_k \LCs_{ijk} L_k$.
The neglected cubic term which is proportional to $(3 k_x^2 k_y - k_y^3) \uptau_2 \Pauli_z = (3 k_x^2 k_y - k_y^3) \iu \gamma_0 \gamma_5$, where $\gamma_5 \defeq \gamma_0 \gamma_1 \gamma_2 \gamma_3 = - \uptau_1 \Pauli_z$, reduces the symmetry group further down to the dihedral group generated by $C_{n z}$ and $C_{2x}$ that we started with.
Note how $\MatU(P) = \gamma_0$ and $\MatTR = - \gamma_1 \gamma_3$, and how $L_3 = K_{12} = \tfrac{1}{2} \uptau_0 \Pauli_z$ and $L_1 = K_{23} = \tfrac{1}{2} \uptau_3 \Pauli_x$ agree with Eqs.~\eqref{eq:eds-J12-exp} and~\eqref{eq:eds-J31-exp}, respectively.

The alternative choice of Eq.~\eqref{eq:eds-J31-exp-v2} for $\MatU(C_{2x})$ would have given us the gamma matrices $\gamma_0' = \uptau_3 \Pauli_0$, $\gamma_1' = \uptau_1 \Pauli_x$, $\gamma_2' = \uptau_1 \Pauli_y$, and $\gamma_3' = \uptau_1 \Pauli_z$.
These are related to the previous ones through $\mathcal{B}^{\dag} \gamma_{\mu}' \mathcal{B} = \gamma_{\mu}$, where $\mathcal{B} = \diag(1,1,-\iu,\iu)$.
All subsequent results rely only on the intrinsic Clifford algebra structure of the gamma matrices and their precise form in no way affects any of our conclusions.

We have thus found that anisotropic gapped Dirac models describe SOC-inverted bands of opposite parities near the $\Gamma$ point.
This is true for other high-symmetry points of the Brillouin zone as well if $P$, $C_{n z}$ with $n\geq 3$, and $C_{2x}$ are symmetry operations (belong to the little group) of these points.
When the high-symmetry points $\vb{k}_{\star}$ have multiplicity higher than one, as happens when not all symmetry operations map $\vb{k}_{\star}$ to $\vb{k}_{\star}$ modulo inverse lattice vectors $\vb{G}$, multiple valleys arise, each described by a Dirac model.
Although effective Dirac models have been found long ago in graphite, bismuth, and \ce{SnTe}~\cite{Wallace1947, Cohen1960, Wolff1964, Rogers1968, Hsieh2012, Ando2013, Ando2015}, and more recently in topological insulators such as \ce{Bi2Se3}, \ce{Bi2Te3}, and \ce{Sb2Te3}~\cite{Zhang2009, Liu2010}, the derivation of this section showcases that this generically holds true for band-inverted systems with SOC.

\subsubsection{Plasmon propagator and electrostatic coupling} \label{sec:el-dip-sc-Dirac-model-int}
In light of the previously derived action~\eqref{eq:eds-HS-int-action}, the part describing the internal dynamics of the plasmon field is given by
\begin{align}
\action_{\Phi}[\Phi] &= \frac{1}{2} \sum_q \Phi_{-q} V^{-1}(q) \Phi_{q},
\end{align}
where in the bare plasmon propagator
\begin{align}
V^{-1}(q) &= \epsilon_{\perp} \mleft(q_x^2 + q_y^2\mright) + \epsilon_z q_z^2 \label{eq:eds-V-1-phi}
\end{align}
we allow for anisotropy between the $xy$ plane and $z$ axis.

Within the Dirac model, electric dipole moments are represented by $\psi^{\dag} \gamma_i \psi = \bar{\psi} \gamma_0 \gamma_i \psi$, where $i \in \{1, 2, 3\}$.
To see why, we note that the $\iu \gamma_0 \gamma_i$ which enter $H_{\vb{k}}$ transform as $\vb{k}$.
Therefore multiplying with $\iu \gamma_0$ will preserve the parity, while inverting the time-reversal sign, to give the unique Hermitian matrices which transform as electric dipoles; see Tab.~\ref{tab:eds-gamma-class}.
Ferroelectric modes couple to Dirac fermions in the same way~\cite{KoziiBiRuhman2019, Kozii2022}, as expected from symmetry.
The electrostatic coupling term thus equals
\begin{align}
\begin{aligned}
\action_c[\psi, \Phi] &= \frac{\iu}{\sqrt{\upbeta L^d}} \sum_{q} \Phi_{-q} \rho_{q} \\
&= - \frac{\iu}{\sqrt{\upbeta L^d}} \sum_{kp} \bar{\psi}_{k} A(k, p) \psi_{p} \Phi_{k-p},
\end{aligned}
\end{align}
where $\upbeta = 1 / (k_B T)$, $L^d$ is the volume, and
\begin{align}
\rho_{q} &= - \sum_k \bar{\psi}_{k} A(k, k+q) \psi_{k+q} \label{eq:eds-rho-psi}
\end{align}
is the density.
In the bare interaction vertex
\begin{align}
A(k, p) &= e \gamma_0 + \iu \eta_{\perp} (k_x - p_x) \gamma_0 \gamma_1 + \iu \eta_{\perp} (k_y - p_y) \gamma_0 \gamma_2 + \iu \eta_z (k_z - p_z) \gamma_0 \gamma_3 \label{eq:eds-A-psi}
\end{align}
we allow for anisotropy between the in-plane $\eta_{\perp}$ and out-of-plane $\eta_z$ electric dipole moments.
For later convenience, we retained the dependence of $A(k, p)$ on both the incoming $p = (\omega_p, \vb{p})$ and outgoing $k = (\omega_k, \vb{k})$ electron four-momenta.

\subsection{Polarization and optical conductivity} \label{sec:el-dip-sc-Dirac-opticond}
The polarization or plasmon self-energy $\Pi(q)$ is defined with the convention
\begin{align}
\mathscr{V}^{-1}(q) &= V^{-1}(q) + \Pi(q),
\end{align}
where
\begin{align}
\mathscr{V}(q) \Kd_{q + q'} = \ev{\Phi_{q} \Phi_{q'}}
\end{align}
is the dressed plasmon propagator.
The small-momentum behavior of the polarization determines the symmetric part of the optical conductivity in the following way:
\begin{align}
\sigma_{ij}(\varpi_q) &= - \iu \frac{\varpi_q}{2} \mleft.\frac{\partial^2 \Pi^R(\varpi_q, \vb{q})}{\partial q_i \partial q_j}\mright|_{\vb{q} = \vb{0}}.
\end{align}
Here, $\Pi^R(q) = \Pi^R(\varpi_q, \vb{q})$ is the retarded real-time polarization which is obtained from $\Pi(q) = \Pi(\omega_q, \vb{q})$ via analytic continuation $\iu \omega_q \to \hbar \varpi_q + \iu 0^+$.

Within RPA, $\Pi(q)$ is given by the fermionic polarization bubble which would have the form
\begin{align}
\Pi(q) \propto \sum_{\vb{k} n n' s s'} \frac{f_{\vb{k} + \vb{q} n} - f_{\vb{k} n'}}{\varepsilon_{\vb{k} + \vb{q} n} - \varepsilon_{\vb{k} n'} + \iu \omega_q} \abs{\braket{u_{\vb{k} + \vb{q} n s}}{u_{\vb{k} n' s'}}}^2
\end{align}
if we ignored the dipolar coupling.
Here, $\varepsilon_{\vb{k} n}$ are the dispersions, $u_{\vb{k} n s}$ the eigenvectors, and $f_{\vb{k} n} = 1 / \mleft(\Elr^{\upbeta \varepsilon_{\vb{k} n}} + 1\mright)$ are the Fermi-Dirac occupation factors.

In most systems, the electric monopole-monopole contribution to $\Pi(q)$, which is schematically written above, is dominant and gives the leading contribution to the optical conductivity.
However, in quasi-2D systems the Hamiltonian $H_{\vb{k}}$ has weak $k_z$-dependence, making both $\varepsilon_{\vb{k} n}$ and $u_{\vb{k} n s}$ weakly dependent on $k_z$, in contrast to the coupling of the $z$-axis electric dipoles $\eta_z$ [Eq.~\eqref{eq:eds-A-psi}].
It then follows that the monopole-monopole contribution to $\sigma_{zz}(\varpi_q)$ is small in quasi-2D systems, whereas the dipolar contributions can be large.
In particular, for the model of the previous section we have evaluated the polarization in the quasi-2D limit:
\begin{align}
v_z &= 0, & \eta_{\perp} &= 0,
\end{align}
which is also of interest for RG reasons discussed in the next section.
The $T = 0$ result is:
\begin{align}
\Pi^R(\varpi_q, 0, 0, q_z) &= \frac{\Lambda_z m^2 \eta_z^2 q_z^2}{\pi^2 v^2 \hbar \varpi_q} \mleft[\log\abs{\frac{2 \upmu + \hbar \varpi_q}{2 \upmu - \hbar \varpi_q}} + \iu \pi \HTh\mleft(\hbar \abs{\varpi_q} - 2 \upmu\mright)\mright], \label{eq:eds-Pi-R-quasi2D}
\end{align}
where $\Lambda_z$ is the $q_z$ cutoff, $q_z \in [- \Lambda_z, \Lambda_z]$, $\upmu = \sqrt{m^2 + v^2 k_F^2}$ is the chemical potential, and $\HTh$ is the Heaviside theta function.
Note that in the no doping limit $k_F \to 0$, $\upmu$ should go to $m$, not $0$, in the above expression.
This result we derive below, in Sec.~\ref{sec:el-dip-sc-polarization}.

The $z$-axis optical conductivity is therefore exclusively given by the $z$-axis dipole fluctuations:
\begin{align}
\sigma_{zz}(\varpi_q) &= \frac{\Lambda_z m^2 \eta_z^2}{\pi^2 v^2 \hbar} \mleft(\pi \HTh\mleft(\hbar \abs{\varpi_q} - 2 \upmu\mright) - \iu \log\abs{\frac{2 \upmu + \hbar \varpi_q}{2 \upmu - \hbar \varpi_q}}\mright).
\end{align}
Due to interband excitations, above the gap we obtain a flat real part of the conductivity, which is very similar to the usual behavior of the in-plane optical conductivity for a two-dimensional Dirac spectrum~\cite{Gorbar2002, Pyatkovskiy2008}.
The surprise is that we obtain this result for the $z$-axis conductivity, even though the band velocity along this direction is zero.
The matrix element responsible for this is exclusively the anomalous dipole element of Eq.~\eqref{eq:eds-dipol-matrix-elem}.
The band gap $m$ and in-plane Fermi velocity $v$ entering $\sigma_{zz}(\varpi_q)$ can both be measured using ARPES.
If one finds weak to no dispersion along the $\vu{e}_z$ direction in ARPES, but nonetheless measures a substantial $z$-axis optical conductivity, then this provides direct evidence for the $z$-axis dipole elements of our theory.

In summary, in quasi-2D Dirac systems the $z$-axis dipole fluctuations that are so important for our pairing mechanism of Sec.~\ref{sec:el-dip-sc-Dirac-pairing} are directly observable in the $z$-axis optical conductivity.

\subsubsection{Evaluation of the polarization bubble} \label{sec:el-dip-sc-polarization}
Here, we evaluate the lowest-order contribution to the polarization $\Pi(q)$.
Because of the RG considerations discussed in the next Sec.~\ref{sec:el-dip-sc-Dirac-RG}, we only consider the quasi-2D limit:
\begin{align}
v_z &= 0, & \eta_{\perp} &= 0.
\end{align}
For quasi-2D geometries, we shall find it convenient to use bolded vectors with $\perp$ subscripts to denote in-plane vectors, as in:
\begin{align}
\begin{aligned}
k &= (\omega_k, \vb{k}_{\perp}, k_z), &\hspace{50pt}
\vb{k}_{\perp} &= (k_x, k_y), \\
q &= (\omega_q, \vb{q}_{\perp}, q_z), &\hspace{50pt}
\vb{q}_{\perp} &= (q_x, q_y).
\end{aligned}
\end{align}
Except the real-time polarization for $\vb{q}_{\perp} = \vb{0}$ that we gave in Eq.~\eqref{eq:eds-Pi-R-quasi2D}, here we also evaluate the imaginary-time polarization for $\upmu = 0$ and for $\omega_q = 0$.
The former we shall use during the RG of the next Sec.~\ref{sec:el-dip-sc-Dirac-RG}, while the latter is employed in Sec.~\ref{sec:el-dip-sc-Dirac-pairing} where we investigate the pairing instabilities of our model.

\begin{figure}[t]
\centering
\includegraphics[width=0.45\textwidth]{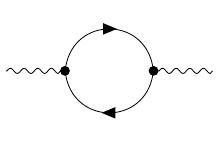}
\captionbelow[The fermionic bubble diagram which gives the leading contribution to the polarization.]{\textbf{The fermionic bubble diagram which gives the leading contribution to the polarization.}
The solid vertices contain both monopole and dipole contributions, as specified in Fig.~\ref{fig:el-dip-sc-plasmon-fermion-vertex}.
Solid lines stand for electrons and wiggly lines for plasmons.
The corresponding expression is given in Eq.~\eqref{eq:eds-bubble-diagram-expression}.}
\label{fig:el-dip-sc-polarization-diagram}
\end{figure}

The polarization is defined with the convention $\Pi(q) = \mathscr{V}^{-1}(q) - V^{-1}(q)$, where $\mathscr{V}(q) \Kd_{q + q'} = \ev{\Phi_{q} \Phi_{q'}}$ is the dressed plasmon propagator.
To lowest order in the coupling, it is given by the fermionic bubble diagram:
\begin{align}
\Pi(q) &= - \int \frac{\dd[4]{k}}{(2\pi)^4} \Tr G(k) A(k, k+q) G(k+q) A(k+q, k), \label{eq:eds-bubble-diagram-expression}
\end{align}
where the thermodynamic and $T = 0$ limits were taken,
\begin{align}
G(k) &= \frac{m \one + \iu \mleft[(\omega_k - \iu \upmu) \gamma_0 + v (k_x \gamma_1 + k_y \gamma_2)\mright]}{m^2 + (\omega_k - \iu \upmu)^2 + v^2 \vb{k}_{\perp}^2} \equiv \frac{\mathcal{X}_k}{\mathcal{Y}_k}
\end{align}
is the bare fermionic Green's function [Eq.~\eqref{eq:eds-G-1-psi}], and
\begin{align}
A(k, p) &= e \gamma_0 + \iu \eta_z (k_z - p_z) \gamma_0 \gamma_3
\end{align}
is the bare vertex in the quasi-2D limit under consideration [Eq.~\eqref{eq:eds-A-psi}].
The corresponding diagram is drawn in Fig.~\ref{fig:el-dip-sc-polarization-diagram}.
$\mathcal{X}_k$ ($\mathcal{Y}_k$) is a shorthand for the numerator (denominator) of $G(k)$.

First, we consider the retarded real-time polarization $\Pi^R(\varpi_q, \vb{q}_{\perp} = \vb{0}, q_z)$ for finite and positive $\upmu \geq m$, finite real-time frequencies $\varpi_q \neq 0$, arbitrary $q_z$, and vanishing $\vb{q}_{\perp} = (q_x, q_y) = \vb{0}$.
Because the dispersion does not depend on $q_z$, it is straightforward to evaluate the frequency integral to get:
\begin{align}
\Pi(\omega_q, \vb{q}_{\perp} = \vb{0}, q_z) &= \frac{\Lambda_z}{\pi} \int_{k_F}^{\infty} \frac{k_{\perp} \dd{k_{\perp}}}{2 \pi} \frac{2 m^2 \eta_z^2 q_z^2}{m^2 + v^2 k_{\perp}^2} \mleft(\frac{1}{2 \sqrt{m^2 + v^2 k_{\perp}^2} + \iu \omega_q} + \frac{1}{2 \sqrt{m^2 + v^2 k_{\perp}^2} - \iu \omega_q}\mright),
\end{align}
where $\Lambda_z$ is the $q_z$ cutoff, $q_z \in [- \Lambda_z, \Lambda_z]$, and $k_F$ is the Fermi wavevector, $\upmu = \sqrt{m^2 + v^2 k_F^2}$.
The retarded real-time polarization is obtained through the substitution $\iu \omega_q \to \hbar \varpi_q + \iu 0^+$.
After applying the Sokhotski-Plemelj formula and evaluating the momentum integral, one obtains the result
\begin{align}
\Pi^R(\varpi_q, \vb{q}_{\perp} = \vb{0}, q_z) &= \frac{\Lambda_z m^2 \eta_z^2 q_z^2}{\pi^2 v^2 \hbar \varpi_q} \mleft[\log\abs{\frac{2 \upmu + \hbar \varpi_q}{2 \upmu - \hbar \varpi_q}} + \iu \pi \HTh\mleft(\hbar \abs{\varpi_q} - 2 \upmu\mright)\mright]
\end{align}
which was provided in Eq.~\eqref{eq:eds-Pi-R-quasi2D}.

For the next two cases, we express the denominator with the help of the Feynman parametrization:
\begin{align}
\frac{1}{\mathcal{Y}_k \mathcal{Y}_{k+q}} &= \int_0^1 \dd{x} \frac{1}{\mleft[(1-x) \mathcal{Y}_k + x \mathcal{Y}_{k+q}\mright]^2} = \int_0^1 \dd{x} \frac{1}{\mleft[\mathcal{Y}_p + x (1-x) \mleft(\omega_q^2 + v^2 \vb{q}_{\perp}^2\mright)\mright]^2},
\end{align}
where $p = k + x \, q$.
In the momentum integral we then switch from $k$ to $p$.
Up to terms which are odd in any component of $p$ and thus vanish under the integral, the numerator trace equals
\begin{align}
- \Tr \mathcal{X}_k A(k, k+q) \mathcal{X}_{k+q} A(k+q, k) &= \mathcal{E}_1 + \mathcal{E}_2 \cdot (1-2x) + \mathcal{E}_3 \cdot x(1-x) + \mathcal{E}_4 \cdot (v^2 \vb{p}_{\perp}^2 - \omega_p^2),
\end{align}
where
\begin{align}
\begin{aligned}
\mathcal{E}_1 &= - 4 (e^2 - \eta_z^2 q_z^2) m^2 - 4 (e^2 + \eta_z^2 q_z^2) \upmu^2, &\hspace{50pt}
\mathcal{E}_2 &= - 4 (e^2 + \eta_z^2 q_z^2) \upmu \iu \omega_q, \\
\mathcal{E}_3 &= 4 (e^2 + \eta_z^2 q_z^2) (v^2 \vb{q}_{\perp}^2 - \omega_q^2), &\hspace{50pt}
\mathcal{E}_4 &= - 4 (e^2 + \eta_z^2 q_z^2).
\end{aligned}
\end{align}

When $\upmu = 0$, there is an $\SO(3)$ symmetry in the $(\omega_p, v \vb{p}_{\perp})$ variables because of which in the numerator $\vb{p}_{\perp}^2 \to \frac{2}{3} p^2$ and $\omega_p^2 \to \frac{1}{3} v^2 p^2$.
The radial integral is then readily evaluated using dimensional regularization:
\begin{align}
\int_0^{\infty} \frac{p^{2+\epsilon} \dd{p}}{(\Delta^2 + v^2 p^2)^2} &= \frac{(1+\epsilon) \pi}{4 v^4 \cos\frac{\epsilon \pi}{2}} \mleft(\Delta / v\mright)^{\epsilon-1} = \begin{cases}
\displaystyle \frac{\pi}{4 v^3 \Delta}, & \text{for $\epsilon = 0$,} \\[8pt]
\displaystyle - \frac{3 \pi \Delta}{4 v^5}, & \text{for $\epsilon \to 2$.}
\end{cases}
\end{align}
The $\epsilon = 2$ case, which arises during the evaluation of the $\mathcal{E}_4$ term contribution, formally diverges.
This divergence is actually spurious.
If instead of radially integrating in frequency and momentum, one first executes the frequency integral and then the momentum integral, one finds a convergent result for the $\mathcal{E}_4$ contribution which agrees with the dimensionally regularized result.
In detail, the integrals
\begin{align}
\int_{-\infty}^{\infty} \frac{v^2 \vb{p}_{\perp}^2 \dd{\omega}}{\mleft(m^2 + v^2 \vb{p}_{\perp}^2 + \omega^2\mright)^2} &= \frac{\pi v^2 \vb{p}_{\perp}^2}{2 (m^2 + v^2 \vb{p}_{\perp}^2)^{3/2}}, \\
\int_{-\infty}^{\infty} \frac{\omega^2 \dd{\omega}}{\mleft(m^2 + v^2 \vb{p}_{\perp}^2 + \omega^2\mright)^2} &= \frac{\pi}{2 \sqrt{m^2 + v^2 \vb{p}_{\perp}^2}}
\end{align}
individually both go like $p_{\perp}^{-1}$ for large $p_{\perp}$.
This makes their in-plane momentum integrals linearly divergent.
However, their sum
\begin{align}
\int_{-\infty}^{\infty} \frac{\mleft(\omega^2 - v^2 \vb{p}_{\perp}^2\mright) \dd{\omega}}{\mleft(m^2 + v^2 \vb{p}_{\perp}^2 + \omega^2\mright)^2} = \frac{\pi m^2}{2 (m^2 + v^2 \vb{p}_{\perp}^2)^{3/2}}
\end{align}
goes like $p_{\perp}^{-3}$ at large $p_{\perp}$, giving a convergent result which agrees with dimensional regularization.
The $x$ integrals can be evaluated through a $x \to y = 4 x (1-x)$ substitution with the help of
\begin{align}
\int_0^1 \frac{\dd{y}}{\sqrt{1-y}} \frac{1}{\sqrt{1 + Q^2 y}} &= \frac{2}{Q} \arccot\frac{1}{Q}, \\
\int_0^1 \frac{y \dd{y}}{\sqrt{1-y}} \frac{1}{\sqrt{1 + Q^2 y}} &= \frac{1}{Q} \mleft[\mleft(1 - \frac{1}{Q^2}\mright) \arccot\frac{1}{Q} + \frac{1}{Q}\mright], \\
\int_0^1 \frac{\dd{y}}{\sqrt{1-y}} \sqrt{1 + Q^2 y} &= Q \mleft[\mleft(1 + \frac{1}{Q^2}\mright) \arccot\frac{1}{Q} + \frac{1}{Q}\mright].
\end{align}
The final result is
\begin{align}
\mleft.\Pi(q)\mright|_{\upmu = 0} &= \frac{\Lambda_z \vb{q}_{\perp}^2 (e^2 + \eta_z^2 q_z^2)}{4 \pi^2 \sqrt{\omega_q^2 + v^2 \vb{q}_{\perp}^2}} \mleft[(1-r_q^2) \arccot r_q + r_q\mright] + \frac{2 \Lambda_z m^2 \eta_z^2 q_z^2}{\pi^2 v^2 \sqrt{\omega_q^2 + v^2 \vb{q}_{\perp}^2}} \arccot r_q, \label{eq:eds-mu0-polarization-previous}
\end{align}
where $q = (\omega_q, \vb{q}_{\perp}, q_z)$, $\vb{q}_{\perp} = (q_x, q_y)$, and $r_q \defeq 2 m / \sqrt{\omega_q^2 + v^2 \vb{q}_{\perp}^2}$.
This $\upmu = 0$ polarization reproduces the polarization of Ref.~\cite{DTSon2007} in the $m \to 0$, $\eta_z \to 0$ limit.

When $\omega_q = 0$, but $\upmu \geq m$ is finite and positive, we proceed by first evaluating the frequency integral.
We write:
\begin{align}
\begin{aligned}
\Pi(\omega_q = 0, \vb{q}) &= \frac{\Lambda_z}{\pi} \int_0^1 \dd{x} \int_0^{\infty} \frac{p_{\perp} \dd{p_{\perp}}}{2 \pi} \\
&\hspace{26pt} \times \int_{-\infty}^{\infty} \frac{\dd{\omega_p}}{2 \pi} \mleft(\frac{4 (e^2 + \eta_z^2 q_z^2)}{\mathcal{Y}_p + x (1-x) v^2 \vb{q}_{\perp}^2} + \frac{- 8 e^2 m^2 - 8 (e^2 + \eta_z^2 q_z^2) v^2 \vb{p}_{\perp}^2}{\mleft[\mathcal{Y}_p + x (1-x) v^2 \vb{q}_{\perp}^2\mright]^2}\mright).
\end{aligned}
\end{align}
Note that during the evaluation of the contour integral, one must not overlook the additional Dirac delta function that appears in the second term:
\begin{align}
\int_{-\infty}^{\infty} \frac{\dd{\omega}}{2 \pi} \frac{1}{\Delta + (\omega - \iu \upmu)^2} &= \frac{1}{2 \sqrt{\Delta}} \HTh(\Delta - \upmu^2), \\
\int_{-\infty}^{\infty} \frac{\dd{\omega}}{2 \pi} \frac{1}{\mleft[\Delta + (\omega - \iu \upmu)^2\mright]^2} &= \frac{1}{2 \sqrt{\Delta}} \mleft(\frac{\HTh(\Delta - \upmu^2)}{2 \Delta} - \Dd(\Delta - \upmu^2)\mright).
\end{align}
The $p_{\perp}$ and $x$ integrals are now readily evaluated.
For $q_{\perp} \leq 2 k_F$, $p_{\perp}$ goes from $\sqrt{k_F^2 - x(1-x) q_{\perp}^2}$ to infinity for all $x$.
For $q_{\perp} > 2 k_F$, one has to separately consider $\abs{x}$ which are smaller and larger than $\frac{1}{2} \mleft(1 - \sqrt{1 - 4 k_F^2 / q_{\perp}^2}\mright)$.
After some lengthy algebra, one finds that
\begin{align}
\frac{\Pi(\omega_q = 0, \vb{q})}{\DOSg_F e^2} &= \begin{cases}
\displaystyle 1 + \frac{\eta_z^2 q_z^2}{e^2}, & \text{for $q_{\perp} \leq 2 k_F$,} \\[16pt]
\displaystyle \begin{aligned}
&\mleft(1 + \frac{\eta_z^2 q_z^2}{e^2}\mright) \mleft(1 - \frac{\sqrt{q_{\perp}^2 - 4 k_F^2}}{2 q_{\perp}}\mright) \\[4pt]
&\hspace{6pt} + \frac{\displaystyle \mleft[1 + \frac{\eta_z^2 q_z^2}{e^2}\mright] v^2 q_{\perp}^2 - 4 m \mleft[1 - \frac{\eta_z^2 q_z^2}{e^2}\mright]}{4 \upmu v q_{\perp}} \arctan\frac{v \sqrt{q_{\perp}^2 - 4 k_F^2}}{2 \upmu},
\end{aligned} & \text{for $q_{\perp} > 2 k_F$,}
\end{cases} \label{eq:eds-static-smallq-pol-previous}
\end{align}
where
\begin{align}
\DOSg_F &= \frac{\Lambda_z \upmu}{\pi^2 v^2}, &
q_{\perp} &= \sqrt{q_x^2 + q_y^2}, &
\upmu &= \sqrt{m^2 + v^2 k_F^2}.
\end{align}
In the $\eta_z \to 0$ limit, this $\Pi(\omega_q = 0, \vb{q})$ reduces to the expression derived in Refs.~\cite{Gorbar2002, Pyatkovskiy2008}.

\subsection{Renormalization group analysis} \label{sec:el-dip-sc-Dirac-RG}
\begin{figure}[t!]
\centering
\includegraphics[width=0.70\textwidth]{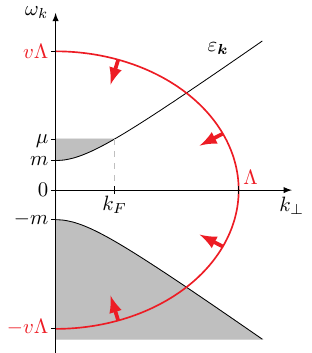}
\captionbelow[A schematic of the renormalization-group procedure.]{\textbf{A schematic of the renormalization-group procedure.}
Here, $\omega_k$ and $k_{\perp} = \sqrt{k_x^2 + k_y^2}$ are the frequency and in-plane momentum, respectively, $2 m$ is the band gap, $\upmu$ is the chemical potential, $k_F$ is the Fermi wavevector, and $\varepsilon_{\vb{k}} = \sqrt{m^2 + v^2 \vb{k}_{\perp}^2}$ is the dispersion.
The occupied states are shaded in grey and the cutoff of Eq.~\eqref{eq:eds-cutoff-def} is highlighted in red.
Arrows indicate the direction of the renormalization-group flow.}
\label{fig:el-dip-sc-RG-sketch}
\end{figure}

Here we study how the fluctuations of high-energy states modify the low-energy physics of our model.
To this end, we first analyze the naive scaling under RG flow, which is depicted in Fig.~\ref{fig:el-dip-sc-RG-sketch}.
We show that the electric dipole coupling is irrelevant in 3D systems, while in quasi-2D systems its out-of-plane component is marginal.
Afterwards, for the quasi-2D case we derive the $1$-loop RG flow equations in the limit of a large number of fermionic flavors $N$.
The technical parts of this calculation we delegate to the end of this section (Secs.~\ref{sec:el-dip-sc-RG-diagrams}).
Using these $1$-loop RG flow equations, we establish that the out-of-plane dipolar coupling $\eta_z$ is marginally relevant (Fig.~\ref{fig:el-dip-sc-RG-etaz-flow}).
Consequently, $\eta_z$ becomes enhanced at low energies.

Cooper pairing, which we study in the next section, takes place only when the screening is strong enough.
The Thomas-Fermi wavevector $k_{\text{TF}} = \sqrt{e^2 \DOSg_F / \epsilon_0}$ thus needs to be larger than the Fermi sea size $k_F$.
Since the density of states $\DOSg_F \propto k_F^2 / (\hbar v_F)$, $k_{\text{TF}} \propto k_F \sqrt{\upalpha}$ where $\upalpha = e^2 / (\hbar v_F \epsilon_0)$ is the monopole coupling constant.
For this reason, throughout this section we focus on the strong-coupling regime $\upalpha \gg 1$.

The strong-coupling regime is not accessible through direct perturbation theory, which is why we use a large-$N$ expansion, $N$ being the number of fermion flavors.
Formally, we modify the model by introducing an additional summation over fermionic flavor indices in Eqs.~\eqref{eq:eds-S-psi} and~\eqref{eq:eds-rho-psi}.
Although in the end we take $N$ to be of order unity, the hope is that by organizing the calculation in orders of $1/N$ we can at least make definite statements about some strongly coupled model that resembles our model.
When the band inversion point is not located at $\vb{k} = 0$, multiple valleys arise, each described by a Dirac model.
This naturally gives larger values for $N$, provided that the intervalley interactions are small.

At the start of the RG procedure, the momentum cutoff $\Lambda$ is initially much larger than the Fermi wave vector $k_F$ and we integrate out high-energy degrees of freedom until $\Lambda$ becomes comparable to $k_F$; see Fig.~\ref{fig:el-dip-sc-RG-sketch}.
To a first approximation, we may thus set the chemical potential mid-gap, i.e., $k_F$ to zero.
Since we are only interested in the low-temperature physics, we may also set $T = 0$.
Throughout this section, we thus set
\begin{align}
\upmu &= 0, & T &= 0. \label{eq:eds-muT0-limit}
\end{align}
Finite $\upmu$ and $T$ are both reintroduced later when we study Cooper pairing given a cutoff $\Lambda \sim k_F$.

\subsubsection{Tree-level scaling} \label{sec:el-dip-sc-Dirac-RG-tree}
First, we study the tree-level scaling (when $\upmu = k_F = T = 0$).
In light of the Dirac form, the cutoff $\Lambda$ we impose on both momenta and frequencies according to
\begin{align}
\norm{k}^2 \defeq \omega_k^2 / v^2 + k_x^2 + k_y^2 + (v_z / v)^2 k_z^2 < \Lambda^2. \label{eq:eds-cutoff-def}
\end{align}
The full action~\eqref{eq:eds-model-action}, with all of its terms spelled out, is given by
\begin{align}
\begin{aligned}
\action[\psi, \Phi] &= \sum_k^{\Lambda} \bar{\psi}_{k} \mleft[m \one - \iu \mleft(\omega_k \gamma_0 + v (k_x \gamma_1 + k_y \gamma_2) + v_z k_z \gamma_3\mright)\mright] \psi_{k} + \frac{1}{2} \sum_q^{\Lambda} \Phi_{-q} \mleft[\epsilon_{\perp} \vb{q}_{\perp}^2 + \epsilon_z q_z^2\mright] \Phi_{q} \\[-4pt]
&\qquad - \frac{\iu}{\sqrt{\upbeta L^d}} \sum_{kpq}^{\Lambda} \Kd_{k-p+q} \Phi_{-q} \bar{\psi}_{k} \mleft[e \gamma_0 - \iu \eta_{\perp} q_x \gamma_0 \gamma_1 - \iu \eta_{\perp} q_y \gamma_0 \gamma_2 - \iu \eta_z q_z \gamma_0 \gamma_3\mright] \psi_{p},
\end{aligned} \label{eq:eds-model-action-again}
\end{align}
where the sum over the $N$ fermionic flavor indices has been suppressed.
The fields $\psi = \psi_< + \psi_>$ and $\Phi = \Phi_< + \Phi_>$ we decompose into slow and fast parts with four-momenta within $0 \leq \norm{k} < \Lambda/b$ and $\Lambda/b < \norm{k} < \Lambda$, respectively; here $b = \Elr^{\ell} > 1$.
\pagebreak The naive slow-field action, which is \linebreak
obtained by substituting the slow fields into the above equation, can be recast into the original action written above [Eq.~\eqref{eq:eds-model-action-again}] through the rescaling:
\begin{align}
\begin{aligned}
k &\mapsto k' = b^1 k, \\
\psi_{k} &\mapsto \psi'_{k'} = b^{-\eta_{\psi}} \psi_{k}, \\
\Phi_{k} &\mapsto \Phi'_{k'} = b^{-\eta_{\Phi}} \Phi_{k}.
\end{aligned}
\end{align}
The $\eta_{\psi}$ and $\eta_{\Phi}$ exponents we choose so that the fermionic frequency $\bar{\psi}_{k} \omega_k \psi_{k}$ and monopole coupling $\Phi_{-q} \psi_{k} e \gamma_0 \psi_{k+q}$ terms are invariant.

In 3+1D, for the naive scaling exponents we find
\begin{align}
\begin{aligned}
m' &= b^{-4} b^{2 \eta_{\psi}} m &&= b^1 m, \\
Z_{\omega}' &= b^{-4} b^{2 \eta_{\psi}} b^{-1} &&\equiv 1 \implies \eta_{\psi} = \frac{5}{2}, \\
v' &= b^{-4} b^{2 \eta_{\psi}} b^{-1} v &&= v, \\
v_z' &= b^{-4} b^{2 \eta_{\psi}} v^{-1} v_z &&= v_z, \\
\epsilon_{\perp}' &= b^{-4} b^{2 \eta_{\Phi}} b^{-2} \epsilon_{\perp} &&= \epsilon_{\perp}, \\
\epsilon_{z}' &= b^{-4} b^{2 \eta_{\Phi}} b^{-2} \epsilon_{z} &&= \epsilon_{z}, \\
e' &= b^{-8} b^{2 \eta_{\psi} + \eta_{\Phi}} e &&\equiv e \implies \eta_{\Phi} = 3, \\
\eta_{\perp}' &= b^{-8} b^{2 \eta_{\psi} + \eta_{\Phi}} b^{-1} \eta_{\perp} &&= b^{-1} \eta_{\perp}, \\
\eta_{z}' &= b^{-8} b^{2 \eta_{\psi} + \eta_{\Phi}} b^{-1} \eta_{z} &&= b^{-1} \eta_{z},
\end{aligned}
\end{align}
where $Z_{\omega}$ is the proportionality constant of the fermionic frequency term $\bar{\psi}_{k} \omega_k \psi_{k}$.
The first $b^{-4}$ and $b^{-8}$ factors come from the rescaling of the four-momentum integral(s), the middle $\sim \eta_{\psi}, \eta_{\Phi}$ factors come from the fields, while the last factor, if present, comes from any additional powers of momentum present in the corresponding term.
If we call $\action'$ the action~\eqref{eq:eds-model-action-again} with $m, v, \ldots$ replaced by the $m', v', \ldots$ from above, but the same cutoff $\Lambda$, then $\action[\psi_<, \Phi_<] = \action'[\psi', \Phi']$.
The coupling constant of a general local momentum-conserving term which we may schematically write as ($M, K, L_1, L_2 \in \N_0$)
\begin{align}
\sim g \sum_{kq} \Kd_{\sum k + \sum q} \psi_k^M \Phi_q^K \abs{k}^{L_1} \abs{q}^{L_2}
\end{align}
scales as
\begin{align}
g' &= b^{4 - 3M/2 - K - L_1 - L_2} g.
\end{align}
The electric dipole couplings $\eta_{\perp}$ and $\eta_{z}$ are thus naively irrelevant, as are all higher-order momentum-conserving local terms in the action which preserve $\Phi \to - \Phi$ symmetry and particle number.
Because the scaling of $\eta_{\perp}$ and $\eta_z$ only receives loop corrections of order $N^{-1}$ or higher, in 3D Dirac systems electric dipole moments become increasingly weak at low energies.

In quasi-2D systems, however, $v_z \approx 0$ and the Fermi surface is cylindrical instead of spherical.
Consequently, during the RG we do not rescale the momenta along $z$.
This changes the naive scaling dimensions to
\begin{align}
\begin{aligned}
m' &= b^{-3} b^{2 \eta_{\psi}} m &&= b^1 m, \\
Z_{\omega}' &= b^{-3} b^{2 \eta_{\psi}} b^{-1} &&\equiv 1 \implies \eta_{\psi} = 2, \\
v' &= b^{-3} b^{2 \eta_{\psi}} b^{-1} v &&= v, \\
\epsilon_{\perp}' &= b^{-3} b^{2 \eta_{\Phi}} b^{-2} \epsilon_{\perp} &&= b^{-1} \epsilon_{\perp}, \\
\epsilon_{z}' &= b^{-3} b^{2 \eta_{\Phi}} \epsilon_{z} &&= b^{1} \epsilon_{z}, \\
e' &= b^{-6} b^{2 \eta_{\psi} + \eta_{\Phi}} e &&\equiv e \implies \eta_{\Phi} = 2, \\
\eta_{\perp}' &= b^{-6} b^{2 \eta_{\psi} + \eta_{\Phi}} b^{-1} \eta_{\perp} &&= b^{-1} \eta_{\perp}, \\
\eta_{z}' &= b^{-6} b^{2 \eta_{\psi} + \eta_{\Phi}} \eta_{z} &&= \eta_{z}, \\
g' & &&= b^{3 - M - K - L_1 - L_2} g.
\end{aligned}
\end{align}
Hence the out-of-plane dipole moment is now marginal, and we shall later see that loop corrections make it marginally relevant.
The monopole coupling $e$ remains marginal.
Intuitively, the reason why the dipolar couplings $\eta_{\perp}$ and $\eta_z$ were previously irrelevant is that they come with an additional power of momentum compared to the charge $e$.
As this momentum becomes smaller because of the restricted momentum range ($\Lambda \to \Lambda/b$), they become increasingly less important, at least in three dimensions for $k_F \ll \Lambda$.
In quasi-2D systems, however, the exchanged momentum along the $\vu{e}_z$ direction is always large (on the order of the Brillouin zone height) and the importance of the $\eta_z$ term is always (naively) the same, which explains its marginality as $\Lambda$ is decreased.
As for $\epsilon_z$, it is relevant, as expected for what is essentially a $z$-dependent mass of the plasmon field.
However, the electrons themselves also gap the plasmon field and in the strong screening limit their contribution is dominant.
This is why we do not consider the flow of $\epsilon_z$ later on.

Given our interest in dipole effects, we focus on quasi-2D systems.
Since $\eta_{\perp}$ is irrelevant, we may set it to zero from the outset.
We therefore consider the regime
\begin{align}
v_z &= 0, & \eta_{\perp} &= 0 \label{eq:eds-quasi-2D-param}
\end{align}
from now on.
In practice, the $z$-axis dispersion and $\eta_{\perp}$ have to be small compared to $m$ and $\eta_z$, respectively, for our calculation to apply.
For quasi-2D geometries, we shall find it convenient to use bolded vectors with $\perp$ subscripts to denote in-plane vectors.
For instance:
\begin{align}
\begin{aligned}
k &= (\omega_k, \vb{k}_{\perp}, k_z), &\hspace{50pt}
\vb{k}_{\perp} &= (k_x, k_y), \\
q &= (\omega_q, \vb{q}_{\perp}, q_z), &\hspace{50pt}
\vb{q}_{\perp} &= (q_x, q_y).
\end{aligned} \label{eq:eds-bold-vec-conv}
\end{align}

\subsubsection{$1$-loop RG flow equations} \label{sec:el-dip-sc-Dirac-RG-flow}
To formulate the RG flow equations, we use the Callan-Symanzik equations~\cite{Zinn-Justin2002}.
Let us assume that we have found how all the states up to the cutoff $\Lambda$ renormalize the fermionic Green's function $G(k)$ of Eq.~\eqref{eq:eds-G-1-psi} into $\mathscr{G}(k) = \ev{\psi_{k} \bar{\psi}_{k}}$:
\begin{align}
\mathscr{G}^{-1}(k) &= Z_m m \one - \iu Z_{\omega} \omega_k \gamma_0 - \iu Z_v v (k_x \gamma_1 + k_y \gamma_2) + \cdots \, ,
\end{align}
and the same for the interaction vertex $A(k, p) \to \mathscr{A}(k, p)$ of Eq.~\eqref{eq:eds-A-psi}:
\begin{align}
\mathscr{A}(k, p) = Z_e e \gamma_0 + \iu Z_{\eta z} \eta_z (k_z - p_z) \gamma_0 \gamma_3 + \cdots \, .
\end{align}
The Callan-Symanzik equations follow from the requirement that this asymptotic behavior for small $k, p$ stays preserved as we change $\Lambda$.
Before imposing this, we need to fix the scale of the fields $\psi$ and $\Phi$ which can in general depend on $\Lambda$.
We choose $\psi \to Z_{\omega}^{-1/2} \psi$, in which case the Callan-Symanzik equations take the form:
\begin{align}
\begin{aligned}
\dv{}{\Lambda} \mleft[\frac{Z_m}{Z_{\omega}} m\mright] &= 0, &\hspace{50pt}
\dv{}{\Lambda} \mleft[\frac{Z_v}{Z_{\omega}} v\mright] &= 0, \\
\dv{}{\Lambda} \mleft[\frac{Z_e}{Z_{\omega}} e\mright] &= 0, &\hspace{50pt}
\dv{}{\Lambda} \mleft[\frac{Z_{\eta z}}{Z_{\omega}} \eta_z\mright] &= 0.
\end{aligned}
\end{align}
Because $\Phi$ couples to the Noether charge of the $\Ugp(1)$ phase rotation symmetry $\psi \to \Elr^{\iu \vartheta} \psi$, there is an exact Ward identity $Z_e = Z_{\omega}$ which implies that the charge $e$ does not flow.
The proof of this important fact we provide later, in Sec.~\ref{sec:el-dip-sc-Ward-id}.
As for the other parameters, the chain rule gives the RG flow equations:
\begin{align}
\sum_j \mleft(\Kd_{ij} + \frac{g_j}{Z_i} \pdvc{Z_i}{g_j}{\Lambda, g_{\ell}}\mright) \frac{\Lambda}{g_j} \dv{g_j}{\Lambda} &= - \frac{\Lambda}{Z_i} \pdvc{Z_i}{\Lambda}{g_{\ell}},
\end{align}
where
\begin{align}
g_i &= \begin{pmatrix}
m \\ v \\ \eta_z
\end{pmatrix}, &
Z_i &= \begin{pmatrix}
{Z_m}/{Z_{\omega}} \\
{Z_v}/{Z_{\omega}} \\
{Z_{\eta z}}/{Z_{\omega}}
\end{pmatrix}.
\end{align}
Since $Z_i = 1 + \bigO(N^{-1})$, as we later show, to $N^{-1}$ order the RG flow equations simplify to:
\begin{align}
\frac{\Lambda}{g_i} \dv{g_i}{\Lambda} &= - \frac{\Lambda}{Z_i} \pdv{Z_i}{\Lambda}. \label{eq:eds-RG-flow-expr}
\end{align}
In these RG flow equations we have not included $\epsilon_{\perp}$ or $\epsilon_z$ because the bare interaction is negligible compared to the polarization in the strong coupling limit.
$\epsilon_{\perp}$ and $\epsilon_z$ we shall therefore keep constant (independent of $\Lambda$) and only include in various expressions to make them dimensionless.

To lowest order in $N$, the plasmon self-energy $\Pi(q)$ is given by the fermionic polarization bubble which is drawn in Fig.~\ref{fig:el-dip-sc-polarization-diagram}.
We have evaluated it in Sec.~\ref{sec:el-dip-sc-polarization}, with the result [Eq.~\eqref{eq:eds-mu0-polarization-previous}]:
\begin{align}
\Pi(q) &= N \frac{\Lambda_z \vb{q}_{\perp}^2 (e^2 + \eta_z^2 q_z^2)}{4 \pi^2 \sqrt{\omega_q^2 + v^2 \vb{q}_{\perp}^2}} \mleft[(1-r_q^2) \arccot r_q + r_q\mright] + N \frac{2 \Lambda_z m^2 \eta_z^2 q_z^2}{\pi^2 v^2 \sqrt{\omega_q^2 + v^2 \vb{q}_{\perp}^2}} \arccot r_q, \label{eq:eds-mu0-polarization}
\end{align}
where $\Lambda_z$ is the $q_z$ cutoff, $q_z \in [- \Lambda_z, \Lambda_z]$, and
\begin{align}
r_q \defeq \frac{2 m}{\sqrt{\omega_q^2 + v^2 \vb{q}_{\perp}^2}}.
\end{align}
Notice how $\Pi(q)$, unlike the bare $V^{-1}(q)$ of Eq.~\eqref{eq:eds-V-1-phi}, is frequency-dependent as well as non-analytic at $q = 0$.

The next step is to evaluate the various renormalization factors $Z_i$, which we do to $N^{-1}$ order.
The relevant self-energy and vertex correction diagrams are standard and the details of their evaluation are delegated to Sec.~\ref{sec:el-dip-sc-RG-diagrams}.
Although the shell integrals cannot be carried out analytically, they can be simplified by introducing the dimensionless parameters:
\begin{align}
\begin{aligned}
\tilde{m} &\defeq \frac{m}{v \Lambda}, &\hspace{80pt}
\upalpha &\defeq \frac{e^2}{\epsilon_{\perp} v}, \\
\tilde{\eta}_z &\defeq \frac{\Lambda_z \eta_z}{e}, &\hspace{80pt}
\tilde{\Lambda} &\defeq \frac{\Lambda}{\Lambda_z},
\end{aligned}
\end{align}
and expressing the shell momentum
\begin{align}
q = (\omega_q, \vb{q}_{\perp}, q_z)
\end{align}
in terms of dimensionless $\tilde{\omega}$ and $\tilde{q}_z$ through
\begin{align}
\omega_q &= v \Lambda \tilde{\omega}, &
\abs{\vb{q}_{\perp}} &= \Lambda \sqrt{1-\tilde{\omega}^2}, &
q_z &= \Lambda_z \tilde{q}_z.
\end{align}

The strong-coupling large-$N$ RG flow equations are to $1$-loop order given by:
\begin{align}
\begin{aligned}
\frac{1}{\tilde{m}} \dv{\tilde{m}}{\ell} &= \beta_{m} = 1 + \frac{4}{\displaystyle \pi^2 N (1+\tilde{m}^2)^2} \int_0^1 \dd{\tilde{q}_z} \int_0^1 \dd{\tilde{\omega}} \frac{\mathscr{B}_{m}(\tilde{\omega}, \tilde{q}_z)}{\mathscr{P}(\tilde{\omega}, \tilde{q}_z)}, \\
\frac{1}{\upalpha} \dv{\upalpha}{\ell} &= \beta_{\upalpha} = - \frac{4}{\displaystyle \pi^2 N (1+\tilde{m}^2)^2} \int_0^1 \dd{\tilde{q}_z} \int_0^1 \dd{\tilde{\omega}} \frac{\mathscr{B}_{\upalpha}(\tilde{\omega}, \tilde{q}_z)}{\mathscr{P}(\tilde{\omega}, \tilde{q}_z)}, \\
\frac{1}{\tilde{\eta}_z} \dv{\tilde{\eta}_z}{\ell} &= \beta_{\eta z} = \frac{4}{\displaystyle \pi^2 N (1+\tilde{m}^2)^2} \int_0^1 \dd{\tilde{q}_z} \int_0^1 \dd{\tilde{\omega}} \frac{\mathscr{B}_{\eta z}(\tilde{\omega}, \tilde{q}_z)}{\mathscr{P}(\tilde{\omega}, \tilde{q}_z)}, \\
\frac{1}{\tilde{\Lambda}} \dv{\tilde{\Lambda}}{\ell} &= -1,
\end{aligned} \label{eq:eds-RG-flow-final}
\end{align}
where $\ell$ determines the cutoff through $\Lambda = \Lambda_0 / b = \Lambda_0 \Elr^{- \ell}$ and
\begin{align}
\begin{aligned}
\mathscr{B}_{m}(\tilde{\omega}, \tilde{q}_z) &= (1-\tilde{\omega}^2) \mleft[1 - (1+\tilde{m}^2) \tilde{\eta}_z^2\mright] \tilde{\Lambda}^2 - (\tilde{\omega}^2 + \tilde{m}^2) \tilde{\eta}_z^2 \tilde{q}_z^2, \\
\mathscr{B}_{\upalpha}(\tilde{\omega}, \tilde{q}_z) &= (1-\tilde{\omega}^2 + 2\tilde{m}^2) (\tilde{\Lambda}^2 + \tilde{\eta}_z^2 \tilde{q}_z^2), \\
\mathscr{B}_{\eta z}(\tilde{\omega}, \tilde{q}_z) &= 2 \tilde{m}^2 (\tilde{\Lambda}^2 + \tilde{\eta}_z^2 \tilde{q}_z^2), \\
\mathscr{P}(\tilde{\omega}, \tilde{q}_z) &= (1-\tilde{\omega}^2) \tfrac{2}{\pi} \mleft[(1-4\tilde{m}^2) \arccot(2\tilde{m}) + 2\tilde{m}\mright] \tilde{\Lambda}^2 \\
&\hspace{4pt} + \mleft\{(1-\tilde{\omega}^2) \tfrac{2}{\pi} \mleft[(1+4\tilde{m}^2) \arccot(2\tilde{m}) + 2\tilde{m}\mright] + 8 \tilde{\omega}^2 \tilde{m}^2 \tfrac{2}{\pi} \arccot(2\tilde{m})\mright\} \tilde{\eta}_z^2 \tilde{q}_z^2.
\end{aligned}
\end{align}
These RG flow equations are the main result of this section and one of the main results of Ref.~\cite{Palle2024-el-dip} on which this chapter is based.

By inspection, one sees that $\mathscr{P}$, $\mathscr{B}_{\upalpha}$, and $\mathscr{B}_{\eta z}$ are strictly positive for all $\tilde{\omega}$ and $\tilde{q}_z$, whereas $\mathscr{B}_{m}$ can be positive or negative.
Consequently, the dimensionless out-of-plane electric dipole moment $\tilde{\eta}_z$ is always marginally relevant, while the effective fine-structure constant $\upalpha$ is always marginally irrelevant.

The flow of the dimensionless gap $\tilde{m}$ is the simplest: it grows with an exponent that approximately equals $+1$ even when we extrapolate $N \to 1$, as the numerical evaluating of the shell integral shows.
Once $\tilde{m}$ becomes on the order of $\sim 1$, the RG flow should be terminated.
Even though large $\tilde{m}$ are thus never reached, let us nonetheless note that all three $\mathscr{B}_i / \mathscr{P} \propto \tilde{m}$ for large $\tilde{m}$ and therefore the flow of both $\upalpha$ and $\tilde{\eta}$ is suppressed as $\tilde{m} \to + \infty$, as expected.
In addition, the RG flow equations are symmetric with respect to $\tilde{m} \to - \tilde{m}$ so we may always choose $\tilde{m} \geq 0$, as we do below.

\begin{figure}[t!]
\centering
\includegraphics[width=0.95\textwidth]{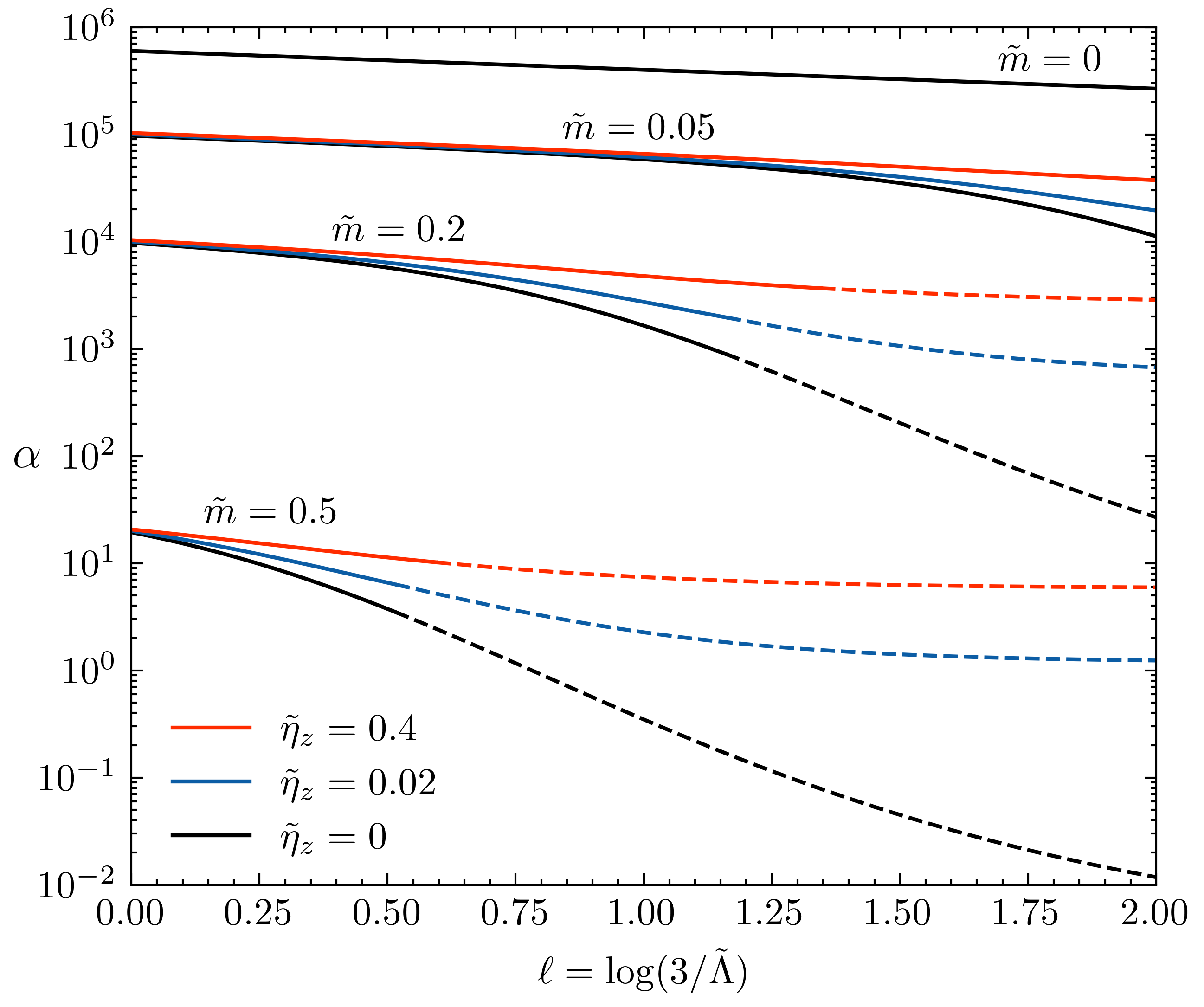}
\captionbelow[The RG flow of $\upalpha$ with $N = 1$ for various initial $\tilde{m}(\ell=0)$ and $\tilde{\eta}_z(\ell=0)$, as indicated on the figure~\cite{Palle2024-el-dip}.]{\textbf{The RG flow of $\upalpha$ with $N = 1$ for various initial $\tilde{m}(\ell=0)$ and $\tilde{\eta}_z(\ell=0)$, as indicated on the figure}~\cite{Palle2024-el-dip}.
Solid lines become dashed when $\tilde{m}(\ell) > 1$.
The $\upalpha(\ell)$ curves associated with different initial masses we have offset relative to each other via multiplication (displacement on a log scale).
We are allowed to do this because $\upalpha(\ell=0)$ enters as a multiplicative factor in the solution of the RG flow equations~\eqref{eq:eds-RG-flow-final}.}
\label{fig:el-dip-sc-RG-alpha-flow}
\end{figure}

The flow of $\upalpha$ for a gapless 2D Dirac system without electric dipoles was analyzed in Ref.~\cite{DTSon2007} and we recover their $\partial_{\ell} \upalpha = - \frac{4}{\pi^2 N} \upalpha$ result when we set $\tilde{m} = 0$.
Our analysis shows that the flow towards small $\upalpha$ persists for finite gaps $\tilde{m}$ and finite $z$-axis dipolar couplings $\tilde{\eta}_z$.
The detailed behavior is shown in Fig.~\ref{fig:el-dip-sc-RG-alpha-flow}, where we plot the flow of $\upalpha$ for different initial values of the mass $\tilde{m}$ and dipole element $\tilde{\eta}_z$.
Notice that $\upalpha$ does not enter any of the beta functions $\beta_i$ in the strong-coupling limit $\upalpha \to \infty$.
Hence, we may offset the solutions via multiplication, as we did in Fig.~\ref{fig:el-dip-sc-RG-alpha-flow} for illustration purposes only.
The suppression of $\upalpha$ is stronger for intermediate $\tilde{m} \sim 1$ than for very small $\tilde{m} \to 0$, and we shall later see that this is accompanied by an enhancement of $\tilde{\eta}$ that also predominantly takes place for $\tilde{m} \sim 1$.
On the other hand, because $\mathscr{B}_{\upalpha} / \mathscr{P} = 1$ when $\tilde{m} = 0$, $\tilde{\eta}_z$ has a negligible effect on the flow of $\upalpha$ for small $\tilde{m}$.
For intermediate $\tilde{m} \sim 1$, small $\tilde{\eta}_z$ are more favorable for the suppression of $\upalpha$ than large $\tilde{\eta}_z$, as can be seen from Fig.~\ref{fig:el-dip-sc-RG-alpha-flow}.
Both positive and negative $\tilde{\eta}_z$ affect $\upalpha$ the same way because of horizontal reflection symmetry $\tilde{\eta}_z \to - \tilde{\eta}_z$, which is respected by Eqs.~\eqref{eq:eds-RG-flow-final}; below we assume $\tilde{\eta}_z \geq 0$.

The dependence of the flow of the dipole strength $\tilde{\eta}_z$ on the mass $\tilde{m}$ is more subtle than that of the monopole coupling $\upalpha$.
Its beta function $\beta_{\eta z}$ vanishes for both small and large $\tilde{m}$.
That large gaps suppress the flow of $\tilde{\eta}_z$ is expected because large gaps suppress the mixing of parities that is needed for high-energy fluctuations to affect electric dipole moments.
Less obvious is the fact that there is a chiral $\Ugp(1)$ symmetry $\psi \to \Elr^{\iu \vartheta \gamma_3} \psi$ in the gapless limit $m \to 0$ (with $k_F = v_z = \eta_{\perp} = 0$) and that the out-of-plane electric dipole moments precisely couple to its charge $\bar{\psi} \gamma_0 \gamma_3 \psi$.
As a result, the associated Ward identity guarantees that $Z_{\eta z} = Z_{\omega}$, precluding any renormalization of $\eta_z$, as we prove in Sec.~\ref{sec:el-dip-sc-Ward-id}.
The largest increase in $\tilde{\eta}_z$ thus happens for moderate $\tilde{m} \sim 1$, and for large $\tilde{\Lambda}$, as follows from the fact that $\mathscr{B}_{\eta z} \propto \tilde{\Lambda}^2$.

\begin{figure}[t!]
\centering
\includegraphics[width=0.95\textwidth]{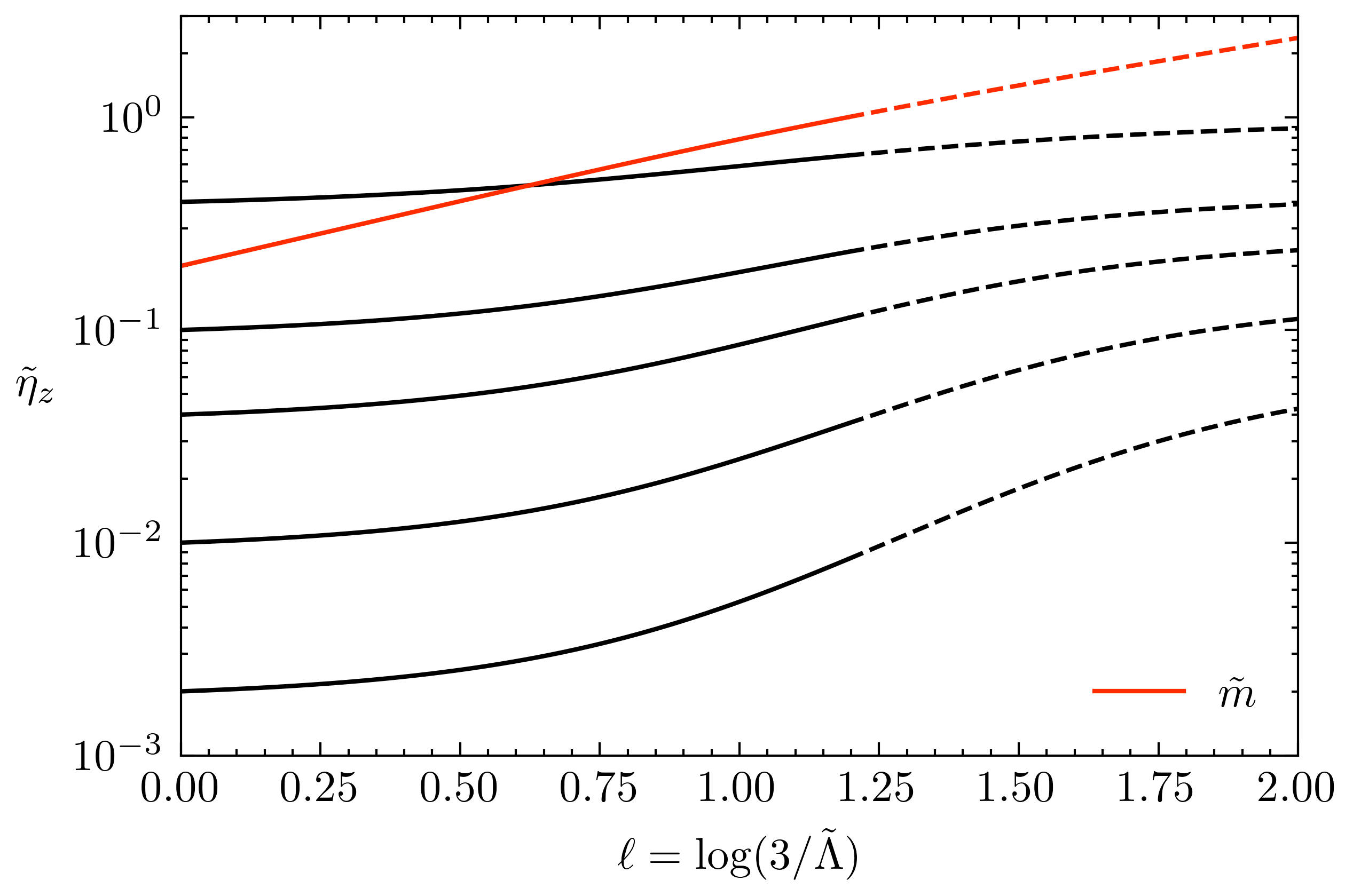}
\captionbelow[The RG flow of $\tilde{\eta}_z$ and $\tilde{m}$ with $N = 1$ for an initial $\tilde{\Lambda}(\ell = 0) = 3$, $\tilde{m}(\ell = 0) = 0.2$, and $\tilde{\eta}_z(\ell = 0) \in \{0.002, 0.01, 0.04, 0.1, 0.4\}$~\cite{Palle2024-el-dip}.]{\textbf{The RG flow of $\tilde{\eta}_z$ and $\tilde{m}$ with $N = 1$ for an initial $\tilde{\Lambda}(\ell = 0) = 3$, $\tilde{m}(\ell = 0) = 0.2$, and $\tilde{\eta}_z(\ell = 0) \in \{0.002, 0.01, 0.04, 0.1, 0.4\}$}~\cite{Palle2024-el-dip}.
Solid lines become dashed when $\tilde{m}(\ell) > 1$.
There are small variations in how $\tilde{m}$ flows, depending on $\tilde{\eta}_z(\ell = 0)$, which we are not shown.
The same scale is used for both $\tilde{m}$ and $\tilde{\eta}_z$.}
\label{fig:el-dip-sc-RG-etaz-flow}
\end{figure}

The numerical results for the flow of the $z$-axis dipole element $\tilde{\eta}_z$ are shown in Fig.~\ref{fig:el-dip-sc-RG-etaz-flow}.
These results depend on the initial values of $\tilde{\Lambda}$, $\tilde{m}$, and $\tilde{\eta}_z$, which are specified below.
Note that they do not depend on $\upalpha$ as long as it is large because $\upalpha(\ell)$ decouples from the rest in the strong-coupling limit described by Eqs.~\eqref{eq:eds-RG-flow-final}.

For $\tilde{\Lambda}$, we assume that initially $\tilde{\Lambda} = 3$, which corresponds to a reasonable amount of anisotropy for a quasi-2D system ($\Lambda = 3 \Lambda_z$).
The RG flow we run until $\ell = 2$, at which point $\tilde{\Lambda} = 3 \Elr^{-2} = 0.41$.
The Fermi radius $k_F$, which we neglected [Eq.~\eqref{eq:eds-muT0-limit}], is thus on the order of $0.2 \Lambda_z$.

Regarding the gap, in Fig.~\ref{fig:el-dip-sc-RG-alpha-flow} we only show the results for an initial $\tilde{m} = 0.2$.
We have explored other initial values as well and we have found that the enhancement of $\tilde{\eta}_z$ is comparable in magnitude to that shown in Fig.~\ref{fig:el-dip-sc-RG-alpha-flow} in the range $\tilde{m} \in \langle 0.05, 1.0\rangle$, whereas outside of this range it is a lot smaller.
As already remarked, the flow of $\tilde{m}$, given an initial value, is not significantly affected by $\tilde{\eta}_z$ so only one curve for $\tilde{m}(\ell)$ is shown in Fig.~\ref{fig:el-dip-sc-RG-alpha-flow}.

The RG flow is given for five different initial values of $\tilde{\eta}_z$, ranging from $0.002$ to $0.4$.
As can be seen in Fig.~\ref{fig:el-dip-sc-RG-alpha-flow}, although smaller $\tilde{\eta}_z$ tend to get more enhanced, sometimes by even two orders of magnitude (if we take $N \to 1$), the final value of $\tilde{\eta}_z(\ell=2)$ declines with decreasing $\tilde{\eta}_z(\ell=0)$.
Larger microscopic electric dipole moments $\tilde{\eta}_z(\ell=0)$ thus always lead to larger effective dipole moments $\tilde{\eta}_z(\ell=2)$.
It is also worth noting that the increase in $\tilde{\eta}_z$ is finite even if we extend $\ell$ to go from $-\infty$ to $+\infty$.
The reason lies in the fact, discussed earlier, that both small and large $\tilde{m}$ suppress the beta function of $\tilde{\eta}_z$: the former because of a chiral symmetry (Sec.~\ref{sec:el-dip-sc-Ward-id}) and the latter because of weak parity mixing.
Hence the dipole matrix element grows only in an intermediate window before $\tilde{m}$ becomes too large.
This should be contrasted to the flow of $\upalpha$ which stops for large $\ell$, but is exponential for small $\ell \to - \infty$.

\subsubsection{Evaluation of $1$-loop self-energy and vertex diagrams} \label{sec:el-dip-sc-RG-diagrams}
Here we evaluate the $1$-loop fermionic self-energy and electron-plasmon vertex diagrams in the quasi-2D limit $v_z = \eta_{\perp} = 0$ [Eq.~\eqref{eq:eds-quasi-2D-param}] with $\upmu = T = 0$ [Eq.~\eqref{eq:eds-muT0-limit}].
These diagrams underlie the RG flow equations~\eqref{eq:eds-RG-flow-final}.

\begin{figure}[t]
\centering
\begin{subfigure}[t]{0.5\textwidth}
\centering
\includegraphics[width=0.80\textwidth]{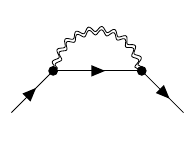}
\subcaption{}
\end{subfigure}%
\begin{subfigure}[t]{0.5\textwidth}
\centering
\includegraphics[width=0.80\textwidth]{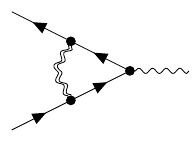}
\subcaption{}
\end{subfigure}
\captionbelow[The diagrams of the leading contributions to the electronic self-energy (a) and electron-plasmon vertex (b).]{\textbf{The diagrams of the leading contributions to the electronic self-energy (a) and electron-plasmon vertex (b).}
In the large-$N$ limit, the plasmon propagators needs to be dressed, as indicated by the double wiggly lines.
Solid lines are fermion propagators.
In all cases, the vertices contain both monopole and dipole contributions, as specified in Fig.~\ref{fig:el-dip-sc-plasmon-fermion-vertex}.}
\label{fig:el-dip-sc-1-loop-diagrams}
\end{figure}

The fermionic self-energy is defined as
\begin{align}
\Sigma(k) = \mathscr{G}^{-1}(k) - G^{-1}(k),
\end{align}
where
\begin{align}
\ev{\psi_{k, \alpha} \bar{\psi}_{p, \beta}} = \mathscr{G}_{\alpha \beta}(k) \Kd_{k - p}.
\end{align}
To lowest order, it is given by the Fock term [Fig.~\ref{fig:el-dip-sc-1-loop-diagrams}(a)]:
\begin{align}
\Sigma(k) &= \int \frac{\dd[4]{q}}{(2\pi)^4} A(k, k+q) G(k+q) A(k+q, k) \cdot \mathscr{V}(-q).
\end{align}
The Hartree term has been omitted because it merely results in an absolute displacement that can be absorbed into the chemical potential.
The bare $G(k)$ and $A(k, p)$ are (Eqs.~\eqref{eq:eds-G-1-psi} and~\eqref{eq:eds-A-psi} in Sec.~\ref{sec:el-dip-sc-Dirac-model}):
\begin{align}
G(k) &= \frac{m \one + \iu \mleft[\omega_k \gamma_0 + v (k_x \gamma_1 + k_y \gamma_2)\mright]}{m^2 + \omega_k^2 + v^2 \vb{k}_{\perp}^2}, \\
A(k, p) &= e \gamma_0 + \iu \eta_z (k_z - p_z) \gamma_0 \gamma_3.
\end{align}
Note that the interaction needs to be dressed ($\mathscr{V}$ appears instead of $V$ in $\Sigma$) with the polarization bubble diagram because of the large-$N$ limit.
In a slight abuse of terminology, we shall still call this diagram ``$1$-loop,'' even though a geometric series of loops has been summed up in the interaction.

When $v_z = \eta_{\perp} = \upmu = 0$, one finds that:
\begin{align}
A(k, k+q) G(k+q) A(k+q, k) &= \frac{\tilde{\mathcal{E}} \cdot \one + \mathcal{E}_0 \cdot \gamma_0 + \mathcal{E}_1 \cdot \gamma_1 + \mathcal{E}_2 \cdot \gamma_2 + \mathcal{E}_3 \cdot \gamma_3}{m^2 + (\omega_k + \omega_q)^2 + v^2 (\vb{k}_{\perp} + \vb{q}_{\perp})^2},
\end{align}
where
\begin{align}
\begin{aligned}
\tilde{\mathcal{E}} &= m (e^2 - q_z^2 \eta_z^2), &\hspace{50pt}
\mathcal{E}_0 &= \iu \mleft(e^2 + q_z^2 \eta_z^2\mright) (\omega_k + \omega_q), \\
\mathcal{E}_1 &= - \iu \mleft(e^2 + q_z^2 \eta_z^2\mright) v (k_x + q_x), &\hspace{50pt}
\mathcal{E}_2 &= - \iu \mleft(e^2 + q_z^2 \eta_z^2\mright) v (k_y + q_y), \\
\mathcal{E}_3 &= 2 \iu m e q_z \eta_z.
\end{aligned}
\end{align}

By expanding in small $k$ and dropping everything odd under $q$, one obtains:
\begin{align}
\mathscr{G}^{-1}(k) &= G^{-1}(k) + \Sigma(k) = Z_m m \one - \iu Z_{\omega} \omega_k \gamma_0 - \iu Z_v v \mleft(k_x \gamma_1 + k_y \gamma_2\mright) + \cdots \, ,
\end{align}
where:
\begin{align}
Z_m &= 1 + \int \frac{\dd[4]{q}}{(2\pi)^4} \frac{e^2 - q_z^2 \eta_z^2}{m^2 + \omega_q^2 + v^2 \vb{q}_{\perp}^2} \cdot \mathscr{V}(-q), \\
Z_{\omega} &= 1 + \int \frac{\dd[4]{q}}{(2\pi)^4} \frac{- (e^2 + q_z^2 \eta_z^2) (m^2 + v^2 \vb{q}_{\perp}^2 - \omega_q^2)}{\mleft[m^2 + \omega_q^2 + v^2 \vb{q}_{\perp}^2\mright]^2} \cdot \mathscr{V}(-q), \\
Z_v &= 1 + \int \frac{\dd[4]{q}}{(2\pi)^4} \frac{(e^2 + q_z^2 \eta_z^2) (m^2 + \omega_q^2)}{\mleft[m^2 + \omega_q^2 + v^2 \vb{q}_{\perp}^2\mright]^2} \cdot \mathscr{V}(-q).
\end{align}

The dressed vertex is defined by amputating the connected fermion-boson Green's function:
\begin{align}
\ev{\psi_{k, \alpha} \bar{\psi}_{p, \beta} \Phi_{q}} - \ev{\psi_{k, \alpha} \bar{\psi}_{p, \beta}} \ev{\Phi_{q}} &= \frac{\iu}{\sqrt{\upbeta L^d}} \mleft[\mathscr{G}(k) \mathscr{A}(k, p) \mathscr{G}(p)\mright]_{\alpha \beta} \mathscr{V}(q) \Kd_{k - p + q}. \label{eq:eds-dressed-vert-def}
\end{align}
Recall that $L^d$ is the volume and $\upbeta = 1 / (k_B T)$.
To lowest order in $N$, it equals [Fig.~\ref{fig:el-dip-sc-1-loop-diagrams}(b)]
\begin{align}
\mathscr{A}(k, p) &= A(k, p) - \int \frac{\dd[4]{q}}{(2\pi)^4} A(k, k+q) G(k+q) A(k+q, p+q) G(p+q) A(p+q, p) \cdot \mathscr{V}(-q),
\end{align}
where the interaction again needs to be dressed with the polarization bubble.

Multiplying out the matrices in the above expression for $\mathscr{A}(k, p)$ results in gamma matrices of all orders, going from $\one$ and $\gamma_{\mu}$ up to $\gamma_{\mu} \gamma_{\nu} \gamma_{\rho} \gamma_{\sigma} = \LCs_{\mu \nu \rho \sigma} \gamma_0 \gamma_1 \gamma_2 \gamma_3$.
At $k = p = 0$, only the $\propto \gamma_0$ term survives, giving a renormalization of the charge $e$.
At linear order in $k$ and $p$, we find terms $\propto (k_i - p_i) \gamma_0 \gamma_i$ which renormalize $\eta_{\perp}$ and $\eta_z$, but also an additional term $\propto (\omega_k + \omega_p) \one$.
This additional term is irrelevant, just like $\eta_{\perp}$, so we shall neglect it.
As for the remaining terms which are higher order in $k$ and $p$, they are also irrelevant under RG flow and we therefore neglect them as well.
After some lengthy algebra, we find that
\begin{align}
\mathscr{A}(k, p) &= Z_e e \gamma_0 + \iu Z_{\eta z} \eta_z (k_z - p_z) \gamma_0 \gamma_3 + \cdots \, ,
\end{align}
where:
\begin{align}
Z_e &= 1 + \int \frac{\dd[4]{q}}{(2\pi)^4} \frac{- (e^2 + q_z^2 \eta_z^2) (m^2 + v^2 \vb{q}_{\perp}^2 - \omega_q^2)}{\mleft[m^2 + \omega_q^2 + v^2 \vb{q}_{\perp}^2\mright]^2} \cdot \mathscr{V}(-q), \\
Z_{\eta z} &= 1 + \int \frac{\dd[4]{q}}{(2\pi)^4} \frac{- (e^2 + q_z^2 \eta_z^2) (- m^2 + v^2 \vb{q}_{\perp}^2 - \omega_q^2)}{\mleft[m^2 + \omega_q^2 + v^2 \vb{q}_{\perp}^2\mright]^2} \cdot \mathscr{V}(-q).
\end{align}
Notice that $Z_e = Z_{\omega}$ and that $Z_{\eta z} = Z_{\omega}$ when $m = 0$.
This is a consequence of exact Ward identities which we prove in the next section.

In all the renormalization factors $Z_i$, the frequency and in-plane momentum integrals go up to $\Lambda$, as specified by $\norm{q}^2 = \omega_q^2 / v^2 + \vb{q}_{\perp}^2 < \Lambda^2$ [Eq.~\eqref{eq:eds-cutoff-def}].
Differentiating by $\Lambda$ in Eq.~\eqref{eq:eds-RG-flow-expr} thus gives the shell integrals we provided in the RG equations~\eqref{eq:eds-RG-flow-final}.

\subsection{Ward identities and a chiral symmetry} \label{sec:el-dip-sc-Ward-id}
Ward identities are exact (non-perturbative) identities which express the ways symmetries constrain the renormalization of the theory~\cite{Zinn-Justin2002, Mahan2000}.
Here we prove two Ward identities for the limit $v_z = \eta_{\perp} = \upmu = 0$.
We are focusing on this limit because of the RG considerations of Sec.~\ref{sec:el-dip-sc-Dirac-RG}.
The first Ward identity follows from charge conservation, while the second one follows from a chiral symmetry which only holds in the massless $m = 0$ limit.

We start by writing the Euclidean action of Eq.~\eqref{eq:eds-model-action} or~\eqref{eq:eds-model-action-again} in real space and imaginary time:
\begin{align}
\action[\psi, \Phi] &= \int_x \bar{\psi}(x) \mleft[m - \sum_{\mu'=0}^{2} \gamma_{\mu'} \partial_{\mu'} - \iu \gamma_0 \mleft(e \Phi(x) + \eta_z \gamma_3 \partial_3 \Phi(x)\mright)\mright] \psi(x) + \frac{1}{2} \epsilon \int_x \sum_{j=1}^{3} (\partial_j \Phi(x))^2.
\end{align}
Here $x = x^{\mu} = (\tau, \vb{r})$, $\int_x = \int \dd{\tau} \dd[3]{r}$, and $\partial_{\mu} = \partial / \partial x^{\mu}$.
Temporarily, we have set $v = 1$ and $\epsilon_{\perp} = \epsilon_z = \epsilon$, which we shall restore later.

Let us recall that within the imaginary-time formalism, averages are defined as
\begin{align}
\ev{\mathcal{F}[\psi, \Phi]} &= \frac{1}{\mathcal{Z}} \int \DD{\psi} \DD{\Phi} \Elr^{- \action[\psi, \Phi]} \mathcal{F}[\psi, \Phi],
\end{align}
where $\mathcal{F}$ is a functional of the fields,
\begin{align}
\mathcal{Z} &= \int \DD{\psi} \DD{\Phi} \Elr^{- \action[\psi, \Phi]}
\end{align}
is the partition function, and the integrals are path integrals which go over all possible field configurations:
\begin{align}
\DD{\psi} &= \prod_{x, \alpha} \dd{\bar{\psi}_{\alpha}(x)} \dd{\psi_{\alpha}(x)}, &
\DD{\Phi} &= \prod_x \dd{\Phi(x)}.
\end{align}
Here $\psi_{\alpha}(x)$ and $\bar{\psi}_{\alpha}(x)$ are Grassmann variables, while $\Phi(x)$ are real variables.
An important property of this path integral measure is that it is affine, i.e., translation invariant.
Thus integrating over all $\psi$ and $\Phi$ should give the same result as integrating over all $\psi' = \psi + \var{\psi}$ and $\Phi' = \Phi + \var{\Phi}$.
For the averages, this implies the following exact Schwinger-Dyson equation:
\begin{align}
- \ev{\var{\action} \cdot \mathcal{F}} + \ev{\var{\mathcal{F}}} &= 0 \label{eq:eds-Schw-Dy-eq}
\end{align}
which is the path-integral equivalent of the Heisenberg equations of motion in the canonical formalism.
Note that the order above matters: $\ev{\var{\action}/\var{\psi} \cdot \mathcal{F}} = - \ev{\mathcal{F} \cdot \var{\action}/\var{\psi}}$ when both the field $\psi$ that we are varying and the functional $\mathcal{F}$ that we are  averaging are Grassmann-odd.
Similarly, during the chain rule the order also matters for Grassmann-even $\mathcal{F}$:
\begin{align}
\var{\mathcal{F}} &= \int_x \mleft[\sum_{\alpha} \var{\psi_{\alpha}(x)} \fdv{\mathcal{F}}{\psi_{\alpha}(x)} + \sum_{\alpha} \var{\bar{\psi}_{\alpha}(x)} \fdv{\mathcal{F}}{\bar{\psi}_{\alpha}(x)} + \var{\Phi(x)} \fdv{\mathcal{F}}{\Phi(x)}\mright].
\end{align}

Under an infinitesimal $\Ugp(1)$ phase rotation $\psi(x) \mapsto \Elr^{\iu \vartheta(x)} \psi(x)$, $\bar{\psi}(x) \mapsto \bar{\psi}(x) \Elr^{- \iu \vartheta(x)}$, $\Phi(x) \mapsto \Phi(x)$, the action changes by
\begin{align}
\var{\action} &= \iu \int_x \vartheta(x) \sum_{\mu'=0}^{2} \partial_{\mu'} \mleft(\bar{\psi}(x) \gamma_{\mu'} \psi(x)\mright).
\end{align}
By applying the Schwinger-Dyson equation~\eqref{eq:eds-Schw-Dy-eq} to the functional $\mathcal{F}[\psi, \Phi] = \psi_{\alpha_2}(x_2) \bar{\psi}_{\alpha_3}(x_3)$, one obtains the Ward-Takahashi identity:
\begin{align}
(\Kd_{x - x_2} - \Kd_{x - x_3}) \ev{\psi(x_2) \bar{\psi}(x_3)} &= \sum_{\mu'=0}^{2} \ev{\partial_{\mu'} \mleft[\bar{\psi}(x) \gamma_{\mu'} \psi(x)\mright] \cdot \psi(x_2) \bar{\psi}(x_3)}.
\end{align}
Physically, this identity expresses the conservation of charge within a four-point thermal average.
In Fourier space, it takes the form:
\begin{align}
\ev{\psi_{k_1+q} \bar{\psi}_{k_2}} - \ev{\psi_{k_1} \bar{\psi}_{k_2-q}} &= \sum_p \sum_{\mu'=0}^{2} \iu q_{\mu'} \ev{\bar{\psi}_{p} \gamma_{\mu'} \psi_{p+q} \cdot \psi_{k_1} \bar{\psi}_{k_2}}. \label{eq:eds-WTI1}
\end{align}

Motivated by the above expression, let us introduce for an arbitrary $4 \times 4$ matrix $\Gamma$ the amputated matrix-fermion vertex:
\begin{align}
\mathscr{W}_{\Gamma}(k, q) &\defeq \sum_p \ev{\bar{\psi}_{p} \Gamma \psi_{p+q} \cdot \mathscr{G}^{-1}(k) \psi_{k} \bar{\psi}_{k+q} \mathscr{G}^{-1}(k+q)}.
\end{align}

The Ward-Takahashi identity~\eqref{eq:eds-WTI1}, with $k_1 = k$ and $k_2 = k + q$, can now be recast into
\begin{align}
\mathscr{G}^{-1}(k+q) - \mathscr{G}^{-1}(k) &= - \iu \omega_q \mathscr{W}_{\gamma_0}(k, q) - \iu v q_x \mathscr{W}_{\gamma_1}(k, q) - \iu v q_y \mathscr{W}_{\gamma_2}(k, q),
\end{align}
where we have restored $v$.
In particular, this means that:
\begin{align}
\mathscr{W}_{\gamma_{0}}(k, q=0) &= \iu \pdv{}{\omega_k} \mathscr{G}^{-1}(k), \\
\mathscr{W}_{\gamma_{1}}(k, q=0) &= \frac{\iu}{v} \pdv{}{k_x} \mathscr{G}^{-1}(k), \\
\mathscr{W}_{\gamma_{2}}(k, q=0) &= \frac{\iu}{v} \pdv{}{k_y} \mathscr{G}^{-1}(k).
\end{align}
Thus if for small $k$
\begin{align}
\mathscr{G}^{-1}(k) &= Z_m m \one - \iu Z_{\omega} \omega_k \gamma_0 - \iu Z_v v \mleft(k_x \gamma_1 + k_y \gamma_2\mright) + \cdots \, ,
\end{align}
it follows that
\begin{align}
\mathscr{W}_{\gamma_{0}}(k, q=0) &= Z_{\omega} \gamma_0, \label{eq:eds-Wgam0-expr} \\
\mathscr{W}_{\gamma_{1}}(k, q=0) &= Z_{v} \gamma_1, \\
\mathscr{W}_{\gamma_{2}}(k, q=0) &= Z_{v} \gamma_2.
\end{align}

The Schwinger-Dyson equation that follows from varying $\Phi_{-q} = \Phi_{q}^{*}$ with $\mathcal{F} = \psi_{k_1,\alpha_1} \bar{\psi}_{k_2,\alpha_2}$ is:
\begin{align}
V^{-1}(q) \ev{\Phi_{q} \psi_{k_1} \bar{\psi}_{k_2}} &= \frac{\iu}{\sqrt{\upbeta L^d}} \sum_p \ev{\bar{\psi}_{p} \gamma_0 (e - \iu \eta_z q_z \gamma_3) \psi_{p+q} \cdot \psi_{k_1} \bar{\psi}_{k_2}}.
\end{align}
After employing Eq.~\eqref{eq:eds-dressed-vert-def} on the left-hand side under the assumption that $\ev{\Phi_{q \neq 0}} = 0$, the above becomes:
\begin{align}
\mathscr{A}(k, k+q) &= V(q) \mathscr{V}^{-1}(q) \cdot \mleft[e \mathscr{W}_{\gamma_0}(k, q) - \iu \eta_{z} q_z \mathscr{W}_{\gamma_0 \gamma_3}(k, q)\mright]. \label{eq:eds-SDeq-phi}
\end{align}
If we now further assume that for small four-momenta $V(q) \mathscr{V}^{-1}(q) = Z_{\epsilon} + \cdots$ and
\begin{align}
\mathscr{A}(k, k+q) &= Z_e e \gamma_0 - \iu Z_{\eta z} \eta_z q_z \gamma_0 \gamma_3 + \cdots \, ,
\end{align}
as well as exploit Eq.~\eqref{eq:eds-Wgam0-expr}, we obtain the Ward identity $Z_e = Z_{\epsilon} Z_{\omega}$.
In Sec.~\ref{sec:el-dip-sc-polarization}, we found that $\Pi(q) = \mathscr{V}^{-1}(q) - V^{-1}(q)$ is non-analytic at $q = 0$, which implies that $\Pi(q)$ cannot be Taylor expanded at $q = 0$.
Moreover, there is no canonical decomposition of $\Pi(q)$ into a non-analytic part and analytic part (which could then be expanded around $q = 0$).
Hence no part of $\Pi(q)$ contributes to the renormalization of the bare plasmon propagator.
Consequently, $Z_{\epsilon} = 1$ and we obtain the Ward identity:
\begin{align}
Z_e = Z_{\omega}.
\end{align}
Physically, this identity expresses the fact that charge does not renormalize, as we explicitly saw on the $1$-loop level in Sec.~\ref{sec:el-dip-sc-RG-diagrams}.

Apart from the $\Ugp(1)$ phase rotation symmetry which is associated with charge conservation, in the massless limit there is an additional $\Ugp(1)$ rotation symmetry of the form $\psi(x) \mapsto \Elr^{\iu \vartheta(x) \gamma_3} \psi(x)$, $\bar{\psi}(x) \mapsto \bar{\psi}(x) \Elr^{\iu \vartheta(x) \gamma_3}$, $\Phi(x) \mapsto \Phi(x)$.
Given that $\gamma_3$ anticommutes with all $\gamma_{\mu}$ just like $\gamma_5$, physically this represents a chiral symmetry of the model.
Analogous manipulations to the previous give the Ward-Takahashi identity
\begin{align}
\gamma_3 \mathscr{G}^{-1}(k+q) + \mathscr{G}^{-1}(k) \gamma_3 &= \iu \omega_q \mathscr{W}_{\gamma_0 \gamma_3}(k, q) + \iu v q_x \mathscr{W}_{\gamma_1 \gamma_3}(k, q) + \iu v q_y \mathscr{W}_{\gamma_2 \gamma_3}(k, q),
\end{align}
which implies
\begin{align}
\mathscr{W}_{\gamma_0 \gamma_3}(k, q=0) &= Z_{\omega} \gamma_0 \gamma_3, \\
\mathscr{W}_{\gamma_1 \gamma_3}(k, q=0) &= Z_{v} \gamma_1 \gamma_3, \\
\mathscr{W}_{\gamma_2 \gamma_3}(k, q=0) &= Z_{v} \gamma_2 \gamma_3.
\end{align}
From Eq.~\eqref{eq:eds-SDeq-phi} we now obtain the Ward identity
\begin{align}
Z_{\eta z} = Z_{\omega},
\end{align}
where we used the fact that $\mathscr{W}_{\gamma_0}(k, q)$ cannot be linear in $q_z$ because of horizontal reflection symmetry.
In the massless limit, the chiral symmetry thus protects the out-of-plane electric dipole moment $\eta_z$ from renormalizing.

\section{Pairing due to electric monopole-dipole interactions} \label{sec:el-dip-sc-pairing}
The strongly repulsive nature of the Coulomb interaction is often one of the biggest obstacles to the formation of Cooper pairs.
Its monopole-monopole part by itself is repulsive and suppresses pairing.
However, the monopole-dipole and dipole-dipole parts can yield unconventional superconductivity (SC) if the screening and dipole moments are strong enough, as we show here.
Although we call this pairing after the monopole-dipole term only, we are not neglecting dipole-dipole interactions in our analysis, but are rather emphasizing the fact that the monopole-dipole coupling is the main source of pairing.
Starting from an effective instantaneous interaction among Fermi-level electrons, such as the one obtained at the end of the RG flow of the previous section, we first summarize the formalism for analyzing SC instabilities in Sec.~\ref{sec:el-dip-sc-lin-gap-eq}.
The expressions that we obtain are very similar to those that we previously had in Sec.~\ref{sec:QCP-model-lin-gap-eq} of Chap.~\ref{chap:loop_currents} for the exchange of order-parameter modes.
We compare and contrast the two in Sec.~\ref{sec:el-dip-sc-comparison-OP-ex}.
Using this formalism, in Sec.~\ref{sec:el-dip-sc-qualitative-pairing} we then study the pairing due to electric monopole-dipole interactions for general systems and we derive a number of its properties.
A toy model is analyzed in the last subsection.
The pairing in quasi-2D Dirac metals, which were the subject of the preceding Sec.~\ref{sec:el-dip-sc-Dirac}, we analyze in the next part of the chapter.

\subsection{Linearized gap equation and formalism} \label{sec:el-dip-sc-lin-gap-eq}
To study Cooper pairing, we use the linearized gap equation that we derived in Appx.~\ref{app:lin_gap_eq}.
If we keep the electron-electron interaction generic for the moment, then we may write it as
\begin{align}
\Haml_{\text{int}} = \frac{1}{4 L^d} \sum \Kd_{\vb{k}_1+\vb{k}_2-\vb{k}_3-\vb{k}_4} U_{\alpha_{1} \alpha_{2} \alpha_{3} \alpha_{4}}(\vb{k}_{1},\vb{k}_{2},\vb{k}_{3},\vb{k}_{4}) \psi_{\vb{k}_1, \alpha_1}^{\dag} \psi_{\vb{k}_2, \alpha_2}^{\dag} \psi_{\vb{k}_4, \alpha_4} \psi_{\vb{k}_3, \alpha_3},
\end{align}
where $U$ is fully antisymmetrized with respect to particle exchange and the sum goes over all four momenta and spin and orbital degrees of freedom.
At leading order in this interaction, in Appx.~\ref{app:lin_gap_eq} we obtained the following linearized gap equation, formulated as an eigenvalue problem [Eq.~\eqref{eq:final-lin-gap-eq}]:
\begin{align}
\sum_n \int\limits_{\varepsilon_{\vb{k} n} = 0} \frac{\dd{S_{\vb{k}}}}{(2\pi)^d} \sum_{A=0}^{3} \PintW_{BA}(\vb{p}_m, \vb{k}_n) d_{A}(\vb{k}_n) = \lambda \, d_{B}(\vb{p}_m). \label{eq:eds-lin-gap-eq}
\end{align}
Here $n, m$ are band indices, $\varepsilon_{\vb{k} n}$ is the band dispersion displaced by the chemical potential, the momenta $\vb{k}_n, \vb{p}_m$ are on the Fermi surfaces which are determined by $\varepsilon_{\vb{k} n} = \varepsilon_{\vb{p} m} = 0$, $\dd{S_{\vb{k}}}$ is a surface element, $A = B = 0$ corresponds to even-parity and $A, B \in \{1, 2, 3\}$ to odd-parity pairing, $d_{A}(\vb{k}_n)$ is the pairing $\vb{d}$-vector, and $\PintW_{BA}(\vb{p}_m, \vb{k}_n)$ is the pairing interaction.
This linearized gap equation applies to spin-orbit-coupled Fermi liquids with space-inversion and time-reversal symmetries whose Fermi surfaces do not touch each other or have Van Hove singularities on them.

Positive pairing eigenvalues $\lambda$ correspond to SC states with transition temperatures:
\begin{align}
k_B T_c &= \frac{2 \Elr^{\upgamma_E}}{\pi} \hbar \omega_c \, \Elr^{- 1 / \lambda} \approx 1.134 \, \hbar \omega_c \, \Elr^{- 1 / \lambda}, \label{eq:eds-BCS-Tc-expr}
\end{align}
where $\upgamma_E$ is the Euler-Mascheroni constant and $\hbar \omega_c$ is the energy cutoff of the theory, which is assumed to be much smaller than the bandwidth.
The leading instability has the largest positive $\lambda$.

The pairing interaction is given by:
\begin{align}
\PintW_{BA}(\vb{p}_m, \vb{k}_n) &= - \sum_{\alpha_1 \alpha_2 \alpha_3 \alpha_4}  \frac{\mleft[\MatTR^{*} \mathcal{P}_{\vb{p} m}^{B}\mright]_{\alpha_2 \alpha_1} \mleft[\mathcal{P}_{\vb{k} n}^{A} \MatTR^{\intercal}\mright]_{\alpha_3 \alpha_4}}{4 \abs{\grad_{\vb{p}} \varepsilon_{\vb{p} m}}^{1/2} \abs{\grad_{\vb{k}} \varepsilon_{\vb{k} n}}^{1/2}} U_{\alpha_1 \alpha_2 \alpha_3 \alpha_4}(\vb{p}, -\vb{p}, \vb{k}, -\vb{k}),
\end{align}
where $\mathcal{P}_{\vb{k} n}^{A}$ are the Pauli-matrix-weighted band projectors:
\begin{align}
\mathcal{P}_{\vb{k} n}^{A} = \sum_{s s'} u_{\vb{k} n s} (\Pauli_A)_{ss'} u_{\vb{k} n s'}^{\dag}.
\end{align}
Here $s, s' \in \{\uparrow, \downarrow\}$ are the pseudospins, $\Pauli_A$ are the Pauli matrices, $\alpha_i$ are combined orbital and spin indices, $u_{\vb{k} n s}$ are the normalized band eigenvectors which diagonalize the one-particle Hamiltonian, $H_{\vb{k}} u_{\vb{k} n s} = \varepsilon_{\vb{k} n} u_{\vb{k} n s}$, and $\MatTR$ is the unitary matrix that determines how single-particle states transform under the antiunitary time-reversal operator, $\SymTR^{-1} \psi_{\vb{k}, \alpha_1} \SymTR = \sum_{\alpha_2} \MatTR_{\alpha_1 \alpha_2}^{*} \psi_{-\vb{k}, \alpha_2}$.
A pseudospins degeneracy requires both space-inversion and time-reversal symmetry, which we henceforth assume.
See Appx.~\ref{app:lin_gap_eq} for further details.

For the plasmon-mediated monopole and dipole interaction of Eq.~\eqref{eq:eds-Hint-vq}, the monopole and dipole fermionic bilinears of Eq.~\eqref{eq:eds-dipole_real_space} we write in the following way:
\begin{align}
\DipD_{\mu \vb{q}} &= - e \sum_{\vb{k}} \psi^{\dag}_{\vb{k}} \Gamma_{\mu \vb{k}, \vb{k}+\vb{q}} \psi_{\vb{k}+\vb{q}}. \label{eq:eds-Pmu-assumed-form}
\end{align}
The interaction now reads:
\begin{align}
\begin{aligned}
U_{\alpha_{1} \alpha_{2} \alpha_{3} \alpha_{4}}(\vb{k}_{1},\vb{k}_{2},\vb{k}_{3},\vb{k}_{4}) &= e^2 \sum_{\mu \nu} V_{\mu \nu}(\vb{k}_1-\vb{k}_3) \mleft[\Gamma_{\mu \vb{k}_1, \vb{k}_3}\mright]_{\alpha_1 \alpha_3} \mleft[\Gamma_{\nu \vb{k}_2, \vb{k}_4}\mright]_{\alpha_2 \alpha_4} \\
&\qquad - \text{(the same with $\alpha_3 \leftrightarrow \alpha_4$ and $\vb{k}_3 \leftrightarrow \vb{k}_4$)}.
\end{aligned}
\end{align}
After some manipulations that exploit the fact that $\DipD_{\mu}$ is even under time reversal, so $\SymTR^{-1} \DipD_{\mu \vb{q}} \SymTR = \DipD_{\mu, -\vb{q}}$ and $\MatTR^{\dag} \Gamma_{\mu \vb{k}, \vb{p}} \MatTR = \Gamma_{\mu, -\vb{k}, -\vb{p}}^{*}$, but also Hermitian, so $\DipD_{\mu \vb{q}}^{\dag} = \DipD_{\mu, -\vb{q}}$ and $\Gamma_{\nu \vb{p}, \vb{k}}^{\dag} = \Gamma_{\nu \vb{k}, \vb{p}}$, for the pairing interaction we obtain:
\begin{align}
\PintW_{BA}(\vb{p}_m, \vb{k}_n) &= - e^2 \frac{\Pintw_{BA}(\vb{p}_m, \vb{k}_n) + \Pintw_{BA}(\vb{p}_m, - \vb{k}_n) p_A}{4 \abs{\grad_{\vb{p}} \varepsilon_{\vb{p} m}}^{1/2} \abs{\grad_{\vb{k}} \varepsilon_{\vb{k} n}}^{1/2}}, \label{eq:eds-W-pairing-def}
\end{align}
where $p_{A=0} = - p_{A=1,2,3} = +1$ and
\begin{align}
\Pintw_{BA}(\vb{p}_m, \vb{k}_n) &= \sum_{\mu \nu} V_{\mu \nu}(\vb{p}-\vb{k}) \Tr \mathcal{P}_{\vb{p} m}^{B} \Gamma_{\mu \vb{p}, \vb{k}} \mathcal{P}_{\vb{k} n}^{A} \Gamma_{\nu \vb{p}, \vb{k}}^{\dag}. \label{eq:eds-w-pairing-def-alt}
\end{align}
The trace arising in $\Pintw_{BA}(\vb{p}_m, \vb{k}_n)$ goes over both spin and orbital degrees of freedom and one can alternatively write it as a pseudospin trace:
\begin{align}
\begin{aligned}
\PintF_{BA}^{\mu \nu}(\vb{p}_m, \vb{k}_n) &\defeq \Tr \mathcal{P}_{\vb{p} m}^{B} \Gamma_{\mu \vb{p}, \vb{k}} \mathcal{P}_{\vb{k} n}^{A} \Gamma_{\nu \vb{p}, \vb{k}}^{\dag} \\
&= \tr_s \Pauli_B \Pintf_{\mu}(\vb{p}_m, \vb{k}_n) \Pauli_A \Pintf_{\nu}^{\dag}(\vb{p}_m, \vb{k}_n),
\end{aligned}
\end{align}
where
\begin{align}
\mleft[\Pintf_{\mu}(\vb{p}_m, \vb{k}_n)\mright]_{s's} &\defeq u_{\vb{p} m s'}^{\dag} \Gamma_{\mu \vb{p}, \vb{k}} u_{\vb{k} n s}.
\end{align}
$\PintF_{BA}^{\mu \nu}$ and $\Pintf_{\mu}$ we shall call pairing form factors.

\subsection{Comparison to pairing due to order-parameter fluctuations} \label{sec:el-dip-sc-comparison-OP-ex}
At this point, a comparison to the analysis of pairing due to order-parameter fluctuations of Chap.~\ref{chap:loop_currents} is instructive.
In Sec.~\ref{sec:QCP-model-lin-gap-eq} of Chap.~\ref{chap:loop_currents}, for the pairing interaction we found [Eqs.~\eqref{eq:lin-gap-eq-boson-exchange-W-mat} and~\eqref{eq:lin-gap-eq-boson-exchange-M-mat}]:
\begin{align}
\PintW_{BA}^{\text{(Ch.\ref{chap:loop_currents})}}(\vb{p}_m, \vb{k}_n) &= p_{\TRop} \, g^2 \frac{\Pintw_{BA}^{\text{(Ch.\ref{chap:loop_currents})}}(\vb{p}_m, \vb{k}_n) + \Pintw_{BA}^{\text{(Ch.\ref{chap:loop_currents})}}(\vb{p}_m, -\vb{k}_n) p_A}{4 \abs{\grad_{\vb{p}} \varepsilon_{\vb{p} m}}^{1/2} \abs{\grad_{\vb{k}} \varepsilon_{\vb{k} n}}^{1/2}}, \\
\Pintw_{BA}^{\text{(Ch.\ref{chap:loop_currents})}}(\vb{p}_m, \vb{k}_n) &= \frac{L^d}{\mathcal{N}} \sum_{ab} \chi^{\text{(Ch.\ref{chap:loop_currents})}}(\vb{p} - \vb{k},0) \Kd_{ab} \Tr \mathcal{P}_{\vb{p} m}^{B} \Gamma_{a \vb{p}, \vb{k}}^{\text{(Ch.\ref{chap:loop_currents})}} \mathcal{P}_{\vb{k} n}^{A} \mleft[\Gamma_{b \vb{p}, \vb{k}}^{\text{(Ch.\ref{chap:loop_currents})}}\mright]^{\dag},
\end{align}
which is formally very similar to what we found in this section.
However, there are a number of important differences:
\begin{enumerate}
\item For TR-even order parameters, $\PintW_{BA}^{\text{(Ch.\ref{chap:loop_currents})}} \propto p_{\TRop} g^2 = + g^2$ is overall attractive.
In contrast, the TR-even electric monopoles and dipoles of this section give an overall repulsive $\PintW_{BA} \propto - e^2$.
Thus it is the TR-odd order parameters which result in Cooper pairing that is analogous to the one considered in this chapter.

\item The order-parameter field $\Phi_a^{\text{(Ch.\ref{chap:loop_currents})}}$ transforms under an arbitrary \emph{irreducible} representation, while the plasmon field $\Phi$ transforms like a TR-even scalar ($A_{1g}^{+}$), just like the electric charge density.
Moreover, the fermion-boson coupling matrices $\Gamma_{a \vb{p}, \vb{k}}^{\text{(Ch.\ref{chap:loop_currents})}}$ transform under the same irreducible representation as $\Phi_a^{\text{(Ch.\ref{chap:loop_currents})}}$, while the components of $\Gamma_{\mu \vb{p}, \vb{k}}$ transform as a scalar ($\mu = 0$) and vector ($\mu = 1, 2, 3$), i.e., its representation is reducible.
Consequently, in $\Pintw_{BA}^{\text{(Ch.\ref{chap:loop_currents})}}$ the sum over the irrep component indices $a, b$ must be $\propto \Kd_{ab}$ (see Sec.~\ref{sec:theory-of-invariants} of Appx.~\ref{app:group_theory}), while in our case the sum over $\mu, \nu$ in $\Pintw_{BA}$ is non-trivial.
In particular, the $V_{\mu j}(\vb{p}-\vb{k})$ need to contract with the dipole matrices $\Gamma_{j \vb{p}, \vb{k}}$ in the right way to give a $A_{1g}^{+}$ density-like object.

\item Relatedly, for higher-dimensional irreps the order parameter of Chap.~\ref{chap:loop_currents} has multiple components and the associated matrix $\chi^{\text{(Ch.\ref{chap:loop_currents})}}(\vb{p} - \vb{k},0) \Kd_{ab}$ is invertible. In our case there is always just one bosonic (plasmon) field and the $4 \times 4$ matrix $V_{\mu \nu}(\vb{p}-\vb{k})$ is non-invertible, with rank $1$.

\item When it comes to the structure of the coupling $\Gamma$ matrices, in Chap.~\ref{chap:loop_currents} we focused on loop currents whose $\Gamma_{a \vb{p}, \vb{k}}^{\text{(Ch.\ref{chap:loop_currents})}}$ are purely orbital.
Electric monopoles and dipoles also have purely orbital $\Gamma_{\mu \vb{p}, \vb{k}}$, but with the notable difference that they are TR-even.
\end{enumerate}

The origin of the first difference is that the Coulomb interaction $\Haml_C = \frac{1}{2} \int_{\vb{r}, \vb{r}'} \rho_e(\vb{r}) V(\vb{r}-\vb{r'}) \rho_e(\vb{r}')$ is repulsive, whereas the exchange of an order-parameter field always gives an attractive interaction of the form $\Haml_{\text{int}} = - \frac{1}{2} g^2 \int_{\vb{r}, \vb{r}'} \sum_a \phi_{a}^{\text{(Ch.\ref{chap:loop_currents})}}(\vb{r}) \chi^{\text{(Ch.\ref{chap:loop_currents})}}(\vb{r}-\vb{r'}) \phi_{a}^{\text{(Ch.\ref{chap:loop_currents})}}(\vb{r'})$ [Eq.~\eqref{eq:effective-multi-exch-int}], at least in the limit of negligible retardation.
With the help of a Hubbard-Stratonovich transformation, the Coulomb interaction can also be recast as an exchange of a bosonic (plasmon) field, as we discussed in Sec.~\ref{sec:el-dip-sc-plasmon-field}.
However, in the resulting $\action_{\text{int}} = \frac{1}{2} \epsilon_0 \int_{x} \mleft(\grad \Phi(x)\mright)^2 + \iu \int_{x} \Phi(x) \rho(x)$ [Eq.~\eqref{eq:eds-HS-int-action}] the coupling between the plasmon and density must be imaginary to ensure that the integral over the real-valued plasmon field $\Phi(x) = \Phi^{*}(x)$ converges, i.e., the $\iu$ cannot be absorbed into $\Phi$.
Among other things, this means that $\ev{\Phi(x)}$, if finite, is imaginary, as follows from the Schwinger-Dyson equation $\nabla^2 \ev{\Phi(x)} = \iu \ev{\rho(x)} / \epsilon_0$.
Evidently, $\Phi$ is just $(-\iu)$ times the scalar potential of the electromagnetic field and its $(-\iu)$ can be understood as arising from the Wick rotation of the electromagnetic four-potential $A_{\mu}$ to Euclidean time.
On the other hand, the order-parameter field $\Phi_a^{\text{(Ch.\ref{chap:loop_currents})}}(x)$, if it condenses, on physical grounds must attain a real value.
This constrains the coupling between the real-valued field $\Phi_a^{\text{(Ch.\ref{chap:loop_currents})}}$ and the Hermitian fermionic bilinear $\phi_a^{\text{(Ch.\ref{chap:loop_currents})}}$ to necessarily be real.
The order-parameter field $\Phi_a^{\text{(Ch.\ref{chap:loop_currents})}}$ can also be formulated as a field operator in the canonical formalism and its coupling to $\phi_a^{\text{(Ch.\ref{chap:loop_currents})}}$ then must be real to ensure that the Hamiltonian is Hermitian (which, in turn, is needed to make time evolution unitary).
Conversely, the plasmon field $\Phi$ (i.e., the scalar potential) does not arise as an operator or a dynamical degree of freedom in the Hamiltonian formalism, but as a Lagrange multiplier that enforces Gauss' law.\footnote{For an interesting recent discussion of Hubbard-Stratonovich transformations in the presence of both attraction and repulsion, see Ref.~\cite{Dalal2023}.}

The second and third differences are self-explanatory.

Regarding the last difference, we have already discussed one important implication of this difference in Sec.~\ref{sec:el-dip-sc-itinerant-dipoles}, namely, that electric dipole moments cannot be carried by electrons in the absence of spin-orbit coupling.
This is equivalent to the statement that the pairing form factor $\Pintf_{\mu}(\vb{p}_n, \vb{k}_n)$ vanishes at forward scattering $\vb{p}_n \to \vb{k}_n$ for the dipolar $\mu = 1, 2, 3$.
For loop currents, due to their opposite time-reversal sign, in Sec.~\ref{sec:Cp-channel-gen-sym-constr} we found the opposite: that even-parity loop-currents decouple at forward scattering.
Both statements follow from oddness under $P \TRop$ symmetry and in both cases we find pseudospin-triplet pairing form factors in the presence of SOC.

Given these similarities, it should come as no surprise that in the next section we shall be able to prove statements that resemble those we proved in Sec.~\ref{sec:TR-positivity} of Chap.~\ref{chap:loop_currents}.

\subsection{Pairing symmetry and upper bounds on the pairing strength} \label{sec:el-dip-sc-qualitative-pairing}
The fact that all interactions between the electric monopoles and dipoles are mediated by the same electrostatic field allows us to make a number of very general statements regarding the pairing.
To encode this fact, we start by writing the $V_{\mu \nu}$ of Eq.~\eqref{eq:eds-VUmklapp} in the following way:
\begin{align}
V_{\mu \nu}(\vb{q}) &= \mathrm{v}_{\mu}(\vb{q}) V(\vb{q}) \mathrm{v}_{\nu}^{*}(\vb{q}), \label{eq:eds-vVv-decomp}
\end{align}
where
\begin{align}
\mathrm{v}_{\mu}(\vb{q}) &= \begin{pmatrix}
1 \\
\iu q_i
\end{pmatrix} = \begin{pmatrix}
1 \\
\iu q_x \\
\iu q_y \\
\iu q_z
\end{pmatrix}.
\end{align}
See also Fig.~\ref{fig:el-dip-sc-plasmon-fermion-vertex}.
After renormalization, only $V(\vb{q}) \to \mathscr{V}(\vb{q})$ changes.
It then follows that
\begin{align}
\Pintw_{00}(\vb{p}_m, \vb{k}_n) &= V(\vb{p}-\vb{k}) \sum_{s's} \abs{\mleft[\bar{\Pintf}(\vb{p}_m, \vb{k}_n)\mright]_{s's}}^2
\end{align}
is strictly positive in the singlet channel, with $\bar{\Pintf}$ given by
\begin{align}
\mleft[\bar{\Pintf}(\vb{p}_m, \vb{k}_n)\mright]_{s's} &\defeq \sum_{\mu=0}^{3} \mathrm{v}_{\mu}(\vb{p}-\vb{k}) \mleft[\Pintf_{\mu}(\vb{p}_m, \vb{k}_n)\mright]_{s's}.
\end{align}
The singlet pairing interaction $\PintW_{00}(\vb{p}_m, \vb{k}_n)$ is therefore negative-definite.
For negative-definite matrices, the Perron-Frobenius theorem~\cite{Berman1994} applies and states that the largest-in-magnitude eigenvalue $\lambda_{\star}$ is negative and that the corresponding eigenvector $d_{\star}(\vb{k}_n)$ has no nodes, i.e., is an $s$-wave SC state.
While $\lambda_{\star}$ and $d_{\star}(\vb{k}_n)$ do not correspond to a SC instability, they are nonetheless a useful reference that bounds the possible pairing instabilities.
In particular, all positive singlet eigenvalues are bounded by $\abs{\lambda_{\star}}$ and, to be orthogonal to $d_{\star}(\vb{k}_n)$, their eigenvectors need to either have nodes or sign changes between Fermi surfaces.
Hence any singlet superconductivity must be unconventional and weaker than $\abs{\lambda_{\star}}$.
Note that extended $s$-wave pairing is still possible.

The triplet eigenvalues are bounded by $\abs{\lambda_{\star}}$ as well.
To show this, consider the eigenvector corresponding to the largest triplet eigenvalue.
Using the $\SU(2)$ local pseudospin gauge freedom, we may always orient this eigenvector along the $\vu{e}_3$ direction.
The corresponding
\begin{align}
\Pintw_{33}(\vb{p}_m, \vb{k}_n) &= V(\vb{p}-\vb{k}) \sum_{s's} (\pm)_{s'} (\pm)_{s} \abs{\mleft[\bar{\Pintf}(\vb{p}_m, \vb{k}_n)\mright]_{s's}}^2,
\end{align}
where $(\pm)_{\uparrow} = - (\pm)_{\downarrow} = +1$, is therefore bounded by $\Pintw_{00}(\vb{p}_m, \vb{k}_n)$, as is $\PintW_{33}(\vb{p}_m, \vb{k}_n)$ by $\abs{\PintW_{00}(\vb{p}_m, \vb{k}_n)}$.
A corollary of the Perron-Frobenius theorem~\cite{Berman1994} then states that the largest-in-magnitude triplet eigenvalue is strictly smaller in magnitude than the largest-in-magnitude singlet eigenvalue $\lambda_{\star}$, which we wanted to show.
That said, the largest \emph{positive} triplet eigenvalue may still be larger than the largest \emph{positive} singlet eigenvalue, resulting in triplet pairing overall.
Clearly, the proofs of these statements are completely analogous to those of Sec.~\ref{sec:TR-positivity} concerning pairing due to order-parameter exchange, which are in turn similar to the results of Ref.~\cite{Brydon2014} concerning phonon-exchange superconductivity.

Although it is, of course, expected that electronic mechanisms can only give superconductivity that is unconventional (not $s$-wave), the arguments of the previous paragraphs show this rigorously.
More interesting is the statement that the Cooper pairing strength is bounded by the strength of the repulsion, as measured by $\lambda_{\star}$.
To get an intuition regarding $\lambda_{\star}$, let us consider the simplest limit where there is only monopole coupling with $\Gamma_{\mu=0, \vb{k}, \vb{k}+\vb{q}} = \one$ in Eq.~\eqref{eq:eds-Pmu-assumed-form}.
We may then schematically write
\begin{align}
\begin{aligned}
\lambda_{\star} &\approx - \frac{1}{2} \DOSg_F \ev{\frac{e^2}{\epsilon_0 \vb{q}^2 + e^2 \DOSg_F}}_{\text{FS}} \\[2pt]
&\sim - \frac{1}{2} \, \frac{e^2 \DOSg_F}{\epsilon_0 k_F^2} \log\mleft(1 + \frac{\epsilon_0 k_F^2}{e^2 \DOSg_F}\mright),
\end{aligned} \label{eq:eds-schem-lambdaStar}
\end{align}
where in the interaction $V(\vb{q})$ we included Thomas-Fermi screening, the average is a Fermi surface average, $k_F$ characterizes the size of the Fermi sea ($\sim k_F^2$ is the area), and the total density of states (DOS) is
\begin{align}
\DOSg_F &= 2 \sum_n \int\limits_{\varepsilon_{\vb{k} n} = 0} \frac{\dd{S_{\vb{k}}}}{(2\pi)^3} \frac{1}{\abs{\grad_{\vb{k}} \varepsilon_{\vb{k} n}}}.
\end{align}
The factor of two comes from the spins.

Hence $\lambda_{\star}$ goes like $\sim \DOSg_F \abs{\log \DOSg_F}$ to zero for small $\DOSg_F$, and to $-1/2$ for large $\DOSg_F$.
Clearly then, a small DOS is unfavorable for superconductivity, as expected.
Less obviously, one cannot make the pairing arbitrarily strong by increasing the DOS because of the DOS-dependent screening.
This is in distinction to other mechanisms, such as pairing due to phonons and, to some extent, also the pairing due to quantum-critical boson exchange~\cite{Millis2001, Chubukov2003, Lederer2015, Lederer2017, Klein2018, Klein2023}, where the DOS can be increased while the pairing interaction changes only moderately.

\begin{figure}[t]
\centering
\begin{subfigure}[t]{0.35\textwidth}
\centering
\includegraphics[width=0.90\textwidth]{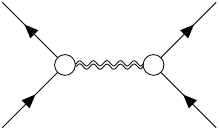}
\subcaption{}
\end{subfigure}%
\begin{subfigure}[t]{0.65\textwidth}
\centering
\includegraphics[width=0.90\textwidth]{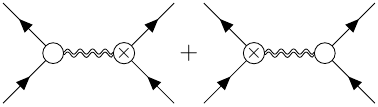}
\subcaption{}
\end{subfigure}
\captionbelow[The leading contributions to the pairing interaction derive from monopole-monopole (a) and monopole-dipole (b) coupling.]{\textbf{The leading contributions to the pairing interaction derive from monopole-monopole (a) and monopole-dipole (b) coupling.}
The double wiggly lines indicates that the plasmon propagator is screened, while the cross indicates a dipole vertex, as defined in Fig.~\ref{fig:el-dip-sc-plasmon-fermion-vertex}.
With sufficient screening, the repulsive contribution from (a) mainly acts in the $s$-wave channel and is orthogonal to the attractive contribution from (b).}
\label{fig:el-dip-sc-monopole-dipole-exchange}
\end{figure}

Finally, we show that our interaction can indeed have positive eigenvalues, resulting in superconductivity, when the screening and dipole moments are strong enough.
Our reasoning is the following:
In the interaction~\eqref{eq:eds-vVv-decomp}, $V(\vb{q})$ decreases with increasing $\vb{q}$, whereas the dipolar part of $\mathrm{v}_{\mu}(\vb{q}) = \mleft(1, \iu q_i\mright)$ increases.
Strong screening means that $V(\vb{q})$ decays weakly with increasing $\vb{q}$.
Thus sufficiently large electric dipole moments can overwhelm this decay to give an interaction that is overall more strongly repulsive at finite $\vb{q}$ than at $\vb{q} = \vb{0}$.
It then follows that pairing eigenvectors which change sign every $\vb{Q}$, where $\vb{Q} \neq \vb{0}$ is the repulsion peak, have positive eigenvalues~\cite{Maiti2013}, which we wanted to show.
A qualitative argument for this statement is given in Sec.~\ref{sec:cuprate-sym-choice-mechanism} of Chap.~\ref{chap:cuprates}.
A similar behavior occurs in the celebrated Kohn-Luttinger mechanism~\cite{Kohn1965, Luttinger1966, Maiti2013, Kagan2014} in which the overscreening of $V(\vb{q})$ is a consequence of the $2 k_F$ non-analyticity of the system.
In our case, the electric dipoles are responsible for this overscreening and formally it develops already in the leading order of the Coulomb interaction (Fig.~\ref{fig:el-dip-sc-monopole-dipole-exchange}).
In particular, to leading order in powers of the electric dipole moment, the interaction that is responsible for the pairing in our mechanism is the screened monopole-dipole interaction which is shown in Fig.~\ref{fig:el-dip-sc-monopole-dipole-exchange}(b).

\subsection{Pairing in a spherical toy model} \label{sec:el-dip-sc-pairing-toy-model}
To illustrate our mechanism, let us consider a Fermi liquid with spherical symmetry and only one Fermi surface.
For the interaction and coupling we assume
\begin{align}
\begin{aligned}
e^2 V(\vb{p}-\vb{k}) &= U_0 + U_1 \vu{p} \vdot \vu{k} + \cdots \, , \\[2pt]
\Pintf_{0}(\vb{p}, \vb{k}) &= \Pauli_0, \\[2pt]
\Pintf_{i}(\vb{p}, \vb{k}) &= - \frac{\eta}{2 e} \mleft[(\vu{p} + \vu{k}) \vcross \vb{\Pauli}\mright]_{i},
\end{aligned}
\end{align}
where $\vu{p} = \vb{p} / \abs{\vb{p}}$ and $\vu{k} = \vb{k} / \abs{\vb{k}}$ are direction unit vectors, while $\abs{\vb{p}} = \abs{\vb{k}} = k_F$.
$U_1 > 0$ quantifies the degree of screening and $\eta$ is the electric dipole moment.
Notice that spin-orbit coupling is needed (Sec.~\ref{sec:el-dip-sc-itinerant-dipoles}) for the dipolar pairing form factors $\Pintf_{i}(\vb{p}, \vb{k})$ to have the form we assumed here.
The $\Pintf_{i}$ that we wrote down is the simplest one that is consistent with symmetries.
To linear order in $U_1$ and $\eta$, we find that:
\begin{gather}
\begin{gathered}
\PintW_{00}(\vb{p}, \vb{k}) = - \frac{U_0}{v_F}, \\[2pt]
\PintW_{ij}(\vb{p}, \vb{k}) = 2 \frac{U_0 k_F \eta}{v_F e} (\hat{p}_i \hat{k}_j - \hat{p}_j \hat{k}_i) - \frac{U_1}{v_F} \vu{p} \vdot \vu{k} \, \Kd_{ij},
\end{gathered}
\end{gather}
where $i, j \in \{1, 2, 3\}$ and $v_F = \abs{\grad_{\vb{k}} \varepsilon_{\vb{k}}}$ is the Fermi velocity at $\abs{\vb{k}} = k_F$.
In the singlet channel we find no pairing, while for the leading instability in the triplet channel we find
\begin{gather}
\begin{gathered}
\lambda_1 = \frac{2}{3} \DOSg_F U_0 k_F \eta / e - \frac{1}{6} \DOSg_F U_1, \\[4pt]
\vb{d}_1(\vb{k}) = \vu{k}
\end{gathered}
\end{gather}
which has pseudoscalar symmetry ($\sim \vu{k} \vdot \vb{\Pauli}$).
There is also a subleading $p$-wave instability with
\begin{gather}
\begin{gathered}
\lambda_2 = \frac{1}{3} \DOSg_F U_0 k_F \eta / e - \frac{1}{6} \DOSg_F U_1, \\[4pt]
\vb{d}_{2, a}(\vb{k}) = \vu{e}_{a} \vcross \vu{k}
\end{gathered}
\end{gather}
which is threefold degenerate; $a \in \{x, y, z\}$ is the degeneracy (irrep) index and $\vu{e}_a$ are Cartesian unit vectors.
Thus if dipole moments are strong compared to the screening, namely $k_F \eta /e > U_1 / (4 U_0)$, the monopole-dipole electrostatic interaction will result in superconductivity of pseudoscalar symmetry.

\section{Cooper pairing in quasi-2D Dirac metals} \label{sec:el-dip-sc-Dirac-pairing}
Here we study the superconducting instabilities of the dipolar Dirac model of Sec.~\ref{sec:el-dip-sc-Dirac-model} in the quasi-2D limit $v_z \approx 0$, that is $v_z \Lambda_z \ll m$.
The starting point our analysis is the effective model that emerges at the end of the RG flow of Sec.~\ref{sec:el-dip-sc-Dirac-RG}.
This effective model has a negligible in-plane dipole coupling $\eta_{\perp} \approx 0$, an enhanced out-of-plane dipole coupling $\eta_z$, and a momentum cutoff $\Lambda \sim k_F$.
Its Cooper pairing we analyze using the linearized gap equation we introduced in Sec.~\ref{sec:el-dip-sc-lin-gap-eq}.
For strong enough screening and $z$-axis dipole moments $\eta_z$, we find that unconventional odd-parity Cooper pairing takes place which has pseudoscalar symmetry $\sim \vb{k} \vdot \vb{\Pauli}$, similar to the superfluid state of \ce{^3He-B}; see Figs.~\ref{fig:el-dip-sc-FS-d-vec-sketch} and~\ref{fig:el-dip-sc-pairing-w-screening}.
In addition, we find a competitive subleading pairing instability of $p$-wave symmetry.

As in our RG treatment, we employ a large-$N$ expansion to analytically access the regime of strong screening.
A slight difference from Sec.~\ref{sec:el-dip-sc-Dirac-RG} is that the cutoff is not imposed on the frequencies [Eq.~\eqref{eq:eds-cutoff-def}], but only on the momenta through their energies $\varepsilon_{\vb{k}} = \sqrt{m^2 + v^2 \vb{k}_{\perp}^2} - \upmu$.
Because we ended the RG flow with a $\Lambda \sim k_F$, our energy cutoff $\hbar \omega_c$ is on the order of the Fermi energy $E_F = \upmu - m = \sqrt{m^2 + v^2 k_F^2} - m$.
Note that the same convention with the energy cutoff was used in the derivation of Eqs.~\eqref{eq:eds-lin-gap-eq} and~\eqref{eq:eds-BCS-Tc-expr} in Appx.~\ref{app:lin_gap_eq}.

Another minor difference from before is that we need to impose periodicity along the $\vu{e}_z$ direction on the model.
Instead of Eqs.~\eqref{eq:eds-V-1-phi} and~\eqref{eq:eds-A-psi}, we thus use
\begin{align}
V^{-1}(\vb{q}) &= \epsilon_{\perp} \vb{q}_{\perp}^2 + \frac{4 \Lambda_z^2 \epsilon_z}{\pi^2} \sin^2\frac{\pi q_z}{2 \Lambda_z}, \\
A(k, p) &= e \gamma_0 + \iu \frac{\Lambda_z \eta_z}{\pi} \sin\frac{\pi (k_z - p_z)}{\Lambda_z} \cdot \gamma_0 \gamma_3.
\end{align}
This is necessary because we are interested in momenta with $\abs{\vb{q}_{\perp}} \sim k_F$ and $q_z \sim \Lambda_z$.
The origin of this periodicity is the Umklapp sum along $\vu{e}_z$, as we discussed after Eq.~\eqref{eq:eds-VUmklapp}.
We only consider quasi-2D systems with $v_z = \eta_{\perp} = 0$ because of the RG considerations of Sec.~\ref{sec:el-dip-sc-Dirac-RG-tree}.

In the limit of strong screening, the interaction is given by the polarization bubble which in the static $\omega_q = 0$ limit for $\abs{\vb{q}_{\perp}} \leq 2 k_F$ equals [Eq.~\eqref{eq:eds-static-smallq-pol-previous}]:
\begin{align}
\Pi(\omega_q = 0, \vb{q}) &= \mathscr{V}^{-1}(\vb{q}) - V^{-1}(\vb{q}) = N \DOSg_F \mleft[e^2 + \frac{\Lambda_z^2 \eta_z^2}{\pi^2} \sin^2\frac{\pi q_z}{\Lambda_z}\mright], \label{eq:eds-static-smallq-pol}
\end{align}
where
\begin{align}
\DOSg_F &= \frac{\Lambda_z \upmu}{\pi^2 v^2}, &
\upmu &= \sqrt{m^2 + v^2 k_F^2}.
\end{align}
Although this was evaluated without a cutoff ($\Lambda \to \infty$), reintroducing it does not significantly influence this expression.

In the $\Pintw_{BA}(\vb{p}, \vb{k})$ pairing interaction of Eq.~\eqref{eq:eds-w-pairing-def-alt}, we therefore use
\begin{align}
V_{\mu \nu}(\vb{q}) &= \begin{pmatrix}
1 & \displaystyle - \iu \frac{\Lambda_z}{\pi} \sin\frac{\pi q_z}{\Lambda_z} \\[10pt]
\displaystyle \iu \frac{\Lambda_z}{\pi} \sin\frac{\pi q_z}{\Lambda_z} & \displaystyle \frac{\Lambda_z^2}{\pi^2} \sin^2\frac{\pi q_z}{\Lambda_z}
\end{pmatrix} \mathscr{V}(\vb{q}), \\[2pt]
\Gamma_{\mu \vb{p}, \vb{k}} &= \begin{pmatrix}
\one \\[2pt]
(\eta/e) \gamma_3
\end{pmatrix} = \begin{pmatrix}
\one \\[2pt]
- (\eta/e) \uptau_2 \Pauli_0
\end{pmatrix},
\end{align}
where $\mu, \nu \in \{0, 3\}$; the $\mu, \nu = 1, 2$ components have been omitted because they vanish.

To calculate the pairing interaction $\PintW_{BA}(\vb{p}, \vb{k})$ of Eq.~\eqref{eq:eds-W-pairing-def}, we need to diagonalize the Dirac Hamiltonian [Eq.~\eqref{eq:eds-Dirac-Haml}]:
\begin{align}
H_{\vb{k}} &= m \uptau_3 \Pauli_0 + v \uptau_2 (k_x \Pauli_y - k_y \Pauli_x) - \upmu \uptau_0 \Pauli_0.
\end{align}
The dispersion of the conduction band is
\begin{align}
\varepsilon_{\vb{k}} &= \sqrt{m^2 + v^2 \vb{k}_{\perp}^2} - \upmu,
\end{align}
and the corresponding conduction band eigenvectors are easily found to be
\begin{align}
u_{\vb{k}\uparrow} &= \frac{1}{\sqrt{\mathcal{N}_{\vb{k}}}} \begin{pmatrix}
m + \sqrt{m^2 + v^2 \vb{k}_{\perp}^2} \\ 0 \\ 0 \\ - v (k_x + \iu k_y)
\end{pmatrix}, \\
u_{\vb{k}\downarrow} &= \frac{1}{\sqrt{\mathcal{N}_{\vb{k}}}} \begin{pmatrix}
0 \\ m + \sqrt{m^2 + v^2 \vb{k}_{\perp}^2} \\ v (k_x - \iu k_y) \\ 0
\end{pmatrix},
\end{align}
where $\vb{k}_{\perp} = (k_x, k_y)$ [Eq.~\eqref{eq:eds-bold-vec-conv}], $\uparrow, \downarrow$ are pseudospins, and
\begin{align}
\mathcal{N}_{\vb{k}} &= 2 \sqrt{m^2 + v^2 \vb{k}_{\perp}^2} \mleft(m + \sqrt{m^2 + v^2 \vb{k}_{\perp}^2}\mright).
\end{align}
In this particular gauge, the symmetry transformation rules of the pseudospins are identical to those of the spins:
\begin{align}
\MatU(g) \begin{pmatrix}
u_{\vb{k}\uparrow} & u_{\vb{k}\downarrow}
\end{pmatrix} &= \begin{pmatrix}
u_{R(g)\vb{k}\uparrow} & u_{R(g)\vb{k}\downarrow}
\end{pmatrix} S(g), \\
\MatTR \begin{pmatrix}
u_{\vb{k}\uparrow}^{*} & u_{\vb{k}\downarrow}^{*}
\end{pmatrix} &= \begin{pmatrix}
u_{-\vb{k}\uparrow} & u_{-\vb{k}\downarrow}
\end{pmatrix} \iu \Pauli_y.
\end{align}
The $\MatU(g)$ and $S(g)$ matrices are given in Tab.~\ref{tab:eds-sym-tranf-matrices}, while $\MatTR = \uptau_3 \iu \Pauli_y$ [Eq.~\eqref{eq:eds-U_P-TH-def}].
At each $\vb{k}$, $u_{\vb{k}\uparrow} = \MatU(P) \MatTR u_{\vb{k}\downarrow}^{*} = \uptau_0 \iu \Pauli_y u_{\vb{k}\downarrow}^{*}$.
The most notable difference from the most general transformation rules we wrote down in Eqs.~\eqref{eq:band-eigenvector-sym-TR-rule} and~\eqref{eq:band-eigenvector-sym-transf-rule} of Sec.~\ref{sec:LC-gen-model-sym} (Chap.~\ref{chap:loop_currents}) is that there is no $\vb{k}$-dependence in the pseudospin rotation matrices, despite the presence of spin-orbit coupling.
This is made possible by the absence of $\vb{k}$-dependence in the $\MatU(g)$.
Because the pseudospins transform like spins, this means that the triplet-channel pairing $\vb{d}$-vectors transform like vectors, as described by Eq.~\eqref{eq:fundamental-triplet-dvec-sym-transf-rule} of Sec.~\ref{sec:lin-gap-eq-spectral-symmetries} with $R_{\vb{p} m}(g) \to R(g)$.

The Fermi surface is a cylinder and the in-plane momenta that are on the cylindrical Fermi surface we shall parameterize with azimuthal angles:
\begin{align}
\begin{aligned}
\vb{p}_{\perp} &= k_F (\cos \theta_p, \sin \theta_p), \\
\vb{k}_{\perp} &= k_F (\cos \theta_k, \sin \theta_k).
\end{aligned}
\end{align}
Now it is a straightforward task to find $\PintW_{BA}(\theta_p, p_z, \theta_k, k_z)$ as given by Eq.~\eqref{eq:eds-W-pairing-def}.
The final expression for $\PintW_{BA}$ that one obtains is fairly complicated, and one cannot diagonalize it [Eq.~\eqref{eq:eds-lin-gap-eq}] analytically for general momentum-dependent interactions $\mathscr{V}(\vb{q})$.
Thus one needs to resort to numerical methods.

\subsection{Analytic solution of the perfect screening limit}
Physically, we are interested in the limit of strong screening in which case the momentum dependence of $\mathscr{V}(\vb{q})$ is weak.
To understand this limit, a good starting point is to consider a constant Hubbard-like interaction
\begin{align}
\mathscr{V}(\vb{q}) &= \frac{1}{\DOSg_F e^2} \equiv U_0 \label{eq:eds-perfect-screening}
\end{align}
which corresponds to the large-$N$ limit [Eq.~\eqref{eq:eds-static-smallq-pol}] with the $q_z$ dependence neglected.
The numerical results, which we present in the next section, can be well understood by analyzing this idealized scenario. 
For a constant interaction, we can exactly diagonalize $\PintW_{BA}$.
The result is~\cite{Palle2024-el-dip}:
\begin{align}
W(\theta_p, p_z, \theta_k, k_z) &= \frac{e^2 U_0}{v} \sum_{n=1}^{12} w_n \sum_{a=1}^{\dim n} d_{n,a}(\theta_p, p_z) d_{n,a}^{\intercal}(\theta_k, k_z), \label{eq:eds-W-eigs-def}
\end{align}
where $w_n$ are dimensionless eigenvalues of degeneracy $\dim n$ and $d_{n,a}(\theta_p, p_z) = d_{n,a}^{*}(\theta_p, p_z)$ are the corresponding eigenvectors, which we made real-valued.
Both are listed in Tab.~\ref{tab:eds-W-eigs}, reproduced from Ref.~\cite{Palle2024-el-dip}.
The eigenvectors are orthogonal and normalized according to
\begin{align}
\int_{-\pi}^{\pi} \frac{\dd{\theta_k}}{2 \pi} \int_{-\Lambda_z}^{\Lambda_z} \frac{\dd{k_z}}{2 \pi} \, d_{n,a}^{\intercal}(\theta_k, k_z) d_{n',a'}(\theta_k, k_z) &= \frac{\Lambda_z}{\pi} \Kd_{nn'} \Kd_{aa'}.
\end{align}
The corresponding pairing eigenvalues $\lambda$ arising in the linearized gap equation~\eqref{eq:eds-lin-gap-eq} therefore equal
\begin{align}
\lambda_n &= \frac{w_n}{2 \sqrt{1 + \hat{m}^2}},
\end{align}
where
\begin{align}
\hat{m} &\defeq \frac{m}{v k_F}, &
\hat{r}_{\pm} &\defeq \frac{\sqrt{1 + \hat{m}^2} \pm \hat{m}}{\sqrt{2 (1 + 2 \hat{m}^2)}}, &
\hat{\eta} &\defeq \frac{\Lambda_z \eta_z}{\pi \, e} \label{eq:eds-new-meta}
\end{align}
are dimensionless measures of the gap and electric dipole coupling.
Given how $\Lambda_z / \pi$ arises in many places, we shall find it convenient to henceforth set the lattice constant along $z$ to unity:
\begin{align}
\Lambda_z &= \pi.
\end{align}

\begin{figure}[t]
\centering
\includegraphics[width=0.5\textwidth]{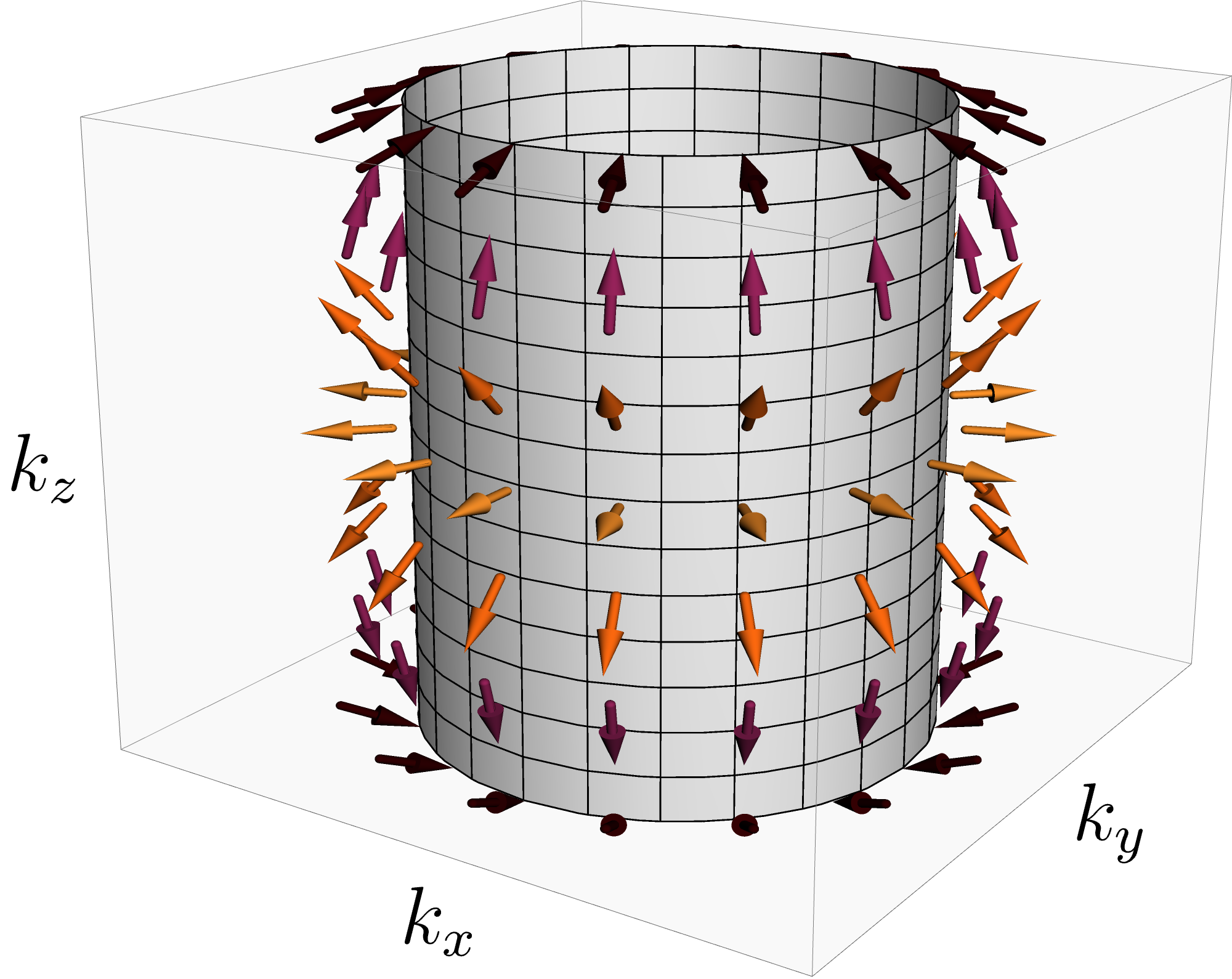}
\captionbelow[The $\vb{d}$-vector of the leading pairing state of a quasi-2D Dirac metal as a function of momentum, assuming perfect screening.]{\textbf{The $\vb{d}$-vector of the leading pairing state of a quasi-2D Dirac metal as a function of momentum, assuming perfect screening.}
The grey cylinder is the Fermi surface.
The arrows indicate the direction of the $\vb{d}$-vector which is given by Eq.~\eqref{eq:eds-leading-pairing} with $\hat{\eta} > 0$.
The overall symmetry of this $\vb{d}$-vector pattern is pseudoscalar.}
\label{fig:el-dip-sc-FS-d-vec-sketch}
\end{figure}

Of the twelve $w_n$, four are positive and give positive $\lambda$ which correspond to superconducting instabilities.
The leading instability among these four is odd-parity and pseudospin-triplet, with ($n = 5, 6$ in Tab.~\ref{tab:eds-W-eigs}):
\begin{gather}
\begin{gathered}
\lambda_{5/6} = \frac{\abs{\hat{\eta}}}{4 \sqrt{1 + \hat{m}^2}}, \\
\vb{d}_{5/6}(\theta_k, k_z) = \begin{pmatrix}
\cos \theta_k \cos k_z \\
\sin \theta_k \cos k_z \\
\sgn \hat{\eta} \, \sin k_z
\end{pmatrix}.
\end{gathered} \label{eq:eds-leading-pairing}
\end{gather}
Since $\vb{d}_{5/6}(\theta_k, k_z) \sim (k_x, k_y, \pm k_z)$, its symmetry is pseudoscalar.
The $\vb{d}$-vector of this solution is depicted in Fig.~\ref{fig:el-dip-sc-FS-d-vec-sketch}.

The subleading pairing instability is also odd-parity and pseudospin-triplet, but has $p$-wave symmetry and is weaker by a factor in between $\sqrt{2}$ and $2$ from the leading instability.
It is a two-component pairing state that may either give rise to time-reversal symmetry breaking or nematic superconductivity, depending on the quartic coefficients in the Ginzburg-Landau expansion (cf.\ Sec.~\ref{sec:SRO-GL-analysis} of the next chapter).
Its pairing eigenvalue equals:
\begin{gather}
\lambda_{7/8} = \frac{\abs{\hat{\eta}} \sqrt{1 + 2 \hat{m}^2}}{8 (1 + \hat{m}^2)}. \label{eq:eds-subleading-pairing}
\end{gather}
The corresponding two degenerate eigenvectors are ($n = 7, 8$ entries of Tab.~\ref{tab:eds-W-eigs}):
\begin{align}
\vb{d}_{7/8,x}(\theta_k, k_z) &= \begin{pmatrix}
0 \\
- \hat{r}_{-} \sgn \hat{\eta} \, \sin 2 \theta_k \sin k_z \\
(- \hat{r}_{+} + \hat{r}_{-} \sin 2 \theta_k) \sgn \hat{\eta} \sin k_z \\
\sqrt{2} \sin \theta_k \cos k_z
\end{pmatrix}, \\
\vb{d}_{7/8,y}(\theta_k, k_z) &= \begin{pmatrix}
0 \\
(\hat{r}_{+} + \hat{r}_{-} \cos 2 \theta_k) \sgn \hat{\eta} \sin k_z \\
\hat{r}_{-} \sgn \hat{\eta} \, \sin 2 \theta_k \sin k_z \\
- \sqrt{2} \cos \theta_k \cos k_z
\end{pmatrix}.
\end{align}

In agreement with our general discussion of Sec.~\ref{sec:el-dip-sc-qualitative-pairing}, the largest-in-magnitude $\lambda$ which bounds all other $\lambda$ is ($n = 1$ in Tab.~\ref{tab:eds-W-eigs})
\begin{align}
\lambda_{\star} &= \lambda_1 = - \frac{1 + 2 \hat{m}^2 + \hat{\eta}^2}{4 (1 + \hat{m}^2)}
\end{align}
and it has an even-parity pseudospin-singlet $s$-wave eigenvector.
Compare with Eq.~\eqref{eq:eds-schem-lambdaStar}.

\newpage
\subsubsection{Table of pairing eigenvalues and eigenvectors}

{\renewcommand{\arraystretch}{1.3}
\renewcommand{\tabcolsep}{10pt}
\begin{longtable}[c]{lNNc}
\caption[The eigenvalues $w_n$ and eigenvectors $d_{n,a}(\theta_k, k_z)$ arising in the eigen-expansion \eqref{eq:eds-W-eigs-def} of the pairing interaction $\PintW_{BA}(\theta_p, p_z, \theta_k, k_z)$ of a quasi-2D Dirac metal with a constant interaction~\cite{Palle2024-el-dip}.]{\textbf{The eigenvalues $w_n$ and eigenvectors $d_{n,a}(\theta_k, k_z)$ arising in the eigen-expansion~\eqref{eq:eds-W-eigs-def} of the pairing interaction $\PintW_{BA}(\theta_p, p_z, \theta_k, k_z)$ of a quasi-2D Dirac metal with a constant interaction}~\cite{Palle2024-el-dip}.
Here $\displaystyle \hat{m} \defeq \frac{m}{v k_F}$, $\displaystyle \hat{r}_{\pm} \defeq \frac{\sqrt{1 + \hat{m}^2} \pm \hat{m}}{\sqrt{2 (1 + 2 \hat{m}^2)}}$, $\displaystyle \hat{\eta} \defeq \frac{\Lambda_z \eta_z}{\pi \, e}$, $\Lambda_z = \pi$, and $\theta_k$ is the azimuthal angle specifying the in-plane position on the cylindrical Fermi surface, $\vb{k}_{\perp} = k_F (\cos \theta_k, \sin \theta_k)$.
The degeneracy $\dim n$ of the $n$-th eigenvalue is either $1$ or $2$, depending on how many eigenvectors are shown in the table.
In cases when $\dim n = 1$, we suppress the $a \in \{1, \ldots, \dim n\}$ index.
For the $p$-wave cases with $\dim n = 2$, we have ensured that the two components transform like $(x|y)$, so sometimes we need to negate and permute the components, like in $(k_y\Pauli_z|-k_x\Pauli_z)$; see also Sec.~\ref{sec:examples-convention-D4h} of Appx.~\ref{app:group_theory}.
For even-parity pseudospin-singlet eigenvectors only the first component is finite, while in odd-parity pseudospin-triplet eigenvectors only the last three components are finite and together constitute the corresponding Balian-Werthamer $\vb{d}$-vector.} \label{tab:eds-W-eigs} \\[206pt] \hline\hline
$n,a$ & $w_n$ & $d_{n,a}(\theta_k, k_z)$ & symmetry \endfirsthead
\caption[]{(continued)} \\[20pt] \hline\hline
$n,a$ & $w_n$ & $d_{n,a}(\theta_k, k_z)$ & symmetry \endhead
\hline
$1$ & $\displaystyle - \frac{1 + 2 \hat{m}^2 + \hat{\eta}^2}{2 \sqrt{1+\hat{m}^2}}$ & $\begin{pmatrix}
1 \\ 0 \\ 0 \\ 0
\end{pmatrix}$ & $s$-wave \\[24pt]
$2$ & $\displaystyle - \frac{1 - \hat{\eta}^2}{2 \sqrt{1+\hat{m}^2}}$ & $\begin{pmatrix}
0 \\ \cos \theta_k \\ \sin \theta_k \\ 0
\end{pmatrix}$ & {\renewcommand{\arraystretch}{0.8}\begin{tabular}{N}
pseudoscalar \\
$k_x \Pauli_x + k_y \Pauli_y$
\end{tabular}} \\[24pt]
$3$ & $\displaystyle - \frac{1 + \hat{\eta}^2}{2 \sqrt{1+\hat{m}^2}}$ & $\begin{pmatrix}
0 \\ \sin \theta_k \\ - \cos \theta_k \\ 0
\end{pmatrix}$ & {\renewcommand{\arraystretch}{0.8}\begin{tabular}{N}
$p$-wave \\
$z$-axis vector \\
$k_x \Pauli_y - k_y \Pauli_x \sim \vu{e}_z$
\end{tabular}} \\[24pt]
$4,x$ & $\displaystyle - \frac{2 - \hat{\eta}^2}{4 \sqrt{1+\hat{m}^2}}$ & $\begin{pmatrix}
0 \\ 0 \\ 0 \\ \sqrt{2} \sin \theta_k
\end{pmatrix}$ & {\renewcommand{\arraystretch}{0.8}\begin{tabular}{N}
$p$-wave \\
$x$-axis vector \\
$k_y \Pauli_z \sim \vu{e}_x$
\end{tabular}} \\[24pt]
$4,y$ & & $\begin{pmatrix}
0 \\ 0 \\ 0 \\ - \sqrt{2} \cos \theta_k
\end{pmatrix}$ & {\renewcommand{\arraystretch}{0.8}\begin{tabular}{N}
$p$-wave \\
$y$-axis vector \\
$- k_x \Pauli_z \sim \vu{e}_y$
\end{tabular}} \\[24pt] \hline\hline
\pagebreak \hline
$5$ & $\displaystyle \frac{\hat{\eta}}{2}$ & $\begin{pmatrix}
0 \\ \cos \theta_k \cos k_z \\ \sin \theta_k \cos k_z \\ \sin k_z
\end{pmatrix}$ & {\renewcommand{\arraystretch}{0.8}\begin{tabular}{N}
pseudoscalar \\
$k_x \Pauli_x + k_y \Pauli_y + k_z \Pauli_z$
\end{tabular}} \\[24pt]
$6$ & $\displaystyle - \frac{\hat{\eta}}{2}$ & $d_{5}(\theta_k, -k_z)$ & {\renewcommand{\arraystretch}{0.8}\begin{tabular}{N}
pseudoscalar \\
$k_x \Pauli_x + k_y \Pauli_y - k_z \Pauli_z$
\end{tabular}} \\[10pt]
$7,x$ & $\displaystyle \frac{\hat{\eta} \sqrt{1 + 2 \hat{m}^2}}{4 \sqrt{1+\hat{m}^2}}$ & $\begin{pmatrix}
0 \\
- \hat{r}_{-} \sin 2 \theta_k \sin k_z \\
(- \hat{r}_{+} + \hat{r}_{-} \sin 2 \theta_k) \sin k_z \\
\sqrt{2} \sin \theta_k \cos k_z
\end{pmatrix}$ & {\renewcommand{\arraystretch}{0.8}\begin{tabular}{N}
$p_x$-wave \\
$k_y \Pauli_z + \cdots \sim \vu{e}_x$
\end{tabular}} \\[24pt]
$7,y$ & & $\begin{pmatrix}
0 \\
(\hat{r}_{+} + \hat{r}_{-} \cos 2 \theta_k) \sin k_z \\
\hat{r}_{-} \sin 2 \theta_k \sin k_z \\
- \sqrt{2} \cos \theta_k \cos k_z
\end{pmatrix}$ & {\renewcommand{\arraystretch}{0.8}\begin{tabular}{N}
$p_y$-wave \\
$- k_x \Pauli_z + \cdots \sim \vu{e}_y$
\end{tabular}} \\[24pt]
$8,x$ & $\displaystyle - \frac{\hat{\eta} \sqrt{1 + 2 \hat{m}^2}}{4 \sqrt{1+\hat{m}^2}}$ & $d_{7,x}(\theta_k, -k_z)$ & $p_x$-wave \\[24pt]
$8,y$ & & $d_{7,y}(\theta_k, -k_z)$ & $p_y$-wave \\[24pt]
$9$ & $\displaystyle \frac{\hat{\eta}^2}{4\sqrt{1+\hat{m}^2}}$ & $\sqrt{2} \cos(2 k_z) \cdot d_{1}(\theta_k, k_z)$ & $s$-wave \\[24pt]
$10$ & $\displaystyle - \frac{\hat{\eta}^2}{4\sqrt{1+\hat{m}^2}}$ & $\sqrt{2} \cos(2 k_z) \cdot d_{2}(\theta_k, k_z)$ & pseudoscalar \\[24pt]
$11$ & $\displaystyle \frac{\hat{\eta}^2}{4\sqrt{1+\hat{m}^2}}$ & $\sqrt{2} \cos(2 k_z) \cdot d_{3}(\theta_k, k_z)$ & $p_z$-wave \\[24pt]
$12,x$ & $\displaystyle - \frac{\hat{\eta}^2}{8\sqrt{1+\hat{m}^2}}$ & $\sqrt{2} \cos(2 k_z) \cdot d_{4,x}(\theta_k, k_z)$ & $p_x$-wave \\[24pt]
$12,y$ & & $\sqrt{2} \cos(2 k_z) \cdot d_{4,y}(\theta_k, k_z)$ & $p_y$-wave
\\[6pt] \hline\hline
\end{longtable}}

\newpage

\subsection{Numerical solutions of the linearized gap equation}
\begin{figure}[t!]
\centering
\includegraphics[width=\textwidth]{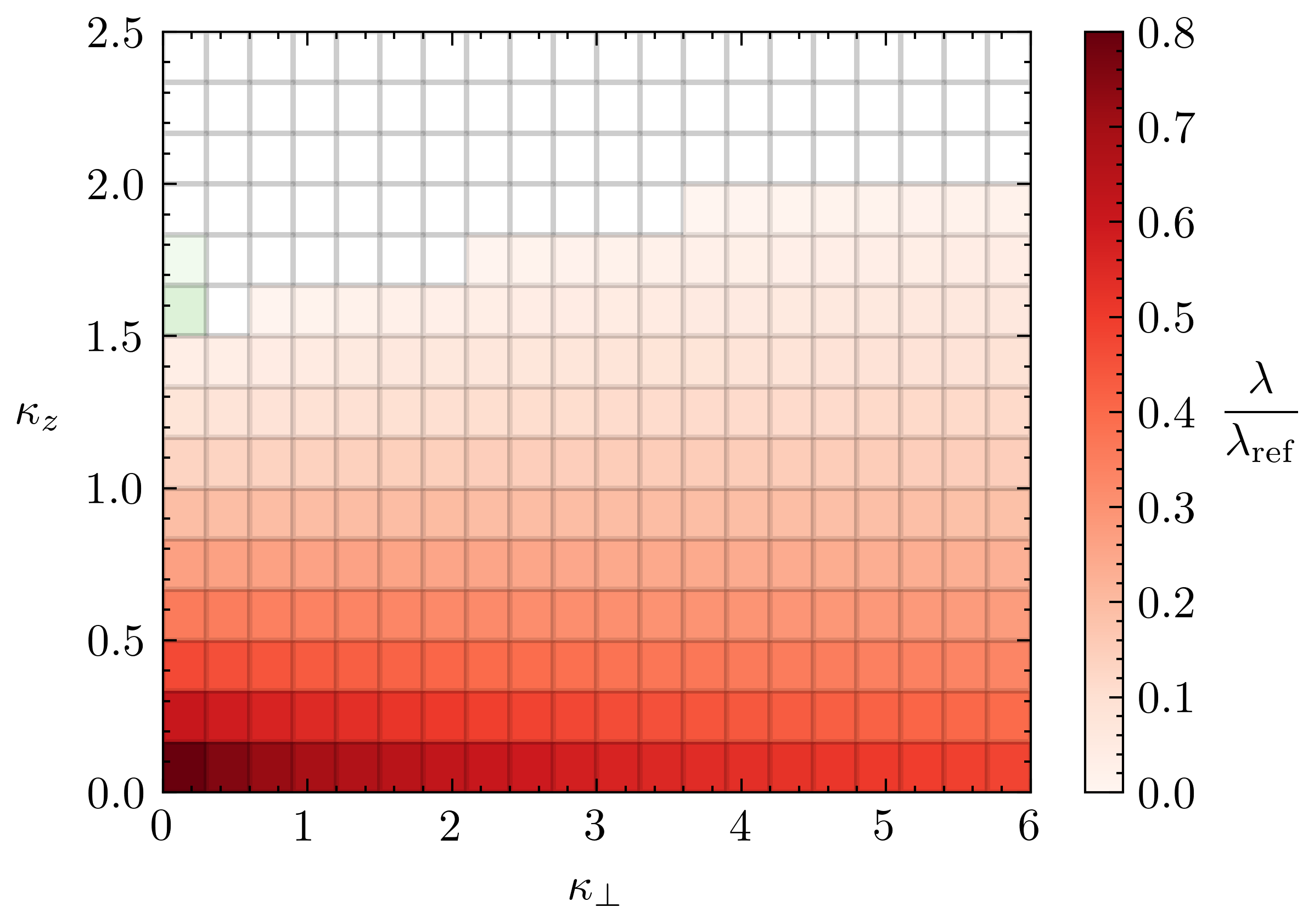} 
\captionbelow[The largest pairing eigenvalue $\lambda$ as a function of the screening parameters $\kappa_{\perp}$ and $\kappa_z$ entering the RPA interaction of Eq.~\eqref{eq:eds-RPA-int} for the case $\hat{m} = 1$ with $\hat{\eta} = 0.3$ {[Eq.~\eqref{eq:eds-new-meta}]}~\cite{Palle2024-el-dip}.]{\textbf{The largest pairing eigenvalue $\lambda$ as a function of the screening parameters $\kappa_{\perp}$ and $\kappa_z$ entering the RPA interaction of Eq.~\eqref{eq:eds-RPA-int} for the case $\hat{m} = 1$ with $\hat{\eta} = 0.3$ [Eq.~\eqref{eq:eds-new-meta}]}~\cite{Palle2024-el-dip}.
$\lambda$ is found by numerically solving Eq.~\eqref{eq:eds-lin-gap-eq} on a dense grid and the reference $\lambda_{\text{ref}} = \abs{\hat{\eta}} / 4 \sqrt{1 + \hat{m}^2} \approx 0.05$ is given in Eq.~\eqref{eq:eds-leading-pairing}.
The leading pairing state has pseudoscalar ($A_{1u}$) symmetry in regions colored red, which is almost everywhere.
Colored white are the regions of large $\kappa_z$ where there is no pairing.
On the two points around $(\kappa_{\perp}, \kappa_{z}) \approx (0, 1.7)$ highlighted green, the leading pairing state is $p$-wave with a small $\lambda / \lambda_{\text{ref}} \approx 0.01$.}
\label{fig:el-dip-sc-pairing-w-screening}
\end{figure}

A more realistic screened interaction is given by RPA (i.e., large-$N$, what we wrote down at the start of this Sec.~\ref{sec:el-dip-sc-Dirac-pairing}):
\begin{align}
\mathscr{V}(\theta_q, q_z) &= \mathscr{V}(\vb{p}-\vb{k}) = \frac{U_0}{\displaystyle 1 + \kappa_{\perp} \sin^2\frac{\theta_q}{2} + \kappa_z \sin^2\frac{q_z}{2} + \hat{\eta}^2 \sin^2 q_z}, \label{eq:eds-RPA-int}
\end{align}
where $\theta_q = \theta_p - \theta_k$, $q_z = p_z - k_z$, $U_0 \defeq 1 / (\DOSg_F e^2)$, and the strength of the screening we specify using the dimensionless parameters:
\begin{align}
\kappa_{\perp} &\defeq \frac{4 k_F^2 \epsilon_{\perp}}{\DOSg_F e^2} = \frac{4 \pi}{\sqrt{1 + \hat{m}^2}} \frac{\pi k_F}{\Lambda_z} \frac{v \epsilon_{\perp}}{e^2}, \\
\kappa_z &\defeq \frac{4 \Lambda_z^2 \epsilon_z}{\pi^2 \DOSg_F e^2} = \frac{4 \pi}{\sqrt{1 + \hat{m}^2}} \frac{\Lambda_z}{\pi k_F} \frac{v \epsilon_z}{e^2}.
\end{align}
For such a $\mathscr{V}(\theta_q, q_z)$, we have numerically investigated the resulting pairing instabilities.
The results for one generic parameter choice, previously presented in Ref.~\cite{Palle2024-el-dip}, are shown in Fig.~\ref{fig:el-dip-sc-pairing-w-screening}.
For general parameter sets, we find that pairing takes place only when $\kappa_{\perp}$ and $\kappa_z$ are sufficiently small compared to $\abs{\hat{\eta}}$.
This agrees with the conclusions drawn from the schematic example we considered in Sec.~\ref{sec:el-dip-sc-pairing-toy-model}.
Moreover, the symmetry of the leading pairing state is robustly pseudoscalar triplet, with essentially the same $\vb{d}$-vector as in Eq.~\eqref{eq:eds-leading-pairing}.
A $p$-wave instability also arises that, although usually weaker by a factor of $\sim \sqrt{2}$ than the leading instability, in a few cases becomes leading.

In many materials, $v$ is on the order of \SI{1}{\electronvolt\angstrom} which gives a small $\upalpha^{-1} = v \epsilon_0 / e^2 \sim 0.006$. 
Hence for $\hat{m} \sim 1$, $k_F / \Lambda_z \sim 1$, and $\epsilon_{\perp} \sim \epsilon_z \sim \epsilon_0$ the screening coefficients $\kappa_{\perp}$ and $\kappa_z$ can be very small, i.e., the screening can be very efficient.
In other words, for physically realistic parameters the momentum-dependence of the screened interaction can be such that it only modestly suppresses the pairing eigenvalue $\lambda$ from its $\kappa_{\perp} = \kappa_z = 0$ value of Eq.~\eqref{eq:eds-leading-pairing}.
That said, one should keep in mind that $\upalpha$ flows toward weak coupling under RG, as shown in Sec.~\ref{sec:el-dip-sc-Dirac-RG}; see Fig.~\ref{fig:el-dip-sc-RG-alpha-flow}.
For the materials that motivated the current study, one finds $v \sim \SI{3}{\electronvolt\angstrom}$ in the case of \ce{Bi2Se3} and \ce{Bi2Te3}~\cite{Liu2010} and $v \sim \SI{1}{\electronvolt\angstrom}$ in the case of \ce{SnTe}~\cite{Hsieh2012}.
The dielectric constants are up to $\sim 10$ in the frequency range of interest for these materials~\cite{Eddrief2016, Lewis1987, Suzuki1995}, giving a small enough $\upalpha^{-1} \sim 0.2$ for our theory to be of relevance.

We have thus found that the leading paring instability is odd-parity and of pseudoscalar ($A_{1u}$) symmetry.
It is interesting to note that states of such symmetry are more robust to disorder than usual~\cite{Dentelski2020}.
As demonstrated in Ref.~\cite{Dentelski2020}, this follows from the fact that the pseudoscalar pairing state transforms like a singlet under the combined application of chiral and time-reversal symmetry, which in turn implies that it is protected by an effective Anderson theorem relative to disorder which respect these symmetries.

If we use $\hat{\eta} \approx 0.5$ as the largest value for the effective dimensionless electric-dipole coupling that follows from the RG treatment of Sec.~\ref{sec:el-dip-sc-Dirac-RG}, we obtain from Eq.~\eqref{eq:eds-leading-pairing} a dimensionless pairing eigenvalue $\lambda \approx 0.1$ which puts the system into the weak-coupling BCS regime.
A quantitative estimate of the transition temperature requires knowledge of the cutoff energy $\hbar \omega_c$.
Using for example $E_F \sim \SI{1}{\electronvolt}$, which is the appropriate energy scale for an electronic mechanism, one gets transition temperatures in the sub-Kelvin regime.
While these transition temperatures are not large, they do give rise to unconventional pairing in materials without strong local electron correlations or quantum-critical fluctuations of any kind.

\subsection{The leading pseudoscalar pairing state is not topological}
Interestingly, the leading pairing state of Eq.~\eqref{eq:eds-leading-pairing} can be interpreted as the quasi-2D solid-state analog of the B phase of superfluid \ce{^3He}~\cite{Leggett1975, Vollhardt1990, Volovik2003}.
In the helium case, it is known that this phase is topological in three dimensions~\cite{Volovik2003, Volovik2009, Mizushima2015}, belonging to the class DIII in the classification of non-interacting gapped topological matter~\cite{Chiu2016, Ludwig2016}.
Hence, it couples to gravitational instantons through a topological $\theta$ term and its boundary contains a Majorana cone of topologically-protected surface Andreev bound states~\cite{Mizushima2015, Ludwig2016}.
To test whether our state is topological, we have evaluated the corresponding topological invariant~\cite{Volovik2009, Mizushima2015}
\begin{align}
\int \frac{\dd[3]{k}}{48 \pi^2} \sum_{ij\ell} \LCs_{ij\ell} \tr\mleft(\upupsilon_2 H_{\text{BdG},\vb{k}}^{-1} \frac{\partial H_{\text{BdG},\vb{k}}}{\partial k_i} H_{\text{BdG},\vb{k}}^{-1} \frac{\partial H_{\text{BdG},\vb{k}}}{\partial k_j} H_{\text{BdG},\vb{k}}^{-1} \frac{\partial H_{\text{BdG},\vb{k}}}{\partial k_{\ell}}\mright) \in \Z
\end{align}
and found that it vanishes.
Here $\upupsilon_{\mu}$ are Pauli matrices in Nambu space, the Nambu spinor is $\mleft(\psi, (\MatTR \psi^{\dag})^{\intercal}\mright)$, where $\MatTR = \uptau_3 \iu \Pauli_y$, and $H_{\text{BdG},\vb{k}} = \upupsilon_3 H_{\vb{k}} + \upupsilon_1 \Delta(\vb{k})$ is the Bogoliubov-de~Gennes Hamiltonian which anticommutes with $\upupsilon_2$, $\{H_{\text{BdG},\vb{k}}, \upupsilon_2\} = 0$.
Hence our pairing state is topologically trivial.
As shown in Ref.~\cite{Fu2010}, fully-gapped odd-parity superconducting states need to have a Fermi surface which encloses an odd number of time-reversal invariant momenta to be topological.
In our case, the cylindrical Fermi surface encloses not only the $\Gamma$ point, but also the $Z$ point $\vb{k} = (0, 0, \pi)$, unlike \ce{^3He-B}, which explains the difference in topology.

\section{Summary, discussion, and comparison to related work} \label{sec:el-dip-sc-final-discussion}
In this chapter, which is based on Ref.~\cite{Palle2024-el-dip}, we developed the theory of electric dipole excitations of electronic states residing near the Fermi level (Sec.~\ref{sec:el-dip-sc-dipole-theory}), we demonstrated that out-of-plane electric dipole fluctuations become enhanced at low energies in spin-orbit-coupled quasi-2D Dirac systems (Sec.~\ref{sec:el-dip-sc-Dirac}), and we showed that electric monopole-dipole interactions induce unconventional low-temperature superconductivity in sufficiently screened systems (Sec.~\ref{sec:el-dip-sc-pairing}).
In quasi-2D Dirac metals in particular, in Sec.~\ref{sec:el-dip-sc-Dirac-pairing} we found that the resulting pairing state is an odd-parity state of pseudoscalar ($A_{1u}$) symmetry, similar to the superfluid phase of \ce{^3He-B}~\cite{Leggett1975, Vollhardt1990, Volovik2003}, with a competitive subleading $p$-wave instability appearing as well.
These are the main results of the current chapter.

In our general treatment of dipole fluctuations of Sec.~\ref{sec:el-dip-sc-dipole-theory}, we made two key observations.
The first one is that intraband electric dipole excitations require spin-orbit coupling to maintain a finite coupling to plasmons in the long-wavelength limit.
The second one is that the same plasmon field mediates all effective electric multipole-multipole interactions that arise from the electron-electron Coulomb interaction.
With these in mind, we then formulated a general theory of itinerant dipole excitations and their electrostatic interactions.
In addition, we related our treatment of dynamically fluctuating dipoles to the modern theory of polarization~\cite{Resta1994, Resta2000} and showed that the King-Smith--Vanderbilt formula~\cite{KingSmith1993} for the (static) polarization acquires an anomalous term within tight-binding descriptions.

When strong spin-orbit coupling inverts bands of opposite parities, dipole fluctuations are especially strong.
The vicinity of such band-inverted points is, moreover, generically described by Dirac models.
Although this has been known in various particular cases~\cite{Cohen1960, Wolff1964, Rogers1968, Zhang2009, Liu2010, Hsieh2012}, in Sec.~\ref{sec:el-dip-sc-Dirac-model-band} of this chapter we presented a general symmetry derivation of this important fact, before turning to the renormalization group analysis of dipole excitations in Dirac systems in Sec.~\ref{sec:el-dip-sc-Dirac-RG}.
Our large-$N$ RG analysis of the strong-screening limit reveled that, although irrelevant in most systems, electric dipole coupling is marginally relevant along the out-of-plane direction in quasi-2D geometries.
Even though the enhancement of the effective $z$-axis (out-of-plane) dipole coupling is limited, it is sufficiently large to imply that electronic dipole interactions cannot be ignored at low energies.
As a concrete experimental footprint, in Sec.~\ref{sec:el-dip-sc-Dirac-opticond} we have found that this $z$-axis dipole coupling gives the dominant contribution to the $z$-axis optical conductivity in quasi-2D Dirac systems.

The electric monopole-dipole coupling between itinerant electrons, introduced in this chapter, causes unconventional superconductivity whenever dipole moments are sufficiently strong compared to the screening, as we established in Sec.~\ref{sec:el-dip-sc-pairing}.
Even when other pairing mechanisms are present, as long as they mostly act in the $s$-wave channel which is suppressed by the electric monopole-monopole repulsion, electric monopole-dipole interactions can still be the main cause of pairing.
Hence, in systems not governed by strong local electronic correlations or nearly critical collective modes, the proposed mechanism is a possible source of unconventional low-temperature superconductivity.
Using arguments similar to those of Ref.~\cite{Brydon2014} and Sec.~\ref{sec:TR-positivity}, we showed that the pairing due to our mechanism is necessarily unconventional, but also that it is not likely to reach high temperatures (strong coupling).
For comparison, the pairing due to the exchange of phonons~\cite{Brydon2014}, ferroelectric modes~\cite{KoziiBiRuhman2019, Enderlein2020, Kozii2022, Volkov2022, Klein2023}, and non-magnetic odd-parity fluctuations~\cite{Kozii2015} robustly favors conventional $s$-wave pairing and is able to reach strong coupling.
Although we included dipole-dipole interactions in our analysis, we found that they give a weaker (subleading) contribution to the Cooper pairing for realistic dipole strengths.
This should be contrasted with pairing in degenerate dipolar Fermi gases~\cite{Baranov2002, Bruun2008, Sieberer2011, Liu2012, Baranov2012}, discussed in more detail below, in which the neutrality of the cold-atom fermions precludes monopole-dipole interactions, rendering dipole-dipole interactions dominant.

Our theory of dipole excitations of Fermi-surface states resembles theories of ferroelectric metals where itinerant electrons couple to ferroelectric modes~\cite{Edge2015, Benedek2016, Chandra2017, KoziiBiRuhman2019, Volkov2020, Enderlein2020, Kozii2022, Volkov2022, Klein2023}, which are usually soft polar phonons~\cite{Cochran1960}.
In both cases, the electrons couple through a fermionic dipole bilinear that is odd under parity and even under time reversal.
Hence this coupling is direct only in the presence of spin-orbit coupling~\cite{KoziiBiRuhman2019, Volkov2020}, as we proved in Sec.~\ref{sec:el-dip-sc-itinerant-dipoles}.
However, in our case there is no independent collective mode associated with this dipole bilinear.
Instead, as we showed in Sec.~\ref{sec:el-dip-sc-plasmon-field}, the dipole bilinear contributes to the total charge density alongside a monopole bilinear, and its fluctuations are mediated by the same plasmon field which mediates all electrostatic interactions.
In contrast, ferroelectric modes propagate separately from plasmons and can thus be tuned to quantum criticality, for instance.
As discussed in Ref.~\cite{Klein2023}, this may even give rise to non-Fermi liquid behavior.
We do not expect that such behavior emerges in our theory as dipolar fluctuations will remain massive due to the screening of the Coulomb interaction.

Another distinction between our problem and ferroelectric metals is that, in the Cooper channel, the Coulomb interaction and its monopole-dipole and dipole-dipole parts are repulsive, whereas the exchange of ferroelectric modes is attractive.
The former can therefore only give unconventional pairing (Sec.~\ref{sec:el-dip-sc-qualitative-pairing}), whereas the latter robustly prefers conventional $s$-wave pairing~\cite{KoziiBiRuhman2019, Enderlein2020, Kozii2022, Volkov2022, Klein2023}, as expected for a type of phonon exchange~\cite{Brydon2014}.
The same distinction applies when comparing our problem to that of metals coupled to more general non-magnetic odd-parity fluctuations~\cite{Kozii2015}.
Apart from this sign difference in the dipole-dipole interaction, a further dissimilarity is that it is the monopole-dipole interaction that is primarily responsible for the pairing in our theory.
This follows from the fact that the dimensionless dipole coupling constant $\tilde{\eta} < 1$ for realistic parameter values so dipole-dipole interactions ($\sim \tilde{\eta}^2$) are weaker than monopole-dipole ones ($\sim \tilde{\eta}$).
For further comparison to pairing mechanisms based on order-parameter exchange, refer to Sec.~\ref{sec:el-dip-sc-comparison-OP-ex}.

In degenerate fermionic gases composed of cold atoms or molecules, electric dipole-dipole interactions have been proposed as a source of pairing in a number of theories~\cite{Baranov2002, Bruun2008, Sieberer2011, Liu2012, Baranov2012} which appear similar to ours.
Further inspection reveals that they are very different.
A comparison is still instructive.
In these theories, the particles are neutral single-component fermions which carry electric dipole moments.
The electric monopole-dipole interaction, which is key to our mechanism, is thus absent, nor is there any need for screening of the monopole-monopole repulsion.
Their dipole-dipole interaction has no internal structure and its momentum dependence solely determines the preferred pairing channel, whereas in our theory the pseudospin structure of the interaction plays an equally important role.
Their dipoles are also aligned along an external field, giving a net polarization.
In contrast, our electric dipole density varies across the Fermi surface, with opposite momenta and opposite pseudospins having opposite dipole densities (Fig.~\ref{fig:el-dip-sc-FS-dip-density}).
Finally, unlike in our theory, the nature of their dipole moments is unimportant and one may exchange electric for magnetic dipoles, as has been done experimentally~\cite{Lu2012}.

The pairing mechanism proposed in this chapter is similar to other electronic mechanism~\cite{Maiti2013, Kagan2014}, which derive in one form or another from the electron-electron Coulomb interaction.
In their pioneering study~\cite{Kohn1965, Luttinger1966}, Kohn and Luttinger showed that the non-analyticity originating from the sharpness of the Fermi surface induces pairing with high orbital angular momentum $\ell$ in isotropic 3D systems, even when the short-ranged bare interaction is repulsive in all channels.
Although non-analyticity has proven to be a negligible source of pairing, giving $T_c \sim \SI{e-11}{\kelvin}$ or smaller~\cite{Kohn1965}, the idea that the overscreening of a bare repulsive interaction can result in pairing has survived and been developed in many ways~\cite{Maiti2013, Kagan2014}.
Subsequent work generalized this mechanism to isotropic 2D systems~\cite{Chubukov1993} and low-density Hubbard models~\cite{Baranov1992, Baranov1992-p2, Kagan2011}, as well as showed that the pairing extends to $\ell = 1$ for a bare repulsive contact interaction in isotropic 3D systems~\cite{Fay1968, Kagan1988, Baranov1992-p2, Efremov2000}, with a $T_c \sim \SI{e-3}{\kelvin}$ when applied to \ce{^3He}~\cite{Kagan1988}.
For repulsive Hubbard models, asymptotically exact weak-coupling solutions were found which gave pairing in both $p$-wave and $d$-wave channels~\cite{Raghu2010, Raghu2011}.

In our mechanism, just like in the Kohn-Luttinger-like mechanisms, an initially repulsive interaction becomes overscreened, resulting in pairing.
Both mechanisms need the interaction to be, or become, nearly momentum-independent.
Because we had started from the long-ranged unscreened Coulomb interaction, to screen it properly we needed to reach the strong-coupling regime of large $\upalpha = e^2 / (\hbar v_F \epsilon_0)$.
Since this regime cannot be analytically treated in the unmodified model~\cite{Chubukov1989, Raghu2012}, we employed a large-$N$ expansion, $N$ being the number of fermion flavors.
In contrast, Kohn-Luttinger-like mechanisms start from a short-ranged repulsive interaction which is readily perturbatively treated.
The origin of the overscreening is different between the two mechanisms as well.
In our mechanism, the electric dipole terms appearing in the bare vertex are responsible, and not perturbative corrections to the Cooper-channel interaction.
Once projected onto the Fermi surface, the dipolar part of the bare vertex acquires a non-trivial structure in pseudospin space which plays an important role in choosing the pairing symmetry.
In Kohn-Luttinger-like mechanisms, on the other hand, the pairing symmetry is essentially chosen by the momentum-dependence of the overscreened interaction.

In light of the strong dipole fluctuations we had found in quasi-2D Dirac systems,  in the penultimate Sec.~\ref{sec:el-dip-sc-Dirac-pairing} we explored their pairing instabilities.
Across most of the parameter range, the dominant pairing state due to electric monopole-dipole interactions has pseudoscalar ($A_{1u}$) symmetry and resembles the Balian-Werthamer state of \ce{^3He-B}~\cite{Leggett1975, Vollhardt1990, Volovik2003} (Figs.~\ref{fig:el-dip-sc-FS-d-vec-sketch} and~\ref{fig:el-dip-sc-pairing-w-screening}).
Since the dimensionless dipole coupling is at best a fraction of the monopole coupling, the pairing problem is expected to be in the weak-coupling regime.
Although we estimated transition temperatures on order of \SI{0.1}{\kelvin}, a detailed prediction of $T_c$ will depend on a number of material parameters, making quantitative predictions rather unreliable.
That said, it is interesting to observe that \ce{SnTe} is well-described by Dirac models~\cite{Rogers1968, Hsieh2012, Ando2015} and that an $A_{1u}$ pairing state is consistent with experiments performed on \ce{In}-doped \ce{SnTe}~\cite{Novak2013, Zhong2013, Smylie2018-p2, Smylie2020, Saghir2014, Maeda2017}.
This suggests that our mechanism could be of relevance.
In the case of doped \ce{Bi2Se3}, which is also well-described by Dirac models~\cite{Zhang2009, Liu2010}, there is strong evidence for nematic $p$-wave pairing~\cite{Yonezawa2019, Matano2016, Pan2016, Yonezawa2017, Asaba2017, Du2017, Smylie2018, Cho2020, Yonezawa2017, Smylie2016, Smylie2017, Fu2010, Venderbos2016, Hecker2018}, which in our mechanism is a competitive subleading instability.
In combination with electron-phonon interactions~\cite{Brydon2014, Wu2017}, it is possible that this subleading $p$-wave state becomes leading.
A symmetry-breaking strain field could have a similar effect, but only if it is sufficiently large.

Despite their unusual superconductivity, neither \ce{SnTe} nor \ce{Bi2Se3} have strong local electronic correlations or nearly critical collective modes, which was one of the motivations for the current work, which is based on Ref.~\cite{Palle2024-el-dip}.
Is parity-mixing and spin-orbit coupling enough to obtain unconventional superconductivity, even in mundane weakly correlated systems?
And can such a mechanism deliver unconventional pairing as the leading instability?
The proposed mechanism answers both in the affirmative.

\chapter{Constraints on the pairing symmetry of strontium ruthenate \ce{Sr2RuO4}}
\label{chap:Sr2RuO4}

The unconventional low-temperature superconductivity of strontium ruthenate \ce{Sr2RuO4} was discovered in 1994~\cite{Maeno1994}.
In the intervening three decades, an impressive array of experiments have been performed on \ce{Sr2RuO4} with high precision and on exceedingly pure samples~\cite{Mackenzie2003, Bergemann2003, Maeno2012, Kallin2012, LiuMao2015, Mackenzie2017, Maeno2024}.
Yet despite this large community effort that has made strontium ruthenate one of the most-studied unconventional superconductors, the high quality of crystal samples that should have made the experiments and their interpretation unambiguous, and the extraordinarily well-characterized and well-understood Fermi liquid normal state that should have made the theoretical understanding of this material within reach, fundamental questions concerning the nature of the unconventional superconductivity (SC) of strontium ruthenate (SRO) remain~\cite{Maeno2024}.
The biggest two are ``What is the pairing symmetry of the SC state?'' and ``What is the pairing mechanism?''
In this chapter, we discuss the recent progress in which the present author has been involved in~\cite{Palle2023-ECE, Jerzembeck2024} that addresses the former question.
Although the text and figures of Refs.~\cite{Palle2023-ECE, Jerzembeck2024} have been recycled in many places in the current chapter, there is also a significant amount of additional material.
Most of it builds and further elaborates upon the results of Refs.~\cite{Palle2023-ECE, Jerzembeck2024}.

For a long time, the leading candidate for the pairing state of SRO was an (odd-parity, spin-triplet) chiral $p$-wave state~\cite{Mackenzie2003, Maeno2012, Kallin2012, LiuMao2015, Mackenzie2017}.
As we shall extensively review in Sec.~\ref{sec:SRO-lit-review}, such a state appeared to be the most consistent with the then-available experiments.
The absence of a change in the NMR Knight shift~\cite{Ishida1998, Murakawa2007} and polarized neutron diffraction (PND)~\cite{Duffy2000} as one entered the SC state suggested spin-triplet pairing, as did the the observation of $\pi$ phase shifts~\cite{Nelson2004} and half-quantum vortices~\cite{Jang2011} indicating odd parity.
Moreover, zero-field muon spin relaxation~\cite{Luke1998, Luke2000} and polar Kerr effect~\cite{Xia2006, Kapitulnik2009} experiments supported time-reversal symmetry-breaking (TRSB) in the SC state, which would imply that SC domains exist, in agreement with what was observed in Josephson junction interference patterns~\cite{Kidwingira2006}.
The simplest state consistent with these experiments is a chiral $p$-wave state (see Tab.~\ref{tab:SRO-SC-state-options}), and indeed influential early theories~\cite{Rice1995, Baskaran1996}, published right after the discovery of SC in SRO~\cite{Maeno1994}, predicted $p$-wave pairing based on an analogy to superfluid \ce{^3He}.

However, even at that time tensions existed in the experimental evidence~\cite{Mackenzie2017}.
A drop in the NMR Knight shift should be visible for some directions of the magnetic field even for triplet SC states, but was not observed for any direction~\cite{Murakawa2007, Maeno2012}.
Likewise, the apparent Pauli limiting~\cite{Mackenzie2017} of the in-plane upper-critical magnetic field is difficult to reconcile with triplet pairing.
Spontaneous magnetization and currents should appear on the surface around defects for TRSB SC states, but have not been observed, despite numerous searches~\cite{Tamegai2003, Bjornsson2005, Kirtley2007, Hicks2010, Curran2011}.
Furthermore, experimentally it was found that $T_c$ depends quadratically on shear strain without any thermodynamically measurable splitting of the transition~\cite{Hicks2014, Steppke2017}, whereas a chiral $p$-wave state should split linearly with shear strain into two measurable transitions.
Finally, multiple experiments have reported low-temperature behavior that is only consistent with nodal SC states~\cite{NishiZaki2000, Deguchi2004, Matsui2001, Lupien2001, Ishida2000, Bonalde2000, Deguchi2004-p2}, in contradiction to chiral $p$-wave pairing which is fully gapped (nodeless).

Five years ago, the paradigm began to shift~\cite{Maeno2024}, with the preponderance of evidence currently standing against odd-parity spin-triplet pairing of any kind.
The key experiment that challenged the old paradigm was a revision of the temperature-dependence of the NMR Knight shift~\cite{Pustogow2019, Ishida2020}.
As they discovered in Ref.~\cite{Pustogow2019}, the Knight shift does, in fact, significantly drop as one enters the SC state of SRO.
This enabled them to rule out chiral $p$-wave pairing whose $\vb{d}$-vector points along the $z$-axis ($\vb{d}(\vb{k}) \sim (k_x \pm \iu \, k_y) \vu{e}_z$, Tab.~\ref{tab:SRO-SC-state-options}).
With later Knight shift measurements~\cite{Chronister2021}, they provided strong evidence against spin-triplet pairing of any kind.
The explanation for why early experiments~\cite{Ishida1998, Murakawa2007} found no changes in the Knight shift at $T_c$ is that, at the $\sim \SI{1}{\kelvin}$ temperatures relevant for SRO ($T_c = \SI{1.5}{\kelvin}$), sufficiently energetic NMR pulses can locally heat up the sample to the normal state~\cite{Pustogow2019, Ishida2020, Chronister2021}, implying that they were not measuring the Knight shift of the SC state.
Moreover, this NMR pulse heat-up effect acts only on time-scales much shorter than the nuclear spin-lattice relaxation time $T_1$, which is why clear features were observed at $T_c$ in the early NMR measurements of $T_1$~\cite{Ishida1997, Ishida2000, Murakawa2007}, but not in the NMR Knight shift~\cite{Ishida1998, Murakawa2007}.
Motivated by this finding, PND measurements have been redone as well~\cite{Petsch2020}, at a smaller magnetic field and with better statistics than before~\cite{Duffy2000}, and they also report a drop in the magnetic susceptibility.

With these discoveries, the study of SRO has been reinvigorated, as has the debate regarding what is the correct pairing symmetry~\cite{Maeno2024}.
In Sec.~\ref{sec:SRO-lit-review}, we review both old and recent experimental studies of SRO and summarize what is currently known about the pairing state.
In brief, we know that the SC state is unconventional, that it has line nodes, at least some of which are vertical, and that it is more likely to be even-parity than odd-parity.
The SC order parameter appears to couple quadratically to all strains, except $\epsilon_{xy}$ shear strain for which there is inconclusive evidence that it couples linearly.
Whether the (homogeneous) SC state breaks time-reversal (TR) symmetry is not clear.
It is worth remarking that the most direct and theoretically minimalistic interpretations of the currently-available experiments are regularly at odds with one another in SRO, like with regard to TRSB.
An open question in the field, which has bearing on the field of unconventional superconductivity more broadly, is whether the interpretation of some of the well-established experimental probes needs to be reexamined.

Without fine-tuning or invoking special mechanisms, it is very challenging to theoretically interpret the superconductivity of SRO in terms of a homogeneous pairing state (described by Ginzburg-Landau theory, etc.).
Developing a theory, even on the phenomenological level, that reconciles the various experimental results is an outstanding open problem of the field.
Many proposals~\cite{Romer2019, Romer2020,
Kivelson2020, Willa2021, Yuan2021, Sheng2022, Yuan2023,
Clepkens2021, Romer2021,
Suh2020, Fukaya2022,
Wagner2021,
Leggett2021, Huang2021,
Huang2021-p2,
Gingras2022, Scaffidi2023} have been put forward in the last few years, but no consensus has formed around which proposal is the correct one.
That said, the focus of the current chapter will not be theories of SRO as such, which we shall only discuss in the passing, but on theoretically analyzing experiments.

Two experimental probes have recently been developed that enable one to significantly narrow down the viable pairing candidates of SRO~\cite{Palle2023-ECE, Jerzembeck2024}.
The first is an apparatus for performing measurements under uniaxial stress~\cite{Hicks2014, Steppke2017}, whether heat capacity, upper-critical magnetic field, nuclear magnetic resonance, muon spin relaxation, or other.
Notably, strain tunes the system without adding disorder, which is known to strongly suppresses $T_c$~\cite{Mackenzie1998, Mao1999, Kikugawa2002, Kikugawa2004}, as expected for an unconventional superconductor.
The second is a method of precisely measuring the elastocaloric effect~\cite{Ikeda2019, Straquadine2020, Ikeda2021}, which is the effect of adiabatic changes in the strain inducing changes in the temperature.

In this chapter, which is based on Refs.~\cite{Palle2023-ECE, Jerzembeck2024}, we discuss the constraints on the pairing symmetry of SRO which follow from recent heat capacity~\cite{Li2021}, magnetic susceptibility~\cite{Jerzembeck2024}, and elastocaloric effect~\cite{Li2022, Jerzembeck2024} measurements performed under in-plane uniaxial stresses.
To be able to explain elastocaloric measurements under $[100]$ stress~\cite{Li2022}, in Sec.~\ref{sec:SRO-analysis-ECE-100} we find that even-parity pairing states must include either large extended $s$-wave, $d_{x^2 - y^2}$-wave, or $(d_{yz} | - d_{xz})$-wave admixtures, where the last possibility arises because of the body-centered lattice of SRO.
These $(d_{yz} | - d_{xz})$-wave admixtures take the form of distinctively body-centered-periodic harmonics that have horizontal line nodes.
Hence $g_{xy(x^2-y^2)}$-wave and $d_{xy}$-wave pairings are excluded as possible dominant even-parity SC states.
The absence of any thermodynamic signatures of transition-splitting under $[110]$ strain~\cite{Jerzembeck2024} furthermore provides strong experimental evidence against bulk two-component SC states of any kind, whether accidental (e.g., $s' + \iu \, d_{xy}$ or $d_{x^2-y^2} + \iu \, g_{xy(x^2-y^2)}$) or symmetry-protected ($d_{xz} + \iu \, d_{yz}$).
As we shall show in Sec.~\ref{sec:SRO-Tc-ECE-analysis-110}, reconciling the measurements of Ref.~\cite{Jerzembeck2024} with related experiments~\cite{Li2021, Benhabib2021, Ghosh2021} requires an extraordinarily high degree of fine-tuning if we assume TRSB.
Given the strong suppression of $T_c$ by non-magnetic impurities~\cite{Mackenzie1998, Mao1999, Kikugawa2002, Kikugawa2004}, the single-component $d_{x^2-y^2}$ pairing state appears to be the simplest one consistent with thermodynamic probes of the SC state, as well as NMR and PND.
The extended $s$-wave pairing is also a viable candidate, although some tuning is needed for it to saturate the Abrikosov-Gor'kov bound regarding $T_c$ suppression by impurities.
That said, neither of these two pairing candidates are without their difficulties.

The chapter is organized as follows.
We start with the fundamentals of strontium ruthenate \ce{Sr2RuO4}.
These are briefly explained at the start of Sec.~\ref{sec:SRO-fundie}, and in more detail in its subsections.
In the first one (Sec.~\ref{sec:SRO-lit-review}) we review all the available experimental investigations of SRO's SC to date and summarize what is currently known about its superconductivity.
In the Sec.~\ref{sec:SRO-cryst-struct} after, we specify SRO's crystal structure and symmetries.
The electronic structure is discussed in Sec.~\ref{sec:SRO-el-struct}, where we also introduce a tight-binding model that we employ in later analyses.
Some basics on the elastic tuning of SRO are recalled in Sec.~\ref{sec:SRO-elastic-tuning}.
In the last subsection~\ref{sec:SRO-SC-construct} that deals with fundamentals, we explain how superconducting states are microscopically constructed and classified in a multiband system such as SRO.
In the remaining Secs.~\ref{sec:SRO-analysis-ECE-100} and \ref{sec:SRO-Tc-ECE-analysis-110}, we present the works of Refs.~\cite{Palle2023-ECE} and~\cite{Jerzembeck2024}, respectively.
In both, the results and derivations of Refs.~\cite{Palle2023-ECE, Jerzembeck2024} are elaborated in more detail than in the published articles.

Sec.~\ref{sec:SRO-analysis-ECE-100} has essentially two parts.
In the first part (Sec.~\ref{sec:SRO-elasto}), we discuss how elastocaloric experiments show that a normal-state entropy maximum becomes a minimum in the SC state (Fig.~\ref{fig:SRO-elasto}) and how this is only possible if there are no vertical line nodes at the Van Hove lines responsible for the normal-state entropy maximum (Fig.~\ref{fig:SRO-bands-DOS}).
In the second part (Sec.~\ref{sec:SRO-behavior-vH}), we exploit the classification of SC states of Sec.~\ref{sec:SRO-SC-construct} to determine which states do not have symmetry-enforced vertical line nodes at the Van Hove lines.
The main result is Tab.~\ref{tab:SRO-main-result}.
As already remarked, we find that only $s$, $d_{x^2 - y^2}$, and $(d_{yz} | - d_{xz})$ dominant pairing states are consistent with the elastocaloric data of Ref.~\cite{Li2022}, where the last $(d_{yz} | - d_{xz})$ state must be made of characteristically body-centered harmonics that have horizontal line nodes.

In the next Sec.~\ref{sec:SRO-Tc-ECE-analysis-110}, we first present the main experimental findings of Ref.~\cite{Jerzembeck2024}: the absence of a cusp in $T_c$ (Fig.~\ref{fig:SRO-J24-Tc}) and the absence of a second anomaly in the elastocaloric data (Fig.~\ref{fig:SRO-J24-ECE}) as $\langle 110 \rangle$ uniaxial stress is applied.
As we explain in the Ginzburg-Landau analysis of the following Sec.~\ref{sec:SRO-GL-analysis}, both a cusp and a second anomaly should take place if the SC state has two components.
This is summarized in Tab.~\ref{fig:schematicPhaseDiagrams}.
Conversely, the reported null-results of Ref.~\cite{Jerzembeck2024}, when combined with the reported jumps in the $c_{66}$ elastic constant~\cite{Benhabib2021, Ghosh2021}, put tight constraints on which two-component states are viable and how finely tuned they must be.
This is the subject of the last Sec.~\ref{sec:SRO-110-implications}.
In particular, we find that TRSB two-component states, both accidental and symmetry-protected, require an implausibly high degree of fine-tuning, which is especially severe in the symmetry-protected case (Fig.~\ref{fig:triplePointTuning}).

\section{Fundamentals of strontium ruthenate} \label{sec:SRO-fundie}
Here we first briefly recollect basics information on strontium ruthenate (SRO) before dwelling into more detail.
In Sec.~\ref{sec:SRO-lit-review} we review the literature on experimental investigations of the pairing state of SRO, with a very brief overview of theories.
After that, in Sec.~\ref{sec:SRO-cryst-struct} we state the crystal structure and symmetries of SRO.
The electronic band structure is explained in Sec.~\ref{sec:SRO-el-struct}, where we also introduce a tight-binding model~\cite{Roising2019} that we later use to study SRO.
The tuning of SRO under external pressure is discussed in Sec.~\ref{sec:SRO-elastic-tuning}.
In the last Sec.~\ref{sec:SRO-SC-construct}, we review how superconducting (SC) states are classified and constructed with the effective three-orbital model of SRO~\cite{Ramires2019, Kaba2019, Huang2019, Palle2023-ECE}.

\begin{figure}[p!]
\centering
\begin{subfigure}{0.46\textwidth}
\centering
\includegraphics[width=\textwidth]{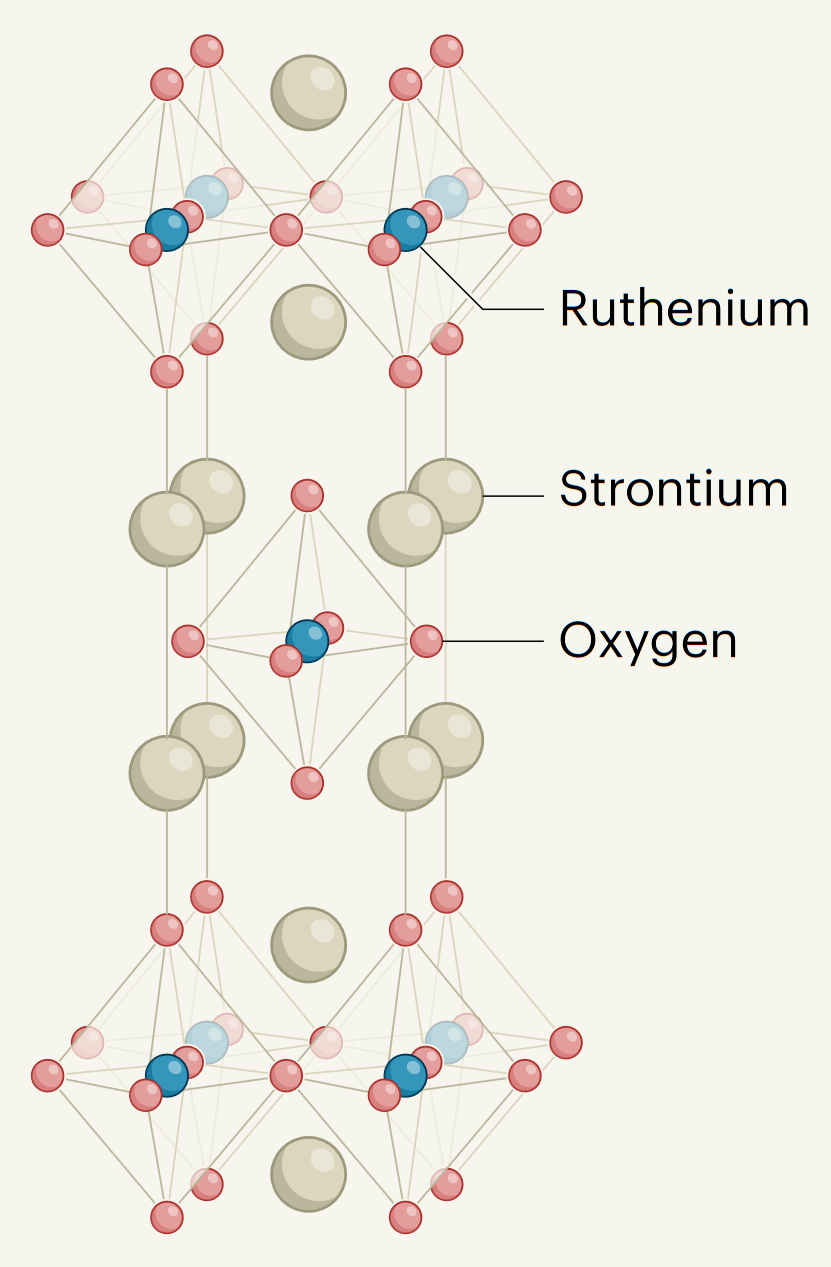}
\subcaption{}
\end{subfigure}%
\begin{subfigure}{0.54\textwidth}
\centering
\includegraphics[width=0.85\textwidth]{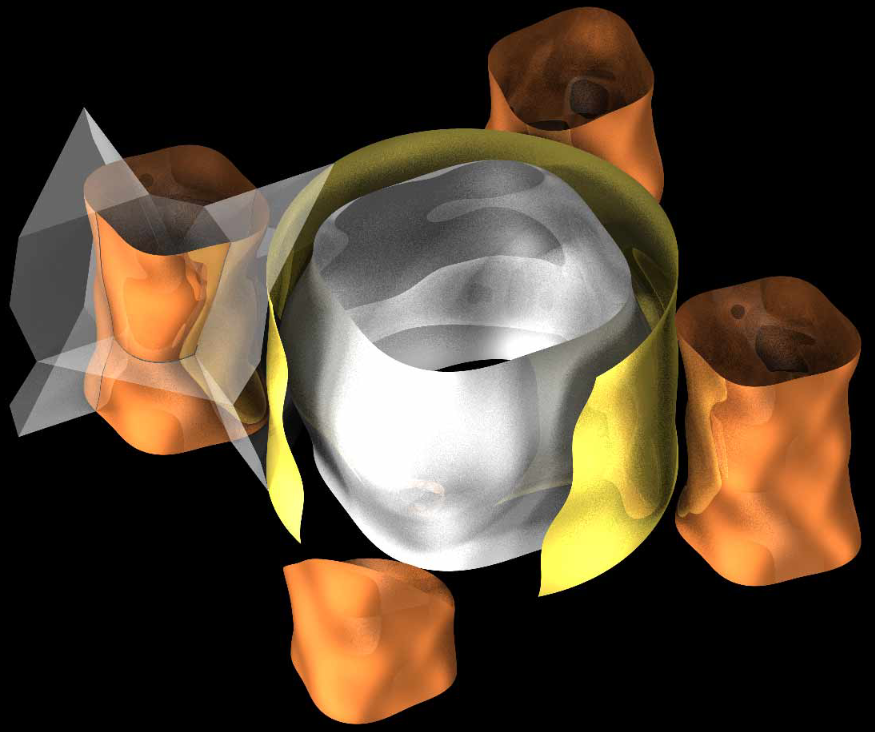}
\subcaption{}
\vspace{12pt}
\raggedleft
\includegraphics[width=0.95\textwidth]{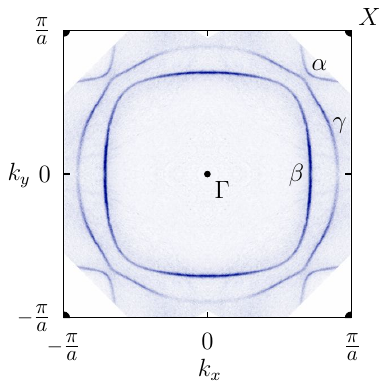} \\[-2pt]
{(c) \hspace{95pt}} 
\end{subfigure}
\captionbelow[Crystal structure of strontium ruthenate \ce{Sr2RuO4}~\cite{Armitage2019} (a), visualization of the three Fermi surfaces of \ce{Sr2RuO4}~\cite{Bergemann2003} (b), and $k_z = 0$ cross-sections of the Fermi surfaces deduced from ARPES~\cite{Tamai2019} (c).]{\textbf{Crystal structure of strontium ruthenate \ce{Sr2RuO4}}~\cite{Armitage2019} \textbf{(a), visualization of the three Fermi surfaces of \ce{Sr2RuO4}}~\cite{Bergemann2003} \textbf{(b), and $k_z = 0$ cross-sections of the Fermi surfaces deduced from ARPES}~\cite{Tamai2019} \textbf{(c).}
The $c$-axis corrugation is exaggerated by a factor of $15$ for clarity under (b), where bronze, silver, and gold stand for the $\alpha$, $\beta$, and $\gamma$ Fermi sheets, respectively.
These Fermi sheets are also denoted under (c).
Figure (a) is reproduced with permission from Springer Nature from Ref.~\cite{Armitage2019}, figure (b) is reproduced with permission from Taylor \& Francis from Ref.~\cite{Bergemann2003}, and figure (c) is reproduced with editing from Ref.~\cite{Tamai2019} (\href{https://creativecommons.org/licenses/by/4.0/deed.en}{CC BY 4.0}).}
\label{fig:SRO-crystal-FS}
\end{figure}

Strontium ruthenate (SRO) is a layered perovskite with chemical composition \ce{Sr2RuO4} and a body-centered tetragonal lattice~\cite{Mackenzie2003, Bergemann2003}.
Its crystal structure is depicted in Fig.~\ref{fig:SRO-crystal-FS}(a), from which one sees that it has the same structure as that of the cuprate superconductor lanthanum barium copper oxide \ce{La_{2-x}Ba_{x}CuO4}, which was previously shown in Fig.~\ref{fig:cuprate-crystal-structure}(a) of Chap.~\ref{chap:cuprates}.
Indeed, this similarity was noticed immediately from the beginning~\cite{Maeno1994}.
The crystal point group is therefore the same, which is namely $D_{4h}$.
The tetragonal $D_{4h}$ point group is discussed at length in Sec.~\ref{sec:tetragonal-group-D4h} of Appx.~\ref{app:group_theory}.
For the reader's convenience, we repeat its character table again in this chapter in Tab.~\ref{tab:D4h-char-tab-again2} of Sec.~\ref{sec:SRO-cryst-struct}, given that the irreducible representations (irreps) of $D_{4h}$ will play an important role in the following discussion.\footnote{For the group theory uninitiated: the simplest way of thinking about irreps is as ways objects can transform under a given point group.
Thus, for instance, when we state that the shear strain component $\epsilon_{xx} - \epsilon_{yy}$ belongs to the irrep $B_{1g}$, we are stating that it transforms the same as the polynomial $x^2-y^2$ (Tab.~\ref{tab:D4h-char-tab-again2}) constructed from the Cartesian coordinates $x$, $y$, and $z$ pointing along the principal axes of the crystal.
In the case of the 2D irreps $E_g$ and $E_u$, the object has two components.
Relatedly, the fact that $x^2+y^2$ and $z^2$ both transform according to $A_{1g}$ means that one cannot tell the two apart purely from symmetries in a tetragonal crystal environment.
For further discussion, see Appx.~\ref{app:group_theory}.}
Despite the structural similarities, the physics of both the normal and the superconducting states could not be more different between the two compounds.

The normal state of SRO below \SI{25}{\kelvin} is a quasi-2D multiband Fermi liquid, as established by numerous experiments~\cite{Mackenzie2003, Bergemann2003}.
This Fermi liquid state is experimentally very well-characterized~\cite{Mackenzie2003, Bergemann2003, Veenstra2014, Tamai2019, Barber2019}.
It has three conduction bands in total, which are conventionally referred to as $\alpha$, $\beta$, and $\gamma$.
All three bands have cylindrical Fermi sheets, as shown in Fig.~\ref{fig:SRO-crystal-FS}(b).
These bands primarily derive from the $t_{2g}$ orbital manifold of the ruthenium atoms, which is made of the $4d_{xz}$, $4d_{yz}$, and $4d_{xy}$ orbitals~\cite{Mackenzie2003, Bergemann2003, Gingras2022}.
In light of the layered highly-anisotropic structure, the \ce{Ru}:$4d_{xz}$ and \ce{Ru}:$4d_{yz}$ orbitals mostly hop along the $x$ and $y$ direction, respectively, and together hybridize into the quasi-1D $\alpha$ and $\beta$ bands.
The middle $\gamma$ band predominantly derives from the \ce{Ru}:$4d_{xy}$ orbital, which hops along both $x$ and $y$ directions, and it is quasi-2D in character, as can be seen from Fig.~\ref{fig:SRO-crystal-FS}(c).
Near the diagonals $k_x = \pm k_y$ where the three Fermi sheets almost touch [Fig.~\ref{fig:SRO-crystal-FS}(c)], there is a large degree of orbital mixing which is partially mediated by spin-orbit coupling.
For further discussion of the normal state, see Sec.~\ref{sec:SRO-el-struct}.

Strontium ruthenate develops superconductivity at stoichiometry, with a low-temperature $T_c$ which reaches \SI{1.5}{\kelvin} in the clean limit~\cite{Mackenzie2003, Maeno2024}.
In contrast to cuprates, adding any doping rapidly suppresses $T_c$ because it adds disorder~\cite{Mackenzie1998, Mao1999, Kikugawa2002, Kikugawa2004} and we shall therefore only discuss pure SRO, here and throughout the chapter.
Some fundamental parameters characterizing the SC state of SRO are provided in Tab.~\ref{tab:SRO-fund-param}.
From the table one sees that the SC is very anisotropic, just like the compound itself [Fig.~\ref{fig:SRO-crystal-FS}(a)].
Phenomenologically, from the Ginzburg-Landau ratios $\kappa = \lambda_L / \chi_0$ it follows that its SC is strongly type~II for in-plane ($\parallel ab$) magnetic fields, but only weakly type~II for magnetic fields pointing along the $c$ axis.
Recently~\cite{Landaeta2023} evidence appeared indicating that the SC state evinces non-local electrodynamics~\cite{Kosztin1997}.

\begin{table}[t]
\centering
\captionabove[Parameters characterizing the superconducting state of \ce{Sr2RuO4} at zero temperature for unstrained and very pure samples.]{\textbf{Parameters characterizing the superconducting state of \ce{Sr2RuO4} at zero temperature for unstrained and very pure samples.}
$T_c$ is the transition temperature, $\Delta_{\text{max}}$ is the gap maximum, $B_c$ is the thermodynamic critical field, $B_{c2}$ is the upper-critical magnetic field,  $\xi_0$ is the Pippard coherence length, and $\lambda_L$ is the London penetration depth.
Starred values were calculated from the others in the following way.
$\xi_{0, ab} = \sqrt{\frac{\Phi_0}{2 \pi B_{c2 \parallel c}}}$, $\Phi_0 = \frac{h}{2 e}$, was determined from orbital limiting, and $\xi_{0, c}$ from the ratio $\frac{\xi_{0, ab}}{\xi_{0, c}}$ measured in Refs.~\cite{Rastovski2013, Kittaka2014}.
$\frac{\lambda_{L, c}}{\xi_{0, c}}$ was calculated from the relation~\cite{Mackenzie2003, Maeno2012} $\sqrt{\frac{\lambda_{L, ab}}{\xi_{0, ab}} \frac{\lambda_{L, c}}{\xi_{0, c}}} = \frac{\tilde{B}_{c2 \parallel ab}}{\sqrt{2} B_c}$ with $\tilde{B}_{c2 \parallel ab} = B_{c2 \parallel c} (\xi_{0, ab} / \xi_{0, c})$ instead of $B_{c2 \parallel ab}$ because the latter is Pauli limited~\cite{Kittaka2009}.
$\lambda_{L, c}$ follows from $\xi_{0, c}$ and $\frac{\lambda_{L, c}}{\xi_{0, c}}$.
Compare with Ref.~\cite{Maeno2024}.}
{\renewcommand{\arraystretch}{1.3}
\renewcommand{\tabcolsep}{10pt}
\hspace*{\stretch{1}} \begin{tabular}{ccc} \hline\hline
parameter & value & Refs. \\ \hline
$T_c$ & \SI{1.5}{\kelvin} & \cite{Mackenzie2003, Maeno2012} \\
$\Delta_{\text{max}}$ & \SI{0.35}{\milli\electronvolt} & \cite{Firmo2013} \\
$B_{c}$ & \SI{23}{\milli\tesla} & \cite{Mackenzie2003, Maeno2012} \\
$B_{c2 \parallel ab}$ & \SI{1.5}{\tesla} & \cite{Mackenzie2003, Maeno2012} \\
$B_{c2 \parallel c}$ & \SI{75}{\milli\tesla} & \cite{Mackenzie2003, Maeno2012} \\
$\xi_{0, ab}$ & \SI{660}{\angstrom} & * \\
$\xi_{0, c}$ & \SI{11}{\angstrom} & * \\
$\lambda_{L, ab}$ & \SI{1900}{\angstrom} & \cite{Maeno2012, Riseman1998} \\
$\lambda_{L, c}$ & \SI{73000}{\angstrom} & *
\\ \hline\hline
\end{tabular} \hspace*{\stretch{1}} \begin{tabular}{ccc} \hline\hline
parameter & value & Refs. \\ \hline \\[-14pt]
$\displaystyle \frac{2 \Delta_{\text{max}}}{k_B T_c}$ & \num{5.4} & * \\[11pt]
$\displaystyle \frac{B_{c2 \parallel ab}}{B_{c2 \parallel c}}$ & \num{20} & * \\[11pt]
$\displaystyle \frac{\lambda_{L, ab}}{\xi_{0, ab}}$ & \num{2.9} & * \\[11pt]
$\displaystyle \frac{\lambda_{L, c}}{\xi_{0, c}}$ & \num{6600} & * \\[11pt]
$\displaystyle \frac{\xi_{0, ab}}{\xi_{0, c}}$ & \num{60} & \cite{Rastovski2013, Kittaka2014}
\\[9pt] \hline\hline
\end{tabular} \hspace*{\stretch{1}}}
\label{tab:SRO-fund-param}
\end{table}

\subsection{Review of experimental investigations of the pairing state} \label{sec:SRO-lit-review}
Despite having been extensively experimentally investigated, the fundamental question of what is the pairing symmetry of SRO remains to this day unanswered~\cite{Maeno2024}.
In Tab.~\ref{tab:SRO-SC-state-options} we list the possible options.
The multiband and spin-orbit-coupled nature of SRO supports a richer set of possible pairing states than single-band superconductivity (SC)~\cite{Ramires2019, Kaba2019, Huang2019}, as we shall explain in Sec.~\ref{sec:SRO-SC-construct}, so the pairing wavefunction that we provide in Tab.~\ref{tab:SRO-SC-state-options} should be understood as schematic examples of pairing states belonging to each symmetry class (irrep).
There have been many reviews of SRO's SC in the past~\cite{Mackenzie2003, Maeno2012, Kallin2012, LiuMao2015, Mackenzie2017, Leggett2021}.
However, given the dramatic change in the experimental outlook, in the introduction of Ref.~\cite{Palle2023-ECE} we have reviewed the literature once more.
Below is an updated version of this review: What do we know about the pairing symmetry of \ce{Sr2RuO4} as of September, 2024?
Recently, a complementary literature review has been published~\cite{Maeno2024} that goes into more details.

The superconductivity of SRO is unconventional.
This has been established early on by the absence of a Hebel-Slichter peak~\cite{Hebel1957, Hebel1959} in the NMR relaxation rate $1/T_1$~\cite{Ishida1997, Ishida2000, Murakawa2007}, and
by the large suppression of the SC transition temperature $T_c$ by non-magnetic impurities~\cite{Mackenzie1998, Mao1999, Kikugawa2002, Kikugawa2004} that saturates the Abrikosov-Gor'kov bound~\cite{Abrikosov1961, Gorkov2008}.
Subsequent experiments have only further confirmed the unconventional character of SRO's SC.

The pairing of SRO is more likely to be even than not.
Recent\footnote{The heating caused by NMR pulses~\cite{Pustogow2019, Ishida2020} has rendered early NMR Knight shift experiments~\cite{Ishida1998}, nicely summarized in Figure~14 of Ref.~\cite{Murakawa2007}, invalid.
The NMR pulse heat-up effect acts on a time-scale much shorter than $T_1$ and has not invalidated the early NMR relaxation rate studies~\cite{Pustogow2019}.
An early polarized neutron scattering study~\cite{Duffy2000} has been superseded by a new one~\cite{Petsch2020} with better statistics, carried out at a smaller magnetic field.
See also the discussed at the start of this chapter.} NMR Knight shift~\cite{Pustogow2019, Ishida2020, Chronister2021} and polarized neutron scattering~\cite{Petsch2020} experiments strongly favor singlet pairing,
as do numerous studies~\cite{Mackenzie2017}\footnote{The evidence for a Pauli-limited $B_{c2 \parallel ab}$ is threefold:
(i) the SC-normal state transition is first-order below $0.5 T_c$, as seen in the hysteresis~\cite{Yonezawa2013, Yonezawa2014, Kittaka2014} and jumps in the specific heat~\cite{Deguchi2002, Yonezawa2014}, thermal conductivity~\cite{Deguchi2002}, magnetocaloric effect~\cite{Yonezawa2013}, ac magnetic susceptibility~\cite{Yaguchi2002}, magnetization~\cite{Kittaka2014}, and Knight shift~\cite{Chronister2021};
(ii) the measured intrinsic SC anisotropy $\xi_{ab} / \xi_c \sim 60$~\cite{Rastovski2013, Kittaka2014} exceeds the critical field anisotropy $B_{c2 \parallel ab} / B_{c2 \parallel c} \sim 20$~\cite{Kittaka2009} by a factor of $3$ at zero temperature in the  absence of strain, and by a factor of $20$ under $\langle 100 \rangle$ uniaxial pressure that maximally enhances $T_c$~\cite{Steppke2017},
whereas for orbitally limited $B_{c2 \parallel ab}$ the two ratios would be comparable; and
(iii) $B_{c2 \parallel ab} \propto \Delta / \mu_B \propto T_c$ under small uniaxial strain~\cite{Jerzembeck2023}, as expected for Pauli limiting.} indicating that the in-plane critical field $B_{c2 \parallel ab}$ is Pauli limited~\cite{Clogston1962}.
Although the observation of $\pi$ phase shifts~\cite{Nelson2004} and half-quantum vortices~\cite{Jang2011, Yasui2017, Cai2022} is at tension with even-parity SC, possible explanations do exist~\cite{Yuan2021, Zutic2005, Lindquist2023}.
Reconciling an \SI{80}{\percent} drop in the in-plane Knight shift~\cite{Chronister2021} with triplet pairing, or a strained critical field anisotropy $B_{c2 \parallel ab} / B_{c2 \parallel c} \sim 3$~\cite{Steppke2017} far below the SC anisotropy $\xi_{ab} / \xi_c \sim 60$~\cite{Rastovski2013, Kittaka2014} without Pauli limiting~\cite{Mackenzie2017}, is significantly more challenging, but perhaps possible~\cite{Ramires2016, Ramires2017}.

The evidence for time-reversal symmetry breaking (TRSB) is mixed.
Zero-field muon spin relaxation (ZF-$\mu$SR)~\cite{Luke1998, Luke2000, Higemoto2014, Grinenko2021-unaxial, Grinenko2021-isotropic} and polar Kerr effect~\cite{Xia2006, Kapitulnik2009} experiments indicate TRSB at a $T_{\text{TRSB}}$ at or very near $T_c$,
yet the current response of micron-sized Josephson junctions~\cite{Saitoh2015, Kashiwaya2019}\footnote{Note: contrary to what is stated in Ref.~\cite{Saitoh2015}, the inversion symmetry $I_c^{+}(H) = - I_c^{-}(-H)$ for which they observe that it becomes restored for small junctions is precisely time-reversal symmetry.} exhibits time-reversal invariance.
Under $\langle 100 \rangle$ uniaxial pressure, ZF-$\mu$SR~\cite{Grinenko2021-unaxial} observes a large splitting between $T_{\text{TRSB}}$ and $T_c$,\footnote{In one sample~\cite{Grinenko2021-unaxial}, $T_{\text{TRSB}}$ and $T_c$ split even without any external pressure.} yet no signatures of a TRSB phase transition below $T_c$ have been found in heat capacity~\cite{Li2021} or elastocaloric~\cite{Li2022} measurements under $\langle 100 \rangle$ strain.
Under disorder and hydrostatic pressure, no splitting between SC and TRSB is observed in ZF-$\mu$SR~\cite{Grinenko2021-isotropic}.
Preliminary ZF-$\mu$SR measurements point towards splitting of SC and TRSB under $\langle 110 \rangle$ uniaxial stress~\cite{Grinenko2023}, but elastocaloric effect measurements performed under the same strain do not find any signatures of a second TRSB transition~\cite{Jerzembeck2024}.
Phenomenologically, TRSB requires a two-component SC order parameter, which is usually taken to couple linearly to $[110]$ stress to explain the jump in the $c_{66}$ elastic coefficient~\cite{Okuda2003, Benhabib2021, Ghosh2021}
However, this linear coupling entails a cusp in $T_c$ as a function of $\epsilon_{110}$ strain that has not been observed~\cite{Jerzembeck2024} and the only way homogeneous TRSB SC states can be reconciled with this absence of a cusp is through delicate fine-tuning~\cite{Jerzembeck2024}.
In the presence of TRSB, spontaneous magnetization and currents are generically expected to appear around domain walls, edges, and defects, yet scanning SQUID and Hall probe microscopy~\cite{Tamegai2003, Bjornsson2005, Kirtley2007, Hicks2010, Curran2011, Curran2014, Curran2023} has failed to find any evidence for them, even though theoretical estimates suggest that they should be measurable if present~\cite{Matsumoto1999, Okuno1999, Kirtley2007, Curran2023}.
Josephson junction experiments~\cite{Saitoh2015, Kidwingira2006, Anwar2013, Anwar2017, Yasui2020} show signs of SC domains in their interference patterns, switching behavior, and size-dependence of their transport properties, but the domains themselves need not be chiral.

The coupling of SC to strain is partially known from measurements of elastic constants.
The main obstacle to making these measurements conclusive is the fact that strain inhomogeneities, such as stacking faults or lattice dislocations, mix elastic waves of different symmetry.\footnote{As pointed out in Ref.~\cite{Willa2021}, dislocations give contributions to elastic constants that are on the order of \SI{1}{\percent}, which is two orders of magnitude larger than the (larger of the two sets of) measured jumps of the elastic constants at $T_c$~\cite{Ghosh2021}.}
That said, according to elastic constant measurements, the SC order appears to couple quadratically to $\epsilon_{xx} - \epsilon_{yy} \in B_{1g}$ strain and possibly linearly to $\epsilon_{xy} \in B_{2g}$ strain.
The evidence for the former is the quadratic dependence of $T_c$ on $\epsilon_{xx} - \epsilon_{yy}$, whether measured globally~\cite{Hicks2014, Steppke2017, Barber2019} or locally~\cite{Watson2018, Mueller2023}, and the absence of a jump at $T_c$ in the shear elastic modulus $c_{B_{1g}} = \tfrac{1}{2} (c_{11} - c_{12})$~\cite{Matsui2001, Benhabib2021, Ghosh2021}.
The evidence for the latter is a jump at $T_c$ in the shear elastic constant $c_{66} \in B_{2g}$~\cite{Okuda2003, Benhabib2021, Ghosh2021}, as measured by ultrasound.
However, the magnitude of this jump varies by a factor of $50$ between the two experimental groups~\cite{Benhabib2021, Ghosh2021} and direct measurements of $T_c$ under $[110]$ strain show linear dependence without any cusp whose magnitude can be fully accounted without linear coupling to $\epsilon_{xy}$~\cite{Jerzembeck2024}.
Moreover, no evidence of transition splitting is found in elastocaloric measurements under $[110]$ strain~\cite{Jerzembeck2024}, as generically expected in the presence of linear coupling to $\epsilon_{xy}$.
This raises the possibility that the observed jump in $c_{66}$ is due to lattice defect effects that, however, need to be channel selective so as to not generate a jump in $c_{B_{1g}}$.
One such proposal~\cite{Willa2021} is that a subleading pairing channel activates near dislocations; the product of the leading and subleading pairing irreps then determines which elastic modulus experiences a jump.
No jump has been observed for the elastic modulus $c_{44} \in E_g$~\cite{Matsui2001, Ghosh2021}, indicating that the coupling to $E_g$ strain is quadratic.
Large jumps in the $A_{1g}$ components of the viscosity tensor have recently been discovered at $T_c$~\cite{Ghosh2022}.

\begin{table}[p!]
\centering
\captionabove[Possible superconducting states of \ce{Sr2RuO4}.]{\textbf{Possible superconducting states of \ce{Sr2RuO4}.}
In the first column are the irreps of the tetragonal point group $D_{4h}$ (Tab.~\ref{tab:D4h-char-tab-again2}) to which the pairing states can belong to, together with the orbital functions often used to specify them ($s$-wave for $A_{1g}$, etc.).
In the middle column are the simplest (lowest order in $\vb{k}$) pairing wavefunction which transform under a given irrep.
In the last column are the orientations of the symmetry-enforced line nodes, which can be vertical (V) or horizontal (H), if present.
Accidental (acc.) line nodes may also arise, as in the case of extended $s$-wave pairing.
The options belonging to the same irrep can be superimposed, in which case only line nodes shared between them survive.
In the case of 2D irreps, their $(\Delta_1|\Delta_2)$ may condense into a time-reversal symmetry-breaking (TRSB) chiral superposition $\Delta_1 \pm \iu \Delta_2$ or rotation symmetry-breaking nematic superposition $\Delta_1$, $\Delta_2$, or $\Delta_1 \pm \Delta_2$, as explained Sec.~\ref{sec:SRO-GL-analysis}.
Only even-frequency pairing without accidental degeneracies between irreps is listed.
See also Sec.~\ref{sec:SRO-SC-construct} and Refs.~\cite{Sigrist1987, Maeno2024}.}
{\renewcommand{\arraystretch}{1.25}
\renewcommand{\tabcolsep}{10pt}
\begin{tabular}{L{80pt}C{180pt}L{105pt}} \hline\hline
\multicolumn{3}{c}{\large Even-parity spin-singlet pairing states:} \\
symmetry & simplest $d_0(\vb{k})$ & line nodes \\ \hline
$A_{1g}(s)$ & $1$ & none \\
$A_{1g}(s)$ & $k_x^2 + k_y^2$ & none \\
$A_{1g}(\text{extended }s)$ & $k_z^2$ & acc.\ horizontal \\
$A_{1g}(\text{extended }s)$ & $k_x^4 + k_y^4 - 6 k_x^2 k_y^2$ & acc.\ vertical \\[5pt]
$A_{2g}(g_{xy(x^2-y^2)})$ & $k_x k_y (k_x^2-k_y^2)$ & vertical \\[5pt]
$B_{1g}(d_{x^2-y^2})$ & $k_x^2-k_y^2$ & vertical \\[5pt]
$B_{2g}(d_{xy})$ & $k_x k_y$ & vertical \\[7pt]
$E_g(d_{yz}|-d_{xz})$ & $(k_y k_z|- k_x k_z)$ & \raisebox{8pt}{$\begin{cases}
\text{H for TRSB}, \\
\text{H\&V for nematic}
\end{cases}$} \\[12pt] \hline \\[-6pt]
\multicolumn{3}{c}{\large Odd-parity spin-triplet pairing states:} \\
symmetry & simplest $\vb{d}(\vb{k})$ & line nodes \\ \hline
$A_{1u}(\text{helical }p)$ & $k_x \vu{e}_x + k_y \vu{e}_y$ & none \\
$A_{1u}(\text{helical }p)$ & $k_z \vu{e}_z$ & horizontal \\[5pt]
$A_{2u}(\text{helical }p)$ & $k_x \vu{e}_y - k_y \vu{e}_x$ & none \\
$A_{2u}(h_{xyz(x^2-y^2)})$ & $k_x k_y k_z (k_x^2-k_y^2) \vu{e}_z$ & H\&V \\[5pt]
$B_{1u}(\text{helical }p)$ & $k_x \vu{e}_x - k_y \vu{e}_y$ & none \\
$B_{1u}(f_{(x^2-y^2)z})$ & $(k_x^2-k_y^2) k_z \vu{e}_z$ & H\&V \\[5pt]
$B_{2u}(\text{helical }p)$ & $k_x \vu{e}_y + k_y \vu{e}_x$ & none \\
$B_{2u}(f_{xyz})$ & $k_x k_y k_z \vu{e}_z$ & H\&V \\[7pt]
$E_u(p_x|p_y)$ & $(k_y \vu{e}_z | - k_x \vu{e}_z)$ & \raisebox{8pt}{$\begin{cases}
\text{none for TRSB}, \\
\text{V for nematic}
\end{cases}$} \\[6pt]
$E_u(p_x|p_y)$ & $(k_z \vu{e}_x | - k_z \vu{e}_y)$ & horizontal
\\ \hline\hline
\end{tabular}}
\label{tab:SRO-SC-state-options}
\end{table}

The preponderance of evidence points towards line nodes.
The expected dependence on temperature is found in the heat capacity~\cite{NishiZaki2000, Deguchi2004, Kittaka2018}, ultrasound attenuation rate~\cite{Matsui2001, Lupien2001}, NMR relaxation rate~\cite{Ishida2000}, and London penetration depth~\cite{Bonalde2000, Landaeta2023, Ferguson2024}.
In weak in-plane fields, the heat capacity~\cite{Deguchi2004-p2, Kittaka2018} and Knight shift~\cite{Chronister2021} obey Volovik scaling ($\propto \sqrt{B / B_{c2}}$) expected of line nodes~\cite{Volovik1993}.
The in-plane thermal conductivity~\cite{Suzuki2002, Hassinger2017} exhibits universal transport, which is a type of transport found only in nodal SC~\cite{Lee1993, Balatsky1995, Sun1995, Graf1996}.
Finally, STM spectroscopy~\cite{Firmo2013, Sharma2020} shows a $V$-shaped conductance minimum,\footnote{One should keep in mind that STM mostly probes the $\alpha, \beta$ bands because of their $d_{xz}, d_{yz}$ orbital characters which make their overlaps with the tip (along $z$) large.}
although this is not completely reproducible~\cite{Lupien2018, Rodriguez2022, Valadkhani2024}.
The only evidence to the contrary is an STM/S study~\cite{Suderow2009} that scanned micron-sized grains ($\sim 10 \, \xi_{0,ab}$) situated on top of SC aluminium and found an implausibly large SC gap $\Delta$ of \SI{3.5}{\kelvin}.
Given that so many studies~\cite{NishiZaki2000, Deguchi2004, Kittaka2018, Matsui2001, Lupien2001, Ishida2000, Bonalde2000, Deguchi2004-p2} found nodal behavior, in some cases down to as low as $\SI{0.04}{\kelvin} \approx T_c / 30$, any fully gapped SC state candidate must have extraordinarily deep minima to be viable.

The line node(s) are more likely to be vertical than horizontal, but this is not completely settled.
If present, the vertical line nodes are most likely located away from the Van Hove points $(\pi, 0)$ and $(0, \pi)$.
Heat capacity~\cite{Kittaka2018} and in-plane thermal conductivity~\cite{Tanatar2001-p2, Izawa2001} both display a fourfold anisotropy in their dependence on the in-plane $\vb{B}$ orientation.\footnote{As pointed out in~\cite{Kittaka2018}, little useful information can be extracted from the out-of-plane field-angle anisotropy.}
Since these anisotropies are small ($\sim \SI{1}{\percent}$), they can be explained by both horizontal and vertical nodes.
That the heat capacity anisotropy has the same sign down to $T_c / 20$ appears to exclude $d_{xy}$-wave pairing~\cite{Kittaka2018}, and perhaps other pairing states too.
A resonance at transfer energy $\approx 2 \Delta$ and momentum with a finite $z$ component was reported below $T_c$ in the inelastic neutron scattering intensity~\cite{Iida2020}, suggesting horizontal line nodes, but was not reproduced in subsequent measurements~\cite{Jenni2021}.
The universal heat transport along $c$ has been found finite with $2 \sigma$ significance~\cite{Hassinger2017}, indicating that nodal quasi-particles have a finite $c$-axis velocity.
If true, this result is strong evidence against symmetry-enforced horizontal line nodes.
Elastocaloric effect measurements under $\langle 100 \rangle$ uniaxial pressure~\cite{Li2022} reveal that the normal-state entropy attains a maximum at the Lifshitz transition strain $\epsilon_{100} = \SI{-0.44}{\percent} \equiv \epsilon_{\text{VH}}$, which becomes a minimum as one enters the SC state~\cite{Palle2023-ECE}.
Further analysis shows that this can only be accounted for if there are no vertical line nodes at the Van Hove points $(\pi, 0)$ and $(0, \pi)$~\cite{Palle2023-ECE}.
Note that these same Van Hove points are responsible for the normal-state entropy maximum~\cite{Steppke2017, Barber2019, Sunko2019}.
From the upper-critical field dependencies on temperature in a very pure sample, in Ref.~\cite{Landaeta2023} they deduced that SRO's SC exhibits non-local electrodynamics~\cite{Kosztin1997}, which is a type of SC response where nodal excitation are important.
They find that the $T$-dependence of the penetration depth is more consistent with vertical than horizontal line nodes~\cite{Landaeta2023}, however further information, like the number or precise locations of the nodes, cannot be inferred~\cite{Roising2024}.

Interface and surface experiments offer limited information.
Josephson junctions to conventional superconductors behave in unusual ways and suffer from irreproducibility~\cite{LiuMao2015, Anwar2021, Maeno2024}, which is one of the reasons these experiments have not been conclusive.
Their unusual behavior (as seen in their interference, switching, and size-dependence) has most often been interpreted as evidence of domains~\cite{Saitoh2015, Kidwingira2006, Anwar2013, Anwar2017, Yasui2020}, but deducing any more precise information on the structure of the SC order parameter has been challenging.
Some experiments have shown signs of $\pi$ shifts~\cite{Nelson2004, Jang2011, Yasui2017, Cai2022}, indicating odd-parity SC, but their interpretation is not clear-cut~\cite{Yuan2021, Zutic2005, Lindquist2023}.
STM tunneling conductance measurements have also been inconsistent~\cite{Upward2002, Firmo2013, Lupien2018, Sharma2020, Rodriguez2022, Valadkhani2024}, likely due to surface reconstruction effects~\cite{Damascelli2000, Lupien2018, Rodriguez2022, Valadkhani2024}.
A $V$-shaped conductance minimum has been reported in Ref.~\cite{Firmo2013}, indicating line nodes.
In another STM study~\cite{Sharma2020}, they considered the Fourier transform of the real-space tunneling conductance and found peaks at nesting vectors expected of $d_{x^2-y^2}$-wave SC.
However, the peaks are not clearly resolved because of noise (see Supplementary Information of Ref.~\cite{Sharma2020}) and when compatibility with other pairings was later investigated~\cite{Bhattacharyya2023}, their measurements were found to be consistent with extended $s$-wave pairing, as well as accidentally degenerate $s' + \iu \, d_{xy}$ and $s' + \iu \, d_{x^2-y^2}$ states.

This concludes the review of the experimental literature concerning strontium ruthenate's superconductivity.
When it comes to theories, many~\cite{Romer2019, Romer2020,
Kivelson2020, Willa2021, Yuan2021, Sheng2022, Yuan2023,
Clepkens2021, Romer2021,
Suh2020, Fukaya2022,
Wagner2021,
Leggett2021, Huang2021,
Huang2021-p2,
Gingras2022, Scaffidi2023} have been developed in the wake of the landmark NMR Knight shift study of Pustogow et al.~\cite{Pustogow2019}.
Although some~\cite{Leggett2021, Huang2021} still explore odd-parity pairing as an option, most recent theories are based on even-parity SC states.
In Tab.~\ref{tab:SRO-SC-state-options} we list the possible SC state which are based on only one irrep.
The most studied of such states are the chiral (TRSB) $E_g$ state $d_{xz} + \iu \, d_{yz}$~\cite{Suh2020, Fukaya2022} and the one-component $B_{1g}$ state $d_{x^2-y^2}$~\cite{Willa2021}.
Most other proposals assume an accidental (fine-tuned) degeneracy between two distinct irreps, which should be contrasted with $d_{xz} + \iu \, d_{yz}$ where the degeneracy is symmetry-enforced.
Such proposals include $s' + \iu \, d_{x^2-y^2}$ pairing~\cite{Romer2019, Romer2020}, $d_{x^2 - y^2} + \iu \, g_{xy(x^2 - y^2)}$ pairing~\cite{Kivelson2020, Willa2021, Yuan2021, Sheng2022, Yuan2023}, and $s' + \iu \, d_{xy}$ pairing~\cite{Clepkens2021, Romer2021}, where $s'$ denotes extended $s$-wave states.
In most of these proposals, the accidentally degenerate pairing state is a proper bulk order, while in others~\cite{Willa2021} the mixing among irreps emerges only near lattice defects.
To explain the puzzling experimental phenomenology of SRO, some have pursued even more exotic ideas, such as mixing of even- and odd-parity SC states~\cite{Scaffidi2023} or mixing of even- and odd-frequency pairing~\cite{Gingras2022}.
For further discussion of theories of SRO, we refer the reader to Ref.~\cite{Maeno2024}.

In Secs.~\ref{sec:SRO-analysis-ECE-100} and~\ref{sec:SRO-Tc-ECE-analysis-110}, we elaborate in more detail how the results of Refs.~\cite{Palle2023-ECE, Jerzembeck2024} were obtained.
These results were already mention during the literature review of this section.

\subsection{Crystal structure and symmetries} \label{sec:SRO-cryst-struct}
\begin{figure}[t]
\centering
\includegraphics[width=0.75\textwidth]{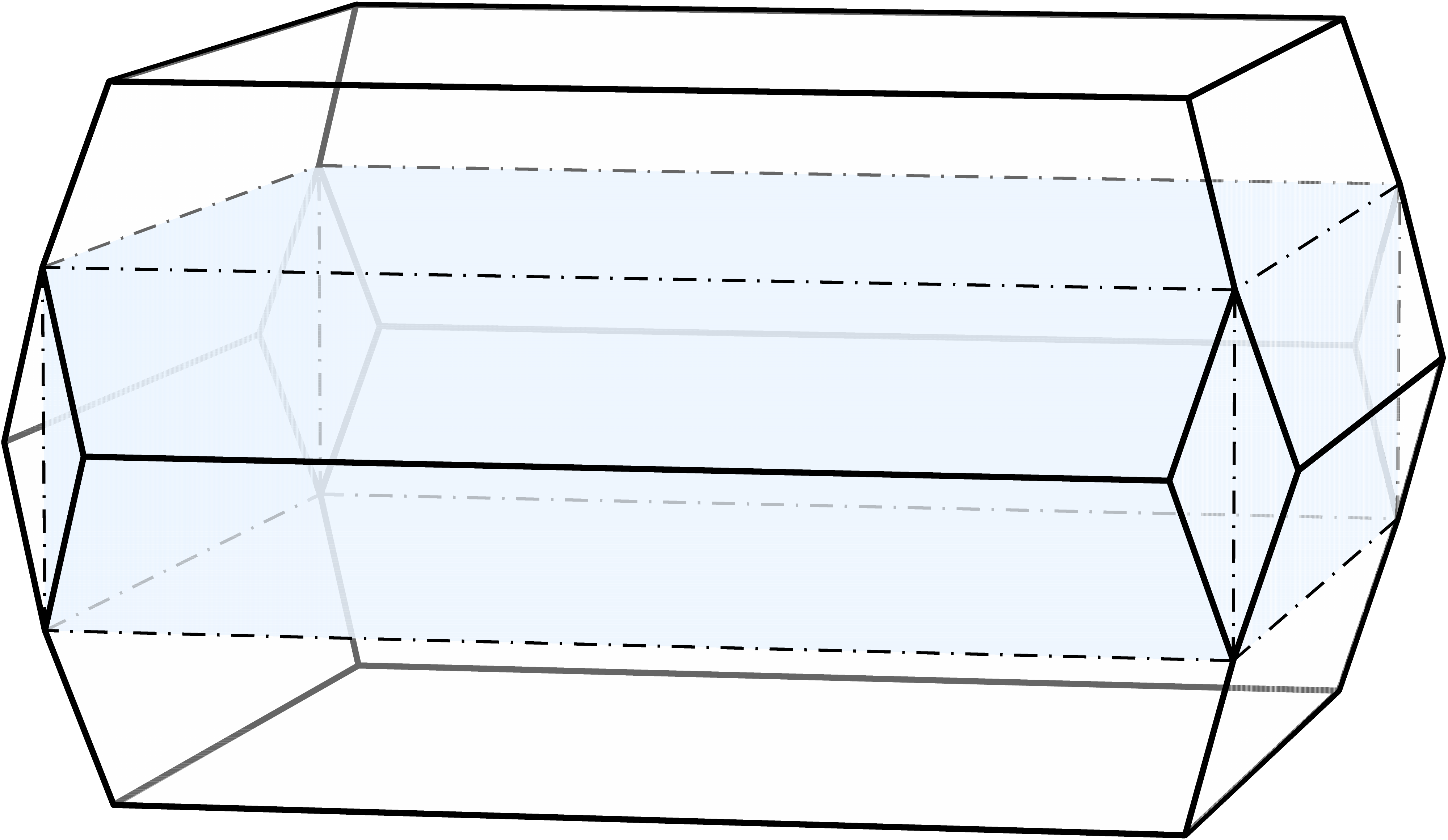}
\captionbelow[The first Brillouin zone of \ce{Sr2RuO4}, drawn in proportion.]{\textbf{The first Brillouin zone of \ce{Sr2RuO4}, drawn in proportion.}
The larger polyhedron that has thick black edges corresponds to the body-centered-tetragonal first Brillouin zone of \ce{Sr2RuO4}.
The smaller rectangular cuboid shaded in blue is the simple-tetragonal first Brillouin zone, shown for reference.
The $c$ axis points upwards.}
\label{fig:SRO-BZ}
\end{figure}

As shown in Fig.~\ref{fig:SRO-crystal-FS}(a), SRO is a layered perovskite with a body-centered tetragonal lattice.
Its lattice constants equal~\cite{Mackenzie2003, Bergemann2003}:
\begin{align}
a = b &= \SI{3.86}{\angstrom}, &
c &= \SI{12.7}{\angstrom}.
\end{align}
The primitive lattice vectors of the body-centered tetragonal lattice of SRO are:
\begin{align}
\begin{aligned}
\vb{a}_1 &= a \vu{e}_x, \\[4pt]
\vb{a}_2 &= a \vu{e}_y, \\[4pt]
\vb{a}_3 &= \frac{a}{2} \vu{e}_x + \frac{a}{2} \vu{e}_y + \frac{c}{2} \vu{e}_z,
\end{aligned}
\end{align}
where the $x, y, z$ Cartesian coordinates have been aligned with the principal $a, b, c$ axes of the lattice.
The corresponding reciprocal lattice is face-centered tetragonal, which is equivalent to body-centered tetragonal for tetragonal systems.
The reciprocal primitive lattice vectors are ($\vb{a}_i \vdot \vb{b}_j = 2 \pi \Kd_{ij}$):
\begin{align}
\begin{aligned}
\vb{b}_1 &= \frac{2 \pi}{a} \vu{e}_x - \frac{2 \pi}{c} \vu{e}_z, \\
\vb{b}_2 &= \frac{2 \pi}{a} \vu{e}_y - \frac{2 \pi}{c} \vu{e}_z, \\
\vb{b}_3 &= \frac{4 \pi}{c} \vu{e}_z.
\end{aligned}
\end{align}
The corresponding first Brillouin zone is draw in Fig.~\ref{fig:SRO-BZ}.
Instead of the crystal momenta
\begin{align}
\vb{k} &= k_x \vu{e}_x + k_y \vu{e}_y + k_z \vu{e}_z = \begin{pmatrix}
k_x \\ k_y \\ k_z
\end{pmatrix},
\end{align}
we shall often use the dimensionless
\begin{align}
\vb{\DimK} &= \DimK_x \vu{e}_x + \DimK_y \vu{e}_y + \DimK_z \vu{e}_z = \begin{pmatrix}
\DimK_x \\
\DimK_y \\
\DimK_z
\end{pmatrix} = \begin{pmatrix}
a k_x \\
a k_y \\
c k_z
\end{pmatrix}.
\end{align}
The lattice constants $a, c$ we retain because the precise geometry of the Brillouin zone and its boundary will be important in Sec.~\ref{sec:SRO-analysis-ECE-100}.

For reference, if the system were simple tetragonal, the reciprocal primitive lattice vectors would equal
\begin{align}
\begin{aligned}
\vb{b}_1' &= \frac{2 \pi}{a} \vu{e}_x, \\
\vb{b}_2' &= \frac{2 \pi}{a} \vu{e}_y, \\
\vb{b}_3' &= \frac{2 \pi}{c} \vu{e}_z.
\end{aligned}
\end{align}
Since $\vb{b}_1 = \vb{b}_1' - \vb{b}_3'$, $\vb{b}_2 = \vb{b}_2' - \vb{b}_3'$, and $\vb{b}_3 = 2 \vb{b}_3'$, it follows that every function which is simple-tetragonal periodic is also body-centered-tetragonal periodic.
The converse is not necessarily true: $f(\vb{k}) = \cos(\tfrac{1}{2} \DimK_x) \cos(\tfrac{1}{2} \DimK_y) \sin(\tfrac{1}{2} \DimK_z)$ is body-centered-periodic [$f(\vb{k} + \vb{b}_i) = f(\vb{k})$], but not simple-periodic [$f(\vb{k} + \vb{b}_i') = - f(\vb{k})$], for instance.
This point will be of significance in Sec.~\ref{sec:SRO-behavior-vH}, during our analysis of which pairing states have vertical line nodes on the Van Hove lines.

\begin{table}[t]
\centering
\captionabove[The character table of the tetragonal point group $D_{4h}$~\cite{Dresselhaus2007}.]{\textbf{The character table of the tetragonal point group $D_{4h}$}~\cite{Dresselhaus2007}.
The irreps are divided according to parity into even (subscript $g$) and odd ($u$) ones.
To the left of the irreps are the simplest polynomials constructed from the coordinates $\vb{r} = (x, y, z)$ that transform according to them.
$C_4$ are \SI{90}{\degree} rotations around $\vu{e}_z$.
$C_2$, $C_2'$, and $C_2''$ are \SI{180}{\degree} rotations around $\vu{e}_z$, $\vu{e}_x$ or $\vu{e}_y$, and the diagonals $\vu{e}_x \pm \vu{e}_y$, respectively.
$P$ is space inversion or parity.
Improper rotations $S_4$ and mirror reflections $\Sigma_h$, $\Sigma_v'$, and $\Sigma_d''$ are obtained by composing $C_4$, $C_2$, $C_2'$, and $C_2''$ with $P$, respectively.}
{\renewcommand{\arraystretch}{1.3}
\renewcommand{\tabcolsep}{8.4pt}
\begin{tabular}{cc|rrrrr|rrrrr} \hline\hline
\multicolumn{2}{c|}{$D_{4h}$} & $E$ & $2 C_4$ & $C_2$ & $2 C_2'$ & $2 C_2''$ & $P$ & $2 S_4$ & $\Sigma_h$ & $2 \Sigma_v'$ & $2 \Sigma_d''$
\\ \hline
$1$, $x^2+y^2$, $z^2$ & $A_{1g}$ & $1$ & $1$ & $1$ & $1$ & $1$ & $1$ & $1$ & $1$ & $1$ & $1$
\\
$xy(x^2-y^2)$ & $A_{2g}$ & $1$ & $1$ & $1$ & $-1$ & $-1$ & $1$ & $1$ & $1$ & $-1$ & $-1$
\\
$x^2-y^2$ & $B_{1g}$ & $1$ & $-1$ & $1$ & $1$ & $-1$ & $1$ & $-1$ & $1$ & $1$ & $-1$
\\
$xy$ & $B_{2g}$ & $1$ & $-1$ & $1$ & $-1$ & $1$ & $1$ & $-1$ & $1$ & $-1$ & $1$
\\
$(yz|-xz)$ & $E_g$ & $2$ & $0$ & $-2$ & $0$ & $0$ & $2$ & $0$ & $-2$ & $0$ & $0$
\\ \hline
$xyz(x^2-y^2)$ & $A_{1u}$ & $1$ & $1$ & $1$ & $1$ & $1$ & $-1$ & $-1$ & $-1$ & $-1$ & $-1$
\\
$z$ & $A_{2u}$ & $1$ & $1$ & $1$ & $-1$ & $-1$ & $-1$ & $-1$ & $-1$ & $1$ & $1$
\\
$xyz$ & $B_{1u}$ & $1$ & $-1$ & $1$ & $1$ & $-1$ & $-1$ & $1$ & $-1$ & $-1$ & $1$
\\
$(x^2-y^2)z$ & $B_{2u}$ & $1$ & $-1$ & $1$ & $-1$ & $1$ & $-1$ & $1$ & $-1$ & $1$ & $-1$
\\
$(x|y)$ & $E_u$ & $2$ & $0$ & $-2$ & $0$ & $0$ & $-2$ & $0$ & $2$ & $0$ & $0$
\\ \hline\hline
\end{tabular}}
\label{tab:D4h-char-tab-again2}
\end{table}

The space group of SRO is $I4/mmm$~\cite{Mackenzie2003, Bergemann2003}.
This space group is symmorphic, i.e., there are no symmetry operations, such as glide plane, screw axis, or others, which include fractional translations.
Hence translations and point group operations can be treated separately.
As previously already mentioned, the point group of SRO is $D_{4h}$ ($4/mmm$ in Hermann-Mauguin notation) and its character table is given in Tab.~\ref{tab:D4h-char-tab-again2}.
This point group is worked out in great detail in Sec.~\ref{sec:tetragonal-group-D4h} of Appx.~\ref{app:group_theory}.
Here, let us just note that $D_{4h}$ is generated by four-fold rotations around the $z$ axis $C_{4z}$, two-fold rotations around the $x$ axis $C_{2x}$, two-fold rotations around the $d_+ = x + y$ diagonal $C_{2d_+}$, and parity $P$.
The center (fixed point) of all of these operations are the ruthenium atoms.
In principle, as the center we could also choose the point $\vb{R} + \tfrac{1}{2} (\vb{a}_1 + \vb{a}_2)$, which is in the middle of the neighboring four ruthenium atoms of a layer.
This latter choice for the center yields point group operations which are equivalent, up to a lattice translation, to the former ones.
We shall always use ruthenium atoms as the center.
By inspecting the crystal structure [Fig.~\ref{fig:SRO-crystal-FS}(a)], one may verify that these operations really are symmetries.
One may also confirm the same for the primitive lattice vectors: $R(g) \vb{a}_i = \vb{a}_i + \vb{R}$ and $R(g) \vb{b}_i = \vb{b}_i + \vb{G}$ for $g \in D_{4h}$ and lattice vectors $\vb{R}, \vb{G}$.

\subsection{Electronic structure and the $t_{2g}$ orbital tight-binding model} \label{sec:SRO-el-struct}
Here we explain the electronic structure of SRO and introduce a tight-binding model for its Fermi liquid phase.

\begin{figure}[t]
\centering
\includegraphics[width=\textwidth]{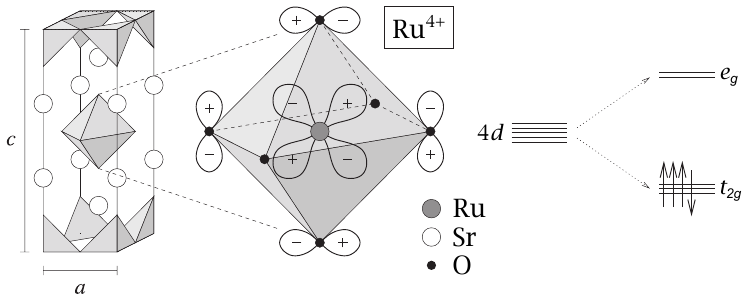}
\captionbelow[The octahedral environment of a ruthenium atom (center) within the layered perovskite crystal lattice of \ce{Sr2RuO4} (left) and the splitting of the $4d$ orbitals cause by such an environment (right)~\cite{Bergemann2003}.]{\textbf{The octahedral environment of a ruthenium atom (center) within the layered perovskite crystal lattice of \ce{Sr2RuO4} (left) and the splitting of the $4d$ orbitals cause by such an environment (right)}~\cite{Bergemann2003}.
The crystal field splits the five degenerate $4d$ levels into a low-lying $t_{2g}$ orbital manifold made of $(d_{yz}, d_{zx}, d_{xy})$ and an elevated $e_g$ orbital manifold made of $(d_{x^2+y^2-2z^2}, d_{x^2-y^2})$.
Reproduced with editing from Ref.~\cite{Bergemann2003}, with permission from Taylor \& Francis.}
\label{fig:SRO-octahedral-Ru}
\end{figure}

The atomic electron configuration of \ce{Ru} is [\ce{Kr}]$5s^{2}4d^{6}$, of \ce{Sr} is [\ce{Kr}]$5s^{2}$, and of \ce{O} is [\ce{He}]$2s^{2}2p^{4}$.
If for the valencies of strontium and oxygen we take the usual \ce{Sr^{2+}} and \ce{O^{2-}} values, then the ruthenium atoms are left in the configuration \ce{Ru^{4+}} = [\ce{Kr}]$5s^{0}4d^{4}$.
In \ce{Sr2RuO4}, each ruthenium atom is surrounded by an octahedron whose vertices are oxygen atoms, as shown in Fig.~\ref{fig:SRO-octahedral-Ru}.
This octahedral environment lifts the degeneracy of the $4d$ orbital, as sketched on the right of Fig.~\ref{fig:SRO-octahedral-Ru}~\cite{Bergemann2003}.
Specifically, the five $d$ orbital fall into the $T_{2g}$ and $E_g$ irreps of the octahedral group $O_h$, from which the corresponding orbital manifolds derive their name: $t_{2g}(d_{yz} | d_{zx} | d_{xy})$ and $e_g(d_{x^2+y^2-2z^2}|\sqrt{3} \, d_{x^2-y^2})$.
The states closest to the Fermi level derive primarily from the partially filled $t_{2g}$ orbitals, with some anti-bonding admixtures coming from the \ce{O}:$2p$ orbitals~\cite{Bergemann2003}.

Among transition metal oxides, metallic behavior is fairly rare because of the small hopping amplitudes, on the one hand, and the large on-site repulsion, on the other, both of which are a consequence of the small radius of the $d$ orbitals~\cite{Bergemann2003}.
The result is usually an insulating magnetic state, as in the cuprates (Sec.~\ref{sec:cuprate-basics}).
In the case of SRO, however, metallic behavior robustly emerges at low temperatures.
More precisely, below around \num{30} Kelvins, SRO settles into a quasi-2D multiband Fermi liquid state~\cite{Mackenzie2003, Bergemann2003}.
There are three conduction bands in SRO, which are conventionally called $\alpha$, $\beta$, and $\gamma$, and their Fermi sheets are cylindrical~\cite{Mackenzie2003, Bergemann2003}.
They are depicted in Fig.~\ref{fig:SRO-crystal-FS}(b).
The $\alpha$ band is hole-like, while the $\beta$ and $\gamma$ bands are electron-like.

\begin{figure}[t]
\centering
\begin{subfigure}[t]{0.5\textwidth}
\raggedright
\includegraphics[width=0.95\textwidth]{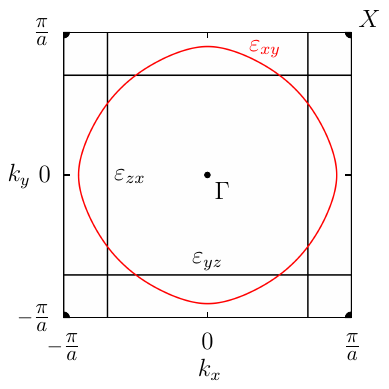}
\subcaption{}
\end{subfigure}%
\begin{subfigure}[t]{0.5\textwidth}
\raggedright
\includegraphics[width=0.95\textwidth]{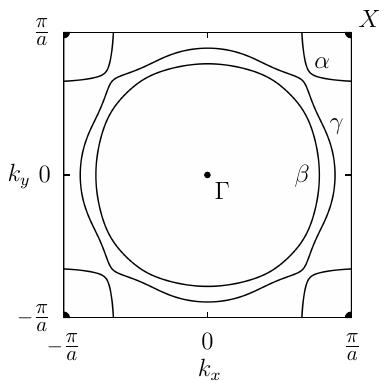}
\subcaption{}
\end{subfigure}
\captionbelow[The Fermi surfaces of \ce{Sr2RuO4}, as determined by the schematic dispersions of Eqs.~(\ref{eq:SRO-schematic-disp1}--\ref{eq:SRO-schematic-disp3}) (a) and the tight-binding model of Eq.~\eqref{eq:SRO-TBA-Haml} (b).]{\textbf{The Fermi surfaces of \ce{Sr2RuO4}, as determined by the schematic dispersions of Eqs.~}(\ref{eq:SRO-schematic-disp1}--\ref{eq:SRO-schematic-disp3}) \textbf{(a) and the tight-binding model of Eq.~\eqref{eq:SRO-TBA-Haml} (b).}
The solid lines are the $k_z = 0$ cross-sections of the cylindrical $\alpha$, $\beta$, and $\gamma$ Fermi sheets of unstrained SRO, shown in Fig.~\ref{fig:SRO-crystal-FS}(b).
The parameters that were used in Eq.~\eqref{eq:SRO-TBA-Haml} are those of Ref.~\cite{Roising2019}.
The $\varepsilon_{xy}(\vb{k}) = 0$ Fermi surface is colored red under (a) for clarity.}
\label{fig:SRO-theoretical-FS}
\end{figure}

The conduction bands of SRO primarily derive from the $t_{2g}(d_{yz} | d_{zx} | d_{xy})$ orbital manifold of the \ce{Ru} atoms~\cite{Mackenzie2003, Bergemann2003, Gingras2022}.
To a first approximation, due to the high anisotropy of SRO ($c / a = 3.3$), $d_{yz}$ and $d_{xz}$ hop along only one in-plane direction and have the following one-dimensional tight-binding dispersions:
\begin{align}
\varepsilon_{yz}(\vb{k}) &= - \upmu - 2 t \cos a k_y, \label{eq:SRO-schematic-disp1} \\
\varepsilon_{zx}(\vb{k}) &= - \upmu - 2 t \cos a k_x. \label{eq:SRO-schematic-disp2}
\end{align}
The $d_{xy}$ hops along both in-plane directions, with the following approximate 2D tight-binding dispersion:
\begin{align}
\varepsilon_{xy}(\vb{k}) &= - \upmu - 2 t (\cos a k_x + \cos a k_y) - 4 t' \cos(a k_x) \cos(a k_y), \label{eq:SRO-schematic-disp3}
\end{align}
where $\upmu \approx \SI{0.35}{\electronvolt}$, $t \approx \SI{0.3}{\electronvolt}$, and $t' \approx \SI{0.1}{\electronvolt}$~\cite{Roising2019, Suh2020}.
The corresponding schematic Fermi surfaces are drawn in Fig.~\ref{fig:SRO-theoretical-FS}(a).
Notice how they already reproduce the broad qualitative shape of the three Fermi surfaces of SRO.
After introducing interorbital mixing and spin-orbit coupling (SOC), $\varepsilon_{yz}(\vb{k})$ and $\varepsilon_{zx}(\vb{k})$ hybridize into the quasi-1D $\alpha$ and $\beta$ bands, while $\varepsilon_{xy}(\vb{k})$ hybridizes into the quasi-2D $\gamma$ band, with the result shown in Fig.~\ref{fig:SRO-theoretical-FS}(b).
Let us note that the $d_{yz}$ and $d_{zx}$ orbitals are even, while $d_{xy}$ is odd, under horizontal reflections, which in turn forbids the mixing of $d_{yz}, d_{zx}$ with $d_{xy}$ for $k_z = 0$ in the absence of SOC.
With SOC, the two may mix, and this mixing is strongest at the diagonals $k_x = \pm k_y$ where the three Fermi sheets almost touch.
Including interlayer hopping adds warping along $k_z$.
It is worth noting that although a Fermi liquid that behaves as if weakly interacting, interactions are significant in SRO and its quasi-particles are strongly renormalized by electronic correlations~\cite{Mackenzie2003, Bergemann2003, Veenstra2014, Tamai2019, Barber2019}.

To describe the Fermi-liquid quasi-particles of the normal state, we shall now introduce a tight-binding model based on the $t_{2g}$ orbitals of ruthenium.
Below \SI{25}{\kelvin}, SRO is well-described by such a tight-binding model~\cite{Roising2019, Suh2020, Zabolotnyy2013, Cobo2016, Burganov2016}.
Within it, the hopping amplitudes $\mathcal{T}_{\vb{\delta}}$ between neighboring lattice sites are significantly constrained by the symmetries of SRO.
In a body-centered lattice, hopping amplitudes along the half-diagonal $\vb{\delta} = \vb{a}_3 = \frac{1}{2} \mleft(a \vu{e}_x + a \vu{e}_y + c \vu{e}_z\mright)$, as well as many other $\vb{\delta}$, are also possible.
However, all such characteristically body-centered hoppings necessarily connect different layers and are thus suppressed by SRO's anisotropy.
For the purpose of making estimates, throughout this chapter we shall employ the tight-binding parameters of Ref.~\cite{Roising2019}, listed in Tab.~\ref{tab:SRO-TBA-params}.

In our definition of the model, we use cyclical ordering of the $t_{2g}$ orbitals.
The column-vector of fermionic annihilation operators (spinor) we define as:
\begin{align}
\psi(\vb{R}) &\defeq \begin{pmatrix}
\ce{Ru}\colon 4d_{yz}(\vb{R}) \\[2pt]
\ce{Ru}\colon 4d_{zx}(\vb{R}) \\[2pt]
\ce{Ru}\colon 4d_{xy}(\vb{R})
\end{pmatrix} = \begin{pmatrix}
\ce{Ru}\colon 4d_{yz}(\vb{R}) \,\otimes \ket{\uparrow} \\[2pt]
\ce{Ru}\colon 4d_{yz}(\vb{R}) \,\otimes \ket{\downarrow} \\[2pt]
\ce{Ru}\colon 4d_{zx}(\vb{R}) \,\otimes \ket{\uparrow} \\[2pt]
\ce{Ru}\colon 4d_{zx}(\vb{R}) \,\otimes \ket{\downarrow} \\[2pt]
\ce{Ru}\colon 4d_{xy}(\vb{R}) \,\otimes \ket{\uparrow} \\[2pt]
\ce{Ru}\colon 4d_{xy}(\vb{R}) \,\otimes \ket{\downarrow}
\end{pmatrix}, \label{eq:SRO-cyclic-spinor}
\end{align}
with the Fourier convention
\begin{align}
\psi_{\vb{k}} &= \frac{1}{\sqrt{\mathcal{N}}} \sum_{\vb{R}} \Elr^{- \iu \vb{k} \vdot \vb{R}} \psi(\vb{R}),
\end{align}
where $\vb{k} = (k_x, k_y, k_z)$ are crystal momenta which always go over the first Brillouin zone only and $\mathcal{N}$ is the number of unit cells.
The subscript ordering does not matter in $d_{yz} = d_{zy}$, $d_{zx} = d_{xz}$, $d_{xy} = d_{yx}$.
This same convention is used in Ref.~\cite{Suh2020}.
When comparing to Refs.~\cite{Zabolotnyy2013, Cobo2016, Burganov2016, Roising2019, Bhattacharyya2023, Roising2024}, among others, one should keep in mind that they use a different ordering of the orbitals.

Symmetries act on fermions in the following way in the cyclically ordered basis:
\begin{align}
\SymU^{\dag}(g) \psi_{\vb{k}} \SymU(g) &= \MatU(g) \psi_{R(g^{-1}) \vb{k}} = O(g) \otimes S(g) \psi_{R(g^{-1}) \vb{k}}, \label{eq:SRO-sym-transf-rule1} \\
\SymTR^{-1} \psi_{\vb{k}} \SymTR &= (\one \otimes \iu \Pauli_{y}) \psi_{- \vb{k}}, \label{eq:SRO-sym-transf-rule2}
\end{align}
where $\SymU(g)$ are the Fock-space point group operators, $g \in D_{4h}$, $\SymTR$ is the Fock-space time-reversal (TR) operator, and $R, O, S$ are unitary representations of $D_{4h}$ whose generators are listed in Tab.~\ref{tab:SRO-transf-mat}.
Here $\one$ is the $3 \times 3$ identity matrix and $\Pauli_{\mu}$ are Pauli matrices.
Because the \ce{Ru} atoms are centered at the Bravais lattice points $\vb{R}$, they map to themselves under all point group operations and the corresponding symmetry matrices therefore do not depend on $\vb{k}$.
This makes \ce{Sr2RuO4} symmetry-wise simpler to treat than cuprates.
Compare with Sec.~\ref{sec:extended-basis-def} of Chap.~\ref{chap:cuprates}.

\begin{table}[t]
\centering
\captionabove[The symmetry transformation matrices of the four generators $g$ of the point group $D_{4h}$ in the cyclically ordered basis~\eqref{eq:SRO-cyclic-spinor}.]{\textbf{The symmetry transformation matrices of the four generators $g$ of the point group $D_{4h}$ in the cyclically ordered basis~\eqref{eq:SRO-cyclic-spinor}.}
$C_{4z}$ is a rotation by $\pi/2$ around the $z$-axis.
$C_{2x}$ and $C_{2d_+}$ are rotations by $\pi$ around $x$ and the diagonal $d_{+} = x+y$, respectively.
$P$ is parity.
$R(g)$, $O(g)$, and $S(g)$ are vector, orbital, and spin transformation matrices, respectively, which enter Eq.~\eqref{eq:SRO-sym-transf-rule1}.
$\Pauli_{\mu}$ are Pauli matrices.}
{\renewcommand{\arraystretch}{1.3}
\renewcommand{\tabcolsep}{10pt}
\begin{tabular}{cNNN} \hline\hline
$g$ & $R(g)$ & $O(g)$ & $S(g)$ 
\\ \hline \\[-13pt]
$C_{4z}$ & $\begin{pmatrix} 0 & -1 & 0 \\ 1 & 0 & 0 \\ 0 & 0 & 1 \end{pmatrix}$ & $\begin{pmatrix} 0 & 1 & 0 \\ -1 & 0 & 0 \\ 0 & 0 & -1 \end{pmatrix}$ & $\displaystyle \frac{\Pauli_{0} - \iu \Pauli_{z}}{\sqrt{2}}$ \\[16pt]
$C_{2x}$ & $\begin{pmatrix} 1 & 0 & 0 \\ 0 & -1 & 0 \\ 0 & 0 & -1 \end{pmatrix}$ & $\begin{pmatrix} 1 & 0 & 0 \\ 0 & -1 & 0 \\ 0 & 0 & -1 \end{pmatrix}$ & $\displaystyle - \iu \Pauli_{x}$ \\[16pt]
$C_{2d_+}$ & $\begin{pmatrix} 0 & 1 & 0 \\ 1 & 0 & 0 \\ 0 & 0 & -1 \end{pmatrix}$ & $\begin{pmatrix} 0 & -1 & 0 \\ -1 & 0 & 0 \\ 0 & 0 & 1 \end{pmatrix}$ & $\displaystyle - \iu \frac{\Pauli_{x} + \Pauli_{y}}{\sqrt{2}}$ \\[16pt]
$P$ & $\begin{pmatrix} -1 & 0 & 0 \\ 0 & -1 & 0 \\ 0 & 0 & -1 \end{pmatrix}$ & $\begin{pmatrix} 1 & 0 & 0 \\ 0 & 1 & 0 \\ 0 & 0 & 1 \end{pmatrix}$ & $\displaystyle \Pauli_{0}$
\\ \\[-13pt] \hline\hline
\end{tabular}}
\label{tab:SRO-transf-mat}
\end{table}

Because there is only one ruthenium atom per a body-centered unit cell, the tight-binding Hamiltonian has the form:
\begin{equation}
\Haml_0 = - \sum_{\vb{R}, \vb{\delta}} \psi_{\vb{R} + \vb{\delta}}^{\dag} \mleft[\mathcal{T}_{\vb{\delta}} \otimes \Pauli_0 + \sum_{i=1}^{3} \mathcal{L}_{\vb{\delta}; i} \otimes \Pauli_{i}\mright] \psi_{\vb{R}},
\end{equation}
where $\vb{R}, \vb{\delta}$ go over the body-centered tetragonal lattice.
The Hamiltonian is Hermitian only when
\begin{align}
\mathcal{T}_{- \vb{\delta}} &= \mathcal{T}_{\vb{\delta}}^{\dag}, &
\mathcal{L}_{- \vb{\delta}; i} &= \mathcal{L}_{\vb{\delta}; i}^{\dag}.
\end{align}
It respects point group symmetries, $\SymU^{\dag}(g) \Haml_0 \SymU(g) = \Haml_0$, only when the following relations which constrain and relate different hopping amplitudes hold:
\begin{align}
O^{\dag}(g) \mathcal{T}_{\vb{\delta}} O(g) &= \mathcal{T}_{R(g^{-1}) \vb{\delta}}, \\
O^{\dag}(g) \mathcal{L}_{\vb{\delta}; i} O(g) &= \det R(g) \sum_{j=1}^{3} R_{ij}(g) \mathcal{L}_{R(g^{-1}) \vb{\delta}; j}.
\end{align}
To ensure time-reversal invariance, the matrix elements of
\begin{align}
\mathcal{T}_{\vb{\delta}}^{*} = \mathcal{T}_{\vb{\delta}}
\end{align}
must be real, while those of
\begin{align}
\mathcal{L}_{\vb{\delta}; i}^{*} = - \mathcal{L}_{\vb{\delta}; i}
\end{align}
must be imaginary.

Symmetries that map $\vb{\delta}$ to itself constrain the forms of the hopping amplitudes.
For the nine closest $\vb{\delta}$ of SRO, we find that
\begin{align}
\begin{aligned}
\mathcal{T}_{\vb{0}} &= \begin{pmatrix} \upmu_{\text{1D}} & 0 & 0 \\ 0 & \upmu_{\text{1D}} & 0 \\ 0 & 0 & \upmu_{\text{2D}} \end{pmatrix}, & \hspace{40pt}
\mathcal{T}_{\vb{a}_1} &= \begin{pmatrix} t_1 & 0 & 0 \\ 0 & t_2 & 0 \\ 0 & 0 & \bar{t}_1 \end{pmatrix}, \\
\mathcal{T}_{\vb{a}_1 + \vb{a}_2} &= \begin{pmatrix} t_3 & t_{i1} & 0 \\ t_{i1} & t_3 & 0 \\ 0 & 0 & \bar{t}_2 \end{pmatrix}, & \hspace{40pt}
\mathcal{T}_{\vb{a}_3} &= \begin{pmatrix} t_4 & t_{i2} & t_{j1} \\ t_{i2} & t_4 & t_{j1} \\ t_{j1} & t_{j1} & \bar{t}_3 \end{pmatrix},
\end{aligned}
\end{align}
and
\begin{align}
\begin{aligned}
\mathcal{T}_{2 \vb{a}_1} &= \begin{pmatrix} t_5 & 0 & 0 \\ 0 & t_6 & 0 \\ 0 & 0 & \bar{t}_4 \end{pmatrix}, & \hspace{40pt}
\mathcal{T}_{2 \vb{a}_1 + \vb{a}_2} &= \begin{pmatrix} t_7 & t_{i3} & 0 \\ t_{i3} & t_8 & 0 \\ 0 & 0 & \bar{t}_5 \end{pmatrix}, \\
\mathcal{T}_{2 (\vb{a}_1 + \vb{a}_2)} &= \begin{pmatrix} t_9 & t_{i4} & 0 \\ t_{i4} & t_9 & 0 \\ 0 & 0 & \bar{t}_6 \end{pmatrix}, & \hspace{40pt}
\mathcal{T}_{3 \vb{a}_1} &= \begin{pmatrix} t_{10} & 0 & 0 \\ 0 & t_{11} & 0 \\ 0 & 0 & \bar{t}_7 \end{pmatrix}, \\
\mathcal{T}_{2\vb{a}_3-\vb{a}_1-\vb{a}_2} &= \begin{pmatrix} t_{12} & 0 & 0 \\ 0 & t_{12} & 0 \\ 0 & 0 & \bar{t}_8 \end{pmatrix}.
\end{aligned}
\end{align}
Among these $\mathcal{T}_{\vb{\delta}}$ for the closest $\vb{\delta}$ whose $\mathcal{T}_{\vb{\delta}}$ are thus also largest, only $\mathcal{T}_{\vb{a}_3}$ and $\mathcal{T}_{2\vb{a}_3-\vb{a}_1-\vb{a}_2}$ connect different layers, reflecting the high anisotropy of SRO.
Here $2\vb{a}_3-\vb{a}_1-\vb{a}_2 =  c \vu{e}_z$.
Moreover, it is only through $\mathcal{T}_{\vb{a}_3}$ that the body-centered periodicity of SRO is felt on the level of the one-particle band structure.
The on-site SOC takes the form:
\begin{align}
\mathcal{L}_{\vb{0}; 1} &= \eta_{\perp} \begin{pmatrix} 0 & 0 & 0 \\ 0 & 0 & - \iu \\ 0 & \iu & 0 \end{pmatrix}, &
\mathcal{L}_{\vb{0}; 2} &= \eta_{\perp} \begin{pmatrix} 0 & 0 & \iu \\ 0 & 0 & 0 \\ - \iu & 0 & 0 \end{pmatrix}, &
\mathcal{L}_{\vb{0}; 3} &= \eta_z \begin{pmatrix} 0 & - \iu & 0 \\ \iu & 0 & 0 \\ 0 & 0 & 0 \end{pmatrix}.
\end{align}
Notice that the $\mathcal{L}_{\vb{0}; i}$ have the same form as the orbital angular momentum of vectors, $(L_i)_{jk} = - \iu \LCs_{ijk}$, which is one of the benefits of using cyclical ordering for the $t_{2g}$ orbitals.
Off-site ($\vb{k}$-dependent) spin-orbit coupling we shall not include, although one should keep in mind that some~\cite{Suh2020} have found that it has a large effect on the preferred Cooper pairing, even when small.

In momentum space, the tight-binding Hamiltonian reads:
\begin{align}
\begin{aligned}
H_{\vb{k}} &= - \sum_{\vb{\delta}} \mleft[\mathcal{T}_{\vb{\delta}} \otimes \Pauli_0 + \sum_{i=1}^{3} \mathcal{L}_{\vb{\delta}; i} \otimes \Pauli_{i}\mright] \Elr^{- \iu \vb{k} \vdot \vb{\delta}} \\
&= \begin{pmatrix}
h_{\text{1D}}(\vb{k}) & h_{i}(\vb{k}) & h_{j}(\vb{k}) \\
& h_{\text{1D}}(\vb{p}) & h_{j}(\vb{p}) \\
\cc && h_{\text{2D}}(\vb{k})
\end{pmatrix} \Pauli_0 + \begin{pmatrix}
0 & \iu \eta_z \Pauli_z & - \iu \eta_{\perp} \Pauli_y \\
- \iu \eta_z \Pauli_z & 0 & \iu \eta_{\perp} \Pauli_x \\
\iu \eta_{\perp} \Pauli_y & - \iu \eta_{\perp} \Pauli_x & 0
\end{pmatrix},
\end{aligned} \label{eq:SRO-TBA-Haml}
\end{align}
where
\begin{align}
\vb{k} &= \begin{pmatrix}
k_x \\
k_y \\
k_z
\end{pmatrix}, &
\vb{p} &= R(C_{2d_+}) \vb{k} = \begin{pmatrix}
k_y \\
k_x \\
k_z
\end{pmatrix}, &
\vb{\DimK} &= \begin{pmatrix}
\DimK_x \\
\DimK_y \\
\DimK_z
\end{pmatrix} = \begin{pmatrix}
a k_x \\
a k_y \\
c k_z
\end{pmatrix},
\end{align}
and
\begin{align}
h_{\text{1D}}(\vb{k}) &= - \upmu_{\text{1D}} - 2 t_1 \cos \DimK_x - 2 t_2 \cos \DimK_y - 4 t_3 \cos \DimK_x \cos \DimK_y \\
&\hspace{18pt} - 8 t_4 \cos \tfrac{1}{2} \DimK_x \cos \tfrac{1}{2} \DimK_y \cos \tfrac{1}{2} \DimK_z - 2 t_5 \cos 2 \DimK_x - 2 t_6 \cos 2 \DimK_y \notag \\
&\hspace{24pt} - 4 t_7 \cos 2 \DimK_x \cos \DimK_y - 4 t_8 \cos \DimK_x \cos 2 \DimK_y - 4 t_9 \cos 2 \DimK_x \cos 2 \DimK_y \notag \\
&\hspace{30pt} - 2 t_{10} \cos 3 \DimK_x - 2 t_{11} \cos 3 \DimK_y - 2 t_{12} \cos \DimK_z, \notag \\[8pt]
h_{\text{2D}}(\vb{k}) &= - \upmu_{\text{2D}} - 2 \bar{t}_1 \mleft(\cos \DimK_x + \cos \DimK_y\mright) - 4 \bar{t}_2 \cos \DimK_x \cos \DimK_y \\
&\hspace{18pt} - 8 \bar{t}_3 \cos \tfrac{1}{2} \DimK_x \cos \tfrac{1}{2} \DimK_y \cos \tfrac{1}{2} \DimK_z - 2 \bar{t}_4 \mleft(\cos 2 \DimK_x + \cos 2 \DimK_y\mright) \notag \\
&\hspace{24pt} - 4 \bar{t}_5 \mleft(\cos 2 \DimK_x \cos \DimK_y + \cos \DimK_x \cos 2 \DimK_y\mright) - 4 \bar{t}_6 \cos 2 \DimK_x \cos 2 \DimK_y \notag \\
&\hspace{30pt} - 2 \bar{t}_7 \mleft(\cos 3 \DimK_x + \cos 3 \DimK_y\mright) - 2 \bar{t}_8 \cos \DimK_z, \notag \\[8pt]
h_{i}(\vb{k}) &= 4 t_{i1} \sin \DimK_x \sin \DimK_y + 8 t_{i2} \sin \tfrac{1}{2} \DimK_x \sin \tfrac{1}{2} \DimK_y \cos \tfrac{1}{2} \DimK_z \\
&\hspace{18pt} + 8 t_{i3} \mleft(\cos \DimK_x + \cos \DimK_y\mright) \sin \DimK_x \sin \DimK_y + 4 t_{i4} \sin 2 \DimK_x \sin 2 \DimK_y, \notag \\[8pt]
h_{j}(\vb{k}) &= 8 t_{j1} \sin \tfrac{1}{2} \DimK_x \cos \tfrac{1}{2} \DimK_y \sin \tfrac{1}{2} \DimK_z.
\end{align}
All the tight-binding parameters appearing in the above expressions are real.

Of the six $t_{2g}$ states (including spin degrees of freedom), four are occupied, as depicted on the right of Fig.~\ref{fig:SRO-octahedral-Ru}.
This means that within the model at zero temperature:
\begin{align}
2 \int\limits_{\text{1\textsuperscript{st}BZ}} \frac{\dd[3]{k}}{V_{\text{BZ}}} \sum_{n=1}^{3} \HTh(- \varepsilon_{\vb{k} n}) &= 4,
\end{align}
where $\HTh$ is the Heaviside theta function, $V_{\text{BZ}} = 2 (2 \pi)^3 / (a^2 c)$, and the integral goes over the body-centered first Brillouin zone shown in Fig.~\ref{fig:SRO-BZ}.
The band energies $\varepsilon_{\vb{k} n}$ of $H_{\vb{k}}$ are numbered in ascending ordering,
\begin{align}
\varepsilon_{\vb{k} 1} \equiv \varepsilon_{\vb{k} \alpha} ~<~ \varepsilon_{\vb{k} 2} \equiv \varepsilon_{\vb{k} \gamma} ~<~ \varepsilon_{\vb{k} 3} \equiv \varepsilon_{\vb{k} \beta},
\end{align}
with the lowest one corresponding to the $\alpha$ band, the highest one to the $\beta$ band, and the middle one to the $\gamma$ band.

\begin{table}[p!]
\centering
\captionabove[The values of our tight-binding model parameters according to various references.]{\textbf{The values of our tight-binding model parameters according to various references.}
Parameters not shown vanish for (or have not been considered in) a given reference.
Refs.~\cite{Zabolotnyy2013, Cobo2016, Burganov2016, Roising2024} obtained their parameter values by fitting to ARPES data, while Refs.~\cite{Roising2019, Suh2020} fitted to the ARPES-based tight-binding $17$-band model of Ref.~\cite{Veenstra2014}.
Refs.~\cite{Suh2020, Roising2024} include a few additional terms whose small hopping parameters are not listed or included in our Hamiltonian~\eqref{eq:SRO-TBA-Haml}.
We use the values shown in the Ref.~\cite{Roising2019} column.}
{\renewcommand{\arraystretch}{1.25}
\renewcommand{\tabcolsep}{7.2pt}
\begin{tabular}{cC{49pt}C{49pt}C{49pt}C{49pt}C{49pt}C{49pt}}  \hline\hline
& \multicolumn{5}{c}{value [\si{\milli\electronvolt}]} \\
parameter & Ref.~\cite{Zabolotnyy2013} & Ref.~\cite{Cobo2016} & Ref.~\cite{Burganov2016} & Ref.~\cite{Roising2019} & Ref.~\cite{Suh2020} & Ref.~\cite{Roising2024} \\ \hline
$\upmu_{\text{1D}}$ & $122$ & $109$ & $178$ & $286.9$ & $443.5$ & $209.9$ \\
$t_1$ & $16$ & $9$ & $13$ & $27.8$ & $134.0$ & $49.95$ \\
$t_2$ & $145$ & $88$ & $165$ & $257.8$ & $362.4$ & $281.35$ \\
$t_3$ & & & & $-22.4$ & $44.0$ & $-11.83$ \\
$t_4$ & & & & $13.6$ & $0.023$ & $12.75$ \\
$t_5$ & & & & $3.2$ & $5.73$ & $0$ \\
$t_6$ & & & & $-35.5$ & $1.02$ & $-87.15$ \\
$t_7$ & & & & $0$ & $7.52$ & $0$ \\
$t_8$ & & & & $-4.7$ & $13.93$ & $-12.95$ \\
$t_9$ & & & & $0$ & $0$ & $0$ \\
$t_{10}$ & & & & $0$ & $0$ & $0$ \\
$t_{11}$ & & & & $-2.4$ & $0$ & $-5.50$ \\
$t_{12}$ & & & & $0$ & $- 2.52$ & $0$ \\ \hline
$\upmu_{\text{2D}}$ & $122$ & $109$ & $176$ & $351.9$ & $212.3$ & $284.2$ \\
$\bar{t}_1$ & $81$ & $80$ & $119$ & $356.8$ & $262.4$ & $229.1$ \\
$\bar{t}_2$ & $39$ & $40$ & $49$ & $126.3$ & $43.73$ & $82.5$ \\
$\bar{t}_3$ & & & & $-1.0$ & $- 1.81$ & $-1.54$ \\
$\bar{t}_4$ & $5$ & $5$ & $0$ & $17.0$ & $- 34.23$ & $-3.75$ \\
$\bar{t}_5$ & & & & $22.3$ & $- 8.07$ & $6.325$ \\
$\bar{t}_6$ & & & & $0$ & $0$ & $8.20$ \\
$\bar{t}_7$ & & & & $0$ & $0$ & $1.75$ \\
$\bar{t}_8$ & & & & $0$ & $3.16$ & $0$ \\ \hline
$t_{i1}$ & $0$ & $0$ & $21$ & $-2.0$ & $-16.25$ & $0$ \\
$t_{i2}$ & & & & $7.8$ & $9.98$ & $-9.05$ \\
$t_{i3}$ & & & & $0$ & $-3.94$ & $0$ \\
$t_{i4}$ & & & & $0$ & $0$ & $0$ \\ \hline
$t_{j1}$ & & & & $2.7$ & $8.30$ & $0$ \\
$\eta_{\perp}$ & $32$ & $35$ & $0$ & $59.2$ & $57.39$ & $81.0$ \\
$\eta_z$ & $32$ & $35$ & $0$ & $59.2$ & $57.39$ & $81.0$
\\ \hline\hline
\end{tabular}}
\label{tab:SRO-TBA-params}
\end{table}

In the remainder of the chapter, whenever we make estimates, we shall employ the tight-binding parameter values of Ref.~\cite{Roising2019}, which they found by fitting to the ARPES-based tight-binding $17$-band model of Ref.~\cite{Veenstra2014}.
Their tight-binding parameter values are reproduced in Tab.~\ref{tab:SRO-TBA-params}, where we also compare them to other references.
The hopping amplitudes of Refs.~\cite{Roising2019} and~\cite{Suh2020} are broadly in agreement, as one would expect given that both were fitted to Ref.~\cite{Veenstra2014}.
High-resolution ARPES measurements have recently been carried out~\cite{Tamai2019} and the fit to the corresponding data~\cite{Roising2024} gives parameter values not too different from Refs.~\cite{Roising2019, Suh2020}.
However, the hoppings of all three~\cite{Roising2019, Suh2020, Roising2024} are by a factor of two or so larger than those of Refs.~\cite{Zabolotnyy2013, Cobo2016, Burganov2016}, which are also ARPES-derived; see Tab.~\ref{tab:SRO-TBA-params}.
Although all these models give the correct shapes for the Fermi sheets, find that the $\gamma$ band is responsible for over \SI{50}{\percent} of the normal-state DOS, and predict a roughly \SI{20}{\percent} increase in the DOS at Van Hove strain (see Sec.~\ref{sec:SRO-elasto}), consistent with the entropy data that we later show (Fig.~\ref{fig:SRO-elasto}), the predicted values for the total DOS differ by a factor of two.
The total DOS $\DOSg_F$ is directly related to the Sommerfeld coefficient $\gamma_N = (\pi^2 / 3) R \DOSg_F$, which is experimentally in between \num{38}~\cite{NishiZaki2000, Deguchi2004, Deguchi2004-p2} and \SI[per-mode=symbol]{40}{\milli\joule\per\square\kelvin\per\mol}~\cite{Kittaka2018} for very pure samples ($T_c \geq \SI{1.48}{\kelvin}$); here $R$ is the molar gas constant.
By a rescaling all hopping parameters, one can preserve the Fermi surface shapes and relative DOS contributions, while increasing or decreasing the Fermi velocities to reproduce the \num{16.5} states per \si{\electronvolt} per body-centered tetragonal unit cell seen in experiment.
The main takeaway is that the various estimates that we make might be off by a factor of two, which is still sufficient for our purposes and does not impact the arguments of Sec.~\ref{sec:SRO-analysis-ECE-100} regarding the elastocaloric effect under $\langle 100 \rangle$ pressure in any way.

The dispersion of the $\gamma$ band near the Van Hove line $\mleft(0, \frac{\pi}{a}, k_z\mright)$, that we later provide in Eqs.~\eqref{eq:SRO-saddle-epsilon} and~\eqref{eq:SRO-VHdispexpansion}, was found by diagonalizing the $H_{\vb{k}}$ of Eq.~\eqref{eq:SRO-TBA-Haml} with the parameter values of Ref.~\cite{Roising2019} (Tab.~\ref{tab:SRO-TBA-params}).

\subsection{Elastic coupling and the $\gamma$ band Lifshitz transition} \label{sec:SRO-elastic-tuning}
With the development~\cite{Hicks2014, Steppke2017} of experimental techniques capable of applying uniaxial stress on SRO in a controlled manner accessible to various probes, many such experiments have been performed on SRO in recent years~\cite{Hicks2014, Taniguchi2015, Steppke2017, Barber2019, Sunko2019, Li2021, Li2022, Pustogow2019, Chronister2021, Jerzembeck2022, Jerzembeck2023, Noad2023, Jerzembeck2024}.
Uniaxial stress applied along the $[100]$ directions in particular has been shown to dramatically influence SRO, in part because the $\gamma$ band experiences a Lifshitz transition at $\epsilon_{100} = \SI{-0.44}{\percent} \equiv \epsilon_{\text{VH}}$ strain~\cite{Steppke2017, Barber2019, Sunko2019}.
Here we describe the elastic coupling of SRO and specify how the tight-binding model of the previous section couples to in-plane strain fields.

The strain and stress tensors we shall denote $\epsilon_{ij}$ and $\sigma_{ij}$, respectively, where $i, j \in \{x, y, z\}$ and the associated $\vu{e}_i$ directions are aligned along the principal axes of the crystal.
Given that $\epsilon_{ij} = \epsilon_{ji}$ and $\sigma_{ij} = \sigma_{ji}$ are symmetric, one conventionally defines~\cite{Lupien2002}:
\begin{align}
\begin{aligned}
\epsilon_1 &\equiv \epsilon_{xx}, &\hspace{50pt}
\epsilon_2 &\equiv \epsilon_{yy}, &\hspace{50pt}
\epsilon_3 &\equiv \epsilon_{zz}, \\
\epsilon_4 &\equiv 2 \epsilon_{yz}, &\hspace{50pt}
\epsilon_5 &\equiv 2 \epsilon_{xz}, &\hspace{50pt}
\epsilon_6 &\equiv 2 \epsilon_{xy},
\end{aligned}
\end{align}
and
\begin{align}
\begin{aligned}
\sigma_1 &\equiv \sigma_{xx}, &\hspace{50pt}
\sigma_2 &\equiv \sigma_{yy}, &\hspace{50pt}
\sigma_3 &\equiv \sigma_{zz}, \\
\sigma_4 &\equiv \sigma_{yz}, &\hspace{50pt}
\sigma_5 &\equiv \sigma_{xz}, &\hspace{50pt}
\sigma_6 &\equiv \sigma_{xy}.
\end{aligned}
\end{align}
This is called Voigt notation.
The factors of two ensure that
\begin{align}
\sum_{i,j=x,y,z} \sigma_{ij} \epsilon_{ij} = \sum_{i=1}^{6} \sigma_i \epsilon_i.
\end{align}

\begin{table}[t]
\centering
\captionabove[Elastic constants of \ce{Sr2RuO4} at $T = \SI{4}{\kelvin}$ temperature and their symmetries.]{\textbf{Elastic constants of \ce{Sr2RuO4} at $T = \SI{4}{\kelvin}$ temperature and their symmetries.}
Ref.~\cite{Benhabib2021} measured them using ultrasound echos, while Ref.~\cite{Ghosh2021} employed resonant ultrasound spectroscopy.
The irreducible representations of $D_{4h}$ shown under the symmetry column are defined in Tab.~\ref{tab:D4h-char-tab-again2}.}
{\renewcommand{\arraystretch}{1.3}
\renewcommand{\tabcolsep}{10pt}
\begin{tabular}{cc|cc} \hline\hline
& & \multicolumn{2}{c}{value [\si{\giga\pascal}]} \\
symmetry & parameter & ~Ref.~\cite{Ghosh2021}~ & ~Ref.~\cite{Benhabib2021}~ \\ \hline
$A_{1g}$ & $\tfrac{1}{2} (c_{11} + c_{12})$ & $190.8$ & $182$ \\
$B_{1g}$ & $\tfrac{1}{2} (c_{11} - c_{12})$ &$53.1$ & $51$ \\
$A_{1g}$ & $c_{13}$ & $85.0$ & \\
$A_{1g}$ & $c_{33}$ & $257.2$ & \\
$E_g$ & $c_{44}$ &$69.5$ & $68.2$ \\
$B_{2g}$ & $c_{66}$ & $65.5$ & $64.3$
\\ \hline\hline
\end{tabular}}
\label{tab:SRO-elastic-constants}
\end{table}

When uniaxial stress of magnitude $\sigma_{\vu{n}}$ is applied along the direction $\vu{n}$, this means that the stress tensor equals
\begin{align}
\sigma_{ij} &= \sigma_{\vu{n}} \hat{n}_i \hat{n}_j.
\end{align}
For small enough stresses, the elastic response is linear and given by
\begin{align}
\epsilon_i &= \sum_{j=1}^{6} c_{ij}^{-1} \sigma_j,
\end{align}
where $c_{ij}$ is the elasticity tensor.
For tetragonal systems such as SRO it has the form~\cite{Lupien2002}:
\begin{align}
c = \begin{pmatrix}
c_{11} & c_{12} & c_{13} & & & \\
c_{12} & c_{11} & c_{13} & & & \\
c_{13} & c_{13} & c_{33} & & & \\
& & & c_{44} & & \\
& & & & c_{44} & \\
& & & & & c_{66}
\end{pmatrix},
\end{align}
where the elements not shown vanish.
The inverse of the elasticity tensor is called the elastic compliance tensor and it has the same form:
\begin{align}
\mathcal{E} \defeq c^{-1} = \begin{pmatrix}
\mathcal{E}_{11} & \mathcal{E}_{12} & \mathcal{E}_{13} & & & \\
\mathcal{E}_{12} & \mathcal{E}_{11} & \mathcal{E}_{13} & & & \\
\mathcal{E}_{13} & \mathcal{E}_{13} & \mathcal{E}_{33} & & & \\
& & & \mathcal{E}_{44} & & \\
& & & & \mathcal{E}_{44} & \\
& & & & & \mathcal{E}_{66}
\end{pmatrix},
\end{align}
where
\begin{align}
\begin{aligned}
\mathcal{E}_{11} &= \frac{c_{11} c_{33} - c_{13}^2}{(c_{11} - c_{12}) \mleft[(c_{11} + c_{12}) c_{33} - 2 c_{13}^2\mright]}, &\quad
\mathcal{E}_{12} &= - \nu_{xy} \mathcal{E}_{11}, &\quad
\mathcal{E}_{13} &= - \nu_{xz} \mathcal{E}_{11}, \\
\mathcal{E}_{33} &= \frac{c_{11} + c_{12}}{(c_{11} + c_{12}) c_{33} - 2 c_{13}^2}, &
\mathcal{E}_{44} &= \frac{1}{c_{44}}, &
\mathcal{E}_{66} &= \frac{1}{c_{66}},
\end{aligned}
\end{align}
and
\begin{align}
\nu_{xy} &= \frac{c_{12} c_{33} - c_{13}^2}{c_{11} c_{33} - c_{13}^2}, &
\nu_{xz} &= \frac{(c_{11} - c_{12}) c_{13}}{c_{11} c_{33} - c_{13}^2}.
\end{align}
$\nu_{xy}$ and $\nu_{xz}$ are called Poisson ratios and they quantify the degree to which $x$-axis stress induces strain along $y$ and $z$.
This is true in general: $\sigma_{ij} = \sigma_{\vu{n}} \hat{n}_i \hat{n}_j$ induces a finite
\begin{align}
\epsilon_{\vu{n}} &\defeq \sum_{i,j=x,y,z} \hat{n}_i \epsilon_{ij} \hat{n}_j,
\end{align}
but also strain components orthogonal to $\hat{n}_i$.
Note on nomenclature: when we write $\sigma_{110}$, for instance, this shall mean that $\sigma_{ij} = \sigma_{110} \hat{n}_i \hat{n}_j$ with $\vu{n} = (1,1,0)/\sqrt{2}$.
On the other hand, $\epsilon_{100}$ will entail that $\epsilon_{xx} = \epsilon_{100}$, but also $\epsilon_{yy} = - \nu_{xy} \epsilon_{100}$ and $\epsilon_{zz} = - \nu_{xz} \epsilon_{100}$.
In experiment, one applies stress, not strain, which is why we use different conventions for strain and stress.

The elastic constants of SRO at $T = \SI{4}{\kelvin}$ are given in Tab.~\ref{tab:SRO-elastic-constants}.
In this chapter, we shall use the values of Ref.~\cite{Ghosh2021} throughout for which $c_{11} = 243.9$, $c_{12} = 137.7$, and:
\begin{align}
\begin{aligned}
\mathcal{E}_{11} &= \frac{1}{\SI{160.1}{\giga\pascal}}, &\hspace{50pt}
\mathcal{E}_{33} &= \frac{1}{\SI{219.3}{\giga\pascal}}, \\[4pt]
\nu_{xy} &= 0.5079, &\hspace{50pt}
\nu_{xz} &= 0.1626.
\end{aligned}
\end{align}

\begin{figure}[t]
\centering
\includegraphics[width=\textwidth]{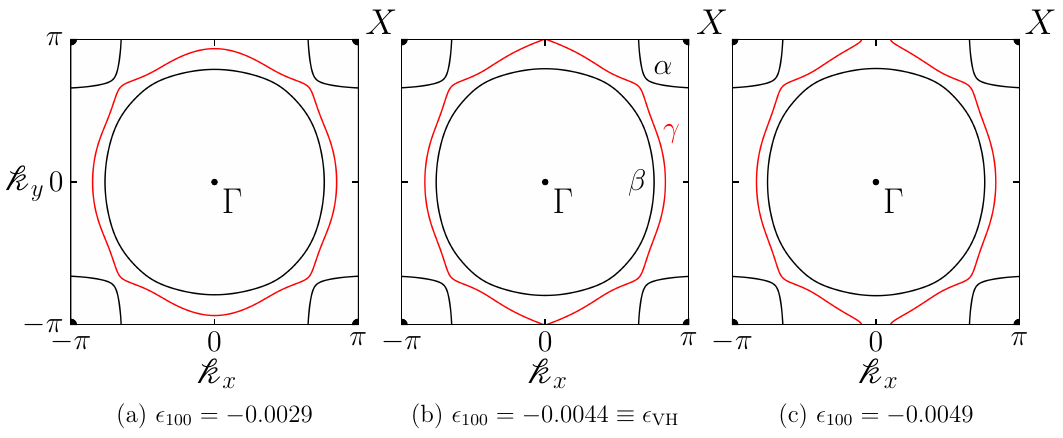}
\captionbelow[Evolution of the three Fermi sheets of \ce{Sr2RuO4} with increasing $\langle 100 \rangle$ uniaxial stress according to our tight-binding model.]{\textbf{Evolution of the three Fermi sheets of \ce{Sr2RuO4} with increasing $\langle 100 \rangle$ uniaxial stress according to our tight-binding model.}
The Hamiltonian is given in Eq.~\eqref{eq:SRO-TBA-Haml}, with the strain coupling specified in Eqs.~\eqref{eq:SRO-TBA-Haml-elasto1}, \eqref{eq:SRO-TBA-Haml-elasto2}, and~\eqref{eq:SRO-TBA-Haml-elasto3}.
The parameter values used are those of Ref.~\cite{Roising2019}, given in Tab.~\ref{tab:SRO-TBA-params}.
The $\epsilon_{xx} = \epsilon_{100}$ strain is given in the caption, while $\epsilon_{yy} = - \nu_{xy} \epsilon_{100}$ with $\nu_{xy} = 0.508$.
The $\gamma$ band which experiences a Lifshitz transition at $\epsilon_{\text{VH}}$ strain is highlighted in red.
$\DimK_x = a k_x$ and $\DimK_y = a k_y$.}
\label{fig:SRO-FS-strain-evolution}
\end{figure}

The strain induced by external stresses modifies the tight-binding model we introduced in the preceding Sec.~\ref{sec:SRO-el-struct}.
The coupling to in-plane strain we adapt from the Supplementary Information of Ref.~\cite{Li2022}.
In particular, given a stress applied along the $\langle 100 \rangle$ direction, this induces an $\epsilon_{xx} = \epsilon_{100}$ strain and a $\epsilon_{yy} = - \nu_{xy} \epsilon_{100}$ strain specified by the Poisson ratio $\nu_{xy} = 0.508$~\cite{Ghosh2021}.
The induced $\epsilon_{zz} = - \nu_{xz} \epsilon_{100}$ strain we neglect because it mainly affects interlayer hopping amplitudes which are small.
These strains modify the tight-binding Hamiltonian~\eqref{eq:SRO-TBA-Haml} by replacing
\begin{align}
h_{\text{1D}}(\vb{k}) &= - \upmu_{\text{1D}} - 2 t_2 \cos \DimK_y + \cdots, \\
h_{\text{1D}}(\vb{p}) &= - \upmu_{\text{1D}} - 2 t_2 \cos \DimK_x + \cdots
\end{align}
with
\begin{align}
h_{\text{1D}}(\vb{k}) &=  - \mleft(1 + \beta_{\upmu} \frac{\epsilon_{xx} + \epsilon_{yy}}{2}\mright) \upmu_{\text{1D}} - 2 t_2 (1 - \beta \epsilon_{yy}) \cos \DimK_y + \cdots, \label{eq:SRO-TBA-Haml-elasto1} \\
h_{\text{1D}}(\vb{p}) &=  - \mleft(1 + \beta_{\upmu} \frac{\epsilon_{xx} + \epsilon_{yy}}{2}\mright) \upmu_{\text{1D}} - 2 t_2 (1 - \beta \epsilon_{xx}) \cos \DimK_x + \cdots, \label{eq:SRO-TBA-Haml-elasto2}
\end{align}
respectively, and
\begin{align}
h_{\text{2D}}(\vb{k}) &= - \upmu_{\text{2D}} - 2 \bar{t}_1 \mleft(\cos \DimK_x + \cos \DimK_y\mright) - 4 \bar{t}_2 \cos \DimK_x \cos \DimK_y + \cdots
\end{align}
with
\begin{align}
\begin{aligned}
h_{\text{2D}}(\vb{k}) &= - \mleft(1 + \alpha_{\upmu} \frac{\epsilon_{xx} + \epsilon_{yy}}{2}\mright) \upmu_{\text{2D}} - 2 \bar{t}_1 (1 - \alpha \epsilon_{xx}) \cos \DimK_x \\
&\hspace{18pt} - 2 \bar{t}_1 (1 - \alpha \epsilon_{yy}) \cos \DimK_y - 4 \bar{t}_2 \mleft(1 - \alpha' \frac{\epsilon_{xx} + \epsilon_{yy}}{2}\mright) \cos \DimK_x \cos \DimK_y + \cdots \, .
\end{aligned} \label{eq:SRO-TBA-Haml-elasto3}
\end{align}
The strain-dependence of the terms not shown has been neglected, as in Ref.~\cite{Li2022}.
We use the values $\alpha = \alpha' = \beta = 15.2$ and $\alpha_{\upmu} = \beta_{\upmu} = 2.7$.
In Ref.~\cite{Li2022} the value $\alpha = \alpha' = 15.62$ with $\beta = \alpha_{\upmu} = \beta_{\upmu} = 0$ was used instead.
Both result in a Lifshitz transition at the $\epsilon_{100} = \SI{-0.44}{\percent} \equiv \epsilon_{\text{VH}}$ strain, as measured in experiment~\cite{Steppke2017, Barber2019, Sunko2019}.
The evolution of the Fermi surfaces with strain is shown in Fig.~\ref{fig:SRO-FS-strain-evolution}.
The Van Hove strain Fermi sheets are also shown in Fig.~\ref{fig:SRO-bands-DOS}
The origin of these changes in the hopping amplitudes are changes in the interatomic distances.
$\upmu$ adjusts to keep the particle number constant.

\subsection{Construction and classification of multiband superconducting states} \label{sec:SRO-SC-construct}
The multiband nature of \ce{Sr2RuO4} (SRO) allows for a richer set of possible superconducting (SC) states than usual~\cite{Ramires2019, Kaba2019, Huang2019}.
Here we detail how the construction of SC states is carried out, following Refs.~\cite{Ramires2019, Kaba2019, Huang2019, Palle2023-ECE}.
The usual pairing wavefunctions (neglecting crystalline periodicity) we listed in Tab.~\ref{tab:SRO-SC-state-options}.

Microscopically, SC is described by a gap matrix $\Delta_{\alpha \beta}(\vb{k})$ that has both momentum and spin-orbit structure.
It is the possibility of a non-trivial orbital structure that sets multiband systems apart from singleband ones.
Thus, for instance, when dealing with even pairings, we cannot simply assume a spin singlet that transforms trivially ($A_{1g}$) under all symmetry operations and equate the irrep of the momentum wavefunction with the irrep of the total gap matrix.
The irrep of the gap matrix is determined by the \emph{product} of the irreps of its momentum and spin-orbit parts, as we explain below (cf.\ Sec.~\ref{sec:cup-ord-param-constr}).
Within the effective tight-binding model of Sec.~\ref{sec:SRO-el-struct}, there are spin-orbit matrices belonging to all possible irreps of $D_{4h}$ for both even- and odd-parity pairings (Tab.~\ref{tab:SRO-Gamma-class}).

Superconductivity emerges from the condensation of an order parameter in the particle-particle sector.
Let us call this complex order-parameter field $\Phi_{a \vb{q}}$.
Its symmetry transformation rules are:
\begin{align}
\SymU^{\dag}(g) \Phi_{a \vb{q}} \SymU(g) &= \sum_{b=1}^{\dim \Phi} \RepM_{ab}^{\Phi}(g) \Phi_{b, R(g^{-1}) \vb{q}}, \\
\Elr^{\iu \vartheta \hat{N}} \Phi_{a \vb{q}} \Elr^{- \iu \vartheta \hat{N}} &= \Elr^{- \iu 2 \vartheta} \Phi_{a \vb{q}}, \\
\SymTR^{-1} \Phi_{a \vb{q}} \SymTR &= p_{\TRop} \Phi_{a, -\vb{q}},
\end{align}
where $a, b \in \{1, \ldots, \dim \Phi\}$ are component indices, $\RepM_{ab}^{\Phi}$ is a representation of the point group, $g$ are point group operations, $\SymTR$ is time reversal (TR), and $\hat{N}$ is the many-body particle-number operator:
\begin{align}
\hat{N} &= \sum_{\vb{k} \alpha} \psi_{\vb{k}, \alpha}^{\dag} \psi_{\vb{k}, \alpha}, &
\Elr^{\iu \vartheta \hat{N}} \psi_{\vb{k}, \alpha} \Elr^{- \iu \vartheta \hat{N}} &= \Elr^{- \iu \vartheta} \psi_{\vb{k}, \alpha}.
\end{align}
Because a simple phase rotation of the complex field $\Phi_{a \vb{q}} \mapsto \iu \Phi_{a \vb{q}}$ changes $p_{\TRop} \mapsto - p_{\TRop}$, we may set $p_{\TRop} = 1$.
Time-reversal symmetry-breaking (TRSB) takes place through the condensation of multiple SC order-parameter components with complex phase differences, as we shall see in Sec.~\ref{sec:SRO-GL-analysis}.

This SC order parameter couples to fermions through
\begin{align}
\Haml_c &= \sum_{a \vb{q}} \Phi_{a, -\vb{q}} \phi_{a \vb{q}} + \Hc,
\end{align}
where
\begin{align}
\phi_{a \vb{q}} &= \frac{1}{\sqrt{\mathcal{N}}} \sum_{\vb{k} \alpha \beta} \psi_{\vb{k}, \alpha}^{\dag} \Delta_{a; \alpha \beta}(\vb{k}, \vb{k}+\vb{q}) \psi_{-\vb{k}-\vb{q}, \beta}^{\dag}.
\end{align}
Here $\alpha, \beta$ are indices which go over both spin and orbital degrees of freedom and $\mathcal{N}$ is the number of unit cells.
Because of the fermionic anticommutation, the SC gap matrix $\Delta_{a; \alpha \beta}(\vb{k}, \vb{p})$ satisfies the particle exchange property:
\begin{align}
\Delta_{a; \beta \alpha}(\vb{k}, \vb{p}) &= - \Delta_{a; \alpha \beta}(-\vb{p}, -\vb{k}).
\end{align}
In principle, we could allow for $\Delta_{a; \alpha \beta}(\vb{k}, \vb{p})$ to have a particle-exchange symmetric part, but once contracted with fermions such a part would vanish identically and not contribute to $\Haml_c$.

Given that the SC order parameter is still a fluctuating field that has not yet condensed, all symmetries must be respected by the coupling $\Haml_c$, i.e.,
\begin{align}
\SymU^{\dag}(g) \Haml_c \SymU(g) &= \Haml_c, &
\Elr^{\iu \vartheta \hat{N}} \Haml_c \Elr^{- \iu \vartheta \hat{N}} &= \Haml_c, &
\SymTR^{-1} \Haml_c \SymTR &= \Haml_c.
\end{align}
Provided that the fermionic symmetry transformation rules have the form~\eqref{eq:SRO-sym-transf-rule1} and~\eqref{eq:SRO-sym-transf-rule2}, for the $\Delta_{a; \alpha \beta}(\vb{k}, \vb{p})$ this implies
\begin{align}
\MatU^{\dag}(g) \Delta_{a}(\vb{k}, \vb{p})  \MatU^{*}(g) &= \sum_{b=1}^{\dim \Phi} \RepM_{ab}^{\Phi}(g) \Delta_{b}\mleft(R(g^{-1}) \vb{k}, R(g^{-1}) \vb{p}\mright), \\
(\one \otimes \iu \Pauli_y)^{\dag} \Delta_{a}(\vb{k}, \vb{p}) (\one \otimes \iu \Pauli_y)^{*} &= \Delta_{a}^{*}(-\vb{k}, -\vb{p}).
\end{align}
Here we have also assumed that $\RepM_{ab}^{\Phi}$ is irreducible, because otherwise additional coefficients specifying the anisotropies could arise, as explained in Sec.~\ref{sec:theory-of-invariants} of Appx.~\ref{app:group_theory}.

After condensation, on the mean-field level the pairing term of the Hamiltonian becomes
\begin{align}
\Haml_{\Delta} &= \sum_{\vb{k} \alpha \beta} \psi_{\vb{k}, \alpha}^{\dag} \Delta_{\alpha \beta}(\vb{k}) \psi_{- \vb{k}, \beta}^{\dag} + \Hc,
\end{align}
where
\begin{align}
\Delta_{\alpha \beta}(\vb{k}) &= \frac{1}{\sqrt{\mathcal{N}}} \sum_a \ev{\Phi_{a, \vb{q}=\vb{0}}} \Delta_{a; \alpha \beta}(\vb{k}, \vb{k}).
\end{align}
Here we only study zero-momentum Cooper pairing, although we should mention that there have been interesting recent experiment on finite-momentum SC in the presence of a magnetic field in SRO~\cite{Kinjo2022}.
Even though we are treating the SC as instantaneous, symmetry-wise these pairing states behave the same as more general even-frequency pairings~\cite{LinderBalatsky2019}.
Hence considering instantaneous SC states is sufficient for the purpose of classifying them.
Odd-frequency pairings we shall not consider, although some~\cite{Komendova2017, Gingras2022} have explored such possibilities.

The SC order parameter has a global $\Ugp(1)$ phase rotation symmetry associated with particle-number conservation.
Because of this, even when $\SymTR^{-1} \Haml_{\Delta} \SymTR$ results in a phase difference compared to $\Haml_{\Delta}$, as long as this phase difference can be absorbed into $\ev{\Phi_{a, \vb{q}=\vb{0}}}$, TR symmetry cannot be said to be broken.
Only when there are unremovable and imaginary relative phase differences between the $\ev{\Phi_{a, \vb{q}=\vb{0}}}$ components does TR symmetry break.

If the pairing were conventional, all point group operations would be preserved and \linebreak $\SymU^{\dag}(g) H_{\Delta} \SymU(g) = H_{\Delta}$ would hold for all $g \in D_{4h}$, giving the constraint $\MatU^{\dag}(g) \Delta(\vb{k})  \MatU^{*}(g) = \Delta\mleft(R(g^{-1}) \vb{k}\mright)$.
Unconventional pairing is classified by the way it breaks this constraint:
\begin{align}
\MatU^{\dag}(g) \Delta_a\mleft(R(g) \vb{k}\mright)  \MatU^{*}(g) &= \sum_{b = 1}^{\dim \zeta} \RepM_{ab}^{\zeta}(g) \Delta_b(\vb{k}). \label{eq:SRO-transf-Delta}
\end{align}
Here, $\zeta$ is an irrep of $D_{4h}$, $a, b$ are indices internal to the irrep, and $\RepM_{ab}^{\zeta}$ are the corresponding matrices.
Only for the 2D irreps $E_{g,u}$ are there multiple possible $\RepM_{ab}^{\zeta}$.
We use the following convention (Eqs.~\eqref{eq:Egu-irrep-mat-conventions1} and~\eqref{eq:Egu-irrep-mat-conventions2}, Sec.~\ref{sec:examples-convention-D4h}):
\begin{align}
\RepM^{E}(C_{4z}) &= \begin{pmatrix}
 0 & -1 \\
 1 & 0 \\
\end{pmatrix}, &
\RepM^{E}(C_{2x}) &= \begin{pmatrix}
 1 & 0 \\
 0 & -1 \\
\end{pmatrix}, &
\RepM^{E}(C_{2d_+}) &= \begin{pmatrix}
 0 & 1 \\
 1 & 0 \\
\end{pmatrix}, \label{eq:SRO-E-rep-rho}
\end{align}
with $\RepM^{E_g}(P) = \Pauli_0$ and $\RepM^{E_u}(P) = - \Pauli_0$.
Fermionic anticommutation and time-reversal symmetry in addition yield:
\begin{align}
(\iu \Pauli_y)^{\dag} \Delta_{a}^{*}(\vb{k}) (\iu \Pauli_y) &= \Delta_{a}(-\vb{k}) = - \Delta_a^{\intercal}(\vb{k}), \label{eq:SRO-exch-prop}
\end{align}
where ${}^{\intercal}$ is transposition and ${}^{*}$ is element-wise complex conjugation.

To construct a $\Delta_a(\vb{k})$ that properly transforms according to Eq.~\eqref{eq:SRO-transf-Delta} and satisfies the constraint~\eqref{eq:SRO-exch-prop}, we need to combine the momentum dependence and spin-orbit structure in just the right way.
This is accomplished~\cite{Ramires2019, Kaba2019, Huang2019, Palle2023-ECE} by first separately classifying pairing wavefunctions and spin-orbit matrices (Tabs.~\ref{tab:SRO-d-func-class} and~\ref{tab:SRO-Gamma-class}), and then combining them according to a set of rules (Tab.~\ref{tab:D4h-irrep-prod-tab}, Appx.~\ref{app:group_theory}).
Let us emphasize that the SC order parameter $\Phi_a$ that enters Ginzburg-Landau theory belongs to the irrep determined by the total SC gap $\Delta_a(\vb{k})$ according to Eq.~\eqref{eq:SRO-transf-Delta}, and not to the irreps of its momentum or spin-orbit parts.

Pairing wavefunctions $f_a(\vb{k})$ are classified according to:
\begin{align}
f_a\mleft(R(g) \vb{k}\mright) &= \sum_{b = 1}^{\dim \zeta} \RepM_{ab}^{\zeta}(g) f_b(\vb{k}). \label{eq:SRO-transf-d}
\end{align}
All $f_a(\vb{k})$ should be made periodic, just like $\Delta_a(\vb{k})$.
If we call $\DimK_x = a k_x$, $\DimK_y = a k_y$, and $\DimK_z = c k_z$, the primitive translations of a body-centered tetragonal lattice map $(\DimK_x, \DimK_y, \DimK_z)$ to $(\DimK_x + 2 \pi, \DimK_y, \DimK_z - 2 \pi)$, $(\DimK_x, \DimK_y + 2 \pi, \DimK_z - 2 \pi)$, and $(\DimK_x, \DimK_y, \DimK_z + 4 \pi)$.
As discussed in Sec.~\ref{sec:SRO-cryst-struct}, some functions can be body-centered-tetragonal periodic, but not simple-tetragonal periodic.
Conventionally, we choose $f_a(\vb{k})$ to always be real,
\begin{align}
f_a^{*}(\vb{k}) &= f_a(\vb{k}).
\end{align}
Examples of pairing wavefunctions are provided in Tab.~\ref{tab:SRO-d-func-class}.
Using the irrep product table~\ref{tab:D4h-irrep-prod-tab}, from these lowest-order lattice harmonics one can systematically construct higher-order ones, as explained in Sec.~\ref{sec:multid-irrep-product}.

When it comes to spin-orbit matrices which we shall denote $\Gamma_a$, notice that $\MatU(P) = \one$ leaves the matrix part of Eq.~\eqref{eq:SRO-transf-Delta} invariant.
This means that all spin-orbit matrices are even.
Odd spin-orbit matrices arise when the conduction bands derive from orbitals of opposite parities, as in the case of cuprates (Sec.~\ref{sec:cuprate-3band-model}) where we indeed found odd-parity orbital matrices (Tab.~\ref{tab:orbital-Lambda-mats}, Sec.~\ref{sec:orbital-Lambda-mats}).
Spin-orbit matrices we classify according to:
\begin{align}
\MatU^{\dag}(g) \Gamma_a \MatU^{*}(g) &= \sum_{b = 1}^{\dim \zeta} \RepM_{ab}^{\zeta}(g) \Gamma_b, \label{eq:SRO-transf-Gamma}
\end{align}
where $\MatU(g) = O(g) \otimes S(g)$ with the $O(g)$ and $S(g)$ provided in Tab.~\ref{tab:SRO-transf-mat}.
Given the transposition appearing in the constraint~\eqref{eq:SRO-exch-prop}, it is natural to further categorize $\Gamma_a$ according to (anti-)symmetry:
\begin{align}
\Gamma_a^{\intercal} &= p_{\Gamma} \Gamma_a, \label{eq:SRO-Gamma-sym}
\end{align}
where $p_{\Gamma} = \pm 1$.
The corresponding irreps we shall denote $\zeta^a$ for $p_{\Gamma} = -1$ and $\zeta^s$ for $p_{\Gamma} = +1$.
We shall also ensure TR invariance:
\begin{align}
(\one \otimes \iu \Pauli_{y})^{\dag} \Gamma_a^{*} (\one \otimes \iu \Pauli_{y}) &= - \Gamma_a^{\intercal}, \label{eq:SRO-Gamma-TRI}
\end{align}
where we have added a minus and a transposition so that we are comparing matrices at the same momentum in Eq.~\eqref{eq:SRO-exch-prop}.

Conventionally~\cite{Balian1963}, the spin-orbit matrices are written in the following way:
\begin{align}
\Gamma_a &= \tilde{\Gamma}_a (\one \otimes \iu \Pauli_y).
\end{align}
Notice that all $O(g)$ are real in Tab.~\ref{tab:SRO-transf-mat} so $O^{*}(g) = O(g)$ in Eq.~\eqref{eq:SRO-transf-Gamma}.
Regarding the spin rotations, their generators $\vb{S} = \tfrac{1}{2} \vb{\Pauli}$ are TR-odd, $(\iu \Pauli_y)^{\dag} \vb{S}^{*} (\iu \Pauli_y) = - \vb{S}$, hence $(\iu \Pauli_y) S^{*}(g) = S(g) (\iu \Pauli_y)$.
Consequently, the transformation rules~\eqref{eq:SRO-transf-Gamma} for $\tilde{\Gamma}_a$ take the form:
\begin{align}
\MatU^{\dag}(g) \tilde{\Gamma}_a \MatU(g) &= \sum_{b = 1}^{\dim \zeta} \RepM_{ab}^{\zeta}(g) \tilde{\Gamma}_b. \label{eq:SRO-transf-Gamma-v2}
\end{align}
As the basis of the orbital part of $\Gamma_a$, we use the following Gell-Mann matrices $\Lambda_{\mu}$ (see also \nameref{app:conventions}):
\begin{align}
\begin{aligned}
\Lambda_0 &= \begin{pmatrix}
1 & 0 & 0 \\
0 & 1 & 0 \\
0 & 0 & 0
\end{pmatrix}, &\qquad \Lambda_1 &= \begin{pmatrix}
0 & 1 & 0 \\
1 & 0 & 0 \\
0 & 0 & 0
\end{pmatrix}, &\qquad \Lambda_2 &= \begin{pmatrix}
0 & -\iu & 0 \\
\iu & 0 & 0 \\
0 & 0 & 0
\end{pmatrix}, \\
\Lambda_3 &= \begin{pmatrix}
1 & 0 & 0 \\
0 & -1 & 0 \\
0 & 0 & 0
\end{pmatrix}, &
\Lambda_4 &= \begin{pmatrix}
0 & 0 & 0 \\
0 & 0 & 0 \\
0 & 0 & \sqrt{2}
\end{pmatrix}, &
\Lambda_5 &= \begin{pmatrix}
0 & 0 & 1 \\
0 & 0 & 0 \\
1 & 0 & 0
\end{pmatrix}, \\
\Lambda_6 &= \begin{pmatrix}
0 & 0 & -\iu \\
0 & 0 & 0 \\
\iu & 0 & 0
\end{pmatrix}, & \Lambda_7 &= \begin{pmatrix}
0 & 0 & 0 \\
0 & 0 & 1 \\
0 & 1 & 0
\end{pmatrix}, & \Lambda_8 &= \begin{pmatrix}
0 & 0 & 0 \\
0 & 0 & -\iu \\
0 & \iu & 0
\end{pmatrix}.
\end{aligned} \label{eq:SRO-GM-mat-list}
\end{align}
They are normalized so that $\tr \Lambda_{A} \Lambda_{B} = 2 \Kd_{AB}$.
The spin-orbit matrices we write in terms of these:
\begin{align}
\Gamma_a &\sim \sum_{A \mu} \Lambda_{A} \otimes \Pauli_{\mu} (\iu \Pauli_{y}).
\end{align}
Given that $\Lambda_{A}^{\dag} = \Lambda_{A}$ for all $A \in \{0, \ldots, 8\}$, written thusly $\Gamma_a$ automatically satisfy time-reversal invariance~\eqref{eq:SRO-Gamma-TRI}.
In three-band systems, there are in total $3^2 \times 4 = 36$ possible $\Gamma_a$, of which $15$ are antisymmetric and $21$ are symmetric.
The categorization of all $\Lambda_{A} \otimes \Pauli_{\mu} (\iu \Pauli_{y})$ is given in Tab.~\ref{tab:SRO-Gamma-class}.

\begin{table}[t]
\centering
\captionabove[A sample of possible pairing wavefunctions $f_a(\vb{k})$, categorized according to the transformation rule of Eq.~\eqref{eq:SRO-transf-d}.]{\textbf{A sample of possible pairing wavefunctions $f_a(\vb{k})$, categorized according to the transformation rule of Eq.~\eqref{eq:SRO-transf-d}.}
The irrep subscripts $g$ and $u$ mean even and odd under parity, respectively.
The two-component $\mleft(f_1(\vb{k}) | f_2(\vb{k})\mright)$ transform according to the $\RepM^{E}(g)$ matrices given in Eq.~\eqref{eq:SRO-E-rep-rho}.
$\vb{k} = (k_x, k_y, k_z)$ and $\DimK_x = a k_x$, $\DimK_y = a k_y$, $\DimK_z = c k_z$.
Highlighted red are those wavefunctions that are periodic under body-centered-tetragonal translations, but not under simple-tetragonal translations (Sec.~\ref{sec:SRO-cryst-struct}).}
{\renewcommand{\arraystretch}{1.3}
\renewcommand{\tabcolsep}{10pt}
\begin{tabular}{c|c} \hline\hline
irrep $\zeta$ & pairing wavefunction $f_a(\vb{k})$ 
\\ \thickhline \\[-16pt]
$A_{1g}$ & $1$,~~ $\cos \DimK_x + \cos \DimK_y$,~~ $\cos \DimK_z$,~~ $\cos \DimK_x \cos \DimK_y$
\\[4pt] \hline \\[-16pt]
$A_{2g}$ & $\mleft(\cos \DimK_x - \cos \DimK_y\mright) \sin \DimK_x \sin \DimK_y$
\\[4pt] \hline \\[-16pt]
$B_{1g}$ & $\cos \DimK_x - \cos \DimK_y$
\\[4pt] \hline \\[-16pt]
$B_{2g}$ & $\sin \DimK_x \sin \DimK_y$,~~ $\textcolor{red}{\sin \tfrac{1}{2} \DimK_x \sin \tfrac{1}{2} \DimK_y \cos \tfrac{1}{2} \DimK_z}$
\\[4pt] \hline \\[-12pt]
$E_g$ & $\begin{pmatrix}
\sin \DimK_y \sin \DimK_z \\
- \sin \DimK_x \sin \DimK_z
\end{pmatrix}$,~~ $\textcolor{red}{\begin{pmatrix}
\cos \tfrac{1}{2} \DimK_x \sin \tfrac{1}{2} \DimK_y \sin\tfrac{1}{2} \DimK_z \\
- \sin\tfrac{1}{2} \DimK_x \cos\tfrac{1}{2} \DimK_y \sin\tfrac{1}{2} \DimK_z
\end{pmatrix}}$
\\[15pt] \thickhline \\[-16pt]
$A_{1u}$ & $\textcolor{red}{\mleft(\cos \DimK_x - \cos \DimK_y\mright) \sin \tfrac{1}{2} \DimK_x \sin \tfrac{1}{2} \DimK_y \sin \tfrac{1}{2} \DimK_z}$
\\[4pt] \hline \\[-16pt]
$A_{2u}$ & $\sin \DimK_z$,~~ $\textcolor{red}{\cos \tfrac{1}{2} \DimK_x \cos \tfrac{1}{2} \DimK_y \sin \tfrac{1}{2} \DimK_z}$
\\[4pt] \hline \\[-16pt]
$B_{1u}$ & $\textcolor{red}{\sin \tfrac{1}{2} \DimK_x \sin \tfrac{1}{2} \DimK_y \sin \tfrac{1}{2} \DimK_z}$
\\[4pt] \hline \\[-16pt]
$B_{2u}$ & $\mleft(\cos \DimK_x - \cos \DimK_y\mright) \sin \DimK_z$
\\[4pt] \hline \\[-12pt]
$E_u$ & $\begin{pmatrix}
\sin \DimK_x \\
\sin \DimK_y
\end{pmatrix}$,~~ $\begin{pmatrix}
(\cos \DimK_x - \cos \DimK_y) \sin \DimK_x \\
(\cos \DimK_y - \cos \DimK_x) \sin \DimK_y
\end{pmatrix}$,~~ $\textcolor{red}{\begin{pmatrix}
\sin \tfrac{1}{2} \DimK_x \cos \tfrac{1}{2} \DimK_y \cos \tfrac{1}{2} \DimK_z \\
\cos \tfrac{1}{2} \DimK_x \sin \tfrac{1}{2} \DimK_y \cos \tfrac{1}{2} \DimK_z
\end{pmatrix}}$
\\[15pt] \hline\hline
\end{tabular}}
\label{tab:SRO-d-func-class}
\end{table}

\begin{table}[t]
\centering
\captionabove[Spin-orbit matrices $\Gamma_a = \tilde{\Gamma}_a (\one \otimes \iu \Pauli_y)$ categorized according to the transformation rule~\eqref{eq:SRO-transf-Gamma} and (anti-)symmetry~\eqref{eq:SRO-Gamma-sym}.]{\textbf{Spin-orbit matrices $\Gamma_a = \tilde{\Gamma}_a (\one \otimes \iu \Pauli_y)$ categorized according to the transformation rule~\eqref{eq:SRO-transf-Gamma} and (anti-)symmetry~\eqref{eq:SRO-Gamma-sym}.}
Only the $\tilde{\Gamma}_a$ parts are shown.
The irrep subscript $g$ means even under parity.
The irrep superscript $s$ ($a$) indicates that $p_{\Gamma} = +1$ ($-1$) in Eq.~\eqref{eq:SRO-Gamma-sym}, i.e., that the corresponding $\Gamma_a$ matrices are (anti-)symmetric under transposition.
The matrices are written in terms of the Gell-Mann matrices $\Lambda_{A}$ listed in Eq.~\eqref{eq:SRO-GM-mat-list} and Pauli matrices $\Pauli_{\mu}$.
The two-component $(\Gamma_1 | \Gamma_2)$ transform according to the $\RepM^{E}(g)$ given in Eq.~\eqref{eq:SRO-E-rep-rho}.
Highlighted blue are the singlet and triplet pairings with trivial orbital structures, typical of one-band Cooper pairing.
Underlined are purely orbital $\tilde{\Gamma}_a$.}
{\renewcommand{\arraystretch}{1.3}
\renewcommand{\tabcolsep}{10pt}
\begin{tabular}{c|c} \hline\hline \\[-16pt]
irrep $\zeta$ & spin-orbit matrix $\tilde{\Gamma}_a = \Gamma_a (\iu \Pauli_y)^{\dag}$ 
\\ \thickhline \\[-16pt]
$A_{1g}^{a}$ & $\underline{\textcolor{blue}{\Lambda_0 \Pauli_0}}$,~~ $\Lambda_2 \Pauli_z$,~~ $\underline{\textcolor{blue}{\Lambda_4 \Pauli_0}}$,~~ $\Lambda_6 \Pauli_y - \Lambda_8 \Pauli_x$
\\[4pt] \hline \\[-16pt]
$A_{2g}^{a}$ & $\Lambda_6 \Pauli_x + \Lambda_8 \Pauli_y$
\\[4pt] \hline \\[-16pt]
$B_{1g}^{a}$ & $\underline{\Lambda_3 \Pauli_0}$,~~ $\Lambda_6 \Pauli_y + \Lambda_8 \Pauli_x$
\\[4pt] \hline \\[-16pt]
$B_{2g}^{a}$ & $\underline{\Lambda_1 \Pauli_0}$,~~ $\Lambda_6 \Pauli_x - \Lambda_8 \Pauli_y$
\\[4pt] \hline \\[-12pt]
$E_g^{a}$ & $\begin{pmatrix}
\Lambda_2 \Pauli_y \\
- \Lambda_2 \Pauli_x
\end{pmatrix}$,~~ $\begin{pmatrix}
\underline{\Lambda_7 \Pauli_0} \\
- \underline{\Lambda_5 \Pauli_0}
\end{pmatrix}$,~~ $\begin{pmatrix}
\Lambda_6 \Pauli_z \\
\Lambda_8 \Pauli_z
\end{pmatrix}$
\\[15pt] \thickhline \\[-16pt]
$A_{1g}^{s}$ & $\Lambda_5 \Pauli_y - \Lambda_7 \Pauli_x$
\\[4pt] \hline \\[-16pt]
$A_{2g}^{s}$ & $\textcolor{blue}{\Lambda_0 \Pauli_z}$,~~ $\underline{\Lambda_2 \Pauli_0}$,~~ $\textcolor{blue}{\Lambda_4 \Pauli_z}$,~~ $\Lambda_5 \Pauli_x + \Lambda_7 \Pauli_y$
\\[4pt] \hline \\[-16pt]
$B_{1g}^{s}$ & $\Lambda_1 \Pauli_z$,~~ $\Lambda_5 \Pauli_y + \Lambda_7 \Pauli_x$
\\[4pt] \hline \\[-16pt]
$B_{2g}^{s}$ & $\Lambda_3 \Pauli_z$,~~ $\Lambda_5 \Pauli_x - \Lambda_7 \Pauli_y$
\\[4pt] \hline \\[-12pt]
$E_g^{s}$ & $\textcolor{blue}{\begin{pmatrix}
\Lambda_0 \Pauli_x \\
\Lambda_0 \Pauli_y
\end{pmatrix}}$,~~ $\begin{pmatrix}
\Lambda_1 \Pauli_y \\
\Lambda_1 \Pauli_x
\end{pmatrix}$,~~ $\begin{pmatrix}
\Lambda_3 \Pauli_x \\
- \Lambda_3 \Pauli_y
\end{pmatrix}$,~~ $\textcolor{blue}{\begin{pmatrix}
\Lambda_4 \Pauli_x \\
\Lambda_4 \Pauli_y
\end{pmatrix}}$,~~ $\begin{pmatrix}
\Lambda_5 \Pauli_z \\
\Lambda_7 \Pauli_z
\end{pmatrix}$,~~ $\begin{pmatrix}
\underline{\Lambda_8 \Pauli_0} \\
- \underline{\Lambda_6 \Pauli_0}
\end{pmatrix}$
\\[15pt] \hline\hline
\end{tabular}}
\label{tab:SRO-Gamma-class}
\end{table}

SC gap matrices $\Delta(\vb{k})$ are constructed by combining pairing wavefunctions $f_a(\vb{k})$ and spin-orbit matrices $\Gamma_a$.
Because of the exchange property $\Delta_{a}(-\vb{k}) = - \Delta_a^{\intercal}(\vb{k})$ [Eq.~\eqref{eq:SRO-exch-prop}], we may only combine even $f_a(\vb{k})$ with antisymmetric $\Gamma_a$, or odd $f_a(\vb{k})$ with symmetric $\Gamma_a$.
Now consider a $f_a(\vb{k}) \in \zeta_{f}$ and $\Gamma_a \in \zeta_{\Gamma}$, where $\zeta_{f}$ and $\zeta_{\Gamma}$ are irreps of the $D_{4h}$ point group.
The composite object
\begin{align}
\Delta_{ab}(\vb{k}) \equiv \Gamma_a f_b(\vb{k})
\end{align}
then transforms according to the direct product representation $\zeta_{\Gamma} \otimes \zeta_{f}$:
\begin{align}
\MatU^{\dag}(g) \Delta_{ab}\mleft(R(g) \vb{k}\mright)  \MatU^{*}(g) &= \sum_{a' = 1}^{\dim \zeta_{\Gamma}} \sum_{b' = 1}^{\dim \zeta_{f}} \RepM_{aa'}^{\zeta_{\Gamma}}(g) \RepM_{bb'}^{\zeta_{f}}(g) \Delta_{a'b'}(\vb{k}).
\end{align}
Since we want to construct SC gap matrices that transform according to \emph{irreducible} representations [Eq.~\eqref{eq:SRO-transf-Delta}], we decomposed $\Delta_{ab}(\vb{k})$ into irreducible parts with the help of Tab.~\ref{tab:D4h-irrep-prod-tab}.
This is explained in more detail in Sec.~\ref{sec:multid-irrep-product} of Appx.~\ref{app:group_theory}.
The most general $\Delta_a(\vb{k})$ belonging to irrep $\zeta_{\Delta}$ is then given by a sum over all possible $f_a(\vb{k}) \in \zeta_{f}$ and $\Gamma_a \in \zeta_{\Gamma}$ such that $\zeta_{\Delta} \in \zeta_{\Gamma} \otimes \zeta_{f}$.

For example, let us construct SC gap matrices belonging to $B_{1g}$.
In Tab.~\ref{tab:D4h-irrep-prod-tab} every row has a $B_1$, meaning antisymmetric $\Gamma_a$ belonging to every irrep could be used.
Combining $\Lambda_0  \Pauli_0 (\iu \Pauli_{y}) \in A_{1g}^{a}$ and $\cos \DimK_x - \cos \DimK_y \in B_{1g}$ gives a $\Delta(\vb{k}) = \Lambda_0 (\iu \Pauli_{y}) \mleft(\cos \DimK_x - \cos \DimK_y\mright) \in B_{1g}$, but so do many others:
\begin{align}
\begin{aligned}
A_{1g}^{a} \otimes B_{1g}\colon &\quad \mleft(\Lambda_6 \Pauli_{y} - \Lambda_8 \Pauli_{x}\mright) (\iu \Pauli_{y}) \mleft(\cos \DimK_x - \cos \DimK_y\mright), \\
A_{2g}^{a} \otimes B_{2g}\colon &\quad \mleft(\Lambda_6 \Pauli_{x} + \Lambda_8 \Pauli_{y}\mright) (\iu \Pauli_{y}) \sin \DimK_x \sin \DimK_y, \\
B_{1g}^{a} \otimes A_{1g}\colon &\quad \Lambda_3 (\iu \Pauli_{y}) \cos \DimK_x \cos \DimK_y, \\
B_{2g}^{a} \otimes A_{2g}\colon &\quad \Lambda_1 (\iu \Pauli_{y}) \mleft(\cos \DimK_x - \cos \DimK_y\mright) \sin \DimK_x \sin \DimK_y, \\
E_g^{a} \otimes E_g\colon &\quad \Lambda_2 \mleft(\Pauli_{x} \sin \DimK_x - \Pauli_{y} \sin \DimK_y\mright) (\iu \Pauli_{y}) \sin \DimK_z,
\end{aligned}
\end{align}
etc.
The most general $\Delta(\vb{k}) \in B_{1g}$ is a linear superposition of all of these options.
The construction for other irreps proceeds analogously.
Refer to Sec.~\ref{sec:cup-ord-param-constr} for a discussion in a formally similar context.

Having constructed the $6 \times 6$ gap matrices $\Delta_{\alpha \beta}(\vb{k})$ which describe SC on the mean-field level, let us compare them to the usual one-band case.
In the one-band case, $\Delta(\vb{k}) = d_0(\vb{k}) \iu \Pauli_y$ for even-parity singlet states and $\Delta(\vb{k}) = \vb{d}(\vb{k}) \vdot \vb{\Pauli} (\iu \Pauli_y)$ for odd-parity triplet states, with the $d_{\mu}(\vb{k})$ as given in Tab.~\ref{tab:SRO-SC-state-options}, for instance.
The analogues of such states are highlighted blue in Tab.~\ref{tab:SRO-Gamma-class}.
In the multiband case, this continues to be true in the sense that, once $\Delta_{\alpha \beta}(\vb{k})$ is projected onto the bands, it is a pseudospin singlet or triplet, depending on the parity.
To be more precise, let us introduce the band-projected SC gap matrix:
\begin{align}
\mleft[\mathscr{d}_{a}(\vb{k}_n)\mright]_{ss'} &\defeq u_{\vb{k} n s}^{\dag} \Delta_{a}(\vb{k}) u_{- \vb{k} n s'}^{*} = \sum_{\mu} \mathscr{d}_{a}^{\mu}(\vb{k}_n) \mleft[\Pauli_{\mu} (\iu \Pauli_{y})\mright]_{ss'},
\end{align}
where the Pauli matrices act in pseudospin (Kramers' degeneracy) space spanned by $s, s' \in \{\uparrow, \downarrow\}$ and $H_{\vb{k}} u_{\vb{k} n s} = \varepsilon_{\vb{k} n} u_{\vb{k} n s}$ diagonalize the Hamiltonian of Eq.~\eqref{eq:SRO-TBA-Haml}.
Since all three $t_{2g}$ orbitals are even, $\MatU(P) = \one$ and we may always locally choose a gauge in which $u_{- \vb{k} n s} = u_{\vb{k} n s}$ so that
\begin{align}
\mathscr{d}_{a}(\vb{k}_n) = p_P \mathscr{d}_{a}(- \vb{k}_n) = - p_P \mathscr{d}_{a}^{\intercal}(\vb{k}_n).
\end{align}
Hence $\mathscr{d}_{a}(\vb{k}_n)$ is a pseudospin-singlet with only the $\mu = 0$ component for even-parity $\Delta(\vb{k})$ ($p_P = +1$), and a pseudospin-triplet with $\mu \in \{x, y, z\}$ components for odd-parity $\Delta(\vb{k})$ ($p_P = -1$).

However, in multiband systems interband coupling is also possible, although it is not expected to be important in a Fermi liquid such as SRO, where SC is essentially a Fermi surface phenomenon.
More interesting is the possibility of having non-trivial orbital structures.
Once projected onto the Fermi surface(s), such orbital structure is expected to modulate the $\mathscr{d}_{a}(\vb{k}_n)$ in the same way a pairing wavefunction $f_a(\vb{k})$ belonging to the same irrep would.
So the way the Fermi surface gets gapped is not qualitatively different.
However, many other quantities (tunneling, spin response, etc.) depend more sensitively on the local spin-orbit structure of the SC gap matrix.
As an extreme example, consider the following state which is spin-singlet, but has odd parity: $\Delta(\vb{k}) = \mleft(\Lambda_8 \sin \DimK_x  - \Lambda_6 \sin \DimK_y\mright) (\iu \Pauli_y) \in A_{1u}$.
Such states are constructed from the $(\Lambda_8 | - \Lambda_6) \in E_g^s$ and $\Lambda_2 \in A_{2g}^s$ Gell-Mann matrices which represent in-plane and $z$-axis orbital angular momentum operators (Tab.~\ref{tab:SRO-Gamma-class}), respectively, and they can be understood as orbital triplets.
Because external probes couple to the physical spin, and not the pseudospin, in this regard such states are expected to behave similarly to even-parity spin-singlet states.
It is worth remarking that these states require spin-orbit coupling if their $\mathscr{d}_{a}(\vb{k}_n)$ are to be finite, because otherwise oddness of orbital angular momentum under $P \TRop$ implies that $\mathscr{d}_{a}(\vb{k}_n) = 0$, as one may show using arguments similar to those of Sec~\ref{sec:Cp-channel-gen-sym-constr} or~\ref{sec:el-dip-sc-itinerant-dipoles}.

\section{Constraints from elastocaloric measurements under~$[100]$~uniaxial stress} \label{sec:SRO-analysis-ECE-100}
As already mentioned in Sec.~\ref{sec:SRO-lit-review}, compelling evidence on the gap structure of \ce{Sr2RuO4} (SRO) has recently emerged from measurements performed under uniaxial pressure.
When $\langle 100 \rangle$ uniaxial pressure is applied on SRO, its superconductivity (SC) is drastically enhanced~\cite{Hicks2014, Taniguchi2015, Steppke2017, Barber2019, Jerzembeck2023}, with $T_c$ increasing from \SI{1.5}{\kelvin} to a maximal \SI{3.5}{\kelvin} before decaying again.
The most likely cause of this enhancement is the Lifshitz transition that occurs at $\epsilon_{100} = \SI{-0.44}{\percent} \equiv \epsilon_{\text{VH}}$ strain~\cite{Steppke2017, Barber2019, Sunko2019} which is accompanied by an increase in the density of states (DOS).
The DOS peaks at $\epsilon_{\text{VH}}$, as does the normal-state entropy~\cite{Li2022}.
In the SC state, however, the entropy becomes a \emph{minimum} at $\epsilon_{\text{VH}}$, as directly measured by the elastocaloric effect~\cite{Li2022}.
As we shown in this section, which is based on Ref.~\cite{Palle2023-ECE}, this is only possible if SRO's SC does not have vertical line nodes at the Van Hove lines that induce the DOS peak at $\epsilon_{\text{VH}}$.
This is a strong constraint on possible pairing states, one whose implications we explore in the current section which reuses much of the text from Ref.~\cite{Palle2023-ECE}.
The final piece of the argument is that these properties of strained SRO carry over to the unstrained SC state, which is supported by the absence of any signatures of a bulk SC state change at finite strain in the heat capacity~\cite{Li2021}, elastocaloric effect~\cite{Li2022}, or NMR Knight shift~\cite{Pustogow2019, Chronister2021}.

The main result of Ref.~\cite{Palle2023-ECE} is that, among even pairings, only $s$-wave ($A_{1g}$), $d_{x^2 - y^2}$-wave ($B_{1g}$), and body-centered periodic $(d_{yz}| - d_{xz})$-wave ($E_g$) pairings gap the Van Hove lines.
Thus the SC state must include admixtures from at least one of these three pairings to be consistent with the elastocaloric experiment of Ref.~\cite{Li2022}.
The logic of our argument does not put any constraints on the subleading channels.
For instance, almost degenerate states like $s' + \iu \, d_{x^2-y^2}$~\cite{Romer2019, Romer2020}, $d_{x^2 - y^2} + \iu \, g_{xy(x^2 - y^2)}$~\cite{Kivelson2020, Willa2021, Yuan2021, Sheng2022, Yuan2023}, and $s' + \iu \, d_{xy}$~\cite{Clepkens2021, Romer2021} are consistent with a dominant $d_{x^2 - y^2}$-wave or $s$-wave state; here $s'$ stands for extended (nodal) $s$-wave states.
Among odd-parity pairings, all irreps can gap the Van Hove lines.
However, $A_{2u}$ and $B_{2u}$ pairings must be made of body-centered periodic wavefunctions, and for the rest we find non-trivial constraints on the orientations of their Balian-Werthamer $\vb{d}$-vectors~\cite{Balian1963}.

This section largely follows the structure of the article itself~\cite{Palle2023-ECE}.
It is organized as follows.
In Sec.~\ref{sec:SRO-elasto}, we explain what has been measured in the elastocaloric experiment~\cite{Li2022} and why these measurements forbid vertical line nodes at the Van Hove lines.
The precise location of the Van Hove lines is the subject of Sec.~\ref{sec:SRO-vH-line}. 
The main results are presented in Sec.~\ref{sec:SRO-behavior-vH}: how the momentum and spin-orbit parts of the SC gap behave near the Van Hove lines and which SC states are excluded by the elastocaloric measurements.
Tab.~\ref{tab:SRO-main-result} is our main result.
In the last Sec.~\ref{sec:SRO-ECE-art-discussion}, we discuss our results.

\subsection{Elastocaloric measurements and the gapping of Van Hove lines} \label{sec:SRO-elasto}
The elastocaloric effect describes the change in the temperature that accompanies an adiabatic change in the strain $\epsilon_{ij}$.
By measuring it, one may determine the dependence of the entropy $S$ on strain.
This is made possible by the thermodynamic identity:
\begin{equation}
\pdvc{T}{\epsilon_{ij}}{S} =  - \frac{T}{C_{\epsilon}(T)} \pdvc{S}{\epsilon_{ij}}{T}, \label{eq:SRO-elasto-id}
\end{equation}
where $C_{\epsilon}(T) = T \mleft(\partial S / \partial T\mright)_{\epsilon}$ is the heat capacity at constant strain.
Recently, important progress has been made in the experimental techniques for measuring the elastocaloric effect and in their analysis for correlated electron systems~\cite{Ikeda2019, Straquadine2020, Ikeda2021}.

\begin{figure}[t!]
\centering
\includegraphics[width=0.90\textwidth]{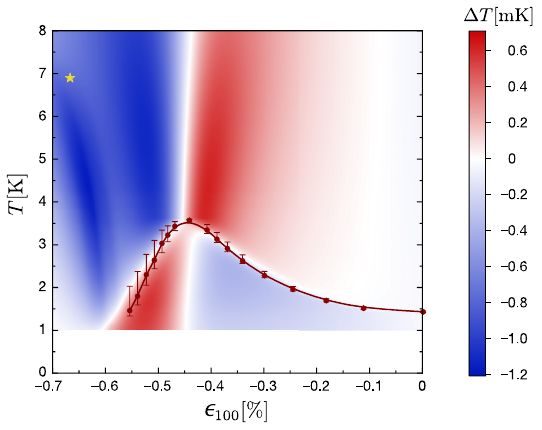}
\captionbelow[Elastocaloric measurements of \ce{Sr2RuO4} as a function of temperature $T$ and compressive uniaxial $\langle 100 \rangle$ strain $\epsilon_{100}$~\cite{Li2022}.]{\textbf{Elastocaloric measurements of \ce{Sr2RuO4} as a function of temperature $T$ and compressive uniaxial $\langle 100 \rangle$ strain $\epsilon_{100}$}~\cite{Li2022}.
The color indicates the measured change in the temperature $\Delta T$ when an ac strain $\Delta \epsilon_{100}$ of a magnitude in between \num{2.9e-6} and \num{3.5e-6} with frequency \SI{1513}{\hertz} is applied on \ce{Sr2RuO4}.
The solid red circles are the superconducting transition temperatures determined from specific heat measurements of Ref.~\cite{Li2021}.
The yellow star indicates the magnetic phase transition temperature deduced from muon spin relaxation in Ref.~\cite{Grinenko2021-unaxial}.
The latter agrees with the phase boundary identified by the dark blue contrast seen in the elastocaloric data for $\epsilon_{100}$ in between \SI{-0.6}{\percent} and \SI{-0.7}{\percent}~\cite{Li2022}.
Reproduced with editing from Ref.~\cite{Li2022} (\href{https://creativecommons.org/licenses/by/4.0/deed.en}{CC BY 4.0}).}
\label{fig:SRO-ECE-data}
\end{figure}

The elastocaloric effect has been measured two years ago for strain applied along the $[100]$ direction~\cite{Li2022}.
Numerical analysis of this dense data set~\cite{Li2022}, which is shown in Fig.~\ref{fig:SRO-ECE-data}, enables the separation of the contribution from $C_{\epsilon}$ and the reconstruction of the dependence of the entropy on strain.
The results of this analysis are plotted in Fig.~\ref{fig:SRO-elasto}.
The data shown in this figure is available in the Supplementary Material of Ref.~\cite{Palle2023-ECE}.

\begin{figure}[p!]
\centering
\includegraphics[width=0.91\textwidth]{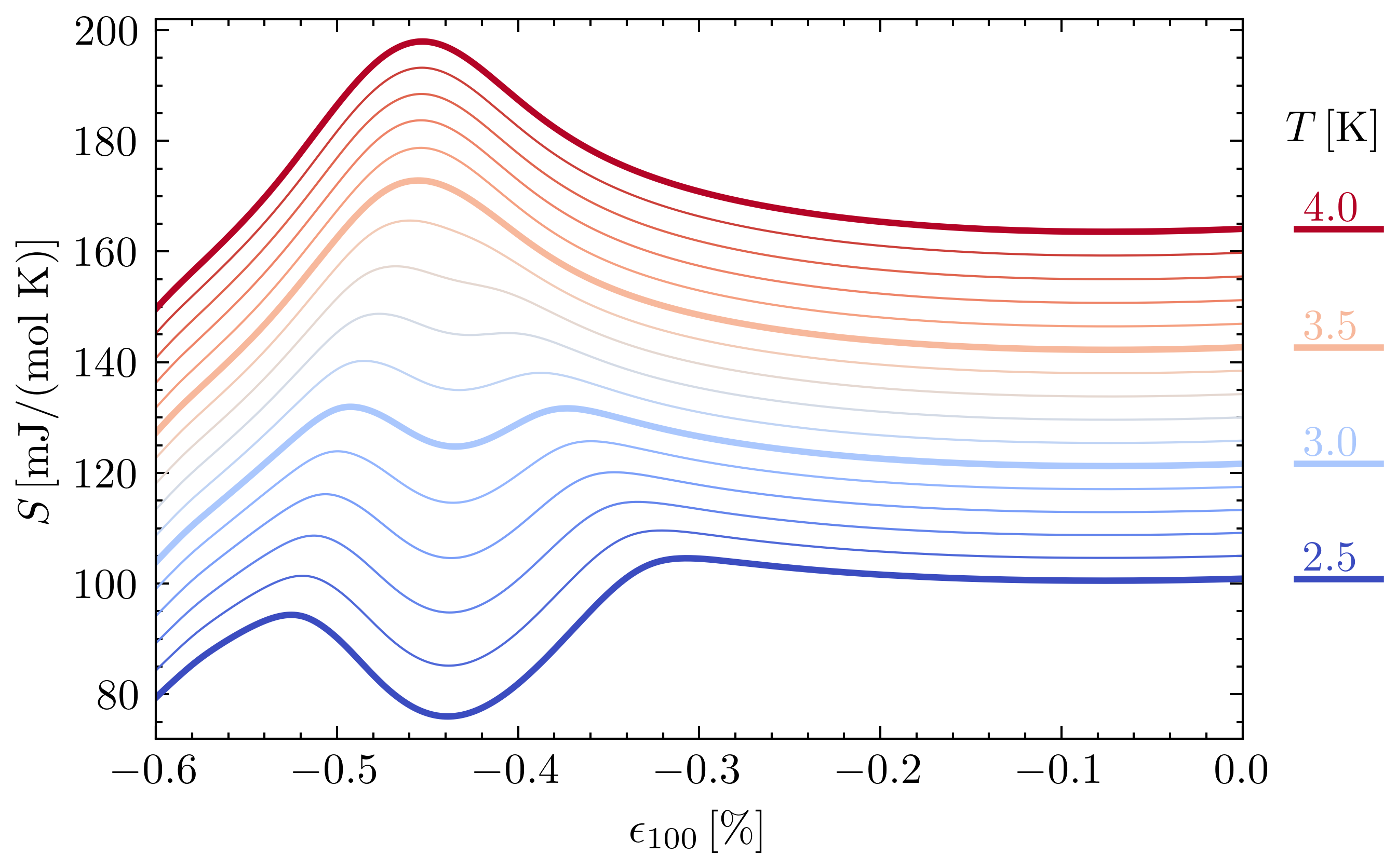} \\
{\includegraphics[width=0.83\textwidth]{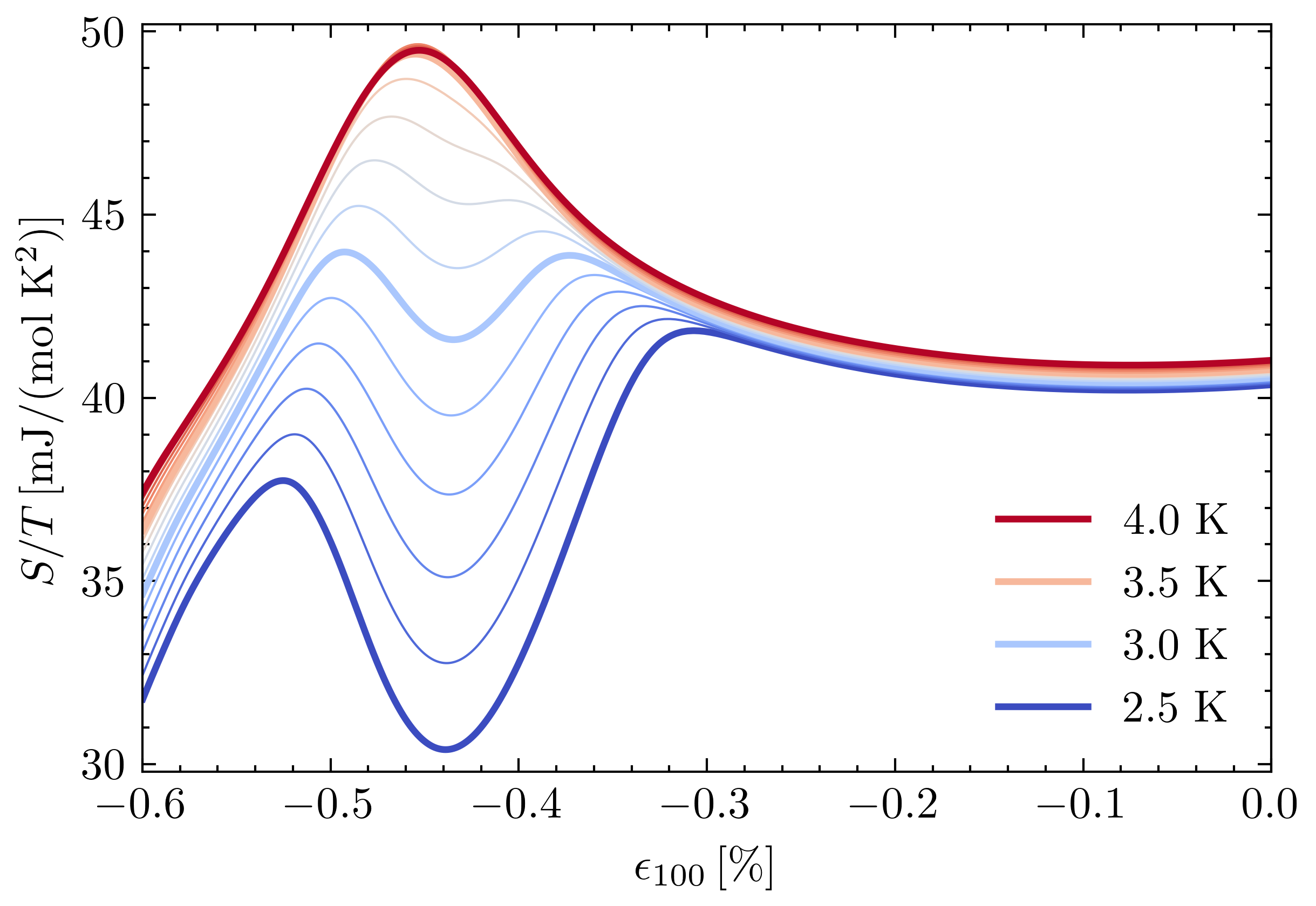}\hspace{26pt}}
\captionbelow[The entropy $S$ (top) and ratio $S / T$ (bottom) of \ce{Sr2RuO4} as a function of applied $\epsilon_{100}$ strain at constant temperatures $T$ ranging from \SI{2.5}{\kelvin} (blue) to \SI{4.0}{\kelvin} (red) in \SI{0.1}{\kelvin} increments~\cite{Palle2023-ECE}.]{\textbf{The entropy $S$ (top) and ratio $S / T$ (bottom) of \ce{Sr2RuO4} as a function of applied $\epsilon_{100}$ strain at constant temperatures $T$ ranging from \SI{2.5}{\kelvin} (blue) to \SI{4.0}{\kelvin} (red) in \SI{0.1}{\kelvin} increments}~\cite{Palle2023-ECE}.
At Van Hove strain $\epsilon_{100} = \SI{-0.44}{\percent} \equiv \epsilon_{\text{VH}}$, $T_c$ attains its maximal value of \SI{3.5}{\kelvin}.
Above (below) \SI{3.5}{\kelvin}, the entropy has a maximum (minimum) at $\epsilon_{\text{VH}}$ strain.
In the top figure, entropies at different temperatures are naturally offset from each other by their temperature dependence.
In the bottom they coalesce because for a Fermi liquid $S$ is linear in $T$.
The entropy has been reconstructed from the elastocaloric measurements of Ref.~\cite{Li2022}, shown in Fig.~\ref{fig:SRO-ECE-data}, using Eq.~\eqref{eq:SRO-elasto-id}.
The plotted data is available in the Supplementary Material of Ref.~\cite{Palle2023-ECE}.}
\label{fig:SRO-elasto}
\end{figure}

As clearly seen in the figure, the normal-state entropy has a maximum at the Van Hove strain $\epsilon_{100} = \SI{-0.44}{\percent} \equiv \epsilon_{\text{VH}}$.
As we enter the SC state, however, this maximum becomes a \emph{minimum} as a function of strain.
To understand this behavior, let us recall that the entropy of a Fermi liquid is given by~\cite{SolyomVol3, Coleman2015}:
\begin{align}
S &= V \frac{\pi^2}{3} k_B^2 T \int \dd{E} \, \Dd_{T}(E) \DOSg(E), \label{eq:SRO-Fermi-liquid-entropy}
\end{align}
where $V$ is the volume, $T$ is the temperature, $E$ is the energy relative to the chemical potential, $\DOSg(E)$ is the DOS, i.e., the number of states (including both spins) per unit cell and energy, and
\begin{align}
\Dd_{T}(E) &\defeq \frac{3}{\pi^2 k_B T} \mleft[- f_E \log f_E - (1-f_E) \log(1-f_E)\mright],
\end{align}
where $f_E = 1/(\Elr^{E/k_B T} + 1)$ is the Fermi-Dirac occupation factor.
Notice that $\Dd_{T}(E) \to \Dd(E)$ as $T \to 0$ so $S \propto T \DOSg(0)$ at low temperature.
This formula applies to both the normal and the SC state.
Thus to understand the entropy, we need to study the DOS near the Fermi level $E = 0$.

In the normal state, at Van Hove strain the $\gamma$ band experiences a Lifshitz transition in which its cylindrical Fermi surface opens at the Van Hove lines $\vb{k}_{\text{VH}} \approx \mleft(0, \pm \frac{\pi}{a}, k_z\mright)$ along the $k_y$ direction~\cite{Steppke2017, Barber2019, Sunko2019}.
This is shown in Fig.~\ref{fig:SRO-FS-strain-evolution}.
Because of the particularly weak $k_z$-dispersion of the $\gamma$ band at $\vb{k}_{\text{VH}}$ ($\sim \SI{1}{\kelvin}$), the Van Hove lines contribute a pronounced peak in the DOS that is only rounded on an energy scale of about one Kelvin~\cite{Li2022}.
It is this peak in the DOS that explains the observed normal-state entropy maximum (Fig.~\ref{fig:SRO-elasto}).

\begin{figure}[t]
\centering
\begin{subfigure}[t]{0.38\textwidth}
\raggedright
\includegraphics[width=\textwidth]{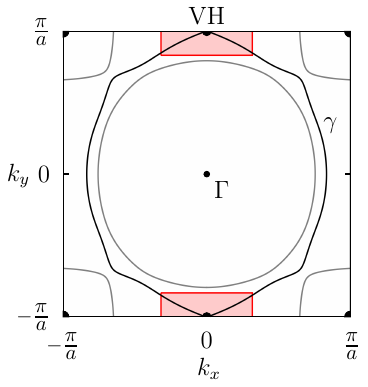}
\subcaption{}
\end{subfigure}%
\begin{subfigure}[t]{0.62\textwidth}
\centering
\includegraphics[width=0.95\textwidth]{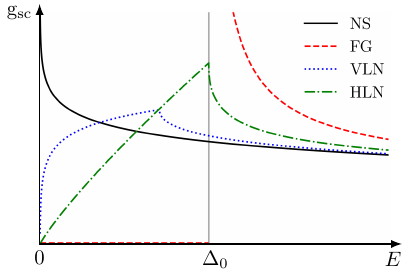}
\subcaption{}
\end{subfigure}
\captionbelow[The Fermi surfaces of \ce{Sr2RuO4} at Van Hove strain $\varepsilon_{100} = \varepsilon_{\text{VH}}$ (a) and how the Van Hove (VH) line $\mleft(0, \frac{\pi}{a}, k_z\mright)$ contribution to the density of states $\DOSg_{\text{sc}}$ depends on the superconducting gapping (b)~\cite{Palle2023-ECE}.]{\textbf{The Fermi surfaces of \ce{Sr2RuO4} at Van Hove strain $\varepsilon_{100} = \varepsilon_{\text{VH}}$ (a) and how the Van Hove (VH) line $\mleft(0, \frac{\pi}{a}, k_z\mright)$ contribution to the density of states $\DOSg_{\text{sc}}$ depends on the superconducting gapping (b)}~\cite{Palle2023-ECE}.
The Fermi sheets shown under (a) are the $k_z = 0$ cross-sections determined by our tight-binding model (Secs.~\ref{sec:SRO-el-struct},~\ref{sec:SRO-elastic-tuning}).
In the density of states $\DOSg_{\text{sc}}(E)$, only the contribution coming from the vicinity of the VH line $\mleft(0, \frac{\pi}{a}, k_z\mright)$ is included.
This region is highlighted red under (a).
Under (b), NS stands for normal state ($\Delta(\vb{k}) = 0$), FG for full gapping of the VH line ($\Delta(\vb{k}) = \Delta_0$), VLN for a vertical line node crossing the VH line ($\Delta(\vb{k}) \propto k_x$), and HLN for a horizontal line node crossing the VH line ($\Delta(\vb{k}) \propto k_z$).
These correspond to Eqs.~\eqref{eq:SRO-VH-DOS-NS}, \eqref{eq:SRO-VH-DOS-FG}, \eqref{eq:SRO-VH-DOS-VLN}, and \eqref{eq:SRO-VH-DOS-HLN}, respectively.
The VLN case (with $\vartheta = \pi/4$) was calculated numerically.
The parameter values $m_{*}^{-1} = \SI{3200}{\kelvin}$, $\Delta_0 = \SI{3}{\kelvin}$, and $\Lambda = 0.5$ were used in all four cases.
Note that the Fermi energy ($E = 0$) is tuned precisely to the saddle point, so this depicts the density of states at the Van Hove strain, shown under (a).}
\label{fig:SRO-bands-DOS}
\end{figure}

To gain a qualitative understanding of what sort of pairings can induce an entropy minimum at $\epsilon_{\text{VH}}$ strain, it is sufficient to consider the $\gamma$ band near the Van Hove lines.
This region is highlighted red in Fig.~\ref{fig:SRO-bands-DOS}(a).
This is justified by the fact that the $\gamma$ band contributes \SI{60}{\percent} of the total DOS (Sec.~\ref{sec:SRO-el-struct}) and is solely responsible for the normal-state peak in the entropy.
For the moment, we shall also neglect the $k_z$-dispersion; we discuss its impact later.

The DOS of a band in 2D with a dispersion $\varepsilon_{\vb{k}}$ and SC gap $\Delta(\vb{k})$ is given by:
\begin{equation}
\DOSg_{\text{sc}}(E) = 2 \int \frac{\dd{\DimK_x} \dd{\DimK_y}}{(2 \pi)^2} \Dd\mleft(E - \xi_{\vb{k}}\mright),
\end{equation}
where the $2$ is due to spin,
\begin{equation}
\xi_{\vb{k}} = \sqrt{\varepsilon_{\vb{k}}^2 + \abs{\Delta(\vb{k})}^2}
\end{equation}
is the Bogoliubov quasi-particle dispersion, and temporarily in this section we define $\DimK$ relative to the Van Hove point $(0, \pi/a)$:
\begin{align}
\vb{\DimK} &= \DimK_x \vu{e}_x + \DimK_y \vu{e}_y = \begin{pmatrix}
\DimK_x \\
\DimK_y
\end{pmatrix} = \begin{pmatrix}
a k_x \\
a k_y - \pi
\end{pmatrix}.
\end{align}
It is often easier to calculate the integrated DOS
\begin{equation}
\mathscr{N}_{\text{sc}}(E) = \int_0^E \dd{E'} \, \DOSg_{\text{sc}}(E') = 2 \int_{\xi_{\vb{k}} \leq E} \frac{\dd{\DimK_x} \dd{\DimK_y}}{(2 \pi)^2}
\end{equation}
instead and then differentiate it to get $\DOSg_{\text{sc}}(E)$.
Near the Van Hove point $\vb{k} = (0, \pi/a)$, the dispersion of the $\gamma$ band is approximately given by (Sec.~\ref{sec:SRO-el-struct}):
\begin{equation}
\varepsilon_{\vb{k}} = \frac{1}{2m_1} \DimK_x^2 - \frac{1}{2m_2} \DimK_y^2 = \frac{1}{m_{*}} \DimQ_{+} \DimQ_{-}, \label{eq:SRO-saddle-epsilon}
\end{equation}
where $\DimQ_{\pm} = \frac{1}{\sqrt{2}} \mleft(r \DimK_x \pm \DimK_y / r\mright)$ and
\begin{align}
m_{*} &= \sqrt{m_1 m_2} = \frac{1}{\SI{3200}{\kelvin}}, &
r &= \sqrt[4]{m_2 / m_1} = 0.59.
\end{align}
The values of $m_{1, 2}$ and $r$ were deduced from the Hamiltonian~\eqref{eq:SRO-TBA-Haml} with the parameter values of Ref.~\cite{Roising2019} (Tab.~\ref{tab:SRO-TBA-params}).
Since this expression for $\varepsilon_{\vb{k}}$ only applies near the Van Hove point, we impose a momentum cutoff $\abs{\DimQ_{\pm}} \leq \Lambda$.
This corresponds to the region highlighted red in Fig.~\ref{fig:SRO-bands-DOS}(a).
(The region depicted in Fig.~\ref{fig:SRO-bands-DOS}(a) has cutoffs imposed on $\DimK_{x,y}$, but this is just for illustration purposes.)

In the normal state (NS),
\begin{equation}
\Delta^{\text{NS}}(\vb{k}) = 0
\end{equation}
and the DOS at the Van Hove strain equals:
\begin{equation}
\DOSg_{\text{sc}}^{\text{NS}}(E) = \frac{8 m_{*}}{(2 \pi)^2} \log\frac{\Lambda^2}{m_{*} E}. \label{eq:SRO-VH-DOS-NS}
\end{equation}
This diverges logarithmically as $E \to 0$.
As we move away from $\epsilon_{100} = \epsilon_{\text{VH}}$, the logarithmic divergence is moved away from the Fermi level $E = 0$, explaining the normal-state entropy maximum~\cite{Li2022}.

If we fully gap (FG) the saddle point like so
\begin{equation}
\Delta^{\text{FG}}(\vb{k}) = \Delta_0,
\end{equation}
then the DOS vanishes up to $\Delta_0$  and diverges above it in the following way:
\begin{equation}
\DOSg_{\text{sc}}^{\text{FG}}(E) = \begin{cases}
0, & \text{when $E \leq \Delta_0$,} \\[4pt]
\displaystyle \frac{8 m_{*}}{(2 \pi)^2} \frac{E}{\sqrt{E^2 - \Delta_0^2}} \log\frac{\Lambda^2}{m_{*} \sqrt{E^2 - \Delta_0^2}}, & \text{when $E > \Delta_0$.}
\end{cases} \label{eq:SRO-VH-DOS-FG}
\end{equation}
Since $\Dd_{T}(E)$ in Eq.~\eqref{eq:SRO-Fermi-liquid-entropy} has a width $\sim k_B T$, for sufficiently large $\Delta_0 / k_B T$ the normal-state entropy maximum can be suppressed so strongly that it becomes a minimum as a function of strain.
Hence fully gapping the Van Hove lines reproduces the features of Fig.~\ref{fig:SRO-elasto}.
Note that a constant gap does not necessarily mean an $s$-wave state, but merely that the gap is finite in the vicinity of the Van Hove point.
For instance, $d_{x^2-y^2}$-wave pairing is finite at the Van Hove point $(0, \pi/a)$ and approximately constant around it.
The same is true for extended $s$-wave pairing which has vertical line nodes away from the Van Hove points.
Our analysis focuses only on the behavior of the pairing gap near the saddle point of the dispersion.

Can pairings with nodal lines at the Van Hove lines also reproduce the SC entropy minimum?
To answer this question, let us calculate the DOS for a vertical and horizontal line node.
For vertical line nodes (VLN), there are two cases to distinguish: when $\Delta(\vb{k})$ is linear and when $\Delta(\vb{k})$ is quadratic in the (displaced) momentum $\vb{\DimK} = (a k_x, a k_y - \pi)$.

In the linear case, we may always write the gap as:
\begin{equation}
\Delta^{\text{VLN}}(\vb{k}) = \Delta_0 \mleft(\DimQ_{+} \cos \vartheta + \DimQ_{-} \sin \vartheta\mright) / \Lambda \equiv \Delta_0 (\DimP_1 / \Lambda).
\end{equation}
In the limit of small $E$, the inequality $\xi_{\vb{k}} \leq E$ that determines $\mathscr{N}_{\text{sc}}(E)$ simplifies to
\begin{equation}
\frac{\Delta_0^2}{\Lambda^2} \DimP_1^2 + \frac{\sin^2(2 \vartheta)}{4 m_{*}^2} \DimP_2^4 \leq E^2,
\end{equation}
where $\DimP_2 = \DimQ_{-} \cos \vartheta - \DimQ_{+} \sin \vartheta$.
The area enclosed by this inequality equals $\pi' \abs{\DimP_{1, \text{max}}}_E \abs{\DimP_{2, \text{max}}}_E$, where $\pi' = 4 \int_0^1 \dd{x} \sqrt{1-x^4} = 3.496...$, and therefore for small $E$:
\begin{equation}
\DOSg_{\text{sc}}^{\text{VLN}}(E \to 0) = \frac{3 \pi'}{(2 \pi)^2} \frac{\Lambda}{\Delta_0} \sqrt{\frac{2 m_{*} E}{\abs{\sin 2 \vartheta}}}. \label{eq:SRO-VH-DOS-VLN}
\end{equation}
This $\DOSg_{\text{sc}}^{\text{VLN}} \propto \sqrt{E}$ behavior persists up to the point where $\DOSg_{\text{sc}}^{\text{VLN}}(E_w) \approx \DOSg_{\text{sc}}^{\text{NS}}(E_w)$.
By solving this equation with $\Delta_0 \sim \SI{3}{\kelvin}$ (the $T_c$ at $\epsilon_{100} = \epsilon_{\text{VH}}$) and $\Lambda \sim 0.5$, one obtains $E_w \sim \SI{0.2}{\kelvin}$.\footnote{The solution of $\sqrt{x} = \frac{1}{2} \delta \log(1/x)$ is $x = \delta^2 W^2(1/\delta)$, where $W(x)$ is the Lambert $W$-function. In our case $x = m_{*} E_w / \Lambda^2$ and $\delta = (8 \sqrt{2} / 3 \pi') (m_{*} \Delta_0 / \Lambda^2)$.}
Exceptionally, when $\vartheta = 0$ or $\pi/2$, one finds a constant DOS up to $\Delta_0$:
\begin{equation}
\DOSg_{\text{sc}}^{\text{VLN}'}(E \leq \Delta_0) = \frac{8 m_{*}}{(2 \pi)^2} \arcsinh\frac{\Lambda^2}{m_{*} \Delta_0}.
\end{equation}
Thus if a single line node cuts through the Van Hove point, the DOS generically vanishes like $\sqrt{E}$ in a very narrow range $E \lesssim \SI{0.2}{\kelvin}$.
If this line node is fine-tuned to coincide with the lines $\DimQ_{+} = 0$ or $\DimQ_{-} = 0$, then the DOS becomes finite and large.

The second case is when $\Delta(\vb{k})$ is quadratic in $\vb{k}$.
Quadratic $\Delta(\vb{k})$ may correspond to a line node with a quadratic orthogonal dispersion, a pair of line nodes that intersect at $\vb{k} = \vb{0}$, or a point node, depending on the eigenvalues of the Hessian at $\vb{k} = (0, \pi/a)$.
The inequality $\xi_{\vb{k}} \leq E$ is in this case invariant under the scaling $\vb{k} \mapsto \sqrt{\alpha} \, \vb{k}$, $E \mapsto \alpha E$.
Hence $\mathscr{N}_{\text{sc}}(E)$ is linear in $E$ for small $E$, yielding a finite $\DOSg_{\text{sc}}^{\text{VLN}''}(E = 0)$ and no opening of a gap.
Exceptionally, when we have two SC line nodes that coincide with the Van Hove strain Fermi surfaces $\DimQ_{\pm} = 0$, the SC gap equals $\Delta(\vb{k}) = \Delta_0 (\DimQ_{+} \DimQ_{-} / \Lambda^2)$, from which we see that $\DOSg_{\text{sc}}^{\text{VLN}'''}$ retains the normal-state logarithmic singularity, albeit with a renormalized $1/m_{*} \mapsto \sqrt{1/m_{*}^2 + \Delta_0^2 / \Lambda^4}$.

Lastly, there's the possibility of a horizontal line node (HLN) crossing the vertical Van Hove line $\mleft(0, \pi/a, k_z\mright)$.
For a schematic
\begin{equation}
\Delta^{\text{HLN}}(\vb{k}) = \Delta_0 (\DimK_z / \pi),
\end{equation}
the 3D DOS can be calculated by averaging Eq.~\eqref{eq:SRO-VH-DOS-FG}:
\begin{equation}
\begin{aligned}
\DOSg_{\text{sc}}^{\text{HLN}}(E) &= \int_{- \pi}^{\pi} \frac{\dd{\DimK_z}}{2 \pi} \mleft.\DOSg_{\text{sc}}^{\text{FG}}(E)\mright|_{\Delta_0 \to \Delta_0 \abs{\DimK_z} / \pi} \\
&= \frac{4 m_{*}}{(2 \pi)^2} \frac{E}{\Delta_0} \mleft[\pi \log\frac{2 \Lambda^2}{m_{*} E} - \mathcal{X}(E)\mright],
\end{aligned} \label{eq:SRO-VH-DOS-HLN}
\end{equation}
where
\begin{equation}
\mathcal{X}(E) = \begin{cases}
\displaystyle 0, &  \quad \text{when $E \leq \Delta_0$,} \\[4pt]
\displaystyle \begin{gathered}
(\pi - 2 \arccos x) \log\frac{\Lambda^2}{m_{*} E} + 2 \log(2 x) \arcsin(x) \\
- 2 \log(x) \arctan\frac{x}{\sqrt{1-x^2}} + \Cl_2(\varphi),
\end{gathered} &  \quad \text{for $E > \Delta_0$,}
\end{cases}
\end{equation}
with
\begin{align}
x &= \sqrt{1 - \Delta_0^2/E^2}, &
\varphi &= \arccos(1 - 2 x^2).
\end{align}
Here $\Cl_2(\varphi) = \sum_{n=1}^{\infty} \sin(n \varphi) / n^2$ is the Clausen function.
$\DOSg_{\text{sc}}^{\text{HLN}}$ is thus roughly linear in $E$ up to $\Delta_0$.

The dependence of the DOS $\DOSg_{\text{sc}}(E)$ for different realizations of the SC gap $\Delta(\vb{k})$ near the saddle point is summarized in Fig.~\ref{fig:SRO-bands-DOS}(b).

Now we come back to the question of whether line nodes at the Van Hove lines are consistent with an entropy minimum.
To clarify this issue, we need to take into account the $k_z$-dispersion, the energy integral in Eq.~\eqref{eq:SRO-Fermi-liquid-entropy}, and the DOS contributions of the other bands.

The $k_z$-dispersion of the $\gamma$ band smears all characteristically 2D features of the DOS by the scale of its energy variation $\var{\varepsilon_{\text{VH}}} \sim \SI{2}{\kelvin}$ [Eq.~\eqref{eq:SRO-VHdispexpansion}].
The normal-state logarithmic singularity becomes a peak.
The $\DOSg_{\text{sc}}^{\text{VLN}} \propto \sqrt{E}$ ascent is cut off to give a finite zero-energy DOS that is because of $E_w / \var{\varepsilon_{\text{VH}}} \ll 1$ of the same magnitude as the normal-state DOS.
Finally, the HLN DOS attains a finite zero-energy DOS that is at most a factor of three or so smaller than the normal-state DOS (since $\var{\varepsilon_{\text{VH}}} / \Delta_0 \sim 1$).
The $\Dd_T(E)$ factor in Eq.~\eqref{eq:SRO-Fermi-liquid-entropy} leads to a temperature smearing that has a similar effect: the ``effective DOS'' that enters the entropy is not $\DOSg_{\text{sc}}(0)$, but $\DOSg_{\text{sc}}(E)$ averaged over $E \sim k_B T$.
All in all, because of these smearing effects, vertical line nodes at the Van Hove lines $\mleft(0, \pm \pi / a, k_z\mright)$ do not suppress the entropy contribution coming from the Van Hove lines, whereas horizontal line nodes can indeed suppress it.

Because of the strain-dependence of $T_c$, the SC gap becomes $\epsilon_{100}$-dependent at constant $T$, peaking at Van Hove strain.
A strong enough gapping of the $\alpha$ and $\beta$ bands could then, in principle, suppress the entropy more than the Van Hove singularities enhance it, resulting in a minimum.
To exclude this scenario, we have calculated the entropy for the case when the $\alpha$, $\beta$, and \SI{80}{\percent} of the $\gamma$ band are fully gapped $\Delta(\vb{k}) = \Delta_0$, while the remaining \SI{20}{\percent} of the $\gamma$ band that includes the Van Hove lines is fully nodal with a vanishing $\Delta(\vb{k}) = 0$.
In particular, for the total DOS we have assumed the form:
\begin{align}
\DOSg_{\text{sc}}^{\text{tot}}(E) &= \DOSg_{\text{VH}} + \HTh(E - \Delta_0) \frac{E}{\sqrt{E^2 - \Delta_0^2}} \DOSg_{\text{rest}},
\end{align}
where $\DOSg_{\text{VH}}$ is the normal-state DOS coming from the parts of the $\gamma$ sheet that are close to the Van Hove lines and $\DOSg_{\text{rest}}$ is the remaining normal-state DOS.
For the temperature-dependence of the SC gap $\Delta_0$ we used the Ansatz
\begin{align}
\Delta_0 &= 1.76 \, k_B T_c \tanh\mleft(1.76 \sqrt{\frac{T_c}{T} - 1}\mright).
\end{align}
Both $\DOSg_{\text{VH}}(\epsilon_{100}) + \DOSg_{\text{rest}}(\epsilon_{100}) \propto \mleft.S(\epsilon_{100}, T) / T\mright|_{T > T_c}$ and $T_c(\epsilon_{100})$ are known experimentally.
Only the ratio $\DOSg_{\text{VH}} / \DOSg_{\text{rest}}$ needs to be calculated, which we have done using the tight-binding model of Sec.~\ref{sec:SRO-el-struct} whose coupling to strain is described in Sec.~\ref{sec:SRO-elastic-tuning}.
One finds that $\DOSg_{\text{rest}}(\epsilon_{100})$ is roughly strain-independent, as expected.
The entropy is calculated by evaluating Eq.~\eqref{eq:SRO-Fermi-liquid-entropy} with $\DOSg_{\text{sc}}^{\text{tot}}(E)$.
The result of this calculation is that a minimum as a function of strain does develop, but the drop in the entropy is \SI{20}{\percent} too small when compared to experiment at $\SI{2.5}{\kelvin}$.
Thus even in this worst-case scenario, where line nodes that are known~\cite{NishiZaki2000, Deguchi2004, Kittaka2018, Matsui2001, Lupien2001, Ishida2000, Bonalde2000, Deguchi2004-p2} to be present in the system are neglected, the Van Hove lines must be gapped in some way to agree with experiment.

The final conclusion that follows from all of these considerations is that the Van Hove lines $\vb{k}_{\text{VH}} \approx \mleft(0, \pm \frac{\pi}{a}, k_z\mright)$ must be either fully gapped or can at most have a horizontal line node crossing them.
Hence, we may exclude vertical line nodes at $\vb{k}_{\text{VH}}$ near Van Hove strain, as previously suggested in Ref.~\cite{Li2022}.
This is one of the main results of Ref.~\cite{Palle2023-ECE}.
That the heat capacity jump is maximal at the Van Hove strain~\cite{Li2021} also supports this conclusion.
Vertical line nodes away from the Van Hove lines are still possible.

To draw conclusions for the unstrained tetragonal system from measurements performed at uniaxial strain $\epsilon_{100} \approx \epsilon_{\text{VH}}$, we rely on the assumption that the pairing states of the strained and unstrained system are adiabatically connected.
Measurements of the highly-sensitive elastocaloric effect~\cite{Li2022} and heat capacity~\cite{Li2021} show no hints of a transition between two different bulk SC states under $[100]$ strain.
By contrast, the onset of spin-density waves, previously found through muon spin relaxation~\cite{Grinenko2021-unaxial}, is clearly visible in the elastocaloric data of Ref.~\cite{Li2022}, shown in Fig.~\ref{fig:SRO-ECE-data}.
So the elastocaloric effect is able to identify a variety of phase transitions, as expected for an indirect probe of the entropy.

We may thus exclude all SC states of the unstrained system that are adiabatically connected to SC states of the $\epsilon_{100}$ strained system which have a vertical line node at $\vb{k}_{\text{VH}} \approx \mleft(0, \pm \frac{\pi}{a}, k_z\mright)$~\cite{Palle2023-ECE}.
Given that $\epsilon_{100}$ strain preserves all the symmetry operations that map the Van Hove lines to themselves, as we shall see in Sec.~\ref{sec:SRO-behavior-vH}, we may conclude that there are no vertical line nodes at either $\mleft(\pm \frac{\pi}{a}, 0, k_z\mright)$ nor $\mleft(0, \pm \frac{\pi}{a}, k_z\mright)$ in the unstrained tetragonal system.
Intuitively, this means that SRO's SC takes full advantage of the enhanced DOS induced by the Van Hove lines.
Indeed, the drastic enhancement of $T_c$ and $B_{c2}$ under uniaxial pressure~\cite{Hicks2014, Taniguchi2015, Steppke2017, Barber2019, Jerzembeck2023} were suggestive of this conclusion long ago, but only with the recent elastocaloric measurements of Ref.~\cite{Li2022} could more conclusive statements be made~\cite{Palle2023-ECE}.

\subsection{Location of and dispersion at the Van Hove lines} \label{sec:SRO-vH-line}
Here we establish that the Van Hove lines are adequately approximated with $\mleft(\pm \frac{\pi}{a}, 0, k_z\mright)$ and $\mleft(0, \pm \frac{\pi}{a}, k_z\mright)$, i.e., with straight vertical lines located at $\mleft(\pm \frac{\pi}{a}, 0\mright)$ and $\mleft(0, \pm \frac{\pi}{a}\mright)$.
For a simple-tetragonal lattice, the Van Hove lines are lines of high symmetry.
However, they are not located precisely on the boundary of the body-centered first Brillouin zone relevant to \ce{Sr2RuO4}, which could in principle allow for large deviations away from $\mleft(\pm \frac{\pi}{a}, 0, k_z\mright)$ and $\mleft(0, \pm \frac{\pi}{a}, k_z\mright)$.
As we shall see, the high anisotropy of SRO makes these deviations negligible, justifying the subsequent analysis.

Van Hove points are points in momentum space where the gradient of the band energy $\varepsilon_{\vb{k}}$ vanishes.
In 3D, the solutions of $\grad_{\vb{k}} \varepsilon_{\vb{k}} = \vb{0}$ are generically isolated points.
However, quasi-2D dispersions may yield Van Hove \emph{lines}, that is, lines on which a number of Van Hove points are situated of similar energy.
The quality of the emergent Van Hove lines is quantified by how well-aligned the Van Hove points are to a line and by how close the energies of the Van Hove points are.

Consider the Van Hove line $\mleft(0, \frac{\pi}{a}, k_z\mright)$.
Then for any two $\vb{k} = \mleft(\var{k_x}, \frac{\pi}{a} + \var{k_y}, k_z\mright)$ and $\vb{k}' = R(g) \vb{k}$ related by a symmetry operation $g \in D_{4h}$, $\varepsilon_{\vb{k}} = \varepsilon_{\vb{k}' + \vb{G}}$ for any reciprocal lattice vector $\vb{G}$.
Applying this to parity gives $\grad_{\vb{k}} \varepsilon_{\vb{k}} = \vb{0}$ at the mid-points of the Brillouin zone faces, which for body-centered tetragonal SRO are $\mleft(0, \frac{\pi}{a}, \pm \frac{\pi}{c}\mright)$.
These are the first two Van Hove points.
The positions of the other two Van Hove points are restricted by symmetry to be at $\mleft(0, \frac{\pi}{a} + \var{k_{\text{VH}, 2}}, 0\mright)$ and $\mleft(0, \frac{\pi}{a} - \var{k_{\text{VH}, 2}}, \pm \frac{2 \pi}{c}\mright)$.
Reflection across the $k_x = 0$ plane implies $\partial_{k_x} \varepsilon_{\vb{k}} = 0$ in the $k_x = 0$ plane and reflection across the $k_z = 0$ plane implies $\partial_{k_z} \varepsilon_{\vb{k}} = 0$ in the planes $k_z = 0, \pm \frac{2 \pi}{c}$.
If the system were simple tetragonal-periodic, then reflection across the $k_y = 0$ plane would imply $\partial_{k_y} \varepsilon_{\vb{k}} = 0$ in the $k_y = \pm \frac{\pi}{a}$ planes, making $\var{k_{\text{VH}, 2}} = 0$.
Because of the smallness of the characteristically body-centered hopping in SRO, which is always between layers (Sec.~\ref{sec:SRO-el-struct}), $\var{k_{\text{VH}, 2}}$ is very close to zero.

\begin{figure}[t!]
\centering
\includegraphics[width=\textwidth]{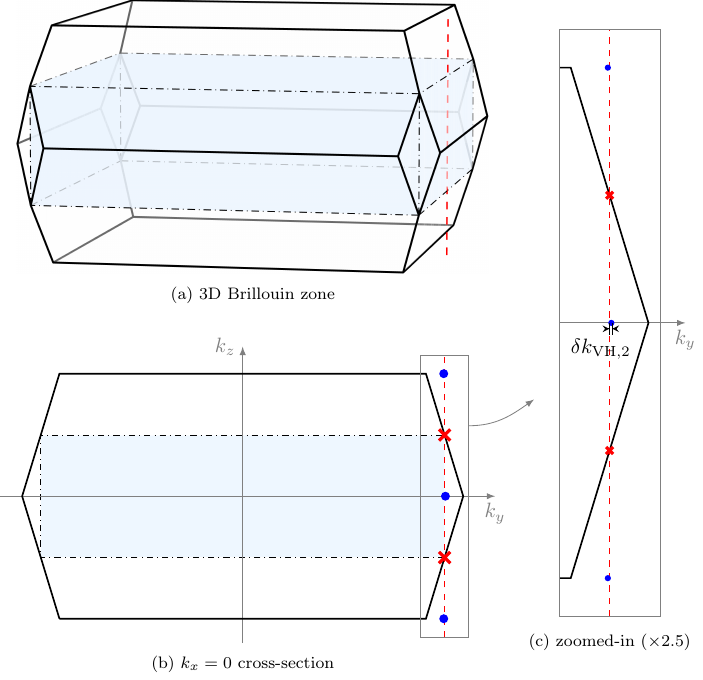}
\captionbelow[The body-centered tetragonal Brillouin zone of SRO (a), its $k_x = 0$ cross-section (b), and the region around the $\mleft(0, \frac{\pi}{a}, k_z\mright)$ Van Hove line (c).]{\textbf{The body-centered tetragonal Brillouin zone of SRO (a), its $k_x = 0$ cross-section (b), and the region around the $\mleft(0, \frac{\pi}{a}, k_z\mright)$ Van Hove line (c).}
Shaded in blue is the simple tetragonal Brillouin zone.
The red crosses are the $\mleft(0, \frac{\pi}{a}, \pm \frac{\pi}{c}\mright)$ Van Hove points.
The blue dots are the $\mleft(0, \frac{\pi}{a} + \var{k_{\text{VH}, 2}}, 0\mright)$ and $\mleft(0, \frac{\pi}{a} - \var{k_{\text{VH}, 2}}, \pm \frac{2 \pi}{c}\mright)$ Van Hove points.
Together they constitute the Van Hove line $\mleft(0, \frac{\pi}{a}, k_z\mright)$, drawn here with a dashed red line.
The displacement length $\var{k_{\text{VH}, 2}} \approx 0.013 / a$ is designated under (c).}
\label{fig:SRO-vanHove}
\end{figure}

From the tight-binding model of Sec.~\ref{sec:SRO-el-struct}, we may extract the following simplified expression for the dispersion of the $\gamma$ band near the Van Hove line $\mleft(0, \frac{\pi}{a}, k_z\mright)$:
\begin{align}
\begin{aligned}
\varepsilon_{\vb{k} \gamma} &= \upmu_{\text{VH}} + \frac{a^2}{2 m_1} k_x^2 - \frac{a^2}{2 m_2} \mleft(k_y - \frac{\pi}{a}\mright)^2 \\
&\hspace{14pt} - \var{\varepsilon_{\text{VH}}} \cos c k_z + \frac{a^2}{m_2} \var{k_{\text{VH}, 2}} \mleft(k_y - \frac{\pi}{a}\mright) \cos\frac{c k_z}{2}.
\end{aligned}  \label{eq:SRO-VHdispexpansion}
\end{align}
Its form follows from symmetry; only the lowest powers in $k_x, k_y$ and lowest harmonics in $k_z$ were retained.
Using the parameters of Ref.~\cite{Roising2019}, we find that
\begin{align}
\begin{aligned}
\upmu_{\text{VH}} &= \SI{54}{\milli\electronvolt}, &\hspace{50pt}
\var{\varepsilon_{\text{VH}}} &= \SI{2.4}{\kelvin}, &\hspace{50pt}
a \var{k_{\text{VH}, 2}} &= 0.013, \\
m_1^{-1} &= \SI{1100}{\kelvin}, &\hspace{50pt}
m_2^{-1} &= \SI{9300}{\kelvin}.
\end{aligned}
\end{align}
While this dispersion was derived from a model of unstrained SRO, it offers a good understanding of the effects of the $k_z$-dispersion on the Van Hove line.
The deviation of the Van Hove points from the $(\frac{\pi}{a}, 0, k_z)$-line is characterized by $\var{k_{\text{VH}, 2}} \ll \frac{2 \pi}{a}$, which is a factor of $500$ smaller than the width of the Brillouin zone.
Furthermore, the difference in the $\gamma$ band energies of the Van Hove points is given by $\var{\varepsilon_{\text{VH}}}$ which is on the order of a few kelvins.
We may thus conclude that the four Van Hove points, illustrated in Fig.~\ref{fig:SRO-vanHove}, together constitute a Van Hove line $\mleft(0, \pi/a, k_z\mright)$ to a high degree of accuracy~\cite{Palle2023-ECE}.
The same is true for the Van Hove lines $\mleft(0, - \pi/a, k_z\mright)$ and $\mleft(\pm \pi/a, 0, k_z\mright)$.

\subsection{Behavior of superconducting states on the Van Hove lines} \label{sec:SRO-behavior-vH}
To see which SC states are excluded by the fact that vertical line nodes on the Van Hove lines are incompatible with the elastocaloric effect data of Ref.~\cite{Li2022}, let us briefly recall which SC states are possible~\cite{Ramires2019, Kaba2019, Huang2019}.

As we discussed at length in Sec.~\ref{sec:SRO-SC-construct}, the multiband nature of SRO allows for a richer set of possible SC states than usual.
The main novelty is that the gap matrix $\Delta_{\alpha \beta}(\vb{k})$ can have non-trivial orbital structure.
As we found in Sec.~\ref{sec:SRO-SC-construct}, for the effective model of SRO based on the $t_{2g}(d_{yz}|d_{zx}|d_{xy})$ orbitals of \ce{Ru}, spin-orbit matrices belonging to all possible irreps of $D_{4h}$ exist, for both even- and odd-parity pairings (Tab.~\ref{tab:SRO-Gamma-class}).
The irrep of the total gap matrix $\Delta(\vb{k})$ is determined by the \emph{product} of the irreps of its momentum and spin-orbit parts.
Thus for all symmetry channels, generic SC states have non-trivial orbital structures and it is not sufficient to just analyze the pairing wavefunctions.
One needs to study the symmetry properties of the spin-orbit matrices as well.

\pagebreak

\begin{figure}[t]
\centering
\includegraphics[width=0.85\textwidth]{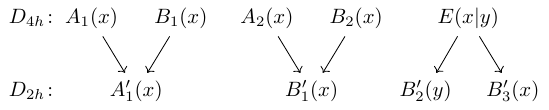}
\captionbelow[How the $D_{4h}$ irreps (top) reduce to $D_{2h}$ irreps (bottom) in the presence of $\epsilon_{100}$ uniaxial strain.]{\textbf{How the $D_{4h}$ irreps (top) reduce to $D_{2h}$ irreps (bottom) in the presence of $\epsilon_{100}$ uniaxial strain.}
Parity stays the same so we have suppressed the $g$ and $u$ subscripts.
The pair $(x|y)$ transforms according to the standard $\RepM^{E}(g)$ matrices of Eq.~\eqref{eq:SRO-E-rep-rho}, Sec.~\ref{sec:SRO-SC-construct}, that we also use elsewhere in the thesis (Sec.~\ref{sec:examples-convention-D4h}).}
\label{fig:SRO-D4h-to-D2h}
\end{figure}

Now we analyze which SC states of the $\epsilon_{100}$-strained system gap the Van Hove lines sufficiently strongly to be able to explain the elastocaloric experiment~\cite{Li2022}.
Viable unstrained SC states must be adiabatically connected to these states.
As we shall see, in the arguments of this section the key symmetry operations are those that map the Van Hove lines $\vb{k}_{\text{VH}} = \mleft(0, \pm \frac{\pi}{a}, k_z\mright)$ to themselves.
As it turns out, although $\epsilon_{100}$ strain reduces the point group from $D_{4h}$ to $D_{2h}$, whose character table is provided in Tab.~\ref{tab:D2h-char-tab-again}, the symmetries that map the Van Hove lines to themselves are the same for both $D_{4h}$ and $D_{2h}$.
They are listed in Tab.~\ref{tab:SRO-rho-kvH}.
Hence we may do the whole analysis either with or without $\epsilon_{100}$ strain.
We have opted for the latter.
Using Fig.~\ref{fig:SRO-D4h-to-D2h}, one may translate all the results for irreps of $D_{4h}$ derived in this section, which is based on Ref.~\cite{Palle2023-ECE}, into results for irreps of $D_{2h}$.
Fig.~\ref{fig:SRO-D4h-to-D2h} also specifies which irreps of $D_{2h}$ are adiabatically connected to which irreps of $D_{4h}$, which brings us back to the initial $D_{4h}$ irreps.

\begin{table}[t]
\centering
\captionabove[The character table of the orthorhombic point group $D_{2h}$~\cite{Dresselhaus2007}.]{\textbf{The character table of the orthorhombic point group $D_{2h}$}~\cite{Dresselhaus2007}.
This is the point group of \ce{Sr2RuO4} when $\langle 100 \rangle$ uniaxial stress is applied on the system.
The point group in the absence of stress is $D_{4h}$ (Tab.~\ref{tab:D4h-char-tab-again2}).
The irreps are divided according to parity into even (subscript $g$) and odd ($u$) ones.
To the left of the irreps are the simplest polynomials constructed from the coordinates $\vb{r} = (x, y, z)$ that transform according to them.
Primes have been added on the irreps to distinguish them from $D_{4h}$ irreps.
$C_{2z}$, $C_{2y}$, and $C_{2x}$ are \SI{180}{\degree} rotations around $\vu{e}_z$, $\vu{e}_y$, and $\vu{e}_x$, respectively.
$P$ is space inversion or parity.
Mirror reflections $\Sigma_z$, $\Sigma_y$, and $\Sigma_x$ are obtained by composing $C_{2z}$, $C_{2y}$, and $C_{2x}$ with $P$, respectively.}
{\renewcommand{\arraystretch}{1.3}
\renewcommand{\tabcolsep}{10pt}
\begin{tabular}{cc|rrrr|rrrr} \hline\hline
\multicolumn{2}{c|}{$D_{2h}$} & $E$ & $C_{2z}$ & $C_{2y}$ & $C_{2x}$ & $P$ & $\Sigma_z$ & $\Sigma_y$ & $\Sigma_x$
\\ \hline
$1$, $x^2$, $y^2$, $z^2$ & $A_{1g}'$ & $1$ & $1$ & $1$ & $1$ & $1$ & $1$ & $1$ & $1$
\\
$xy$ & $B_{1g}'$ & $1$ & $1$ & $-1$ & $-1$ & $1$ & $1$ & $-1$ & $-1$
\\
$xz$ & $B_{2g}'$ & $1$ & $-1$ & $1$ & $-1$ & $1$ & $-1$ & $1$ & $-1$
\\
$yz$ & $B_{3g}'$ & $1$ & $-1$ & $-1$ & $1$ & $1$ & $-1$ & $-1$ & $1$
\\ \hline
$xyz$ & $A_{1u}'$ & $1$ & $1$ & $1$ & $1$ & $-1$ & $-1$ & $-1$ & $-1$
\\
$z$ & $B_{1u}'$ & $1$ & $1$ & $-1$ & $-1$ & $-1$ & $-1$ & $1$ & $1$
\\
$y$ & $B_{2u}'$ & $1$ & $-1$ & $1$ & $-1$ & $-1$ & $1$ & $-1$ & $1$
\\
$x$ & $B_{3u}'$ & $1$ & $-1$ & $-1$ & $1$ & $-1$ & $1$ & $1$ & $-1$
\\ \hline\hline
\end{tabular}}
\label{tab:D2h-char-tab-again}
\end{table}

Let us consider the Van Hove line ($k_z \in \R$):
\begin{equation}
\vb{k}_{\text{VH}} = \begin{pmatrix}
0 \\ \pi/a \\ k_z
\end{pmatrix}.
\end{equation}
For a SC gap matrix $\Delta_a(\vb{k})$ to be able to gap the $\gamma$ band at $\vb{k}_{\text{VH}}$, both its pairing wavefunction $f_a(\vb{k})$ and the projection of its spin-orbit matrix $\Gamma_a$ onto the $\gamma$ band must be finite there.

The only point group symmetries $g \in D_{4h}$ that constrain $f_a\mleft(\vb{k}_{\text{VH}}\mright)$ or the band projections of $\Gamma_a$ are those that map the $\mleft(0, \frac{\pi}{a}, k_z\mright)$ line to itself, modulo body-centered reciprocal lattice vectors.
One readily find that these are
\begin{align}
\begin{aligned}
\Sigma_x\colon k_z &\mapsto k_z, \\
C_{2z}, \Sigma_y\colon k_z &\mapsto k_z + \frac{2 \pi}{c}, \\
C_{2y}, \Sigma_h\colon k_z &\mapsto - k_z, \\
C_{2x}, P\colon k_z &\mapsto - k_z + \frac{2 \pi}{c}.
\end{aligned} \label{eq:SRO-how-g-maps-kZ}
\end{align}
Here, $C_{2x}$, $C_{2y}$, $C_{2z}$ are rotations by $\pi$ around $x$, $y$, and $z$, respectively, and $\Sigma_x = P C_{2x}$, $\Sigma_y = P C_{2y}$, $\Sigma_h = P C_{2z}$ are reflections.
Given that $C_{2z} = \Sigma_x \Sigma_y$, $C_{2y} = \Sigma_x \Sigma_h$, and $P = \Sigma_x C_{2x}$, we may focus solely on the reflections and $P$ (or $C_{2x}$).
The other point group operations do not yield any additional constraints.
Their matrices are listed in Tab.~\ref{tab:SRO-rho-kvH}.
The strongest constraints follow from $\Sigma_x$ because it maps $k_z \mapsto k_z$.
In the simple-tetragonal limit, $k_z \cong k_z + \frac{2 \pi}{c}$ so $\vb{k}_{\text{VH}}$ are on the Brillouin zone boundary and $\Sigma_y, C_{2z}$ give strong constraints too.

\begin{table}[t]
\centering
\captionabove[The symmetry transformation matrices $\RepM^{\zeta}(g)$ for all the irreps $\zeta$ of the tetragonal point group $D_{4h}$ and for all the point group operations $g \in D_{4h}$ which map the Van Hove line $\mleft(0, \frac{\pi}{a}, k_z\mright)$ to itself.]{\textbf{The symmetry transformation matrices $\RepM^{\zeta}(g)$ for all the irreps $\zeta$ of the tetragonal point group $D_{4h}$ and for all the point group operations $g \in D_{4h}$ which map the Van Hove line $\mleft(0, \frac{\pi}{a}, k_z\mright)$ to itself.}
Highlighted red are the negative elements which constrain various things to vanish during symmetry arguments.
In the last column is how the the $k_z$ coordinate of $\mleft(0, \frac{\pi}{a}, k_z\mright)$ gets mapped to itself under $g$, $R(g) \mleft(0, \frac{\pi}{a}, k_z\mright)^{\intercal} + \vb{G} \equiv \mleft(0, \frac{\pi}{a}, g \cdot k_z\mright)^{\intercal}$, modulo body-centered-tetragonal inverse lattice vectors $\vb{G}$.}
{\renewcommand{\arraystretch}{1.3}
\renewcommand{\tabcolsep}{5.5pt}
\begin{tabular}{c|rrrrNrrrrN|c} \hline\hline
& \multicolumn{10}{c|}{$\RepM^{\zeta}(g)$} & \\
$g$
& $A_{1g}$ & $A_{2g}$ & $B_{1g}$ & $B_{2g}$ & $E_{g}$
& $A_{1u}$ & $A_{2u}$ & $B_{1u}$ & $B_{2u}$ & $E_{u}$
& $g \cdot k_z$
\\ \hline 
&&&&&&&&&&& \\[-15pt]
$\one$
& $1~$ & $1~$ & $1~$ & $1~$ & $\begin{pmatrix} 1 & 0 \\ 0 & 1 \end{pmatrix}$
& $1~$ & $1~$ & $1~$ & $1~$ & $\begin{pmatrix} 1 & 0 \\ 0 & 1 \end{pmatrix}$
& $k_z$
\\[9pt]
$C_{2x}$
& $1~$ & $\textcolor{red}{-1}~$ & $1~$ & $\textcolor{red}{-1}~$ & $\begin{pmatrix} 1 & 0 \\ 0 & \textcolor{red}{-1} \end{pmatrix}$
& $1~$ & $\textcolor{red}{-1}~$ & $1~$ & $\textcolor{red}{-1}~$ & $\begin{pmatrix} 1 & 0 \\ 0 & \textcolor{red}{-1} \end{pmatrix}$
& $- k_z + \frac{2 \pi}{c}$
\\[9pt]
$C_{2y}$
& $1~$ & $\textcolor{red}{-1}~$ & $1~$ & $\textcolor{red}{-1}~$ & $\begin{pmatrix} \textcolor{red}{-1} & 0 \\ 0 & 1 \end{pmatrix}$
& $1~$ & $\textcolor{red}{-1}~$ & $1~$ & $\textcolor{red}{-1}~$ & $\begin{pmatrix} \textcolor{red}{-1} & 0 \\ 0 & 1 \end{pmatrix}$
& $- k_z$
\\[9pt]
$C_{2z}$
& $1~$ & $1~$ & $1~$ & $1~$ & $\begin{pmatrix} \textcolor{red}{-1} & 0 \\ 0 & \textcolor{red}{-1} \end{pmatrix}$
& $1~$ & $1~$ & $1~$ & $1~$ & $\begin{pmatrix} \textcolor{red}{-1} & 0 \\ 0 & \textcolor{red}{-1} \end{pmatrix}$
& $k_z + \frac{2 \pi}{c}$
\\[9pt]
$P$
& $1~$ & $1~$ & $1~$ & $1~$ & $\begin{pmatrix} 1 & 0 \\ 0 & 1 \end{pmatrix}$
& $\textcolor{red}{-1}~$ & $\textcolor{red}{-1}~$ & $\textcolor{red}{-1}~$ & $\textcolor{red}{-1}~$ & $\begin{pmatrix} \textcolor{red}{-1} & 0 \\ 0 & \textcolor{red}{-1} \end{pmatrix}$
& $- k_z + \frac{2 \pi}{c}$
\\[9pt]
$\Sigma_x$
& $1~$ & $\textcolor{red}{-1}~$ & $1~$ & $\textcolor{red}{-1}~$ & $\begin{pmatrix} 1 & 0 \\ 0 & \textcolor{red}{-1} \end{pmatrix}$
& $\textcolor{red}{-1}~$ & $1~$ & $\textcolor{red}{-1}~$ & $1~$ & $\begin{pmatrix} \textcolor{red}{-1} & 0 \\ 0 & 1 \end{pmatrix}$
& $k_z$
\\[9pt]
$\Sigma_y$
& $1~$ & $\textcolor{red}{-1}~$ & $1~$ & $\textcolor{red}{-1}~$ & $\begin{pmatrix} \textcolor{red}{-1} & 0 \\ 0 & 1 \end{pmatrix}$
& $\textcolor{red}{-1}~$ & $1~$ & $\textcolor{red}{-1}~$ & $1~$ & $\begin{pmatrix} 1 & 0 \\ 0 & \textcolor{red}{-1} \end{pmatrix}$
& $k_z + \frac{2 \pi}{c}$
\\[9pt]
$\Sigma_h$
& $1~$ & $1~$ & $1~$ & $1~$ & $\begin{pmatrix} \textcolor{red}{-1} & 0 \\ 0 & \textcolor{red}{-1} \end{pmatrix}$
& $\textcolor{red}{-1}~$ & $\textcolor{red}{-1}~$ & $\textcolor{red}{-1}~$ & $\textcolor{red}{-1}~$ & $\begin{pmatrix} 1 & 0 \\ 0 & 1 \end{pmatrix}$
& $- k_z$
\\[8pt] \hline\hline
\end{tabular}}
\label{tab:SRO-rho-kvH}
\end{table}

Consider one of the point group elements $g \in D_{4h}$ listed in Tab.~\ref{tab:SRO-rho-kvH} and a $k_{z,\star}$ that $g$ maps to itself, modulo $\frac{4 \pi}{c}$.
As written in Tab.~\ref{tab:SRO-rho-kvH}, this means that $g \cdot k_{z,\star} = k_{z,\star} \mod \frac{4 \pi}{c}$.
In light of Eq.~\eqref{eq:SRO-how-g-maps-kZ}, $k_{z,\star}$ may take the following values, depending on $g$:
\begin{itemize}
\item For $g = \Sigma_x$, all $k_{z,\star} \in \R$ are allowed.
\item For $g = \Sigma_y$ or $C_{2z}$, no $k_{z,\star}$ maps to itself in a body-centered system like SRO. In the simple-tetragonal limit, all $k_{z,\star} \in \R$ map to themselves and are thus allowed.
\item For $g = \Sigma_h$ or $C_{2y}$, only $k_{z,\star} \in \{0, \pm \frac{2 \pi}{c}\}$ are allowed.
\item For $g = P$ or $C_{2x}$, only $k_{z,\star} \in \{\pm \frac{\pi}{c}\}$ are allowed.
\end{itemize}

For such $k_{z,\star}$, periodicity and the symmetry transformation rule of pairing wavefunctions (Eq.~\eqref{eq:SRO-transf-d}, Sec.~\ref{sec:SRO-SC-construct}) give the following symmetry constraint:
\begin{align}
f_a\mleft(0, \tfrac{\pi}{a}, k_{z,\star}\mright) &= \sum_{b = 1}^{\dim \zeta} \RepM_{ab}^{\zeta}(g) f_b\mleft(0, \tfrac{\pi}{a}, k_{z,\star}\mright).
\end{align}
Because all $\RepM^{\zeta}(g)$ are diagonal (see Tab.~\ref{tab:SRO-rho-kvH}), the above constrains each component of $f_a$ individually.
In particular, notice that whenever $\RepM_{aa}^{\zeta}(g) = -1$, this constrains the pairing wavefunctions $f_a\mleft(0, \tfrac{\pi}{a}, k_{z,\star}\mright)$ to vanish identically by symmetry.
These negative elements are highlighted red in Tab.~\ref{tab:SRO-rho-kvH}.
By going through all the irreps and point group operations, we find the following symmetry-enforced behavior of $f_a\mleft(0, \frac{\pi}{a}, k_z\mright)$, depending on its irrep and $k_z = k_{z,\star}$:
\begin{itemize}
\item $f$ belonging to $A_{2g}$, $B_{2g}$, $A_{1u}$, and $B_{1u}$ vanish for all $k_z$.
\item For $(f_1|f_2) \in E_g$, $f_2$ vanishes for all $k_z$, whereas $f_1$ vanishes only at $k_z = 0, \pm \frac{2 \pi}{c}$.
\item For $(f_1|f_2) \in E_u$, $f_1$ vanishes for all $k_z$, whereas $f_2$ vanishes only at $k_z = \pm \frac{\pi}{c}$.
\item For those $(f_1|f_2) \in E_{g/u}$ that are periodic under simple tetragonal translations ($k_z \cong k_z + \frac{2 \pi}{c}$), both components vanish for all $k_z$.
\item $f$ from irreps $A_{2u}$ and $B_{2u}$ vanish only at $k_z = 0$, $\pm \frac{\pi}{c}$, and $\pm \frac{2 \pi}{c}$, but are otherwise unconstrained.
\item $f$ from $A_{1g}$ and $B_{1g}$ are completely unconstrained for all $k_z$.
\end{itemize}

Next we study the spin-orbit matrices $\Gamma_a$.
We do so by considering the pairing of the band eigenstates of the problem and focus on intraband pairing.
To explore it, we need to project the $\Gamma_a$ onto the bands.
Call $u_{\vb{k} \gamma s}$ the eigenvectors of the $\gamma$ band: $H_{\vb{k}} u_{\vb{k} \gamma s} = \varepsilon_{\vb{k} \gamma} u_{\vb{k} \gamma s}$ with the $H_{\vb{k}}$ given in Eq.~\eqref{eq:SRO-TBA-Haml}.
The projection is then given by:
\begin{align}
\mleft[\mathscr{P}_{a}(\vb{k})\mright]_{ss'} &\defeq u_{\vb{k} \gamma s}^{\dag} \Gamma_{a} u_{- \vb{k} \gamma s'}^{*} = \sum_{\mu} \mathscr{P}_{a}^{\mu}(\vb{k}) \mleft[\Pauli_{\mu} (\iu \Pauli_{y})\mright]_{ss'},
\end{align}
where $s, s' \in \{\uparrow, \downarrow\}$ are the pseudospins.
Since all three $t_{2g}$ orbitals are even, $\MatU(P) = \one$ and we may always locally choose a gauge in which $u_{- \vb{k} \gamma s} = u_{\vb{k} \gamma s}$ so that $\mathscr{P}_{a}(\vb{k}) = \mathscr{P}_{a}(- \vb{k}) = p_{\Gamma} \mathscr{P}_{a}^{\intercal}(\vb{k})$, where $\Gamma_a^{\intercal} = p_{\Gamma} \Gamma_a$.
In turn this implies that $\mathscr{P}_{a}(\vb{k})$ has only the $\mu = 0$ component for antisymmetric $\Gamma_a$ ($p_{\Gamma} = -1$) and only the $\mu \in \{x, y, z\}$ components for symmetric $\Gamma_a$ ($p_{\Gamma} = +1$).

Whenever a $g \in D_{4h}$ maps a $\vb{k}_{\star}$ to itself modulo periodicity, its symmetry transformation matrix $\MatU(g) = O(g) \otimes S(g)$ (Sec.~\ref{sec:SRO-el-struct}, Tab.~\ref{tab:SRO-transf-mat}) commutes with the normal-state Hamiltonian $H_{\vb{k}}$:
\begin{align}
\MatU^{\dag}(g) H_{\vb{k}_{\star}} \MatU(g) &= H_{R(g^{-1}) \vb{k}_{\star}} = H_{\vb{k}_{\star}}.
\end{align}
This means that the interband parts of $\MatU(g)$ vanish.
Here we are assuming that $H_{\vb{k}+\vb{G}} = H_{\vb{k}}$ is periodic, which entails a periodic momentum-space gauge.\footnote{This point is of more significance in systems where some of the orbitals have non-trivial Wyckoff positions, as in the cuprates.
See Secs.~\ref{sec:extended-basis-def} and~\ref{sec:ASV-conventions} in particular.
The gauges used for SRO are always periodic.}
As for the intraband part, we may choose a basis for the Kramers' degenerate subspace such that it takes a spin-like form:
\begin{align}
u_{\vb{k}_{\star} \gamma s'}^{\dag} \MatU(g) u_{\vb{k}_{\star} \gamma s} &= \mleft[S(g)\mright]_{s's},
\end{align}
or equivalently:
\begin{align}
\MatU(g) u_{\vb{k}_{\star} \gamma s} &= \sum_{s'} u_{\vb{k}_{\star} \gamma s'} \mleft[S(g)\mright]_{s's}.
\end{align}
Although such transformation rules do not apply to general spin-orbit-coupled systems (see Sec.~\ref{sec:LC-gen-model-sym}) even at high-symmetry points $\vb{k}_{\star}$, one may verify that they hold for the Van Hove lines in the effective $t_{2g}$ model of strontium ruthenate of Sec.~\ref{sec:SRO-el-struct}.
Notice also that $\MatU(P) = \one$ so rotations and reflections act in the same way on the eigenvectors and Hamiltonian.

The symmetry transformation rule of spin-orbit matrices (Eq.~\eqref{eq:SRO-transf-Gamma}, Sec.~\ref{sec:SRO-SC-construct}) now gives the following constraint on the spin-orbit matrix projections:
\begin{align}
S^{\dag}(g) \mathscr{P}_{a}(\vb{k}_{\star}) S^{*}(g) &= \sum_{b = 1}^{\dim \zeta} \RepM_{ab}^{\zeta}(g) \mathscr{P}_{b}(\vb{k}_{\star}).
\end{align}
For $\vb{k}_{\star}$ on the Van Hove line $\mleft(0, \frac{\pi}{a}, k_z\mright)$, the $g$ from Tab.~\ref{tab:SRO-rho-kvH} constrain certain $\mathscr{P}_{a}^{\mu}(\vb{k}_{\star})$ to vanish, depending on the (anti-)symmetry, irrep, and $k_z = k_{z,\star}$.
To write down the constraints more explicitly, let us note that all $\RepM^{\zeta}(g)$ are diagonal, that $(\iu \Pauli_y) S^{*}(g) = S(g) (\iu \Pauli_y)$, and also that $\Pauli_{0}$ is a scalar, while $\Pauli_{i}$ transforms like a pseudovector ($E_g \oplus A_{2g}$).
Hence for antisymmetric $\Gamma_a^{\intercal} = - \Gamma_a$:
\begin{align}
\mathscr{P}_{a}^{0}(\vb{k}_{\star}) &= \RepM_{aa}^{\zeta}(g) \mathscr{P}_{a}^{0}(\vb{k}_{\star}),
\end{align}
while for symmetric $\Gamma_a^{\intercal} = + \Gamma_a$:
\begin{align}
\begin{aligned}
\mathscr{P}_{a}^{x}(\vb{k}_{\star}) &= \RepM_{11}^{E_g}(g) \RepM_{aa}^{\zeta}(g) \mathscr{P}_{a}^{x}(\vb{k}_{\star}), \\
\mathscr{P}_{a}^{y}(\vb{k}_{\star}) &= \RepM_{22}^{E_g}(g) \RepM_{aa}^{\zeta}(g) \mathscr{P}_{a}^{y}(\vb{k}_{\star}), \\
\mathscr{P}_{a}^{z}(\vb{k}_{\star}) &= \RepM^{A_{2g}}(g) \RepM_{aa}^{\zeta}(g) \mathscr{P}_{a}^{z}(\vb{k}_{\star}).
\end{aligned}
\end{align}
As previously, when $\mathscr{P}_{a}^{\mu}$ is equal to minus itself due to some symmetry, it vanishes.
The (anti-)symmetry $\Gamma_a^{\intercal} = p_{\Gamma} \Gamma_a$ we shall denote with an irrep superscript $s$ ($a$) when $p_{\Gamma} = +1$ ($p_{\Gamma} = -1$).
Thus, for instance, $\Gamma \in A_{1g}^{a}$ are antisymmetric under transposition, whereas $\Gamma \in B_{1g}^{s}$ are symmetric under transposition.
The symmetry-enforced behavior of $\mathscr{P}_{a}^{\mu}\mleft(0, \frac{\pi}{a}, k_z\mright)$ we may summarize as follows:
\begin{itemize}
\item $\Gamma$ belonging to $A_{2g}^{a}$ and $B_{2g}^{a}$ have $\mathscr{P}^{0} = 0$ for all $k_z$.
\item $(\Gamma_1|\Gamma_2) \in E_g^{a}$ have $\mathscr{P}_{2}^{0} = 0$ for all $k_z$, whereas $\mathscr{P}_{1}^{0} = 0$ only at $k_z = 0, \pm \frac{2 \pi}{c}$.
\item $\Gamma \in A_{1g}^{s}$ and $B_{1g}^{s}$ have $\mathscr{P}^{y} = \mathscr{P}^{z} = 0$ for all $k_z$, and $\mathscr{P}^{x} = 0$ only at $k_z = 0, \pm \frac{2 \pi}{c}$.
\item $\Gamma \in A_{2g}^{s}$ and $B_{2g}^{s}$ have $\mathscr{P}^{x} = 0$ for all $k_z$, and $\mathscr{P}^{y} = 0$ only at $k_z = 0, \pm \frac{2 \pi}{c}$. $\mathscr{P}^{z}$ is unconstrained.
\item $(\Gamma_1|\Gamma_2) \in E_g^{s}$ have $\mathscr{P}_{1}^{y} = \mathscr{P}_{1}^{z} = \mathscr{P}_{2}^{x} = 0$ for all $k_z$, and $\mathscr{P}_{2}^{z} = 0$ only at $k_z = 0, \pm \frac{2 \pi}{c}$. The remaining $\mathscr{P}_{1}^{x}$ and $\mathscr{P}_{2}^{y}$ are unconstrained.
\item The $\mathscr{P}^{0}$ of $\Gamma$ from $A_{1g}^{a}$ and $B_{1g}^{a}$ are completely unconstrained for all $k_z$.
\end{itemize}
In the limit of vanishing body-centered tetragonal hopping, the following $\mathscr{P}_{a}^{\mu}$ vanish in addition:
\begin{itemize}
\item For $(\Gamma_1|\Gamma_2) \in E_g^{a}$, $\mathscr{P}_{1}^{0}$ vanishes for all $k_z$ so both $\mathscr{P}_{a}^{0}$ are zero.
\item For $\Gamma \in A_{1g}^{s}$ and $B_{1g}^{s}$, $\mathscr{P}^{\mu}$ completely vanish for all $k_z$.
\item For $\Gamma \in A_{2g}^{s}$ and $B_{2g}^{s}$, $\mathscr{P}^{y} = 0$ for all $k_z$, but $\mathscr{P}^{z}$ is still unconstrained.
\item For $(\Gamma_1|\Gamma_2) \in E_g^{s}$, $\mathscr{P}_{2}^{z} = 0$ for all $k_z$, but $\mathscr{P}_{1}^{x}$ and $\mathscr{P}_{2}^{y}$ are still unconstrained.
\end{itemize}
Owning to the fact that all characteristically body-centered hopping is necessarily between layers and that these hoppings are very small in SRO because of its high anisotropy, the vanishing $\mathscr{P}_{a}^{\mu}$ listed above are very small for SRO, although not precisely zero.
Using the tight-binding model of Ref.~\cite{Roising2019}, described in Sec.~\ref{sec:SRO-el-struct}, we have quantified their smallness: the vanishing $\mathscr{P}_{a}^{\mu}$ listed above are by a factor of $50$ or more smaller than the largest possible $\mathscr{P}_{a}^{\mu} \sim 1$, where all $\Gamma_a$ have been normalized to $\tr \Gamma_a^{\dag} \Gamma_a = 1$ for a fair comparison.
Note that we did not analyze odd-parity spin-orbit matrices because they do not arise in the $t_{2g}$ model of SRO, as follows from the fact that all orbitals are even; see Tab.~\ref{tab:SRO-Gamma-class}.

Unlike the above anisotropy argument, arguments based on the $d_{xy}$ orbital character of the $\gamma$ band do not suppress any irreps, but only inform us on which $\Gamma_a$ from within a given irrep have large $\mathscr{P}_{a}^{\mu}$.

\begin{table}[t!]
\centering
\captionabove[Even-parity and odd-parity superconducting states that do not have a vertical line node at the Van Hove line $\mleft(0, \frac{\pi}{a}, k_z\mright)$~\cite{Palle2023-ECE}.]{\textbf{Even-parity and odd-parity superconducting states that do not have a vertical line node at the Van Hove line $\mleft(0, \frac{\pi}{a}, k_z\mright)$}~\cite{Palle2023-ECE}.
These states are constructed by combining pairing wavefunctions $f_a(\vb{k})$ with spin-orbit matrices $\Gamma_a$ according to the multiplication table of $D_{4h}$ irreps provided in Tab.~\ref{tab:D4h-irrep-prod-tab} of Sec.~\ref{sec:multid-irrep-product}.
An $s$ superscript on a spin-orbit matrix irrep means that the matrices are symmetric ($\Gamma^{\intercal} = + \Gamma$), whereas an $a$ superscript indicates antisymmetry under transposition.
A zero component of $E_{g/u}$ indicates that it vanishes identically on $\mleft(0, \frac{\pi}{a}, k_z\mright)$.
Highlighted red are those $f_a$ that must be periodic under body-centered translations, but not under simple tetragonal translations, to be finite on $\mleft(0, \frac{\pi}{a}, k_z\mright)$.
For examples, see Tab.~\ref{tab:SRO-d-func-class} from Sec.~\ref{sec:SRO-SC-construct}.
Such $f_a$ have horizontal line nodes at $k_z = 0, \pm \frac{2 \pi}{c}$.
Highlighted blue are those $\Gamma_a$ whose projections onto the $\gamma$ band are suppressed by two orders of magnitude because of the weakness of body-centered interlayer hopping.
Such $\Gamma_a$ are unable to account for the elastocaloric experiment of Ref.~\cite{Li2022}, but are listed for the sake of completeness.}
{\renewcommand{\arraystretch}{1.3}
\renewcommand{\tabcolsep}{10pt}
\begin{tabular}{|c"c|c|c|}
\multicolumn{4}{c}{\large Even-parity pairings that are finite on $\mleft(0, \frac{\pi}{a}, k_z\mright)$:} \\[2pt] \hline
$\otimes$ & $A_{1g}(f)$ & $B_{1g}(f)$ & $E_g(\textcolor{red}{f_1} | 0)$
\\ \thickhline
$A_{1g}^{a}(\Gamma)$ & $A_{1g}(\Gamma f)$ & $B_{1g}(\Gamma f)$ & $E_g(\Gamma \textcolor{red}{f_1} | 0)$
\\ \hline
$B_{1g}^{a}(\Gamma)$ & $B_{1g}(\Gamma f)$ & $A_{1g}(\Gamma f)$ & $E_g(\Gamma \textcolor{red}{f_1} | 0)$
\\ \hline
$E_g^{a}(\textcolor{blue}{\Gamma_1} | 0)$
& $E_g(\textcolor{blue}{\Gamma_1} f | 0)$
& $E_g(\textcolor{blue}{\Gamma_1} f | 0)$
& $A_{1g}(\textcolor{blue}{\Gamma_1} \textcolor{red}{f_1} + 0) \oplus B_{1g}(\textcolor{blue}{\Gamma_1} \textcolor{red}{f_1} - 0)$
\\ \hline
\multicolumn{4}{c}{} \\[-2pt]
\multicolumn{4}{c}{\large Odd-parity pairings that are finite on $\mleft(0, \frac{\pi}{a}, k_z\mright)$:} \\[2pt] \hline
$\otimes$ & $A_{2u}(f)$ & $B_{2u}(f)$ & $E_u(0 | \textcolor{red}{f_2})$
\\ \thickhline
$A_{1g}^{s}(\textcolor{blue}{\Gamma})$ & $A_{2u}(\textcolor{blue}{\Gamma} f)$ & $B_{2u}(\textcolor{blue}{\Gamma} f)$ & $E_u(0 | \textcolor{blue}{\Gamma} \textcolor{red}{f_2})$
\\ \hline
$A_{2g}^{s}(\Gamma)$ & $A_{1u}(\Gamma f)$ & $B_{1u}(\Gamma f)$ & $E_u(\Gamma \textcolor{red}{f_2} | 0)$
\\ \hline
$B_{1g}^{s}(\textcolor{blue}{\Gamma})$ & $B_{2u}(\textcolor{blue}{\Gamma} f)$ & $A_{2u}(\textcolor{blue}{\Gamma} f)$ & $E_u(0 | - \textcolor{blue}{\Gamma} \textcolor{red}{f_2})$
\\ \hline
$B_{2g}^{s}(\Gamma)$ & $B_{1u}(\Gamma f)$ & $A_{1u}(\Gamma f)$ & $E_u(\Gamma \textcolor{red}{f_2} | 0)$
\\ \hline
$E_g^{s}(\Gamma_1 | \Gamma_2)$
& $E_u(\Gamma_2 f | - \Gamma_1 f)$
& $E_u(\Gamma_2 f | \Gamma_1 f)$
& $\begin{matrix}
A_{1u}(0 + \Gamma_2 \textcolor{red}{f_2}) \oplus A_{2u}(\Gamma_1 \textcolor{red}{f_2} - 0) \\
{} \oplus B_{1u}(0 - \Gamma_2 \textcolor{red}{f_2}) \oplus B_{2u}(\Gamma_1 \textcolor{red}{f_2} + 0)
\end{matrix}$
\\ \hline
\end{tabular}}
\label{tab:SRO-main-result}
\end{table}

Finally, we synthesize the results found for $f_a$ and $\Gamma_a$.
This is done by going through the multiplication table of $D_{4h}$ irreps (Tab.~\ref{tab:D4h-irrep-prod-tab} in Sec.~\ref{sec:multid-irrep-product} of Appx.~\ref{app:group_theory}) and seeing which entries yield a $\Delta_a(\vb{k})$ with a finite $\gamma$ band projection.
The results are summarized in Tab.~\ref{tab:SRO-main-result}, which is reproduced from Ref.~\cite{Palle2023-ECE}.
Tab.~\ref{tab:SRO-main-result} is the main result of the current analysis (and Ref.~\cite{Palle2023-ECE}).
As mentioned, SRO's anisotropy suppresses the blue entries of the table by two orders of magnitude.
This means that a $\Delta$ with a maximal value $\sim k_B T_c$ is way too small on the Van Hove lines to explain the observed entropy quenching~\cite{Li2022}.
Hence the blue entries of Tab.~\ref{tab:SRO-main-result} are excluded as possible leading SC states as well.

From Tab.~\ref{tab:SRO-main-result} we see that, among even-parity pairings, only $A_{1g}$, $B_{1g}$, and $E_g$ irreps have SC states that do not have symmetry-enforced vertical line nodes on the Van Hove lines.
Thus even-parity pairings must have admixtures from one of these three irreps to be able to explain the elastocaloric experiment of Ref.~\cite{Li2022}.
It is worth noting that within these three irreps, pairings with symmetry-enforced vertical line nodes on $\vb{k}_{\text{VH}}$ do exist, like for instance $\Delta(\vb{k}) = \Lambda_1 (\iu \Pauli_{y}) \sin a k_x \sin a k_y \in B_{2g}^{a} \otimes B_{2g} = A_{1g}$; the Gell-Mann matrix $\Lambda_1$ is defined in Eq.~\eqref{eq:SRO-GM-mat-list} of Sec.~\ref{sec:SRO-SC-construct}.
So Tab.~\ref{tab:SRO-main-result} also yields non-trivial information on the spin-orbit and momentum structure of these Van Hove line-gapping admixtures.

One such piece of information is that $E_g$ pairing must be made of wavefunctions $f_a$ that are body-centered periodic, but not simple tetragonal periodic.
The lowest order such $(d_{yz} | - d_{xz}) \in E_g$ is (Tab.~\ref{tab:SRO-d-func-class}):
\begin{equation}
\mleft(\cos\frac{a k_x}{2} \sin\frac{a k_y}{2} \sin\frac{c k_z}{2} \middle| - \sin\frac{a k_x}{2} \cos\frac{a k_y}{2} \sin\frac{c k_z}{2}\mright).
\label{eq:SRO-nastyone}
\end{equation}
It is this pairing state, only allowed because of the body-centered tetragonal structure of SRO, that opens a gap at the Van Hove line and that we cannot exclude based on the elastocaloric data.
In Ref.~\cite{Suh2020} it was shown that such a pairing state can be stabilized by a strongly momentum-dependent spin-orbit coupling.
A better understanding of the origin of such momentum dependence might help elucidate whether this state is a viable option for SRO's SC.
In distinction, the $E_g$ pairing state
\begin{equation}
(\sin a k_y  \sin c k_z | - \sin a k_x  \sin c k_z), \label{eq:SRO-usualone}
\end{equation}
which would be the only allowed one for simple-tetragonal lattices, cannot be the only pairing state as it does not open a gap on the Van Hove line.
An important difference between these two types of states [Eq.~\eqref{eq:SRO-nastyone} vs.~\eqref{eq:SRO-usualone}] is that the former always have horizontal line nodes at $k_z = 0, \pm \frac{2 \pi}{c}$.

\begin{figure}[p!]
\centering
\includegraphics[width=0.62\columnwidth]{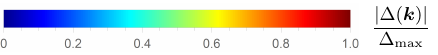}\\[4pt]
\begin{subfigure}{\textwidth}
\centering
\includegraphics[width=\textwidth]{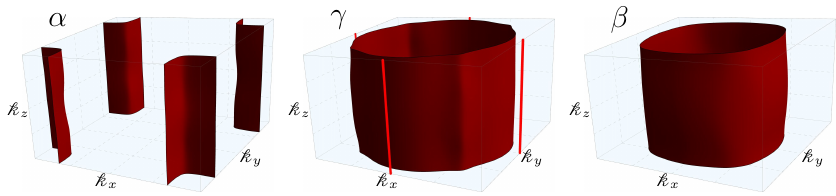}
\subcaption{$\Delta(\vb{k}) = \mleft(\sqrt{2} \Lambda_0 + \Lambda_4\mright) (\iu \Pauli_{y}) \in A_{1g}$}
\end{subfigure}\\[8pt]
\begin{subfigure}{\textwidth}
\centering
\includegraphics[width=\textwidth]{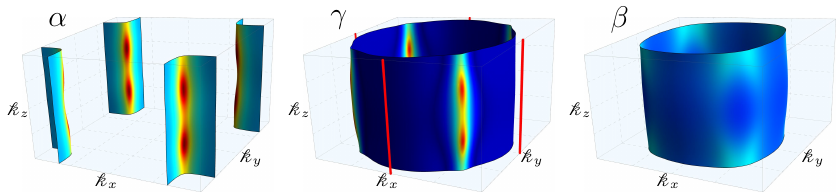}
\subcaption{$\Delta(\vb{k}) = \Lambda_2 \Pauli_z (\iu \Pauli_{y}) (\iu \Pauli_{y}) \in A_{1g}$}
\end{subfigure}\\[8pt]
\begin{subfigure}{\textwidth}
\centering
\includegraphics[width=\textwidth]{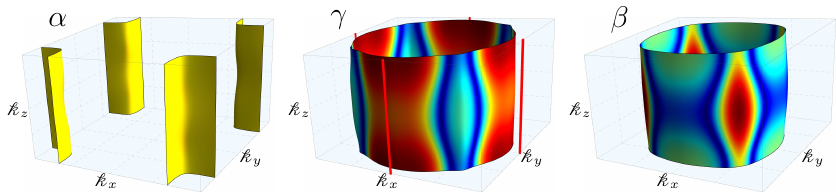}
\subcaption{$\Delta(\vb{k}) = \mleft(\Lambda_0 - \sqrt{2} \Lambda_4\mright) (\iu \Pauli_{y}) \in A_{1g}$}
\end{subfigure}\\[8pt]
\begin{subfigure}{\textwidth}
\centering
\includegraphics[width=\textwidth]{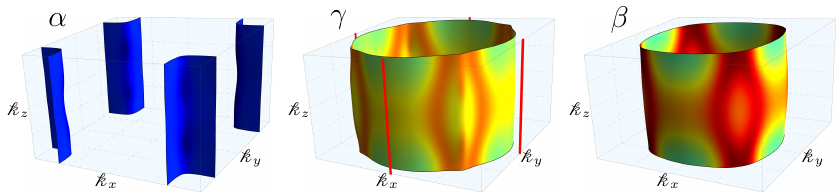}
\subcaption{$\Delta(\vb{k}) = \mleft(\Lambda_6 \Pauli_{y} - \Lambda_8 \Pauli_{x}\mright) (\iu \Pauli_{y}) \in A_{1g}$}
\end{subfigure}
\captionbelow[Projections onto the Fermi sheets of Van Hove line-gapping superconducting states.]{\textbf{Projections onto the Fermi sheets of Van Hove line-gapping superconducting states.}
$\DimK_x = a k_x \in [- \pi, \pi]$, $\DimK_y = a k_y \in [- \pi, \pi]$, and $\DimK_z = c k_z \in [- 2 \pi, 2 \pi]$.
In the $\gamma$ sheet plots, the Van Hove lines $\mleft(\pm \frac{\pi}{a}, 0, k_z\mright)$ and $\mleft(0, \pm \frac{\pi}{a}, k_z\mright)$ are highlighted red.}
\label{fig:SRO-SCamps1}
\end{figure}

\begin{figure}[p!]
\centering
\includegraphics[width=0.62\columnwidth]{SRO-SCamps/legend.pdf}\\[4pt]
\begin{subfigure}{\textwidth}
\centering
\includegraphics[width=\textwidth]{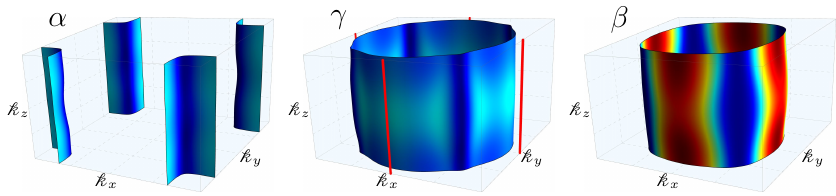}
\subcaption{$\Delta(\vb{k}) = \Lambda_3 (\iu \Pauli_{y}) \mleft(\cos \DimK_x - \cos \DimK_y\mright) \in A_{1g}$}
\end{subfigure}\\[8pt]
\begin{subfigure}{\textwidth}
\centering
\includegraphics[width=\textwidth]{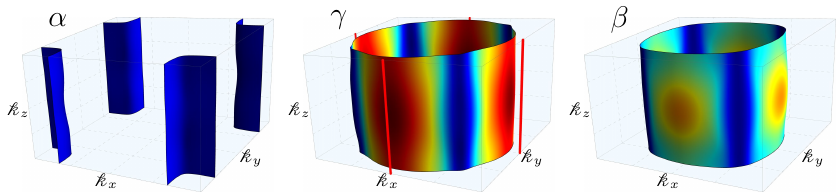}
\subcaption{$\Delta(\vb{k}) = \mleft(\Lambda_6 \Pauli_{y} + \Lambda_8 \Pauli_{x}\mright) (\iu \Pauli_{y}) \mleft(\cos \DimK_x - \cos \DimK_y\mright) \in A_{1g}$}
\end{subfigure}\\[8pt]
\begin{subfigure}{\textwidth}
\centering
\includegraphics[width=\textwidth]{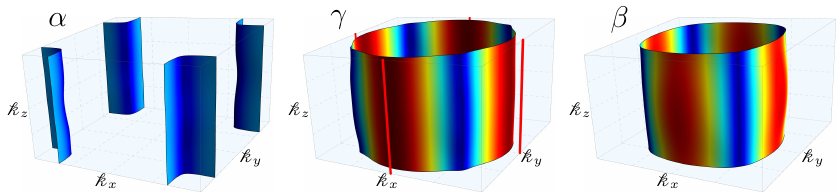}
\subcaption{$\Delta(\vb{k}) = \mleft(\sqrt{2} \Lambda_0 + \Lambda_4\mright) (\iu \Pauli_{y}) \mleft(\cos \DimK_x - \cos \DimK_y\mright) \in B_{1g}$}
\end{subfigure}\\[8pt]
\begin{subfigure}{\textwidth}
\centering
\includegraphics[width=\textwidth]{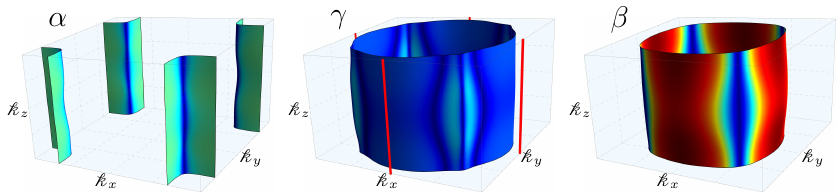}
\subcaption{$\Delta(\vb{k}) = \Lambda_2 \Pauli_z (\iu \Pauli_{y}) \mleft(\cos \DimK_x - \cos \DimK_y\mright) \in B_{1g}$}
\end{subfigure}
\captionbelow[Projections onto the Fermi sheets of Van Hove line-gapping superconducting states (continued).]{\textbf{Projections onto the Fermi sheets of Van Hove line-gapping superconducting states (continued).}
$\DimK_x = a k_x \in [- \pi, \pi]$, $\DimK_y = a k_y \in [- \pi, \pi]$, and $\DimK_z = c k_z \in [- 2 \pi, 2 \pi]$.
In the $\gamma$ sheet plots, the Van Hove lines $\mleft(\pm \frac{\pi}{a}, 0, k_z\mright)$ and $\mleft(0, \pm \frac{\pi}{a}, k_z\mright)$ are highlighted red.}
\label{fig:SRO-SCamps2}
\end{figure}

\begin{figure}[p!]
\centering
\includegraphics[width=0.62\columnwidth]{SRO-SCamps/legend.pdf}\\[4pt]
\begin{subfigure}{\textwidth}
\centering
\includegraphics[width=\textwidth]{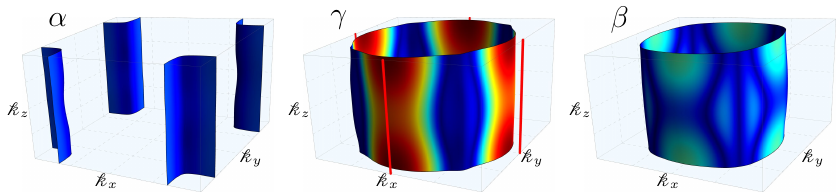}
\subcaption{$\Delta(\vb{k}) = \mleft(\Lambda_0 - \sqrt{2} \Lambda_4\mright) (\iu \Pauli_{y}) \mleft(\cos \DimK_x - \cos \DimK_y\mright) \in B_{1g}$}
\end{subfigure}\\[8pt]
\begin{subfigure}{\textwidth}
\centering
\includegraphics[width=\textwidth]{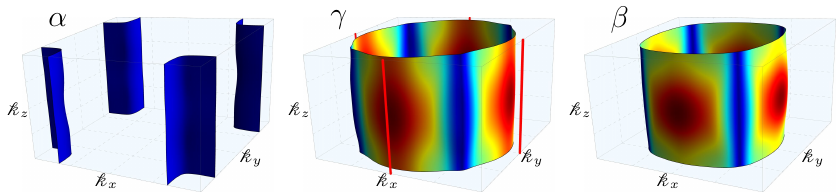}
\subcaption{$\Delta(\vb{k}) = \mleft(\Lambda_6 \Pauli_{y} - \Lambda_8 \Pauli_{x}\mright) (\iu \Pauli_{y}) \mleft(\cos \DimK_x - \cos \DimK_y\mright) \in B_{1g}$}
\end{subfigure}\\[8pt]
\begin{subfigure}{\textwidth}
\centering
\includegraphics[width=\textwidth]{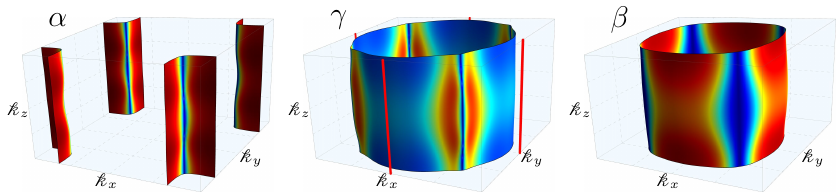}
\subcaption{$\Delta(\vb{k}) = \Lambda_3 (\iu \Pauli_{y}) \in B_{1g}$}
\end{subfigure}\\[8pt]
\begin{subfigure}{\textwidth}
\centering
\includegraphics[width=\textwidth]{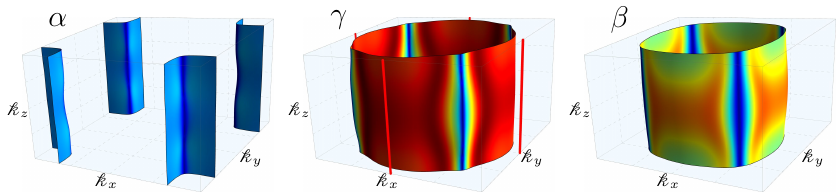}
\subcaption{$\Delta(\vb{k}) = \mleft(\Lambda_6 \Pauli_{y} + \Lambda_8 \Pauli_{x}\mright) (\iu \Pauli_{y}) \in B_{1g}$}
\end{subfigure}
\captionbelow[Projections onto the Fermi sheets of Van Hove line-gapping superconducting states (continued).]{\textbf{Projections onto the Fermi sheets of Van Hove line-gapping superconducting states (continued).}
$\DimK_x = a k_x \in [- \pi, \pi]$, $\DimK_y = a k_y \in [- \pi, \pi]$, and $\DimK_z = c k_z \in [- 2 \pi, 2 \pi]$.
In the $\gamma$ sheet plots, the Van Hove lines $\mleft(\pm \frac{\pi}{a}, 0, k_z\mright)$ and $\mleft(0, \pm \frac{\pi}{a}, k_z\mright)$ are highlighted red.}
\label{fig:SRO-SCamps3}
\end{figure}

\begin{figure}[p!]
\centering
\includegraphics[width=0.62\columnwidth]{SRO-SCamps/legend.pdf}\\[4pt]
\begin{subfigure}{\textwidth}
\centering
\includegraphics[width=\textwidth]{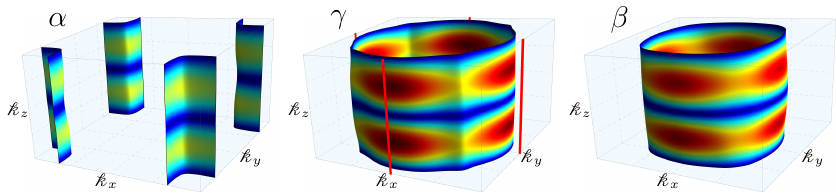}
\subcaption{$\Delta(\vb{k}) = \mleft(\sqrt{2} \Lambda_0 + \Lambda_4\mright) (\iu \Pauli_{y}) \mleft(\sin\frac{1}{2}\DimK_x \cos\frac{1}{2}\DimK_y \pm \iu \cos\frac{1}{2}\DimK_x \sin\frac{1}{2}\DimK_y\mright) \sin\frac{1}{2}\DimK_z \in E_{g}$}
\end{subfigure}\\[8pt]
\begin{subfigure}{\textwidth}
\centering
\includegraphics[width=\textwidth]{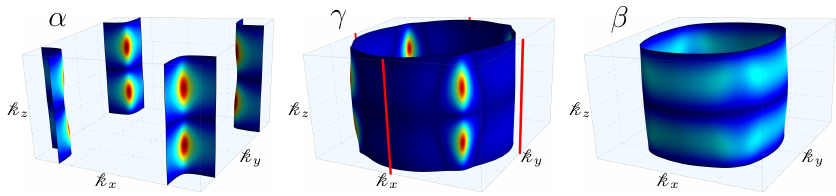}
\subcaption{$\Delta(\vb{k}) = \Lambda_2 \Pauli_z (\iu \Pauli_{y}) \mleft(\sin\frac{1}{2}\DimK_x \cos\frac{1}{2}\DimK_y \pm \iu \cos\frac{1}{2}\DimK_x \sin\frac{1}{2}\DimK_y\mright) \sin\frac{1}{2}\DimK_z \in E_{g}$}
\end{subfigure}\\[8pt]
\begin{subfigure}{\textwidth}
\centering
\includegraphics[width=\textwidth]{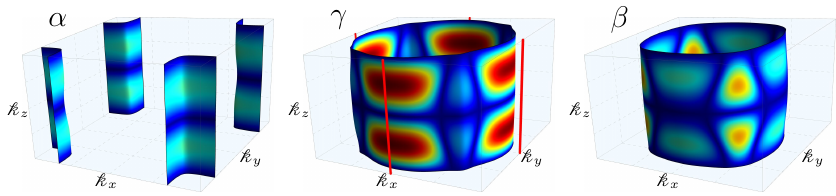}
\subcaption{$\Delta(\vb{k}) = \mleft(\Lambda_0 - \sqrt{2} \Lambda_4\mright) (\iu \Pauli_{y}) \mleft(\sin\frac{1}{2}\DimK_x \cos\frac{1}{2}\DimK_y \pm \iu \cos\frac{1}{2}\DimK_x \sin\frac{1}{2}\DimK_y\mright) \sin\frac{1}{2}\DimK_z \in E_{g}$}
\end{subfigure}\\[8pt]
\begin{subfigure}{\textwidth}
\centering
\includegraphics[width=\textwidth]{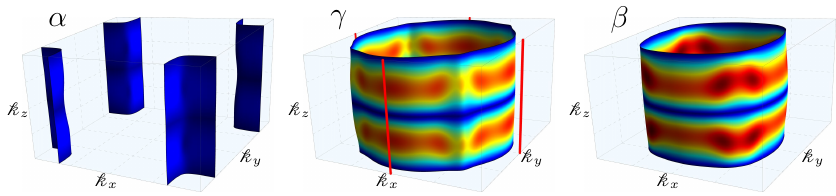}
\subcaption{$\Delta(\vb{k}) = \mleft(\Lambda_6 \Pauli_{y} - \Lambda_8 \Pauli_{x}\mright) (\iu \Pauli_{y}) \mleft(\sin\frac{1}{2}\DimK_x \cos\frac{1}{2}\DimK_y \pm \iu \cos\frac{1}{2}\DimK_x \sin\frac{1}{2}\DimK_y\mright) \sin\frac{1}{2}\DimK_z \in E_{g}$}
\end{subfigure}
\captionbelow[Projections onto the Fermi sheets of Van Hove line-gapping superconducting states (continued).]{\textbf{Projections onto the Fermi sheets of Van Hove line-gapping superconducting states (continued).}
$\DimK_x = a k_x \in [- \pi, \pi]$, $\DimK_y = a k_y \in [- \pi, \pi]$, and $\DimK_z = c k_z \in [- 2 \pi, 2 \pi]$.
In the $\gamma$ sheet plots, the Van Hove lines $\mleft(\pm \frac{\pi}{a}, 0, k_z\mright)$ and $\mleft(0, \pm \frac{\pi}{a}, k_z\mright)$ are highlighted red.}
\label{fig:SRO-SCamps4}
\end{figure}

\begin{figure}[t!]
\centering
\includegraphics[width=0.62\columnwidth]{SRO-SCamps/legend.pdf}\\[4pt]
\begin{subfigure}{\textwidth}
\centering
\includegraphics[width=\textwidth]{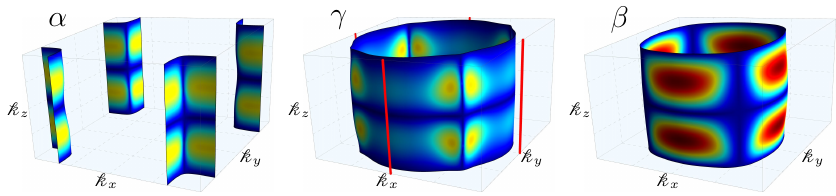}
\subcaption{$\Delta(\vb{k}) = \Lambda_3 (\iu \Pauli_{y}) \mleft(\sin\frac{1}{2}\DimK_x \cos\frac{1}{2}\DimK_y \pm \iu \cos\frac{1}{2}\DimK_x \sin\frac{1}{2}\DimK_y\mright) \sin\frac{1}{2}\DimK_z \in E_{g}$}
\end{subfigure}\\[8pt]
\begin{subfigure}{\textwidth}
\centering
\includegraphics[width=\textwidth]{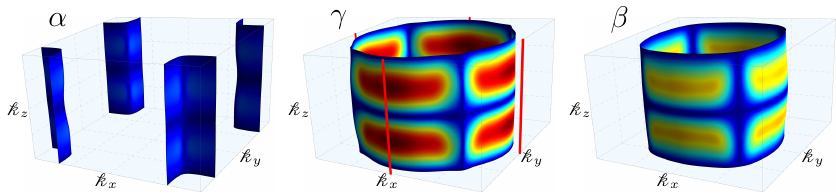}
\subcaption{$\Delta(\vb{k}) = \mleft(\Lambda_6 \Pauli_{y} + \Lambda_8 \Pauli_{x}\mright) \mleft(\sin\frac{1}{2}\DimK_x \cos\frac{1}{2}\DimK_y \pm \iu \cos\frac{1}{2}\DimK_x \sin\frac{1}{2}\DimK_y\mright) \sin\frac{1}{2}\DimK_z \in E_{g}$}
\end{subfigure}
\captionbelow[Projections onto the Fermi sheets of Van Hove line-gapping superconducting states (continued).]{\textbf{Projections onto the Fermi sheets of Van Hove line-gapping superconducting states (continued).}
$\DimK_x = a k_x \in [- \pi, \pi]$, $\DimK_y = a k_y \in [- \pi, \pi]$, and $\DimK_z = c k_z \in [- 2 \pi, 2 \pi]$.
In the $\gamma$ sheet plots, the Van Hove lines $\mleft(\pm \frac{\pi}{a}, 0, k_z\mright)$ and $\mleft(0, \pm \frac{\pi}{a}, k_z\mright)$ are highlighted red.}
\label{fig:SRO-SCamps5}
\end{figure}

In Figs.~\ref{fig:SRO-SCamps1} to \ref{fig:SRO-SCamps5}, we have plotted the Fermi surface-projections of a number of Van Hove line-gapping even-parity SC states from Tab.~\ref{tab:SRO-main-result}.
These have been constructed by combining the six $A_{1g}^{a}$ and $B_{1g}^{a}$ spin-orbit matrices (Tab.~\ref{tab:SRO-Gamma-class}) with the lowest order $A_{1g}$, $B_{1g}$, and $E_{g}$ pairing wavefunctions (Tab.~\ref{tab:SRO-d-func-class}).
Note that $\sqrt{2} \Lambda_0 + \Lambda_4 = \sqrt{2} \one$.
$\Delta(\vb{k})$ constructed from the highly suppressed $E_g^{a}$ spin-orbit matrices (blue in Tab.~\ref{tab:SRO-Gamma-class}) are not shown.
Of all the possible superpositions in the case of $E_g$ pairing (see Sec.~\ref{sec:SRO-GL-analysis}), we have plotted the chiral ones as they are the most interesting because of the various evidence~\cite{Luke1998, Luke2000, Higemoto2014, Grinenko2021-unaxial, Grinenko2021-isotropic, Xia2006, Kapitulnik2009} indicating TRSB.
The most general Van Hove line-gapping $\Delta(\vb{k})$ belonging to $A_{1g}$, $B_{1g}$, or chiral $E_g$ is a superposition of the depicted ones, plus higher order harmonics.
In the figures $\DimK_x = a k_x \in [- \pi, \pi]$, $\DimK_y = a k_y \in [- \pi, \pi]$, and $\DimK_z = c k_z \in [- 2 \pi, 2 \pi]$.
In the middle $\gamma$ sheet plots, the Van Hove lines $\mleft(\pm \frac{\pi}{a}, 0, k_z\mright)$ and $\mleft(0, \pm \frac{\pi}{a}, k_z\mright)$ have been highlighted red.
Even though the projections of some $\Delta(\vb{k})$ onto the $\gamma$ band might be small (shaded blue) near the Van Hove lines [Fig.~\ref{fig:SRO-SCamps1}(b), Fig.~\ref{fig:SRO-SCamps2}(a)\&(d), Fig.~\ref{fig:SRO-SCamps3}(c), Fig.~\ref{fig:SRO-SCamps4}(b), Fig.~\ref{fig:SRO-SCamps5}(a)], they are only exactly zero at a certain $\DimK_z$ for the $\Delta(\vb{k}) \in E_g$ that have horizontal nodes at $\DimK_z = 0, \pm 2 \pi$.

\newpage

Among odd pairings, all irreps have pairings without symmetry-enforced vertical line nodes on $\vb{k}_{\text{VH}}$.
However, the orientations of the Balian-Werthamer $\vb{d}$-vectors~\cite{Balian1963} are non-trivially restricted and the non-suppressed $A_{2u}$ and $B_{2u}$ pairings are necessarily made of characteristically body-centered periodic wavefunctions $f_a$.
They thus have horizontal line nodes.

In multiband systems with spin-orbit coupling, a $\vb{d}$-vector is associated with each band in its pseudospin (Kramers') space.
It is defined through:
\begin{align}
u_{\vb{k} n s'}^{\dag} \Delta(\vb{k}) u_{- \vb{k} n s}^{*} &\equiv \mleft[\vb{\mathscr{d}}_{\vb{k} n} \vdot \vb{\Pauli_{}} (\iu \Pauli_{y})\mright]_{s's},
\end{align}
where $u_{\vb{k} n s}$ are the Kramers-degenerate eigenvectors of the $n$-th band.
They satisfy $u_{- \vb{k} n s} = u_{\vb{k} n s}$ because $\MatU(P) = \one$.
We make the following gauge choice for the pseudospins:
\begin{align}
\begin{aligned}
u_{\vb{k} n s'}^{\dag} (\one \otimes \iu \Pauli_{y}) u_{\vb{k} n s}^{*} &= \mleft[\iu \Pauli_{y}\mright]_{s's}, \\
u_{\vb{k} n s'}^{\dag} (\one \otimes \Pauli_{z}) u_{\vb{k} n s} &= \mleft[\iota_z \Pauli_{z}\mright]_{s's}, \\
u_{\vb{k} n s'}^{\dag} (\one \otimes \Pauli_{x}) u_{\vb{k} n s} &= \mleft[\iota_x \Pauli_{x} + \delta_{xz} \Pauli_{z}\mright]_{s's},
\end{aligned} \label{eq:SRO-pseudo-gauge}
\end{align}
where $\iota_z, \iota_x, \delta_{xz} \in \R$.
This is the closest one can make the pseudospins look like spins for general momenta $\vb{k}$.
In general $\delta_{xz}$ is not zero, nor are the $\delta_{yx}, \delta_{yz}$ from $u_{\vb{k} n s'}^{\dag} (\one \otimes \Pauli_{y}) u_{\vb{k} n s} = \mleft[\iota_y \Pauli_{y} + \delta_{yx} \Pauli_{x} + \delta_{yz} \Pauli_{z}\mright]_{s's}$.
However, in SRO the only regions where $\delta_{xz}, \delta_{yx}, \delta_{yz}$ are substantially different from zero is at the nesting of the $\alpha$, $\beta$, and $\gamma$ bands at $k_x = \pm k_y$ (see Fig.~\ref{fig:SRO-theoretical-FS}).
The explanation for this is the fact that spin-orbit coupling most strongly affects the band structure there, as we discussed in Sec.~\ref{sec:SRO-el-struct}.

Using the $t_{2g}$ orbital-based tight-binding model of SRO that we introduces in Sec.~\ref{sec:SRO-el-struct} [Eq.~\eqref{eq:SRO-TBA-Haml}], we have explored the orientation of the $\vb{\mathscr{d}}_{\vb{k} n}$-vectors on the $\alpha$, $\beta$, and $\gamma$ Fermi sheets.
Everywhere except near the $k_x = \pm k_y$ nesting of the sheets, we find that symmetric spin-orbit matrices from 1D irreps have $\vb{\mathscr{d}}_{\vb{k} n}$ pointing along $\pm \vu{e}_z$, whereas $(\Gamma_1|\Gamma_2)$ from $E_g^{s}$ always have in-plane $\vb{\mathscr{d}}_{\vb{k} n}$.
So the non-suppressed $A_{2u}$ and $B_{2u}$ from Tab.~\ref{tab:SRO-main-result}(b) have $\vb{\mathscr{d}}_{\vb{k} n} \parallel \vu{e}_z$.
Moreover, among odd-parity pairings not made of body-centered $f_a(\vb{k})$, $A_{1u}$ and $B_{1u}$ pairings have $\vb{\mathscr{d}}_{\vb{k} n} \parallel \vu{e}_z$ and $E_u$ pairings have in-plane $\vb{\mathscr{d}}_{\vb{k} n}$.
Given that body-centered $(f_1|f_2) \in E_u$ have horizontal line nodes, on the one hand, and that the spin susceptibility is intimately related to the orientation of the Balian-Werthamer $\vb{d}$-vector, on the other, this information may prove to be useful in further narrowing down the odd-pairing SC candidates.

\subsection{Discussion} \label{sec:SRO-ECE-art-discussion}
The article~\cite{Palle2023-ECE} on which the current section is based was motivated by the measurements of the elastocaloric effect of \ce{Sr2RuO4} under strain which were reported in Ref.~\cite{Li2022} (Fig.~\ref{fig:SRO-ECE-data}).
The elastocaloric effect measures, with high accuracy, the entropy derivative $\partial S(\epsilon, T) / \partial \epsilon$.
Above $T_c$, the elastocaloric effect revealed a pronounced maximum in the entropy as function of $\langle 100 \rangle$ strain $\epsilon_{100}$.
As demonstrated in Ref.~\cite{Li2022}, this maximum of $S(\epsilon)$ can be fully accounted for by the DOS enhancement that occurs when the Fermi energy crosses the Van Hove points near the lines $\mleft(0, \pm \frac{\pi}{a}, k_z\mright)$.
Below $T_c$, the entropy maximum was found to transform into a minimum (Fig.~\ref{fig:SRO-elasto}).
This is only possible if the states near the saddle points of the electronic dispersion open a gap as one enters the SC state.
Hence, with rather minimal modeling, it is possible to obtain information about the momentum-space structure of the SC gap from a thermodynamic measurement.

In order to draw more detailed conclusions about the allowed pairing states, we performed a symmetry analysis for a three-dimensional, three-band description of SRO.
Here we focus primarily on even-parity states, given the strong evidence for even parity in NMR measurements~\cite{Pustogow2019, Ishida2020, Chronister2021}.
From a simple two-dimensional perspective, one would conclude that the SC state must open a gap at the Van Hove points $\mleft(\pm \frac{\pi}{a}, 0\mright)$ and $\mleft(0, \pm\frac{\pi}{a}\mright)$.
However, to distinguish the relevant pairing states, in particular those of the 2D irreducible representation $E_g$ that transform like $(d_{yz} | - d_{xz})$, we must include the third momentum direction.
It is well known that the energy dispersion of SRO is strongly anisotropic.
Indeed, our analysis shows that the energy scale below which the three-dimensionality of the Fermi surface becomes important is about one kelvin [Eq.~\eqref{eq:SRO-VHdispexpansion}], fully consistent with magneto-oscillation experiments~\cite{Mackenzie2003}.
We also show that the saddle points deviate by very small amounts $\var{k_{\text{VH}, 2}} \ll \frac{2\pi}{a}$ from the lines $\mleft(\pm \frac{\pi}{a}, 0, k_z\mright)$ and $\mleft(0, \pm \frac{\pi}{a}, k_z\mright)$.
However, this need not be the case for the SC state.
While the single particle spectrum of SRO is highly anisotropic, it is possible that many-body interactions that are responsible for the SC pairing couple different layers more efficiently.
Hence, at least in principle, one should not exclude a strong dependence of the gap function on $k_z$; such dependence is crucial for the $(d_{yz} | - d_{xz})$-wave pairing states.

With these insights, we then turned to the symmetry analysis of potential pairing states.
If one assumes for a moment that the crystal structure of SRO is simple tetragonal, one is left with only two possible even pairing states, namely, the $s$-wave state of $A_{1g}$ symmetry and the $d_{x^2-y^2}$-wave state of $B_{1g}$ symmetry.
Given that fine-tuning is required for $s$-wave pairing to be consistent with the pair-breaking role of impurities~\cite{Mackenzie1998, Mao1999, Kikugawa2002, Kikugawa2004}, $d_{x^2-y^2}$-wave pairing would then appear to be the only natural pairing candidate.
However, \ce{Sr2RuO4} is a body-centered tetragonal compound.
The corresponding symmetry analysis now allows, in addition to $d_{x^2-y^2}$-wave pairing, for a $(d_{yz} | - d_{xz})$-wave state of $E_g$ symmetry like the one given in Eq.~\eqref{eq:SRO-nastyone}.

Our analysis does, however, allow us to exclude $d_{xy}$-wave pairing states that transform like $B_{2g}$ and $g_{xy(x^2-y^2)}$-wave pairing states that transform like $A_{2g}$ as sole pairing states.
Such states may at best be subleading contenders that could be added to the pairing wavefunction at fine-tuned points of accidental degeneracy.
In addition, we can exclude $(d_{yz} | - d_{xz})$-wave pairing that is exclusively of the type given in Eq.~\eqref{eq:SRO-usualone}.
The nature of our argument does not allow us to more precisely quantify how large these subleading $d_{xy}$-wave or $g_{xy(x^2-y^2)}$-wave contributions are because they vanish precisely where the elastocaloric experiment is most sensitive: at the Van Hove lines.
Thus, while the elastocaloric measurements do not allow for a unique determination of the superconducting order parameter symmetry, they do constrain the available options.
To finally resolve the nature of superconductivity in \ce{Sr2RuO4} requires a better understanding of the origin of time-reversal symmetry-breaking and of the orientation of line nodes.

In the next section, we discuss a subsequent work~\cite{Jerzembeck2024} which reported strong evidence against homogeneous time-reversal symmetry-breaking, and two-component superconducting order parameters more broadly.

\section{Constrains from $T_c$ and elastocaloric measurements under~$[110]$~uniaxial stress} \label{sec:SRO-Tc-ECE-analysis-110}
A significant number of experiments performed on strontium ruthenate (SRO) suggest that its superconducting (SC) order parameter has two components.
On the one hand, there is the old evidence indicating time-reversal symmetry-breaking (TRSB) in the SC state, as seen in muon spin relaxation ($\mu$SR)~\cite{Luke1998, Luke2000, Higemoto2014}, polar Kerr effect~\cite{Xia2006, Kapitulnik2009}, and Josephson junction~\cite{Kidwingira2006} experiments.
As we shall explain here, TRSB in the SC state necessitates two components with a complex phase difference.
On the other hand, there is an old ultrasound study~\cite{Okuda2003} that found a jump in the $c_{66} \in B_{2g}$ elastic coefficient.
Such a jump can only take place when the SC order parameter couples linearly to $\epsilon_{xy} \in B_{2g}$ strain, which is in turn only possible for two-component SC.
This we shall explain in more detail in Sec.~\ref{sec:SRO-GL-analysis}.
More recently, in the wake of the landmark Knight shift study of Pustogow et al.~\cite{Pustogow2019}, the TRSB signal in the $\mu$SR rate has been reproduced~\cite{Grinenko2021-unaxial, Grinenko2021-isotropic, Grinenko2023} and found to split from the SC transition under $[100]$ and $[110]$ uniaxial stress.
Regarding $c_{66}$, from the data of Ref.~\cite{Okuda2003} it is not entirely clear that the sharp feature at $T_c$ is a jump.
Three years ago this has been confirmed~\cite{Benhabib2021, Ghosh2021}, albeit with estimates for the $\Delta c_{66}$ jump that differ by a factor of $50$ between the two ultrasound measurement techniques.
Taken together, these two sets of experiments strongly suggest that SRO exhibits a chiral two-components SC that couples linearly to $\epsilon_{xy}$ strain.
Among even-parity SC states, this leaves only three options: $s' + \iu \, d_{xy}$, $g_{xy(x^2-y^2)} + \iu \, d_{x^2-y^2}$, and $d_{xz} + \iu \, d_{yz}$.\footnote{For a list of all options, excluding accidental degeneracies, see Tab.~\ref{tab:SRO-SC-state-options}.}
The degeneracy between the two components is accidental in the former two and symmetry-enforced in the latter ($d_{xz} + \iu \, d_{yz} \in E_g$).

However, not all evidence is consistent with a two-component SC state, as we already remarked during our literature review of Sec.~\ref{sec:SRO-lit-review}.
On the one hand, numerous experiments~\cite{Saitoh2015, Kashiwaya2019, Tamegai2003, Bjornsson2005, Kirtley2007, Hicks2010, Curran2011, Curran2014, Curran2023} have searched for TRSB and found no evidence for it.
On the other, cross-checking against thermodynamic measurements~\cite{Hicks2014, Steppke2017, Barber2019, Li2021, Li2022, Jerzembeck2024} reveals inconsistencies with TRSB or linear coupling to $\epsilon_{xy} \in B_{2g}$ strain, especially if these two phenomena are to be interpreted in terms of a homogeneous SC state.
It is this cross-checking that has motivated the study~\cite{Jerzembeck2024} whose results I present in the current section.
Much of the text of the current section has been recycled from Ref.~\cite{Jerzembeck2024}.

As often happens when a large number of experiments are performed on a single material, the results and/or interpretations of some experiments disagree.
While it is appropriate for theory to attempt to reconcile apparently contradictory results, the possibility of experimental error must also be kept in mind.
In the context of SRO, a noted example of the latter are early NMR Knight shift measurements~\cite{Ishida1998, Murakawa2007}.
As we reviewed in the introduction of this chapter, a reduction in the Knight shift was measured at $T_c$ only after a subtle systematic error was uncovered~\cite{Pustogow2019, Ishida2020, Chronister2021}.
Notably, this development was preceded by experimental contradictions: Pauli limiting was observed~\cite{Yonezawa2013, Yonezawa2014, Kittaka2014} which is at tension with the absence of a reduction in the Knight shift~\cite{Mackenzie2017}.
It is therefore important to cross-check experiments to see whether a coherent picture of SRO's remarkable SC can be attained.
In this regard, thermodynamic experiments hold a privileged position which rests on their unambiguous interpretation and well-developed measuring techniques.

When it comes to cross-checking, positive results are always more helpful as guides than negative ones.
Recently, Ref.~\cite{Grinenko2021-unaxial} reported that the transition temperature of TRSB $T_{\text{TRSB}}$, as seen in non-thermodynamic $\mu$SR measurements, splits from $T_c$ under $\langle 100 \rangle$ stress.
However, high-resolution heat capacity~\cite{Li2021} and elastocaloric~\cite{Li2022} measurements performed under $\langle 100 \rangle$ stress failed to resolve any anomaly at the reported~\cite{Grinenko2021-unaxial} TRSB temperature.
A Ginzburg-Landau analysis of TRSB SC states (Sec.~\ref{sec:SRO-GL-analysis}) moreover demonstrates that their $T_c$ should develop a cusp in its dependence on shear strain~\cite{SigristUeda1991, Walker2002}.
Yet this cusp has not been observed for uniaxial $[100]$ stress (which induces $\epsilon_{xx} - \epsilon_{yy} \in B_{1g}$ shear strain), despite several searches~\cite{Hicks2014, Steppke2017, Barber2019, Watson2018, Mueller2023}.
Reconciling the two within a Ginzburg-Landau description requires considerable fine-tuning.

The subject of the current section is the cross-checking of the results of Refs.~\cite{Benhabib2021, Ghosh2021, Grinenko2023} which has been carried out in Ref.~\cite{Jerzembeck2024}.
Two main results were reported in Refs.~\cite{Benhabib2021, Ghosh2021, Grinenko2023}.
First, a jump in the $c_{66} \in B_{2g}$ elastic modulus at $T_c$ was reported in ultrasound echo measurements~\cite{Benhabib2021} and resonant ultrasound spectroscopy~\cite{Ghosh2021}.
Second, $\mu$SR measurements~\cite{Grinenko2023} found that the TRSB transition temperature $T_{\text{TRSB}}$ splits from $T_c$ under $[110]$ pressure with a $T_{\text{TRSB}} < T_c$.
Through Ehrenfest relations which we derive in Sec.~\ref{sec:SRO-GL-analysis}, these two results imply that the cusp of $T_c(\sigma_{110})$ and that the splitting of $T_{\text{TRSB}}$ away from $T_c$ under $\langle 110 \rangle$ stress $\sigma_{110}$ should be easily observable, if their results are taken at face value.

In Ref.~\cite{Jerzembeck2024}, high-resolution measurements have been carried out of both the magnetic susceptibility ($T_c$) and the elastocaloric effect under $[110]$ uniaxial pressure.
Within tight limits, neither a cusp nor transition splitting is resolved in the data.
As we show in Sec.~\ref{sec:SRO-110-implications}, these results cannot be plausibly reconciled with the observed jumps in $c_{66}$ under the assumption of a homogeneous SC state -- the level of tuning implied is implausibly fine.
The data is also not consistent with the transition splitting seen in $\mu$SR~\cite{Grinenko2023}.
In contrast, the data of Ref.~\cite{Jerzembeck2024} is in agreement with previous work under $[001]$ and hydrostatic pressure~\cite{Forsythe2002, Jerzembeck2022}, confirming thermodynamic consistency.

In the remainder of this section, we first we present the main experimental findings of Ref.~\cite{Jerzembeck2024}.
Then we carry out a general Ginzburg-Landau analysis of two-component SC states which couple linearly to $\sigma_6 \in B_{2g}$ stress.
Using the results of this analysis, in the final Sec.~\ref{sec:SRO-110-implications} we examine the consistency and fine-tuning that is needed for the ultrasound~\cite{Benhabib2021, Ghosh2021} and thermodynamic~\cite{Li2021, Jerzembeck2024} experiments to be in agreement under the assumption of a homogeneous two-component SC state.

\subsection{Experimental findings: no indications of a cusp or transition splitting} \label{sec:SRO-Tc-ECE-110-findings}
\begin{figure}[p!]
\centering
\includegraphics[width=0.84\textwidth]{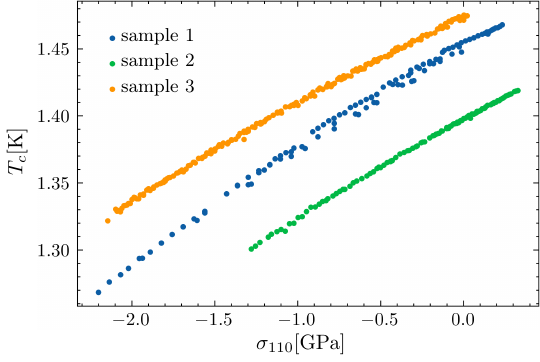} \\[10pt]
{\includegraphics[width=0.884\textwidth]{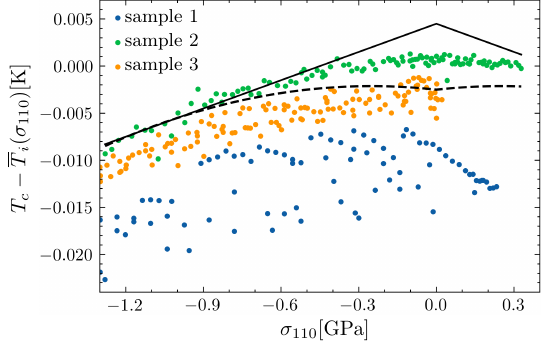} \hspace{14pt}}
\captionbelow[Dependence of the superconducting transition temperature $T_c$ on $\langle 110 \rangle$ uniaxial stress $\sigma_{110}$, as determined by magnetic susceptibility measurements~\cite{Jerzembeck2024}.]{\textbf{Dependence of the superconducting transition temperature $T_c$ on $\langle 110 \rangle$ uniaxial stress $\sigma_{110}$, as determined by magnetic susceptibility measurements}~\cite{Jerzembeck2024}.
In the bottom panel, $T_c$ is displaced by linear fits $\overline{T}_i(\sigma_{110}) = \overline{T}_{0,i} + \varrho_i \sigma_{110}$ with $\overline{T}_{0,i} = (1.464, 1.397, 1.477)~\si{\kelvin}$ and $\varrho_i = (0.0719, 0.0679, 0.0586)~\si{\kelvin\per\giga\pascal}$ for samples $i = (1, 2, 3)$, respectively.
$\overline{T}_{0,i}$ have been intentionally chosen to vertically offset the different samples for clarity.
The solid black line is the curve $\SI{0.0045}{\kelvin} - \SI{0.01}{\kelvin\per\giga\pascal} \abs{\sigma_{110}}$, while the dashed black line is the curve  $\SI{-0.0025}{\kelvin} + \SI{0.003}{\kelvin\per\giga\pascal} \abs{\sigma_{110}} - \SI{0.006}{\kelvin\per\giga\pascal\squared} (\sigma_{110})^2$.
The cusps of these two curves give estimates for cusps in $T_c$ below experimental resolution.
The plotted data is available at~\cite{Jerzembeck2024}.}
\label{fig:SRO-J24-Tc}
\end{figure}

\begin{figure}[p!]
\centering
\includegraphics[width=\textwidth]{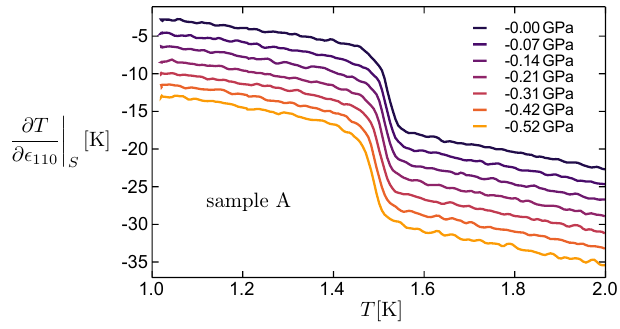} \\[4pt]
\includegraphics[width=\textwidth]{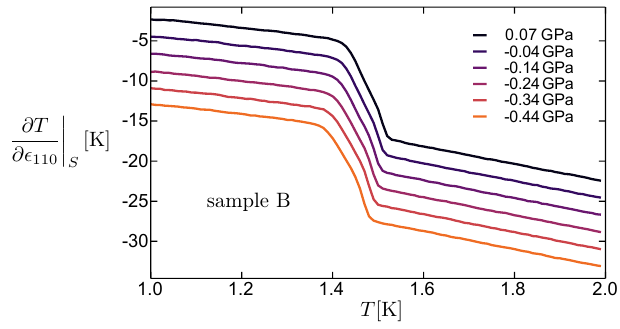}
\captionbelow[Elastocaloric measurements of \ce{Sr2RuO4} as a function of temperature $T$ as a small $\epsilon_{110}$ strain is adiabatically varied along the $\langle 110 \rangle$ direction~\cite{Jerzembeck2024}.]{\textbf{Elastocaloric measurements of \ce{Sr2RuO4} as a function of temperature $T$ as a small $\epsilon_{110}$ strain is adiabatically varied along the $\langle 110 \rangle$ direction}~\cite{Jerzembeck2024}.
For clarity, all curves apart from the black ones have been shifted vertically with respect to each other.
The colors of the curves indicate the average $\langle 110 \rangle$ uniaxial pressure $\sigma_{110}$ that is applied on the sample, as designated in the plots.
The plotted data is available at~\cite{Jerzembeck2024}.}
\label{fig:SRO-J24-ECE}
\end{figure}

The main experimental findings of Ref.~\cite{Jerzembeck2024} are shown in Figs.~\ref{fig:SRO-J24-Tc} and~\ref{fig:SRO-J24-ECE}.

As can be seen in Fig.~\ref{fig:SRO-J24-Tc}, overall $T_c$ depends linearly on $\sigma_{110}$.
This is expected because uniaxial $\sigma_{110}$ stress implies $\sigma_{xx} = \sigma_{yy} = \sigma_{xy} = \sigma_{yx} = \sigma_{110}/2$ stresses, which in turn induce not only the shear $\epsilon_6 = 2 \epsilon_{xy} \in B_{2g}$ component of the strain tensor $\epsilon_{ij}$, but also $A_{1g}$ components (Sec.~\ref{sec:SRO-elastic-tuning}).
By symmetry, $\epsilon_{A_{1g}}$ are always allowed to couple linearly to the SC order parameter.
If a coupling $\propto \epsilon_{A_{1g}} \abs{\Phi}^2$ is present in the free energy, this means that the temperature $T_c$ at which the quadratic coefficient of the Ginzburg-Landau expansion becomes negative changes linearly with $\epsilon_{A_{1g}}$.
By using the relations of Sec.~\ref{sec:SRO-elastic-tuning}, one may show that the measured $\dd{T_c}/\dd{\sigma_{110}} = \SI[per-mode=symbol, parse-numbers=false]{64 \pm 7}{\milli\kelvin\per\giga\pascal}$ (Fig.~\ref{fig:SRO-J24-Tc}) and the previously measured $\dd{T_c}/\dd{\sigma_{001}} = \SI[per-mode=symbol, parse-numbers=false]{76 \pm 5}{\milli\kelvin\per\giga\pascal}$~\cite{Jerzembeck2022} imply that $\dd{T_c}/\dd{\sigma_{\text{hyd}}} = \SI[per-mode=symbol, parse-numbers=false]{202 \pm 12}{\milli\kelvin\per\giga\pascal}$, which agrees with the $\dd{T_c}/\dd{\sigma_{\text{hyd}}} = \SI[per-mode=symbol, parse-numbers=false]{220 \pm 20}{\milli\kelvin\per\giga\pascal}$ of Ref.~\cite{Forsythe2002}.
The measurements are thus thermodynamically consistent.

More significant is the fact that no cusp is resolved in $T_c$ at $\sigma_{110} = 0$.
From the lower panel of Fig.~\ref{fig:SRO-J24-Tc}, we see that within experimental resolution $T_c$ depends quadratically on $\sigma_{110}$, after the linear dependence is subtracted.
For comparison, if we had a two-component SC order parameter $(\Phi_1|\Phi_2)$, it would be able to couple linearly to $\epsilon_6 \in B_{2g}$ through a term of the form:
\begin{align}
\epsilon_6 \mleft(\Phi_1^{*} \Phi_2 + \Phi_2^{*} \Phi_1\mright) &= \epsilon_6 \mleft(\abs{\Phi_+}^2 - \abs{\Phi_-}^2\mright),
\end{align}
where $\Phi_{\pm} = (\Phi_1 \pm \Phi_2) / \sqrt{2}$.
Hence the quadratic coefficients of $\Phi_+$ and $\Phi_-$, which are equal at $\epsilon_6 = 0$, would be offset in opposite directions and only one of them would become negative at $T_c$ in the presence of finite $\epsilon_6$ strain.
The associated transition temperature would therefore grow linearly for both positive and negative $\epsilon_6$, with the following dependence near $\sigma_{110} = 0$ ($\sigma_6 = \sigma_{110}/2 = c_{66} \epsilon_6$):
\begin{align}
T_c(\sigma_{110}) &= T_{c0} + \dv{T_c}{\sigma_{110}} \sigma_{110} + \abs{\dv{T_c}{\sigma_6}} \cdot \abs{\sigma_6} + \cdots \, .
\end{align}
In contrast, in Fig.~\ref{fig:SRO-J24-Tc} we find that the the quadratic dependence goes downwards.
Even if we imagine that internal strain inhomogeneities smear the cusp $\propto \abs{\sigma_6}$, and they would have to be very large, on the order of $\sim \SI{0.3}{\giga\pascal}$, to do so, we should still see some dip in $T_c - \overline{T}_i$ at $\sigma_{110} = 0$.
None is observed, and if we try to fit one (dashed black line in the lower panel of Fig.~\ref{fig:SRO-J24-Tc}), we obtain the following upper bound on the cusp:
\begin{align}
\abs{\dv{T_c}{\epsilon_6}} = 2 c_{66} \abs{\dv{T_c}{\sigma_{110}}} \leq 2 \cdot \SI{65.5}{\giga\pascal} \cdot \SI{0.003}{\kelvin\per\giga\pascal} = \SI{0.4}{\kelvin}. \label{eq:SRO-my-tighted-bnd}
\end{align}
Here we used the $c_{66}$ value of Ref.~\cite{Ghosh2021}, listed in Tab.~\ref{tab:SRO-elastic-constants}.
In the article itself~\cite{Jerzembeck2024}, a different procedure was used for estimating the upper bound that gave a more conservative upper bound:
\begin{align}
\abs{\dv{T_c}{\epsilon_6}} \leq \SI{1.3}{\kelvin}. \label{eq:SRO-article-Tc-bound}
\end{align}
From the lower panel of Fig.~\ref{fig:SRO-J24-Tc}, we see that this is roughly the bound that one can infer from sample~3.
Sample~2 gives a tighter bound [Eq.~\eqref{eq:SRO-my-tighted-bnd}], while sample~1 gives a looser bound.
In the remainder, we use the bound~\eqref{eq:SRO-article-Tc-bound}.

As already discussed in Sec.~\ref{sec:SRO-elasto}, the elastocaloric effect is the effect of adiabatic changes in the strain $\epsilon_{ij}$ inducing changes in the temperature.
More importantly for the current discussion, the associated quantity [Eq.~\eqref{eq:SRO-elasto-id}]
\begin{equation}
\pdvc{T}{\epsilon_{ij}}{S} =  - \frac{T}{C_{\epsilon}(T)} \pdvc{S}{\epsilon_{ij}}{T}
\end{equation}
is sensitive to phase transitions and it can be measured with a higher signal-to-noise ratio than heat capacity~\cite{Li2021, Li2022}.
The results are shown in Fig.~\ref{fig:SRO-J24-ECE}, reproduced from Ref.~\cite{Jerzembeck2024}.
The details of how these results were obtained from the measured thermocouple voltage can be found in the article~\cite{Jerzembeck2024}.
Evidently, only one main transition is observed in the elastocaloric effect, with apparently no visible sign of uniaxial pressure-dependent splitting of the
main transition.
Any structure in the transition that is seen at zero pressure (likely due to slight inhomogeneity of the strain field and/or defect density) remains the same at non-zero pressure.
Thus no evidence of a second transition is present in the elastocaloric effect~\cite{Jerzembeck2024}.
With additional assumptions, one can make statements on how finely tuned the second transition would have to be to evade detection.
We refer the interested reader to the article~\cite{Jerzembeck2024} for this.
Below we mainly analyze the implications of the bound~\eqref{eq:SRO-article-Tc-bound} when compared to the observed jumps in the $c_{66}$ elastic modulus~\cite{Benhabib2021, Ghosh2021}.

\subsection{Ginzburg-Landau analysis of two-component superconducting states} \label{sec:SRO-GL-analysis}
In this section, we analyze the response of a two-component order parameter $\vb{\Phi} = (\Phi_{1}, \Phi_{2})^{\intercal}$ to $\sigma_6 = \sigma_{xy}$ shear stress within the Ginzburg-Landau framework, under the assumptions of homogeneous strain and superconductivity.
While the analysis of a symmetry-protected two-component order parameter had already been done for the $D_{4h}$ point group~\cite{Benhabib2021, Ghosh2021}, the case of accidental degeneracy has not been analyzed in the literature to the degree of detail required for the analysis of Ref.~\cite{Jerzembeck2024}.
Here we reproduce the analysis of Ref.~\cite{Jerzembeck2024} with more elaborations and with a more elegant parametrization.

The case of a symmetry-protected two-component order parameter corresponds to the two-dimensional irreducible representations $E_g$ and $E_u$ whose wavefunctions we may write as $(d_{yz} | - d_{xz})$ and $(p_x | p_y)$, respectively.
The unusual ordering for the $E_g$ irreducible representation (irrep) is to ensure that the two components transform under the conventional transformation matrices which we consistently use through the thesis; see Eqs.~\eqref{eq:Egu-irrep-mat-conventions1} and~\eqref{eq:Egu-irrep-mat-conventions2} of Sec.~\ref{sec:examples-convention-D4h} or the Eq.~\eqref{eq:SRO-E-rep-rho} of Sec.~\ref{sec:SRO-SC-construct}.

Accidental degeneracy could, in principle, be between any pair of one-dimensional irreps.
Because of ultrasound experiments~\cite{Benhabib2021, Ghosh2021}, we consider only those degenerate pairs that couple linearly to $\sigma_6 \in B_{2g}$, which are namely $A_{1g}(s) \oplus B_{2g}(d_{xy})$ and $B_{1g}(d_{x^2-y^2}) \oplus A_{2g}\mleft(g_{xy(x^2-y^2)}\mright)$.
Odd-parity 1D irrep pairs, such as $A_{1u}(k_x \vu{e}_x + k_y \vu{e}_y) \oplus B_{2u}(k_x \vu{e}_y + k_y \vu{e}_x)$ and $B_{1u}(k_x \vu{e}_x - k_y \vu{e}_y) \oplus A_{2u}(k_x \vu{e}_y - k_y \vu{e}_x)$, are also in principle possible, but are not deemed likely due to NMR Knight shift~\cite{Pustogow2019, Ishida2020, Chronister2021} and Pauli limiting~\cite{Yonezawa2013, Yonezawa2014, Kittaka2014} experiments, as we discussed in Sec.~\ref{sec:SRO-lit-review}.
Formally, the analysis is identical for even- and odd-parity SC states, and precisely which pair of accidentally degenerate 1D irreps we consider does not matter, as long as the product of their two irreps is $B_{2g}$.
Quadratic coupling to $\sigma_6$  does not induce a jump in the shear elastic modulus $c_{66}$ nor does it split the transition.

\begin{table}[t]
\captionabove[Irreducible representations (irreps) of the $D_{4h}$ point group under which the bilinear forms $\Upsilon_{\mu} \defeq \vb{\Phi}^{\dag} \Pauli_{\mu} \vb{\Phi}$ transform.]{\textbf{Irreducible representations (irreps) of the $D_{4h}$ point group under which the bilinear forms $\Upsilon_{\mu} \defeq \vb{\Phi}^{\dag} \Pauli_{\mu} \vb{\Phi}$ transform.}
The bilinears are constructed from a two-component order parameter $\vb{\Phi} = (\Phi_{1}, \Phi_{2})^{\intercal}$ which belongs to the 2D irreps $E_{g,u}$ on the left, whereas on the right $\Phi_{1,2}$ belong to two distinct 1D irreps  $\zeta_{1,2}$, respectively.
The $+$ ($-$) irrep superscript indicates evenness (oddness) under time reversal.
We only analyze accidentally degenerate pairs whose $\zeta_{1} \otimes \zeta_{2} = B_{2g}$.}
{\renewcommand{\arraystretch}{1.3}
\renewcommand{\tabcolsep}{10pt}
\hspace*{\stretch{1}}
\begin{tabular}{|c|c|} \hline\hline
\multicolumn{2}{|c|}{$\vb{\Phi} \in E_{g,u}$} \\
bilinear & irrep \\ \hline
$\Upsilon_{0}$ & $A_{1g}^{+}$ \\
$\Upsilon_{x}$ & $B_{2g}^{+}$ \\
$\Upsilon_{y}$ & $A_{2g}^{-}$ \\
$\Upsilon_{z}$ & $B_{1g}^{+}$ \\[2pt]
\hline\hline
\end{tabular} \hspace*{\stretch{1}}
\begin{tabular}{|c|c|} \hline\hline
\multicolumn{2}{|c|}{$\Phi_1 \in \zeta_{1}$, $\Phi_2 \in \zeta_{2}$} \\
bilinear & irrep \\\hline
$\Upsilon_{0}$ & $A_{1g}^{+}$ \\
$\Upsilon_{x}$ & $(\zeta_{1} \otimes \zeta_{2})^{+}$ \\
$\Upsilon_{y}$ & $(\zeta_{1} \otimes \zeta_{2})^{-}$ \\
$\Upsilon_{z}$ & $A_{1g}^{+}$ \\[2pt]
\hline\hline
\end{tabular} \hspace*{\stretch{1}}}
\label{tab:SRO-phi-irreps}
\end{table}

\begin{figure}[t!]
\centering
\includegraphics[width=0.95\textwidth]{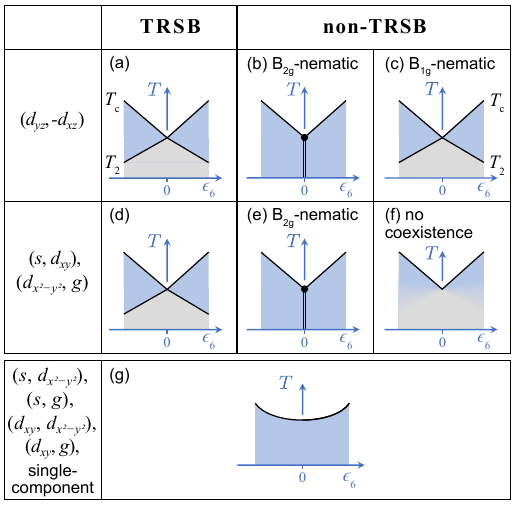}
\captionbelow[Temperature $T$ vs.\ shear strain $\epsilon_6 = 2 \epsilon_{xy}$ phase diagrams for various possible superconducting order parameters~\cite{Jerzembeck2024}.]{\textbf{Temperature $T$ vs.\ shear strain $\epsilon_6 = 2 \epsilon_{xy}$ phase diagrams for various possible superconducting order parameters}~\cite{Jerzembeck2024}.
The first column specifies the two order parameter components $(\Phi_1, \Phi_2)$.
At $\epsilon_6 = 0$, they condense into $\Phi_1 \pm \iu \Phi_2$ for the time-reversal symmetry-breaking (TRSB) case, into $\Phi _1 \pm \Phi_2$ for the $B_{2g}$-nematic case, and into $\Phi_1$ or $\Phi_2$ for the no coexistence/$B_{1g}$-nematic case.
In the last row the precise ordering does not matter because there is no linear coupling to $\epsilon_6$ strain.
For $(d_{yz}, - d_{xz}) \in E_g$ the two components are degenerate by symmetry, while for the other cases the are degenerate by accident.
$g$ is a shorthand for $g_{xy(x^2-y^2)} \in A_{2g}$.
In all panels, single black lines indicate second-order transitions, double lines indicate first-order transitions, and color gradients indicate crossovers.}
\label{fig:schematicPhaseDiagrams}
\end{figure}

Before we proceed with the Ginzburg-Landau analysis, let us first broadly sketch how the SC transition is expected to split under $\epsilon_6$ strain, depending on the symmetries.
This is summarized in Fig.~\ref{fig:schematicPhaseDiagrams}.
Let us introduce the bilinear forms:
\begin{equation}
\Upsilon_{\mu} \defeq \vb{\Phi}^{\dag} \Pauli_{\mu} \vb{\Phi},
\end{equation}
where $\Pauli_0$ is the $2 \times 2$ identity matrix and $\Pauli_{x, y, z}$ are Pauli matrices.
The transformation properties of $\Upsilon_{\mu}$ are are easily deduced with the help of Tab.~\ref{tab:D4h-irrep-prod-tab} from Appx.~\ref{app:group_theory} and we have summarized them in Tab.~\ref{tab:SRO-phi-irreps}.
A sufficient condition for a cusp in $T_c(\sigma_6)$ is that there exists a $\Upsilon_{\mu}$ that transforms like the shear strain $\sigma_6 \in B_{2g}^{+}$, where the $+$ superscript indicates evenness under time reversal (TR).
In our case, this is only possible for $\Upsilon_x$.
If $\Upsilon_x$ acquires a non-zero expectation value below $T_c$ at $\epsilon_6 = 0$, then $\epsilon_6$ strain acts like a conjugate field that lifts the degeneracy between $\pm \ev{\Upsilon_x}$, and only one transition takes place since the symmetry associated with $\Upsilon_x$ is already broken.
Moreover, the transition between the $\pm \ev{\Upsilon_x}$ states as a function of $\epsilon_6$ at fixed $T < T_c$ is first-order.
This corresponds to the $B_{2g}$-nematic column of Fig.~\ref{fig:schematicPhaseDiagrams}.
If, on the other hand,  $\Upsilon_x$ is not the bilinear that acquires a finite expectation value below $T_c$ at $\epsilon_6 = 0$, an additional symmetry can still break, resulting in a second transition.
This second transition can be a nematic one, as for the $B_{1g}$-nematic states of $E_{g}$ or $E_u$, or a TRSB one.
In the $B_{1g}$-nematic case, in going from the $\Phi_1 \pm \Phi_2$ states at finite $\epsilon_6$ above $T_2$ into the $\Phi_1 + c \Phi_2$ with $c \neq \pm 1$ states below $T_2$, it is the diagonal rotation symmetry $C_{2d_+}\colon \Phi_{1,2} \mapsto \Phi_{2,1}$ that breaks down.
Notice that this symmetry is present even when $\sigma_{xy}$ or $\sigma_{110}$ stress is applied on the system.
This symmetry does not mix the two components in the accidentally degenerate case, $C_{2d_+}\colon \Phi_{1,2} \mapsto \pm \Phi_{1,2}$, and is therefore always broken, which explains the crossover shown in Fig.~\ref{fig:schematicPhaseDiagrams}(f).
Quadratic coupling to $\epsilon_6$ is incapable of splitting the transition for the simple reason that $(\epsilon_6)^2 \in A_{1g}^{+}$ transforms trivially under symmetries.
Hence no splitting or cusp is found for SC order parameter which can only couple quadratically to $\epsilon_6$ [Fig.~\ref{fig:schematicPhaseDiagrams}(g)].

The Ginzburg-Landau expansion of the free energy in the absence of stress is given by
\begin{align}
F = F_n &+ \frac{a}{2} \Upsilon_0 + \frac{\tilde{a}}{2} \Upsilon_z + \sum_{\mu=0,x,y,z} \frac{v_{\mu}}{4} \Upsilon_{\mu}^2 + \frac{\tilde{v}}{4} \Upsilon_{0} \Upsilon_{z}, \label{eq:SRO-GL-GLexp}
\end{align}
where $F_n$ is the normal-state free energy.
From Tab.~\ref{tab:SRO-phi-irreps} and the irrep product Tab.~\ref{tab:D4h-irrep-prod-tab}, it is straightforward to confirm that this is the most general form of an invariant ($A_{1g}^{+}$) function that is quadratic in $\Upsilon_{\mu}$ (quartic in $\Phi$).
Due to the Fierz identity
\begin{equation}
\Upsilon_0^2 = \sum_{i=x,y,z} \Upsilon_{i}^2, \label{eq:SRO-GL-Fierz}
\end{equation}
there is a redundancy between the $v_{\mu} \Upsilon_{\mu}^2$ terms that we eliminate by setting
\begin{equation}
v_0 = 0.
\end{equation}
Note that this is different from Appendix~E of Ref.~\cite{Jerzembeck2024} where $v_z$ was set to zero.
As it turns out, setting $v_0$ to zero results in simpler and more symmetric expressions.

In the case of a symmetry-protected degeneracy, $\Upsilon_{z}$ transforms under $B_{1g}$ and therefore
\begin{align}
\tilde{a} &= \tilde{v} = 0 \qquad\text{for symmetry-protected $\vb{\Phi} \in E_{g,u}$.}
\end{align}
Below the transition temperature $T_{c0}$, the quadratic coefficient changes sign.
To leading order in temperature, $a(T)$ is thus linear in $T$ with a positive slope $\dot{a} > 0$:
\begin{equation}
a(T) = (T - T_{c0}) \times \dot{a}, \label{eq:SRO-GL-aofT}
\end{equation}
whereas the quartic coefficients are $T$-independent.

When $\Phi_{1, 2}$ belong to two 1D irreps, $\Upsilon_{z}$ transforms trivially and both $\tilde{a}$ and $\tilde{v}$ are allowed to be finite.
However, since $\Phi_1$ and $\Phi_2$ are unrelated by symmetry, we may rescale them $(\Phi_1, \Phi_2) \mapsto (s \Phi_1, s^{-1} \Phi_2)$ by a factor $s = \mleft(v_z - \tilde{v}\mright)^{1/8} / \mleft(v_z + \tilde{v}\mright)^{1/8}$ so that after the rescaling
\begin{align}
\tilde{v} = 0,
\end{align}
which we henceforth assume.
Regarding $\tilde{a}$, in the expansion $F = \dot{a}_1 \mleft(T-T_{\text{c0},1}\mright) \abs{\Phi_{1}}^{2} + \linebreak \dot{a}_2 \mleft(T-T_{\text{c0},2}\mright) \abs{\Phi_{2}}^{2} + \cdots$ the fine-tuning of the two transition temperatures corresponds to the requirement that $T_{\text{c0},1} = T_{\text{c0},2} \equiv T_{c0}$.
Hence $a(T)$ is given by Eq.~\eqref{eq:SRO-GL-aofT} with $\dot{a} = \dot{a}_1 + \dot{a}_2$, while
\begin{equation}
\tilde{a}(T) = \alpha \times a(T)
\end{equation}
for a $T$-independent coefficient $\alpha = (\dot{a}_1 - \dot{a}_2) / (\dot{a}_1 + \dot{a}_2)$.
$\alpha$ can take any value in between $-1$ and $1$ and reflects the absence of a symmetry transformation connecting $\Phi_1$ and $\Phi_2$.
Thus in the symmetry-protected case the only formal difference is that $\alpha = 0$, given that $\tilde{v} = 0$ in both cases.

Let us now include elasticity.
When strains $\epsilon_i$ are present in the system, they couple to the superconductivity via
\begin{equation}
F_{c} = \sum\limits_{i=1}^6 \sum\limits_{a,b=1}^2 \lambda_{iab} \epsilon_i \Phi_a^* \Phi_b, \label{eq:SRO-GL-lambda-coupling-constants}
\end{equation}
where $\lambda_{iab}$ are the coupling constants and $\epsilon_i$ are in Voigt notation (Sec.~\ref{sec:SRO-elastic-tuning}).
As it turns out, when the elastic free energy is quadratic in $\epsilon_i$,
\begin{equation}
F_{\epsilon} = \frac{1}{2}\sum\limits_{i,j=1}^6 c_{ij,0} \epsilon_i \epsilon_j,
\end{equation}
one may decouple the elastic and superconducting parts of the free energy, greatly simplifying the free energy minimization problem.
Here $c_{ij,0}$ is the elastic tensor in the absence of superconductivity.
This decoupling is accomplished by introducing the ``external'' strain
\begin{equation}
\epsilon_{i,0} \defeq \epsilon_i + \sum\limits_{j=1}^6 \sum\limits_{a,b=1}^2 c_{ij,0}^{-1} \lambda_{jab} \Phi_a^* \Phi_b, \label{eq:SRO-GL-externalStrain}
\end{equation}
which is decoupled from $\vb{\Phi}$ and directly related to the external stress:
\begin{align}
\epsilon_{i,0} = \sum_{j=1}^6 c_{ij,0}^{-1} \sigma_j.
\end{align}
It is the strain that would be obtained under a given set of stresses in the absence of superconductivity.

In practice, the difference between $\epsilon_{i,0}$ and the total strain $\epsilon_i$ is negligible for \ce{Sr2RuO4}: the larger of the two reported values of $\Delta c_{66}$ is $\sim 10^{-5} c_{66,0}$~\cite{Benhabib2021, Ghosh2021}, and the experimental upper limit on any spontaneous nematic strain is on the order of $10^{-8}$ [Eq.~\eqref{eq:SRO-GL-nematicUpperLimit}], far smaller than the scale of the strains applied during experiments.
For this reason, during our presentation of the experimental results of Ref.~\cite{Jerzembeck2024} (Sec.~\ref{sec:SRO-Tc-ECE-110-findings}) we made no distinction between $\epsilon_{i,0}$ and $\epsilon_i$, nor shall we distinguish the two during our analysis of Sec.~\ref{sec:SRO-110-implications}.
Here, we retain this distinction to be able to calculate the jump in the shear modulus $\Delta c_{66}$.

In the presence of $\sigma_6$ external shear stress, the total free energy after decoupling therefore equals
\begin{equation}
F = F_n + F_{\epsilon 0} + F_{\Phi 0},
\end{equation}
where the elastic part is
\begin{equation}
F_{\epsilon 0} = \frac{1}{2} c_{66,0} \epsilon_{6,0}^2 - \sigma_6 \epsilon_{6,0}
\end{equation}
and the superconducting part is
\begin{equation}
F_{\Phi 0} = \frac{a}{2} \Upsilon_0 + \alpha \frac{a}{2} \Upsilon_z + \frac{v_x}{4} \Upsilon_x^2 + \frac{v_y}{4} \Upsilon_y^2 + \frac{v_z}{4} \Upsilon_z^2 + \frac{\sigma_6 \lambda_6}{c_{66,0}} \Upsilon_x. \label{eq:SRO-GL-F_Delta}
\end{equation}
The form of the coupling to $\sigma_6 \in B_{2g}^{+}$ follows from Tab.~\ref{tab:SRO-phi-irreps}.
For the accidentally degenerate case, here we assumed that $\zeta_{1} \otimes \zeta_{2} = B_{2g}$.
As already remarked, $\alpha = 0$ in the symmetry-protected case, while for accidental degeneracies $\alpha$ can take any value in between $-1$ and $1$.
By enacting $(\Phi_1, \Phi_2) \mapsto (\Phi_1, -\Phi_2)$, $\Upsilon_x \mapsto - \Upsilon_x$ so we can always make
\begin{align}
\lambda_6 > 0,
\end{align}
which we henceforth assume.
In shifting from $\epsilon_i$ to $\epsilon_{i,0}$, the quartic coefficients $v_x, v_y, v_z$ have been renormalized.

The minimum of the elastic free energy is $F_{\epsilon 0} = - \frac{1}{2} c_{66,0} \epsilon_{6,0}^2$ with $\epsilon_{6,0} = \sigma_6 / c_{66,0}$.

To find the minimum of $F_{\Phi 0}$, we use the spherical parametrization
\begin{align}
\begin{pmatrix}
\Phi_1 \\
\Phi_2
\end{pmatrix} &= \Phi_0 \begin{pmatrix}
\cos \tfrac{\vartheta}{2} \\[2pt]
\sin \tfrac{\vartheta}{2} \, \Elr^{\iu \varphi}
\end{pmatrix}
\end{align}
in terms of which
\begin{align}
\vb{\Upsilon} &= \begin{pmatrix}
\Upsilon_x \\
\Upsilon_y \\
\Upsilon_z
\end{pmatrix} = \Phi_0^2 \begin{pmatrix}
\sin \vartheta \cos \varphi \\
\sin \vartheta \sin \varphi \\
\cos \vartheta
\end{pmatrix}.
\end{align}
Evidently, all three $v_{x,y,z}$ must be positive if the free energy $F_{\Phi 0}$ is to be bounded from below because otherwise we could orient $\vb{\Upsilon}$ along the negative direction to get $F_{\Phi 0} \to - \infty$ as $\Phi_0 \to + \infty$.
For later convenience, the $v_x$, $v_y$, and $v_z$ parameters we write in the following symmetric manner:
\begin{align}
\begin{aligned}
v_x &= (1 + \kappa + \sqrt{3} \, \kappa') w, \\
v_y &= (1 + \kappa - \sqrt{3} \, \kappa') w, \\
v_z &= (1 - 2 \kappa) w.
\end{aligned}
\end{align}
The $u$, $\gamma$, and $\gamma'$ parameters previously employed in Ref.~\cite{Jerzembeck2024} are related to our parameters through $v_x = (1 + \gamma + \gamma') u$, $v_y = (1 + \gamma - \gamma') u$, and $v_z = u$, which is less symmetric.
The condition that $v_{x,y,z} > 0$ is equivalent to $w > 0$ with $(\kappa, \kappa')$ constrained to lie within an equilateral triangle centered at zero.
This physical phase space of the Ginzburg-Landau theory is drawn in Fig.~\ref{fig:SRO-GL-phase-space}.

The SC free energy in spherical coordinates attains the form:
\begin{equation}
F_{\Phi 0} = A(\vartheta, \varphi) \frac{a}{2} \Phi_0^2 + W(\vartheta, \varphi) \frac{w}{4} \Phi_0^4, \label{eq:SRO-GL-F_reduced}
\end{equation}
where
\begin{align}
A(\vartheta, \varphi) &= 1 + \alpha \cos(\vartheta) + \beta \sin(\vartheta)\cos(\varphi), \label{eq:SRO-GL-A} \\
W(\vartheta, \varphi) &= 1 - 2 \kappa + K(\varphi) \sin^2(\vartheta), \\
K(\varphi) &= 3 \kappa + \sqrt{3} \, \kappa' \cos(2\varphi).
\end{align}
Here we have introduced the shorthand:
\begin{equation}
\beta \defeq \frac{2 \lambda_6 \epsilon_{6,0}}{a} = \frac{2 \lambda_6 \sigma_6}{c_{66,0} (T - T_{c0}) \dot{a}}.
\end{equation}
$\beta$ is the main parameter through which the temperature $T$ and external strain $\epsilon_{6,0}$ enter the analysis.
The saddle point equations for the non-trivial solution whose
\begin{align}
\Phi_0^2 &= - \frac{a A(\vartheta, \varphi)}{w W(\vartheta, \varphi)} > 0
\end{align}
are given by:
\begin{align}
\begin{aligned}
0 & = \sin(\varphi) \sin(\vartheta) \mleft[\sqrt{3} \, \kappa' \cos(\varphi) \sin(\vartheta) - \frac{\beta W(\vartheta, \varphi)}{2 A(\vartheta, \varphi)}\mright], \\
0 & = \sin( \vartheta) \cos(\vartheta) K(\varphi) + \mleft[\alpha \sin (\vartheta) - \beta \cos(\varphi) \cos(\vartheta)\mright] \frac{W(\vartheta, \varphi)}{A(\vartheta, \varphi)}.
\end{aligned} \label{eq:SRO-GL-saddlePoint}
\end{align}

In light of the Fierz identity~\eqref{eq:SRO-GL-Fierz}, the saddle point equations can also be formulated in terms of the $\vb{\Upsilon}$ bilinears directly:
\begin{align}
\begin{aligned}
(a + v_x \Upsilon_0) \Upsilon_x &= - a \beta \Upsilon_0, \\
(a + v_y \Upsilon_0) \Upsilon_y &= 0, \\
(a + v_z \Upsilon_0) \Upsilon_z &= - a \alpha \Upsilon_0,
\end{aligned}
\end{align}
where $\Upsilon_0 = \sqrt{\Upsilon_x^2 + \Upsilon_y^2 + \Upsilon_z^2} = \Phi_0^2 > 0$.

\subsubsection{Solutions in the absence of $B_{2g}$ stress ($\sigma_6 = 0$)} \label{sec:SRO-GL-noStrain-sols}
In the absence of applied stress ($\beta = 0$), these saddle point equations are easily solved.
They give three classes of solutions.
\begin{itemize}
\item No coexistence ($\Phi_{1,2}$ only) solutions whose $\vb{\Phi}$ has only one finite component:
\begin{align}
\vb{\Phi} &= \Phi_0 \begin{pmatrix}
1 \\ 0
\end{pmatrix} \quad\text{or}\quad \Phi_0 \begin{pmatrix}
0 \\ 1
\end{pmatrix}. \label{eq:SRO-GL-noStrain-sols-B1g}
\end{align}
I.e., $\vartheta = 0$ or $\pi$ and $\vb{\Upsilon} = \pm \, \Phi_0^2 \, \vu{e}_z$.
In the case of symmetry-protected degeneracy ($\alpha = 0$), the $\Phi_1$ only and $\Phi_2$ only ground states are degenerate because of the diagonal rotation symmetry $C_{2d_+}\colon \Phi_{1,2} \mapsto \Phi_{2,1}$ which continues to be a symmetry in the presence of $\sigma_6$ stress.
For $\vb{\Phi} \in E_{g,u}$, the two solutions we may thus identify with $B_{1g}$-nematic order.
\item $B_{2g}$-nematic solutions whose
\begin{align}
\vb{\Phi} &= \Phi_0 \begin{pmatrix}
\cos \tfrac{\vartheta}{2} \\[2pt]
\pm \sin \tfrac{\vartheta}{2}
\end{pmatrix} \label{eq:SRO-GL-noStrain-sols-B2g}
\end{align}
with a $\displaystyle \vartheta = \arccos\mleft(\frac{\alpha v_x}{v_z - v_x}\mright)$ and $\varphi = 0$ or $\pi$. $\vb{\Upsilon} = \Phi_0^2 (\pm \sin \vartheta \, \vu{e}_x + \cos \vartheta \, \vu{e}_z)$.
Here the relevant symmetry operations are \SI{180}{\degree} rotations around the $x$ and $y$ axes which act according to $(\Phi_1, \Phi_2) \mapsto \pm (\Phi_1, - \Phi_2)$.
\item Time-reversal symmetry-breaking (TRSB) solutions whose
\begin{align}
\vb{\Phi} &= \Phi_0 \begin{pmatrix}
\cos \tfrac{\vartheta}{2} \\[2pt]
\pm \, \iu \sin \tfrac{\vartheta}{2}
\end{pmatrix} \label{eq:SRO-GL-noStrain-sols-TRSB}
\end{align}
with a $\displaystyle \vartheta = \arccos\mleft(\frac{\alpha v_y}{v_z - v_y}\mright)$ and $\displaystyle \varphi = \pm \frac{\pi}{2}$. $\vb{\Upsilon} = \Phi_0^2 (\pm \sin \vartheta \, \vu{e}_y + \cos \vartheta \, \vu{e}_z)$.
Time-reversal acts on $\vb{\Phi}$ through complex conjugation: $(\Phi_1, \Phi_2) \mapsto (\Phi_1^{*}, \Phi_2^{*})$
\end{itemize}

\begin{figure}[t!]
\centering
\includegraphics[width=0.85\textwidth]{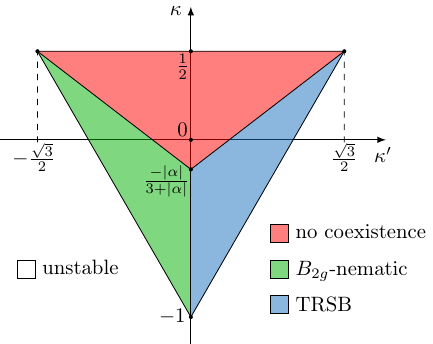}
\captionbelow[The phase space of the Ginzburg-Landau theory for accidentally degenerate two-component superconducting order parameters ($\abs{\alpha} > 0$) in the absence of strain ($\sigma_6 = \beta = 0$).]{\textbf{The phase space of the Ginzburg-Landau theory for accidentally degenerate two-component superconducting order parameters ($\abs{\alpha} > 0$) in the absence of strain ($\sigma_6 = \beta = 0$).}
The color indicates the global ground state, as specified by Eqs.~\eqref{eq:SRO-GL-noStrain-sols-B1g}, \eqref{eq:SRO-GL-noStrain-sols-B2g}, and~\eqref{eq:SRO-GL-noStrain-sols-TRSB}.
The region outside the equilateral triangle is unstable on the quartic level.
In the no coexistence region, $\Phi_1$ ($\Phi_2$) is preferred for $\alpha > 0$ ($\alpha < 0$).
For symmetry-protected order parameters, $\alpha = 0$ so the triple point $(\kappa', \kappa) = \mleft(0, \frac{-\abs{\alpha}}{3 + \abs{\alpha}}\mright)$ coincides with the origin $\vb{0}$ and the $\Phi_{1,2}$ solutions of the no coexistence region become degenerate and can be identified with $B_{1g}$-nematic order.}
\label{fig:SRO-GL-phase-space}
\end{figure}

The free energy values for these solutions are:
\begin{align}
F_{\Phi 0} &= - \frac{a^2}{4} \begin{cases}
\displaystyle \frac{(1 + \alpha)^2}{v_z}, & \text{for $\Phi_1$ only,} \\[8pt]
\displaystyle \frac{(1 - \alpha)^2}{v_z}, & \text{for $\Phi_2$ only,} \\[8pt]
\displaystyle \frac{1}{v_x} + \frac{\alpha^2}{v_z - v_x}, & \text{for $B_{2g}$-nematic,} \\[8pt]
\displaystyle \frac{1}{v_y} + \frac{\alpha^2}{v_z - v_y}, & \text{for TRSB.}
\end{cases} \label{eq:SRO-GL-app-free-energy}
\end{align}
For the preferred global minimum, $\vb{\Upsilon}$ points in the ``softest'' direction whose quartic coefficients $v_{x,y,z}$ are the smallest, as one would intuitively expect.
To be more precise, introduce the set
\begin{align}
\mathscr{V} = \mleft(v_x, ~v_y, ~\frac{v_z}{1 + \abs{\alpha}}\mright) = w \times \mleft(1 + \kappa + \sqrt{3} \, \kappa', ~1 + \kappa - \sqrt{3} \, \kappa', ~\frac{1 - 2 \kappa}{1 + \abs{\alpha}}\mright).
\end{align}
Then the global minimum is
\begin{itemize}
\item $\Phi_1$ only when $\displaystyle \min \mathscr{V} = \frac{v_z}{1 + \abs{\alpha}}$ and $\alpha > 0$,
\item $\Phi_2$ only when $\displaystyle \min \mathscr{V} = \frac{v_z}{1 + \abs{\alpha}}$ and $\alpha < 0$,
\item $B_{2g}$-nematic when $\displaystyle \min \mathscr{V} = v_x$, and
\item TRSB when $\displaystyle \min \mathscr{V} = v_y$.
\end{itemize}
The corresponding phase diagram is shown in Fig.~\ref{fig:SRO-GL-phase-space}.
For vanishing $\alpha$, the triple point is moved to the origin and the no coexistence region attains two degenerate  $B_{1g}$-nematic solutions.

\subsubsection{Solutions in the presence of $B_{2g}$ stress ($\sigma_6 \neq 0$)}
First, let us consider $T > T_{c0}$.
In this case, given that $a = (T - T_{c0}) \dot{a} > 0$, a non-trivial solution with $F_{\Phi 0} < 0$ is only obtained when $A(\vartheta, \varphi) < 0$.
By minimizing Eq.~\eqref{eq:SRO-GL-A}, we see that the minimum of $A(\vartheta, \varphi)$ is $1 - \sqrt{\alpha^2 + \beta^2}$ and has $\varphi = 0$ or $\pi$ with $\vartheta \neq 0$.
This corresponds to $B_{2g}$-nematic order.
Hence the upper transition occurs for
\begin{equation}
\abs{\beta} = \beta_c = \sqrt{1 - \alpha^2}, \label{eq:SRO-beta_c-eq}
\end{equation}
which translates to
\begin{equation}
T_c = T_{c0} + \frac{\lambda_6 \abs{\epsilon_{6,0}}}{\dot{a}} \frac{2}{\sqrt{1-\alpha^2}}
\end{equation}
and the symmetry of the state is $B_{2g}$-nematic.
In the symmetry-protected case ($\alpha = 0$), $\vartheta = - \tfrac{1}{2} \pi \sgn \beta = - \tfrac{1}{2} \pi \sgn \sigma_6$, while for $\alpha \neq 0$ the angle $\vartheta$ takes values in between $0$ and $- \pi \sgn \beta = - \pi \sgn \sigma_6$.

Now consider reducing $T$ below $T_c$.
As illustrated in Fig.~\ref{fig:schematicPhaseDiagrams}, a second transition takes place when the ground state breaks time-reversal symmetry, whether the degeneracy is symmetry-protected or not, and when the ground state is $B_{1g}$-nematic.
In the latter case, the degeneracy must be symmetry-protected because only then is the $(\Phi_1, \Phi_2) \mapsto (\Phi_2, \Phi_1)$ diagonal rotation symmetry present which forbids a smooth crossover between $B_{1g}$ and $B_{2g}$-nematic states.

To determine the lower transition temperature $T_2$, we need to solve the saddle point equations~\eqref{eq:SRO-GL-saddlePoint} and figure out which solution yields the smallest free energy.

\begin{figure}[t]
\centering
\begin{subfigure}[t]{0.5\textwidth}
\raggedright
\includegraphics[width=0.98\textwidth]{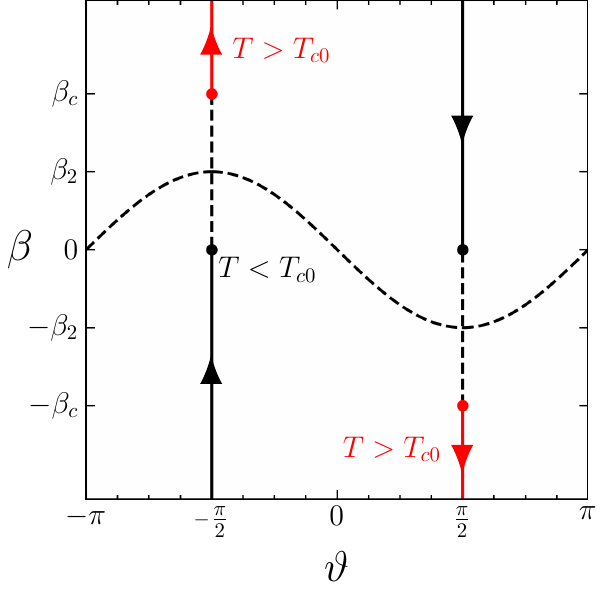}
\subcaption{}
\end{subfigure}%
\begin{subfigure}[t]{0.5\textwidth}
\raggedright
\includegraphics[width=0.98\textwidth]{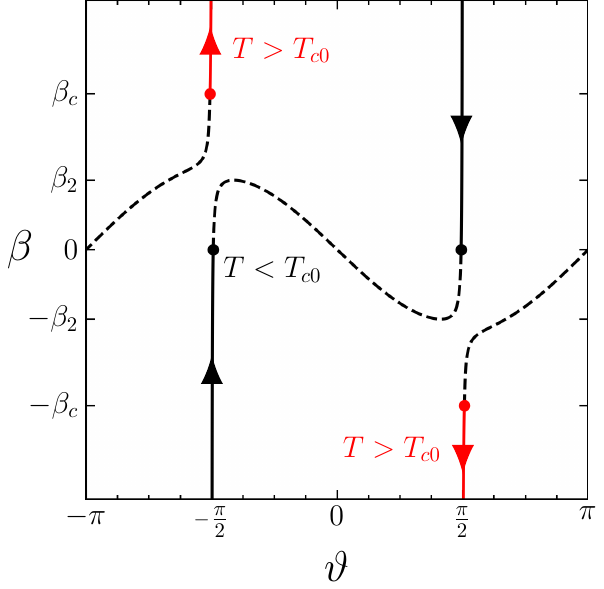}
\subcaption{}
\end{subfigure}
\captionbelow[The evolution of nematic ($\varphi = 0$) saddle point solutions as a function of $\beta = (2 \lambda_6 / \dot{a}) \cdot \epsilon_{6,0} / (T - T_{c0})$ when $B_{2g}$-nematic states are the ground state for $v_x / v_z = 0.5$ and $\alpha = 0$ (a) and for $v_x / v_z = 0.5$ and $\alpha = 0.02$ (b).]{\textbf{The evolution of nematic ($\varphi = 0$) saddle point solutions as a function of $\beta = (2 \lambda_6 / \dot{a}) \cdot \epsilon_{6,0} / (T - T_{c0})$ when $B_{2g}$-nematic states are the ground state for $v_x / v_z = 0.5$ and $\alpha = 0$ (a) and for $v_x / v_z = 0.5$ and $\alpha = 0.02$ (b).}
Solid lines are stable solutions, while dashed lines are unstable solutions, of Eq.~\eqref{eq:SRO-GL-nematicSaddlePointEq}.
Black (red) indicates the portion where $T < T_{c0}$ ($T > T_{c0}$).
The $\vartheta \in \langle -\pi, 0 \rangle$ part corresponds to $\epsilon_{6,0} > 0$, whereas for strain $\epsilon_{6,0} < 0$ the angle $\vartheta \in \langle 0, \pi \rangle$.
The arrows indicate the direction of the evolution as the temperature is lowered.
The red dots are the initial solutions at the upper transition $T = T_c$, while black dots are the final solutions in the absence of strain $\epsilon_{6,0} = 0$ (formally $T \to - \infty$).
The $\beta_c$ and $\beta_2$ are provided in Eqs.~\eqref{eq:SRO-beta_c-eq} and~\eqref{eq:SRO-beta_2-eq}.}
\label{fig:SRO-bifurcation-B2g}
\end{figure}

\begin{figure}[p!]
\centering
\begin{subfigure}[t]{0.5\textwidth}
\raggedright
\includegraphics[width=0.98\textwidth]{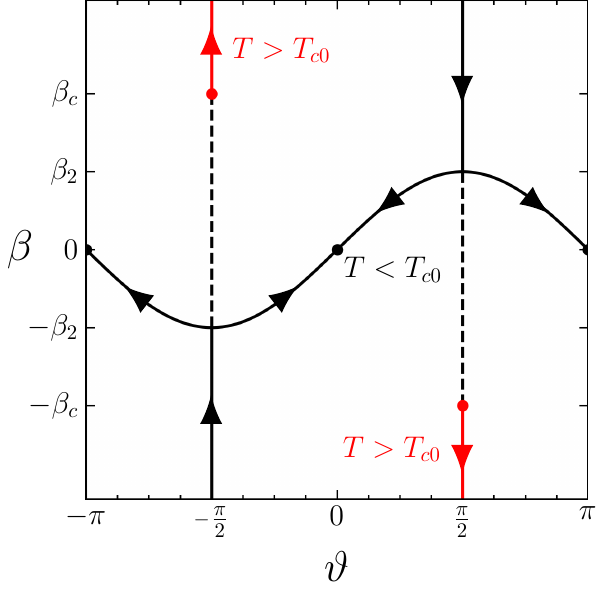}
\subcaption{}
\end{subfigure} \\[20pt]
\begin{subfigure}[t]{0.5\textwidth}
\raggedright
\includegraphics[width=0.98\textwidth]{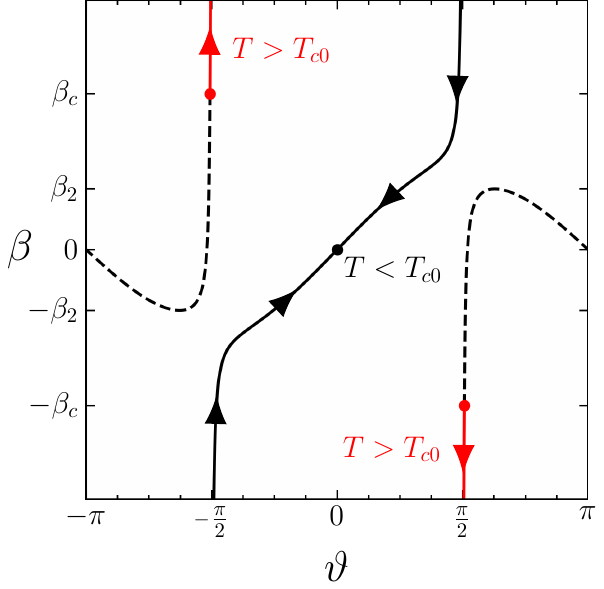}
\subcaption{}
\end{subfigure}%
\begin{subfigure}[t]{0.5\textwidth}
\raggedright
\includegraphics[width=0.98\textwidth]{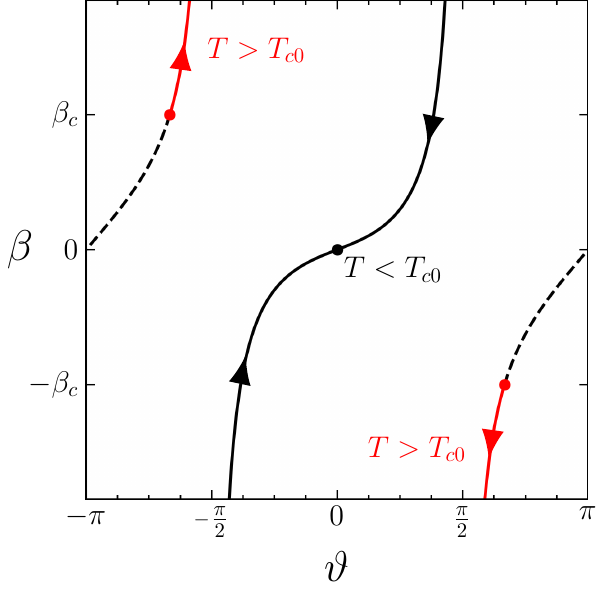}
\subcaption{}
\end{subfigure}
\captionbelow[The evolution of nematic ($\varphi = 0$) saddle point solutions as a function of $\beta = (2 \lambda_6 / \dot{a}) \cdot \epsilon_{6,0} / (T - T_{c0})$ when no coexistence ($B_{1g}$-nematic) states are the ground state for $v_x / v_z = 1.5$ and $\alpha = 0$ (a), for $v_x / v_z = 1.5$ and $\alpha = 0.02$ (b), and for $v_x / v_z = 0.8$ and $\alpha = 0.5$ (c).]{\textbf{The evolution of nematic ($\varphi = 0$) saddle point solutions as a function of $\beta = (2 \lambda_6 / \dot{a}) \cdot \epsilon_{6,0} / (T - T_{c0})$ when no coexistence ($B_{1g}$-nematic) states are the ground state for $v_x / v_z = 1.5$ and $\alpha = 0$ (a), for $v_x / v_z = 1.5$ and $\alpha = 0.02$ (b), and for $v_x / v_z = 0.8$ and $\alpha = 0.5$ (c).}
Solid lines are stable solutions, while dashed lines are unstable solutions, of Eq.~\eqref{eq:SRO-GL-nematicSaddlePointEq}.
Black (red) indicates the portion where $T < T_{c0}$ ($T > T_{c0}$).
The $\vartheta \in \langle -\pi, 0 \rangle$ part corresponds to $\epsilon_{6,0} > 0$, whereas for strain $\epsilon_{6,0} < 0$ the angle $\vartheta \in \langle 0, \pi \rangle$.
The arrows indicate the direction of the evolution as the temperature is lowered.
The red dots are the initial solutions at the upper transition $T = T_c$, while black dots are the final solutions in the absence of strain $\epsilon_{6,0} = 0$ (formally $T \to - \infty$).
The $\beta_c$ and $\beta_2$ are provided in Eqs.~\eqref{eq:SRO-beta_c-eq} and~\eqref{eq:SRO-beta_2-eq}.}
\label{fig:SRO-bifurcation-B1g}
\end{figure}

We start with the nematic case.
Its $\varphi = 0$, while its $\vartheta$ is determined by the transcendental equation:
\begin{equation}
\beta \cos(\vartheta) = \mleft(\frac{v_x}{v_z} - 1\mright) \cos(\vartheta) \sin(\vartheta) + \alpha \frac{v_x}{v_z} \sin(\vartheta). \label{eq:SRO-GL-nematicSaddlePointEq}
\end{equation}
For $\alpha \neq 0$, this equation cannot be inverted to get $\vartheta(\beta)$ in closed form.
However, plotting $\beta$ as a function of $\vartheta$ is just as instructive, as we have done in Figs.~\ref{fig:SRO-bifurcation-B1g} and~\ref{fig:SRO-bifurcation-B2g}.
By inspecting this equation (see figures), one may readily confirm that it has two solutions for large $\abs{\beta}$.
When $\abs{\beta}$ becomes smaller than
\begin{equation}
\beta_2 = \frac{\abs{v_x - v_z}}{v_z} \mleft[1 - \abs{\mathcal{X}}^{2/3}\mright]^{3/2}, \label{eq:SRO-beta_2-eq}
\end{equation}
two additional solutions may appear if
\begin{equation}
\mathcal{X} \equiv \frac{\abs{\alpha} v_x}{v_z - v_x}
\end{equation}
is smaller than $1$, $\abs{\mathcal{X}} < 1$.

As can be seen from Fig.~\ref{fig:SRO-bifurcation-B2g}, when the $\sigma_6 = \beta = 0$ ground state is $B_{2g}$-nematic, $\vartheta$ of the global minimum changes smoothly with temperature at a fixed $\sigma_6$ and there is no second transition.
The same happens when $\Phi_1$ or $\Phi_2$ are the ground states and $\alpha \neq 0$ [Fig.~\ref{fig:SRO-bifurcation-B1g}(b)\&(c)]: we have a smooth crossover.
This follows from the fact there is no symmetry which would prevent such a crossover.

When the ground state is $B_{1g}$-nematic and $\alpha = 0$ [Fig.~\ref{fig:SRO-bifurcation-B1g}(a)], $B_{1g}$-nematic solutions overtake the $B_{2g}$-nematic solutions below $\abs{\beta} = \beta_2 = (v_x - v_z) / v_z$, yielding
\begin{equation}
T_2 = T_{c0} - \frac{\lambda_6 \abs{\epsilon_{6,0}}}{\dot{a}} \frac{2 v_z}{v_x - v_z}.
\end{equation}

When the ground state is TRSB with symmetry-protected degeneracy:
\begin{equation}
T_2 = T_{c0} - \frac{\lambda_6 \abs{\epsilon_{6,0}}}{\dot{a}} \frac{2 v_y}{v_x - v_y}. \label{eq:SRO-GL-T2_symmProtectedTRSB}
\end{equation}
Along the line $v_y = v_z$ ($\kappa' = \sqrt{3} \, \kappa$) that is the boundary between the $B_{1g}$ and TRSB regions of the $\alpha=0$ parameter space, these two expressions for $T_2$ agree.
When the ground state is TRSB with accidental degeneracy:
\begin{equation}
T_2 = T_{c0} - \frac{\lambda_6 \abs{\epsilon_{6,0}}}{\dot{a}} \frac{2 v_y}{v_x - v_y} \frac{v_z - v_y}{\sqrt{(v_z - v_y)^2 - \alpha^2 v_y^2}}. \label{eq:SRO-GL-T2_accidentalTRSB}
\end{equation}
In the TRSB case, one may solve the saddle point equations analytically in closed form:
\begin{align}
\vartheta & = \arccos\mleft(\frac{\alpha v_y}{v_z - v_y}\mright), \\
\varphi & = \pm \arccos\mleft(\frac{\lambda \epsilon_{6,0}}{a} \frac{2 v_y}{v_x - v_y} \frac{v_z - v_y}{\sqrt{(v_z - v_y)^2 - \alpha^2 v_y^2}}\mright), \\[2pt]
F_{\Phi 0} & = -\frac{a^2}{4} \mleft(\frac{1}{v_y} + \frac{\alpha^2}{v_z - v_y}\mright) - \frac{\lambda_6^2 \epsilon_{6,0}^2}{v_x - v_y}. \label{eq:SRO-GL-TRSBFreeEnergy}
\end{align}

\subsubsection{Ehrenfest relations} \label{sec:SRO-Ehrenfest_relations_appendix}
The jump in the heat capacity across the superconducting transition is given by:
\begin{equation}
\frac{\Delta C_0}{T_{c0}} = - \mleft.\frac{\partial^2 F_{\Phi 0}}{\partial T^2}\mright|_{T = T_{c0}, \sigma_6 = 0}.
\end{equation}
From the free energy expressions of Eq.~\eqref{eq:SRO-GL-app-free-energy}:
\begin{align}
\frac{\Delta C_0}{T_{c0}} &= \frac{\dot{a}^2}{2} \begin{cases}
\displaystyle \frac{(1 + \abs{\alpha})^2}{v_z}, & \text{for $\Phi_1$ or $\Phi_2$ only,} \\[8pt]
\displaystyle \frac{1}{v_x} + \frac{\alpha^2}{v_z - v_x}, & \text{for $B_{2g}$-nematic} \\[8pt]
\displaystyle \frac{1}{v_y} + \frac{\alpha^2}{v_z - v_y}, & \text{for TRSB.}
\end{cases}
\end{align}

The shear elastic modulus $c_{66}$ below $T_c$ is given by
\begin{equation}
\frac{1}{c_{66}} = \frac{1}{c_{66,0}} - \mleft.\frac{\partial^2 F_{\Phi 0}}{\partial \sigma_6^2}\mright|_{T, \sigma_6 = 0}.
\end{equation}
The jump $\Delta c_{66} = c_{66,0} - c_{66}|_{T=T_{c0}}$ is the difference between $c_{66}$ just above $T_{c0}$ and that just below it.
Since $\Delta c_{66}$ is so small, below we use $1/c_{66} = 1/c_{66,0} + \Delta c_{66} / c_{66,0}^2$.

When the ground state is $\Phi_1$ or $\Phi_2$ only,
\begin{equation}
\Delta c_{66} = 2 \lambda_6^2 \frac{1 + \abs{\alpha}}{(1 + \abs{\alpha}) v_x - v_z}.
\end{equation}
This is derived by solving Eq.~\eqref{eq:SRO-GL-nematicSaddlePointEq} for small $\beta$.
In the case of symmetry-enforced degeneracy ($\alpha = 0$), that is, $B_{1g}$-nematic ground states, one obtains the following Ehrenfest relation:
\begin{align}
\Delta c_{66} &= \frac{\Delta C_0}{T_{c0}} \abs{\dv{T_c}{\epsilon_{6,0}}} \abs{\dv{T_2}{\epsilon_{6,0}}}. \label{eq:SRO-B1g-Ehrenfest-rel}
\end{align}
In the general $\alpha \neq 0$ case, we could try using $T_c$ instead of $T_2$ above, but the corresponding dimenionless ratio
\begin{align}
\frac{\Delta c_{66}}{\displaystyle \frac{\Delta C_0}{T_{c0}} \abs{\dv{T_c}{\epsilon_{6,0}}} \abs{\dv{T_c}{\epsilon_{6,0}}}} &= \frac{(1 - \alpha^2) (1 + \mathcal{X} / \abs{\alpha})}{(1 + \abs{\alpha}) (\mathcal{X} - 1)}
\end{align}
can be any positive real number, depending on the values of $\alpha$ and $\mathcal{X} = \abs{\alpha} v_x / (v_z - v_x) \in \langle - \infty, - \abs{\alpha} \rangle \cup \langle 1, + \infty \rangle$ which we do not know.

When the ground state is $B_{2g}$-nematic, by solving Eq.~\eqref{eq:SRO-GL-nematicSaddlePointEq} one finds that
\begin{equation}
\Delta c_{66} = 2 \lambda_6^2 \frac{1 + \mathcal{X}^3 / \abs{\alpha}}{v_x (1 - \mathcal{X}^2)},
\end{equation}
and therefore
\begin{equation}
\frac{\Delta c_{66}}{\displaystyle \frac{\Delta C_0}{T_{c0}} \abs{\dv{T_c}{\epsilon_{6,0}}} \abs{\dv{T_c}{\epsilon_{6,0}}}} = \frac{(1 - \alpha^2) (1 + \mathcal{X}^3 / \abs{\alpha})}{(1 + \abs{\alpha} \mathcal{X}) (1 - \mathcal{X}^2)}, \label{eq:SRO-B2g-Ehrenfest-rel}
\end{equation}
where
\begin{align}
0 ~<~ \mathcal{X} \equiv \frac{\abs{\alpha} v_x}{v_z - v_x} ~<~ 1.
\end{align}
When $\alpha = \mathcal{X} = 0$, this expression reduces to the standard Ehrenfest relation.
The stability condition for $B_{2g}$-nematic order corresponds to $0 < \mathcal{X} < 1$ indicated above and the right-hand side can equal any number between $(1 - \alpha^2)$ and $+ \infty$ for $\alpha \neq 0$ and $\mathcal{X}$ in this range.

When the ground state is TRSB, the second derivative of Eq.~\eqref{eq:SRO-GL-TRSBFreeEnergy} with respect to $\epsilon_{6,0}$ yields
\begin{equation}
\Delta c_{66} = 2 \lambda_6^2 \frac{1}{v_x - v_y}.
\end{equation}
The corresponding Ehrenfest relation takes the form:
\begin{equation}
\frac{\Delta c_{66}}{\displaystyle \frac{\Delta C_0}{T_{c0}} \abs{\dv{T_c}{\epsilon_{6,0}}} \abs{\dv{T_2}{\epsilon_{6,0}}}} = \frac{\sqrt{1-\alpha^2}\sqrt{1 - \mathcal{Y}^2}}{1 + \abs{\alpha} \mathcal{Y}} \leq 1, \label{eq:SRO-TRSB-Ehrenfest-rel}
\end{equation}
where
\begin{align}
0 ~<~ \mathcal{Y} \equiv \frac{\abs{\alpha} v_y}{v_z - v_y} ~<~ 1.
\end{align}
In the $0 < \mathcal{Y} < 1$ region where TRSB is the ground state, the right-hand side of Eq.~\eqref{eq:SRO-TRSB-Ehrenfest-rel} takes values in between $0$ and $1$, and for $\alpha = 0$ equals $1$.

\subsubsection{Ratio relations} \label{sec:SRO-ratio-relations}
Here we show that the ratios of the jumps at the upper and lower transitions are related.
These relations hold only for the symmetry-protected case ($\alpha = 0$).
The heat capacity and elastic modulus jumps we shall denote $\Delta C_c$ and $\Delta c_{66,\text{c}}$ at the upper transition ($T = T_c$), and $\Delta C_2$ and $\Delta c_{66,2}$ at the lower transition ($T = T_2$), respectively.

The jumps at the upper transition are ($\alpha = 0$, $\sigma_6 \neq 0$):
\begin{align}
\frac{\Delta C_c}{T_c} &= \frac{\dot{a}^2}{2 v_x}, \\
\Delta c_{66,\text{c}} &= \frac{2 \lambda_6^2}{v_x}.
\end{align}
The jumps at the lower transition are ($\alpha = 0$, $\sigma_6 \neq 0$):
\begin{align}
\frac{\Delta C_2}{T_2} &= \frac{\dot{a}^2}{2} \begin{cases}
\displaystyle \frac{v_x - v_z}{v_x v_z}, & \text{for $B_{1g}$-nematic,} \\[8pt]
\displaystyle \frac{v_x - v_y}{v_x v_y}, & \text{for TRSB,}
\end{cases} \\
\Delta c_{66,2} &= 2 \lambda_6^2 \begin{cases}
\displaystyle \frac{v_z}{v_x (v_x - v_z)}, & \text{for $B_{1g}$-nematic,} \\[8pt]
\displaystyle \frac{v_y}{v_x (v_x - v_y)}, & \text{for TRSB.}
\end{cases}
\end{align}
To find these expressions, we had to solve Eq.~\eqref{eq:SRO-GL-nematicSaddlePointEq} around the $\beta$ at which the solutions bifurcate.
Note that $\Delta C_c/T_c + \Delta C_2/T_2$ and $\Delta c_{66,\text{c}} + \Delta c_{66,2}$ reproduce the previous $\Delta C_0/T_{c0}$ and $\Delta c_{66}$ with $\alpha = 0$.
Combining, we obtain the ratio relations:
\begin{equation}
\frac{\displaystyle \abs{\dv{T_2}{\epsilon_{6,0}}}}{\displaystyle \abs{\dv{T_c}{\epsilon_{6,0}}}} = \frac{\displaystyle \frac{\Delta C_c}{T_c}}{\displaystyle \frac{\Delta C_2}{T_2}} = \frac{\Delta c_{66,2}}{\Delta c_{66,\text{c}}} = \begin{cases}
\displaystyle \frac{v_z}{v_x - v_z}, & \text{for $B_{1g}$-nematic,} \\[8pt]
\displaystyle \frac{v_y}{v_x - v_y}, & \text{for TRSB.}
\end{cases} \label{eq:SRO-GL-ratioRelationsAppendix}
\end{equation}
Thus small second-transition heat capacity jumps $\Delta C_2$ imply large cusps for $T_2$, but also large elastic modulus jumps $\Delta c_{66,2}$ at the second transition.

With some work, one can also derive the ratio relation for the accidentally-degenerate TRSB case ($\alpha \neq 0$):
\begin{align}
\begin{aligned}
\frac{\sqrt{1 - \mathcal{Y}^2}}{\sqrt{1 - \alpha^2}} \cdot \frac{\displaystyle \abs{\dv{T_2}{\epsilon_{6,0}}}}{\displaystyle \abs{\dv{T_c}{\epsilon_{6,0}}}} &= \frac{(1 - \mathcal{Y}^2)^2 (1 + \abs{\alpha} \mathcal{X})}{(1 - \alpha^2)^2 (1 - \mathcal{Y}^3 / \mathcal{X})} \cdot \frac{\displaystyle \frac{\Delta C_c}{T_c}}{\displaystyle \frac{\Delta C_2}{T_2}} \\
&= \frac{(1 - \alpha^2) (1 - \mathcal{Y}^3 / \mathcal{X})}{(1 - \mathcal{Y}^2) (1 + \abs{\alpha} \mathcal{X})} \cdot \frac{\Delta c_{66,2}}{\Delta c_{66,\text{c}}} = \frac{v_y}{v_x - v_y}.
\end{aligned} \label{eq:SRO-GL-ratioRelationsAppendix-attempt}
\end{align}
TRSB is the ground state at $\sigma_6 = 0$ when $\mathcal{Y} \in \langle 0, 1 \rangle$ and $\mathcal{X} \in \langle - \infty, - \abs{\alpha} \rangle \cup \langle \mathcal{Y}, + \infty \rangle$.
By varying $\mathcal{X}$, $\mathcal{Y}$, and $\alpha \in \langle -1, 1 \rangle$ within their allowed ranges, one can make the prefactors arising in the above equation take any value.
This makes the TRSB ratio relation for accidentally degenerate states of little practical use.

Unlike the Ehrenfest relations, these ratio relations tie together properties at finite strain.
If we are applying $\sigma_{110}$ stress, for instance, this will induce not only $\epsilon_{6,0}$ strain, but also $A_{1g}$ strain components.
Since $\dot{a}$, $\lambda_6$, $v_{\mu}$, etc., can all depend linearly on $A_{1g}$ strain, the ratio relation~\eqref{eq:SRO-GL-ratioRelationsAppendix} formally holds only as $\epsilon_{6,0} \to 0$.
More precisely, the ratio of the cusps in Eq.~\eqref{eq:SRO-GL-ratioRelationsAppendix} is evaluated at $\epsilon_{A_{1g}} = 0$, while the ratios of the heat capacity and elastic modulus jumps are evaluated at a finite $\epsilon_{A_{1g}} \neq 0$, which means that the $v_{\mu}$ ratios that they equal [rightmost part of Eq.~\eqref{eq:SRO-GL-ratioRelationsAppendix}] are suppose to be evaluated at different $A_{1g}$ strains.
That said, the Ginzburg-Landau expansion of the free energy only holds in the vicinity of $T_c$, so $\abs{T_c - T_2}$ needs to be small anyway.
The domain of validity of Eq.~\eqref{eq:SRO-GL-ratioRelationsAppendix} is thus not any smaller or larger than that of the Ginzburg-Landau analysis as a whole.

As an aside, let us note that the reason why non-trivial Ehrenfest and ratio relations can be derived in the first place is because the \emph{cusp-like part} of the slope $\abs{\dd{T_c}/\dd{\epsilon_{6,0}}}$, the \emph{jump} in the heat capacity $\Delta C$, the \emph{jump} in the $c_{66}$ elastic coefficient, etc., all isolate only one coupling constant: $\lambda_6$.
One may thus relate the corresponding dimensionless, experimentally-measurable quantities to the Ginzburg-Landau expansion coefficients.
In contrast, if we were to look at the total $T_c$ slope, total heat capacity, and so on, because of the contributions from other $\lambda_{iab}$ in Eq.~\eqref{eq:SRO-GL-lambda-coupling-constants}, it is difficult to make similar statements.

\subsubsection{Bounds on the nematic strain}
The second term in Eq.~\eqref{eq:SRO-GL-externalStrain} defines the ``internal'' strain, which is the strain generated by the superconducting order parameter:
\begin{align}
\begin{aligned}
\epsilon_6^\text{nem} &= -\frac{\lambda_6}{c_{66}}\big(\Phi_1^* \Phi_2 + \Phi_2^* \Phi_1\big) \\
&= -\frac{\lambda_6}{c_{66}}\Phi_0^2 \sin \vartheta \cos \varphi.
\end{aligned}
\end{align}
Due to the proportionality to $\cos \varphi$, when $\sigma_6 = 0$ only the $B_{2g}$-nematic states generate a non-zero $\epsilon_6$.
Its value is bounded from above through
\begin{equation}
\frac{c_{66,0} \abs{\epsilon_6^{\text{nem}}}}{\displaystyle \frac{\Delta C_0}{T_{c0}} \abs{\dv{T_c}{\epsilon_{6,0}}} \abs{T - T_{c0}}} = \frac{\sqrt{1 - \alpha^2} \sqrt{1 - \mathcal{X}^2}}{1 + \abs{\alpha} \mathcal{X}} \leq 1, \label{eq:SRO-GL-nematicUpperLimit}
\end{equation}
where the right-hand side is in between $0$ and $1$ in the range $0 < \mathcal{X} = \frac{\abs{\alpha} v_x}{v_z - v_x} < 1$ where $B_{2g}$-nematic order is preferred and for $\alpha = 0$ equals $1$.

\subsubsection{The case of $B_{1g}$ stress} \label{sec:B1g_relations_appendix}
As we shall see in the next Sec.~\ref{sec:SRO-110-implications}, if one combines the measurements of Ref.~\cite{Jerzembeck2024} with those performed under $[100]$ uniaxial stress~\cite{Li2021}, one can put tight constraints on where precisely \ce{Sr2RuO4} must be in the phase diagram of Fig.~\ref{fig:SRO-GL-phase-space}.
See Fig.~\ref{fig:triplePointTuning} in particular.

To make contact with the measurements under $\langle 100 \rangle$ uniaxial stress, here we briefly summarize the results of the Ginzburg-Landau analysis for $B_{1g}$ stress $\sigma_{B_{1g}} = \tfrac{1}{2} (\sigma_1 - \sigma_2) = \sigma_{100} / 2$.
Superconductivity couples linearly to $B_{1g}$ stress only in the case of symmetry-protected degeneracy
\begin{equation}
\alpha = 0,
\end{equation}
which we henceforth consider.

In light of Tab.~\ref{tab:SRO-phi-irreps}, the coupling to $B_{1g}$ stress takes the form
\begin{align}
F_{\Phi 0} = \cdots + \sigma_{B_{1g}} c_{B_{1g}}^{-1} \lambda_{B_{1g}} \Upsilon_z,
\end{align}
where $c_{B_{1g}} \defeq \tfrac{1}{2} (c_{11} - c_{12})$; see Tabs.~\ref{tab:SRO-elastic-constants} and~\ref{tab:SRO-phi-irreps}.
By a rotation
\begin{equation}
\tilde{\vb{\Phi}} = \frac{1}{\sqrt{2}} \begin{pmatrix}
1 & -1 \\
1 & 1
\end{pmatrix} \vb{\Phi}
\end{equation}
and reparametrization
\begin{align}
\begin{aligned}
\tilde{v}_x &= v_z, \\
\tilde{v}_y &= v_y, \\
\tilde{v}_z &= v_x,
\end{aligned}
\end{align}
one obtains a free energy identical in form to Eq.~\eqref{eq:SRO-GL-F_Delta}.
Hence all the previous formulas carry over if we replace $v_x, v_y, v_z, \lambda_6$ with $\tilde{v}_x, \tilde{v}_y, \tilde{v}_z, \lambda_{B_{1g}}$, and exchanges what one identifies as $B_{1g}$ with $B_{2g}$, and vice versa.

The upper transition temperature is given by:
\begin{align}
T_c &= T_{c0} + \frac{2 \lambda_{B_{1g}} \abs{\epsilon_{B_{1g},0}}}{\dot{a}}.
\end{align}
At finite $B_{1g}$ stress, the superconductivity is $B_{1g}$-nematic slightly below $T_c$.
When $B_{1g}$-nematic pairing is the ground state, there is no second transition.
For the other two cases:
\begin{align}
T_2 &= T_{c0} - \frac{2 \lambda_{B_{1g}} \abs{\epsilon_{B_{1g},0}}}{\dot{a}} \begin{cases}
\displaystyle \frac{v_x}{v_z - v_x}, & \text{for $B_{2g}$-nematic,} \\[8pt]
\displaystyle \frac{v_y}{v_z - v_y}, & \text{for TRSB.}
\end{cases}
\end{align}

The heat capacity jumps:
\begin{align}
\frac{\Delta C_c}{T_c} &= \frac{\dot{a}^2}{2 u}, \\
\frac{\Delta C_2}{T_2} &= \frac{\dot{a}^2}{2 u} \begin{cases}
\displaystyle \frac{v_z - v_x}{v_x}, & \text{for $B_{2g}$-nematic,} \\[8pt]
\displaystyle \frac{v_z - v_y}{v_y}, & \text{for TRSB.}
\end{cases}
\end{align}
The jumps in the $B_{1g}$ elastic constants:
\begin{align}
\Delta c_{B_{1g},\text{c}} &= \frac{2 \lambda_{B_{1g}}^2}{u}, \\
\Delta c_{B_{1g},2} &= \frac{2 \lambda_{B_{1g}}^2}{u} \begin{cases}
\displaystyle \frac{v_x}{v_z - v_x}, & \text{for $B_{2g}$-nematic,} \\[8pt]
\displaystyle \frac{v_y}{v_z - v_y}, & \text{for TRSB.}
\end{cases}
\end{align}
The total jumps are obtained by summing the jumps at the upper and lower transition, if it takes place.

The Ehrenfest relation for $B_{1g}$-nematic states:
\begin{align}
\Delta c_{B_{1g}} &= \frac{\Delta C_0}{T_{c0}} \abs{\dv{T_c}{\epsilon_{B_{1g},0}}} \abs{\dv{T_c}{\epsilon_{B_{1g},0}}}.
\end{align}
The Ehrenfest relation when $B_{2g}$-nematic or TRSB pairing is preferred in the absence of stress:
\begin{align}
\Delta c_{B_{1g}} &= \frac{\Delta C_0}{T_{c0}} \abs{\dv{T_c}{\epsilon_{B_{1g},0}}} \abs{\dv{T_2}{\epsilon_{B_{1g},0}}}.
\end{align}
Ratio relations:
\begin{equation}
\frac{\displaystyle \abs{\dv{T_2}{\epsilon_{B_{1g},0}}}}{\displaystyle \abs{\dv{T_c}{\epsilon_{B_{1g},0}}}} = \frac{\displaystyle \frac{\Delta C_c}{T_c}}{\displaystyle \frac{\Delta C_2}{T_2}} = \frac{\Delta c_{B_{1g},2}}{\Delta c_{B_{1g},\text{c}}} = \begin{cases}
\displaystyle \frac{v_x}{v_z - v_x}, & \text{for $B_{2g}$-nematic,} \\[8pt]
\displaystyle \frac{v_y}{v_z - v_y}, & \text{for TRSB.}
\end{cases} \label{eq:SRO-GL-ratioRelationsAppendixB1g}
\end{equation}

\subsection{Theoretical implications: quantifying consistency and fine-tuning} \label{sec:SRO-110-implications}
One of the motivations for the measurements reported in Ref.~\cite{Jerzembeck2024} was to cross-check recent ultrasound experiments~\cite{Benhabib2021, Ghosh2021} which resolved jumps in the $c_{66}$ elastic constant at $T_c$.
As we have seen in the previous section, a jump in $c_{66} \in B_{2g}$ implies that the SC order parameter has two components which couple linearly to $\sigma_6 \in B_{2g}$ stress.
However, two-component SC that couples linearly to $\sigma_6$ should also exhibit transition splitting, as summarized in Tab.~\ref{fig:schematicPhaseDiagrams}, which has not been observed in $T_c$ or elastocaloric measurements~\cite{Jerzembeck2024}, as we reviewed in Sec.~\ref{sec:SRO-Tc-ECE-110-findings}.
The two are clearly at odds with one another.
Using the results of the Ginzburg-Landau analysis of the preceding section, here we examine the degree of fine-tuning that is needed for SRO's SC to be consistent with both experiments.
We do so under the assumption of a homogeneous SC order.
In other words, we shall suppose that all invoked experiments are giving information on bulk, homogeneous thermodynamic phases.

The following jumps in $c_{66}$ have been reported by Benhabib et al.~\cite{Benhabib2021}:
\begin{align}
\Delta c_{66} &= \begin{cases}
\SI{0.026}{\mega\pascal}, & \text{at \SI{169}{\mega\hertz},} \\
\SI{0.13}{\mega\pascal}, & \text{at \SI{201}{\mega\hertz} (not used).}
\end{cases} \label{eq:Benhabib2021-jump}
\end{align}
More precisely, they reported jumps in the ultrasound speed $\var{v_s} / v_s$ of magnitude \SI{0.2}{ppm} and \SI{1.0}{ppm} that are related to $c_{66}$ through $c_{66} = \rho v_s^2$.
For the elastic constants needed during various conversions, we employ those reported in Ref.~\cite{Ghosh2021}, which are listed in Tab.~\ref{tab:SRO-elastic-constants}.
These two$\Delta c_{66}$ were measured with two separate apparatuses using ultrasound pulse echos.
The difference between the two pulse-echo results has been attributed to possible mode mixing in the \SI{201}{\mega\hertz} experiment~\cite{Benhabib2021}.
We shall therefore use the value measured obtained at \SI{169}{\mega\hertz}.
From resonant ultrasound spectroscopy performed at much lower frequencies of approximately \SI{2}{\mega\hertz}, Ghosh et al.~\cite{Ghosh2021} deduced a larger value for the jump ($\Delta c_{66} / c_{66} = \SI{17.5}{ppm}$):
\begin{align}
\Delta c_{66} &= \SI{1.15}{\mega\pascal}.
\end{align}
It has been suggested that the difference between the pulse-echo and resonant ultrasound results is a consequence of the very different measurement frequencies~\cite{Ghosh2021}, with the higher frequencies thought to suppress the jump from its intrinsic thermodynamic value~\cite{Benhabib2021}.
Below we compare our results with both values reported for $\Delta c_{66}$.

On the basis of magnetic susceptibility measurements (Fig.~\ref{fig:SRO-J24-Tc}), in Sec.~\ref{sec:SRO-Tc-ECE-110-findings} we established that any putative cusp is smaller than~\cite{Jerzembeck2024}:
\begin{align}
\abs{\dv{T_c}{\epsilon_6}} \leq \SI{1.3}{\kelvin}. \label{eq:J24-bound-again}
\end{align}
In addition, we shall find it interesting to compare our results to experiments performed under $[100]$ uniaxial stress.
Because $E_g$ (and $E_u$) SC states couple linearly to both $B_{2g}$ and $B_{1g}$ strains (Tab.~\ref{tab:SRO-phi-irreps}), transition splitting, cusps in $T_c$, and jumps in elastic moduli should develop for both $[110]$ and $[100]$ stress directions.
However, neither a $T_c(\epsilon_{B_{1g}})$ cusp~\cite{Hicks2014, Steppke2017, Barber2019, Watson2018, Mueller2023} nor a $\Delta c_{B_{1g}}$ jump~\cite{Benhabib2021, Ghosh2021} has been resolved so the Ehrenfest relations of Sec.~\ref{sec:B1g_relations_appendix} cannot be exploited to make any strong statements.
The ratio relations of Sec.~\ref{sec:B1g_relations_appendix} prove to be more useful because of recent high-resolution heat capacity measurements~\cite{Li2021}.
Although a second transition has not been resolved~\cite{Li2021}, the tight bound
\begin{align}
\mleft.\frac{\displaystyle \frac{\Delta C_2}{T_2}}{\displaystyle \frac{\Delta C_c}{T_c}}\mright|_{\epsilon_{B_{1g}}} \leq 0.05 \label{eq:Li2021-bound}
\end{align}
that they put on the anomaly of any putative lower transition $T_2 < T_c$ can be used to make non-trivial statements.
For reference, the heat capacity anomaly $\Delta C_0$ in the absence of strain or magnetic fields has been measured to be \SI[per-mode=symbol]{40}{\milli\joule\per\mol\per\square\kelvin}~\cite{Kittaka2018} and \SI[per-mode=symbol]{41}{\milli\joule\per\mol\per\square\kelvin}~\cite{NishiZaki2000, Deguchi2004, Deguchi2004-p2} for high-quality samples with $T_c = \SI{1.505}{\kelvin}$ and \SI{1.48}{\kelvin}, respectively.
By using the molar mass \SI[per-mode=symbol]{340.3}{\gram\per\mol} and mass density \SI[per-mode=symbol]{5954}{\kilo\gram\per\cubic\meter}~\cite{Lupien2002}, this translates to
\begin{align}
\frac{\Delta C_0}{T_{c0}} &= \SI[per-mode=symbol]{470}{\joule\per\cubic\meter\per\square\kelvin}
\end{align}
up to a $\pm 10$ uncertainty that we shall suppress.

\begin{figure}[t!]
\centering
{\includegraphics[width=0.90\textwidth]{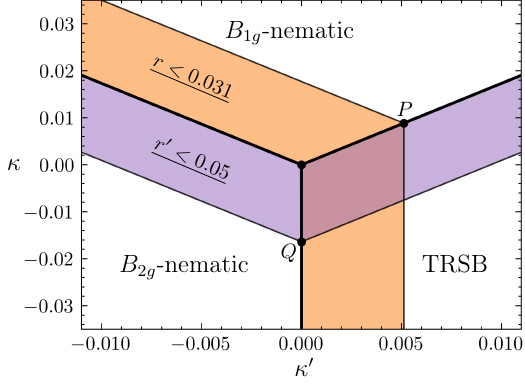} \hspace{16pt}}
\captionbelow[Regions consistent with the absence of a $T_c(\epsilon_6)$ cusp~\cite{Jerzembeck2024} (orange, $r < 0.031$) and the absence of a second heat capacity anomaly $\mleft.\Delta C_2\mright|_{\epsilon_{B_{1g}}}$~\cite{Li2021} (purple, $r' < 0.05$) for symmetry-protected $E_g(d_{yz}|-d_{xz})$ and $E_u(p_x|p_y)$ pairing.]{\textbf{Regions consistent with the absence of a $T_c(\epsilon_6)$ cusp}~\cite{Jerzembeck2024}\textbf{ (orange, $r < 0.031$) and the absence of a second heat capacity anomaly $\mleft.\Delta C_2\mright|_{\epsilon_{B_{1g}}}$}~\cite{Li2021}\textbf{ (purple, $r' < 0.05$) for symmetry-protected $E_g(d_{yz}|-d_{xz})$ and $E_u(p_x|p_y)$ pairing.}
The global minimum is $B_{1g}$-nematic for $\kappa > \sqrt{3} \abs{\kappa'}$, $B_{2g}$-nematic for $\kappa < - \sqrt{3} \, \kappa'$ and $\kappa' < 0$, and time-reversal symmetry-breaking (TRSB) for $\kappa < \sqrt{3} \, \kappa'$ and $\kappa' > 0$.
These global minima are divide with thick black lines (cf.\ Fig.~\ref{fig:SRO-GL-phase-space}).
The $T_c$ cusp bound~\eqref{eq:J24-bound-again} together with the conservative $\Delta c_{66}$ value~\eqref{eq:Benhabib2021-jump} implies $r < 0.031 = r_{\star}$, which is completely inconsistent with $B_{2g}$-nematic states whose $r = 1$, while for $B_{1g}$-nematic and TRSB it is only consistent in the orange region, which is specified by Eqs.~(\ref{eq:SRO-req1},~\ref{eq:SRO-req2}).
The heat capacity bound~\eqref{eq:Li2021-bound} implies $r' < 0.05 = r_{\star}'$, which is consistent with $B_{1g}$-nematic states whose $r' = 0$, while for $B_{2g}$-nematic and TRSB it is only consistent in the purple region, which is specified by Eqs.~(\ref{eq:SRO-rPrime-eq1},~\ref{eq:SRO-rPrime-eq2}).
The lines emanating from $P = \mleft(\frac{\sqrt{3} r_{\star}}{2 (3 + r_{\star})}, \frac{r_{\star}}{2 (3 + r_{\star})}\mright)$ and $Q = \mleft(0, - \frac{r_{\star}'}{3 + r_{\star}'}\mright)$ connect to the outer vertexes of Fig.~\ref{fig:SRO-GL-phase-space}.}
\label{fig:triplePointTuning}
\end{figure}

Now we go through the various possible two-component SC states and discuss the implications of the experimentally reported values that we provided above.
We start with the symmetry-protected $B_{2g}$-nematic SC since this is the simplest one to analyze.
According to the associated Ehrenfest relation [Eq.~\eqref{eq:SRO-B2g-Ehrenfest-rel} with $\alpha = \mathcal{X} = 0$], it follows that:
\begin{align}
\abs{\dv{T_c}{\epsilon_{6}}} &= \sqrt{\frac{\Delta c_{66}}{\displaystyle {\Delta C_0} / {T_{c0}}}} = \begin{cases}
\SI{7.4}{\kelvin}, & \text{for $\Delta c_{66}$ of Ref.~\cite{Benhabib2021},} \\
\SI{49}{\kelvin}, & \text{for $\Delta c_{66}$ of Ref.~\cite{Ghosh2021}.}
\end{cases}
\end{align}
Thus there is a discrepancy between a factor of $5.7$ and $38$ between our bound~\eqref{eq:J24-bound-again} and the ultrasound experiments.
We can therefore rule out bulk $B_{2g}$-nematic SC of the form $d_{xz} \pm d_{yz} \in E_g$ as the origin of the observed jumps in $c_{66}$.
As an aside, even before $T_c$ measurements~\cite{Jerzembeck2024} established the bound~\eqref{eq:J24-bound-again}, from the heat capacity bound~\eqref{eq:Li2021-bound} of Ref.~\cite{Li2021} it was evident that a high degree of fine-tuning is necessary for $B_{2g}$-nematic SC to be viable.
To be more precise, introduce
\begin{align}
r' &\defeq \frac{\displaystyle \abs{\dv{T_c}{\epsilon_{B_{1g}}}}}{\displaystyle \abs{\dv{T_2}{\epsilon_{B_{1g}}}}} = \frac{v_z - v_x}{v_x} = \frac{- 3 \kappa - \sqrt{3} \, \kappa'}{1 + \kappa + \sqrt{3} \, \kappa'}.
\end{align}
Then from the bound~\eqref{eq:Li2021-bound} and the ratio relation~\eqref{eq:SRO-GL-ratioRelationsAppendixB1g} we may deduce that
\begin{align}
r' \leq r_{\star}' = 0.05
\end{align}
and therefore
\begin{align}
- \frac{r_{\star}' + (3 + r_{\star}') \kappa}{(1 + r_{\star}') \sqrt{3}} ~\leq~ \kappa' ~\leq~ - \sqrt{3} \, \kappa. \label{eq:SRO-rPrime-eq1}
\end{align}
Within the Ginzburg-Landau phase space of Fig.~\ref{fig:SRO-GL-phase-space}, this puts any presumed $B_{2g}$-nematic state to be right on the border to the $B_{1g}$-nematic phase.
This region is highlighted purple in Fig.~\ref{fig:triplePointTuning}.

Regarding accidentally degenerate $B_{2g}$-nematic SC, no similarly definite statements can be made because the corresponding Ehrenfest relation~\eqref{eq:SRO-B2g-Ehrenfest-rel}, derived in Sec.~\ref{sec:SRO-Ehrenfest_relations_appendix}, contains two free tuning parameters: $\alpha$ and $\mathcal{X}$.
The only thing we can say is that some degree of fine-tuning is necessary for the accidentally degenerate $B_{2g}$-nematic states (namely $s' \pm d_{xy}$ and $d_{x^2-y^2} \pm g_{xy(x^2-y^2)}$) to be measurable in ultrasound, but not give a visible cusp in $T_c$.

Next, we discuss $B_{1g}$-nematic states.
Under $[100]$ strain, the degeneracy of these states is lifted and no second transition takes places [cf.\ Fig.~\ref{fig:schematicPhaseDiagrams}(b)].
Thus they are automatically consistent with the absence of a heat capacity anomaly [Eq.~\eqref{eq:Li2021-bound}].
However, from our bound~\eqref{eq:J24-bound-again} and the Ehrenfest relation~\eqref{eq:SRO-B1g-Ehrenfest-rel} it follows that that slope of the second transition would have to be enormous to be consistent with the observed jumps in ultrasound:
\begin{align}
\abs{\dv{T_2}{\epsilon_{6}}} &= \frac{\displaystyle \Delta c_{66}}{\displaystyle \frac{\Delta C_0}{T_{c0}} \abs{\dv{T_c}{\epsilon_{6}}}} \geq \begin{cases}
\SI{43}{\kelvin}, & \text{for $\Delta c_{66}$ of Ref.~\cite{Benhabib2021},} \\
\SI{1880}{\kelvin}, & \text{for $\Delta c_{66}$ of Ref.~\cite{Ghosh2021}.}
\end{cases} \label{eq:SRO-B1g-Ehrenfest-rel-again}
\end{align}
Presumingly, such a large change coming from small increases in $\epsilon_6$ should be visible in the elastocaloric data of Fig.~\ref{fig:SRO-J24-ECE}.
No signatures of a second transition are apparent, however.
We can quantify the necessary degree of fine-tuning by considering the dimensionless ratio
\begin{align}
r &\defeq \frac{\displaystyle \abs{\dv{T_c}{\epsilon_{6}}}}{\displaystyle \abs{\dv{T_2}{\epsilon_{6}}}} = \frac{v_x - v_z}{v_z} = \frac{3 \kappa + \sqrt{3} \, \kappa'}{1 - 2 \kappa}
\end{align}
which is directly related to the Ginzburg-Landau coefficients, as we demonstrated in Sec.~\ref{sec:SRO-ratio-relations}.
From the Ehrenfest relation~\eqref{eq:SRO-B1g-Ehrenfest-rel-again}, it now follows that
\begin{align}
r = \frac{\displaystyle \frac{\Delta C_0}{T_{c0}} \abs{\dv{T_c}{\epsilon_{6}}}^2}{\displaystyle \Delta c_{66}} \leq r_{\star} = \begin{cases}
0.031, & \text{for $\Delta c_{66}$ of Ref.~\cite{Benhabib2021},} \\
0.00069, & \text{for $\Delta c_{66}$ of Ref.~\cite{Ghosh2021}.}
\end{cases}
\end{align}
Hence only a small region on the cusp of the $B_{1g}$-$B_{2g}$-nematic boundary is allowed:
\begin{align}
- \sqrt{3} \, \kappa ~\leq~ \kappa' ~\leq~ \frac{r_{\star}}{\sqrt{3}} - \mleft(\sqrt{3} + \frac{2 r_{\star}}{\sqrt{3}}\mright) \kappa. \label{eq:SRO-req1}
\end{align}
This region is colored orange in Fig.~\ref{fig:triplePointTuning}.

For the no coexistence (only $\Phi_1$ or only $\Phi_2$) instances of accidentally degenerate states, little can be inferred because the corresponding Ehrenfest relation has two additional free tuning parameters.
Moreover, the second transition under $\sigma_6$ stress is replaced by a crossover, as depicted in Fig.~\ref{fig:schematicPhaseDiagrams}(f).
This agrees with absence of any additional sharp features in the elastocaloric data of Fig.~\ref{fig:SRO-J24-ECE}.
Let us also note that these accidentally degenerate states couple quadratically to $\epsilon_{B_{1g}}$ strain so no second transition is expected, in agreement with Eq.~\eqref{eq:Li2021-bound}.

Finally, we come to the most interesting case of TRSB.
As previously discussed, a number of non-thermodynamic experiments support TRSB~\cite{Luke1998, Luke2000, Higemoto2014, Grinenko2021-unaxial, Grinenko2021-isotropic, Xia2006, Kapitulnik2009}.
Let us start with the symmetry-protected state $d_{xz} \pm \iu \, d_{yz} \in E_g$.
Such a state should split under both $[100]$ and $[110]$ strain.
Neither has been observed in thermodynamic measurements.
To quantify the degree of fine-tuning necessary to avoid detection, we use the ratio relations~\eqref{eq:SRO-GL-ratioRelationsAppendix} and~\eqref{eq:SRO-GL-ratioRelationsAppendixB1g} to express dimensionless experimentally-bounded quantities in terms of Ginzburg-Landau coefficients:
\begin{align}
r &\defeq \frac{\displaystyle \abs{\dv{T_c}{\epsilon_{6}}}}{\displaystyle \abs{\dv{T_2}{\epsilon_{6}}}} = \frac{v_x - v_y}{v_y} = \frac{2 \sqrt{3} \, \kappa'}{1 + \kappa - \sqrt{3} \, \kappa'}, \\
r' &\defeq \frac{\displaystyle \abs{\dv{T_c}{\epsilon_{B_{1g}}}}}{\displaystyle \abs{\dv{T_2}{\epsilon_{B_{1g}}}}} = \frac{v_z - v_y}{v_y} = \frac{- 3 \kappa + \sqrt{3} \, \kappa'}{1 + \kappa - \sqrt{3} \, \kappa'}.
\end{align}
From Eq.~\eqref{eq:J24-bound-again} and the Ehrenfest relation~\eqref{eq:SRO-TRSB-Ehrenfest-rel} for $\epsilon_6$ strain (with $\alpha = \mathcal{Y} = 0$), it follows that
\begin{align}
r = \frac{\displaystyle \frac{\Delta C_0}{T_{c0}} \abs{\dv{T_c}{\epsilon_{6}}}^2}{\displaystyle \Delta c_{66}} \leq r_{\star} = \begin{cases}
0.031, & \text{for $\Delta c_{66}$ of Ref.~\cite{Benhabib2021},} \\
0.00069, & \text{for $\Delta c_{66}$ of Ref.~\cite{Ghosh2021},}
\end{cases}
\end{align}
which in turn implies
\begin{align}
0 ~\leq~ \kappa' ~\leq~ \frac{(1 + \kappa) r_{\star}}{(2 + r_{\star}) \sqrt{3}}. \label{eq:SRO-req2}
\end{align}
By furthermore exploiting the ratio relation~\eqref{eq:SRO-GL-ratioRelationsAppendixB1g} for $\epsilon_{B_{1g}}$ strain, we find that
\begin{align}
r' \leq r_{\star}' = 0.05,
\end{align}
and therefore
\begin{align}
\sqrt{3} \, \kappa ~\leq~ \kappa' ~\leq~ \frac{r_{\star}' + (3 + r_{\star}') \kappa}{(1 + r_{\star}') \sqrt{3}}. \label{eq:SRO-rPrime-eq2}
\end{align}
The $r_{\star}$ upper bound tells us that the TRSB state must be near the $B_{2g}$-nematic transition, while the $r_{\star}'$ upper bound constrains the SC to the cusp of the TRSB-$B_{1g}$-nematic boundary.
Thus any bulk symmetry-protected TRSB SC state must be doubly fine-tuned to the triplet point $\kappa = \kappa' = 0$ of the phase space, as depicted in Fig.~\ref{fig:triplePointTuning}.
Note the scale in Fig.~\ref{fig:triplePointTuning}, as compared to the total phase space of Fig.~\ref{fig:SRO-GL-phase-space}.
The total stable phase space of Fig.~\ref{fig:SRO-GL-phase-space} has an area $3 \sqrt{3} / 4 = 1.30$, while the allowed region of Fig.~\ref{fig:triplePointTuning} has an area $\frac{\sqrt{3} \, r_{\star} r_{\star}'}{2 (3+r_{\star}) (3+r_{\star}')} = \num{1.5e-4}$.
Evidently, the level of required fine-tuning is extraordinary high.

In case of TRSB SC order constructed from accidentally-degenerate components, the Ehrenfest relation~\eqref{eq:SRO-TRSB-Ehrenfest-rel} becomes an inequality.
It entails the following lower bound:
\begin{align}
\abs{\dv{T_{\text{TRSB}}}{\epsilon_{6}}} \geq \frac{\Delta c_{66}}{\displaystyle \frac{\Delta C_0}{T_{c0}} \abs{\dv{T_c}{\epsilon_{6}}}} \geq \begin{cases}
\SI{43}{\kelvin}, & \text{for $\Delta c_{66}$ of Ref.~\cite{Benhabib2021},} \\
\SI{1880}{\kelvin}, & \text{for $\Delta c_{66}$ of Ref.~\cite{Ghosh2021}.}
\end{cases}
\end{align}
This bound on $\abs{\dd{T_{\text{TRSB}}} / \dd{\epsilon_{6,0}}}$ holds for both the symmetry-protected and accidentally-degenerate case.
One would expect such strong splitting of the transition to be visible in the elastocaloric data of Fig.~\ref{fig:SRO-J24-ECE}, yet no second anomaly was found.
On the other hand, a recent muon spin relaxation experiment reported splitting of $T_{\text{TRSB}} = T_2$ from $T_c$ with the following dependence~\cite{Grinenko2023}:
\begin{align}
\dv{T_{\text{TRSB}}}{\epsilon_{6}} &= \SI[parse-numbers=false]{-90 \pm 30}{\kelvin}.
\end{align}
This agrees with the smaller~\cite{Benhabib2021} of the two reported $c_{66}$ jumps, but not with the larger one~\cite{Ghosh2021}.
Let us observe that the larger one~\cite{Ghosh2021} has been measured at a two orders of magnitude smaller frequency and thus likely reflects the intrinsic thermodynamic value more accurately.
Regarding the heat capacity bound~\eqref{eq:Li2021-bound}, the ratio relation~\eqref{eq:SRO-GL-ratioRelationsAppendixB1g} cannot be generalized to the case of accidentally-degenerate order parameters.
If one attempts to do so [Eq.~\eqref{eq:SRO-GL-ratioRelationsAppendix-attempt}], additional free tuning parameters appear that make the relation uninformative.
Nonetheless, some degree of fine-tuning is still needed if TRSB states such as $s' + \iu \, d_{xy}$ or $d_{x^2-y^2} + \iu \, g_{xy(x^2-y^2)}$ are to be measurable in ultrasound, but not in heat capacity or elastocaloric experiments.

In conclusion, something is amiss with either some of the experiments which explore the superconductivity of strontium ruthenate, or with our theoretical understanding of how to interpret these experiments.
The most straightforward interpretation in terms of a homogeneous bulk superconductivity, as describe by Ginzburg-Landau theory, is filled with tensions.
Depending on what pairing state we presume, we either find outright contradictions or high levels of fine-tuning, which are at times implausibly high.
Given these tensions, it is highly desirable to (re)establish the interpretation and consistency of the fundamental probes used to study unconventional superconductivity, both thermodynamic and non-thermodynamic.
For instance, domains and inhomogeneities might play a more important role in ultrasound, muon spin relaxation, and other probes than was previously appreciated.

The difficulty in obtaining clear thermodynamic evidence for two-component superconductivity, both in the results covered here and in previous measurements under $[100]$ uniaxial stress, suggests that the possibility of single-component pairing in \ce{Sr2RuO4} should be seriously considered, even though it cannot break time-reversal symmetry homogeneously in the bulk.
Thus strontium ruthenate might not be a two-component superconductor.
Yet there is no doubt that a large number of experiments exhibit highly unusual behavior even for an unconventional single-component pairing state, as we reviewed in great detail in Sec.~\ref{sec:SRO-lit-review}.
In circumstances like these, it is particularly important to cross-check and verify experimental results using different methods, and further experiments like those of Ref.~\cite{Jerzembeck2024} hold the promise of pointing the way towards a final understanding of the enigmatic superconductivity of strontium ruthenate.

\appendix

\chapter{Derivation of the linearized gap equation}
\label{app:lin_gap_eq}

Here we derive the linearized version of the BCS gap equation for general systems.
Its solution determines the transition temperature and symmetry of the superconducting state.
Although variants of the linearized gap equation are available for various special cases~\cite{Schrieffer1999, Sigrist2005, Leggett2006}, starting with the original article by Bardeen, Cooper, and Schrieffer~\cite{BCS}, the following derivation is more streamlined and general than what I found elsewhere in the literature.

\section{Hamiltonian of itinerant fermions with instantaneous interactions}
Let us consider a general fermionic system whose one-particle Hamiltonian is given by:
\begin{align}
\Haml_0 &= \sum_{\vb{k}} \psi_{\vb{k}}^{\dag} H_{\vb{k}} \psi_{\vb{k}},
\end{align}
where
\begin{align}
\psi_{\vb{k}} &= \begin{pmatrix}
\psi_{\vb{k}, 1} \\
\vdots \\
\psi_{\vb{k}, 2M}
\end{pmatrix}
\end{align}
is a spinor of the fermionic annihilation operators. They satisfy the usual anticommutation rules
\begin{align}
\begin{aligned}
\{\psi_{\vb{k}, \alpha}, \psi_{\vb{p}, \beta}\} &= 0, \\
\{\psi_{\vb{k}, \alpha}, \psi_{\vb{p}, \beta}^{\dag}\} &= \Kd_{\vb{k} \vb{p}} \Kd_{\alpha \beta}, \\
\{\psi_{\vb{k}, \alpha}^{\dag}, \psi_{\vb{p}, \beta}^{\dag}\} &= 0.
\end{aligned}
\end{align}
Here, $\vb{k}$ and $\vb{p}$ refer to the crystal momentum and their summations go over the first Brillouin zone only.
The $2M$ components go over both spin and orbital degrees of freedom and we index them with lowercase Greek letters $\alpha, \beta \in \{1, \ldots, 2M\}$.
$M$ is the number of orbital or internal degrees of freedom.

For the Bloch Hamiltonian $H_{\vb{k}}$, which is a $2M \times 2M$ Hermitian matrix, the eigen-representation written in the following form
\begin{align}
H_{\vb{k}} &= \sum_{n} \varepsilon_{\vb{k} n} \mathcal{P}_{\vb{k} n} \label{eq:Hk-eigen-rep}
\end{align}
will be useful.
Here, $n$ is the band index, $\varepsilon_{\vb{k} n}$ are the band energies or dispersions, and the band projectors are defined as
\begin{align}
\mathcal{P}_{\vb{k} n} &= \sum_s u_{\vb{k} n s} u_{\vb{k} n s}^{\dag} = \sum_s \dyad{u_{\vb{k} n s}},
\end{align}
where $u_{\vb{k} n s} = \ket{u_{\vb{k} n s}}$ are the band eigenvectors and $s$ is the band degeneracy index.
The band eigenvectors and projectors satisfy:
\begin{align}
\begin{aligned}
H_{\vb{k}} \ket{u_{\vb{k} n s}} &= \varepsilon_{\vb{k} n} \ket{u_{\vb{k} n s}}, &\hspace{50pt}
H_{\vb{k}} \mathcal{P}_{\vb{k} n} = \mathcal{P}_{\vb{k} n} H_{\vb{k}} &= \varepsilon_{\vb{k} n} \mathcal{P}_{\vb{k} n}, \\
\braket{u_{\vb{k} n s}}{u_{\vb{k} m s'}} &= \Kd_{nm} \Kd_{ss'}, &\hspace{50pt}
\mathcal{P}_{\vb{k} n} \mathcal{P}_{\vb{k} m} &= \Kd_{nm} \mathcal{P}_{\vb{k} n}.
\end{aligned}
\end{align}
In systems with parity and time-reversal symmetry, $s \in \{\uparrow, \downarrow\}$ is the Kramers' degeneracy index or pseudospin.
In the absence of spin-orbit coupling, the pseudospin reduces to the physical spin and the eigenvectors and projectors factorize into orbital and spin parts:
\begin{align}
u_{\vb{k} n s} &= u_{\vb{k} n} \otimes \ket{s}, &
\mathcal{P}_{\vb{k} n} &= u_{\vb{k} n} u_{\vb{k} n}^{\dag} \otimes \Pauli_0.
\end{align}
$\Pauli_0$ is the $2 \times 2$ identity matrix.

In addition, let us assume that the fermions interact through an instantaneous momentum-conserving four-fermion interaction:
\begin{align}
\Haml_{\text{int}} = \frac{1}{4 L^d} \sum_{1234} \Kd_{\vb{k}_1 + \vb{k}_2 - \vb{k}_3 - \vb{k}_4} U_{1234} \psi_1^{\dag} \psi_2^{\dag} \psi_4 \psi_3,
\end{align}
where $L^d$ is the volume in $d$ spatial dimensions, $1 \equiv (\vb{k}_1, \alpha_1)$, $2 \equiv (\vb{k}_2, \alpha_2)$, etc., are particle indices, and
\begin{align}
U_{1234} \equiv U_{\alpha_{1} \alpha_{2} \alpha_{3} \alpha_{4}}(\vb{k}_{1},\vb{k}_{2},\vb{k}_{3},\vb{k}_{4})
\end{align}
is the interaction.
Due to the anticommutation of $\psi_{\vb{k}, \alpha}$, the interaction is antisymmetric under particle exchange:
\begin{align}
U_{1234} = - U_{2134} = - U_{1243} = U_{2143}.
\end{align}

The total Hamiltonian is the sum of the one-particle and interacting parts:
\begin{align}
\Haml &= \Haml_0 + \Haml_{\text{int}}.
\end{align}

\section{BCS gap equation and the instability towards Cooper pairing}
To assess the Cooper pairing instability, we decouple the interaction in the Cooper or Bogoliubov channel, i.e., we write
\begin{align}
\psi_1^{\dag} \psi_2^{\dag} \psi_4 \psi_3 = \langle{\psi_1^{\dag} \psi_2^{\dag}}\rangle \psi_4 \psi_3 + \psi_1^{\dag} \psi_2^{\dag} \langle{\psi_4 \psi_3}\rangle + \cdots \, .
\end{align}
This generates a pairing term in the Hamiltonian
\begin{align}
\Haml_{\Delta} &= \frac{1}{2} \sum_{\vb{k} \alpha \beta} \psi_{\vb{k}, \alpha}^{\dag} \Delta_{\alpha \beta}(\vb{k}) \psi_{- \vb{k}, \beta}^{\dag} + \Hc
\end{align}
plus a remainder $(\Haml_{\text{int}} - \Haml_{\Delta})$ that describes fluctuations.
The superconducting gap matrix $\Delta(\vb{k})$ is determined by demanding that
this pairing term coincides with thermally averaged Bogoliubov part of the decoupled interaction.
The resulting self-consistency equation is the BCS gap equation~\cite{Schrieffer1999, Leggett2006}:
\begin{align}
\Delta_{\alpha \beta}(\vb{p}) = - \frac{1}{2 L^d} \sum_{\vb{k} \alpha' \beta'} U^{\text{(Cp.)}}_{\alpha \beta \alpha' \beta'}(\vb{p}, \vb{k}) \ev{\psi_{\vb{k}, \alpha'} \psi_{- \vb{k}, \beta'}},
\end{align}
where the Cooper-channel interaction is defined as
\begin{align}
U^{\text{(Cp.)}}_{\alpha \beta \alpha' \beta'}(\vb{p}, \vb{k}) \defeq U_{\alpha \beta \alpha' \beta'}(\vb{p}, - \vb{p}, \vb{k}, - \vb{k}).
\end{align}
Due to antisymmetry under particle exchange, the gap matrix satisfies
\begin{align}
\Delta_{\alpha \beta}(\vb{k}) &= - \Delta_{\beta \alpha}(-\vb{k}), &
\Delta^{\intercal}(\vb{k}) &= - \Delta(-\vb{k}).
\end{align}

Given that we are only interested in the onset of superconductivity, next we linearize the BCS gap equation.

For weak interactions, fluctuations are negligible and the anomalous average $\ev{\psi_{\vb{k}, \alpha} \psi_{- \vb{k}, \beta}}$ is performed relative to the mean-field Bogoliubov-de~Gennes Hamiltonian
\begin{align}
\Haml_{\text{mf}} &= \Haml_0 + \Haml_{\Delta}.
\end{align}
Using the general inversion formula
\begin{align}
\begin{pmatrix}
A & B \\
C & D
\end{pmatrix}^{-1} &= \begin{pmatrix}
\mleft(A - B D^{-1} C\mright)^{-1} & \mleft(C - D B^{-1} A\mright)^{-1} \\
\mleft(B - A C^{-1} D\mright)^{-1} & \mleft(D - C A^{-1} B\mright)^{-1}
\end{pmatrix}
\end{align}
on the mean-field Euclidean action~\cite{Altland2010, Coleman2015}
\begin{align}
\action_{\text{mf}}[\psi] &= \frac{1}{2} \sum_{\omega_{\ell} \vb{k}} \begin{pmatrix}
\psi_{\vb{k}}^{\dag}(\iu \omega_{\ell}) & \big[\psi_{-\vb{k}}(-\iu \omega_{\ell})\big]^{\intercal}
\end{pmatrix} \begin{pmatrix}
G_{\vb{k}}^{-1}(\iu \omega_{\ell}) & \Delta(\vb{k}) \\
\Delta^{\dag}(\vb{k}) & - \mleft[G_{-\vb{k}}^{\intercal}(-\iu \omega_{\ell})\mright]^{-1}
\end{pmatrix} \begin{pmatrix}
\psi_{\vb{k}}(\iu \omega_{\ell}) \\
\big[\psi_{-\vb{k}}^{\dag}(-\iu \omega_{\ell})\big]^{\intercal}
\end{pmatrix},
\end{align}
one finds that to linear order in $\Delta(\vb{k})$:
\begin{align}
\begin{aligned}
\ev{\psi_{\vb{k}} \psi_{- \vb{k}}^{\intercal}}_{\text{mf}} &= \frac{1}{\upbeta} \sum_{\ell=-\infty}^{\infty} \ev{\psi_{\vb{k}}(\iu \omega_{\ell}) \big[\psi_{- \vb{k}}(-\iu \omega_{\ell})\big]^{\intercal}}_{\text{mf}} \\
&= \frac{1}{\upbeta} \sum_{\ell=-\infty}^{\infty} \mleft(\mleft[G_{- \vb{k}}^{\intercal}(- \iu \omega_{\ell})\mright]^{-1} \Delta^{-1}(\vb{k}) G_{\vb{k}}^{-1}(\iu \omega_{\ell}) + \Delta^{\dag}(\vb{k})\mright)^{-1} \\
&= \frac{1}{\upbeta} \sum_{\ell=-\infty}^{\infty} G_{\vb{k}}(\iu \omega_{\ell}) \Delta(\vb{k}) G_{- \vb{k}}^{\intercal}(- \iu \omega_{\ell}) + \mathcal{O}(\Delta^2),
\end{aligned} \label{eq:anomalous-mf-avg}
\end{align}
where $\upbeta = 1 / (k_B T)$, $\omega_{\ell} = (2 \ell + 1) \pi / \upbeta$ are the fermionic Matsubara frequencies, and $G_{\vb{k}}(\iu \omega_{\ell})$ is the normal-state imaginary-time single-particle propagator:
\begin{align}
G_{\vb{k}}(\iu \omega_{\ell}) &= \frac{1}{- \iu \omega_{\ell} + H_{\vb{k}}} = \sum_n \frac{\mathcal{P}_{\vb{k} n}}{- \iu \omega_{\ell} + \varepsilon_{\vb{k} n}},
\end{align}
which we expressed with the aid of Eq.~\eqref{eq:Hk-eigen-rep}.
Note that $\varepsilon_{\vb{k} n}$ are measured relative to the chemical potential.

Performing the Matsubara summation in Eq.~\eqref{eq:anomalous-mf-avg} using
\begin{align}
\frac{1}{\upbeta} \sum_{\ell=-\infty}^{\infty} \frac{1}{(- \iu \omega_{\ell} + \varepsilon) (\iu \omega_{\ell} + \varepsilon')} &= \frac{\tanh\tfrac{1}{2} \upbeta \varepsilon + \tanh\tfrac{1}{2} \upbeta \varepsilon'}{2 (\varepsilon + \varepsilon')}
\end{align}
gives the linearized gap equation:
\begin{align}
\Delta_{\alpha \beta}(\vb{p}) &= - \frac{1}{2 L^d} \sum_{\vb{k} \alpha' \beta'} U^{\text{(Cp.)}}_{\alpha \beta \alpha' \beta'}(\vb{p}, \vb{k}) \sum_{n m} \frac{\tanh\tfrac{1}{2} \upbeta \varepsilon_{\vb{k} n} + \tanh\tfrac{1}{2} \upbeta \varepsilon_{-\vb{k} m}}{2 (\varepsilon_{\vb{k} n} + \varepsilon_{- \vb{k} m})} \mleft[\mathcal{P}_{\vb{k} n} \Delta(\vb{k}) \mathcal{P}_{- \vb{k} m}^{\intercal}\mright]_{\alpha' \beta'}.
\end{align}

At weak coupling, the pairing instability is dominated by the Cooper logarithm.
Assuming no accidental or near-accidental degeneracies (Kramers' degeneracies are taken care of through the projectors) and time-reversal symmetry ($\varepsilon_{-\vb{k} n} = \varepsilon_{\vb{k} n}$), this means that the $n = m$ terms dominate the above summation.
We thus drop the $n \neq m$ terms.

Next, into the momentum summation we insert
\begin{align}
1 = \int_{- \hbar \omega_c}^{\hbar \omega_c} \dd{\epsilon'} \Dd(\varepsilon_{\vb{k} n} - \epsilon')
\end{align}
where $\hbar \omega_c$ is the energy cutoff of the theory.
After that we neglect the dependence of $\Dd(\varepsilon_{\vb{k} n} - \epsilon')$, \linebreak $U^{\text{(Cp.)}}$, $\mathcal{P}_{\vb{k} n}$, and $\Delta(\vb{k})$ on the direction orthogonal to the Fermi surface $\varepsilon_{\vb{k} n} = 0$, retaining only \linebreak $\tanh(\tfrac{1}{2} \upbeta \epsilon') / (2 \epsilon')$.
This allows us to perform the energy integral, which we can do analytically in the low-temperature limit $X \equiv \tfrac{1}{2} \upbeta \hbar \omega_c \gg 1$ by applying the standard partial integration trick:
\begin{align}
\begin{aligned}
\int_0^{X} \dd{(\log x)} \tanh x &= \log X \tanh X - \int_0^{X} \frac{\dd{x} \log x}{\cosh^2(x)} \\
&\approx \log X + \log \frac{4 \Elr^{\upgamma_E}}{\pi},
\end{aligned}
\end{align}
where $\upgamma_E = 0.5772...$ is the Euler-Mascheroni constant.
The last step is to take the thermodynamic limit of $L^{-d} \sum_{\vb{k}} \Dd(\varepsilon_{\vb{k} n})$, resulting in an integral over the Fermi surface.

The final result is the following linearized gap equation, formulated as an eigenvalue problem:
\begin{align}
- \frac{1}{2} \sum_n \int\limits_{\varepsilon_{\vb{k} n} = 0} \frac{\dd{S_{\vb{k}}}}{(2 \pi)^d} \sum_{\alpha' \beta'} \frac{U^{\text{(Cp.)}}_{\alpha \beta \alpha' \beta'}(\vb{p}, \vb{k})}{\abs{\grad_{\vb{k}} \varepsilon_{\vb{k} n}}} \mleft[\mathcal{P}_{\vb{k} n} \Delta(\vb{k}) \mathcal{P}_{-\vb{k} n}^{\intercal}\mright]_{\alpha' \beta'} &= \lambda \, \Delta_{\alpha \beta}(\vb{p}). \label{eq:lin-gap-eq}
\end{align}
Here, $\dd{S_{\vb{k}}}$ are infinitesimal area elements of the Fermi surface specified by $\varepsilon_{\vb{k} n} = 0$, $d$ is the spatial dimension, and $\vb{p}$ is on the Fermi surface specified by $\varepsilon_{\vb{p} m} = 0$.
The eigenvalue $\lambda$, if positive, determines the transition temperature according to:
\begin{align}
k_B T_c &= \frac{2 \Elr^{\upgamma_E}}{\pi} \hbar \omega_c \, \Elr^{- 1 / \lambda} \approx 1.134 \, \hbar \omega_c \, \Elr^{- 1 / \lambda}.
\end{align}
The leading instability is determined by the largest positive eigenvalue.

Although $k_B T_c$ seemingly depends on the arbitrary cutoff $\hbar \omega_c$, note that the effective interaction of the theory also depends on the cutoff.
This dependence of the effective interaction turns out the be just right to make $k_B T_c$ cutoff-independent~\cite{Leggett2006}.
Under the change of cutoff (i.e., renormalization group flow)
\begin{align}
\frac{1}{\omega_c} \dv{\omega_c}{\ell} &= - 1,
\end{align}
one may show that the pairing eigenvalue flows as a marginally relevant parameter~\cite{Shankar1994}:
\begin{align}
\frac{1}{\lambda} \dv{\lambda}{\ell} &= \lambda,
\end{align}
thereby ensuring that $\dd{T_c}/\dd{\ell} = 0$.

\section{Fermi surface projection and final form of the linearized gap equation} \label{app:lin_gap_eq-fin-form}
Assuming parity and time-reversal symmetries of the most general form
\begin{align}
\SymU^{\dag}(P) \psi_{\vb{k}} \SymU(P) &= \MatU_{\vb{k}}(P) \psi_{- \vb{k}}, \\
\SymTR^{-1} \psi_{\vb{k}} \SymTR &= \MatTR_{\vb{k}}^{*} \psi_{-\vb{k}},
\end{align}
we can further simplify Eq.~\eqref{eq:lin-gap-eq} by introducing Balian-Werthamer $\vb{d}$-vectors~\cite{Balian1963} for each band:
\begin{gather}
\Delta(\vb{p}) = \abs{\grad_{\vb{p}} \varepsilon_{\vb{p} m}}^{1/2} \sum_{B=0}^{3} d_{B}(\vb{p}_m) \mathcal{P}_{\vb{p} m}^{B} \MatTR_{-\vb{p}}^{\intercal}, \label{eq:Balian-Werthamer} \\
\mathcal{P}_{\vb{p} m}^{B} \defeq \sum_{s s'} u_{\vb{p} m s} (\Pauli_B)_{ss'} u_{\vb{p} m s'}^{\dag}.
\end{gather}
Here, $\Pauli_B$ are the Pauli matrices in pseudospin space, $A, B \in \{0, 1, 2, 3\}$, and the subscript on $\vb{p}_m$ indicates on which Fermi surface the momentum lies.
The reason why the TR matrix $\MatTR_{\vb{k}}$ appears here is because Cooper pairing naturally couples states with their time-inverted pairs.
One important implication of this principle is Anderon's theorem~\cite{Abrikosov1959, Abrikosov1959-p2, Anderson1959, Gorkov2008, Andersen2020}.

Note that we have rescaled $d_{B}(\vb{p}_m)$ by the square root of the Fermi velocity $\abs{\grad_{\vb{p}} \varepsilon_{\vb{p} m}}$ to ensure that the matrix which we diagonalize (i.e., the $\PintW_{BA}$ pairing interaction introduced below) is explicitly Hermitian.
The alternative is to have a matrix which we diagonalize relative to a non-trivial scalar product which includes the Fermi velocity as a weight.

If the gap matrix has a well-defined parity eigenvalue,
\begin{align}
\MatU_{\vb{k}}^{\dag}(P) \Delta(\vb{k}) \MatU_{-\vb{k}}^{*}(P) &= p_P \Delta(- \vb{k}),
\end{align}
then by using the relations
\begin{align}
\Delta(- \vb{k}) &= - \Delta^{\intercal}(\vb{k}), \\
\MatTR_{\vb{k}} \MatTR_{-\vb{k}}^{*} &= - \one, \\
\MatU_{-\vb{k}}(P) &= \MatU_{\vb{k}}^{\dag}(P), \\
\MatU_{\vb{k}}(P) \MatTR_{-\vb{k}} u_{\vb{k} n s}^{*} &= \MatTR_{\vb{k}} \MatU_{-\vb{k}}^{*}(P) u_{\vb{k} n s}^{*} = \sum_{s'} u_{\vb{k} n s'} (\iu \Pauli_y)_{s's},
\end{align}
one can show that $B = 0$ corresponds to even-parity ($p_P = +1$) pairing and $B \in \{1, 2, 3\}$ to odd-parity ($p_P = -1$) pairing.
By plugging~\eqref{eq:Balian-Werthamer} into~\eqref{eq:lin-gap-eq}, we obtain the final form of the linearized gap equation that we employ in this thesis:
\begin{gather}
\sum_n \int\limits_{\varepsilon_{\vb{k} n} = 0} \frac{\dd{S_{\vb{k}}}}{(2 \pi)^d} \sum_{A=0}^{3} \PintW_{BA}(\vb{p}_m, \vb{k}_n) \, d_{A}(\vb{k}_n) = \lambda \, d_{B}(\vb{p}_m), \label{eq:final-lin-gap-eq} \\
\PintW_{BA}(\vb{p}_m, \vb{k}_n) \defeq - \sum_{\alpha \beta \alpha' \beta'} \frac{\mleft[\MatTR_{-\vb{p}}^{*} \mathcal{P}_{\vb{p} m}^{B}\mright]_{\beta \alpha} \mleft[\mathcal{P}_{\vb{k} n}^{A} \MatTR_{-\vb{k}}^{\intercal}\mright]_{\alpha' \beta'}}{4 \abs{\grad_{\vb{p}} \varepsilon_{\vb{p} m}}^{1/2} \abs{\grad_{\vb{k}} \varepsilon_{\vb{k} n}}^{1/2}} U^{\text{(Cp.)}}_{\alpha \beta \alpha' \beta'}(\vb{p}, \vb{k}).
\end{gather}
Its solutions fall into irreducible representations of the point group of the system.
In particular, one may show that $\PintW_{0A'} = \PintW_{A'0} = 0$ for $A' \in \{1, 2, 3\}$ so there is no parity mixing.
The Hermitian matrix $\PintW_{BA}(\vb{p}_m, \vb{k}_n)$ we identify as the pairing interaction.
In a few places we shall also employ the following unsymmetrized variant of this linearized gap equation:
\begin{gather}
\sum_n \int\limits_{\varepsilon_{\vb{k} n} = 0} \frac{\dd{S_{\vb{k}}}}{(2 \pi)^d \abs{\grad_{\vb{k}} \varepsilon_{\vb{k} n}}} \sum_{A=0}^{3} \PintV_{BA}(\vb{p}_m, \vb{k}_n) \, \Delta_{A}(\vb{k}_n) = \lambda \, \Delta_{B}(\vb{p}_m), \label{eq:final-lin-gap-eq-unsym} \\
\PintV_{BA}(\vb{p}_m, \vb{k}_n) \defeq - \frac{1}{4} \sum_{\alpha \beta \alpha' \beta'} \mleft[\MatTR_{-\vb{p}}^{*} \mathcal{P}_{\vb{p} m}^{B}\mright]_{\beta \alpha} \mleft[\mathcal{P}_{\vb{k} n}^{A} \MatTR_{-\vb{k}}^{\intercal}\mright]_{\alpha' \beta'} U^{\text{(Cp.)}}_{\alpha \beta \alpha' \beta'}(\vb{p}, \vb{k}),
\end{gather}
where $\Delta_{B}(\vb{p}_m) \defeq \abs{\grad_{\vb{p}} \varepsilon_{\vb{p} m}}^{1/2} d_{B}(\vb{p}_m)$.

These linearized gap equations apply to spin-orbit-coupled Fermi liquids with space-inversion and time-reversal symmetry whose Fermi surfaces do not touch each other or have Van Hove singularities on them.
The interactions that enter them are the effective instantaneous interactions that one obtains by integrating out all states outside of a thin shell, specified by the energy cutoff $\hbar \omega_c$, around the Fermi surface(s).

\chapter{Elements of group and representation theory}
\label{app:group_theory}

Here we recapitulate some elements of group and representation theory that are relevant to our work.
We start with the fundamentals of group and representation theory.
After that, in Sec.~\ref{sec:theory-of-invariants}, we discuss how representation theory can be used to construct invariants.
In Sec.~\ref{sec:SU2-SO3-conventions}, we recall some elementary facts on the vector $\SO(3)$ and spin $\SU(2)$ rotation groups, as well as parity, and state the conventions we use for these two groups throughout the thesis.
The structure of the tetragonal group $D_{4h}$, which is the point group of both many cuprates (Chap.~\ref{chap:cuprates}) and strontium ruthenate (Chap.~\ref{chap:Sr2RuO4}), is reviewed in Sec.~\ref{sec:tetragonal-group-D4h}.
Finally, in Sec.~\ref{sec:multid-irrep-product}, we discuss how to decompose composite objects into irreducible parts.
We also provide an irreducible representation product table for the tetragonal point group $D_{4h}$ (Tab.~\ref{tab:D4h-irrep-prod-tab}) which enables quick decomposing.
For the reader's convenience, we bold group-theoretic terms when we first define them.

The material covered here is standard.
A great book on group and representation theory as it applies to condensed matter physics is Dresselhaus et al.~\cite{Dresselhaus2007}.
The unpublished lecture notes by Arovas~\cite{ArovasUnpublished} are also recommended.
We refer the reader to both for further reading.

\section{Fundamentals} \label{sec:grp-rep-theory-basics}
\subsection{Group theory} \label{sec:grp-rep-theory-basics-grp}
Group theory is the natural mathematical language of symmetries.
The idea behind introducing groups is to abstract away the notion of a symmetry operation away from the precise object on which it acts.
Let us recall how a group is defined mathematically:
\begin{definition}
A \textbf{group} $(G, \circ)$ is a set of transformations, operations, or group elements $g \in G$ that can be composed or multiplied using $\circ\colon G \times G \to G$. Group multiplication, moreover, must satisfy:
\begin{itemize}
\item closure: composing any transformations $g_1, g_2 \in G$ results in a another transformation $g_1 \circ g_2 \in G$,
\item associativity: $(g_1 \circ g_2) \circ g_3 = g_1 \circ (g_2 \circ g_3)$ for all transformations $g_1, g_2, g_3 \in G$,
\item there exists an identity $\one$ such that $\one \circ g = g \circ \one = g$ leaves all $g \in G$ invariant, and
\item every transformation $g \in G$ has an inverse $g^{-1} \in G$ such that $g \circ g^{-1} = g^{-1} \circ g = \one$.
\end{itemize}
One often writes $G$ instead of $(G, \circ)$ and uses juxtaposition instead of $\circ$ to denote group multiplication.
\end{definition}
\noindent Groups in which multiplication is in addition commutative, $g_1 \circ g_2 = g_2 \circ g_1$, are said to be \textbf{Abelian}.
In physical applications, $g$ are operations such as rotations, reflections, or translations.
Group theory allows us to study the structure of such operations abstractly, without committing to any particular object or system on which they act.

Some examples of groups are the trivial group which is made of only the identity $\{\one\}$, the group $\Z_2$ made of $\{+1, -1\}$ with multiplication $\times$, and the cyclic group $\Z_n = \{0, 1, \ldots, n-1\}$ with addition modulo $n$ as the group multiplication.
These are examples of \textbf{finite groups}, i.e., groups with a finite number of elements.
Groups can also have a continuum of elements, such as real numbers $\R$ under addition or phases $\Ugp(1) = \{\Elr^{\iu \vartheta} \mid \vartheta \in \R\}$ under multiplication.
The latter groups are called Lie groups.
More precisely, \textbf{Lie groups} are groups whose set $G$ is a manifold and whose group multiplication and inversion are smooth.
Notable examples are groups of invertible matrices, such as the general linear group $\GL(n)$, unitary group $\Ugp(n)$, and orthogonal group $\Ogp(n)$.
The general linear group $\GL(n)$ is the group of $n \times n$ invertible matrices, which can be either real or complex, with matrix multiplication as the group composition.
$\Ugp(n)$ is made of unitary ($U^{-1} = U^{\dag}$ for $U \in \Ugp(n)$) complex $n \times n$ matrices, while $\Ogp(n)$ is made of orthogonal ($O^{-1} = O^{\intercal}$ for $O \in \Ogp(n)$) $n \times n$ real matrices, again with matrix multiplication as the group multiplication.
The elements of the special linear group $\SL(n)$, special unitary group $\SU(n)$, and special orthogonal group $\SO(n)$ are special compared to $\GL(n)$, $\Ugp(n)$, and $\Ogp(n)$, respectively, in the sense that their matrix determinant is equal to unity.

Two groups $(G, \circ)$ and $(H, \cdot)$ are essentially the same if there exists a mapping $\Upsilon\colon G \to H$, called an isomorphism, that is bijective and preserves multiplication in the sense that
\begin{align}
\Upsilon(g_1 \circ g_2) = \Upsilon(g_1) \cdot \Upsilon(g_2)
\end{align}
for all $g_1, g_2 \in G$.
From this condition, it follows that $\Upsilon(g^{-1}) = [\Upsilon(g)]^{-1}$ and $\Upsilon(\one) = \one$.
Such groups are said to be \textbf{isomorphic} to each other.
A group $H$ is a \textbf{subgroup} of $G$ if all its elements are contained in $G$ and if multiplication acts in the same way for both.
$\Ogp(n)$ is a subgroup of $\GL(n)$, for instance, but so is $\SU(n)$ a subgroup of $\Ugp(n)$.

Two elements $g_1$ and $g_2$ of $G$ are \textbf{conjugate} to each other if there is a $\tilde{g} \in G$ such that
\begin{align}
g_2 = \tilde{g}^{-1} g_1 \tilde{g}.
\end{align}
Conceptually, elements are conjugate if they are in some sense similar, without being outright equal (except in the case of Abelian groups).
``Two elements are equivalent if they are conjugate to each other'' defines an equivalence relation which partitions the group into conjugacy classes.\footnote{Recall that an equivalence relation $\sim$ is a way of formally identifying elements. Equivalence relations are by definition reflexive ($g \sim g$), symmetric ($g_1 \sim g_2 \iff g_2 \sim g_1$), and transitive ($g_1 \sim g_2$ and $g_2 \sim g_3 \implies g_1 \sim g_3$). These three properties are enough to show that equivalence classes (sets of mutually equivalent elements) constitute a partition of the set over which the equivalence relation is defined.}
In other words, every group can be written as a union of disjoint conjugacy classes.
\textbf{Conjugacy classes} are sets of mutually conjugate elements.

In this thesis, we predominantly study crystalline systems.
On the one hand, these systems are symmetric under discrete translations.
On the other hand, they are also symmetric under various operations, such as rotations and reflections, that leave a point fixed.
Symmetry operations which keep a point invariant together constitute the (crystallographic) \textbf{point group} of the crystalline system.
The (crystallographic) \textbf{space group} of the crystalline system is made of all symmetry transformations, without any restrictions on the transformations.
The space group includes lattice-commensurate translations, point group operations, their compositions, but sometimes also additional symmetry operations in which a fractional translation\footnote{Fractional translations are translations which move a fraction of the distance between unit cells, in contrast to lattice-commensurate translations which move by a multiple of the distance. By themselves fractional translations are not symmetries, whereas lattice-commensurate translations are.} is composed with a reflection or rotation.
The corresponding space groups are called non-symmorphic and they are somewhat complicated to treat.
All the systems studied in this thesis have \textbf{symmorphic} space groups, meaning there are no symmetries involving fractional translations.
In symmorphic systems, translations and point group operations can be separately analyzed.
More formally, symmorphic space groups are semidirect products of the group of translations and the point group.

The possible crystals and their space groups and point groups have been classified by crystalographers a long time ago.
In three dimensions, there are symmetry-wise fourteen different ways one can arrange identical point into a periodic lattice.
Such lattices are known as Bravais lattices and given how some of these fourteen Bravais lattices look similar, one speaks of 7 different crystal systems, which are namely: cubic, tetragonal, orthorhombic, hexagonal, trigonal, monoclinic, and triclinic.
Depending on how the atoms are positioned within the Bravais lattice, multiple point groups and space groups are possible for each Bravais lattice type.
In total, there are 32 crystallographic point groups, 73 symmorphic space groups, and 
157 non-symmorphic space groups.
We refer the reader to the book by Dresselhaus et al.~\cite{Dresselhaus2007} and to the Bilbao crystallographic server~\cite{BilbaoCryst1, BilbaoCryst2} for details.
Here we shall only list the notation that we use for point-group symmetry operations throughout the thesis:
\begin{itemize}
\item $E = \one$ is the identity.
\item $P$ is space inversion or parity; $P^2 = \one$.
\item $\mathscr{C}$ is the rotation by $2 \pi$, which can be non-trivial for fermions and half-integer spin; $\mathscr{C}^2 = \one$. The axis of rotation does not matter.
\item $C_n$ are $n$-fold rotations, i.e., rotations by $2 \pi / n$ around some axis; $(C_n)^n = \mathscr{C}$. Conventionally, the $z$ axis is chosen to be along the axis of highest rotational symmetry. When we want to be specific about the rotation axis, we shall usually add the subscripts $x$, $y$, or $z$ for the principal axes, $d_{\pm} = x \pm y$ for the in-plane diagonals, or $D = x + y + z$ for the space diagonal.
\item $\Sigma$ is a reflection or mirroring across some plane; $\Sigma^2 = \mathscr{C}$. When a \SI{180}{\degree} rotation along some axis $\vu{n}$, $C_{2\vu{n}}$, is composed with parity $P$, the result is a reflection across the plane orthogonal to $\vu{n}$.
\item $\Sigma_h = P C_{2z}$ is a reflection across a horizontal plane, which is by definition perpendicular to the axis of highest rotational symmetry.
\item $\Sigma_v = P C_{2x}$ is a reflection across a vertical plane, which by definition contains the axis of highest rotational symmetry.
\item $\Sigma_d = P C_{2d}$ is a reflection across a diagonal plane, which is diagonal relative to the some principal symmetry axes.
\item $S_n = \Sigma_h C_n$ is an improper rotation by $2 \pi / n$ around some axis, which by definition is an $n$-fold rotation around the axis followed by a reflection perpendicular to the axis.
\end{itemize}
This is a slight variation on the Schönflies notation~\cite{Dresselhaus2007}. See also \nameref{app:conventions}.

\subsection{Representation theory} \label{sec:rep-theory-basics}
Having abstracted transformations such as rotations, reflections, etc., into groups, we may now systematically study how these transformations act on different objects.
This is the subject of representation theory.
The objects of prime interest in physics are vectors, which in the abstract sense of linear algebra are simply objects which can by added together and multiplied by scalars.
Representations are defined in the following way:
\begin{definition}
A (linear) \textbf{representation} of a group $G$ over the vector space $V$ is a mapping $\RepM\colon G \to \GL(V)$ in which to each group element $g \in G$ we attribute a linear transformation $\RepM(g)\colon V \to V$ in such a way that both group multiplication and group inversion are respected:
\begin{align}
\RepM(g_1 \circ g_2) &= \RepM(g_1) \RepM(g_2), &
\RepM(g^{-1}) &= \mleft[\RepM(g)\mright]^{-1}.
\end{align}
From this it immediately follows that $\RepM(\one) = \one$.
\end{definition}
\noindent In more concrete settings, $V = \R^n$ or $\C^n$ and $\GL(V)$ is the corresponding group of $n \times n$ (real or complex) matrices $\GL(n)$.
A representation is called real or complex depending on whether its matrices are real or complex.
A representation is unitary when $\RepM(g) \in \Ugp(n)$ are unitary and therefore $\RepM(g^{-1}) = \RepM^{\dag}(g)$.
Similarly, a representation is orthogonal when $\RepM(g) \in \Ogp(n)$ are orthogonal and therefore $\RepM(g^{-1}) = \RepM^{\intercal}(g)$.

In physical applications, $G$ is usually the group of symmetry operations, while the vectors $\in V$ can be Cartesian coordinates of position or momentum, spinors, quantum-mechanical states, multi-component order parameters, sets of operators which transform into each other under symmetries, and many other things.
An important result in this context is Wigner's theorem~\cite{Weinberg1995} which states that symmetries act on quantum-mechanical states through linear operators that are either unitary or antiunitary.
Apart from time reversal, which is represented through an antiunitary operator, it thus follows that representation theory is the natural mathematical language of how symmetries act in quantum mechanics.

One of the main goals of representation theory is to simplify representations.
In general, the $n \times n$ matrices $\RepM(g)$ are quite complicated.
In linear algebra, the main way square matrices are simplified is through diagonalization, i.e., by changing into a basis made of eigenvectors in which the matrix is diagonal.
Within representation theory, we can also enact changes of basis.
Representations which differ by a change of basis are said to be equivalent.
To be more specific, two representations $\RepM$ and $\RepM'$ are \textbf{equivalent} if there is an invertible change-of-basis matrix $\mathcal{B} \in \GL(V)$ such that
\begin{align}
\RepM'(g) = \mathcal{B}^{-1} \RepM(g) \mathcal{B}
\end{align}
for all group elements $g \in G$.
The main task is thus to see to what extent can we use the \emph{same} $\mathcal{B}$ to \emph{simultaneously} diagonalize all $\RepM(g)$.

An important result from linear algebra is that commuting matrices can be simultaneously diagonalized, while non-commuting matrices cannot.
Hence if we have an Abelian group, $[\RepM(g_1), \RepM(g_2)] = 0$ and we can simultaneously diagonalize all $\RepM(g)$.
For more general groups this is not the case.
Instead, the best we can do is to ensure that the $\RepM(g)$ matrices become block-diagonal in the new basis in the sense that:
\begin{align}
\begin{aligned}
\RepM'(g) &= \mathcal{B}^{-1} \RepM(g) \mathcal{B} = \begin{pmatrix}
\RepM_1(g) & & & \\
& \RepM_2(g) & & \\
& & \RepM_3(g) & \\
& & & \ddots~{}
\end{pmatrix} \\
&\equiv (\RepM_1 \oplus \RepM_2 \oplus \RepM_3 \oplus \cdots)(g),
\end{aligned} \label{eq:rho-irrep-decomposition}
\end{align}
where $\RepM_1, \RepM_2, \RepM_3, \ldots$ are the smallest possible representations and $\oplus$ is the direct sum operation.
Clearly, a necessary condition for the existence of such smaller representations is that there exists a vector subspace $V' < V$ which is invariant under all $\RepM(g)$, i.e., $\RepM(g) v' \in V'$ for all $v' \in V'$.
A representation which has a non-trivial\footnote{Non-trivial in the sense that the invariant subspace is neither zero nor the whole space.} invariant subspace is called a \textbf{reducible} representation.
An \textbf{irreducible representation} (\textbf{irrep}) is a representation which is not reducible.
Irreps can also be characterized in an affirmative way as representations for which the set $\{\RepM(g) v\}_{g \in G}$ for any non-zero $v \in V$ always spans the whole space.
The $\RepM_1, \RepM_2, \RepM_3, \ldots$ representations appearing in Eq.~\eqref{eq:rho-irrep-decomposition} are irreps.
Irreps thus constitute elementary building blocks from which all representations are constructed.
In linear algebra, multiplication by scalars (eigenvalues) plays the same role.

Now we state a few fundamental results concerning representations and irreps.
For proofs, see Refs.~\cite{Dresselhaus2007, ArovasUnpublished, Cornwell1984}.
\begin{theorem}[Schur's first lemma]
Consider two irreducible representations $\RepM$ and $\RepM'$ of a group $G$ over the same vector space $V$.
If a linear operator $\mathcal{B}\colon V \to V$ satisfies
\begin{align}
\mathcal{B} \RepM'(g) = \RepM(g) \mathcal{B} \label{eq:Schur1relation}
\end{align}
for all $g \in G$, then either (i) $\mathcal{B} = 0$ or (ii) $\mathcal{B}$ is invertible and $\RepM$ and $\RepM'$ are equivalent.
A non-zero and non-invertible $\mathcal{B}$ is not possible.
\end{theorem}
\noindent The intuition behind this lemma is that, for non-zero $\mathcal{B}$ and $v$, both $\{\RepM'(g) v\}_{g \in G}$ and $\{\RepM(g) \mathcal{B} v\}_{g \in G}$ span the whole space (since both $\RepM$ and $\RepM'$ are irreps) so only an invertible $\mathcal{B}$ is consistent with Eq.~\eqref{eq:Schur1relation}.
\begin{theorem}[Schur's second lemma]
Consider an irreducible representation $\RepM$ of a group $G$ over $V$.
If a linear operator $\mathcal{B}\colon V \to V$ satisfies
\begin{align}
\mathcal{B} \RepM(g) = \RepM(g) \mathcal{B} \label{eq:Schur2relation}
\end{align}
for all $g \in G$, then it is proportional to the identity, i.e., $\mathcal{B} = \lambda \, \one$ for some scalar $\lambda$.
\end{theorem}
\noindent To understand this result, suppose you are given an eigenvector $v$ of $\mathcal{B}$, $\mathcal{B} v = \lambda v$.
Then Eq.~\eqref{eq:Schur2relation} tells us that $\RepM(g) v$ is also an eigenvector with the same eigenvalue $\lambda$.
Since $\RepM$ is an irrep, $\{\RepM(g) v\}_{g \in G}$ spans the whole space and $\mathcal{B}$ must be proportional to the identity.

Notice that Eq.~\eqref{eq:rho-irrep-decomposition} requires that $\RepM(g)$ not only has one invariant subspace $V_1$, but also that it has a complementary invariant subspace $V_{1}^{\perp} = V_2 \oplus V_3 \oplus \cdots$ such that the total space $V = V_1 \oplus V_{1}^{\perp}$.
Here $V_n$ are the subspaces on which $\RepM_n$ act.
Otherwise mixing of the form
\begin{align}
\begin{pmatrix}
\RepM_1(g) & \tilde{\RepM}(g) \\
0 & \RepM_2(g)
\end{pmatrix}
\end{align}
cannot be excluded.
Such $\RepM$ which have complementary invariant subspaces are called \textbf{completely reducible}.
The following theorem clarifies when representations are completely reducible~\cite{Cornwell1984}:
\begin{theorem}[Maschke]
Consider a reducible representation $\RepM$ of a group $G$ over the vector space $V$.
Then this representation is completely reducible if any one of the following three conditions is true:
\begin{itemize}
\item $\RepM$ is a unitary representation,
\item $G$ is a finite or compact group,
\item $G$ is connected, not compact, and semisimple.
\end{itemize}
\end{theorem}
\noindent In the case of unitary $\RepM$, the theorem follows from the fact that the orthogonal complement of an invariant subspace is also invariant.
For finite or compact $G$, the idea is to use an arbitrary scalar product $\langle ~\,, ~ \rangle$ to construct the following scalar product
\begin{align}
\braket{v}{v'} \defeq \frac{1}{\abs{G}} \sum_{g \in G} \langle \RepM(g) v, \RepM(g) v' \rangle
\end{align}
with respect to which $\RepM$ is unitary; here $\abs{G}$ is the number of group elements and the sum over $g \in G$ becomes an integral for continuous compact groups.
After this, the proof proceeds in the same way as for unitary $\RepM$.
Completely reducible representations can always be decomposed into irreps, as written in Eq.~\eqref{eq:rho-irrep-decomposition}.

A subject that is very important, but has not yet been covered, is that of representation characters.
Since this is best explained in the context of an example, we discuss characters in Sec.~\ref{sec:character-theory-D4h} after introducing the $D_{4h}$ point group.

\section{Construction of invariants} \label{sec:theory-of-invariants}
As a simple application of group and representation theory, we shall now prove the following important result on how to construct invariants~\cite{Dresselhaus2007, ArovasUnpublished, Cornwell1984}:
\begin{theorem}[Fundamental Theorem of the Theory of Invariants]
Consider two objects $\vb{v} = (v_1, \ldots, v_{N})^{\intercal}$ and $\vb{u} = (u_1, \ldots, u_{M})^{\intercal}$ whose transformation under the group $G$ is described by the unitary irreducible representations $\RepM_{v}$ and $\RepM_{u}$, respectively.
Then a non-zero bilinear invariant $\vb{v}^{\dag} \Gamma \vb{u}$ specified by the $N \times M$ matrix $\Gamma$ exist if and only if $\RepM_{v}$ and $\RepM_{u}$ are equivalent.
Furthermore, when it exists, $\Gamma$ is unique up to a constant.
In the basis in which $\RepM_{v} = \RepM_{u}$, $\Gamma$ is proportional to the identity, $\Gamma \propto \one$, and the (up to a constant) unique bilinear invariant that one may construct takes the form $\vb{v}^{\dag} \vb{u}$.
\end{theorem}
\begin{proof}[Proof]
The condition that the bilinear $\vb{v}^{\dag} \Gamma \vb{u}$ is invariant in the sense that
\begin{align}
\vb{v}^{\dag} \mleft[\RepM_{v}(g)\mright]^{\dag} \Gamma \RepM_{u}(g) \vb{u} = \vb{v}^{\dag} \Gamma \vb{u}
\end{align}
for all $\vb{v}$, $\vb{u}$, and $g \in G$ is, due to unitary of $\RepM_{v}$, equivalent to the requirement that
\begin{align}
\Gamma \RepM_{u}(g) = \RepM_{v}(g) \Gamma
\end{align}
for all group elements $g \in G$.
This requirement is the same one from Schur's first lemma [Eq.~\eqref{eq:Schur1relation}].
Hence $\Gamma$ can be non-zero only if it is invertible, which implies that $\RepM_{v}$ and $\RepM_{u}$ are equivalent.
Since they are equivalent, we may always switch to a basis in which $\RepM_{v}$ and $\RepM_{u}$ are equal.
Schur's second lemma now tells us that in this basis $\Gamma = \lambda \, \one$ for some scalar $\lambda$.
\end{proof}
\noindent This theorem underlies a great many applications of group theory in physics.
The Hamiltonian, the action, and the free energy are all examples of important operators and scalars which must be invariant under all symmetry operations and their construction is aided by the above theorem.

In practice, one is usually given objects which transform under reducible representations, in which case some work needs to be done to obtain the \emph{irreducible} parts of the objects to which the theorem applies.
In case we want to combine more than two objects into an invariant (e.g.\ $\sum \Gamma_{abc} v_a^{*} u_b w_c$), one does so by first decomposing composite objects (e.g.\ $\{u_b w_c\}$) into irreducible parts and then only later applying the theorem.
The decomposition of composite objects is discussed in Sec.~\ref{sec:multid-irrep-product}.

\section{Rotations, reflections, and parity} \label{sec:SU2-SO3-conventions}
Rotations act on three-dimensional vectors $\vb{v} \in \R^3$ via multiplication with special orthogonal $3 \times 3$ matrices $R \in \SO(3)$:
\begin{align}
\vb{v} \mapsto R \vb{v}.
\end{align}
Rotations act on spinors $\psi \in \C^2$ via multiplication with special unitary $2 \times 2$ matrices $S \in \SU(2)$:
\begin{align}
\psi \mapsto S \psi.
\end{align}
Recall that:
\begin{align}
\SO(3) &\defeq \mleft\{\text{real $3 \times 3$ matrices $R$} \mid R^{-1} = R^{\intercal}, \det R = 1\mright\}, \\
\SU(2) &\defeq \mleft\{\text{complex $2 \times 2$ matrices $S$} \mid S^{-1} = S^{\dag}, \det S = 1\mright\}.
\end{align}
Parity or spatial inversion $P$ inverts vectors,
\begin{align}
P\colon \vb{v} \mapsto - \vb{v},
\end{align}
but acts trivially on spinors:
\begin{align}
P\colon \psi \mapsto \psi.
\end{align}
On the other hand, rotations by $2 \pi$ (around any axis) invert spinors,
\begin{align}
\mathscr{C}\colon \psi \mapsto - \psi,
\end{align}
but acts trivially on vectors:
\begin{align}
\mathscr{C}\colon \vb{v} \mapsto \vb{v}.
\end{align}
The operation of rotating by $2 \pi$ is conventionally denoted $\mathscr{C}$.

By composing parity with rotations, we obtain the orthogonal group
\begin{align}
\Ogp(3) &\defeq \SO(3) \times \{\one, P\} = \mleft\{\text{real $3 \times 3$ matrices $R$} \mid R^{-1} = R^{\intercal}\mright\}
\end{align}
which is the point group of isotropic systems.
In addition to rotations, it includes parity, reflections, and improper rotations.
Its elements have $\det R = \pm 1$.
In crystal systems, this group is broken down to finite subgroups.
To emphasize the fact that the elements of $\Ogp(3)$ represent physical point group transformations, which can act in a variety of ways on different objects, we shall denote them abstractly as $g \in \Ogp(3)$ through the thesis.
The corresponding matrices we shall denote $R(g)$.

When dealing with spinors, one has to allow for $2 \pi$ rotations $\mathscr{C}$ which are non-trivial.
This is done by formally enlarging the point group ($\SO(3) \times \{\one, P\}$ in isotropic systems) to the so-called \textbf{double group of the point group}, which for isotropic systems equals $\SU(2) \times \{\one, P\}$.
In crystals, this isotropic double group of the point group is broken down to finite subgroups.
For the same reasons as for $\Ogp(3)$, elements of $\SU(2) \times \{\one, P\}$ we shall denote abstractly as $g$ and the corresponding matrices as $S(g)$ throughout the thesis.
Notice that $S(P) = \one$ and $R(\mathscr{C}) = \one$.
Just like parity $P$, $\mathscr{C}$ commutes with all other point group operations.

General rotations can be parameterized using Euler angles and the composition of rotations can be understood as a mapping from two sets of Euler angles into a new set of Euler angles.
However, studying $\SU(2)$ and $\SO(3)$ rotations, and infinitely-dimensional Lie groups in general, through their group multiplication turns out to be quite complicated.
Instead, what one does is study infinitesimal rotations, i.e., Lie group elements that are close to the identity.
This is a lot simpler because infinitesimal group elements have a linear ``Lie algebra'' structure which is easier to analyze.
More importantly, the Lie algebra contains almost\footnote{The global structure/topology of the group is not contained. E.g., the generators of $\SO(3)$ and $\SU(2)$ satisfy the same algebra, even though $\SO(3) \neq \SU(2)$.} the same information as group multiplication.

The \textbf{infinitesimal generators} of vector $\SO(3)$ rotations are
\begin{align}
L_x &= \begin{pmatrix}
0 & 0 & 0 \\
0 & 0 & - \iu \\
0 & \iu & 0
\end{pmatrix}, &
L_y &= \begin{pmatrix}
0 & 0 & \iu \\
0 & 0 & 0 \\
- \iu & 0 & 0
\end{pmatrix}, &
L_z &= \begin{pmatrix}
0 & - \iu & 0 \\
\iu & 0 & 0 \\
0 & 0 & 0
\end{pmatrix},
\end{align}
or more compactly $(L_i)_{jk} = - \iu \LCs_{ijk}$, while the generators of spin $\SU(2)$ rotations are
\begin{align}
S_x &= \frac{1}{2} \begin{pmatrix}
0 & 1 \\
1 & 0
\end{pmatrix}, &
S_y &= \frac{1}{2} \begin{pmatrix}
0 & -\iu \\
\iu & 0
\end{pmatrix}, &
S_z &= \frac{1}{2} \begin{pmatrix}
1 & 0 \\
0 & -1
\end{pmatrix},
\end{align}
or more compactly $S_i = \tfrac{1}{2} \Pauli_i$; here $\LCs_{ijk}$ is the Levi-Civita symbol and $\Pauli_i$ are Pauli matrices.
The generators satisfy the spin algebra:
\begin{align}
[L_i, L_j] &= \iu \sum_{k=1}^{3} \LCs_{ijk} L_k, &
[S_i, S_j] &= \iu \sum_{k=1}^{3} \LCs_{ijk} S_k,
\end{align}
which is the Lie algebra of $\SO(3)$ and $\SU(2)$.

In terms of these generators, a rotation by an angle $\vartheta$ around an axis specified by the unit-vector $\vu{n}$ is given by
\begin{align}
R\mleft(g = C_{\vartheta \vu{n}}\mright) &= \exp\mleft(- \iu \, \vartheta \, \vu{n} \vdot \vb{L}\mright), \label{eq:Rrep-exp-expr} \\
S\mleft(g = C_{\vartheta \vu{n}}\mright) &= \exp\mleft(- \iu \, \vartheta \, \vu{n} \vdot \vb{S}\mright), \label{eq:Srep-exp-expr}
\end{align}
where $\vb{S} = (S_x, S_y, S_z)$ and $\vb{L} = (L_x, L_y, L_z)$.
In conjunction with $R(P) = - \one$ and $S(P) = \one$, this completely specifies the representations $R$ and $S$ for $g \in \SU(2) \times \{\one, P\}$.
The two representations are, moreover, related through
\begin{align}
S^{\dag}(g) \Pauli_i S(g) &= \det R(g) \sum_{j=1}^{3} R_{ij}(g) \Pauli_j \label{eq:Pauli-mat-tranfs-expr}
\end{align}
for all $g \in \SU(2) \times \{\one, P\}$.
This relation can be alternatively read as the statement that $\vb{\Pauli} = (\Pauli_x, \Pauli_y, \Pauli_z)$ transforms as a pseudovector.

$R(g)$ is the canonical representation of $\SO(3)$ and $S(g)$ is the canonical representation of $\SU(2)$.
However, one can also consider how rotations act on other objects as well, such as tensors.
Mathematically, this is described by linear representations.
The possible irreducible representations can be derived by introducing the raising and lower operators $J_{\pm} = J_x \pm \iu J_y$, where $\vb{J} = \vb{L}$ or $\vb{S}$.
This is explained in all quantum mechanics textbooks~\cite{Sakurai2017}, albeit without stating that mathematically this amounts to finding irreps of $\SU(2)$ and $\SO(3)$.
The result is well-known.

Irreps of $\SU(2)$ are specified by a non-negative half-integers $j \in \{0, \tfrac{1}{2}, 1, \ldots\}$ called the spin.
The basis of the irrep vector space is made of states $\ket{m}$, $m \in \{-j, -j+1, \ldots, j-1, j\}$, which are eigenstates of $J_z$, but get mixed under $J_x$ and $J_y$, according to:
\begin{align}
J_x^{(j)} \ket{m} &= \frac{1}{2} \mleft[\sqrt{j(j+1) - m(m+1)} \ket{m+1} + \sqrt{j(j+1) - m(m-1)} \ket{m-1}\mright], \\
J_y^{(j)} \ket{m} &= - \frac{\iu}{2} \mleft[\sqrt{j(j+1) - m(m+1)} \ket{m+1} - \sqrt{j(j+1) - m(m-1)} \ket{m-1}\mright], \\
J_z^{(j)} \ket{m} &= m \ket{m}.
\end{align}
The irrep matrices we obtain by exponentiating the generators:
\begin{align}
D^{(j)}\mleft(g = C_{\vartheta \vu{n}}\mright) &= \exp\mleft(- \iu \, \vartheta \, \vu{n} \vdot \vb{J}^{(j)}\mright).
\end{align}
Since $m$ are integer or half-integer, depending on $j$, it follows that $D^{(j)}(\mathscr{C}) = (-1)^{2j} \one$.
Regarding parity, $D^{(j)}(P)$ can be set to either $+ \one$ (for even-parity irreps) or $- \one$ (for odd-parity irreps) independently of $j$ because $P$ commutes with all rotations.
The irreps of $\SO(3)$ are the same as for $\SU(2)$, except for the fact that they can only have integer spin $j = \ell \in \{0, 1, \ldots\}$.
The canonical representations are obtained by setting $j = \tfrac{1}{2}$ and $\ell = 1$:
\begin{align}
S(g) &= D^{(j=1/2)}(g) \text{ with $S(P) = + \one$,} \\
R(g) &= D^{(\ell=1)}(g) \text{ with $R(P) = - \one$.}
\end{align}

There is an infinite number of possible irreps because $\SU(2)$ is an infinitely-dimensional Lie group.
Moreover, this infinite set of irreps is discrete (not a continuum) because $\SU(2)$ is compact.
For finite  groups (which are always compact), the number of possible irreps is finite and equals the number of conjugacy classes.
As we shall see in the next section, when the isotropic group of rotations $\SU(2)$ gets broken down to a finite subgroup, such as the tetragonal point group $D_{4h}$, there will be only a few irreps which conceptually correspond to the lowest-spin irreps of $\SU(2)$.
The high-spin irreps break down into smaller parts because, for finite point groups, there is simply not enough symmetry operations to generate from the state $\ket{m = j}$ all the $2j+1$ states $\{\ket{m}\}_{\abs{m} \leq j}$ when $j$ is large.

\section{The tetragonal point group $D_{4h}$} \label{sec:tetragonal-group-D4h}
The tetragonal point group $D_4$ is a subgroup of $\SO(3)$ which is generated by the following three operations:
\begin{itemize}
\item four-fold rotations around the $z$ axis $C_{4z}$,
\item two-fold rotations around the $x$ axis $C_{2x}$, and
\item two-fold rotations around the $d_{+} = x + y$ diagonal $C_{2d_{+}}$.
\end{itemize}
If we add the fourth generator,
\begin{itemize}
\item parity $P$,
\end{itemize}
we obtain the tetragonal point group $D_{4h}$ which is a subgroup of $\Ogp(3)$.
By multiplying and inverting these finite-group generators in all possible ways, we obtain the whole group.
In principle, we should also state how the different generators compose and commute, but for rotations this is implicitly known since they inherit the group structure from $\SO(3)$.

For the tetragonal point groups, we find that:
\begin{align}
D_{4h} &= D_4 \times \{\one, P\}, &
D_4 &= \mleft\{\one, C_{4z}, C_{2z}, C_{-4z}, C_{2x}, C_{2y}, C_{2d_{+}}, C_{2d_{-}}\mright\},
\end{align}
where $d_{-} = x - y$ and
\begin{align}
\begin{aligned}
C_{2z} &= (C_{4z})^2, &\hspace{50pt} C_{-4z} &= (C_{4z})^{-1} = (C_{4z})^3, \\
C_{2y} &= C_{4z} C_{2x} (C_{4z})^{-1}, &\hspace{50pt} C_{2d_{-}} &= (C_{4z})^{-1} C_{2d_{+}} C_{4z}.
\end{aligned}
\end{align}
By composing with parity, we obtain improper rotations and reflections:
\begin{gather}
\begin{gathered}
S_{-4z} = P C_{4z} = \Sigma_h C_{-4z}, \hspace{30pt}
\Sigma_h = P C_{2z}, \hspace{30pt}
S_{4z} = P C_{-4z} = \Sigma_h C_{4z}, \\
\Sigma_x = P C_{2x}, \hspace{30pt}
\Sigma_y = P C_{2y}, \hspace{30pt}
\Sigma_{d_+} = P C_{2d_+}, \hspace{30pt}
\Sigma_{d_-} = P C_{2d_-}.
\end{gathered}
\end{gather}
These are included in $D_{4h}$:
\begin{align}
D_{4h} = D_4 \cup \mleft\{P, S_{-4z}, \Sigma_h, S_{4z}, \Sigma_x, \Sigma_y, \Sigma_{d_+}, \Sigma_{d_-}\mright\}.
\end{align}
$\Sigma_x$ and $\Sigma_y$ are vertical reflections and $\Sigma_{d_{\pm}}$ are diagonal ones.

Normally, for application purposes, it is not necessary to work out the multiplication of all group elements.
We nonetheless do so here for pedagogical purposes.
The group multiplication table of the $D_4$ point group is provided in Tab.~\ref{tab:D4-mult-table}.
Since parity commutes with everything, that is $[P, g] = 0$ for all $g \in D_4$, as well as $P^2 = \one$, it follows that for $g_1, g_2 \in D_4$:
\begin{gather}
(P g_1) g_2 = g_1 (P g_2) = P (g_1 g_2), \\
(P g_1) (P g_2) = g_1 g_2.
\end{gather}
Thus Tab.~\ref{tab:D4-mult-table} also gives the multiplication rules for $D_{4h} = D_4 \times \{\one, P\}$.
Two notable features of the multiplication Tab.~\ref{tab:D4-mult-table} are (i) there are $g_1$ and $g_2$ for which $g_1 g_2 \neq g_2 g_1$, i.e., group multiplication is not commutative in general, nor in this case in particular, and (ii) every column and row has only one appearance of each number.
The latter is a consequence of the invertibility of group multiplication.
Associativity $(g_1 g_2) g_3 = g_1 (g_2 g_3)$ is not obvious from the table and, in general, one has to verify it.

\begin{table}[t]
\centering
\captionabove[The group multiplication table of the tetragonal point group $D_4$.]{\textbf{The group multiplication table of the tetragonal point group $D_4$.}
For the row with group element $g_1$ and column with group element $g_2$, the table entry gives the result of group multiplication $g_1 g_2$ (in this order).
The group elements have been colored according to conjugacy class to highlight the group structure.}
{\renewcommand{\arraystretch}{1.3}
\renewcommand{\tabcolsep}{10pt}
\begin{tabular}{|c"c|c|c|c|c|c|c|c|}
\hline
& $\textcolor{black}{\one}$ & $\textcolor{red}{C_{4z}}$ & $\textcolor{black}{C_{2z}}$ & $\textcolor{red}{C_{-4z}}$ & $\textcolor{green!70!black}{C_{2x}}$ & $\textcolor{green!70!black}{C_{2y}}$ & $\textcolor{blue}{C_{2d_+}}$ & $\textcolor{blue}{C_{2d_-}}$
\\ \thickhline
$\textcolor{black}{\one}$ & $\textcolor{black}{\one}$ & $\textcolor{red}{C_{4z}}$ & $\textcolor{black}{C_{2z}}$ & $\textcolor{red}{C_{-4z}}$ & $\textcolor{green!70!black}{C_{2x}}$ & $\textcolor{green!70!black}{C_{2y}}$ & $\textcolor{blue}{C_{2d_+}}$ & $\textcolor{blue}{C_{2d_-}}$
\\ \hline
$\textcolor{red}{C_{4z}}$ & $\textcolor{red}{C_{4z}}$ & $\textcolor{black}{C_{2z}}$ & $\textcolor{red}{C_{-4z}}$ & $\textcolor{black}{\one}$ & $\textcolor{blue}{C_{2d_+}}$ & $\textcolor{blue}{C_{2d_-}}$ & $\textcolor{green!70!black}{C_{2y}}$ & $\textcolor{green!70!black}{C_{2x}}$
\\ \hline
$\textcolor{black}{C_{2z}}$ & $\textcolor{black}{C_{2z}}$ & $\textcolor{red}{C_{-4z}}$ & $\textcolor{black}{\one}$ & $\textcolor{red}{C_{4z}}$ & $\textcolor{green!70!black}{C_{2y}}$ & $\textcolor{green!70!black}{C_{2x}}$ & $\textcolor{blue}{C_{2d_-}}$ & $\textcolor{blue}{C_{2d_+}}$
\\ \hline
$\textcolor{red}{C_{-4z}}$ & $\textcolor{red}{C_{-4z}}$ & $\textcolor{black}{\one}$ & $\textcolor{red}{C_{4z}}$ & $\textcolor{black}{C_{2z}}$ & $\textcolor{blue}{C_{2d_-}}$ & $\textcolor{blue}{C_{2d_+}}$ & $\textcolor{green!70!black}{C_{2x}}$ & $\textcolor{green!70!black}{C_{2y}}$
\\ \hline
$\textcolor{green!70!black}{C_{2x}}$ & $\textcolor{green!70!black}{C_{2x}}$ & $\textcolor{blue}{C_{2d_-}}$ & $\textcolor{green!70!black}{C_{2y}}$ & $\textcolor{blue}{C_{2d_+}}$ & $\textcolor{black}{\one}$ & $\textcolor{black}{C_{2z}}$ & $\textcolor{red}{C_{-4z}}$ & $\textcolor{red}{C_{4z}}$
\\ \hline
$\textcolor{green!70!black}{C_{2y}}$ & $\textcolor{green!70!black}{C_{2y}}$ & $\textcolor{blue}{C_{2d_+}}$ & $\textcolor{green!70!black}{C_{2x}}$ & $\textcolor{blue}{C_{2d_-}}$ & $\textcolor{black}{C_{2z}}$ & $\textcolor{black}{\one}$ & $\textcolor{red}{C_{4z}}$ & $\textcolor{red}{C_{-4z}}$
\\ \hline
$\textcolor{blue}{C_{2d_+}}$ & $\textcolor{blue}{C_{2d_+}}$ & $\textcolor{green!70!black}{C_{2x}}$ & $\textcolor{blue}{C_{2d_-}}$ & $\textcolor{green!70!black}{C_{2y}}$ & $\textcolor{red}{C_{4z}}$ & $\textcolor{red}{C_{-4z}}$ & $\textcolor{black}{\one}$ & $\textcolor{black}{C_{2z}}$
\\ \hline
$\textcolor{blue}{C_{2d_-}}$ & $\textcolor{blue}{C_{2d_-}}$ & $\textcolor{green!70!black}{C_{2y}}$ & $\textcolor{blue}{C_{2d_+}}$ & $\textcolor{green!70!black}{C_{2x}}$ & $\textcolor{red}{C_{-4z}}$ & $\textcolor{red}{C_{4z}}$ & $\textcolor{black}{C_{2z}}$ & $\textcolor{black}{\one}$
\\ \hline
\end{tabular}}
\label{tab:D4-mult-table}
\end{table}

The conjugacy classes of the group are much more important during practical applications of group theory in condensed matter.
Conjugacy classes are defined by identifying group elements $g$ and $g'$, $g \sim g'$, whenever there exists a group element $\tilde{g}$ such that $g' = \tilde{g}^{-1} g \tilde{g}$.
There are five conjugacy classes of $D_4$:
\begin{itemize}
\item $E = \{\one\}$,
\item $C_4 = \{C_{4z}, C_{-4z}\}$,
\item $C_2 = \{C_{2z}\}$,
\item $C_2' = \{C_{2x}, C_{2y}\}$, and
\item $C_2'' = \{C_{2d_+}, C_{2d_-}\}$.
\end{itemize}
It is a good exercise to show this.
The identity is always its own conjugacy class.
Notice how the conjugacy classes are made of conceptually similar elements.
Given that parity $P$ commutes with everything, there are ten conjugacy classes of $D_{4h}$, which are those of $D_4$ plus
\begin{itemize}
\item $P = \{P\}$,
\item $S_4 = \{S_{4z}, S_{-4z}\}$,
\item $\Sigma_h = \{\Sigma_h\}$,
\item $\Sigma_v' = \{\Sigma_x, \Sigma_y\}$, and
\item $\Sigma_d'' = \{\Sigma_{d_+}, \Sigma_{d_-}\}$.
\end{itemize}

\subsection{Character theory} \label{sec:character-theory-D4h}
Let us now discuss the representations of the tetragonal point groups $D_4$ and $D_{4h}$.
The completely reducible representations of any point group can always be decomposed into irreps, which are the elementary building blocks of representations.
Finding the irreps of a given finite group, however, is a bit involved and we refer the reader to Refs.~\cite{Dresselhaus2007, ArovasUnpublished, Cornwell1984} for the details on how this is done.
For the point groups which appear in condensed matter physics applications, the results are well-known and tabulated in the form of \textbf{character tables}.
The character tables can be found in books such as Dresselhaus et al.~\cite{Dresselhaus2007} or online on websites such as that of the Bilbao crystallographic server~\cite{BilbaoCryst1, BilbaoCryst2}.
Here we explain how to read and use these tables.

As we have seen in Sec.~\ref{sec:rep-theory-basics}, irreps can come in a number of different forms which are all equivalent in the sense that they differ by a change of basis.
We do not want to be distracted by the detailed way a group is represented in a certain basis, however, because this it not universal.
Instead, we want to characterize the linear operators of the representation $\RepM(g)$ in terms of invariants which \emph{are} universal (basis-independent).
The most important of such invariants in representation theory is the trace, which defines the so-called character.
\begin{definition}
The \textbf{character of a group element} $g \in G$ with respect to a representation $\RepM\colon G \to \GL(V)$ is the trace of $\RepM(g)$:
\begin{align}
\chi_{\RepM}(g) \defeq \Tr \RepM(g).
\end{align}
The \textbf{character of a representation} $\RepM\colon G \to \GL(V)$ is the set of all group element characters:
\begin{align}
\vec{\chi}_{\RepM} \defeq \mleft\{\chi_{\RepM}(g) \mid g \in G\mright\}.
\end{align}
\end{definition}
\noindent An important result from representation theory is that representations $\RepM$ and $\RepM'$ are equivalent for a finite (or compact) group $G$ if and only if the characters of the two representations $\vec{\chi}_{\RepM}$ and $\vec{\chi}_{\RepM'}$ are identical.
Thus characters completely characterize the representations of finite groups.

Here are a few important properties of characters and irreducible representations (irreps)~\cite{Dresselhaus2007, ArovasUnpublished, Cornwell1984}:
\begin{itemize}
\item The character of the identity gives the dimension of the representation since $\Tr \RepM(\one) = \Tr \one = \dim \RepM$.
\item The trivial representation, usually denoted $A$, $A_1$, or $A_{1g}$, in which all $g$ are mapped to plus one, $\RepM_{A_{1g}}(g) = +1$, is always an irrep. All its characters equal $+1$.
\item The character is the same for all elements of the same conjugacy class. This follows from $\Tr \RepM(\tilde{g}^{-1} g \tilde{g}) = \Tr \RepM^{-1}(\tilde{g}) \RepM(g) \RepM(\tilde{g}) = \Tr \RepM(g)$. Thus one may speak of characters of conjugacy classes.
\item For finite groups, the number of conjugacy classes equals the number of irreps.
\item The characters of the irreps are orthogonal in the sense that
\begin{align}
\frac{1}{\abs{G}} \sum_{g \in G} \chi_{\RepM_{\zeta}}(g^{-1}) \chi_{\RepM_{\xi}}(g) &= \frac{1}{\abs{G}} \sum_{\mathcal{C}_n \in G/{\sim}} \abs{\mathcal{C}_n} \chi_{\RepM_{\zeta}}(\mathcal{C}_n^{-1}) \chi_{\RepM_{\xi}}(\mathcal{C}_n) = \Kd_{\zeta \xi}, \label{eq:Schur-orthogonality1}
\end{align}
where $\zeta$ and $\xi$ denote the irreps, $\mathcal{C}_n$ goes over the conjugacy classes $G/{\sim}$, and $\abs{\mathcal{C}_n}$ is the number of elements within the conjugacy class $\mathcal{C}_n$.
\item The characters of the conjugacy classes are orthogonal in the sense that
\begin{align}
\frac{1}{\abs{G}} \sum_{\zeta} \abs{\mathcal{C}_n} \chi_{\RepM_{\zeta}}(\mathcal{C}_n^{-1}) \chi_{\RepM_{\zeta}}(\mathcal{C}_m) &= \Kd_{n m}, \label{eq:Schur-orthogonality2}
\end{align}
where $\zeta$ goes over all irreps of $G$ and $\mathcal{C}_{n,m}$ are conjugacy classes.
\item Irreps are complete in the sense that every representation can be written as a direct sum of irreps.
\item The characters of direct sums of representations add up: $\chi_{\RepM_1 \oplus \RepM_2}(g) = \chi_{\RepM_1}(g) + \chi_{\RepM_2}(g)$.
\item The characters of direct products of representations get multiplied: $\chi_{\RepM_1 \otimes \RepM_2}(g) = \chi_{\RepM_1}(g) \linebreak \times \chi_{\RepM_2}(g)$.
\end{itemize}

The character tables of $\{\one, P\}$ and $D_4$ are provided in Tab.~\ref{tab:D4-P-char-tab}.
From these two tables, the character table of the $D_{4h}$ point group is easily constructed. 
It is shown in Tab.~\ref{tab:D4h-char-tab}.
In character tables, columns correspond to conjugacy classes, which are denote on the top together with their size if larger than $1$.
E.g., the conjugacy class $E$ has only the identity $\{\one\}$, while $2 C_2'$ has two elements which are, namely, $C_{2x}$ and $C_{2y}$.
Rows correspond to irreps whose names are given at the leftmost end.
The entries of the table are the characters $\chi_{\RepM_{\zeta}}(\mathcal{C}_n)$, where the irrep $\zeta$ and conjugacy class $\mathcal{C}_n$ correspond to the given row and column.

\begin{table}[t!]
\centering
\captionabove[The character tables of the (triclinic) point group $S_2 = \{\one, P\}$ and tetragonal point group $D_{4}$~\cite{Dresselhaus2007}.]{\textbf{The character tables of the (triclinic) point group $S_2 = \{\one, P\}$ and tetragonal point group $D_{4}$}~\cite{Dresselhaus2007}.
$P$ is space inversion or parity.
$C_4$ are \SI{90}{\degree} rotations around $\vu{e}_z$.
$C_2$, $C_2'$, and $C_2''$ are \SI{180}{\degree} rotations around $\vu{e}_z$, $\vu{e}_x$ or $\vu{e}_y$, and the diagonals $\vu{e}_x \pm \vu{e}_y$, respectively.}
\begin{subfigure}[t]{0.4\textwidth}
\centering
{\renewcommand{\arraystretch}{1.3}
\renewcommand{\tabcolsep}{10pt}
\hspace{20pt}\begin{tabular}{c|rr} \hline\hline
$S_2$ & $E$ & $P$
\\ \hline
$A_g$ & $1$ & $1$
\\
$A_u$ & $1$ & $-1$
\\ \hline\hline
\end{tabular}}
\end{subfigure}%
\begin{subfigure}[t]{0.6\textwidth}
\centering
{\renewcommand{\arraystretch}{1.3}
\renewcommand{\tabcolsep}{10pt}
\begin{tabular}{c|rrrrr} \hline\hline
$D_4$ & $E$ & $2 C_4$ & $C_2$ & $2 C_2'$ & $2 C_2''$
\\ \hline
$A_{1}$ & $1$ & $1$ & $1$ & $1$ & $1$
\\
$A_{2}$ & $1$ & $1$ & $1$ & $-1$ & $-1$
\\
$B_{1}$ & $1$ & $-1$ & $1$ & $1$ & $-1$
\\
$B_{2}$ & $1$ & $-1$ & $1$ & $-1$ & $1$
\\
$E$ & $2$ & $0$ & $-2$ & $0$ & $0$
\\ \hline\hline
\end{tabular}\hspace{20pt}}
\end{subfigure}
\label{tab:D4-P-char-tab}
\end{table}

\begin{table}[t!]
\centering
\captionabove[The character table of the tetragonal point group $D_{4h}$~\cite{Dresselhaus2007}.]{\textbf{The character table of the tetragonal point group $D_{4h}$}~\cite{Dresselhaus2007}.
The irreps are divided according to parity into even (subscript $g$) and odd ($u$) ones.
$C_4$ are \SI{90}{\degree} rotations around $\vu{e}_z$.
$C_2$, $C_2'$, and $C_2''$ are \SI{180}{\degree} rotations around $\vu{e}_z$, $\vu{e}_x$ or $\vu{e}_y$, and the diagonals $\vu{e}_x \pm \vu{e}_y$, respectively.
$P$ is space inversion or parity.
Improper rotations $S_4$ and mirror reflections $\Sigma_h$, $\Sigma_v'$, and $\Sigma_d''$ are obtained by composing $C_4$, $C_2$, $C_2'$, and $C_2''$ with $P$, respectively.
Notice how the four quadrants have the same structure as the $S_2 = \{\one, P\}$ character table of Tab.~\ref{tab:D4-P-char-tab}, as follows from the commutativity of parity.}
{\renewcommand{\arraystretch}{1.3}
\renewcommand{\tabcolsep}{10pt}
\begin{tabular}{c|rrrrr|rrrrr} \hline\hline
$D_{4h}$ & $E$ & $2 C_4$ & $C_2$ & $2 C_2'$ & $2 C_2''$ & $P$ & $2 S_4$ & $\Sigma_h$ & $2 \Sigma_v'$ & $2 \Sigma_d''$
\\ \hline
$A_{1g}$ & $1$ & $1$ & $1$ & $1$ & $1$ & $1$ & $1$ & $1$ & $1$ & $1$
\\
$A_{2g}$ & $1$ & $1$ & $1$ & $-1$ & $-1$ & $1$ & $1$ & $1$ & $-1$ & $-1$
\\
$B_{1g}$ & $1$ & $-1$ & $1$ & $1$ & $-1$ & $1$ & $-1$ & $1$ & $1$ & $-1$
\\
$B_{2g}$ & $1$ & $-1$ & $1$ & $-1$ & $1$ & $1$ & $-1$ & $1$ & $-1$ & $1$
\\
$E_g$ & $2$ & $0$ & $-2$ & $0$ & $0$ & $2$ & $0$ & $-2$ & $0$ & $0$
\\ \hline
$A_{1u}$ & $1$ & $1$ & $1$ & $1$ & $1$ & $-1$ & $-1$ & $-1$ & $-1$ & $-1$
\\
$A_{2u}$ & $1$ & $1$ & $1$ & $-1$ & $-1$ & $-1$ & $-1$ & $-1$ & $1$ & $1$
\\
$B_{1u}$ & $1$ & $-1$ & $1$ & $1$ & $-1$ & $-1$ & $1$ & $-1$ & $-1$ & $1$
\\
$B_{2u}$ & $1$ & $-1$ & $1$ & $-1$ & $1$ & $-1$ & $1$ & $-1$ & $1$ & $-1$
\\
$E_u$ & $2$ & $0$ & $-2$ & $0$ & $0$ & $-2$ & $0$ & $2$ & $0$ & $0$
\\ \hline\hline
\end{tabular}}
\label{tab:D4h-char-tab}
\end{table}

The properties we just listed for the characters and irreps are all reflected in Tabs.~\ref{tab:D4-P-char-tab} and~\ref{tab:D4h-char-tab}
The first column under $E$ gives the dimension of the irreps.
The first rows has only plus ones because the corresponding representation is trivial.
One may also verify that the rows and columns are orthogonal in the precise way described by Eqs.~\eqref{eq:Schur-orthogonality1} and~\eqref{eq:Schur-orthogonality2}.

\subsection{Examples and conventions for irreducible representations of $D_{4h}$} \label{sec:examples-convention-D4h}
In the case of 1D irreps, the character table explicitly gives us the irrep $\chi_{\RepM_{\zeta}}(g)$, which is actually unique since there is no such thing as changing the basis of a 1D vector space.
In the case of multidimensional irreps, such as $E_g$ or $E_u$ of Tab.~\ref{tab:D4h-char-tab}, however, one has to explicitly specify the basis which one uses and the precise form of the irrep matrices.
Throughout the thesis, whenever we say that an object transform under the irreps $E_g$ or $E_u$ of $D_{4h}$, we shall entail that the transformation matrices have the form:
\begin{align}
\RepM_{E_{g/u}}(C_{4z}) &= \begin{pmatrix}
 0 & -1 \\
 1 & 0 \\
\end{pmatrix}, &
\RepM_{E_{g/u}}(C_{2x}) &= \begin{pmatrix}
 1 & 0 \\
 0 & -1 \\
\end{pmatrix}, &
\RepM_{E_{g/u}}(C_{2d_+}) &= \begin{pmatrix}
 0 & 1 \\
 1 & 0 \\
\end{pmatrix}, \label{eq:Egu-irrep-mat-conventions1}
\end{align}
with
\begin{align}
\RepM_{E_{g}}(P) &= \begin{pmatrix}
1 & 0 \\
0 & 1
\end{pmatrix}, &
\RepM_{E_{u}}(P) &= \begin{pmatrix}
-1 & 0 \\
0 & -1
\end{pmatrix}. \label{eq:Egu-irrep-mat-conventions2}
\end{align}
Note that to specify the $\RepM_{E_{g/u}}(g)$ for all $g \in D_{4h}$, it is sufficient to specify how the matrices look like for the four group generators of $D_{4h}$.
Parity is diagonal for multidimensional irreps because it commutes with all group elements, as follows from Schur's second lemma (Sec.~\ref{sec:rep-theory-basics}).

The best way to get an intuition regarding the various irreps is to think of them in terms of elementary objects which transform under them.
The most basic objects are the real-space coordinates $\vb{r} = (x, y, z)$ and polynomials can be constructed from these coordinates so that they transform under all irreps of the point group.\footnote{In the case of irreps of the \emph{double group} of the point group which are odd under $2 \pi$ rotations, spinors need to be used to represent them.}
Such polynomials are often provided alongside the character table; see Ref.~\cite{Dresselhaus2007}, for example.
In Tab.~\ref{tab:D4h-example-tab}, we have listed the lowest-order coordinate polynomials which transform according to the various irreps of $D_{4h}$.

In the example $(zy|- xz) \in E_g$ of Tab.~\ref{tab:D4h-example-tab}, the peculiar-looking ordering and minus sign are necessary to ensure that the corresponding transformation matrices are the ones given in Eqs.~\eqref{eq:Egu-irrep-mat-conventions1} and~\eqref{eq:Egu-irrep-mat-conventions2}.
For instance, $C_{2x}$ maps $(x, y, z) \mapsto (x, -y, -z)$ hence $(zy|- xz) \mapsto (zy|xz)$, in agreement with Eq.~\eqref{eq:Egu-irrep-mat-conventions1}, only if we flip the places of $xz$ and $yz$.
Similarly, the relative minus sign is needed so that $C_{4z}$ acts through the matrix given in Eq.~\eqref{eq:Egu-irrep-mat-conventions1}.

\begin{table}[t]
\centering
\captionabove[Examples of coordinate polynomials transforming according to the irreps of the tetragonal point group $D_{4h}$~\cite{Palle2023-ECE}.]{\textbf{Examples of coordinate polynomials transforming according to the irreps of the tetragonal point group $D_{4h}$}~\cite{Palle2023-ECE}.
As discussed in the text, $D_{4h}$ is generated by fourfold rotations around $z$, twofold rotations around $x$ and $y$, twofold rotations around the diagonals $x \pm y$, and parity.
It has five even ($A_{1g}$, $A_{2g}$, $B_{1g}$, $B_{2g}$, $E_g$) and five odd ($A_{1u}$, $A_{2u}$, $B_{1u}$, $B_{2u}$, $E_u$) irreps, of which $E_g$ and $E_u$ are two-dimensional.
The character table is given in Tab.~\ref{tab:D4h-char-tab}.}
{\renewcommand{\arraystretch}{1.3}
\renewcommand{\tabcolsep}{10pt}
\begin{tabular}{c|c|c|c|c} \hline\hline
$A_{1g}$ & $A_{2g}$ & $B_{1g}$ & $B_{2g}$ & $E_g$ \\
$1$, $x^2+y^2$, $z^2$ & $xy(x^2-y^2)$ & $x^2-y^2$ & $xy$ & $(yz | -xz)$ \\ \hline
$A_{1u}$ & $A_{2u}$ & $B_{1u}$ & $B_{2u}$ & $E_u$ \\
$xyz(x^2-y^2)$ & $z$ & $xyz$ & $(x^2-y^2)z$ & $(x | y)$
\\ \hline\hline
\end{tabular}}
\label{tab:D4h-example-tab}
\end{table}

\begin{figure}[t]
\centering
\includegraphics[width=0.95\textwidth]{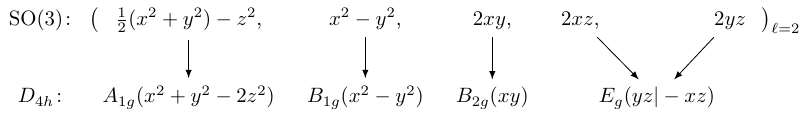}
\captionbelow[The splitting of $d$-wave ($\ell = 2$) spherical harmonics (top) into irreps of $D_{4h}$ (bottom) in the presence of a tetragonal crystal environment.]{\textbf{The splitting of $d$-wave ($\ell = 2$) spherical harmonics (top) into irreps of $D_{4h}$ (bottom) in the presence of a tetragonal crystal environment.}}
\label{fig:spher-harm-split}
\end{figure}

In the case of the isotropic point group $\SO(3)$, the coordinate polynomials which fall into the various irreps are the spherical harmonics.
It is insightful to compare them with Tab.~\ref{tab:D4h-example-tab}.
The $s$-wave ($\ell = 0$) constant wavefunction $1$ belongs to $A_{1g}$, but so does the $d$-wave ($\ell = 2$) wavefunction $x^2 + y^2 - 2 z^2$.
The two $d$-wave functions $x^2 - y^2$ and $2 x y$, although related by a \SI{45}{\degree} rotation around $z$, belong to different irreps because the symmetry operation which relates them is not an element of the point group $D_{4h}$.
For the same reason, the $p$-wave ($\ell = 1$) functions $(x | y)$ and $z$ belong to different irreps of $D_{4h}$.
Indeed, this one may explicitly see by evaluating the vector representation $R(g)$ using Eq.~\eqref{eq:Rrep-exp-expr}:
\begin{align}
\begin{aligned}
R(C_{4z}) &= \begin{pmatrix}
0 & -1 & 0 \\
1 & 0 & 0 \\
0 & 0 & 1
\end{pmatrix}, &\hspace{50pt}
R(C_{2x}) &= \begin{pmatrix}
1 & 0 & 0 \\
0 & -1 & 0 \\
0 & 0 & -1
\end{pmatrix}, \\
R(C_{2d_+}) &= \begin{pmatrix}
0 & 1 & 0 \\
1 & 0 & 0 \\
0 & 0 & -1
\end{pmatrix}, &\hspace{50pt}
R(P) &= \begin{pmatrix}
-1 & 0 & 0 \\
0 & -1 & 0 \\
0 & 0 & -1
\end{pmatrix}.
\end{aligned}
\end{align}
None of the $D_{4h}$ group generators mix $(x | y)$ and $z$.
Hence the two belong to different irreps, $E_u$ and $A_{2u}$, as follows from $R(g) = \RepM_{E_u}(g) \oplus \RepM_{A_{2u}}(g)$.
The broad patter is therefore that spherical harmonics, which are degenerate under $\SO(3)$, have their degeneracy lifted in crystal environments.
The splitting of the degeneracy is depicted in Fig.~\ref{fig:spher-harm-split} for the $\ell = 2$ spherical manifold.

Of course, coordinate polynomials are not the only thing that transforms under irreps.
Matrices, operators, field, etc., can all be decompose into parts which transform according to irreps of the point group of the problem.
For instance, the magnetic field $\vb{B} = (B_x, B_y, B_z)$ decomposes into $(B_x | B_y) \in E_g$ and $B_z \in A_{2g}$.
Similarly, the Pauli (or spin) matrices transform according to:
\begin{align}
\begin{aligned}
S^{\dag}(g) \Pauli_a S(g) &= \sum_{b=1}^{2} \mleft[\RepM_{E_{g}}(g)\mright]_{ab} \Pauli_b, \\
S^{\dag}(g) \Pauli_3 S(g) &= \RepM_{A_{2g}}(g) \Pauli_3,
\end{aligned}
\end{align}
for $g \in D_{4h}$.
Thus $(\Pauli_1 | \Pauli_2) \in E_g$ and $\Pauli_3 \in A_{2g}$.
This is a special case of Eq.~\eqref{eq:Pauli-mat-tranfs-expr}.
Here, $S(g)$ is the spin representation of Sec.~\ref{sec:SU2-SO3-conventions}.
From Eq.~\eqref{eq:Srep-exp-expr}, it follows that for the generators of $D_{4h}$:
\begin{align}
\begin{aligned}
S(C_{4z}) &= \frac{\Pauli_0 - \iu \Pauli_z}{\sqrt{2}}, &\hspace{50pt}
S(C_{2x}) &= - \iu \Pauli_x, \\
S(C_{2d_+}) &= - \iu \frac{\Pauli_x + \Pauli_y}{\sqrt{2}}, &\hspace{50pt}
S(P) &= \Pauli_0.
\end{aligned}
\end{align}
Here $\Pauli_0$ is the $2 \times 2$ identity matrix.

\section{Decomposition of composite objects} \label{sec:multid-irrep-product}
Suppose we are given two vectors $\vb{v} = (v_1, \ldots, v_{N})^{\intercal}$ and $\vb{u} = (u_1, \ldots, u_{M})^{\intercal}$ which transform under the representations $\RepM_{v}$ and $\RepM_{u}$ of a finite group $G$, respectively.
Then the composite object $\{v_a u_b\}$ transforms like
\begin{align}
g\colon v_a u_b &\mapsto \sum_{c=1}^{N} \sum_{d=1}^{M} \mleft[\RepM_{v}(g)\mright]_{ac} \mleft[\RepM_{u}(g)\mright]_{bd} v_c u_d, \label{eq:composite-vu-transf-rule}
\end{align}
which is fairly complicated.
We want to simplify this.

The first step is to introduce the direct-product vector
\begin{align}
\vb{v} \otimes \vb{u} &= \begin{pmatrix}
v_1 u_1 \\ \vdots \\ v_1 u_{M} \\ v_2 u_1 \\ \vdots \\ v_{N} u_{M}
\end{pmatrix}.
\end{align}
Then Eq.~\eqref{eq:composite-vu-transf-rule} can be recast into matrix multiplication with the direct-product matrix $\RepM_{v}(g) \otimes \RepM_{u}(g)$:
\begin{align}
g\colon \vb{v} \otimes \vb{u} &\mapsto \mleft[\RepM_{v}(g) \otimes \RepM_{u}(g)\mright] \mleft(\vb{v} \otimes \vb{u}\mright) = \mleft[\RepM_{v}(g) \vb{v}\mright] \otimes \mleft[\RepM_{u}(g) \vb{u}\mright].
\end{align}
Hence $\vb{v} \otimes \vb{u}$ transforms under $\RepM_{v} \otimes \RepM_{u}$.

The components of $\vb{v}$ and $\vb{u}$ can be scalars or operators.
In the case of operators, there is usually a representation of $G$ on the operator space, call it $\SymU$, which is related to the representations $\RepM_{v}$ and $\RepM_{u}$ through:
\begin{align}
\SymU^{-1}(g) v_a \SymU(g) &= \sum_{b=1}^{N} \mleft[\RepM_{v}(g)\mright]_{ab} v_b, \\
\SymU^{-1}(g) u_a \SymU(g) &= \sum_{b=1}^{M} \mleft[\RepM_{u}(g)\mright]_{ab} u_b.
\end{align}
Notice how these relations are consistent with composition ($g \to g_1 g_2$) and how
\begin{align}
\SymU^{-1}(g) v_a u_b \SymU(g) &= \SymU^{-1}(g) v_a \SymU(g) \SymU^{-1}(g) u_b \SymU(g) = \sum_{c=1}^{N} \sum_{d=1}^{M} \mleft[\RepM_{v}(g)\mright]_{ac} \mleft[\RepM_{u}(g)\mright]_{bd} v_c u_d.
\end{align}

The next step in the simplification is to decompose $\RepM_{v} \otimes \RepM_{u}$ into irreps.
More explicitly, we want to change into a basis in which $\RepM_{v} \otimes \RepM_{u}$ is block diagonal [cf.\ Eq.~\eqref{eq:rho-irrep-decomposition}]:
\begin{align}
\begin{aligned}
\mathcal{B}^{-1} \RepM_{v}(g) \otimes \RepM_{u}(g) \mathcal{B} &= \begin{pmatrix}
\RepM_{\zeta_1}(g) & & & \\
& \RepM_{\zeta_2}(g) & & \\
& & \RepM_{\zeta_3}(g) & \\
& & & \ddots~{}
\end{pmatrix} \\
&= (\RepM_{\zeta_1} \oplus \RepM_{\zeta_2} \oplus \RepM_{\zeta_3} \oplus \cdots)(g).
\end{aligned}
\end{align}
Although finding $\mathcal{B}$ is a bit involved, finding out which $\zeta_1, \zeta_2, \ldots$ irreps appear on the right-hand side is more straightforward since it can be deduced from the characters alone.
By taking the trace of the above, one finds that
\begin{align}
\chi_{\RepM_{v} \otimes \RepM_{u}}(g) &= \chi_{\RepM_{v}}(g) \chi_{\RepM_{u}}(g) = \chi_{\RepM_{\zeta_1}}(g) + \chi_{\RepM_{\zeta_2}}(g) + \chi_{\RepM_{\zeta_3}}(g) + \cdots \, .
\end{align}
Given that we know the characters of $\RepM_{v}$, $\RepM_{u}$, and all the irreps, the above is readily solved to find which irreps appear in the decomposition of $\RepM_{v} \otimes \RepM_{u}$.
The orthogonality of irreps [Eq.~\eqref{eq:Schur-orthogonality1}] is very useful in this context.

Let us now consider the tetragonal point group $D_{4h}$ whose character table is given in Tab.~\ref{tab:D4h-char-tab}.
Introduce the character vectors:\footnote{Improper rotations and reflections need not be included in the vector because parity commutes with everything.}
\begin{align}
\vec{\chi}_{\RepM} = \begin{pmatrix}
\chi_{\RepM}(E), & \chi_{\RepM}(C_4), & \chi_{\RepM}(C_2), & \chi_{\RepM}(C_2'), & \chi_{\RepM}(C_2''), & \chi_{\RepM}(P)
\end{pmatrix}.
\end{align}
By employing the character table, one can now easily find the irrep decompositions of direct products, like for instance:
\begin{align}
\vec{\chi}_{A_{2g} \otimes B_{1u}} &= \begin{pmatrix}
1, & -1, & 1, & -1, & 1, & -1
\end{pmatrix} = \vec{\chi}_{B_{2u}}, \\
\vec{\chi}_{E_g \otimes B_{2g}} &= \begin{pmatrix}
2, & 0, & -2, & 0, & 0, & 2
\end{pmatrix} = \vec{\chi}_{E_g}, \\
\vec{\chi}_{E_g \otimes E_u} &= \begin{pmatrix}
4, & 0, & 4, & 0, & 0, & -4
\end{pmatrix} = \vec{\chi}_{A_{1u}} + \vec{\chi}_{A_{2u}} + \vec{\chi}_{B_{1u}} + \vec{\chi}_{B_{2u}},
\end{align}
and so forth.
In the case of 1D irreps, the above completely answers what we get after a direct product.
In the case of 2D irreps, however, special care needs to be taken to ensure that the 2D vectors transform under the same 2D irrep matrices which were given in Eqs.~\eqref{eq:Egu-irrep-mat-conventions1} and~\eqref{eq:Egu-irrep-mat-conventions2}:
\begin{align}
\RepM_{E}(C_{4z}) &= \begin{pmatrix}
 0 & -1 \\
 1 & 0 \\
\end{pmatrix}, &
\RepM_{E}(C_{2x}) &= \begin{pmatrix}
 1 & 0 \\
 0 & -1 \\
\end{pmatrix}, &
\RepM_{E}(C_{2d_+}) &= \begin{pmatrix}
 0 & 1 \\
 1 & 0 \\
\end{pmatrix}. \label{eq:E-irrep-mat-conventions-again}
\end{align}
Regarding parity, because it commutes with everything, in the direct product one can treat it separately from the rotational part $D_4$ of $D_{4h}$.
After going through all the possible irreps of the $D_{4h}$ point group, one obtains the irrep product table~\ref{tab:D4h-irrep-prod-tab}, which was previously also provided in Ref.~\cite{Palle2023-ECE}.
Let us note that in the case of $E(\vb{v}) \otimes E(\vb{u})$, it is convenient to write the result in terms of Pauli matrices:
\begin{align}
\begin{aligned}
\vb{v}^{\intercal} \Pauli_0 \vb{u} &\in A_{1}, &\hspace{50pt}
\vb{v}^{\intercal} \Pauli_x \vb{u} &\in B_{2}, \\
\vb{v}^{\intercal} \Pauli_y \vb{u} &\in A_{2}, &\hspace{50pt}
\vb{v}^{\intercal} \Pauli_z \vb{u} &\in B_{1}.
\end{aligned}
\end{align}

Analogous irrep product tables can be derived for other point groups.
When these point groups have multidimensional irreps, as it the case, e.g., for the hexagonal point group $D_{6h}$ and cubic point group $O_h$, special care needs to be taken to ensure that the components of the multidimensional irreps consistently transform under the same irrep matrices.
For the respective irrep product tables and a discussion of their derivation, we refer the reader to the doctoral thesis of Charles Steward~\cite{CharlieThesis}.

\begin{landscape}
\begin{table}
\centering
\captionabove[The product table(s) for irreducible representations of the tetragonal point group $D_{4h} = \{\one, P\} \times D_4$~\cite{Palle2023-ECE}.]{\textbf{The product table(s) for irreducible representations of the tetragonal point group $D_{4h} = \{\one, P\} \times D_4$}~\cite{Palle2023-ECE}.
The upper table is the product table for $\{\one, P\}$, which has an even ($g$) and odd ($u$) irrep, while the lower table is the product table for $D_4$.
Both tables are symmetric in the sense that $\zeta(\vb{v}) \otimes \xi(\vb{u}) = \xi(\vb{u}) \otimes \zeta(\vb{v})$ for irreps $\zeta, \xi$.
For $D_4$, notice how the products between 1D irreps have the structure of the $\Z_2 \times \Z_2$ group, with the first $\Z_2 = \{A, B\}$ and the second one corresponding to the subscripts $\{1, 2\}$.
In the case of the 2D irrep $E$, we have ensured that the two components always transform under the same set of matrices of Eq.~\eqref{eq:E-irrep-mat-conventions-again}.
In particular, the ordering is important since $E(u_1|u_2)$ and $E(u_2|u_1)$ imply different transformation rules for $\vb{u} = (u_1, u_2)$.
Thus, when multiplied with a 1D irrep, the vector components sometimes need to be permuted or negated to ensure that the transformation matrices stay the same.}
{\renewcommand{\arraystretch}{1.3}
\renewcommand{\tabcolsep}{10pt}
\begin{tabular}{|c"c|c|} \hline
$\otimes$ & $g$ & $u$ \\ \thickhline
$g$ & $g$ & $u$ \\ \hline
$u$ & $u$ & $g$ \\ \hline
\end{tabular} \\[14pt]
\begin{tabular}{|c"c|c|c|c|c|} \hline
$\otimes$ & $A_1(u)$ & $A_2(u)$ & $B_1(u)$ & $B_2(u)$ & $E(u_1|u_2)$
\\ \thickhline
$A_1(v)$ & $A_1(v u)$ & $A_2(v u)$ & $B_1(v u)$ & $B_2(v u)$ & $E(v u_1|v u_2)$
\\ \hline
$A_2(v)$ & $A_2(v u)$ & $A_1(v u)$ & $B_2(v u)$ & $B_1(v u)$ & $E(v u_2|- v u_1)$
\\ \hline
$B_1(v)$ & $B_1(v u)$ & $B_2(v u)$ & $A_1(v u)$ & $A_2(v u)$ & $E(v u_1|- v u_2)$
\\ \hline
$B_2(v)$ & $B_2(v u)$ & $B_1(v u)$ & $A_2(v u)$ & $A_1(v u)$ & $E(v u_2|v u_1)$
\\ \hline
$E(v_1|v_2)$ & $E(v_1 u|v_2 u)$ & $E(v_2 u|- v_1 u)$ & $E(v_1 u|- v_2 u)$ & $E(v_2 u|v_1 u)$ & $\begin{matrix}
A_1(v_1 u_1 + v_2 u_2) \\
A_2(v_1 u_2 - v_2 u_1) \\
B_1(v_1 u_1 - v_2 u_2) \\
B_2(v_1 u_2 + v_2 u_1)
\end{matrix}$ \\ \hline
\end{tabular}}
\label{tab:D4h-irrep-prod-tab}
\end{table}
\end{landscape}

\backmatter
\let\cleardoublepage\clearpage

\chapter{Notation and Conventions}
\label{app:conventions}

Here we summarize the notation and conventions that we employ throughout the thesis.

We use SI units without exception, with the standard notation for the fundamental constants and units.
Both the reduced Planck constant $\hbar = h / (2 \pi)$ (the Planck constant $h$ is never used) and the Boltzmann constant $k_B$ are retained, i.e., not set to unity.
The elementary charge $e$  one can always tell apart from the Euler constant $\Elr$ from context.
$\upgamma_E = 0.5772...$ is the Euler-Mascheroni constant.

All the systems considered in this thesis are crystalline.
Periodic boundary conditions are always used, unless explicitly stated otherwise.
In Chap.~\ref{chap:cuprates} we have in addition set the lattice constant to unity so $L^d = \mathcal{N}$, $\vb{a}_i = \vu{e}_i$, etc.
The Fourier normalization factors are symmetric [$f(\vb{R}) = \mathcal{N}^{-1/2} \sum_{\vb{k}} \Elr^{\iu \vb{k} \vdot \vb{R}} f_{\vb{k}}$] for fields ($\psi$, $\Psi$, $\phi$, $\Phi$) and are asymmetric [$f(\vb{R}) = \mathcal{N}^{-1} \sum_{\vb{k}} \Elr^{\iu \vb{k} \vdot \vb{R}} f_{\vb{k}}$] for everything else ($\Gamma$, $\rho$, $\DipD$, $\Delta$, $H_{\vb{k}}$).
If not explicitly stated, the Fourier conventions can be easily deduced.

To avoid confusion, we never use the Einstein summation convention, i.e., all summations are explicit.
We employ the Euclidean signature for everything so there are no differences between lower and upper indices ($x_{\mu} = x^{\mu}$, $k_{\mu} = k^{\mu}$, etc.), the Dirac matrices satisfy $\{\gamma_{\mu}, \gamma_{\nu}\} = 2 \Kd_{\mu \nu}$, and so on.
All calculations are performed in imaginary (Euclidean) time, with the only exception being Sec.~\ref{sec:el-dip-sc-Dirac-opticond} where we analytically continue to real time. 

Vectors are bolded and have hats if they are normalized to unity.
We use hats for operators only in a few instances where we wish to distinguish them from their matrices.
Occasionally, Dirac braket notation is used.
\\

\textbf{Basic quantities:}
\vspace{-10pt}
{\renewcommand{\arraystretch}{1.7}
\renewcommand{\tabcolsep}{10pt}
\begin{longtable}[c]{L{0.08\textwidth}L{0.76\textwidth}}
$T$ & temperature \\
$\upbeta$ & thermodynamic beta, $\upbeta \defeq 1/(k_B T)$ \\ 
$T_c$ & superconducting transition temperature \\
$T^{*}$ & pseudogap onset temperature (of cuprates, Chap.~\ref{chap:cuprates}) \\
$\upmu$ & chemical potential \\
$\DOSg$ & density of states \\
$p$ & hole doping \\
$r$ & general tuning parameter or the quantum-critical boson softness parameter \\
$\epsilon_{ij}$ ($\epsilon_i$) & strain tensor components (in Voigt notation) \\
$\sigma_{ij}$ ($\sigma_i$) & stress tensor components (in Voigt notation) \\
$c_{ij}$ & elasticity tensor, in Voigt notation \\
$C$ & heat capacity \\
$S$ & entropy \\
$F$ & free energy \\
$\mathcal{Z}$ & partition function \\
$\action$ & Euclidean (imaginary-time) action \\
$\Haml$ & many-body (Fock-space, second-quantized) Hamiltonian \\
$d$ & spatial dimension \\
$L$ & linear size of the system \\
$L^d$ & total volume of the system \\
$\mathcal{N}$ & total number of unit cells \\
$M$ & number of orbitals per unit cell included in the tight-binding model \\
$\psi$ & column-vector/spinor of fermionic annihilation operators or Grassmann-odd fields, $\psi \equiv (\psi_{1,\uparrow}, \psi_{1,\downarrow}, \ldots, \psi_{M,\uparrow}, \psi_{M,\downarrow})^{\intercal}$ \\
$\Psi$ & extended-basis (Chap.~\ref{chap:cuprates}) or continuum (Chap.~\ref{chap:el_dip_SC}) fermionic field operator \\
$N$ & number of fermionic flavor components (during large-$N$ expansion, Chap.~\ref{chap:el_dip_SC}) \\
$\dim \Phi$ & number of order parameter components \\
$\Phi_a$ & order parameter or the corresponding fluctuating bosonic Grassmann-even fields (or field operators), $a \in \{1, \ldots, \dim \Phi\}$ \\
$\Upsilon_{\mu}$ & bilinear constructed from $\Phi$, $\sim \Phi^{\dag} \Pauli_{\mu} \Phi$ \\
$\phi_a$ & fermionic bilinear conjugate to $\Phi_a$ \\
$\Gamma$ & fermion-boson coupling matrix ($\sim \Phi \, \psi^{\dag} \Gamma \psi$, $\Phi \, \Psi^{\dag} \Gamma \Psi$) or general spin-orbit matrix \\
$\Lambda$ & orbital matrices of various types ($\Gamma \sim \Lambda \otimes \Pauli$) or momentum-space cutoff \\
$\Phi$ & plasmon field \\
$\rho$ & electric charge density \\
$\DipD$ & electric dipole operator (Chap.~\ref{chap:el_dip_SC}) \\
$\Delta$ & superconducting gap matrix \\
$\lambda$ & pairing eigenvalue, $T_c \propto \Elr^{-1/\lambda}$ \\
$\PintW_{AB}$, $\PintV_{AB}$ & pairing interactions which enter the linearized gap equation (Sec.~\ref{app:lin_gap_eq-fin-form}) \\
$\PintF_{AB}$, $\Pintf_a$ & pairing form factors, $\PintF_{AB} = \sum_a \tr_s \Pauli_A \Pintf_a \Pauli_B \Pintf_a^{\dag}$ (Sec.~\ref{sec:LC-prop-Cp-ch-int}) \\
$H_{\vb{k}}$ & band Hamiltonian, including the displacement by the chemical potential $\upmu$ \\
$\varepsilon_{\vb{k} n}$ & dispersion of the $n$-th band with the Fermi level set to zero \\
$u_{\vb{k} n s}$ & band eigenvector of the $n$-th band, $H_{\vb{k}} u_{\vb{k} n s} = \varepsilon_{\vb{k} n} u_{\vb{k} n s}$ \\
$\mathcal{P}_{\vb{k} n}$ & band projector of the $n$-th band, $\mathcal{P}_{\vb{k} n} \defeq \sum_s u_{\vb{k} n s} u_{\vb{k} n s}^{\dag}$ \\
$P$ & parity \\
$C_{m \vu{n}}$ & $m$-fold ($\frac{2\pi}{m}$) rotation around $\vu{n}$ \\
$\Sigma$ & reflections \\
$\SymU(g)$ & many-body unitary symmetry operator \\
$\MatU_{\vb{k}}(g)$ & matrix describing the action of $\SymU(g)$ on fermions $\psi_{\vb{k}}$ in $\vb{k}$-space \\
$\SymTR$ & many-body antiunitary time-reversal operator \\
$\MatTR_{\vb{k}}$ & matrix describing the action of $\SymTR$ on fermions $\psi_{\vb{k}}$ in $\vb{k}$-space \\
$R(g)$ & 3D vector transformation matrices, $R(g) \in \Ogp(3)$ (Appx.~\ref{sec:SU2-SO3-conventions}) \\
$S(g)$ & 2D spinor transformation matrices, $S(g) \in \SU(2)$ (Appx.~\ref{sec:SU2-SO3-conventions}) \\
$\RepM(g)$ & (irreducible) representation matrices, for $D_{4h}$ irreps see Sec.~\ref{sec:examples-convention-D4h}
\end{longtable}}

\textbf{Unit vectors, components, and crystal notation:}
\vspace{-10pt}
{\renewcommand{\arraystretch}{1.7}
\renewcommand{\tabcolsep}{10pt}
\begin{longtable}[c]{L{0.08\textwidth}L{0.76\textwidth}}
$\vu{e}_i$ & Cartesian unit vectors $\{\vu{e}_x, \vu{e}_y, \vu{e}_z\}$, $\vu{e}_i \vdot \vu{e}_j = \Kd_{ij}$. \\
$r_i$ & Cartesian components of $\vb{r}$, $r_i \defeq \vu{e}_i \vdot \vb{r}$. Individually, we shall usually denote them $x$, $y$, $z$ instead of $r_x$, $r_y$, $r_z$ or $r_1$, $r_2$, $r_3$. \\
$k_i$ & Cartesian components of $\vb{k}$, $k_i \defeq \vu{e}_i \vdot \vb{k}$. We shall always denote them \newline $k_x$, $k_y$, $k_z$ instead of $k_1$, $k_2$, $k_3$ because in a few cases the latter denote four-momenta, and not components. Same goes for $p_i$ and $q_j$. \\
$\vb{a}_i$ & Primitive vectors of the real-space Bravais lattice of the system. \\
$\vb{b}_i$ & Primitive vectors of the reciprocal lattice, $\vb{a}_i \vdot \vb{b}_j = 2 \pi \Kd_{ij}$. \\
$[hk\ell]$, $\langle hk\ell \rangle$ & Miller indices describing the direction $\vb{R}_{[hk\ell]} = h \vb{a}_1 + k \vb{a}_2 + \ell \vb{a}_3$, modulo point group symmetries for $\langle hk\ell \rangle$. Bars denote negative integers, $\bar{1} = - 1$, etc. \\
$(hk\ell)$, $\{hk\ell\}$ & Miller indices describing the plane $\vb{r} \vdot \vb{G}_{(hk\ell)} = 0$, where $\vb{G}_{(hk\ell)} = h \vb{b}_1 + k \vb{b}_2 + \ell \vb{b}_3$, modulo point group symmetries for $\{hk\ell\}$.
\end{longtable}}

\textbf{Variables and their domains:}
\vspace{-10pt}
{\renewcommand{\arraystretch}{1.7}
\renewcommand{\tabcolsep}{10pt}
\begin{longtable}[c]{L{0.08\textwidth}L{0.76\textwidth}}
$g$ & Elements of the point group of the system. For naming of individual point group elements, see the end of Sec.~\ref{sec:grp-rep-theory-basics-grp} of Appx.~\ref{app:group_theory}. \\
$\tau$ & Imaginary time, $\in [0, \upbeta]$ and $\int_{\tau} = \int_0^{\upbeta} \dd{\tau}$. \\
$\vb{r}$ & Continuous spatial positions, $\in \R^d$ and $\int_{\vb{r}} = \int \dd[d]{r} = L^d$. These integrals always go over the whole space, and not just one unit cell, unless explicitly stated otherwise. \\
$\vb{R}, \vb{\delta}$ & Direct lattice vectors, $\in \Z \vb{a}_1 + \Z \vb{a}_2 + \Z \vb{a}_3$ and $\sum_{\vb{R}} 1 = \mathcal{N}$. Their sums always go over the whole lattice. $\vb{R}$ vs.\ $\vb{\delta}$ is used to emphasize whether we are dealing with an absolute position or relative displacement, respectively. $\sum_{\vb{\delta}}$ is a sum over lattice neighbors, both close and distant, including $\vb{\delta} = \vb{0}$. \\
$\vb{x}_{\alpha}$ & Relative position of the center of the $\alpha$ orbital within a unit cell, with respect to the Bravais lattice. $\vb{R} + \vb{x}_{\alpha}$ are the absolute positions. \\
$x$ & Spacetime four-vectors, $\in \R^{d+1}$. They can either equal $x = (\tau, \vb{r})$ with $\int_x = \int_0^{\upbeta} \dd{\tau} \int \dd[d]{r}$, or $x = (\tau, \vb{R})$ with $\int_x = \int_0^{\upbeta} \dd{\tau} \sum_{\vb{R}}$, depending on whether we are dealing with a continuum or lattice model. \\
$\omega_{\ell}$ & Matsubara frequencies, can be either bosonic $\omega_{\ell} = 2 \ell \pi / \upbeta$ or fermionic $\omega_{\ell} = (2 \ell + 1) \pi / \upbeta$; should be obvious from context which ones. Matsubara sums $\sum_{\omega_{\ell}}$ always go over all frequencies. \\
$\vb{k}, \vb{p}, \vb{q}$ & Wavevectors/crystal momenta, $\sum_{\vb{k}} 1 = \mathcal{N}$. Their sums and integrals always go over only the first Brillouin zone, unless explicitly stated otherwise. \\
$\vb{G}$ & Reciprocal lattice vectors,  $\in \Z \vb{b}_1 + \Z \vb{b}_2 + \Z \vb{b}_3$, $\Elr^{\iu \vb{G} \vdot \vb{R}} = 1$, and $\sum_{\vb{G}} 1 = \mathcal{N}$. Their sums always go over the whole reciprocal lattice. \\
$k, p, q$ & Crystal momentum four-vectors, $k = (\omega_k \equiv \omega_{\ell}, \vb{k})$ and $\sum_k = \sum_{\omega_{\ell}} \sum_{\vb{k}}$. The Matsubara frequencies corresponding to $k$, $p$, $q$ we usually denote $\omega_k$, $\omega_p$, $\omega_q$ instead of $\omega_{\ell}$, respectively.
\end{longtable}}

\textbf{Indices and their spans:}
\vspace{-10pt}
{\renewcommand{\arraystretch}{1.7}
\renewcommand{\tabcolsep}{10pt}
\begin{longtable}[c]{L{0.08\textwidth}L{0.76\textwidth}}
$\zeta, \xi$ & Denote irreducible representations (irreps). For the tetragonal point group $D_{4h}$ (Sec.~\ref{sec:tetragonal-group-D4h}), $\zeta, \xi \in \{A_{1g}, \, A_{2g}, \, B_{1g}, \, B_{2g}, \, E_g, \, A_{1u}, \, A_{2u}, \, B_{1u}, \linebreak B_{2u}, \, E_u\}$. \\
$a, b$ & Order parameter component indices $\in \{1, \ldots, \dim \Phi\}$, irrep components indices $\in \{1, \ldots, \dim \zeta\}$, or just general matrix indices. $\dim \zeta$ is the dimension of the irrep. This index we suppress for 1D irreps ($\dim \zeta = 1$). \\
$\mu, \nu$ & Spacetime component indices, $\in \{0, 1 \equiv x, 2 \equiv y, 3 \equiv z\}$, or Pauli matrix indices including $\Pauli_0 = \one$. \\
$i, j, k$ & Spatial component or direction indices, $\in \{1 \equiv x, 2 \equiv y, 3 \equiv z\}$, or spin-like Pauli matrix indices (excluding $\Pauli_0$). Also used as the Voigt notation indices, $\in \{1 \equiv xx, 2 \equiv yy, 3 \equiv zz, 4 \equiv yz, 5 \equiv zx, 6 \equiv xy\}$. \linebreak $k$ we use sparingly and, when we do, it should be obvious from context that $k$ isn't a four-momentum. \\
$\alpha, \beta$ & Fermion component indices, covering both orbital and spin degrees of freedom, $\in \{1, 2, \ldots, 2M\}$. In a few instances they go only over orbital degrees of freedom, in which case they span $1, \ldots, M$. \\
$n, m$ & Band indices, $\in \{1, \ldots, M\}$. All systems under consideration have both parity and time-reversal symmetry so their bands are doubly degenerate. Also used as general enumeration indices $\in \{1, 2, \ldots\}$. \\
$s$ & Spin or pseudospin (Kramers' degeneracy) indices, $\in \{\uparrow, \downarrow\}$. \\
$A, B$ & Pauli $\in \{0, 1, 2, 3\}$ or Gell-Mann $\in \{0, 1, \ldots, 8\}$ matrix indices.
\end{longtable}}

\textbf{Special notations:}
\vspace{-10pt}
{\renewcommand{\arraystretch}{1.7}
\renewcommand{\tabcolsep}{10pt}
\begin{longtable}[c]{L{0.12\textwidth}L{0.72\textwidth}}
$\vb{k}_n, \vb{p}_m$ & This means that the wavevector $\vb{k}$ is on the Fermi surface of the $n$-th band, i.e., it satisfies $\varepsilon_{\vb{k} n} = 0$, where $\varepsilon_{\vb{k} n}$ is the dispersion of the $n$-th band displaced by the chemical potential. Likewise, $\vb{p}_m \iff \varepsilon_{\vb{p} m} = 0$. \\
$\vb{k}_{\perp}, \vb{p}_{\perp}, \vb{q}_{\perp}$ & Denotes the in-plane components of $\vb{k}, \vb{p}, \vb{q}$ is quasi-2D systems, \newline i.e., $\vb{k}_{\perp} \equiv (k_x, k_y)$, $\vb{k} = (\vb{k}_{\perp}, k_z) = (k_x, k_y, k_z)$, $\vb{p}_{\perp} \equiv (p_x, p_y)$, etc.
\end{longtable}}

\pagebreak

\textbf{Matrices:}
\begin{itemize}
\item The Pauli matrices are the usual ones:
\begin{align*}
\Pauli_0 = \uptau_0 &= \begin{pmatrix} 1 & 0 \\ 0 & 1 \end{pmatrix}, &
\Pauli_1 \equiv \Pauli_x = \uptau_x &= \begin{pmatrix} 0 & 1 \\ 1 & 0 \end{pmatrix}, \\
\Pauli_2 \equiv \Pauli_y = \uptau_y &= \begin{pmatrix} 0 & -\iu \\ \iu & 0 \end{pmatrix}, &
\Pauli_3 \equiv \Pauli_z = \uptau_z &= \begin{pmatrix} 1 & 0 \\ 0 & -1 \end{pmatrix}.
\end{align*}
$\Pauli_A$ are used for Pauli matrices in spin or pseudospin space, while $\uptau_A$ are used for Pauli matrices in orbital or flavor space.
The tensor product $\otimes$ between $\uptau_A$ and $\Pauli_B$ is usually suppressed.
The $\uptau_A$ only arise in Chap.~\ref{chap:el_dip_SC}.

\item The Dirac gamma matrices employed in Chap.~\ref{chap:el_dip_SC} are:
\begin{align*}
\gamma_0 &= \uptau_3 \Pauli_0 = \begin{pmatrix}
1 & 0 & 0 & 0 \\
0 & 1 & 0 & 0 \\
0 & 0 & -1 & 0 \\
0 & 0 & 0 & -1
\end{pmatrix}, &
\gamma_1 &= \uptau_1 \Pauli_y = \begin{pmatrix}
0 & 0 & 0 & -\iu \\
0 & 0 & \iu & 0 \\
0 & -\iu & 0 & 0 \\
\iu & 0 & 0 & 0
\end{pmatrix}, \\
\gamma_2 &= - \uptau_1 \Pauli_x = \begin{pmatrix}
0 & 0 & 0 & -1 \\
0 & 0 & -1 & 0 \\
0 & -1 & 0 & 0 \\
-1 & 0 & 0 & 0
\end{pmatrix}, &
\gamma_3 &= - \uptau_2 \Pauli_0 = \begin{pmatrix}
0 & 0 & \iu & 0 \\
0 & 0 & 0 & \iu \\
-\iu & 0 & 0 & 0 \\
0 & -\iu & 0 & 0
\end{pmatrix}, \\
\gamma_5 &\defeq \gamma_0 \gamma_1 \gamma_2 \gamma_3 = - \uptau_1 \Pauli_z = \begin{pmatrix}
0 & 0 & \iu & 0 \\
0 & 0 & 0 & \iu \\
-\iu & 0 & 0 & 0 \\
0 & -\iu & 0 & 0
\end{pmatrix}.
\end{align*}
Note that they are Hermitian, $\gamma_{\mu}^{\dag} = \gamma_{\mu}$, and in Euclidean signature, $\{\gamma_{\mu}, \gamma_{\nu}\} = 2 \Kd_{\mu \nu}$.

\item In Chap.~\ref{chap:Sr2RuO4}, we use the following unconventional choice for the nine Gell-Mann matrices:
\begin{align*}
\Lambda_0 &= \begin{pmatrix}
1 & 0 & 0 \\
0 & 1 & 0 \\
0 & 0 & 0
\end{pmatrix}, & \Lambda_1 &= \begin{pmatrix}
0 & 1 & 0 \\
1 & 0 & 0 \\
0 & 0 & 0
\end{pmatrix}, \\
\Lambda_2 &= \begin{pmatrix}
0 & -\iu & 0 \\
\iu & 0 & 0 \\
0 & 0 & 0
\end{pmatrix}, & \Lambda_3 &= \begin{pmatrix}
1 & 0 & 0 \\
0 & -1 & 0 \\
0 & 0 & 0
\end{pmatrix}, \\
\Lambda_4 &= \begin{pmatrix}
0 & 0 & 0 \\
0 & 0 & 0 \\
0 & 0 & \sqrt{2}
\end{pmatrix}, \\
\Lambda_5 &= \begin{pmatrix}
0 & 0 & 1 \\
0 & 0 & 0 \\
1 & 0 & 0
\end{pmatrix}, & \Lambda_6 &= \begin{pmatrix}
0 & 0 & -\iu \\
0 & 0 & 0 \\
\iu & 0 & 0
\end{pmatrix}, \\
\Lambda_7 &= \begin{pmatrix}
0 & 0 & 0 \\
0 & 0 & 1 \\
0 & 1 & 0
\end{pmatrix}, & \Lambda_8 &= \begin{pmatrix}
0 & 0 & 0 \\
0 & 0 & -\iu \\
0 & \iu & 0
\end{pmatrix}.
\end{align*}
These $3 \times 3$ Gell-Mann $\Lambda_{A}$ matrices, used in Chap.~\ref{chap:Sr2RuO4}, should not be conflated with the $5 \times 5$ extended-basis orbital $\Lambda^{\zeta}_{n,a}$ matrices, introduced in Sec.~\ref{sec:orbital-Lambda-mats} of Chap.~\ref{chap:cuprates}.
\end{itemize}

\vspace{10pt}

\textbf{Various:}
\vspace{-10pt}
{\renewcommand{\arraystretch}{1.7}
\renewcommand{\tabcolsep}{10pt}
\begin{longtable}[c]{L{0.17\textwidth}L{0.67\textwidth}}
$\const$ & constant \\
$z^{*}$ & complex conjugate of $z \in \C$ (the notation $\bar{z}$ is never used) \\
$A^*$ & element-wise complex conjugate of $A$, $(A^*)_{ab} = (A_{ab})^*$ \\
$\cc$ & complex conjugate \\
$A^{\intercal}$ & transpose of $A$ \\
$A^{\dag}$ & Hermitian conjugate of $A$, $A^{\dag} = (A^*)^{\intercal}$ \\
$\Hc$ & Hermitian conjugate \\
$\one$ & identity operator or matrix \\
$\diag(x_1, x_2, \ldots)$ & diagonal matrix with $x_1$, $x_2$, \ldots on the diagonal \\
$[A, B]$ & commutator, $[A, B] \defeq A B - B A$ \\
$\{A, B\}$ & anticommutator, $\{A, B\} \defeq A B + B A$ \\
$\tr$, $\Tr$ & trace \\
$\ket{v}$ & a ``ket,'' i.e., a column-vector $v$ in Dirac notation \\
$\bra{v}$ & a ``bra,'' i.e., a conjugated and transposed vector $v^{\dag}$ in Dirac notation \\
$\braket{v}{u}$ & a ``braket,'' i.e., a scalar product between $v$ and $u$ in Dirac notation \\
$\Kd_{ij}$ & Kronecker delta symbol \\
$\LCs_{ijk}$ & Levi-Civita symbol \\
$\Dd(x)$ & Dirac delta function \\
$\HTh(x)$ & Heaviside step function \\
$\sgn(x)$ & sign function \\
$\log$ & natural base-$\Elr$ logarithm (the notation $\ln$ is never used) \\
$\erf(x)$ & error function, $\erf(x) \defeq \frac{2}{\sqrt{\pi}} \int_0^{x} \dd{t} \, \Elr^{-t^2}$ \\
$\Cl_2(x)$ & Clausen function, $\Cl_2(x) \defeq \sum_{n=1}^{\infty} \sin(n x)/n^2$ \\[4pt]
$\grad$ & real-space nabla operator, $\displaystyle \grad \defeq \frac{\partial}{\partial \vb{r}}$ \\[4pt]
$\grad_{\vb{k}}$ & momentum-space nabla operator, $\displaystyle \grad_{\vb{k}} \defeq \frac{\partial}{\partial \vb{k}}$ \\[4pt]
$\bigO(x^n)$ & big O notation
\end{longtable}}

\chapter{List of Abbreviations}
\label{app:abbreviations}

{\large
\renewcommand{\arraystretch}{1.3}
\renewcommand{\tabcolsep}{10pt}
\begin{longtable}[c]{L{0.12\textwidth}L{0.72\textwidth}}
\textbf{$n$D} & $n$ spatial dimensions/$n$-dimensional (for integer $n$) \\
\textbf{AF} & antiferromagnet/antiferromagnetic/antiferromagnetism \\
\textbf{ARPES} & angle-resolved photoemission spectroscopy \\
\textbf{ASV} & Aji, Shekhter, and Varma (authors of Ref.~\cite{Aji2010}) \\
\textbf{BCS} & Bardeen-Cooper-Schrieffer \\
\textbf{BZ} & Brillouin zone \\
\textbf{CDW} & charge-density wave \\
\textbf{DFT} & density functional theory \\
\textbf{DOS} & density of states \\
\textbf{irrep} & irreducible representation \\
\textbf{IUC} & intra-unit-cell (synonymous with homogeneous $\vb{q} = \vb{0}$ order) \\
\textbf{LC} & loop current (synonymous with orbital magnetism) \\
\textbf{$\mu$SR} & muon spin spectroscopy/muon spin rotation/muon spin relaxation \\
\textbf{NMR} & nuclear magnetic resonance \\
\textbf{PND} & polarized neutron diffraction \\
\textbf{QCP} & quantum-critical point \\
\textbf{RG} & renormalization group \\
\textbf{RPA} & random phase approximation \\
\textbf{SC} & superconductor/superconducting/superconductivity \\
\textbf{SDW} & spin-density wave \\
\textbf{SOC} & spin-orbit coupling \\
\textbf{SQUID} & superconducting quantum interference device \\
\textbf{SRO} & strontium ruthenate \ce{Sr2RuO4} \\
\textbf{STM} & scanning tunneling microscopy \\
\textbf{TR} & time reversal \\
\textbf{TRSB} & time-reversal symmetry-breaking
\end{longtable}}

{
\hypersetup{hidelinks}
\listoffigures
\listoftables
}
\chapter{List of Publications}

Here I list all the peer-reviewed publications I have been involved in up to the date of my thesis defense, in reverse chronological order:
\begin{enumerate}
\item \textbf{Grgur Palle} \textcolor{black}{and Jörg Schmalian,} ``{Unconventional superconductivity from electronic dipole fluctuations}'', \href{https://doi.org/10.1103/PhysRevB.110.104516}{Phys.\ Rev.\ B \textbf{110}, 104516 (2024)}. Ref.~\cite{Palle2024-el-dip} in the Bibliography.

\item \textcolor{black}{Fabian Jerzembeck, You-Sheng Li,} \textbf{Grgur Palle}\textcolor{black}{, Zhenhai Hu, Mehdi Biderang, Naoki Kikugawa, Dmitry A.\ Sokolov, Sayak Ghosh, Brad J.\ Ramshaw, Thomas Scaffidi, Michael Nicklas, Jörg Schmalian, Andrew P.\ Mackenzie, and Clifford W.\ Hicks,} ``{$T_c$ and the elastocaloric effect of \ce{Sr2RuO4} under $\langle 110 \rangle$ uniaxial stress: No indications of transition splitting}'', \href{https://doi.org/10.1103/PhysRevB.110.064514}{Phys.\ Rev.\ B \textbf{110}, 064514 (2024)}. Ref.~\cite{Jerzembeck2024} in the Bibliography.

\item \textbf{Grgur Palle}\textcolor{black}{,} ``{Comment on "Towards exact solutions for the superconducting $T_c$ induced by electron-phonon interaction"}'', \href{https://doi.org/10.1103/PhysRevB.110.026501}{Phys.\ Rev.\ B \textbf{110}, 026501 (2024)}. Not cited in the thesis.

\item \textbf{Grgur Palle}\textcolor{black}{, Risto Ojajärvi, Rafael M.\ Fernandes, and Jörg Schmalian,} ``{Superconductivity due to fluctuating loop currents}'', \href{https://doi.org/10.1126/sciadv.adn3662}{Sci.\ Adv.\ \textbf{10}, eadn3662 (2024)}. Ref.~\cite{Palle2024-LC} in the Bibliography.

\item \textbf{Grgur Palle}\textcolor{black}{, Clifford Hicks, Roser Valentí, Zhenhai Hu, You-Sheng Li, Andreas Rost, Michael Nicklas, Andrew P.\ Mackenzie, and Jörg Schmalian,} ``{Constraints on the superconducting state of \ce{Sr2RuO4} from elastocaloric measurements}'', \href{https://doi.org/10.1103/PhysRevB.108.094516}{Phys.\ Rev.\ B \textbf{108}, 094516 (2023)}. Ref.~\cite{Palle2023-ECE} in the Bibliography.

\item \textcolor{black}{Jonas Gaa,} \textbf{Grgur Palle}\textcolor{black}{, Rafael M.\ Fernandes, and Jörg Schmalian,} ``{Fracton-elasticity duality in twisted moiré superlattices}'', \href{https://doi.org/10.1103/PhysRevB.104.064109}{Phys.\ Rev.\ B \textbf{104}, 064109 (2021)}. Not cited in the thesis.

\item \textbf{Grgur Palle} \textcolor{black}{and Denis K.\ Sunko,} ``{Physical limitations of the Hohenberg-Mermin-Wagner theorem}'', \href{https://doi.org/10.1088/1751-8121/ac0a9d}{J.\ Phys.\ A: Math.\ Theor.\ \textbf{54}, 315001 (2021)}. Ref.~\cite{Palle2021-HMW} in the Bibliography.

\item \textbf{Grgur Palle} \textcolor{black}{and Owen Benton,} ``{Exactly solvable spin-$\frac{1}{2}$ XYZ models with highly degenerate partially ordered ground states}'', \href{https://doi.org/10.1103/PhysRevB.103.214428}{Phys.\ Rev.\ B \textbf{103}, 214428 (2021)}. Not cited in the thesis.

\item \textbf{Grgur Palle}\textcolor{black}{,} ``{Fizikalna ograničenja Hohenberg-Mermin-Wagnerovog argumenta}'' (in English ``{Physical limitations of the Hohenberg-Mermin-Wagner argument}''). This is my Master Thesis (2020), which is available online at \url{https://urn.nsk.hr/urn:nbn:hr:217:080906} in Croatian. Not cited in the thesis.

\item \textbf{Grgur Palle}\textcolor{black}{, Luka Bakrač, and Aleksandar Opančar,} ``{A walking roller chain}'', \href{https://doi.org/10.1051/emsci/2018005}{Emergent Scientist \textbf{2}, 6 (2018)}. Not cited in the thesis.
\end{enumerate}

\chapter{Acknowledgments}

First and foremost, I would like to express my sincere gratitude to my advisor Prof.~Dr.~Jörg Schmalian for offering me the opportunity to work with him and for supporting and supervising my doctorate.
It is often said that your doctoral advisor is the most important person of your academic career, and I could not be more grateful for having him in this role.
His patience, sharp insights and foresight, gentle words of encouragement, and cheerful attitude throughout our four years together are all greatly appreciated, as is the freedom he granted me in pursuing topics I am interested in.
I wish to thank him for the time he devoted to me and for the many lessons he passed onto me, on physics and beyond.
It has been a pleasure working together.

Secondly, I would like to thank Prof.~Dr.~Markus Garst for the insightful discussions and collaborations, but also for agreeing to be the second referee of my doctoral defense.
I hope that his reading of my doctoral thesis has been enjoyable.

I am indebted to my many collaborators, previous and ongoing, without whom only a fraction of my doctoral work would have been possible:
Risto Ojajärvi,
Charles Steward,
Iksu Jang,
Tamaghna Hazra,
Jonas Gaa,
Rafael M.\ Fernandes,
Andrey V.\ Chubukov,
Andrew P.\ Mackenzie,
Clifford Hicks,
Fabian Jerzembeck,
You-Sheng Li,
Zhenhai Hu,
Andreas Rost,
Michael Nicklas,
Hilary Noad,
Roser Valentí,
Thomas Scaffidi,
Owen Benton,
Sayak Ghosh,
Brad J.\ Ramshaw,
Lichen Wang, and
Bernhard Keimer.
In particular, I would like to thank Risto Ojajärvi for helping me with a number of figures of Chap.~\ref{chap:cuprates}.
It has been a pleasure working with all of you.
I have learned immensely from our scientific discussions and I look forward to any future collaborations.

I am grateful to the many people that made my stay at KIT so enjoyable and welcoming:
Veronika Stangier,
Davide Valentinis,
Thibault Scoquart,
Hugo Perrin,
Gian-Andrea Inkof,
Roland Willa,
Matthias Hecker,
Egor Kiselev,
Luis Filsinger,
Hanna Ziegler,
Romy Morin,
Michael Rampp,
Adrian Reich,
Lindsay Orr,
Jinhong Park,
Andrei Pavlov,
Paul Pöpperl,
Jonas Karcher,
Vanessa Gall,
Dmitriy Shapiro,
Christian Spanslätt,
Kyrylo Snizhko,
Daniel Hauck,
Konrad Scharff,
Sopheak Sorn,
Lars Franke, 
Safa Lamia Ahmed, and
Paolo Battistoni, not to mention
Risto, Charlie, Iksu, and Tamaghna again.
Special thanks go to the always-helpful Sonja König for her administrative support, to Andreas Poenicke for his technical support, and to master-chef Waldemar Langosch for his daily delicious cooking.
I am also grateful to
Juraj Krsnik,
Eugen Dizer,
Anzumaan Chakraborty,
Henrik S.\ Guttesen, and
Aiman Al-Eryani
for the stimulating discussions on physics and more,
and I would like to thank Aiman for inviting me to Bochum.
It can be easy to forget the many people you cross paths with, but I like to believe that each one of them contributed in their little (or big) way to my success, for which I am grateful.
I hope that the many individuals inevitably left unmentioned can still feel my gratitude.

The support by the KIT Graduate School of Quantum Matter (KSQM) is greatly appreciated, as is the hard work by Yu Goldscheider and others that went into organizing it.
The KSQM Retreats have been some of the more enjoyable parts of my doctorate and the soft skills workshops offered within KSQM have proven to be invaluable to me.

I have had the great fortune of being part of the Elasto-Q-Mat initiative (Collaborative Research Center TRR288), which not only funded my doctoral research, but also connected me to cutting-edge research, like-minded doctoral students, and a welcoming and vibrant scholarly community.
I have greatly profited from this experience, for which I am thankful.

Lastly, I wish to thank my family and dear ones for their endless encouragement and unconditional support.
Such things are easily taken for granted, even though they shouldn't be.
Special thanks go to my uncle Ivo Batistić for his careful reading and helpful recommendations on my thesis.

\printbibliography[heading=bibintoc]

\end{document}